\documentclass[ALICE,manyauthors]{cernphprep}
\usepackage[comma,square,numbers,sort&compress]{natbib}
\usepackage{lineno}
\usepackage{xspace}
\usepackage{xcolor,soul}
\usepackage{hyperref}
\usepackage{xparse}
\usepackage{xstring}
\usepackage{euscript}
\usepackage{enumitem}
\usepackage[utf8]{inputenc}
\usepackage[english]{babel}
\usepackage{graphicx}
\usepackage{blindtext}
\usepackage{float}
\usepackage{adjustbox}
\newsavebox{\measurebox}
\usepackage{pdfpages}
\usepackage[T1]{fontenc}
\usepackage{orcidlink}

\usepackage{datetime} 

\newcommand{\version}[2]{ [\today, \currenttime\ {\scriptsize GMT}] }

\usepackage{subfiles} 

\sethlcolor{green}
\begin{document}

\includepdf{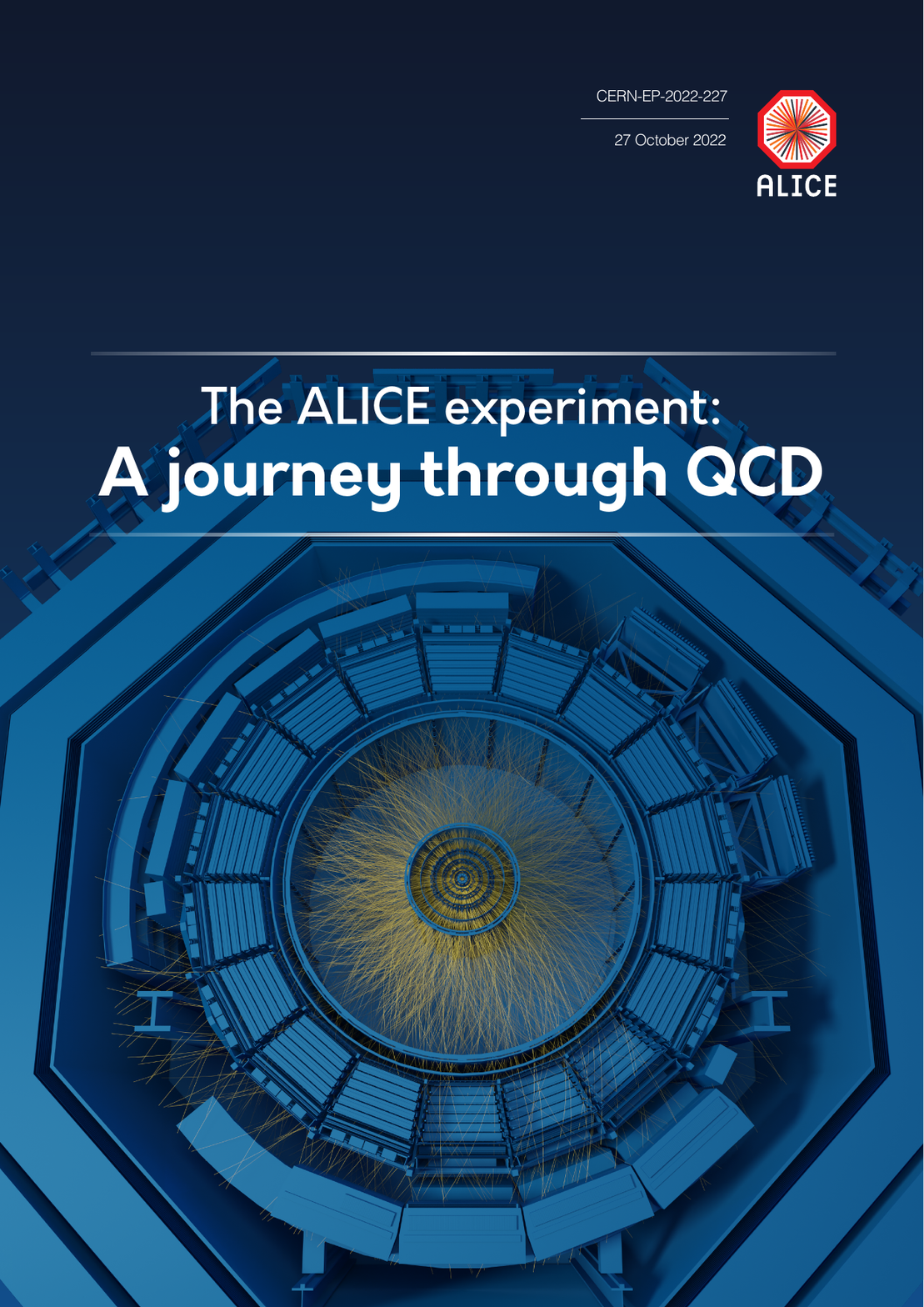}
\begin{titlepage}
\PHyear{2022}       
\PHnumber{227}      
\PHdate{27 October}  

\title{The ALICE experiment: a journey through QCD}
\ShortTitle{The ALICE experiment: a journey through QCD}   

\Collaboration{ALICE Collaboration\thanks{See Appendix~\ref{app:collab} for the list of collaboration members}
}
\ShortAuthor{ALICE Collaboration} 

\begin{abstract}
The ALICE experiment was proposed in 1993, to study strongly-interacting matter at extreme energy densities and temperatures. This proposal entailed a comprehensive investigation of nuclear collisions at the LHC. Its physics programme initially focused on the determination of the properties of the quark--gluon plasma (QGP), a deconfined state of quarks and gluons, created in such collisions. The ALICE physics programme has been extended to cover a broader ensemble of observables related to Quantum Chromodynamics (QCD), the theory of strong interactions.
The experiment has studied Pb--Pb, Xe--Xe, p--Pb and pp collisions in the multi-TeV centre of mass energy range, during the Run 1--2 data-taking periods at the LHC (2009--2018). The aim of this review is to summarise the key ALICE physics results in this endeavor, and to discuss their implications on
the current understanding of the macroscopic and microscopic properties of strongly-interacting matter at the highest temperatures reached in the laboratory. It will review the latest findings on the properties of the QGP created by heavy-ion collisions at LHC energies, and describe the surprising QGP-like effects in pp and p--Pb collisions.
Measurements of few-body QCD interactions, and their impact in unraveling the structure of hadrons and hadronic interactions, will be discussed. ALICE results relevant for physics topics outside the realm of QCD will also be touched upon.
Finally, prospects for future measurements with the ALICE detector in the context of its planned upgrades will also be briefly described.

\end{abstract}
\end{titlepage}

\setcounter{page}{2} 

\tableofcontents

\newpage

\newcommand{\pp}           {pp\xspace}
\newcommand{\ppbar}        {\mbox{$\mathrm {p\overline{p}}$}\xspace}
\newcommand{\XeXe}         {\mbox{Xe--Xe}\xspace}
\newcommand{\PbPb}         {\mbox{Pb--Pb}\xspace}
\newcommand{\pA}           {\mbox{pA}\xspace}
\newcommand{\pPb}          {\mbox{p--Pb}\xspace}
\newcommand{\AuAu}         {\mbox{Au--Au}\xspace}
\newcommand{\dAu}          {\mbox{d--Au}\xspace}

\newcommand{\s}            {\ensuremath{\sqrt{s}}\xspace}
\newcommand{\snn}          {\ensuremath{\sqrt{s_{\mathrm{NN}}}}\xspace}
\newcommand{\pt}           {\ensuremath{p_\mathrm T\,}\xspace}
\newcommand{\meanpt}       {$\langle p_{\mathrm{T}}\rangle$\xspace}
\newcommand{\minv}         {\ensuremath{m_{\rm inv.}}\xspace}
\newcommand{\ycms}         {\ensuremath{y_{\rm CMS}}\xspace}
\newcommand{\ylab}         {\ensuremath{y_{\rm lab}}\xspace}
\newcommand{\etarange}[1]  {\mbox{$\left | \eta \right |~<~#1$}}
\newcommand{\yrange}[1]    {\mbox{$\left | y \right |~<~#1$}}
\newcommand{\dndy}         {\ensuremath{\mathrm{d}N_\mathrm{ch}/\mathrm{d}y}\xspace}
\newcommand{\dndeta}       {\ensuremath{\mathrm{d}N_\mathrm{ch}/\mathrm{d}\eta}\xspace}
\newcommand{\avdndeta}     {\ensuremath{\langle\dndeta\rangle}\xspace}
\newcommand{\dNdy}         {\ensuremath{\mathrm{d}N_\mathrm{ch}/\mathrm{d}y}\xspace}
\newcommand{\Npart}        {\ensuremath{N_\mathrm{part}}\xspace}
\newcommand{\Ncoll}        {\ensuremath{N_\mathrm{coll}}\xspace}
\newcommand{\avNpart}{\ensuremath{\langle N_\mathrm{part} \rangle}\xspace}
\newcommand{\avNcoll}{\ensuremath{\langle N_\mathrm{coll} \rangle}\xspace}
\newcommand{\avTAA}        {\ensuremath{\langle T_\mathrm{AA} \rangle}\xspace}
\newcommand{\avTpPb}        {\ensuremath{\langle T_\mathrm{pPb} \rangle}\xspace}
\newcommand{\Naa}       {\ensuremath{N_{\mathrm{AA}}}\xspace}
\newcommand{\NpPb}   {\ensuremath{N_{\mathrm{pPb}}}\xspace}
\newcommand{\dEdx}         {\ensuremath{\textrm{d}E/\textrm{d}x}\xspace}
\newcommand{\RpPb}         {\ensuremath{R_{\rm pPb}}\xspace}
\newcommand{\mpt}          {\ensuremath{\langle p_{\rm T}\rangle}\xspace}
\newcommand{\mptsquared}   {\ensuremath{\langle p^2_{\rm T}\rangle}\xspace}
\newcommand{\vn}           {\ensuremath{\rm{v}_{n}}\xspace}
\newcommand{\RAA}          {\ensuremath{R_{\rm AA}}\xspace}
\newcommand{\RpA}          {\ensuremath{R_{\rm pA}}\xspace}

\newcommand{\nineH}        {$\sqrt{s}~=~0.9$~Te\kern-.1emV\xspace}
\newcommand{\seven}        {$\sqrt{s}~=~7$~Te\kern-.1emV\xspace}
\newcommand{\twoH}         {$\sqrt{s}~=~0.2$~Te\kern-.1emV\xspace}
\newcommand{\twosevensix}  {$\sqrt{s}~=~2.76$~Te\kern-.1emV\xspace}
\newcommand{\five}         {$\sqrt{s}~=~5.02$~Te\kern-.1emV\xspace}
\newcommand{\thirteen}     {$\sqrt{s}~=~13$~Te\kern-.1emV\xspace}
\newcommand{\twosevensixnn}{$\sqrt{s_{\mathrm{NN}}}~=~2.76$~Te\kern-.1emV\xspace}
\newcommand{\fivenn}       {$\sqrt{s_{\mathrm{NN}}}~=~5.02$~Te\kern-.1emV\xspace}
\newcommand{\eightnn}       {$\sqrt{s_{\mathrm{NN}}}~=~8.16$~Te\kern-.1emV\xspace}
\newcommand{\LT}           {L{\'e}vy-Tsallis\xspace}
\newcommand{\GeVc}         {Ge\kern-.1emV/$c$\xspace}
\newcommand{\MeVc}         {Me\kern-.1emV/$c$\xspace}
\newcommand{\TeV}          {Te\kern-.1emV\xspace}
\newcommand{\GeV}          {Ge\kern-.1emV\xspace}
\newcommand{\MeV}          {Me\kern-.1emV\xspace}
\newcommand{\GeVmass}      {Ge\kern-.2emV/$c^2$\xspace}
\newcommand{\MeVmass}      {Me\kern-.2emV/$c^2$\xspace}
\newcommand{\lumi}         {\ensuremath{\mathcal{L}}\xspace}

\newcommand{\ITS}          {\rm{ITS}\xspace}
\newcommand{\TOF}          {\rm{TOF}\xspace}
\newcommand{\ZDC}          {\rm{ZDC}\xspace}
\newcommand{\ZDCs}         {\rm{ZDCs}\xspace}
\newcommand{\ZNA}          {\rm{ZNA}\xspace}
\newcommand{\ZNC}          {\rm{ZNC}\xspace}
\newcommand{\SPD}          {\rm{SPD}\xspace}
\newcommand{\SDD}          {\rm{SDD}\xspace}
\newcommand{\SSD}          {\rm{SSD}\xspace}
\newcommand{\TPC}          {\rm{TPC}\xspace}
\newcommand{\TRD}          {\rm{TRD}\xspace}
\newcommand{\VZERO}        {\rm{V0}\xspace}
\newcommand{\VZEROA}       {\rm{V0A}\xspace}
\newcommand{\VZEROC}       {\rm{V0C}\xspace}
\newcommand{\Vdecay} 	   {\ensuremath{V^{0}}\xspace}

\newcommand{\jpsi}         {\ensuremath{\text{J}/\psi}\xspace}
\newcommand{\psiprime}     {\ensuremath{\psi(2\text{S})}\xspace}
\newcommand{\ccbar}        {\ensuremath{\text{c}\overline{\text{c}}}\xspace}
\newcommand{\bbbar}        {\ensuremath{\text{b}\overline{\text{b}}}\xspace}
\newcommand{\ee}           {\ensuremath{\text{e}^{+}\text{e}^{-}}\xspace} 
\newcommand{\pip}          {\ensuremath{\pi^{+}}\xspace}
\newcommand{\pim}          {\ensuremath{\pi^{-}}\xspace}
\newcommand{\kap}          {\ensuremath{\rm{K}^{+}}\xspace}
\newcommand{\kam}          {\ensuremath{\rm{K}^{-}}\xspace}
\newcommand{\pbar}         {\ensuremath{\rm\overline{p}}\xspace}
\newcommand{\kzero}        {\ensuremath{{\rm K}^{0}_{\rm{S}}}\xspace}
\newcommand{\lmb}          {\ensuremath{\Lambda}\xspace}
\newcommand{\almb}         {\ensuremath{\overline{\Lambda}}\xspace}
\newcommand{\Om}           {\ensuremath{\Omega^-}\xspace}
\newcommand{\Mo}           {\ensuremath{\overline{\Omega}^+}\xspace}
\newcommand{\X}            {\ensuremath{\Xi^-}\xspace}
\newcommand{\Ix}           {\ensuremath{\overline{\Xi}^+}\xspace}
\newcommand{\Xis}          {\ensuremath{\Xi^{\pm}}\xspace}
\newcommand{\Oms}          {\ensuremath{\Omega^{\pm}}\xspace}
\newcommand{\degree}       {\ensuremath{^{\rm o}}\xspace}
\newcommand{\upsone}         {\ensuremath{\Upsilon(1\text{S})}\xspace}
\newcommand{\rmJpsi}    {\mbox{$\mathrm{J\kern-0.05em /\kern-0.05em\psi}$}}

\newcommand{\note}[1]{{\color{red}\textbf{#1}}}

\section*{Preface}
\label{sec:preface}

\begin{figure}[!b]
\begin{center}
\includegraphics[width = 0.9\textwidth]{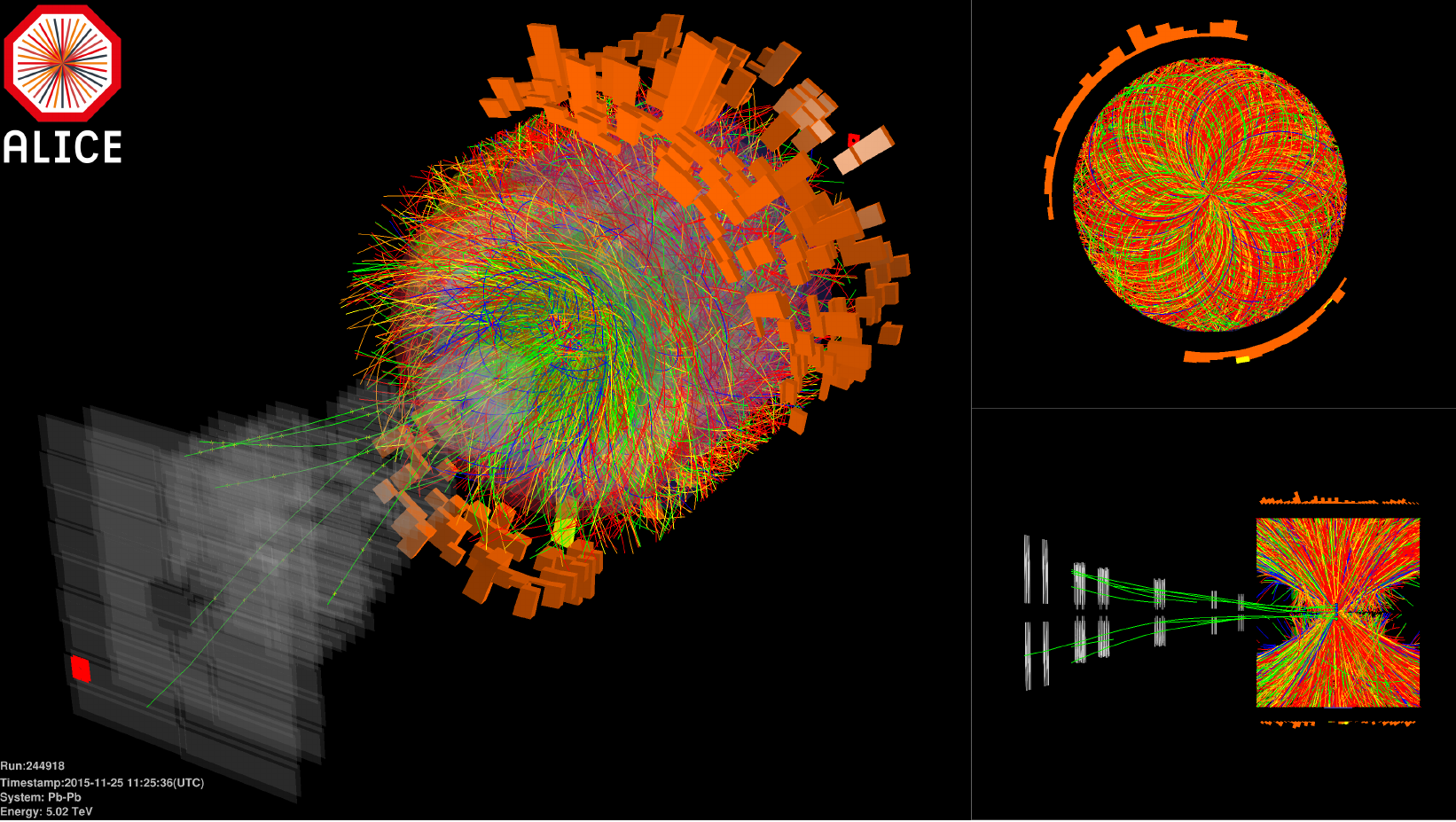}
\end{center}
\caption{An ALICE event display of detected particles created in a \PbPb \fivenn collision.}
\label{fig:ALICEED}
\end{figure}

Relativistic heavy-ion collisions at the LHC create the quark--gluon plasma (QGP): the hottest and densest fluid ever studied in the laboratory on Earth. In contrast to normal nuclear matter, the QGP is a state where quarks and gluons are not confined inside hadrons. It is speculated that the early Universe existed in such a state around $\sim 10^{-6}$ seconds after the Big Bang. The strong force, which principally governs behaviour of the quarks and gluons in such a plasma, and all nuclear matter, is responsible for the vast majority of the visible mass in the Universe. The ALICE detector (Fig.~\ref{fig:ALICEED}) was specifically designed to study the QGP created at LHC energies, and this review is dedicated to summarising what has been learnt in the last decade. In particular, we will address the following questions. What does the Standard Model have to say about the plausibility of the existence of the QGP? Are there emergent phenomena that arise from high density QCD? Does the QGP behave as a gas or a liquid? How quickly does it expand? What happens to the QGP when it is excited by the presence of large momentum/mass quarks or gluons? How do QGP constituents react to the enormous electromagnetic fields created by heavy-ion collisions? Are the extremely high energies at the LHC sufficient for the formation of the QGP in proton--proton collisions?  How do the hadrons that emerge in the dense medium after the transition from QGP to normal matter interact between themselves? And what can we learn from these studies regarding the properties of the extremities in the Universe, such as the core of neutron stars? This review intends to answer these questions in the context of the ALICE measurements performed in the last decade. We will begin by an introduction discussing the theoretical groundings in Quantum Chromodynamics (QCD) for the existence of the QGP, and what heavy-ion collisions can teach us (and have already taught us) about QGP properties. At the end of the introduction, the above questions will be reformulated in the context of what we can learn about QGP formation at LHC energies, and what advantages the ALICE detector brings to this endeavour. We will then discuss how our experimentally based conclusions constrain QGP properties from heavy-ion collisions, and after that we will review measurements of QGP-like signatures in proton--proton (pp) and proton--lead (p--Pb) collisions. We will also discuss the contributions of ALICE measurements in pushing QCD to its limits in the context of few-body interactions.
We will finish by discussing connections to other fields of physics from our studies, and conclude by summarising what we have learnt so far, and what we look forward to for the future phases of ALICE in the 2020s and 2030s.

\newpage

\section{Introduction}
\label{ch:Introduction}

\subsection{Emergent phenomena in QCD: the quark--gluon plasma} 
\label{sec:emergQGP}

Understanding nature is one of the main goals of science. In the reductionist approach, one tries to infer all manifestations of reality using a compact set of relations~\cite{Chibbaro:2014,Weinberg:1987}. Such an approach has proved to be very fruitful in pinning down fundamental interactions, and discovering the basic building blocks of the universe. It has led to a framework, the Standard Model of particle physics~\cite{Glashow:1961tr,Weinberg:1967tq,Higgs:1964pj,Gross:1973id}, that describes the electromagnetic, weak and strong forces. It also allows for quantitative predictions of phenomena involving elementary particles, using the mathematical language of quantum field theory~\cite{Hooft:2015wra}.
The strong interaction, the major topic of ALICE's research, is responsible for the very existence of atomic nuclei. Broadly speaking, it is also responsible for $\sim $95\% of the visible mass of the Universe. 

Quantum Chromodynamics (QCD) is one of the pillars of the Standard Model of particle physics. It is a gauge field theory describing the strong interaction~\cite{Wilczek:1998ma}. One of its distinctive features is asymptotic freedom, a consequence of the non-Abelian nature of the SU(3) group on which QCD is based. Asymptotic freedom results in the interaction between quarks and gluons (partons), the elementary particles that experience strong interaction, becoming weaker when their momentum exchange increases. Therefore, the coupling constant $\alpha_{\rm S}$ of the strong interaction depends on the momentum scale of the interaction. Processes corresponding to large momentum transfers between partons~\cite{Gross:2005kv}, where $\alpha_{\rm S}$ is small, can be investigated via a perturbative approach (pQCD). This involves an expansion to higher orders in $\alpha_{\rm S}$~\cite{Politzer:1973fx}, with only lower orders contributing significantly. For calculations concerning low-momentum processes, the perturbative approach of QCD breaks down, as higher orders become dominant~\cite{Ioffe:2010zz}. Such a non-perturbative regime, where the coupling constant becomes large, is of the utmost interest for the description of several essential features of the strong interaction. In particular, in this regime confinement is observed~\cite{Greensite:2011zz}. This leads to the fact that quarks and gluons, the elementary particles carrying the ``charge'' of the strong interaction, known as colour, cannot be isolated, and therefore not directly observed. Only composite objects, hadrons, which do not possess a net colour charge, can be detected. Another key feature of strong interaction in the non-perturbative regime is the spontaneous breaking of chiral symmetry~\cite{Goldstone:1962es,Koch:1997ei,Fukaya:2009fh}. This  gives rise to the predominant fraction of the mass of hadrons, for example protons and neutrons. Theoretical approaches such as Lattice QCD~\cite{Gupta:1997nd,Muroya:2003qs,Ratti:2018ksb} and Effective Field Theories~\cite{Burgess:2007pt,Bodwin:1994jh,Brambilla:1999xf} represent viable solutions for the description of the non-perturbative regime of strong interactions. In this domain, guidance provided by experimental findings considerably helps the progress of the theory.

The behaviour of extended systems subject to the strong interaction provides a unique avenue to understand the strong interaction further.
For the electromagnetic interaction, condensed matter physics~\cite{Cohen:2006} studies phenomena as magnetism and superconductivity. These are prime examples of emergent behaviour~\cite{OConnor1994-OCOEP}, in the sense that their manifestations do not directly arise from the laws governing microscopic interactions. The study of ``QCD condensed matter'' can be performed by producing a many body system of quarks and gluons, under the conditions of large energy density.  Heating such a system, with zero (or very small) net baryon density, to a temperature exceeding 150--160~MeV (equivalent to more than $10^{12}$~K), leads to the creation of the quark--gluon plasma (QGP). This is a state of matter where two of the basic features of low-temperature QCD, confinement and chiral symmetry breaking, are no longer present~\cite{Cabibbo:1975ig,Shuryak:1977ut,Bazavov:2014pvz}.

\begin{figure}[ht!]
\begin{center}
\includegraphics[width = 0.55\textwidth]{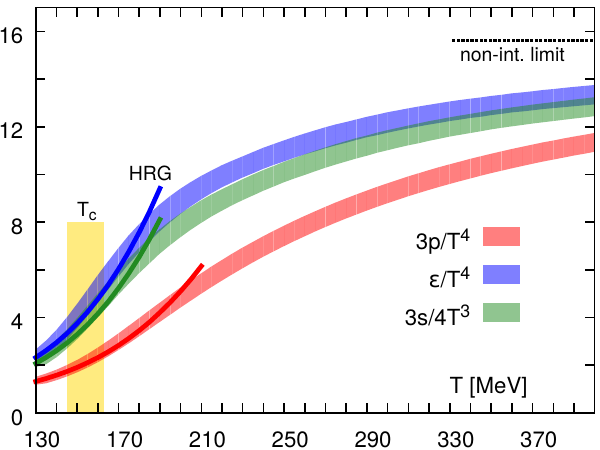}
\end{center}
\caption{Pressure, energy density and entropy density normalised to the 4$^{\rm th}$ (3$^{\rm rd}$ for the latter) power of the temperature, from the Lattice QCD calculations of the HotQCD Collaboration, see Ref.~\cite{Bazavov:2014pvz}. The dark lines show the prediction of the Hadron Resonance Gas model, the horizontal line corresponds to the ideal gas limit for the energy density. The vertical band indicates the cross-over transition region. Corresponding results from the Wuppertal-Budapest Collaboration can be found in Ref.~\cite{Borsanyi:2013bia}.}
\label{fig:latticecrossover}
\end{figure}

A primary role for the study of the QGP is played by Lattice QCD. Lattice QCD uses continuous space-time discretised on a lattice of finite size, with quarks defined at lattice sites, and gluons for the links connecting those sites. This approach can be used for QCD regarding static problems in the non-perturbative domain. Therefore, it is well suited for the study of the thermodynamic properties of a partonic system, such as the QGP. The thermodynamic properties can be estimated more precisely with increasing computational power. In this respect, reliable estimates of the properties of strongly interacting matter via Lattice QCD calculations require a number of important aspects. These include the use of realistic masses~\cite{Bazavov:2018omf} for the dominant quark flavours (the so-called (2+1)-flavour QCD, corresponding to up, down and strange quarks), and a controlled extrapolation to the continuum limit~\cite{Borsanyi:2013bia}. Indeed the charm, bottom and top quarks are too heavy to significantly add to the dynamics of the system, and it is the dynamics of the strange and, most importantly, of the light up and down quarks to determine the rich phase structure of QCD. Once the equation of state of QCD matter is evaluated with this technique, it is then possible to extract the temperature dependence of various thermodynamic quantities, such as the pressure, the energy and entropy density. The main results of such calculations demonstrate that a strongly interacting system with zero net baryon density evolves smoothly from a confined (hadronic) towards a deconfined (quarks and gluons) state, when its temperature is increased up to $T\sim 155$ MeV~\cite{Bazavov:2014pvz,Borsanyi:2013bia}. Since there is no discontinuity in thermodynamic variables, a crossover transition occurs, where de-confined and confined hadronic matter can co-exist and no latent heat is involved. The corresponding temperature is commonly indicated as ``pseudocritical temperature'' $T_{\rm pc}$. Moreover, it is found that such a cross-over occurs is in coincidence with the restoration of the chiral symmetry. The liberation of many new degrees of freedom is indicated by a strong increase in the energy density ($\epsilon$) normalised to the fourth power of temperature (Stefan-Boltzmann law) around the deconfinement temperature. This can be clearly seen in Fig.~\ref{fig:latticecrossover}, where the extrapolation to the continuum of lattice simulations of (2+1)-flavour QCD is shown~\cite{Bazavov:2014pvz}. The temperature dependence of other thermodynamical quantities as pressure ($p$) and entropy density ($s$) are also reported.

Figure~\ref{fig:latticecrossover} also shows that even at $T\sim 400$ MeV, the energy density is still $\sim 20$\% lower than the non-interacting ideal-gas limit. It is expected that this limit will be reached at very large $T$, due to the asymptotic freedom of QCD~\cite{Gross:2005kv}, i.e. the vanishing strength of the strong interaction for increasingly large momentum transfers. This residual difference has profound consequences on the properties of the QGP, with the remaining coupling responsible for various QGP features. In particular, below the ideal gas limit, the QGP manifests itself as a strongly interacting system of quarks and gluons. It behaves as an almost perfect liquid~\cite{Kolb:2003dz}, with the presence of bound states of quasi-particles (qq, gg, qg)~\cite{Shuryak:2004tx}. These properties have furnished in the literature to the definition of sQGP, with the letter ``s'' underlining its strong coupling~\cite{Nagle:2006cj}. Strongly coupled systems like the QGP also occur in the domain of electromagnetic interactions. For example, ultracold atomic Fermi gases in the extremely small pico-eV temperature range are the subject of an intense research activity~\cite{Adams:2012th}.

It is worth stressing that the result shown in Fig.~\ref{fig:latticecrossover} is obtained for a QCD medium with zero net baryon number. This is a particularly relevant configuration as it corresponds to that of the early universe, where deconfined quarks and gluons also hadronised around $T_{\rm c}$. A complementary situation, corresponding to large baryon density and relatively low temperature, may also lead to the creation of the QGP. This could be present in the core of neutron stars~\cite{Annala:2019puf}. In the laboratory, a situation corresponding to that of the early universe can be obtained by colliding heavy ions in the energy range accessible to hadron colliders. The CERN Large Hadron Collider (LHC)~\cite{Evans:2008zzb} is currently the facility where, starting from 2010, the highest collision energies are attained. The values of  the center-of-mass energies per nucleon pair have reached $\sqrt{s_{\rm NN}}\sim 5$ TeV. The main scope of this review is to analyse and discuss the results obtained by ALICE (A Large Ion Collider Experiment)~\cite{Aamodt:2008zz} in order to study QGP properties.

In the following sections of this Introduction, a discussion of the general aspects of experimental and theoretical investigations of high temperature and energy QCD using the ALICE detector will occur (Sec.~\ref{sec:observ}). Then, a brief review of the history and the main results obtained from  lower energy facilities regarding QGP formation will be given (Sec.~\ref{sec:QGPhistory}). A concise introduction of the ALICE detector, with key details on its design, realisation and performances will follow (Sec.~\ref{sec:ALICEExp}). Finally, an overview of the key scientific questions addressed by ALICE, in light of QGP studies, but also exploring other areas of QCD, will be described (Sec.~\ref{sec:KeyQuestions}). In that last Section the organisation of the following chapters will also be discussed.

\subsection{Experimental investigations of the QGP and QCD}
\label{sec:observ}

The purpose of this section is three-fold. It will start by summarising the conceptual framework 
that has arisen in the field of heavy-ion physics over a number of decades. It will then introduce commonly measured observables, which will be shown and elaborated on in subsequent chapters. It will then finish by a brief description of the main theoretical approaches that attempt to use such observables to further investigate the QGP, and more generally QCD. Additional theoretical approaches in the context of comparisons to ALICE data will also be explored in subsequent chapters.

\subsubsection{The evolution of a heavy-ion collision}
\label{sec:STMHI}

\begin{figure}[ht]
\begin{center}
\includegraphics[width = 1\textwidth]{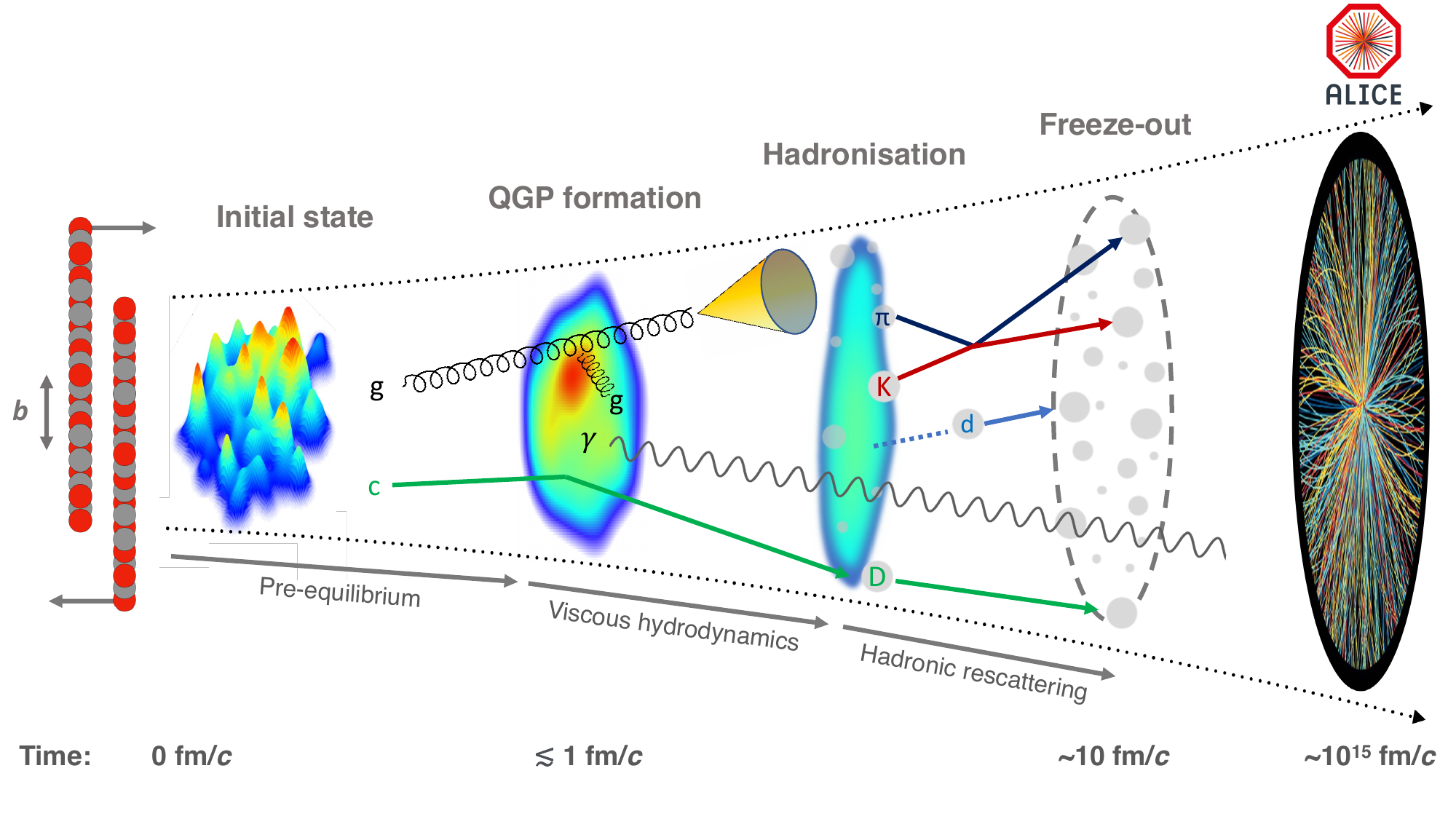}
\end{center}
\caption[Evolution of a heavy-ion collision.]{The evolution of a heavy-ion collision at LHC energies.}
\label{fig:EvolHI}
\end{figure}

Collisions of heavy ions with ultra-relativistic energies are used to create the QGP in the laboratory. The evolution of a heavy-ion collision is commonly described in terms of a series of stages, which can in principle be factorised. They include: (i) an initial state, defined by the wave-functions of the projectiles, which are universal and independent of any specific scattering process; (ii) large-$Q^2$ interactions of partons drawn from the projectiles\footnote{$Q^2$ being the 4 momentum transfer squared}; (iii) smaller-$Q^2$ interactions generating a pre-equilibrated parton gas; (iv) equilibration and expansion of the QGP; (v) hadron formation; (vi) chemical freeze-out of hadrons; (vii) hadronic interactions that subsequently freeze-out kinetically; (viii) free-streaming of stable particles to the detector.\footnote{For a more detailed conceptual description of heavy-ion collisions, see Ref.~\cite{Busza:2018rrf}.} This evolution is illustrated schematically in Fig.~\ref{fig:EvolHI}\footnote{This figure was inspired by illustrations from ~\href{https://u.osu.edu/vishnu/2014/08/06/sketch-of-relativistic-heavy-ion-collisions}{Chun Shen}.}. As the heavy ions collide, an extremely dense region of partons is excited and deposits energy and entropy in the overlap region of the collision. Before they interact, the nuclei at the LHC will be highly Lorentz contracted, as indicated in Fig.~\ref{fig:EvolHI}. The impact parameter $b$ is the distance between the centres of the colliding nuclei. It very closely related to the number of nucleons in the nuclei that participate in an inelastic interaction (at least once), referred to as $N_{\rm part}$. It also controls the volume of the collision region. The total number of inelastic nucleon--nucleon collisions is referred to as $N_{\rm coll}$. When $b$ is small, $N_{\rm part}$ and $N_{\rm coll}$ will be large, and vice-versa. The largest number $N_{\rm part}$ can reach is 2$A$ (if each colliding nuclei have the same number $A$ of nucleons) 
 and for $N_{\rm coll}$ the respective number is $\sim A^{4/3}$ (which can reach values of $\sim$ 2000 for Pb--Pb at LHC energies). Nucleons not participating in the collision are defined as spectators, and continue travelling approximately along the beam direction after the collision. In the inelastic nucleon--nucleon interactions, a variety of QCD processes occur, which, as mentioned, involve a range of $Q^2$ momentum transfers, and each range plays a distinct role. 

The partons within the nuclei that are involved in the smaller-$Q^2$ interactions determine the overall energy density and entropy deposition in the initial state, and their interaction rate is largely driven by $N_{\rm part}$. Such parton interactions lead to a ``lumpiness" of the initial density profile, which is the result of fluctuations in the distribution of nuclear matter, and is depicted in Fig.~\ref{fig:EvolHI}. Immediately after the collision, the smaller-$Q^2$ interactions occur in the context of a weakly coupled pre-equilibrium phase. This is followed by the creation of even softer partons in these processes, which enable the formation of a strongly coupled QGP phase. The hard processes from large-$Q^2$ interactions, with their rate driven by $N_{\rm coll}$, enable the creation of high momentum gluons and high momentum/mass quarks, as indicated in Fig.~\ref{fig:EvolHI} via the gluon and charm quark trajectories. As they have short wavelengths, they will interact with other quarks and gluons on a microscopic level, leading to energy loss effects (the energy being transferred to the medium), and therefore they offer information on the opaqueness of the QGP. The interactions of high-momentum partons with the QGP can be radiative as indicated in Fig.~\ref{fig:EvolHI} for a gluon, as well as elastic, as indicated by the change in direction of the charm quark. The amount of energy loss will depend on the colour charge, momentum, mass, type of process (inelastic or elastic), the distance traversed (path length) of the hard scattered parton, and is subject to stochastic processes. The heavy quarks produced via hard processes can also form quarkonia (bound heavy quark-antiquark states), with their production rate being suppressed because the binding force between the quark and anti-quark is weakened (screened) by the presence of the colour charge of quarks and gluons. That suppression is closely related to the temperature of the QGP, and can be counterbalanced by a regeneration process that recombines heavy quarks participating in the medium interactions, depending on the abundance of heavy quarks. In addition, the parton fragmentation processes (indicated by the yellow cone) lead to jets, partonic showers that arise from these high energy partons, and that fragment into experimentally observable hadrons once the shower components reach low virtuality. That fragmentation pattern in the medium can be altered compared to vacuum-like conditions, e.g. e$^+$e$^-$ collisions.

The evolution of the QGP for most processes involved in soft interactions after $\sim 1$~fm/$c$ can be understood as follows. Since the mean free path of the vast majority of QGP constituents is expected to be much smaller than the size of the QGP formed (assuming these constituents are strongly coupled), multiple interactions drive the expansion. This expansion is highly influenced by the non-uniform energy distribution in the initial state as a function of space, which creates pressure gradients in the QGP stage, that act to cool the system and to smooth out the lumps as time progresses. If the length scales of these gradients are larger than the mean free path, such an evolution exhibits long-wavelength behaviour, which can be described as a liquid in the context of hydrodynamics. A radial flow occurs due to a greater pressure at the centre of the QGP compared to the outskirts, and this leads to a common velocity field outwards. The rate of the hydrodynamic expansion is influenced by the QGP's bulk viscosity, which is its resistance to volume growth. Anisotropic flow is the result of a directional dependence to these pressure gradients. This occurs due to spatial anisotropies in the initial state. These arise if the collision zone is almond shaped (at $b>0$), or due to the lumpiness of the initial state. Such spatial anisotropies are converted to momentum anisotropies via the hydrodynamic response. This is influenced by the QGP's shear viscosity, which quantifies the resistance to fluid deformation. 

Since the nuclei are charged, the movement of the incoming beams sets up a very large magnetic field (from the protons in the nuclei with relativistic energies), which can also have an influence on motion of quarks during the QGP stage. The Chiral Magnetic Effect (CME), which is purported to result from strong parity violation, could lead to a splitting among positive and negative quarks in the QGP along the direction of the extremely large magnetic fields produced from the colliding ions. The decaying magnetic field will also induce an electric field, which can move the electrically charged quarks accordingly. Such an effect prolongs the decay time of the magnetic field (Lenz's law). The high-temperature QGP also produces thermal radiation, in the form of both photons and lepton-antilepton pairs. The electromagnetic radiation does not interact via the strong force in the QGP, and can therefore be used to gain information about the temperature from early to late times in the collision processes. In addition, strange quarks, which have masses below the deconfinement temperature (therefore the QGP temperature) can also be produced easily in the QGP stage, with the dominant mechanism coming from gluon interactions.

During the evolution, the parts of the QGP that cool below the transition temperature $T_{\rm pc}$ will hadronise. %
Since it is likely that the temperature and energy density of the medium will diminish with increasing distance from the collision centre, and, as indicated by Lattice QCD, the transition to normal nuclear matter is a smooth cross-over, it is possible that hadronisation will happen at different times at different places in the phase-space. Nonetheless, the formation of hadrons from the QGP can proceed as follows. Hard partons in a jet will fragment and hadronise in the similar manner as in elementary collisions. For partons at lower momenta, if they share a similar space and momenta as other partons, they can combine into hadrons via coalescence. Heavy (charm or beauty) quarks can also combine with heavy antiquarks, thus forming quarkonia and give rise to an additional production mechanism for closed heavy-flavour hadrons, which can compensate the deficit due to melting in the QGP. 

After the hadrons materialise out of quarks and gluons, primarily pions ($\pi$), kaons (K) or protons (p), the energy density may be large enough to allow for inelastic interactions, with a consequent evolution of their “chemical” composition,  in terms of particle species.  A loosely bound state such as the deuteron (d) shown in Fig.~\ref{fig:EvolHI} is particularly sensitive to such interactions, as it can be easily formed or destroyed. Such interactions cease at the chemical freeze-out temperature $T_{\rm chem}$, fixing the particle composition. Elastic interactions can still continue, demonstrated by the $\pi$--K interaction in Fig.~\ref{fig:EvolHI}, and halt at the kinetic freeze-out temperature $T_{\rm kin}$, which is achieved at the time of $\sim 10$~fm/$c$. At this point, the particle momenta are fixed. These particles travel towards the ALICE detector, where they will be measured $\sim 10^{15}$~fm/$c$ after the initial collision. The high-energy beams of the LHC provide an unprecedented opportunity to study the QGP in the laboratory. The highest centre of mass energy per nucleon pair ($\sqrt{s_{\rm NN}}$) achieved at the LHC in Pb--Pb collisions has been 5.02 TeV, which is $\sim$25 times higher than the top energy available at the Relativistic Heavy-Ion Collider (which began taking data in 2000). This in principle allows for the hottest, densest, and longest ever lived QGP formed in the laboratory to be probed using the ALICE detector.

\subsubsection{Observables in heavy-ion collisions}
\label{sec:expobservables}

There are various experimental probes used to investigate all phases of heavy-ion collisions: the initial state, the QGP phase, and the final hadronic phase. Each of these probes has varying sensitivities to each phase. To start, a fundamental quantity for many of these probes is the Lorentz-invariant differential yield of final state particles, given by:
\begin{equation}
E\frac{{\rm d}^{3}N}{{\rm d}p^3} = \frac{1}{2\pi p_{\rm T}} \frac{{\rm d}^{2}N}{{\rm d}p_{\rm T}{\rm d}y},
\label{eq:Spectra}
\end{equation}
which is the number density of the particle three-momentum scaled by the particle energy ($E$). This will depend on the measured particle species in question, the transverse momentum, \pt, and rapidity, $y$. When the particle species is not known, pseudorapidity $\eta$ is used instead~\footnote{Defined as $\eta=-\ln[\tan{(\theta/2)}]$, $\theta$ being the polar emission angle of the particle. A more complete description of kinematic variables can be found in a Review of Particle Physics by the Particle Data Group~\cite{ParticleDataGroup:2022pth}}, and both are equivalent when the particle energy is much greater than its mass. Different \pt ranges will probe different physical processes, and for clarity in this review, one can define low-\pt as $p_{\rm T} \lesssim 2$~GeV/$c$, intermediate-\pt as $2 \lesssim p_{\rm T} \lesssim 8$~GeV/$c$, and high-\pt as $p_{\rm T} \gtrsim 8$~GeV/$c$. This classification is not intended to provide a rigid distinction regarding these processes, but to aid the reader regarding references to momentum ranges.

\paragraph{Initial state.} Regarding the initial state, for a given collision, the multiplicity can be determined, which is an addition of the number of charged hadrons in a broad momentum range. It plays a critical role in providing a selection on a range of impact parameters, $b$, for heavy-ion collisions. It will be large when $b$ is small, which leads to large numbers of $N_{\rm part}$ that will correspondingly produce large numbers of particles. Such small-$b$ collisions are referred to as central i.e. head-on, whereas collisions with large impact parameters and small numbers of $N_{\rm part}$ (and fewer produced particles) are referred to as peripheral. The multiplicity therefore provides an experimental handle on the centrality of a collision. Such a handle is extremely useful, as many of the system properties such as the energy density or lifetime depend upon the centrality. The multiplicity can also provide a measure of the initial state entropy. If the system hydrodynamically evolves without internal resistance i.e. viscous effects are minimal, the multiplicity in the final state can be used to determine the initial state entropy, as entropy is conserved in this case. If viscous effects become appreciable, entropy is created during the evolution, and the multiplicity then serves as an upper limit to the initial state entropy directly. 

Ultra peripheral collisions (UPCs) provide another handle on the initial state. These are heavy-ion collisions that typically have very large $b$ values (greater than the nuclear diameter e.g. $\sim$ 15 fm for Pb). The nuclei in these collisions are sufficiently separated such that short-range strong interactions are highly suppressed (and therefore no QGP formation), yet the huge electromagnetic fields generated in the vicinity of the colliding nuclei lead to photon mediated interactions. Measurements of the differential cross sections of light (e.g.\,$\rho^{0}$) and heavy (e.g. \jpsi) vector mesons from these collisions are especially sensitive to the momentum distribution of partons (i.e. quarks and gluons) inside the nucleus, parameterised as nuclear parton distribution functions (nPDFs). These distributions are modified compared with proton--proton collisions parton distribution functions (PDFs), and nPDFs play a critical role for understanding the production rate of hard processes in the initial stages of heavy-ion collisions.

\paragraph{QGP phase.} The amount of the hydrodynamic collective motion developed in the QGP phase can be explored using measurements sensitive to radial and anisotropic flow~\cite{Heinz:2013th}, which develop transverse to the beam direction. In the low-\pt region, radial flow causes an increase in the transverse momentum (\pt) of higher-mass hadrons. Measurements of the mean transverse momentum \mpt of identified particles can therefore be used to extract information on the amount of radial flow that has developed. In order to investigate anisotropic flow, Eq.~\ref{eq:Spectra} can be expanded into the azimuthal angular dependence of the transverse momentum vector direction. The most widely used decomposition expresses the $\varphi$ (the angular component of this vector) dependence of the produced particle density as a Fourier series as follows~\cite{Voloshin:1994mz}:
\begin{equation} 
\frac{{\rm d}N}{{\rm d}\varphi} \propto 1 + 2\sum_{n=1}^{\infty} v_{n} \cos [n(\varphi-\Psi_n)],
\label{eq:v_n}
\end{equation}  
The $v_n$ terms, which can be determined experimentally, are referred to as anisotropic flow coefficients. They depend on the particle species, $p_{\rm T}$ and $y$. The term $n$ corresponds to the order of anisotropic flow, and $\Psi_n$ is the corresponding symmetry plane angle, which is the angular direction of anisotropic flow for the order $n$ of interest. For the vast majority of produced hadrons, non-zero $v_n$ coefficients arise mainly from the QGPs hydrodynamic response (with a relatively small contribution from the hadronic state). Regarding the coupling of the created electromagnetic fields with quarks, the CME effect introduces charge-dependent sine terms in  Eq.~\ref{eq:v_n} for the produced hadrons with corresponding coefficients. These are small relative to the $v_n$ coefficients, but can also be explored experimentally. Any motion due to an electric field in the QGP will lead to charge-dependent $v_{1}$ coefficients.

Hard probes, as mentioned, are produced in the earliest times of the collision. One of the ways they can be explored is by extracting their nuclear modification factor $\RAA$. This observable is constructed to be sensitive to changes of the dynamics of hard processes in heavy-ion collisions with respect to expectations from elementary \pp collisions. Within a $y$ or $\eta$ interval, $\RAA$ as a function of transverse momentum is defined as follows:
\begin{equation}
\RAA (\pt) =  \frac{1}{\avTAA} \, \frac{{\rm d}\Naa (\pt)/ {\rm d}\pt}{{\rm d} \sigma _{\rm pp}(\pt) / {\rm d}\pt},
\label{equ:RAA}
\end{equation}

The average nuclear overlap function \avTAA is obtained from the average number of \Ncoll divided by the inelastic nucleon--nucleon cross section for the centrality range of interest, and its estimation is discussed in Sec.~\ref{sec:theotools}. For hard processes, the yield \Naa in heavy-ion collisions is expected to scale with the average nuclear overlap function \avTAA when compared to the production cross section $\sigma _{\rm pp}$ in \pp collisions, in the absence of any QGP or initial state nuclear effects. That being the case, if $R_{\rm AA}(p_{\rm T})=1$, production from heavy-ion collisions can be considered as a superposition of nucleon--nucleon collisions. Any departure from unity reveals how these processes are modified in heavy-ion collisions. In particular, $R_{\rm AA}$ is expected to be below unity at high-\pt for inclusive hadrons from partons undergoing in-medium energy loss. Jet-finding algorithms can also be applied in heavy-ion collisions. The internal structure of jet showers is governed by quantum interference effects, resulting in the phenomenon of “angle-ordering”, whereby the highest \pt hadrons in the shower are on average most closely aligned with the nominal jet axis. A key parameter in jet measurements is the ``jet radius" or ``resolution parameter” $R$, which is effectively the size of the aperture through which the jet shower is viewed. The measured jet yields can be used to determine $\RAA$, while the jet radius dependence of the $R_{\rm AA}$ or jet substructure measurements provide information about the medium modifications of the quark and gluon radiation patterns. The heavy-ion jets can be compared to jets in pp collisions, where no such modifications are expected. Finally, measurements of $v_n$ at high-\pt can be used to explore the path length dependence of energy loss. Rather than resulting from the hydrodynamic response, finite values can arise as large momentum partons lose less energy in the $\Psi_n$ direction as the path length is smaller, compared to the perpendicular direction where this is larger, within the elongated nuclear overlap region. This leads to a larger abundance of high-\pt hadrons in the $\Psi_n$ direction, which, in turn, results in an observed non-zero $v_n$.

A specific class of hard probes, heavy quarks, gives access to several QGP features that can be accessed via differential measurements of $R_{\rm AA}$ and $v_n$. 
A key difference is that the corresponding hadrons are associated with early-stage probes across their entire \pt range. 
D mesons, which carry the vast majority of the charm quarks produced, can be investigated differentially as function of \pt via measurements of $v_n$ and \RAA. Measurements of $v_n$ at low-\pt are sensitive to degree to which heavy quarks take part in the collective expansion of the QGP and approach thermalisation. Measurements of \RAA at high-\pt provide insights on the energy loss processes of heavy quarks, which maybe smaller than for light quarks or gluons, due to the dead cone effect~\cite{Dokshitzer:2001zm}. Measurements of quarkonia (e.g.\,\jpsi or \upsone) \RAA and $v_n$ are also carried out as a function of \pt. The \RAA studies allow the investigation of the aforementioned suppression and regeneration processes, with the latter expected to be dominant in the low-\pt region due to larger heavy quark multiplicity in that kinematic range. Anisotropic flow measurements give complementary information to those from open heavy flavours, with the observed effect not involving the contribution of the flow of light quarks.

Electroweak probes, e.g.\,photons, leptons or Z/W bosons, provide another class of measurements. If they arise from hard processes in the initial stages, measurements of $R_{\rm AA}$ at high-\pt are expected to be unity, and any deviations from this reveal the influences of non-QGP processes that affect this measurement e.g.\,isospin effects or differences in nPDFs and PDFs. On the other hand, measurements of direct photon (i.e.\,not from hadron decays) spectra at low-\pt are expected to arise mainly from the softer processes in the QGP involving charged quarks, and their slope can be related to the temperature when they are created. Direct photons can also be produced from hadronic interactions, so that the observed yield is a convolution of their emission along the whole collision history. Since photons are produced in an expanding QGP medium, they will be blue-shifted, which has to be taken into account regarding their connection to the temperature of the QGP.

\paragraph{Hadronic phase.} A variety of methods are available to investigate the hadronic phase of the heavy-ion collisions, and will be outlined in the following. Measurements of identified hadron spectra and ${v}_n$ at intermediate-\pt are sensitive to the coalescence of quarks upon hadronisation from the QGP phase, which will compete with hadronisation from the well known fragmentation processes observed in QCD interactions. Such a coalescence mechanism enhances baryon production rates and ${v}_n$ values compared to mesons in this \pt range. 
The chemical freeze-out temperature $T_{\rm chem}$ can be explored by measuring the total  yields of specific particles ($\rm{d}\it{N}/\rm{d}\it{y}$) via an integration over \pt in Eq.~\ref{eq:Spectra} - for hadrons that do not decay quickly via the strong or electromagnetic interactions (e.g protons, kaons, charged pions, hyperons etc). Measurements of the production of strongly decaying resonances can offer insight into the duration of the hadronic phase. Their decay products for temperatures both above and below $T_{\rm chem}$ are subject to elastic scattering. If these effects are strong enough, they will lead to a depletion in the resonance yields, as the kinematics of the particles resulting from a resonance decay are altered in such a way that the resonance cannot be reconstructed anymore. On the other hand, regeneration processes from hadron--hadron interactions may form resonances during this phase. Final state hadron--hadron interactions can also be studied via the pair-wise femtoscopic correlation function, which can be extracted experimentally as:
\begin{equation} 
C(k^{*}) = \frac{N_{\rm pairs}\text{ (same event)}}{N_{\rm pairs}\text{ (background)}},
\label{eq:cffem}
\end{equation} 
where $k^{*}$ is the invariant momentum difference between the pairs. The numerator is extracted from pairs produced in the same collision, and the denominator is obtained from a background hypothesis that assumes no correlated pairs, and is normalised such that $C(k^{*})=1$ in the absence of same-event correlations. Femtoscopic correlations arise from quantum correlations which are sensitive to the size of the system at freeze-out, or final state hadron--hadron interactions, both of which predominately occur at low-$p_{\rm T}$. Finally, the slope of hadron spectra can be used to infer the kinetic freeze-out temperature $T_{\rm kin}$. Such an approach relies on modelling to disentangle the contribution from radial flow, and will be discussed in the next chapter.

\subsubsection{Small systems and searches for thresholds of QGP formation}
\label{sec:introsmallsystem}

Proton--proton collisions, where initial state nuclear effects are not relevant, were envisioned to provide a crucial reference to ultrarelativistic heavy-ion collisions. Collisions of p--Pb also have a further complementary role, by providing a reference where cold nuclear matter (CNM) effects are expected, but in principle, no QGP effects. CNM effects include modifications of the parton distribution functions in the nucleus compared to the proton, multiple scatterings in nucleons that collide with more than one other nucleon, the Cronin Effect~\cite{Cronin:1974zm}, parton energy loss in cold nuclear matter, and absorption of the produced hadrons by the nucleus. One way to study CNM effects independently of the formation of the QGP (in principle), is by measuring \RpPb, which is an analogous application of Eq.~\ref{equ:RAA} for p--Pb collisions. As CNM effects are also expected to occur in heavy-ion collisions, such measurements are very important for disentangling CNM effects from QGP effects, and are of interest in their own right. Proton--proton collisions, and to some extent proton--nucleus collisions, have provided opportunities for studying QCD interactions in few-body systems.
For example, inclusive and heavy-flavour jet cross sections provide stringent tests on perturbative QCD (pQCD), while measurements of the production of identified hadrons can be used to test the universality (collision-system independence) of parton-to-hadron fragmentation functions (FFs). The use of femtoscopic measurements to study final state hadron--hadron interactions, which occur for any collision system producing hadrons, is another example. \\ \\ 
All things considered, pp and p--Pb collisions therefore provide an opportunity to study QCD at the LHC in a more dilute environment. Studying the multiplicity dependence of observables described in the previous section for pp and p--Pb collisions provides a means to explore the thresholds for QGP formation. Unlike the case for heavy-ion collisions, high multiplicity events in pp and p--Pb collisions are not expected to result from a trivial increase in the amount of colliding matter: this is obviously the case for pp collisions as $N_{\rm part}$ is always 2. Rather, given the initial system volume is somewhat fixed, high multiplicity events may be associated with collisions that have energy densities exceeding the values required for QGP formation. Indeed, the highest number of particles produced in such collisions are comparable to peripheral heavy-ion collisions at lower energies, where QGP formation is established. Even if the energy densities in pp and p--Pb collisions reach values needed for QGP formation, it is not clear a priori if and which typical QGP effects can be observed in their study. A detailed investigation in this sense, using measurements developed in heavy-ion collisions to explore QGP-like behaviour - such as anisotropic flow coefficients, the production rate of hadrons from hard processes, or changes in hadronisation mechanisms as a function of multiplicity, represents a fascinating avenue of study.

\subsubsection{Theoretical tools}
\label{sec:theotools}

A wide variety of theoretical tools are employed to investigate the aforementioned processes in heavy-ion and light system collisions. They range from fundamental QCD calculations to effective theories and phenomenological models. Some of the main goals of comparing the output of theory to the previously described measurements are as follows:

\begin{enumerate}[label=\roman*.]
 \item Map the spatial and momentum distributions of the initial state.
 \item Provide quantitative constraints on QGP properties and probe QGP behaviour.
 \item Understand the contribution of hadronic interactions to final state measurements and explore the global properties of the hadronic phase.
 \item Assess the degree QGP effects or few body QCD interactions describe the data in heavy and small systems.
\end{enumerate}

\paragraph {Initial state.} The use of initial state modeling is an essential prerequisite to determine QGP properties, and is of vast interest in and of itself. The Monte Carlo (MC) Glauber approach can be used to infer \avNpart, \avNcoll, and \avTAA within a particular centrality/multiplicity range for Pb--Pb and p--Pb collisions~\cite{Miller:2007ri,dEnterria:2020dwq}. This approach assumes the nucleons in the nucleus are positioned according to the Woods-Saxon density distribution. The nucleons travel in an unperturbed trajectory irrespective of whether they interact with other nucleons. The criterion for an inelastic nucleon--nucleon interaction then depends on the cross section of a single interaction, which can be inferred from experimental measures in pp collisions. These assumptions are particularly relevant for heavy-ion collisions at the LHC that are in transparency regime, which implies the nuclear remnants after a collision travel in a straight line. 
\begin{figure}[ht]
\includegraphics[width = 1\textwidth]{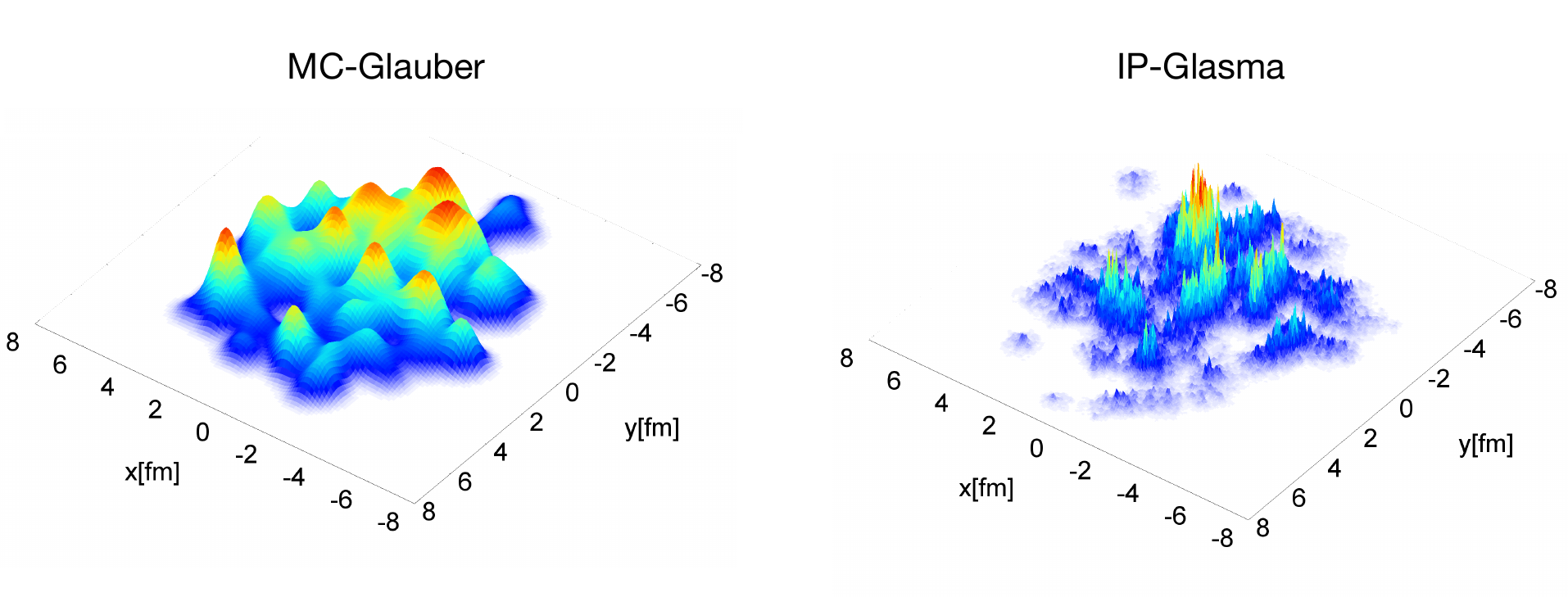}
\caption{The $xy$ distributions of the initial energy density (arbitrary units) from the MC-Glauber and IP-Glasma models for a heavy-ion collision~\cite{Schenke:2012wb}.}
\label{fig:IPMC}
\end{figure}
Another output of such an approach is the initial state spatial eccentricity, $\varepsilon_{\rm{n}}$, in the (transverse) $xy$ plane. This depends on how the matter in the initial state is distributed. It is an essential input for the hydrodynamic model chains (which will be described shortly), as non zero values of $\varepsilon_{\rm{n}}$ generate anisotropic flow. Other initial state models go further. For example, the IP-Glasma model uses MC Glauber assumptions~\cite{Schenke:2012wb}, but then determines the spatial gluon distributions in a nucleon using the QCD based {\it Colour Glass Condensate} ansatz~\cite{PhysRevD.49.2233}. It attempts to model these distributions and how they overlap in a heavy-ion collision in the context of nuclear saturation effects, which control the high number of low-$x$ gluons populating the transverse extent of nucleons in a high-energy collision. Saturation effects dictate the population of the vast majority of partons at low-$x$, and their corresponding spatial distributions. This additional feature influences the value $\varepsilon_{\rm{n}}$ compared to the MC Glauber approach, and IP-Glasma generally predicts higher values of $\varepsilon_{2}$ in a particular centrality range. A comparison of the energy densities from MC-Glauber (derived from nucleons) and IP-Glasma (gluons) is shown in Fig.~\ref{fig:IPMC} in the $xy$ plane. For the study of hard processes in both elementary and nuclear collisions, knowledge of the initial momentum distributions of partons also is an essential prerequisite. These distributions can be constrained using measurements from UPC collisions with the appropriate theoretical comparisons. Parameterisations of PDFs in protons, expressed as a function of the fractional momentum $x$ of the parton and the momentum transfer $Q^2$ are modeled generally starting from the results of deep inelastic scattering experiments. In the nucleus, typical depletion effects (i.e.\,a reduction in the nPDF) such as shadowing are prominent at low-$x$ ($\ll 0.1$), and can have notable consequences on the production of hard processes. Anti-shadowing effects lead to parton excesses relative to pp collisions at high-$x$ ($\sim 0.1$). Recent parameterisations of nPDFs including these effects via fits the data have been investigated elsewhere~\cite{Kovarik:2015cma,Eskola:2016oht}.

\begin{figure}[ht]
\includegraphics[width = 1\textwidth]{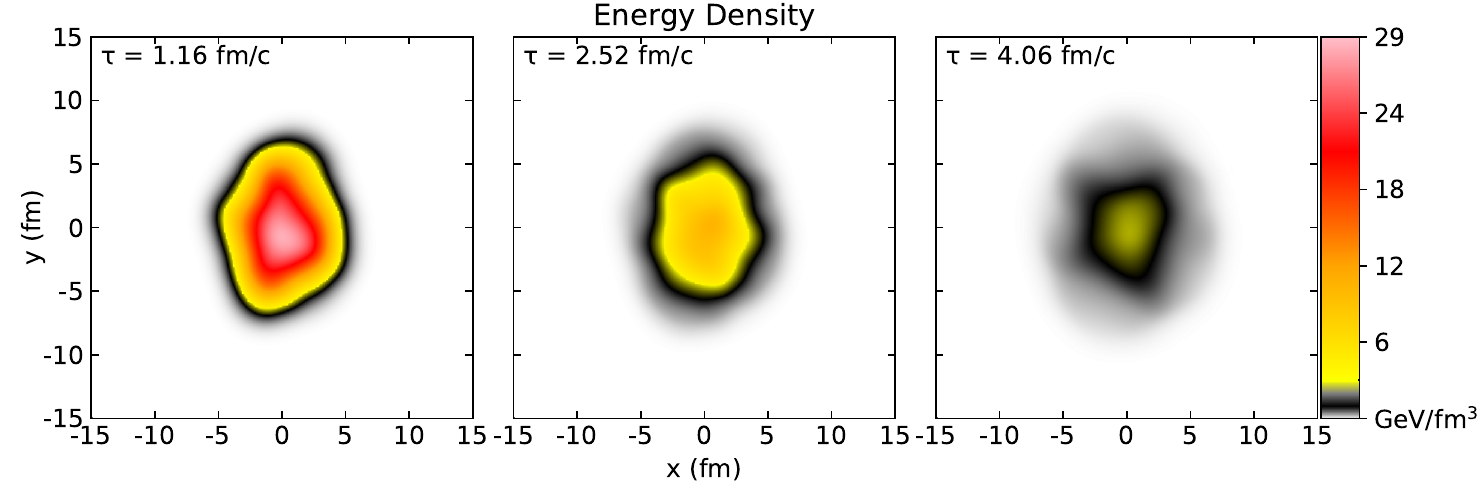}
\caption{A mapping of the energy density in the QGP phases vs time and space for a mid-central ($b=7.5$ fm) Pb-- \fivenn collision using the T$_{\rm{R}}$ENTo-VISHNU model chain~\cite{Bernhard:2019bmu}.}
\label{fig:hydroED}
\end{figure}

\paragraph {QGP properties and behaviour.} Viscous hydrodynamic theory is one of the main methods employed to investigate the dynamical evolution of the QGP. Lattice QCD, being a static theory in space and time, cannot be used for this purpose. Viscous hydrodynamic theory is used in the context of a hydrodynamic model chain e.g. IP-Glasma+MUSIC\cite{Schenke:2020mbo} or T$_{\rm{R}}$ENTo-VISHNU~\cite{Bernhard:2019bmu}, with components that attempt to describe each stage of a collision. These components include the energy/entropy density from an initial state model at the moment of the collision, a pre-equilibrium phase that describes the weakly coupled dynamics prior to the hydrodynamic phase, an implementation of the viscous hydrodynamics to describe the evolution of the QGP when it is formed, a particlisation scheme to create hadrons when the deconfinement temperature is reached, and finally a model to describe hadronic interactions. The models used for hadronic interactions, often called hadronic afterburners, will be described later in this subsection. The viscous hydrodynamic stage itself uses a Lattice QCD equation of state to evolve a strongly interacting system via energy and momentum conservation equations. As the final output of these hydrodynamic model chains are the momentum distributions of hadrons, this facilitates direct comparisons to data; in particular measurements sensitive to radial and anisotropic flow. An example of an intermediate output of such of a model is shown in Fig.~\ref{fig:hydroED}, which demonstrates the evolution of the energy density as function of space (in the transverse plane) and time. 

For anisotropic flow orders of $n=$ 2, or 3, hydrodynamic models exhibit the following factorisation relation:

\begin{equation} 
v_n \approx \kappa_n \, \varepsilon_n
\label{eq:vnen}
\end{equation} 
where the variable $\kappa_n$ encodes the efficiency of the QGP in converting the spatial anisotropy $\varepsilon_n$ to a momentum anisotropy $v_n$. That efficiency is dictated by two prominent QGP transport properties, which are needed as inputs to hydrodynamic models, and are $\eta/s$ and $\zeta/s$. The terms $\eta$ and $\zeta$ are the shear and bulk viscosities respectively, and as mentioned, they influence the amount of anisotropic and radial flow respectively, while $s$ is the entropy density. The $\eta/s$ and $\zeta/s$ terms can be deduced from different coupling regimes that can be explored for the QGP, and are ultimately constrained by measurements sensitive to radial and anisotropic flow. These regimes range from beyond Standard Model approaches that include infinite coupling regimes from string theory techniques (AdS), to Standard Model QCD descriptions in the strong and weak coupling limits~\cite{Arnold:2003zc, Arnold:2006fz, Meyer:2007ic, Denicol:2009am, Rougemont:2017tlu}. 

The production of hard probes and their interactions with the QGP offers distinct information regarding QGP properties. The description of the hadron production from these processes relies on the application of the QCD Collinear Factorisation theorem~\cite{Collins:1989gx} to heavy-ion collisions. As an example, it can be applied as follows to determine the differential hadron ($h$) production cross section~\cite{Qin:2015srf}:
\begin{equation} 
{\rm d}{\sigma}_{AB \to hX} \approx \sum_{abjj'd} f_{a/A}(x_a) \otimes f_{b/B}(x_b) \otimes {\rm d}\sigma_{ab\to jd} \otimes P_{j\to j'} \otimes D_{h/j'}(z_{j'}).
\label{eq:ft}
\end{equation} 
The $A$ and $B$ terms correspond to the colliding nuclei. The $f$ terms represent the parton distribution functions of those nuclei at longitudinal momentum fractions $x_a$, $x_b$ for incoming partons $a$ and $b$. The term ${\rm d}\sigma_{ab\to jd}$ is the cross section of a hard scattered parton $j$, while $P_{j \rightarrow j'}$ describes the effect of the parton $j$ interacting with the coloured medium before fragmenting into hadrons. The term $D$ is the fragmentation function in terms of the fractional momentum $z$ of the hadron from the modified parton $j'$. As mentioned, the hard-scattered partons interact with the colour-charged constituents of the QGP through elastic collisions~\cite{Bjorken:1982tu} and inelastic scatterings, with the latter giving rise to medium-induced gluon radiation~\cite{Gyulassy:1993hr,Baier:1994bd}.
Such interactions cause a modification of the parton shower, usually denoted as ``jet quenching''~\cite{Gyulassy:1990ye}.
The simplest manifestation of jet quenching is a reduction of the energy/momentum of the jet (and of the leading hadron emerging from the jet fragmentation) due to the transport of the initial parton energy to large angles from the jet direction (``energy loss''~\cite{Wang:1998bha}).
In the high-momentum regime, the influence and the effects of the QGP, mainly the collisional and radiative energy loss, can be calculated in the context of pQCD utilising either (semi)analytical calculations of in-medium energy loss~\cite{Gyulassy:2000fs,Djordjevic:2003zk} or Monte Carlo simulations of the microscopic interactions that occur  in an expanding QGP~\cite{Zapp:2008gi,Cao:2017zih}. 
The main energy loss mechanism for high momentum partons is expected to be radiative gluon emission~\cite{Baier:1996kr}. In the framework of the Factorisation theorem, the medium-induced energy degradation of the leading parton can be described through medium-modified parton fragmentation functions~\cite{Salgado:2003gb}.
These models for QGP interactions explore the energy loss of gluons (which are the majority of hard partons produced at LHC energies), light quarks, and heavy quarks, which are influenced differently in the QGP due to the different colour charges and masses~\cite{Dokshitzer:2001zm}. The parameter that is typically used to characterise the medium-induced energy loss in such models is the transport coefficient $\hat{q}$, which can be considered a quantification of the opacity of the QGP. It is defined as the average of the squared transverse momentum transfer per mean free path of a parton due to energy loss in the QGP, and like $\eta/s$, is directly linked to the coupling regime of the QGP. Amongst other things, these models provide as an output $R_{\rm AA}$, which can be directly compared to experimental measurements of jets, or hadrons at high-\pt. They can also be compared to $v_n$ for hadrons from hard processes, in order to test the assumptions of the path-length dependent energy loss used in these approaches. In addition, Monte Carlo based models provide predictions for medium induced modifications of jet properties (such as radial energy profile, jet substructure, and fragmentation functions) in heavy-ion collisions compared to pp interactions.

The study of the transport of heavy quarks (charm and bottom) represents another key topic for the characterisation of QGP properties~\cite{Andronic:2015wma}. In this respect transport models, which attempt to model the heavy quark-parton interactions on a microscopic level as a function of space and time, provide valuable information. The production and transport of heavy quarks themselves in heavy-ion collisions can be modeled under a couple of advantageous assumptions. Firstly, the heavy quark production rates in the initial stages can also be inferred from pQCD calculations, or direct measurements from pp collisions, and secondly, since the heavy-quark masses are considerably higher than the expected QGP temperatures at the LHC, further production of heavy quarks is expected to be small in the QGP stage. In addition, as the temperature of the medium is smaller than the heavy quark mass, the typical momentum exchanges of heavy quarks with the medium are small. The propagation of heavy quarks in the QGP can therefore be described as a “Brownian motion” with many small-momentum kicks.
A key variable to characterise how heavy quarks interact in the QGP is the heavy-quark spatial diffusion coefficient, $D_{\rm s}$, which is related to the relaxation time of heavy quarks in the QGP~\cite{Moore:2004tg}. This time defines the rate at which the initial momentum distribution of the heavy quarks (which are produced far out of equilibrium with the QGP) approaches a thermal spectrum. If this is small in relation to the medium lifetime and the expansion rate of the system, heavy quarks participate in the collective motion of the QGP. It can also be determined for the infinite, strong, and weak coupling regimes ~\cite{Dong:2019byy}, like $\hat{q}$ and $\eta/s$ . Measurements of heavy quark hadron $v_{n}$ and $R_{\rm AA}$ at low and intermediate-\pt provides constraints for the values of $D_{\rm s}$ used in these approaches. Heavy quark transport models can also investigate the production of heavy quarkonium states in the QGP, which is strongly affected by a medium with free colour charges. Typically, one solves a rate equation including suppression and regeneration effects. The former are evaluated starting from the modifications of the quarkonium spectral functions in the QGP, constrained by potential models validated by Lattice QCD inputs,  while the latter are tuned from the measured heavy quark multiplicity, which is very large ($>10^2$) for charm quarks in central Pb--Pb collisions at LHC energies.

\paragraph{Hadronic interactions and the hadronic phase.} Hadronic transport and statistical models play a crucial role for the description of this phase. Regarding the transport models (e.g. UrQMD~\cite{Bleicher:1999xi}), they attempt to describe microscopic hadron--hadron interactions in space and time. They require hadronic cross sections as inputs, and as mentioned are employed as a hadronic afterburner for hydrodynamic model chains, in order to account for final state interactions. The cross sections required for various hadron species interactions are subject to experimental uncertainties, and in some cases, in particular for rarely produced hadrons, have no experimental constraints and therefore require a theoretical input. The application of statistical models on the other hand assumes that the system is thermalised at the hadronic stage, which allows for macroscopic thermodynamic properties to be determined based on fits to measurements of hadron abundances~\cite{BraunMunzinger:2003zd}. The chemical freeze-out temperature is a prime example, as these hadron abundances are fixed at this temperature. The baryochemical potential $\mu_{\rm B}$, which is related to the net baryon density i.e.\,the matter-antimatter asymmetry of the system, is another example. Statistical models are also particularly useful as Lattice QCD calculations predict the deconfinement temperature, therefore the chemical freeze-out temperature obtained from statistical models provides a lower limit for such a prediction. Lattice QCD or chiral effective theories can also be used to explore the strong potential between produced hadrons that undergo low $Q^2$ interactions, in both heavy-ion and light system collisions. It can prescribe potentials between the copiously produced short-lived hadrons (such as $\Lambda$ or $\Xi$) and protons, which can be tested via femtoscopic measurements.

\paragraph{QGP effects and few body QCD interactions in heavy and small systems.} A number of event generators are used in the modelling of heavy-ion or light system collisions, which incorporate QGP effects to a varying degree. The AMPT model uses phenomenological strings to determine the QCD fields created in the initial collision~\cite{Lin:2004en}. These strings break to form quarks, which subsequently interact in accordance with transport equations, and then hadronise to produce particle momentum distributions, which can be compared to data. It has an afterburner, similar to UrQMD, called ART, that operates in a similar way. The EPOS model also attempts to model QGP behaviour~\cite{Pierog:2009zt}, and like AMPT it utilises QCD strings, but differs in its use of a core-corona approach. In the core, the energy density of the strings is sufficient to invoke the QGP description, which is subject to a hydrodynamic evolution. In the corona nucleon--nucleon collisions are treated as independent entities regarding hadron production from such collisions.  The HIJING model is somewhat simpler~\cite{Gyulassy:1994ew}: it assumes nucleon--nucleon interactions produce hadrons independently, and is therefore used as a non-QGP reference with respect to comparisons to heavy-ion data (although it has a jet-quenching option). As mentioned, QCD can be tested directly in the more dilute environment of pp, p--Pb and ultra peripheral Pb--Pb collisions. Such tests take advantage of the more standard tools used in high energy physics. An event generator that does so is PYTHIA~\cite{Sjostrand:2006za}, which provides a well established model for pp collisions. PYTHIA, like HIJING, AMPT and EPOS, describes the QCD field using phenomenological strings which break to produce low-$p_{T}$ hadrons, and also incorporates hard scatterings coupled with event by event parton showers and hadronisation models to describe the production of high-$p_{\rm T}$ hadrons. Other Monte Carlo generators such as MC@NLO~\cite{Frixione:2002ik,Frixione:2003ei} and POWHEG~\cite{Frixione:2007nw} are available to study heavy flavour production. They couple the modeling of the hadronic final state with a perturbative description of the hard scattering process, and are useful to model complex observables involving correlations or heavy-flavour jets. PYTHIA has also been recently extended in the PYTHIA/Angantyr format to model heavy-ion collisions~\cite{Bierlich:2018xfw}, without including a hydrodynamic evolution.

\subsection{Evolution of the field: from first studies to ALICE and beyond}
\label{sec:QGPhistory}

Before the LHC start-up, 
the search for the existence of the quark--gluon plasma and the study of its properties were already the quintessential reason for the existence of a broad community of physicists, with a background ranging from traditional nuclear physics to high-energy particle physics. Such a community had a relevant fraction of its roots in experiments performed with ion beams in the few GeV energy range, starting at BEVALAC in the `70s~\cite{Stock:1982qx}, with strong theory groups devoted to the study of nuclear matter under extreme conditions being also formed in those years~\cite{Hagedorn:1980kb,VanHove:1982qz}. A decisive boost for the field occurred with the availability of nuclear collisions at the BNL AGS ($\sqrt{s_{\rm NN}}\sim 5$ GeV)~\cite{Ogilvie:2001cx} and CERN SPS ($\sqrt{s_{\rm NN}}\sim 20$ GeV)~\cite{Heinz:2000bk}, both starting in 1986. On the theory side a whole series of observables that could be considered as signatures of the formation of the QGP were proposed. Prime examples of these signals included the suppression of the J/$\psi$ yield induced by colour screening in the QGP~\cite{Matsui:1986dk} and the enhancement of strangeness production due to the reduction from constituent to current quark mass in a chirally symmetric and deconfined medium~\cite{Rafelski:1982pu}, leading to a strong excess of multi-strange baryons relative to pp collisions. Enhancements of the dilepton spectrum due to the modification of the $\rho$ spectral function~\cite{Brown:1991kk,Rapp:1999ej} and the production of a thermal signal from the deconfined state~\cite{Kapusta:1991qp,Ruuskanen:1991au} were also among the predicted signatures. All of these signals were observed to various extents in Pb--Pb collision studies performed by several experiments at the SPS in the `90s~\cite{Abreu:2000ni,Andersen:1999ym,Agakishiev:1997au,Afanasiev:2002mx,Arnaldi:2008er}. Although alternative explanations not involving deconfinement could be formulated for some of them, the multiple simultaneous observations of the various proposed signatures was considered by the community as a compelling evidence for the production of a new state of matter with the characteristics of the QGP and a press release in that sense was issued by CERN in 2000~\cite{Heinz:2000bk}.

A few months later (June 2000) the first heavy-ion collider, RHIC, entered operation at Brookhaven National Laboratory, increasing the available collision energy by about one order of magnitude. With this step, the study of the production of hard probes of the QGP received an extraordinary boost and only a few months later some new discoveries came to light in this sector, with the first observation of the suppression in Au--Au collisions at ${\ensuremath{\sqrt{s_{\mathrm{NN}}}}\xspace}$ $= 130$ GeV of high-$p_{\rm T}$ particle production~\cite{Adcox:2001jp,Adler:2002xw}. This signal, much awaited for, was immediately connected to the energy loss of hard partons in the QGP~\cite{Wang:1991xy,Baier:1996sk} and initiated an entire field of investigation, extended through the years to the study of the modification of jet properties in the medium~\cite{Salgado:2003rv,Adams:2006yt,Adler:2005ee}. At the same time, the increase of the hadron multiplicity up to pseudorapidity densities ${\rm d}N_{\rm ch}/{\rm d}\eta \sim$650 for central Au--Au collisions~\cite{Back:2001ae} made accurate studies of the global characteristics of the events possible. It was shown, from the measurements of $v_{2}$ at low-\pt and $R_{AA}$ at high-\pt from charged hadrons that the medium created in nuclear collisions at RHIC behaves like a nearly ideal fluid~\cite{Shuryak:2003xe}. Such a medium implies a thermal equilibration at early times and with constituent interactions that have a very short mean free path, somewhat superseding the early notion of a weakly-coupled, gas-like QGP~\cite{PerfectLiquid:2005}. The values of $\eta/s$ implied from these measurements were close to those expected from the creation of an infinitely coupled fluid prescribed from AdS/CFT: $1/4\pi$~\cite{Kovtun:2004de}. In the later RHIC years (circa 2009), a claim of the observation of a CME-like signal was announced~\cite{Abelev:2009ac}.

The first round of RHIC results~\cite{Arsene:2004fa,Back:2004je,Adams:2005dq,Adcox:2004mh} was an important input for the shaping of the physics program for heavy-ion collisions at the LHC. Studies were carried out mainly in the environment of the ALICE Collaboration, but also by groups participating in the ATLAS and CMS experiments. The further jump in energy by a factor $>20$ ($\sqrt{s_{\rm NN}}=5.5$ TeV for Pb--Pb collisions for the nominal LHC running conditions) creates ideal conditions for the study of hard processes, thanks to the increase in the corresponding production cross sections, both in the strong (charm, beauty) and electroweak (W, Z) sector. At the same time, the further logarithmic but significant increase in the hadronic multiplicity created an obvious interest in the possibility of precise studies of the chemical composition and of collective effects of the medium, also addressing such quantities and their fluctuations on an event-by-event basis.
In this situation, and considering the fact that the lifetime of the  QGP was now expected to be much larger than that of the hadronic phase, the LHC was clearly considered as the ultimate machine for the quantitative study of the deconfined phase~\cite{Schukraft:2001vg,BraunMunzinger:2007zz}. The ALICE experiment and its physics program were designed to address the study of particle production in the widest possible momentum range, from the soft to the hard sector, and to have the capability of answering a number of fundamental questions on the behaviour of the QGP~\cite{Carminati:2004fp,Cortese:2005qfz}. Corresponding physics programs for the ATLAS~\cite{Jenni:721909} and CMS~\cite{dEnterria:2007iyi} experiments were also developed in a later phase, with a stronger emphasis on the detection of high-$p_{\rm T}$ probes, due to the specific design of those experiments, mainly conceived for the study of high-luminosity pp collisions. During the first decade of LHC operation, also the LHCb experiment developed a program to study proton---nucleus and peripheral nucleus--nucleus collisions. 

\begin{figure}[ht!]
\begin{center}
\includegraphics[width = 0.8\textwidth]{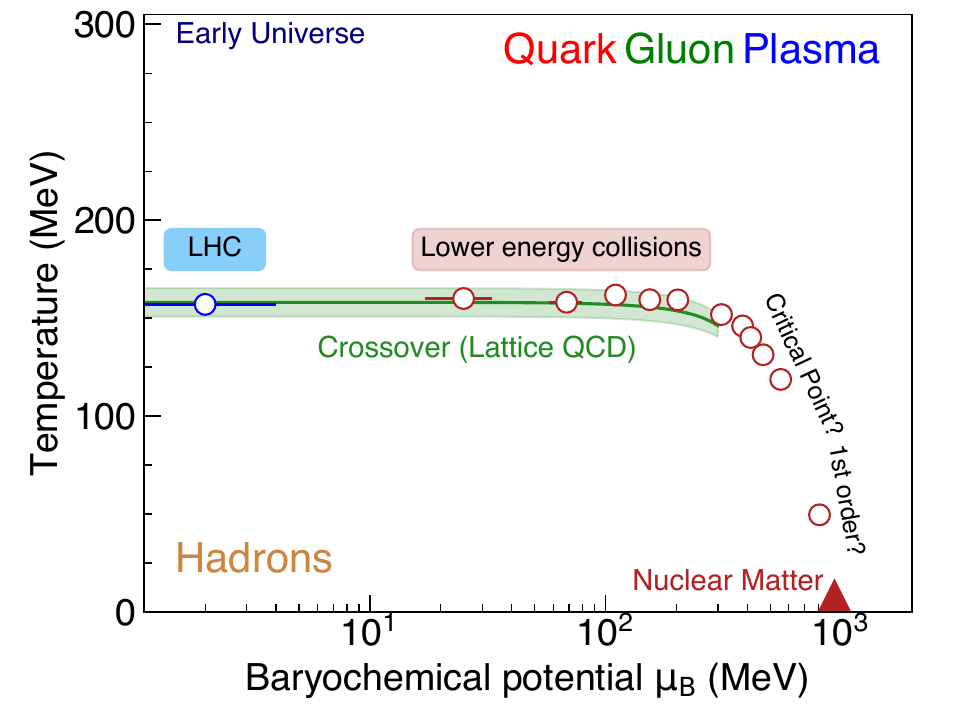}
\end{center}
\vskip -0.5cm 
\caption{A schematic representation of the QCD phase diagram. The green line and band shows the $\mu_{\rm B}$ region accessible to Lattice QCD calculations ~\cite{Borsanyi:2020fev}. The line shows the pseudocritical temperature, whereas the band represents the half-width of the crossover transition i.e. the temperatures where the QGP and hadrons can co-exist. The open points show experimental results for the determination of the chemical freeze-out parameters~\cite{Andronic:2005yp, Cleymans:2005xv, Andronic:2017pug, Flor:2020fdw}. The location of atomic nuclei is also shown, as well as conjectured regions for the presence of a first order phase transition and of a critical point.}
\label{fig:QCDphasediagram}
\end{figure} 

As mentioned in Sec.~\ref{sec:emergQGP}, the QGP created at LHC energies is expected to have a vanishing net baryon-number density. Indeed, when the collision energy becomes very high a ``transparency'' regime sets in, with the baryon number carried out by the colliding nuclei being located outside the centre-of-mass rapidity region ($y_{\rm CM}=0$). The latter is the 
hottest region, where particle production is maximal and the highest temperatures are reached~\cite{Bass:1998ca,Busza:1983rj}. In these kinematic conditions the transition between deconfined and hadronic phase is characterised by a rapid cross-over at the pseudocritical temperature $T_{\rm pc}$~\cite{Bazavov:2014pvz,Borsanyi:2013bia}. 

In a wider perspective, by studying heavy-ion collisions over extended energy ranges, it is possible to obtain information on the QCD phase diagram, usually shown in terms of temperature and baryochemical potential $\mu_{\rm B}$. The latter quantity  corresponds to the energy needed to increase the baryon number by one unit at fixed volume and entropy of the system and is proportional to the net baryon density of the system. A conceptual representation of the QCD phase diagram is illustrated in Fig.~\ref{fig:QCDphasediagram}, where the region explored by LHC experiments is approximately shown. Moving towards lower collision energies implies a lower initial temperature of the QGP phase and an increase in the net baryon density of the system due to an increased ``stopping'' of the baryonic number in a region closer to $y_{\rm CM}=0$. The open points depicted in Fig.~\ref{fig:QCDphasediagram} show the values $T_{\rm chem}$ and $\mu_{\rm B, ch}$ corresponding to the chemical freeze-out of the system at LHC energy (blue) and for lower collision energies (red). In the region where reliable Lattice QCD calculations exist, indicated as a green band, the confinement temperature coincides, within uncertainties, with $T_{\rm chem}$, showing that the chemical equilibration of the system occurs at, or shortly after, hadronisation. At large $\mu_{\rm B}$, a well known technical issue of Lattice QCD calculations, the sign problem~\cite{deForcrand:2010ys},   forces to rely on approximate methods for the calculation of the thermodynamical quantities~\cite{Gavai:2003nn,DElia:2002tig}. These methods predict a first-order phase transition at large baryon density that should end in a critical point when decreasing $\mu_{\rm B}$~\cite{Gavai:2004sd,Stephanov:2004wx}. Studies of this region of the QCD phase diagram are extremely relevant also for our understanding of astrophysical objects as compact neutron stars~\cite{Baym:2017whm}. They represent a field of investigation which is being developed in these years with experiments at existing (RHIC low beam-energy scans, SPS) and forthcoming (FAIR, NICA) facilities~\cite{Bzdak:2019pkr,Friman:2011zz,Ablyazimov:2017guv,Kapishin:2016ojm,Golovatyuk:2016zps,Dahms:2673280}. It is clearly complementary to the one accessible at top RHIC and LHC collider energies where, as already discussed in Sec.~\ref{sec:emergQGP}, a hot and long-lived QGP phase, similar to that of the early Universe, is produced. The program at the highest collider energies will continue at RHIC in the 2020s with the sPHENIX~\cite{PHENIX:2015siv} detector and at the LHC in the 2020s and 2030s with the participation of all four large experiments~\cite{Citron:2018lsq}.

\subsection{ALICE: design considerations, implementation, operation}
\label{sec:ALICEExp}
ALICE (A Large Ion Collider Experiment) was proposed in March 1993 as ``a dedicated heavy-ion experiment'' aimed at studying nuclear collisions at the LHC, in order to analyse the existence of QCD bulk matter and the QGP~\cite{CERN-LHCC-93-016,CERN-LHCC-95-71}. Being the only detector specifically devoted to QGP studies, it was designed to access a large number of specific observables in a wide transverse momentum range, in order to shed light on the various stages of the evolution of the heavy-ion collisions, from the initial state to the QGP phase and to the transition to hadronic matter. At the same time, specific aspects of pp physics were part of the ALICE program from the beginning, and gained progressively more importance.

\begin{figure}[ht!]
\begin{center}
\includegraphics[width=1\linewidth]{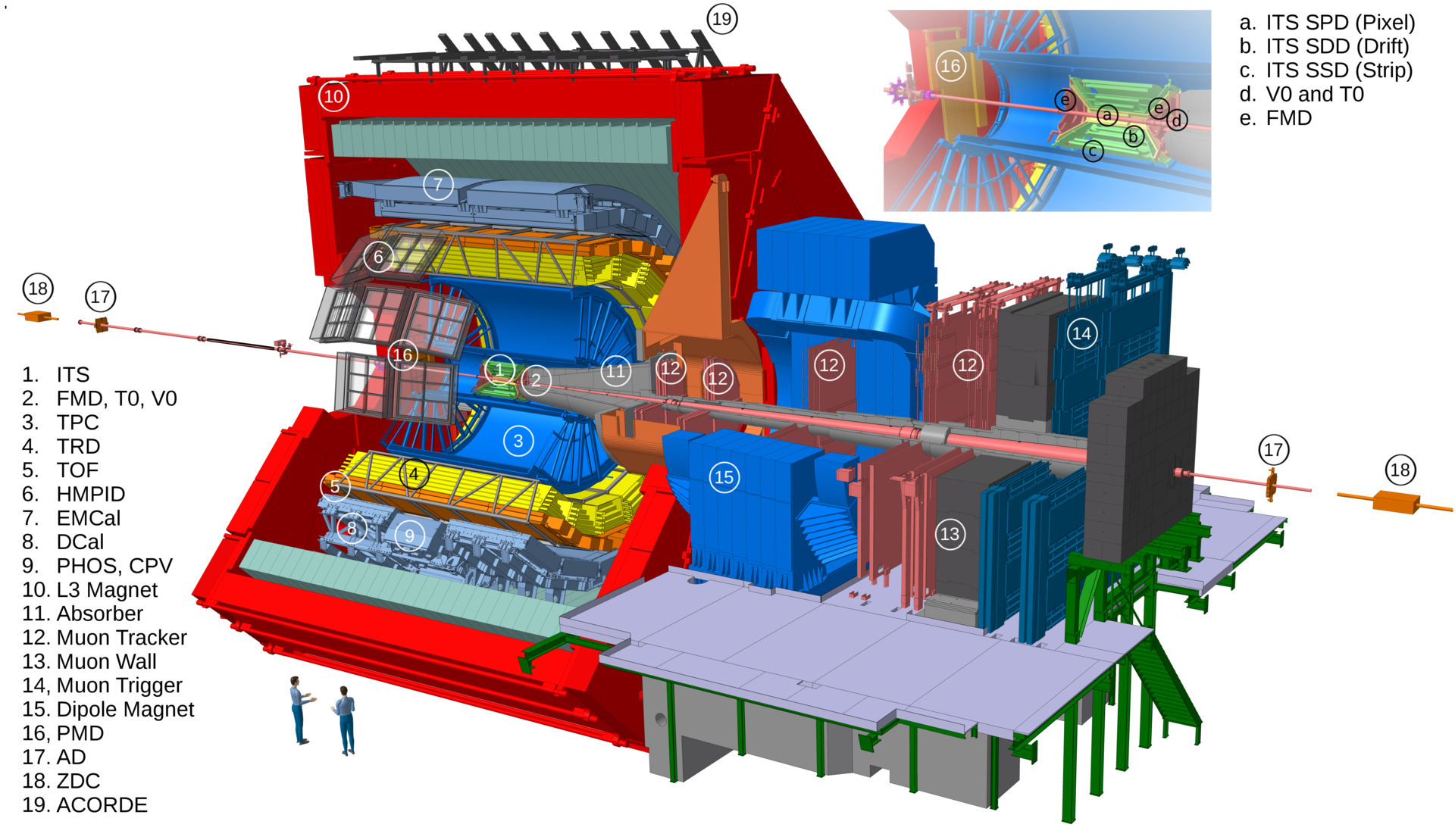}  
\caption{The ALICE detector. A short description of the various subdetectors, as well as information on their kinematic coverage, is given in the text.}
\label{fig:ALICEsetup} 
\end{center}
\end{figure} 

\begin{figure}[ht!]
\centering
\includegraphics[width=0.7\textwidth]{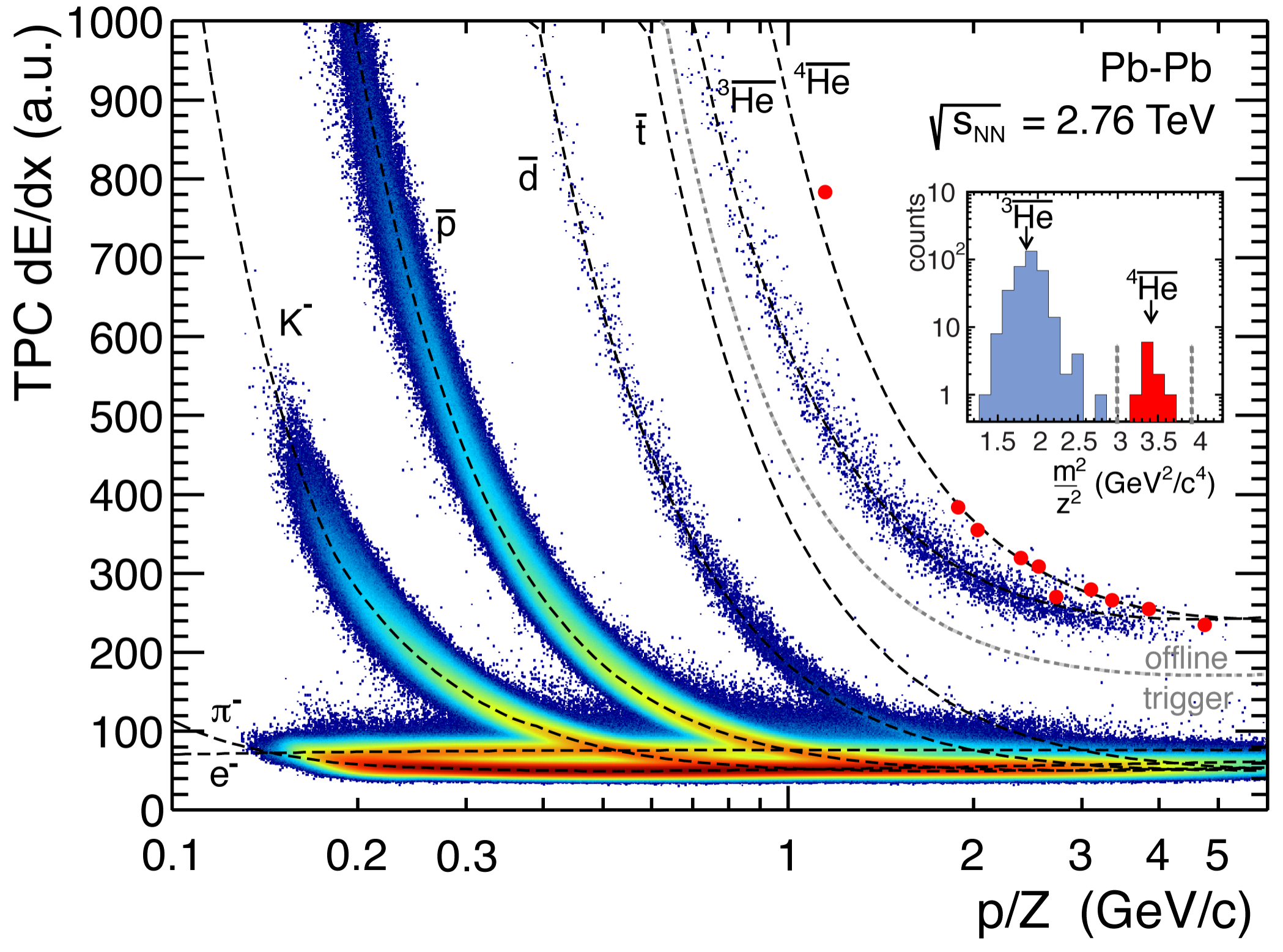} \hbox{\hspace{0.4cm}\includegraphics[width=0.7\textwidth]{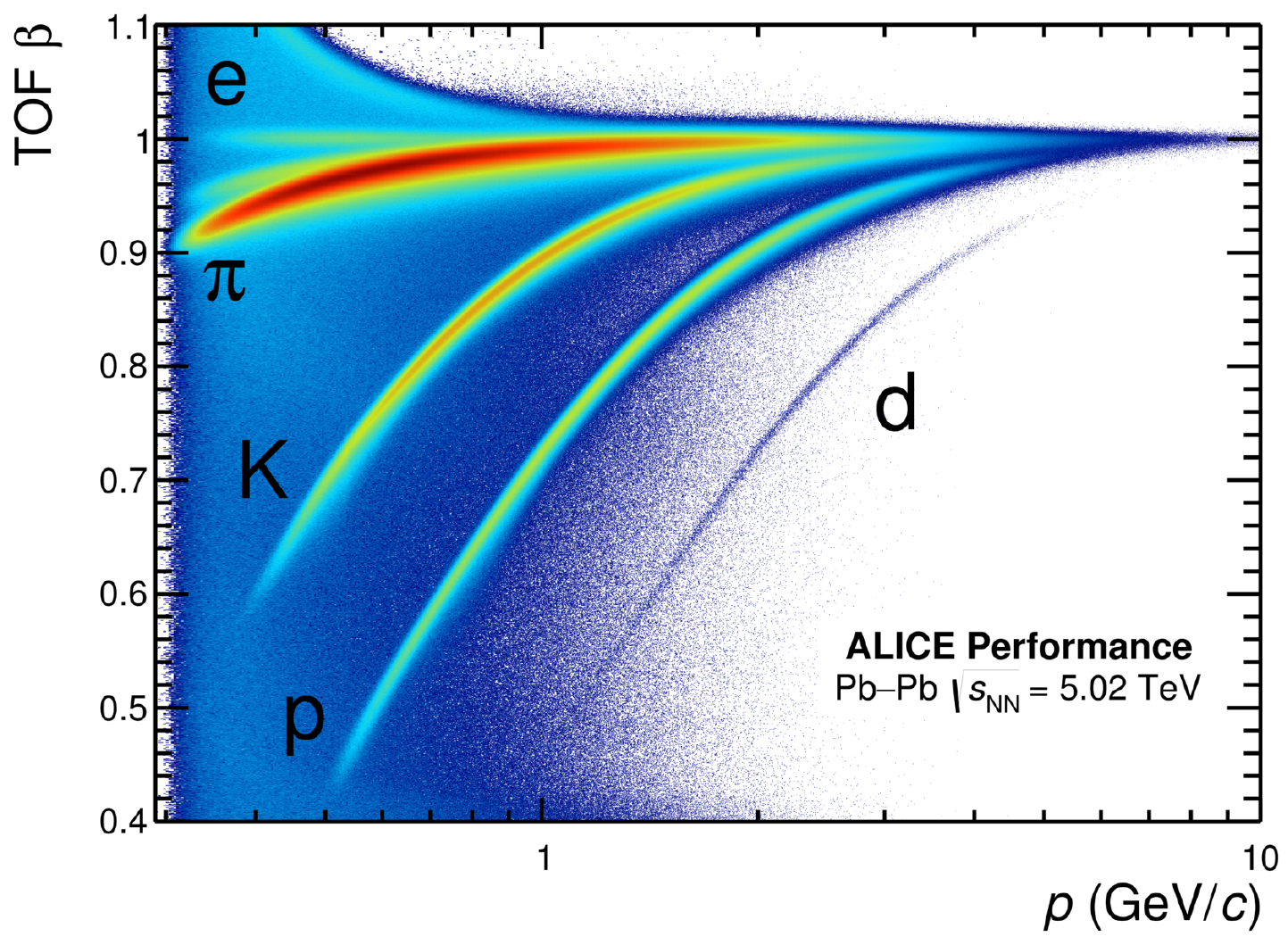}}
\caption{(Top) The ${\rm d}E/{\rm d}x$ signal in the ALICE TPC as a function of magnetic rigidity. The expected curves for various
 particle species are also shown, with the inset panel showing the TOF mass measurement providing additional separation for helium isotopes when $p/Z>$2.3~GeV/$c$. (Bottom) The Time-of-Flight measured in the TOF system as a function of the particle momentum. Tracks are selected with standard cuts inside the pseudorapidity region $|\eta| < 0.5$.}
\label{fig:barrelperf2}
\end{figure}

\begin{figure}[ht!]
\centering
\hbox{\hspace{2.4cm} \includegraphics[width=0.8\textwidth]{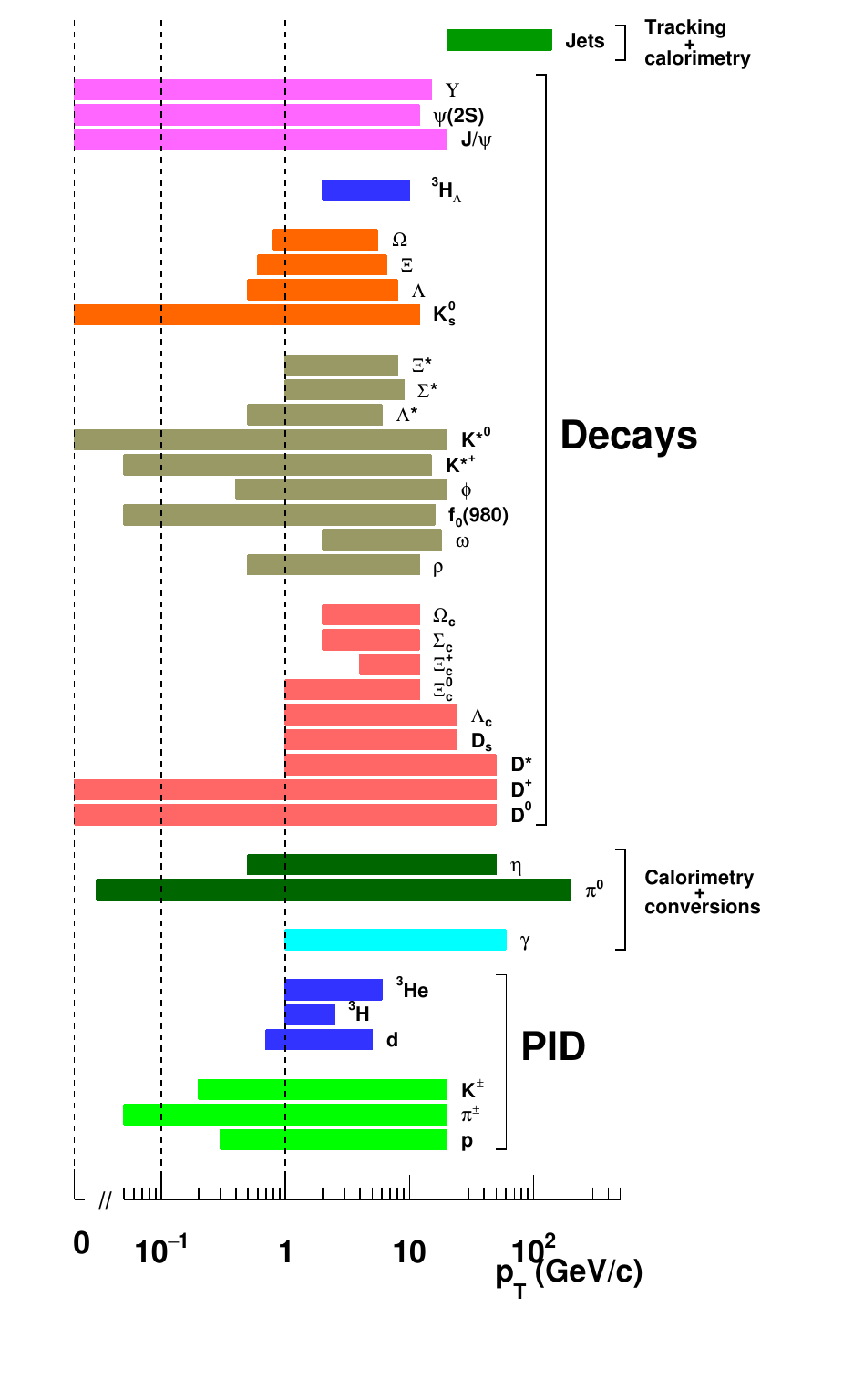}}
\caption{ALICE particle identification and reconstruction capabilities, with the $p_{\rm T}$ coverage corresponding to the published measurements based on pp or Pb--Pb data samples. For W (Z) bosons, the selection $p_{\rm T}>10$ (20) GeV/$c$ is performed on the transverse momentum of the decay muon, without defining a sharp $p_{\rm T}$ range.}
\label{fig:barrelperf1}
\end{figure}

The ALICE detector is situated at the interaction point IP2 of the LHC. The apparatus that was operated in the LHC Runs 1 and 2 (`ALICE\,1', shown in Fig.~\ref{fig:ALICEsetup}) is described in detail in Ref.~\cite{Aamodt:2008zz} and its performance 
is discussed in Ref.~\cite{Abelev:2014ffa} and will be shortly reported hereafter.
The detector is based on a central barrel, covering full azimuth and the pseudorapidity region $|\eta|<0.9$. It provides a robust particle identification up to $p_{\rm T}\sim$~20 GeV/$c$, together with a very low momentum cut-off ($p_{\rm T}\sim 0.15$~GeV/$c$, and even lower for specific reconstruction settings and collision systems) and excellent capabilities for primary and secondary vertex reconstruction (i.e.\,$\sim 100 \, \mu$m resolution for the study of the 2- and 3-prong decays of D mesons). The main charged-particle tracking detectors of ALICE are the Inner Tracking System (ITS)~\cite{Aamodt:2010aa}, composed of six tracking layers, two Silicon Pixel Detectors (SPD), two Silicon Drift Detectors (SDD), and two Silicon Strip Detectors (SSD), and a large Time Projection Chamber (TPC)~\cite{Alme:2010ke}. The latter, together with the SSD and SDD, provide charged-particle identification via measurement of the specific ionisation energy loss ${\rm d}E/{\rm  d}x$. External to the TPC, the tracking is complemented by a Transition Radiation Detector (TRD)\cite{Acharya:2017lco} and a Time Of Flight system (TOF)~\cite{Akindinov:2013tea}. These two detectors also provide electron identification and charged particle identification at intermediate-momenta, respectively. Outside the TOF, the azimuthal region is shared by two electromagnetic calorimeters, with a thickness of about 20 radiation lengths: (i) the high-resolution PHOton Spectrometer (PHOS)~\cite{Dellacasa:1999kd}, based on PbWO$_{\rm 4}$ crystals, (ii) the EMCal~\cite{Abeysekara:2010ze,ALICE:2022qhn}, based on a layered Pb-scintillator sampling technique.
The former is used for photon and neutral meson detection, while the latter due to its significantly larger acceptances is also employed for electron and (di)jet measurements. Finally a High Momentum Particle Identification Detector (HMPID)~\cite{Beole:1998yq}, provides, thanks to the RICH technique, further PID capability at intermediate-$p_{\rm T}$. The central barrel detectors are embedded in the L3 solenoid magnet which delivers a magnetic field up to $B=0.5$~T. Figure~\ref{fig:barrelperf2} illustrates, as an example, the PID performance of the TPC which allows excellent separation for various hadrons and light nuclei at low-$p_{\rm T}$. A good separation between protons, kaons and pions is also achieved in the region of the relativistic rise of ${\rm d}E/{\rm d}x$, up to $p_{\rm T}\sim 20$~GeV/$c$. The PID performance of the TOF system is shown in Fig.~\ref{fig:barrelperf2}. Furthermore, an overview of the ALICE capabilities for the measurement of various hadrons is shown in both Figs.~\ref{fig:barrelperf2} and ~\ref{fig:barrelperf1}, where the approximate coverage, extending to very low-$p_{\rm T}$, for different meson and baryon species is shown, together with the corresponding detection techniques.

The ALICE coverage extends outside the midrapidity region by means of a muon spectrometer and of various sets of smaller forward detectors. The muon spectrometer~\cite{ALICE:1999aa} covers the pseudorapidity region $-4<\eta<-2.5$, it detects charmonia and bottomonia, via their dimuon decay, with a mass resolution of $\sim 70$ and $\sim 150$ MeV/$c^2$, respectively, Z bosons and low-mass hadrons ($\omega$, $\phi$). It also measures single-muon production from decays of heavy-flavour hadrons and W$^{\pm}$ bosons. 
It is based on a dipole magnet with a 3 Tm bending power, and on sets of tracking (Cathode Pad Chambers, CPC) and triggering (Resistive Plate Chambers, RPC) detectors. A hadron absorber, with a thickness of 10 interaction lengths ($\lambda_{\rm i}$) separates the muon spectrometer from the interaction point (IP), and a second absorber (7.2 $\lambda_{\rm i}$-thick) is positioned between the tracking and the trigger chambers. The other forward detectors include a silicon-strip Forward Multiplicity Detector~\cite{Cortese:2004aa} (FMD, on both sides of the IP at  $-3.4<\eta<-1.7$ and $1.7<\eta<5$) and a preshower/gas counter Photon Multiplicity Detector~\cite{Dellacasa:1999ar} (PMD, at $2.3<\eta<3.9$) for the measurement of charged particles and photons. Furthermore, the V0~\cite{Abbas:2013taa}, composed of two sets of scintillator detectors covering $-3.7<\eta<-1.7$ (V0C) and $2.8<\eta<5.1$ (V0A), defines the trigger for minimum-bias collisions and is also used to classify events based on multiplicity or centrality.
The T0~\cite{Cortese:2004aa}, two quartz Cerenkov detectors ($-3.3<\eta<-3$ and $4.6<\eta<4.9$), are used to determine the timing and longitudinal position of the interaction. Finally two sets of Zero-Degree Calorimeters~\cite{ALICE:2012aa} (ZDC) detect spectator neutrons and protons emitted at very small angle and are used mainly for the determination of the centrality of the collisions and to reject out-of-bunch interactions.

For a large experiment as ALICE, the triggering scheme is based on various decision levels~\cite{Fabjan:684651,ALICE:2019jgt}. The Level 0 trigger decision (L0) is made $\sim 0.9 \mu{\rm s}$ after the collision using V0, T0, EMCal, PHOS, and the muon spectrometer trigger chambers. The events accepted at L0 are further evaluated by the Level 1 (L1)
trigger algorithm, which also involves TRD and ZDC, and the corresponding decision is made $\sim 6.5 \mu{\rm s}$ after L0. The Level 2 (L2) decision, taken after about 100 $\mu$s corresponding to the drift time of the TPC, triggers the sending of the event data to DAQ and, in parallel, to the High Level Trigger system (HLT), which also plays an important role in data compression.
In physics data taking a mixture of so-called ``minimum bias'' and rare triggers is used. In Pb--Pb collisions, a typical definition of the minimum bias trigger is a coincidence of signals on the two V0 detectors, that can be complemented by the requirement of signals in both sets of ZDCs, to suppress electromagnetic interactions. Additional selections on the centrality of the event, based on the V0 signal, can also be included at the trigger level. Several rare triggers, designed to enhance the statistical significance for specific physics processes, are also implemented. As an example, the experiment can trigger on the production of single muon and muon pairs in the forward spectrometer~\cite{Bossu:2012jt}, with a programmable $p_{\rm T}$ threshold which can be as low as 0.5 GeV/$c$, and on the deposition of a certain amount of energy in a number of adjacent cells in the EMCal and/or PHOS. Similar trigger conditions are implemented for the pp data taking, with the addition of a specific set of triggers involving the detection of one or more electrons in the TRD, and with the definition of a high-multiplicity trigger based on appropriate thresholds on the V0 signals or on the number of hits on the SPD. Details on the triggering scheme of the experiment can be found in Ref.~\cite{Abelev:2014ffa}. 

ALICE has started physics data taking in 2009 with the first LHC pp collisions at $\sqrt{s}=0.9$~TeV, and since then has collected data with all the collision systems and energies that became available during Run 1 (2009--2013) and Run 2 (2015--2018). In particular, Pb--Pb data taking occurred in 2010 and 2011 at $\sqrt{s_{\rm NN}}=2.76$~TeV, and in 2015 and 2018 at $\sqrt{s_{\rm NN}}=5.02$~TeV. A short run with Xe--Xe collisions at  $\sqrt{s_{\rm NN}}=5.44$~TeV was also performed in 2017. Proton--proton data taking at the same energies as Pb--Pb, serving as a reference for nucleus--nucleus reactions but also for specific QCD studies, were also performed along the years. For the study of cold nuclear matter effects, after a short pilot run in 2012, p--Pb collisions were studied in 2013 and 2016, at $\sqrt{s_{\rm NN}}=5.02$ and 8.16~TeV. In addition to these combinations of collision system and energy, ALICE has taken pp data at all the other collision energies that became available along the years, up to $\sqrt{s}=13$~TeV. 

The data samples collected by ALICE can be characterised by two quantities. The first is the number of recorded minimum bias events $N_{\rm MB}$, which is relevant for all the measurements which involve low-$p_{\rm T}$ particle production. The minimum-bias event rate  is limited essentially by the readout time of the detectors and their occupancy, and when necessary, in particular for pp and p--Pb collisions, was tuned by acting on the beam optics (up to values of $\sim 200$~kHz for pp data taking at $\sqrt{s}=13$~TeV). A typical feature of ALICE data taking is the possibility of operating with a low pile-up rate in pp collisions, allowing as an example a detailed study of the dependence of particle production on the charged hadron multiplicity.  Typical values for $N_{\rm MB}$ in the LHC Run 2 are $\sim 4\times10^9$ for pp collisions at $\sqrt{s}=13$ TeV, complemented by $\sim 2\times 10^9$ high-multiplicity events. For p--Pb, about $\sim 8\times 10^8$ minimum-bias events were collected, mainly at $\sqrt{s_{\rm NN}}=5.02$~TeV. The corresponding number for Pb--Pb collisions at the same energy was about $3\times 10^8$ minimum-bias events, plus a similar number of recorded events with a further selection on centrality. The second relevant quantity is the recorded integrated luminosity $L_{\rm int}$ for rare triggers, with the largest values corresponding typically to (di)muon triggers. For this specific class of triggers, in the LHC Run 2, about 36~pb$^{-1}$ were collected for $\sqrt{s}=13$~TeV pp collisions and 25~nb$^{-1}$ for p--Pb collisions at $\sqrt{s_{\rm NN}}=8.16$~TeV. The corresponding number for Pb--Pb collisions at $\sqrt{s_{\rm NN}}=5.02$~TeV was about 800~$\mu{\rm b}^{-1}$. In Table~\ref{tab:datsamples} the collision systems, center-of-mass energies, integrated luminosity and number of minimum-bias events are listed.

\begin{table}[!ht]
	\begin{center}
		\caption{Collision systems studied by ALICE during LHC run 1 and 2. The center of mass energy per nucleon--nucleon collision, the integrated luminosity for rare triggers and the number of collected minimum bias events are shown. The latter quantity includes samples where selections on hadronic multiplicity (pp) or collision centrality (Pb--Pb) were performed.}\label{tab:datsamples}
        \vskip 1cm
		\begin{tabular}{c|c|c|c|c}
			System  & Year(s) & $\sqrt{s_{\rm NN}}$ (TeV) & $L_{\rm int}$ & $N_{\rm MB}$\\
			\hline
            Pb--Pb & 2010, 2011 & 2.76 & 75 $\mu{\rm b}^{-1}$ & 1.3$\times$10$^8$ \\
            Pb--Pb & 2015, 2018 & 5.02 & 800 $\mu{\rm b}^{-1}$ & 6$\times$10$^8$ \\
            \hline
            Xe--Xe & 2017 & 5.44 & 0.3 $\mu{\rm b}^{-1}$ & 1.1$\times$10$^6$ \\
            \hline
            p--Pb & 2013, 2016 & 5.02 & 18 nb$^{-1}$ & 8$\times$10$^8$ \\
            p--Pb & 2016 & 8.16 & 25 nb$^{-1}$ & 1.3$\times$10$^8$\\
            \hline
            pp & 2009 & 0.9 & 200 $\mu{\rm b}^{-1}$ & 0.5$\times$10$^6$ \\
            pp & 2011 & 2.76 & 100 nb$^{-1}$ & 1.3$\times$10$^8$ \\
            pp & 2010, 2011 & 7 & 1.5 pb$^{-1}$ & 1.6$\times$10$^9$ \\
            pp & 2012 & 8 & 2.5 pb$^{-1}$ & 3.1$\times$10$^8$ \\
            pp & 2015, 2017 & 5.02 & 1.3 pb$^{-1}$ & 10$^9$ \\
            pp & 2015 - 2018 & 13 & 36 pb$^{-1}$ & 6$\times$10$^9$ \\
            
		\end{tabular}
	\end{center}
\end{table}

The data size corresponding to event samples collected in each year of data taking can easily exceed tens of PetaBytes. The large computing and storage power required to process such amounts in order to perform the reconstruction of the physics objects from the raw data is ensured by the Worldwide LHC Computing Grid (WLCG) infrastructure, based on about 200 computing centres located all around the world~\cite{Bird:1447125}. 

In ALICE, the track reconstruction scheme is based on the Kalman filter~\cite{Fruhwirth:1987fm,Ivanov:2006yra,Abelev:2014ffa}. In the central barrel, after a preliminary interaction vertex finding based on SPD information, track finding in the TPC is performed and track candidates are then matched to clusters in the ITS layers. Tracks are extrapolated  to their point of closest approach to the preliminary interaction vertex and are then refitted in the outward direction, including the more external detectors and evaluating quantities relevant for PID. A final refit, starting from the external radius of the TPC is then performed and a final primary interaction vertex is computed. Secondary vertices from particle decays and photon conversions are subsequently searched, pairing tracks exceeding a defined distance of closest approach to the primary vertex and applying several quality cuts. More complex secondary vertex reconstruction (for example for heavy-flavour decays) is performed later at the analysis stage.
In the muon spectrometer a Kalman-based algorithm is also adopted, and the track candidates are extrapolated to the SPD-based vertex position and finally corrected for the non-negligible effect of muon energy loss and multiple scattering in the hadron absorber~\cite{ALICE:2011zqe}. Photon identification in the calorimeters is based on the detection of clusters in those detectors. The electron/positron contribution is rejected by checking the absence of reconstructed charged tracks in the vicinity~\cite{Abelev:2014ffa,Cortese:2005qfz}. Cuts on various shower shape parameters allow rejection of hadronic showers. A complementary detection technique via study of photon conversions to e$^+$e$^-$ pairs in the TPC is also employed~\cite{Abelev:2012cn}. Finally, jet measurements, particularly delicate in the complex heavy-ion collision environment, are performed combining information from charged particle tracking in ITS, TPC and EMCal information for the neutral energy carried by photons from hadronic decays~\cite{Abelev:2013fn,Adam:2015ewa,ALICE:2022qhn}.

The ALICE apparatus has now undergone a major upgrade (see Sec.~\ref{ch:ALICE2}) and the ‘ALICE\,2’ version has recently started data taking in the LHC Run 3~\cite{Abelevetal:2014cna,ALICE:2023udb}. A completely new detector ‘ALICE 3’~\cite{ALICE:2022wwr} is proposed for the LHC Run 5 and beyond (see Sec.~\ref{ch:ALICE3}).

\subsection{Overview of the key scientific questions addressed by ALICE}
\label{sec:KeyQuestions}

High-energy nuclear collisions in the multi-TeV energy range at the LHC provide ideal conditions to produce a long-lived QGP as well as abundant hard probes of such a state up to high $p_{\rm T}$. An accurate and quantitative characterisation of the QGP state and its related phenomenology, via the study of the observables discussed in Sec.~\ref{sec:observ}, is the subject of the ALICE experimental program and has inspired the design of its detector. In addition to this, owing to its high-precision tracking and excellent PID performance down to very low $p_{\rm T}$  ALICE has contributed to addressing further fundamental aspects of the strong interaction and QCD. In particular, more general insights on the manifestations of hadronic interactions can be carried out by studying observables like the production of light anti- and hyper-nuclei, the kaon-nucleon and the hyperon-nucleon interaction. Due to the difficulty in obtaining hyperon beams, such studies can be efficiently performed in the ALICE environment, either with Pb--Pb or high-multiplicity pp collisions. Another set of physics topics related to the strong interaction and not directly connected to QGP, where ALICE gives an important contribution, is the study of specific high-$Q^2$ QCD processes in pp collisions. These include the measurement of inclusive and heavy-flavour jet production, and the study of their substructure. Studies of open heavy flavour, quarkonia and high-$p_{\rm T}$ hadrons also represent an area of interest for the experiment.

In the following, we briefly introduce the main  topics investigated by ALICE, in the form of physics questions. The corresponding answers will be the object of the following chapters of this review.

\begin{enumerate}[leftmargin=*]
    \item {\it What are the thermodynamic and global properties of the QGP produced at the LHC?}
    
    the QGP can be considered as an extended system of strongly interacting quarks and gluons in thermal equilibrium and as such can be characterised by a set of intensive quantities like temperature and energy density. Also, its size and lifetime are essential parameters for its characterisation. Various observables, such as hadronic multiplicity, transverse energy distributions, and direct photon observables are known to be related to these quantities and have been thoroughly investigated in the past. At LHC energies, which provide the highest temperatures to explore the QGP, they fully retain their importance for the evaluation of the thermodynamic properties of the QGP produced in Pb–Pb collisions and, once the temperature of the system decreases below the pseudocritical temperature, also of the subsequent hadronic phase.  The analysis of the evolution of these quantities from small to large collision systems is a field of investigation that has received much attention at the LHC and is crucial for the understanding of the emergence of QGP-related features. Results on these studies are reported in Sec.~\ref{sec:MacroscopicProperties} and Sec.~\ref{section:3.1} for Pb--Pb and smaller collision systems, respectively.
        
    \item {\it What are the hydrodynamic and transport properties of the QGP?}
    
    As the highest QGP temperatures are expected to be achieved at the LHC, that positions heavy-ion collisions at the LHC as paramount when investigating how hydrodynamic collective motion develops in such an extreme environment. In particular, the temperature dependence of $\eta/s$ and $\zeta/s$ can be studied to the fullest extent, via measurements sensitive to anisotropic and radial flow respectively, as the largest possible ranges of temperatures of the QGP can be probed. The temperature dependence of these parameters were not established prior to first heavy-ion collisions at the LHC. Given the discovery of a perfect liquid at RHIC, constraining the temperature dependence of $\eta/s$ and $\zeta/s$ is crucial in assessing how ``perfect" the QGP liquid created at the LHC is, as a weak coupling picture is expected at higher temperatures, due to the scale dependence of the QCD coupling constant. Finally, we will investigate, via measurements of charmed hadrons, whether low-momentum charm quarks produced out of equilibrium in the initial stages are able to participate in the QGP's collective motion. Such an observation was not established prior to the LHC, and will help to constrain the temperature dependence of the diffusion coefficient $D_{\rm s}$ via comparisons to transport models. The ALICE detector is particularly well suited to this task at the LHC due to its large \pt and $\eta$ acceptance, given that hydrodynamic and transport models provide predictions over the full phase space. These topics will be addressed in Sec.~\ref{sec:QGPevolution} and Sec.~\ref{sec:PartonInteractions}.

    \item {\it How does the QGP affect the formation of  hadrons?}
    
    As mentioned, the hadronisation of the QGP can be studied via measurements of hadron production, which provides information on how hadronic states emerge from the bulk of thermalised partons, when the expanding medium approaches the pseudocritical temperature.
    The abundances of different hadron species were observed at AGS, SPS and RHIC energies to closely follow the expected equilibrium populations of a hadron-resonance gas at the temperature at which the chemical freeze-out of the system occurs. This thermal origin of particle production allows for a macroscopic description of the properties of the hadron gas in terms of thermodynamic variables via statistical-hadronisation models and provides access to the chemical freeze-out parameters relevant for the QCD phase diagram.  At the microscopic level, hadron formation from the QGP medium can be influenced by the presence of the bulk of thermalised partons.
    Recombination models were proposed after the first measurements of hadron production at RHIC and provided a natural explanation for some unexpected results of baryon and meson production and $v_{\rm 2}$ measurements in the intermediate-\pt region.
    In these models, the quarks of the bulk, which are close to each other in phase space, can combine into hadrons. The recombination process is expected to dominate for partons at low and intermediate \pt, whereas energetic partons generally escape from the QGP and hadronise outside the medium via fragmentation. High precision measurements of production of different hadron species, including strange and charm mesons and baryons, as well as light nuclei and hypernuclei, are possible with the ALICE detector at the LHC in different collision systems with improved precision, especially at intermediate- and high-\pt, compared to results at RHIC energies. This can provide new insights into the hadronisation mechanisms and the properties of the hadronic phase of the system evolution. 
    The results of studies on QGP hadronisation will be discussed in Sec.~\ref{sec:QGPHadronization}.

    \item {\it How does the QGP affect the propagation of energetic partons?}
    
    The evidence of strong energy loss of high-momentum partons, via single-hadron $R_{\rm AA}$ measurements, was one of the main discoveries at RHIC energies. This led to the first quantitative investigations for key energy loss parameters, such as $\hat{q}$. At high $p_{\rm T}$, jet production is the main observable regarding the propagation of energetic quarks/gluons, both in vacuum and in a medium. The LHC, thanks to the abundant production of such probes, represents a unique laboratory for the study of their coupling to the QGP. Jet energy loss, jet substructure modification, jet deflection and the emission of large-angle radiation are powerful handles for a more detailed phenomenological understanding and for constraining the key theoretical parameters. Specific studies as the colour-charge (quarks vs gluons) and the quark-mass dependence (beauty vs charm vs light partons) of energy loss offer the possibility of a detailed testing on QCD-based predictions. Also, jets probe the QGP at short distances, as the QGP cannot be strongly coupled at all scales. When probed at short distances, its quasi-particle structure, i.e., the free quarks and gluons, should emerge, and their signature is expected to be a large momentum transfer to the jet or its constituents. These topics will be addressed in Sec.~\ref{sec:PartonInteractions}.

\newpage
    \item {\it How does deconfinement in the QGP affect the QCD force?}
    
    The strong interaction binding a heavy quark-antiquark pair is expected to be strongly modified in the presence of a medium like the QGP, characterised by a high density of (deconfined) colour charges. Results at SPS and RHIC energies showed a significant suppression of charmonium states, depending on their binding energy and clearly related to the formation of a strongly interacting system. At the same time they also demonstrated that a quantitative understanding of the observation needs to consider the modification of spectral properties of quarkonia in the medium as they can be inferred from Lattice QCD and implemented in effective theory calculations. The LHC is ideally placed for a decisive step forward in these studies, thanks to the availability of high-statistics results in both the charmonium and, for the first time, bottomonium sector. Specific open points concern the existence and the characteristics of the various bound states in the QGP and the possibility of a (re)generation of quarkonia at the confinement threshold, which constitutes in itself an evidence for the production of a deconfined state. The results on quarkonium-related observables will be described and discussed in Sec.~\ref{sec:Quarkonium}. 

    \item {\it Can the QGP lead to discovery of novel QCD effects?}
    
    In addition to forming the necessary conditions to study the deconfined QCD matter in a controlled way, heavy-ion collisions provide also the possibility to probe novel QCD phenomena that are associated with a chirality imbalance, and thus to parity violation in strong interactions. This imbalance is a consequence of topological transitions between the different vacuum states of QCD, yet another reflection of the non-Abelian nature of the theory, via a mechanism similar to the one responsible for the creation of baryon asymmetry in the early universe. The coupling of these effects with the extremely large magnetic field produced in peripheral heavy-ion collisions is expected to lead to an electric dipole moment of QCD matter, imprinted in the motion of final state particles. This is known as the Chiral Magnetic Effect (CME) which is believed to have a signal within experimental reach. The associated motion of deconfined quarks can only occur in the QGP. The CME is of particular interest at the LHC as heavy-ion collisions at these enegies are expected to provide the largest ever magnetic field created in the laboratory. 
    The ALICE results will be discussed in Sec.~\ref{sec:NovelQCD}, also in relation with the general  status of the field in studies of the early stage electromagnetic fields and on probing anomalous QCD effects.
    
    \item {\it What are the minimal conditions of QGP formation?}
    
    Several signatures of a strongly interacting medium, such as a relatively large strangeness production and anisotropic flow, appear prominently in mid-central and central nucleus--nucleus collisions. In this context, a natural question is if these phenomena also appear in very peripheral nucleus--nucleus collisions or, alternatively, in high-multiplicity proton--proton or proton--nucleus collisions. The ALICE detector is able to conduct a comprehensive study regarding the limits of QGP formation by studying the multiplicity dependence of several observables in pp and p--Pb collisions, which was made possible due to the unprecedented amount of data from these smaller collision systems delivered by the LHC. Advances in several key topics, including strangeness production, the onset of collective behaviour and energy loss phenomena, were achieved, leading to a better understanding of the microscopic origin of many of these phenomena and bringing forth new theoretical challenges. These results will be discussed in Chapter~\ref{ch:SmallSystems}.
    
     \item {\it What is the nature of the initial state of heavy-ion collisions?}
     
    An accurate description of the initial state of colliding nuclei is crucial for the interpretation of many of the measurements carried out in heavy-ion collisions at the LHC, and ultimately for the extraction of the relevant QGP parameters. In particular, the characterisation of the spatial distribution of nuclear matter in terms of energy and density profiles before the QGP formation has an important influence on the quantitative evaluation of its fundamental parameters (e.g. viscosity). Various approaches were developed along the years, considering either nucleons, constituent quarks, or  gluons as the fundamental degrees of freedom. The Colour Glass Condensate framework implies the spatial extent of low-$x$ gluons at LHC energies is determined by the saturation scale. This should have a discernible impact on anisotropic flow coefficients. The corresponding measurements therefore provide an opportunity to test nuclear gluon saturation approaches using the ALICE detector. At a complementary level, a quantitative description of hard processes requires the knowledge of the modifications of the parton longitudinal momentum distribution in the nuclei. Here, a strong suppression (nuclear shadowing) was previously observed at low parton momentum fractions ($x<10^{-2}$). The measurement of such modifications to the parton distribution functions (PDF), which can be seen as a consequence of saturation effects in the low-$x$ region, suffers from significant uncertainties. The availability of results with high statistical precision on electroweak boson production in p--Pb and Pb--Pb, as well as on vector-meson photoproduction in ultraperipheral Pb--Pb collisions, represent the main source of improvement that becomes available at LHC energies. The corresponding results will be described in Chapter~\ref{ch:InitialState}. 
    
    \item{\it What is the nature of hadron--hadron interactions?}
    
    The residual strong interaction among stable and unstable hadrons can be studied using femtoscopic correlation techniques at the LHC in pp and p--Pb collisions, investigated by ALICE with unprecedented precision. This is due to the large acceptance provided by ALICE, and the large integrated luminosities for light systems provided by the LHC. These measured observables can be directly connected to the relative wave function among the hadrons of interest. The small size of the source created in these light collision systems (about 1 fm) provides a unique environment to test the short-range strong interactions among hadron pairs. As mentioned, the measurements sensitive to these interactions can be compared with Lattice QCD or chiral effective theory predictions of hadron--hadron potentials, and the success of these comparisons, in particular for the rarely produced hadrons, may have implications for the modelling of neutron star cores. 
    Hadron--hadron interactions can also be studied by measuring the binding energies of light antinuclei and hypernuclei produced in pp, p--Pb and Pb--Pb collisions. A microscopic understanding of the formation process of light nuclei can be achieved by employing the measurement of the source size extracted from femtoscopy and testing whether hadron interactions can explain nuclei formation via the coalescence of nucleons. These topics are treated in Chapter~\ref{ch:NuclPhysLHC}.

    \item{\it Can ALICE elucidate specific aspects of perturbative QCD and of related ``long distance'' QCD interactions?}
    
    Precise perturbative QCD (pQCD) 
    calculations serve as a baseline for searches for physics Beyond the Standard Model. Moreover, observables with a well understood production cross section in proton--proton collisions are essential to interpret the additional complexity of nonzero temperature QCD in heavy-ion collisions. Measurements of high-$Q^2$ processes typically involve both perturbatively calculable cross sections, as well as long-distance QCD interactions described by nonperturbative objects such as parton distribution functions and fragmentation functions which are commonly assumed to be universal, i.e. independent of the collision system, energy, and observable. Jet measurements, for example, test the importance of higher order terms in state-of-the-art pQCD calculations, and identified particle measurements constrain nonperturbative objects and test the universality of their fragmentation functions. The ALICE  high-precision tracking and particle identification capabilities, with a focus on low to moderate transverse momentum, are unique at the LHC and complementary to the other LHC detectors. The moderate collision rates at the ALICE interaction point provide a clean, low-pileup environment for a variety of high-precision QCD studies. Results of QCD studies with high-$Q^2$ processes in pp collisions will be discussed in Chapter~\ref{ch:QCDpp}.

\end{enumerate}
\newpage    
The next chapters of this review are organised as follows: Chapter~\ref{ch:QGPproperties} contains an extended description of the quantitative studies of the QGP properties, as investigated by ALICE mainly in Pb--Pb collisions. Chapter~\ref{ch:SmallSystems} aims at connecting observations obtained for large and small collision systems with the scope of analyzing the emergence of collective effects related to the strong interaction and possibly to QGP formation. Chapter~\ref{ch:InitialState} will describe the contributions of ALICE to the determination of the properties of the initial state of the high-energy collision systems, which are relevant both for the interpretation of final state quantities and for the study of specific QCD aspects (parton saturation). 
Chapter~\ref{ch:NuclPhysLHC} will address studies of the interaction between hadrons, manifested in the formation of nuclear systems also including strange quarks. Chapter~\ref{ch:QCDpp} deals with selected studies of high-$Q^2$ processes in pp collisions where ALICE has given significant contributions in the frame of LHC experimental results. 
Chapter~\ref{ch:ConnectionsOtherFields}
presents ALICE contributions to fields not directly related with LHC physics.  Chapter~\ref{ch:Conclusions} is devoted to the conclusions. Chapter~\ref{ch:Outlook} 
discusses future prospects with the upgraded `ALICE\,2' (Run\,3--4), the proposed new detector `ALICE\,3' for Run\,5 and beyond, and the open questions that will be addressed by future measurements.

\newpage

\setcounter{section}{1}
\section{The quark--gluon plasma and its properties}
\label{ch:QGPproperties}

The quantitative study of the properties of the quark--gluon plasma produced in nuclear collisions at the LHC represents the core of the ALICE physics program. The possibility of colliding Pb ions at the highest energy presently available at a particle accelerator, with significant further increase only foreseen in a relatively distant future (HE-LHC, FCC), implies that many of the results that will be described in this chapter will represent for a long time the state-of-the-art of our comprehension of low(zero)-$\mu_{\rm B}$ deconfined matter. Due to the large initial temperature, the conditions are favourable for the formation of a long-lived system which quickly reaches thermalisation, allowing a precise estimate of its intensive properties. At the same time, the abundant production of hard partons at LHC energy, constitutes an intense source of probes of the QGP phase, allowing a test of its transport properties and micro-structure. 

This chapter is organized in sections, each one dealing with specific physics aspects of the QGP. After a short introduction to the corresponding topics, the main results obtained by ALICE are described and discussed, with  reference to theoretical models that are found to be relevant for the understanding of the results and the extraction of a physics message. 

Specifically, characterisation of the Pb--Pb events in terms of centrality selection will be discussed first, followed by estimates of the attained energy density and temperature of the QGP, and by a discussion on the spatial extension of the strongly interacting system (Sec.~\ref{sec:MacroscopicProperties}). Then, a discussion on the dynamical properties of the QGP will be presented, mainly dealing with the study of its collective motion that offers insight on the coupling strength of the system (Sec.~\ref{sec:QGPevolution}). In the next step, a discussion of the phenomena related to the (pseudo)critical transition of the QGP towards the hadronic phase is carried out, by describing results on the chemical equilibration and kinetic thermalisation of the system  and investigating the duration of the hadronic phase (Sec.~\ref{sec:QGPHadronization}). Then, Sec.~\ref{sec:PartonInteractions} deals with the use of high-momentum transfer processes as a probe of the QGP, their in-medium  interactions being  used  to quantitatively understand dynamical features such as energy transport and equilibration. A specific set of hard probes, quarkonia, exhibits a strong sensitivity to deconfinement and is more generally sensitive to the modification of the QCD force in the medium, with the corresponding results shown in Sec.~\ref{sec:Quarkonium}. Then, Sec.~\ref{sec:NovelQCD} deals with novel QCD phenomena that can be catalysed by the presence of very strong magnetic fields generated in heavy-ion collisions, as local P and CP violating effects in the strong interaction. 

A brief and schematic summary of the main findings completes each section, and the chapter is then concluded in Sec.~\ref{sec:QGPsummary}, which presents a discussion of the extracted values of QGP-related quantities.

\newpage

\newcommand {\stat}     {({\it stat.})~}
\newcommand {\syst}     {({\it syst.})~}

\newcommand{\ie}        {$i.e.$~}
\newcommand{\eg}        {$e.g.$~}
\newcommand{\Eg}        {$E.g.$~}
\newcommand{\Fig}       {Fig.~}
\newcommand{\fig}       {\Fig}
\newcommand{\Figs}      {Figs.~}
\newcommand{\figs}      {\Figs}
\newcommand{\Figure}    {Figure~}
\newcommand{\Figures}   {Figures~}
\newcommand{\Tab}       {Tab.~}
\newcommand{\tab}       {\Tab}
\newcommand{\Table}     {Table~}
\newcommand{\Tables}    {Tables~}
\newcommand{\chap}      {Chap.~}
\newcommand{\parag}     {Par.~}
\newcommand{\appdx}     {App.~}
\newcommand{\eq}        {Eq.~}
\newcommand{\opt}        {Opt.~}

\newcommand{\tune}      {\emph{tune}}
\newcommand{\tunes}     {\emph{tunes}\xspace}

\newcommand{\orderOf}[1]{\ensuremath{\mathcal{O}}(#1)}

\newcommand{\CheckGr}   {\textcolor{Green}{\normalsize \CheckmarkBold}} 
\newcommand{\SurpriseGr}{\textcolor{Green}{\textbf{!}{\scriptsize !}}}
\newcommand{\Wtf}       {\textcolor{IndianRed}{\textbf{?}!}}
\newcommand{\NB}        {\textcolor{IndianRed}{\HandRight}}
\newcommand{\Caution}   {\textcolor{IndianRed}{\scriptsize \textlhdbend}}
\newcommand{\NoWay}     {\textcolor{IndianRed}{\normalsize \XSolidBrush}}
\newcommand{\UnderWork} {\textcolor{Blue}{\Large \dsagricultural}}
\newcommand{\ToDo}      {\textcolor{Gray}{\scriptsize \textsc{ToDo}}}

\newcommand {\pT}           {\ensuremath{p_{\text{\textsc{t}}}}}
\newcommand {\pt}{\pT}
\newcommand {\pTlw}         {\ensuremath{p_{\text{\textsc{t}}}^{lw} }}
\newcommand {\pTIdx}[1]     {\ensuremath{p_{\text{\textsc{t},#1}}}}
\newcommand {\pTExp}[1]     {\ensuremath{p_{\text{\textsc{t}}}^{#1}}}
\newcommand {\pTIdxExp}[2]  {\ensuremath{p_{\text{\textsc{t},#1}}^{#2}}}
\newcommand {\pTof}[1]      {\ensuremath{p_{\text{\textsc{t}}} \text{(#1)}}}
\newcommand {\pTchJet}      {\ensuremath{p_{\text{\textsc{t},jet}}^{\text{ch}}}}
\newcommand {\pTchEmJet}    {\ensuremath{p_{\text{\textsc{t},jet}}^{\text{ch+em}}}}
\newcommand {\meanpT}       {\ensuremath{\langle p_{\textsc{t}} \kern-0.1em\rangle}}
\newcommand {\mean}[1]      {\ensuremath{\langle #1 \kern-0.1em\rangle}} 
\newcommand {\sqrtSnn}      {\ensuremath{\sqrt{s_\text{\textsc{nn}}}}}
\newcommand {\sqrtS}        {\ensuremath{\sqrt{s}}}
\newcommand {\vTwo}         {\ensuremath{v_{\text{2}}}}
\newcommand {\vThree}       {\ensuremath{v_{\text{3}}}}
\newcommand {\vFour}        {\ensuremath{v_{\text{4}}}}
\newcommand {\vFive}        {\ensuremath{v_{\text{5}}}}
\newcommand {\vSix}         {\ensuremath{v_{\text{6}}}}
\newcommand {\vN}           {\ensuremath{v_{\text{n}}}}
\newcommand {\eT}           {\ensuremath{E_{\text{\textsc{t}}}}}
\newcommand {\mT}           {\ensuremath{m_{\text{\textsc{t}}}}}
\newcommand {\mTmZero}      {\ensuremath{m_{\text{\textsc{t}}} - m_0}}
\newcommand {\minv}         {\mbox{$m_{\ee}$}}
\newcommand {\rap}          {\mbox{$y$}}
\newcommand {\rapLab}       {\mbox{$y_{\text{lab}}$}}
\newcommand {\rapCms}       {\mbox{$y_{\text{\textsc{cms} }}$}}
\newcommand {\absrap}       {\mbox{$\left | y \right | $}}
\newcommand {\rapPart}[1]   {\mbox{$\left | y\text{(#1)} \right | $}}
\newcommand {\rapXi}        {\mbox{$\left | y(\rmXi) \right | $}}
\newcommand {\rapJpsi}      {\mbox{$y_{\tiny \rmJpsi}$}}
\newcommand {\abspseudorap} {\mbox{$\left | \eta \right | $}}
\newcommand {\pseudorap}    {\mbox{$\eta$}}
\newcommand {\pseudorapLab} {\mbox{$\eta_{\,\text{lab}}$}}
\newcommand {\pseudorapCms} {\mbox{$\eta_{\text{\textsc{cms} }}$}}
\newcommand {\cTau}         {\ensuremath{c.\tau}}
\newcommand {\sigee}        {$\sigma_E$/$E$}
\newcommand {\AxEff}        {\ensuremath{\mathscr{A}.\varepsilon}}

\newcommand{\dd}{\mathop{}\!\mathrm{d}}
\newcommand {\oneOverpipT}  {\ensuremath{1/2\pi\pT}}
\newcommand {\crossSec}[1]  {\mbox{$\sigma_{\scriptsize \rm #1}$}}
\newcommand {\visCrossSec}[1]  {\mbox{$\sigma_{\scriptsize \rm #1}^{visible}$}}
\newcommand {\dsigmady}     {\ensuremath{\mathrm{d}\sigma/\mathrm{d}y}}
\newcommand {\dsigmadpt}    {\ensuremath{\mathrm{d}^{2}\sigma/\mathrm{d}\pT}}
\newcommand {\dsigmadptdy}  {\ensuremath{\mathrm{d}^{2}\sigma/\mathrm{d}\pT\mathrm{d}y}}
\newcommand {\dsigmadptdeta}{\ensuremath{\mathrm{d}^{2}\sigma/\mathrm{d}\pT\mathrm{d}\eta}}
\newcommand {\dsigmaXdy}[1] {\ensuremath{\mathrm{d}\sigma\text{(#1)}/\mathrm{d}y}}
\newcommand {\dsigmaXdpt}[1]{\ensuremath{\mathrm{d}\sigma\text{(#1)}/\mathrm{d}\pT}}
\newcommand {\dsigmaXdptdy}[1]  {\ensuremath{\mathrm{d}^{2}\sigma\text{(#1)}/\mathrm{d}\pT\mathrm{d}y}}
\newcommand {\dNdy}         {\ensuremath{\mathrm{d}N/\mathrm{d}y}}
\newcommand {\dNXdy}[1]     {\ensuremath{\mathrm{d}N\text{(#1)}/\mathrm{d}y}}
\newcommand {\dNJpsidy}     {\dNXdy{\rmJpsi}}
\newcommand {\dNdpt}        {\ensuremath{\mathrm{d}N/\mathrm{d}\pT }}
\newcommand {\dNXdptdy}[1]  {\ensuremath{\mathrm{d}^{2}N\text{(#1)}/\mathrm{d}\pT\mathrm{d}y}}
\newcommand {\dNdptdy}      {\ensuremath{\mathrm{d}^{2}N/\mathrm{d}\pT\mathrm{d}y }}
\newcommand {\dNdptdeta}    {\ensuremath{\mathrm{d}^{2}N/\mathrm{d}\pT\mathrm{d}\eta }}
\newcommand {\fracdsigmadptdy}  {\ensuremath{ \frac{\mathrm{d}^{2}\sigma}{\mathrm{d}\pT\mathrm{d}y}}}
\newcommand {\fracdNdptdy}  {\ensuremath{ \frac{\mathrm{d}^{2}N}{\mathrm{d}\pT\mathrm{d}y } }}
\newcommand {\fracdNdy}     {\ensuremath{ \frac{\dN}{\dy}}}
\newcommand {\fracdNdyBold} {\ensuremath{ \frac{\bm{\dN}}{\bm{\dy}}}}

\newcommand {\dNdmtdy}      {\ensuremath{\mathrm{d}^{2}N/\mathrm{d}\mT\mathrm{d}y }}
\newcommand {\dN}           {\ensuremath{\mathrm{d}N }}
\newcommand {\Npp}          {\ensuremath{N_{\textsc{\pp}}}}
\newcommand {\dNsquared}    {\ensuremath{\mathrm{d}^{2}N }}
\newcommand {\dsquared}     {\ensuremath{\mathrm{d}^{2} }}
\newcommand {\dpT}          {\ensuremath{\mathrm{d}\pT }}
\newcommand {\dy}           {\ensuremath{\mathrm{d}y}}
\newcommand {\dNdyBold}     {\ensuremath{\bm{\dN/\dy}}}
\newcommand {\dNchdy}       {\ensuremath{\mathrm{d}N_\text{ch}/\mathrm{d}y }}
\newcommand {\dNchdeta}     {\ensuremath{\mathrm{d}N_\text{ch}/\mathrm{d}\eta }}
\newcommand {\dNchdptdeta}  {\ensuremath{\mathrm{d}^{2}N_\text{ch}/\mathrm{d}\pT\mathrm{d}\eta }}
\newcommand {\RAA}          {\ensuremath{R_\text{AA}}}
\newcommand {\RpA}          {\ensuremath{R_\text{pA}}}
\newcommand {\RpPb}         {\ensuremath{R_\text{pPb}}}
\newcommand {\RPbp}         {\ensuremath{R_\text{Pbp}}}
\newcommand {\RAuAu}        {\ensuremath{R_\text{AuAu}}}
\newcommand {\RPbPb}        {\ensuremath{R_\text{PbPb}}}
\newcommand {\Rcp}          {\ensuremath{R_\text{CP}}}
\newcommand {\hPMVzsCorrel} {\ensuremath{(\text{\hPM-V0})}}
\newcommand {\hVzsCorrel}   {\ensuremath{(\text{h-V0})}}
\newcommand {\mPDG}[1]      {\ensuremath{m_{\text{pdg}}(#1)}}

\newcommand {\Nevt}         {\ensuremath{N_\text{evt}}}
\newcommand {\NevtINEL}     {\ensuremath{N_\text{evt}(\textsc{inel})}}
\newcommand {\NevtNSD}      {\ensuremath{N_\text{evt}(\textsc{nsd})}}
\newcommand {\INEL}         {\ensuremath{\textsc{inel}}}
\newcommand {\INELZero}     {\ensuremath{\textsc{inel}>0}}
\newcommand {\NSD}          {\ensuremath{\textsc{nsd}}}
\newcommand {\dEdx}         {\ensuremath{\textup{d}E/\textup{d}x }}

\newcommand {\bsTe}      {\ensuremath{\bm{T_e}}}
\newcommand {\bsTb}      {\ensuremath{\bm{T_b}}}
\newcommand {\bsTt}      {\ensuremath{\bm{T_T}}}
\newcommand {\bsCt}      {\ensuremath{\bm{C_T}}}
\newcommand {\bsNt}      {\ensuremath{\bm{n}}}
\newcommand {\bsQt}      {\ensuremath{\bm{q}}}
\newcommand {\bsVt}      {\ensuremath{\bm{V}}}
\newcommand {\bsgVt}     {\ensuremath{\bm{g.V}}}
\newcommand {\bsNb}      {\ensuremath{\bm{n}}}
\newcommand {\bsFp}      {\ensuremath{\bm{f_P}}}
\newcommand {\bsCp}      {\ensuremath{\bm{C_P}}}

\newcommand {\Lint}         {\ensuremath{L_{\text{int}}}}

\newcommand {\rphi}         {\mbox{\ensuremath{(r,\varphi)}}}
\newcommand {\alphaS}       {\ensuremath{ \alpha_s}}
\newcommand {\chLeptonAsymm}{\ensuremath{ A_{\ell\ell} }}

\newcommand {\MeanNpart}    {\mbox{\ensuremath{\langle\kern-0.05em N_{\rm part} \kern-0.05em \rangle}}}
\newcommand {\MeanNcoll}    {\mbox{\ensuremath{\langle\kern-0.05em N_{\rm coll} \kern-0.05em \rangle}}}
\newcommand {\sigmaBarlow}  {\ensuremath{\sigma_{Barlow}}}
\newcommand {\sigmaStat}    {\ensuremath{\sigma_{stat}}}
\newcommand {\sigmaSyst}    {\ensuremath{\sigma_{syst}}}
\newcommand {\sigmaTot}     {\ensuremath{\sqrt{\sigmaStat^2 + \sigmaSyst^2}}}

\newcommand {\xXzero}     {\ensuremath{\textsc{x}/X_0}}

\newcommand{\GeVc}         {Ge\kern-.1emV/$c$\xspace}
\newcommand{\MeVc}         {Me\kern-.1emV/$c$\xspace}
\newcommand{\TeV}          {Te\kern-.1emV\xspace}
\newcommand{\GeV}          {Ge\kern-.1emV\xspace}
\newcommand{\MeV}          {Me\kern-.1emV\xspace}

\newcommand{\ycms}         {\ensuremath{y_{\rm CMS}}\xspace}
\newcommand{\ylab}         {\ensuremath{y_{\rm lab}}\xspace}
\newcommand{\etarange}[1]  {\mbox{$\left | \eta \right |~<~#1$}}
\newcommand{\yrange}[1]    {\mbox{$\left | y \right |~<~#1$}}
\newcommand{\dndy}         {\ensuremath{\mathrm{d}N_\mathrm{ch}/\mathrm{d}y}\xspace}
\newcommand{\dndeta}       {\ensuremath{\mathrm{d}N_\mathrm{ch}/\mathrm{d}\eta}\xspace}
\newcommand{\Npart}        {\ensuremath{N_\mathrm{part}}\xspace}
\newcommand{\Ncoll}        {\ensuremath{N_\mathrm{coll}}\xspace}
\newcommand{\avdndeta}     {\ensuremath{\langle\dndeta\rangle}\xspace}
\newcommand{\avNpart}      {\MeanNpart}
\newcommand{\avNcoll}      {\MeanNcoll}
\newcommand{\dNdetape}     {\ensuremath{\frac{2}{\avNpart}\avdndeta}}
\newcommand{\dndetalab}    {\ensuremath{\mathrm{d}N_\mathrm{ch}/\mathrm{d}\eta_\mathrm{lab}}}

\newcommand{\Teff}{\ensuremath{T_{\mathrm{eff}}}}
\newcommand{\temp}{\ensuremath{T}}
\newcommand{\Tchem}{\ensuremath{T_{\mathrm{chem}}}}
\newcommand{\Tkin}{\ensuremath{T_{\mathrm{kin}}}}

\newcommand{\Rgamma}        {\ensuremath{\mathrm{R}_{\gamma}}\xspace}
\newcommand{\vtwogammadir} {\ensuremath{v_{2}^{\gamma, {\rm dir}}}\xspace}
\newcommand{\vtwogammadec} {\ensuremath{v_{2}^{\gamma, {\rm dec}}}\xspace}
\newcommand{\vtwogammainc} {\ensuremath{v_{2}^{\gamma, {\rm inc}}}\xspace}
\newcommand{\gammainc} {\ensuremath{\gamma_{\mathrm{ inc}}}\xspace}
\newcommand{\gammadir} {\ensuremath{\gamma_{\mathrm{ dir}}}\xspace}
\newcommand{\gammadec} {\ensuremath{\gamma_{\mathrm{ dec}}}\xspace}
\newcommand{\npiparam} {\ensuremath{\pi^{0}_{\mathrm{param}}}\xspace}

\newcommand{\Rlong}{\ensuremath{R_{\mathrm{long}}}\xspace}
\newcommand{\Rside}{\ensuremath{R_{\mathrm{side}}}\xspace}
\newcommand{\Rout}{\ensuremath{R_{\mathrm{out}}}\xspace}
\newcommand {\kT}           {\ensuremath{k_{\text{\textsc{t}}}}\xspace}

\newcommand{\ST}{\ensuremath S_{\scriptscriptstyle\mathrm{T}}}
\newcommand{\ET}{\ensuremath E_{\scriptscriptstyle\mathrm{T}}}

\newcommand {\eX}    {\ensuremath{\vec{e}_{\textsc{x}}}}
\newcommand {\eY}    {\ensuremath{\vec{e}_{\textsc{y}}}}
\newcommand {\eZ}    {\ensuremath{\vec{e}_{\textsc{z}}}}

\newcommand {\ePhi}  {\ensuremath{\vec{e}_{\varphi}}}
\newcommand {\eR}    {\ensuremath{\vec{e}_{r}}}

\newcommand{\Sherpa}        {\textsc{Sherpa}\xspace}
\newcommand{\Herwig}        {\textsc{Herwig++}\xspace}
\newcommand{\Epos}          {\textsc{Epos}\xspace}
\newcommand{\Pythia}        {\textsc{Pythia}\xspace}
\newcommand{\Pythiaeight}   {\Pythia~8\xspace}
\newcommand{\Rivet}         {\textsc{Rivet}\xspace}
\newcommand{\HepMC}         {\textsc{HepMc}\xspace}

\newcommand{\Root}          {\textsc{Root}\xspace}
\newcommand{\GeantFour}     {\textsc{Geant4}\xspace}
\newcommand{\Fluka}         {\textsc{Fluka}\xspace}

\newcommand {\pp}        {\ensuremath{\mbox{\text {p\kern-0.05em p}}}\xspace}
\newcommand {\rmpp}      {\ensuremath{\text{p\kern-0.05em p} }}
\newcommand {\ppbar}     {\mbox{$\text{p}\overline{\text{p}}$}}
\newcommand {\PbPb}      {\ensuremath{\mbox{\text{Pb--Pb}} }}
\newcommand {\rmPbPb}    {\ensuremath{\text{PbPb} }}
\newcommand {\AuAu}      {\ensuremath{\mbox{\text{Au--Au}} }}
\newcommand {\ArAr}      {\ensuremath{\mbox{\text{Ar--Ar}} }}
\newcommand {\CuCu}      {\ensuremath{\mbox{\text{Cu--Cu}} }}
\newcommand {\UU}        {\ensuremath{\mbox{\text{U--U}}   }}
\newcommand {\XeXe}      {\ensuremath{\mbox{\text{Xe--Xe}} }}
\renewcommand {\AA}      {\ensuremath{\text{A--A}          }} %
\newcommand {\rmAA}      {\ensuremath{\text{AA}            }} %
\newcommand {\pA}        {\ensuremath{\mbox{\text{p--A}}   }}
\newcommand {\rmpA}      {\ensuremath{\text{pA}            }} %
\newcommand {\dA}        {\ensuremath{\mbox{\text{d--A}}   }}
\newcommand {\pPb}       {\ensuremath{\mbox{\text{p--Pb}}  }}
\newcommand {\Pbp}       {\ensuremath{\mbox{\text{Pb--p}}  }}
\newcommand {\dAu}       {\ensuremath{\mbox{\text{d--Au}}  }}
\newcommand {\pAu}       {\ensuremath{\mbox{\text{p--Au}}  }}
\newcommand {\EplusEminus}      {\ee}
\newcommand {\ee}               {\mbox{$\text{e}^+\text{e}^-$}}
\newcommand {\Eplus}            {\mbox{$\text{e}^+$}}
\newcommand {\Eminus}           {\mbox{$\text{e}^-$}}
\newcommand {\MuPlusMuMinus}    {\mbox{$\mu^+\mu^-$}}
\newcommand {\MuPlus}           {\mbox{$\mu^+$}}
\newcommand {\MuMinus}          {\mbox{$\mu^-$}}

\newcommand {\massStyle}[1] {\mbox{\ensuremath{\text{#1}\kern-0.1em /\kern-0.12em c^2}}}
\newcommand {\mass}     {\massStyle{MeV}}
\newcommand {\kmass}    {\massStyle{keV}}
\newcommand {\mmass}    {\massStyle{MeV}}
\newcommand {\gmass}    {\massStyle{GeV}}

\newcommand {\unitStyle}[1] {\mbox{\ensuremath{\text{#1}}}}
\newcommand {\tev}      {\unitStyle{TeV}}
\newcommand {\gev}      {\unitStyle{GeV}}
\newcommand {\mev}      {\unitStyle{MeV}}
\newcommand {\kev}      {\unitStyle{keV}}
\newcommand {\J}        {\unitStyle{J}}

\newcommand {\momStyle}[1] {\mbox{\ensuremath{\text{#1}\kern-0.1em /\kern-0.12em c}}}
\newcommand {\mmom}     {\momStyle{MeV}}
\newcommand {\gmom}     {\momStyle{GeV}}

\newcommand {\fsec}       {\unitStyle{fs}}
\newcommand {\psec}       {\unitStyle{ps}}
\newcommand {\nsec}       {\unitStyle{ns}}
\newcommand {\musec}      {\mbox{$\mu\unitStyle{s}$}}
\newcommand {\millisec}   {\unitStyle{ms}}
\newcommand {\second}     {\unitStyle{s}}

\newcommand {\MHz}     {\unitStyle{MHz}}
\newcommand {\kHz}     {\unitStyle{kHz}}

\newcommand {\fmC}      {\mbox{$\unitStyle{fm}/\kern-0.12em c$}}

\newcommand {\fm}       {\unitStyle{fm}}
\newcommand {\nm}       {\unitStyle{nm}}
\newcommand {\mum}      {\mbox{$\mu\unitStyle{m}$}}
\newcommand {\mm}       {\unitStyle{mm}}
\newcommand {\cm}       {\unitStyle{cm}}
\newcommand {\m}        {\unitStyle{m}}

\newcommand {\cmq}      {\mbox{$\cm^2$}}
\newcommand {\mmq}      {\mbox{$\mm^2$}}
\newcommand {\mumq}     {\mbox{$\mum^2$}}

\newcommand {\fmCube}   {\mbox{\unitStyle{fm}$^3$}}

\newcommand {\mug}      {\mbox{$\mu\unitStyle{g}$}}
\newcommand {\mg}       {\unitStyle{mg}}
\newcommand {\gram}     {\unitStyle{g}}
\newcommand {\kg}       {\unitStyle{kg}}

\newcommand {\dens}     {\mbox{$\unitStyle{g}/\unitStyle{cm}^{3}$}}

\newcommand {\dg}       {\mbox{$\kern+0.1em ^\circ$}}

\newcommand {\lumi}     {\mbox{$\cm^{-2}\second^{-1}$}}

\newcommand {\barn}     {\unitStyle{b}}
\newcommand {\fb}       {\unitStyle{fb}}
\newcommand {\pb}       {\unitStyle{pb}}
\newcommand {\nb}       {\unitStyle{nb}}
\newcommand {\mub}      {\mbox{$\mu\unitStyle{b}$}}
\newcommand {\mb}       {\unitStyle{mb}}
\newcommand {\kb}       {\unitStyle{kb}}

\newcommand {\invmub}   {\mbox{$\mub^{-1}$}}
\newcommand {\invnb}    {\mbox{$\nb^{-1}$}}
\newcommand {\invpb}    {\mbox{$\pb^{-1}$}}
\newcommand {\invfb}    {\mbox{$\fb^{-1}$}}

\newcommand{\hPM}           {\ensuremath{h^{\pm}}}
\newcommand{\ePlusMinus}    {\mbox{$\mathrm {e^{\pm}}$}}
\newcommand{\muPlusMinus}   {\mbox{$\mathrm {\mu^{\pm}}$}}

\newcommand{\piZero}        {\mbox{$\mathrm {\pi^0}$}}
\newcommand{\piMinus}       {\mbox{$\mathrm {\pi^-}$}}
\newcommand{\piPlus}        {\mbox{$\mathrm {\pi^+}$}}
\newcommand{\piPlusMinus}   {\mbox{$\mathrm {\pi^{\pm}}$}}
\newcommand{\rmPiPlusMinus} {\piPlusMinus}
\newcommand{\rmPiPM}        {\piPlusMinus}
\newcommand{\rmPiMinus}     {\piMinus}
\newcommand{\rmPiPlus}      {\piPlus}

\newcommand{\Kzs}        {\mbox{$\mathrm {K^0_S}$}}
\newcommand{\rmKzero}    {\Kzs}
\newcommand{\Kzl}        {\mbox{$\mathrm {K^0_L}$}}
\newcommand{\rmKzeroL}   {\Kzl}
\newcommand{\Kminus}     {\mbox{$\mathrm {K^-}$}}
\newcommand{\rmKminus}   {\Kminus}
\newcommand{\Kplus}      {\mbox{$\mathrm {K^+}$}}
\newcommand{\rmKplus}    {\Kplus}
\newcommand{\Kplusmin}   {\mbox{$\mathrm {K^{\pm}}$}}
\newcommand{\rmKpm}      {\Kplusmin}
\newcommand{\rmKstar}    {\mbox{$\mathrm{K}^*\mathrm{(892)}^0$}}
\newcommand{\rmPhiMes}   {\mbox{$\mathrm {\phi(1020)}$}}

\newcommand{\proton}    {\mbox{$\mathrm {p}$}}
\newcommand{\pbar}      {\mbox{$\mathrm {\overline{p}}$}}
\newcommand{\pOrPbar}   {\mbox{$\mathrm {p^{\pm}}$}}

\newcommand{\rmLambdaZ}         {\mbox{$\mathrm {\Lambda}$}}
\newcommand{\rmAlambdaZ}        {\mbox{$\mathrm {\overline{\Lambda}}$}}
\newcommand{\rmLambda}          {\mbox{$\mathrm {\Lambda}$}}
\newcommand{\rmAlambda}         {\mbox{$\mathrm {\overline{\Lambda}}$}}
\newcommand{\rmLambdas}         {\mbox{$\mathrm {\Lambda \kern-0.2em + \kern-0.2em \overline{\Lambda}}$}}
\newcommand{\ratioLamOverKzs}   {\rmLambda/\rmKzero}

\newcommand{\rmSigma}       {\mbox{$\mathrm {\Sigma}$}}
\newcommand{\rmSigmaM}      {\mbox{$\mathrm{\Sigma}^{-}$}}
\newcommand{\rmSigmaP}      {\mbox{$\mathrm{\Sigma}^{+}$}}
\newcommand{\rmSigmaZero}   {\mbox{$\mathrm{\Sigma}^{0}$}}
\newcommand{\rmSigmaMres}   {\mbox{$\mathrm{\Sigma(1385)}^{-}$}}
\newcommand{\rmSigmaPres}   {\mbox{$\mathrm{\Sigma(1385)}^{+}$}}

\newcommand{\rmXi}      {\mbox{$\mathrm{\Xi}$}}
\newcommand{\rmXiM}     {\mbox{$\mathrm{\Xi}^{-}$}}
\newcommand{\rmXiPM}    {\mbox{$\mathrm {\Xi^{\pm}}$}}

\newcommand{\rmAxiP}    {\mbox{$\mathrm {\overline{\Xi}^{+}}$}}
\newcommand{\rmXis}     {\mbox{$\mathrm {\Xi^{-} \kern-0.3em + \kern-0.1em \overline{\Xi}^{+}}$}}
\newcommand{\rmXiZero}  {\mbox{$\mathrm {\Xi^{0}}$}}
\newcommand{\rmXiZ}     {\rmXiZero}
\newcommand{\rmXiZres}  {\mbox{$\mathrm {\Xi (1530)^{0}}$}}
\newcommand{\rmAxiZres} {\mbox{$\mathrm {\overline{\Xi} (1530)^{0}}$}}
\newcommand{\rmXiMres}  {\mbox{$\mathrm {\Xi (1530)^{-}}$}}
\newcommand{\rmAxiPres} {\mbox{$\mathrm {\overline{\Xi}(1530)^{+}}$}}

\newcommand{\rmOmega}   {\mbox{$\mathrm {\Omega}$}}
\newcommand{\rmOmegaM}  {\mbox{$\mathrm {\Omega^{-}}$}}
\newcommand{\rmAomegaP} {\mbox{$\mathrm {\overline{\Omega}^{+}}$}}
\newcommand{\rmOmegas}  {\mbox{$\mathrm {\Omega^{-} \kern-0.3em +  \kern-0.1em \overline{\Omega}^{+}}$}}
\newcommand{\rmOmegaPM} {\mbox{$\mathrm {\Omega^{\pm}}$}}

\newcommand{\rmDeuton}   {\mbox{$\mathrm {d}$}}
\newcommand{\rmDeutonPM} {\mbox{$\mathrm {d}^{\pm}$}}
\newcommand{\rmTriton}   {\mbox{$\mathrm {t}$}}
\newcommand{\rmTritonPM} {\mbox{$\mathrm {t}^{\pm}$}}
\newcommand{\rmHeThree}  {\mbox{$\mathrm {^3He}$}}
\newcommand{\rmHeThreePM}{\mbox{$\mathrm {^3He^{2\pm}}$}}
\newcommand{\rmHeFour}   {\mbox{$\mathrm {^4He}$}}
\newcommand{\rmHeFourPM} {\mbox{$\mathrm {^4He^{2\pm}}$}}

\newcommand{\rmHypertriton}  {\mbox{$^{3}_{\Lambda}\mathrm{H}$}}

\newcommand{\rmJpsi}    {\mbox{$\mathrm{J\kern-0.05em /\kern-0.05em\psi}$}}
\newcommand{\rmPsiTwoS} {\mbox{$\mathrm {\psi(2S)}$}}
\newcommand{\rmChicZero}{\mbox{$\mathrm {\chi_{c_0}}$}}
\newcommand{\rmChicOne} {\mbox{$\mathrm {\chi_{c_1}}$}}
\newcommand{\rmChicTwo} {\mbox{$\mathrm {\chi_{c_2}}$}}
\newcommand{\rmChicJ}   {\mbox{$\mathrm {\chi_{c_J}}$}}

\newcommand{\rmLambdaC}         {\mbox{$\mathrm {\Lambda}_{c}^{+}$}}
\newcommand{\rmXiCplus}         {\mbox{$\mathrm {\Xi}_{c}^{+}$}}
\newcommand{\rmXiCzero}         {\mbox{$\mathrm {\Xi}_{c}^{0}$}}
\newcommand{\rmOmegaCzero}      {\mbox{$\mathrm {\Omega}_{c}^{0}$}}
\newcommand{\rmXiCCtwoPlus}     {\mbox{$\mathrm {\Xi}_{cc}^{2+}$}}
\newcommand{\rmOmegaCCtwoPlus}  {\mbox{$\mathrm {\Omega}_{ccc}^{2+}$}}

\newcommand{\rmDzero}   {\mbox{$\mathrm {D}^{0}$}}
\newcommand{\rmDzeroBar}{\mbox{$\mathrm {\overline{D}}^{0}$}}
\newcommand{\rmDplus}   {\mbox{$\mathrm {D}^{+}$}}
\newcommand{\rmDminus}  {\mbox{$\mathrm {D}^{+}$}}
\newcommand{\rmDpm}     {\mbox{$\mathrm {D}^{\pm}$}}
\newcommand{\rmDstar}   {\mbox{$\mathrm{D}^*\mathrm{(2010)}^+$}}
\newcommand{\rmDs}      {\mbox{$\mathrm {D}^{+}_{s}$}}

\newcommand{\rmBzero}       {\mbox{$\mathrm {B^{0}}$}}
\newcommand{\rmBplus}       {\mbox{$\mathrm {B^{+}}$}}
\newcommand{\rmBminus}      {\mbox{$\mathrm {B^{-}}$}}
\newcommand{\rmBplusMinus}  {\mbox{$\mathrm {B^{\pm}}$}}
\newcommand{\rmBzeroS}      {\mbox{$\mathrm {B^{0}_s}$}}

\newcommand{\rmUpsOneS}         {\mbox{$\mathrm {\Upsilon(1S)}$}}
\newcommand{\rmUpsTwoS}         {\mbox{$\mathrm {\Upsilon(2S)}$}}
\newcommand{\rmUpsThreeS}       {\mbox{$\mathrm {\Upsilon(3S)}$}}
\newcommand{\rmUpsTwoThreeS}    {\mbox{$\mathrm {\Upsilon(2S,3S)}$}}
\newcommand{\rmUpsnS}           {\mbox{$\mathrm {\Upsilon(nS)}$}}
\newcommand{\rmUpsOneTwoThreeS} {\mbox{$\mathrm {\Upsilon(1S,2S,3S)}$}}

\newcommand{\rmPhoton}      {\mbox{$\mathrm {\gamma}$}}
\newcommand{\rmWplus}       {\mbox{$\mathrm {W^{+}}$}}
\newcommand{\rmWminus}      {\mbox{$\mathrm {W^{-}}$}}
\newcommand{\rmWPlusMinus}  {\mbox{$\mathrm {W^{\pm}}$}}
\newcommand{\rmZzero}       {\mbox{$\mathrm {Z}$}}

\newcommand{\qqbar}             {\mbox{$q\overline{q}$}}
\newcommand{\ccbar}             {\mbox{$c\overline{c}$}}
\newcommand{\bbbar}             {\mbox{$b\overline{b}$}}
\newcommand{\DDbar}             {\mbox{$\mathrm {D\overline{D}}$}}

\newcommand {\qbar}     {\mbox{$\overline{\text{q}}$}}
\renewcommand {\qqbar}     {\mbox{$\text{q}\overline{\text{q}}$}}

\newcommand{\VZERO}        {\rm{V0}\xspace}
\newcommand{\VZEROA}       {\rm{V0A}\xspace}
\newcommand{\VZEROC}       {\rm{V0C}\xspace} 

\subsection{Macroscopic system properties and QGP thermodynamics}
\label{sec:MacroscopicProperties}

The first question to answer is whether a quark--gluon plasma is formed in \PbPb\ collisions at the LHC and how we can characterise it based on its thermodynamic properties. 
As discussed in Chap.~\ref{ch:Introduction}, crossing the QCD phase transition boundary and forming the QGP require that the initial energy density and temperature reached in the collision are larger than the critical values. 
For these, the most recent Lattice QCD calculations report for the energy density $\epsilon_c$~=~(0.42~$\pm$~0.06)~\GeV/fm$^3$~\cite{Bazavov:2018mes}, and two slightly different values of pseudocritical temperature, $T_{\rm pc}$~=~(156.5~$\pm$~1.5)~\MeV~\cite{Bazavov:2018mes} and $T_{\rm pc}$~=~(158~$\pm$~0.6)~\MeV\cite{Borsanyi:2020fev}. 
As the system expands and cools down, a second transition from a deconfined quark and gluon medium to hadrons takes place.
Therefore, the first quantities we are interested in determining to characterise the system are the initial energy density and the temperature.  
The total transverse energy of the particles produced in the final state is a measure of the energy density, whereas multiplicity in the final state is closely related to entropy production in the collision. 
As in the rapid expansion of the fireball both the total energy and the entropy are expected to be approximately conserved, the use of final state quantities to estimate the initial energy and entropy density is justified.
This is strictly valid in a non-viscous hydrodynamics scenario.
The amount of energy concentrated in the collision zone can be controlled experimentally by selecting events based on  centrality, as discussed in Sec.~\ref{sec:TG1centrality}. 
Results on the charged-particle multiplicity density measured in \PbPb\ collisions at the LHC are summarised and presented in comparison to those at lower energies as well as in \pPb\ and \pp\ collisions in Sec.~\ref{sec:TG1multiplicity}. In Sec.~\ref{sec:TG1energydensity}, we provide an estimate of the energy density as a function of centrality and discuss its relation to hydrodynamic calculations. 
While hadrons are produced in the freeze-out phase of the collision, photons and dileptons are emitted during the entire evolution of the fireball and provide an independent measure of the temperature. Therefore, hadrons and electromagnetic radiation measure different temperatures, corresponding to different stages of the system evolution. This is discussed in Sec.~\ref{sec:TG1directphotons}. 
Besides these thermodynamic properties, the system produced in the collision can be further characterised in terms of its spatial extension and duration by using femtoscopic techniques. The measurement of momentum correlations among particles produced in the final state (i.e.\,at the freeze-out of the system) provides important information on the particle emitting source and on the underlying dynamics. The size and lifetime of the system are discussed in Sec.~\ref{sec:TG1sizelifetime}.

\subsubsection{Centrality of nucleus--nucleus collisions}
\label{sec:TG1centrality}

Nuclei are extended objects and the degree of geometrical overlap between them in the collision, expressed in terms of the impact parameter ($b$), varies. Since $b$ is not directly measurable, an experimental proxy, centrality, is used to characterise the amount of nuclear overlap in the collisions, as anticipated in Sec.~\ref{sec:theotools}.
Centrality is commonly expressed in percentiles of the total nucleus-nucleus cross section ($\sigma_{PbPb}~=~(7.67~\pm~0.16^{syst})$~b for \PbPb\ collisions at \sqrtSnn~=~5.02~TeV~\cite{ALICE-PUBLIC-2018-011}) and, conventionally, small (large) percentiles as for instance 0-5\% (80-90\%) correspond to central (peripheral) collisions~\cite{Abelev:2013qoq}. 

\begin{figure}[htb]
    \begin{center}
    \hbox{\hspace{2.6cm} \includegraphics[width = 0.7\linewidth]{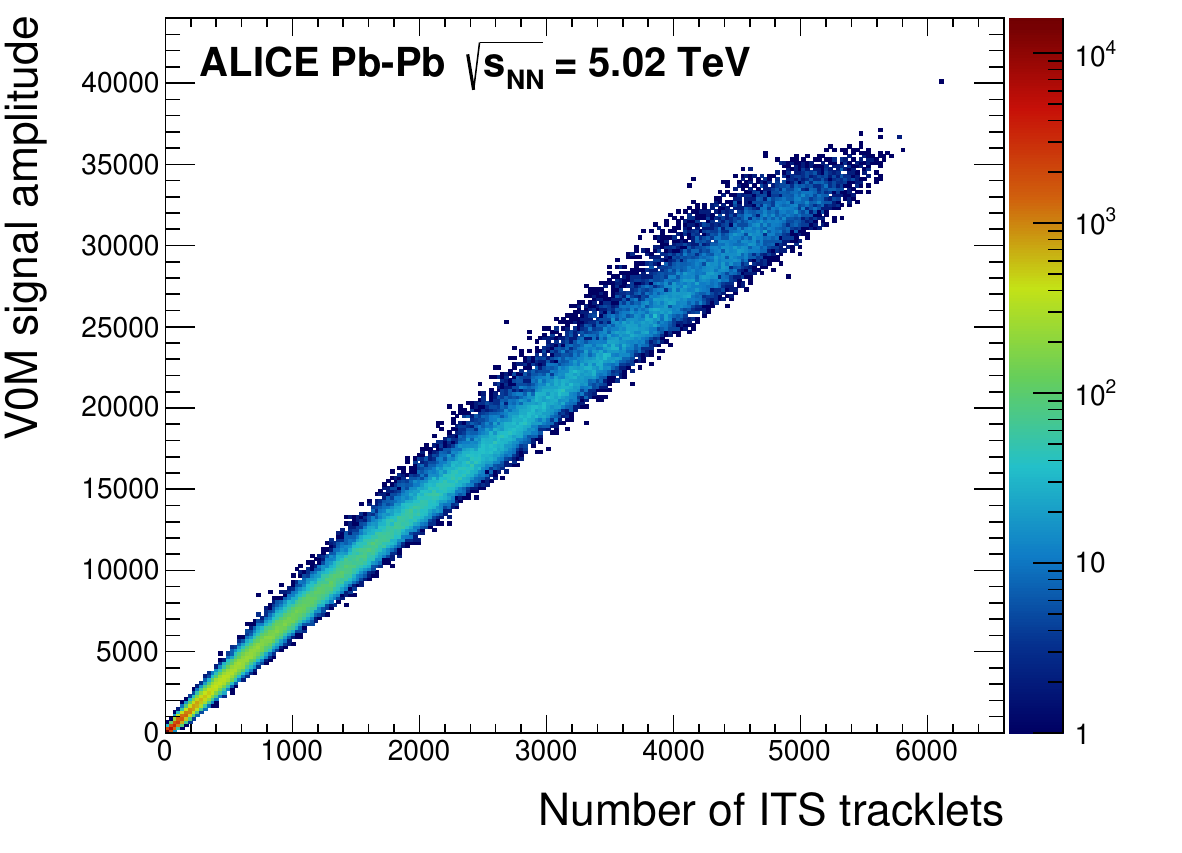}} %
    \includegraphics[width = 0.65\linewidth]{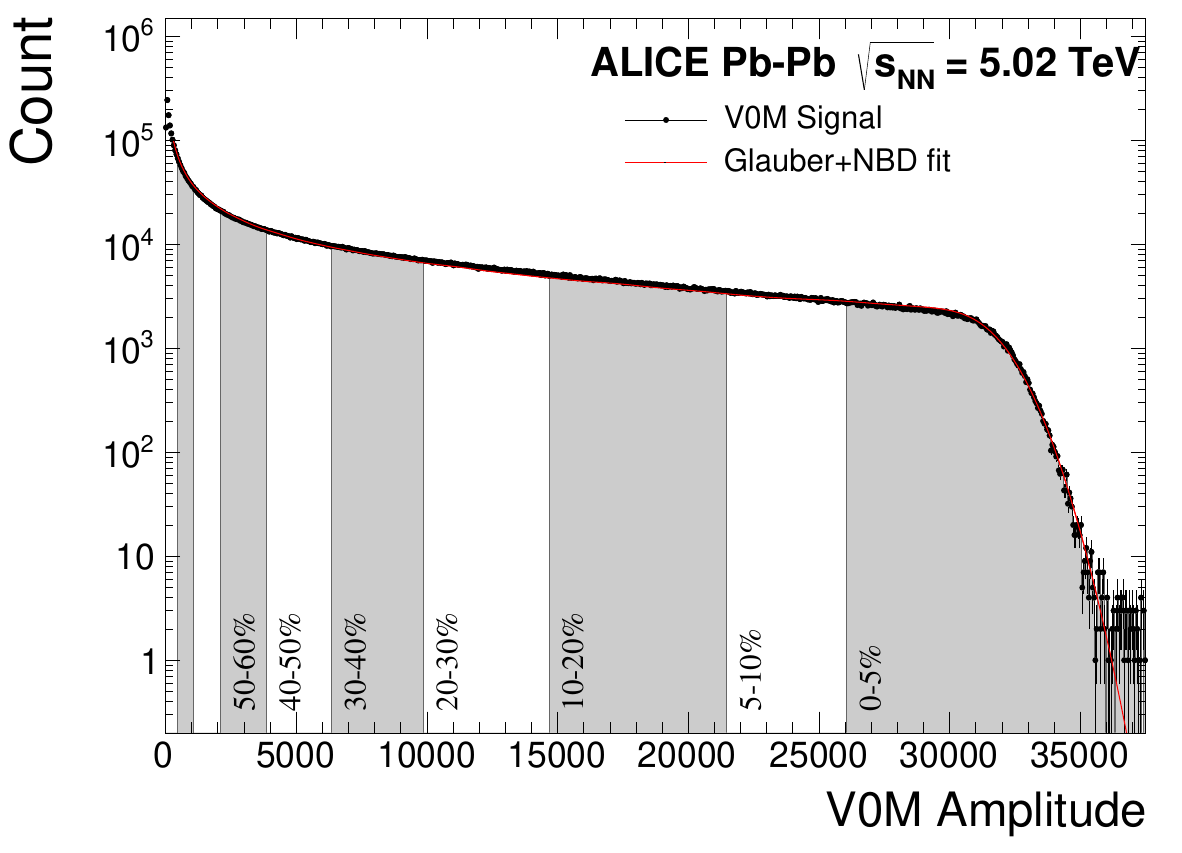}     %
    \end{center}
    \caption{Summed signal amplitude in the V0 scintillators in triggered \PbPb~collisions at \sqrtSnn~=~5.02~\TeV.
    (Top) Correlation between the V0M amplitude, at forward rapidity, and the number of charged-particle track segments (tracklets) reconstructed in the ITS at midrapidity ($\vert\eta\vert<1.4$). (Bottom) Fit (red line) to the V0M distribution with a Glauber model coupled to a negative binomial distribution (NBD). Gray and white bands indicate the classification of the events into centrality classes, with centrality expressed in percentile of the hadronic cross section.}
    \label{fig:centrality}
\end{figure}

In ALICE, centrality is determined by measuring the signal in the V0 scintillator arrays, which is proportional to the number of particles that strike them. 
The signal amplitude measured at forward rapidity in the V0 is observed to be strongly correlated with the multiplicity of charged particles measured at midrapidity. 
This is shown in the top panel of \Fig~\ref{fig:centrality}, which illustrates the correlation between the summed \VZEROA and \VZEROC (the sum is referred to as V0M) signal amplitudes, and the number of charged-particle track segments reconstructed using the ITS at midrapidity ($\vert\eta\vert<1.4$). 
The observed relation of proportionality between the particle multiplicity at midrapidity and the particle multiplicity at forward rapidity, which are two causally-distant observables, constitutes a purely data-driven justification to our centrality classification of events based on the V0 information. 
Alternatively, centrality can be measured using the particle multiplicity in the ITS or the energy deposited by the spectator nucleons in the ZDC, as discussed in detail in Ref.~\cite{Abelev:2013qoq}. The V0 estimator remains the most common choice for centrality classification for the purpose of studying particle production at midrapidity, due to the rapidity separation. 
As illustrated in the bottom panel of \Fig~\ref{fig:centrality} for a sample of 5.02~\TeV~\PbPb~collision events, the centrality classes are defined by slicing the V0M signal amplitude distribution.
The latter is fitted with a Glauber model~\cite{Loizides:2017ack} (see also Sec.~\ref{sec:theotools}) coupled to a model
for particle production based on a negative binomial distribution (NBD), to obtain \avNcoll~and \avNpart~for each centrality class, as detailed in~\cite{Abelev:2013qoq, ALICE-PUBLIC-2018-011}. The NBD is used to reproduce the amplitude for peripheral events where the multiplicity approaches the one observed in \pp\ collisions~\cite{Abelev:2013qoq}.
In peripheral nucleus--nucleus collisions, the selection of the centrality classes on the basis of charged-particle multiplicity, leads to a bias in the determination of the collision geometry parameters~\cite{Loizides:2017sqq}, which has to be taken into account when interpreting measurements that use these parameters, like the nuclear modification factor (see Sec.~\ref{sec:ElossHadrons}).
The event classification based on multiplicity in pp as well as \pPb\ collisions is discussed later, in Sec.~\ref{section:3.1}.

\subsubsection{Charged-particle multiplicity density at midrapidity}
\label{sec:TG1multiplicity}

In heavy-ion collisions, the charged-particle multiplicity per unit of (pseudo)rapidity (\dndeta) is studied as a function of centrality as well as in comparison to different collision systems and at various collision energies. This allows one to investigate the role of the initial energy density and the mechanisms responsible for particle production.

The primary charged-particle multiplicity density at midrapidity ($\vert\eta\vert < 2$) can be obtained from the number of short track segments formed using the position of the primary vertex and all possible combinations of hits between the two SPD layers (the two innermost layers of the ITS). The deposited energy signal in the FMD is used to measure the charged-particle pseudorapidity density in the forward regions ($-3.5 < \eta < -1.8$ and $1.8 < \eta < 5$). In both cases, the multiplicity is corrected for the background from secondary particles as well as for efficiency and detector acceptance. Details on the methods can be found in~\cite{Adam:2016ddh}, for instance.
The number of particles produced at midrapidity increases from peripheral to central collisions, from few tens of particles per unit of rapidity to few thousands. In central (0-5$\%$) \PbPb\ collisions at \sqrtSnn~= 5.02 TeV and in the range $\vert \eta \vert < 0.5$, \avdndeta = 1943 $\pm$ 54 and \avNpart = 385 $\pm$ 3, compared to \avdndeta = 17.52 $\pm$ 1.84 and \avNpart = 7.3 $\pm$ 0.1 in peripheral (80-90$\%$) collisions~\cite{Adam:2015ptt, Adam:2016ddh}.

\begin{figure}[htb]
    \begin{center}
    \includegraphics[width = 0.6\textwidth]{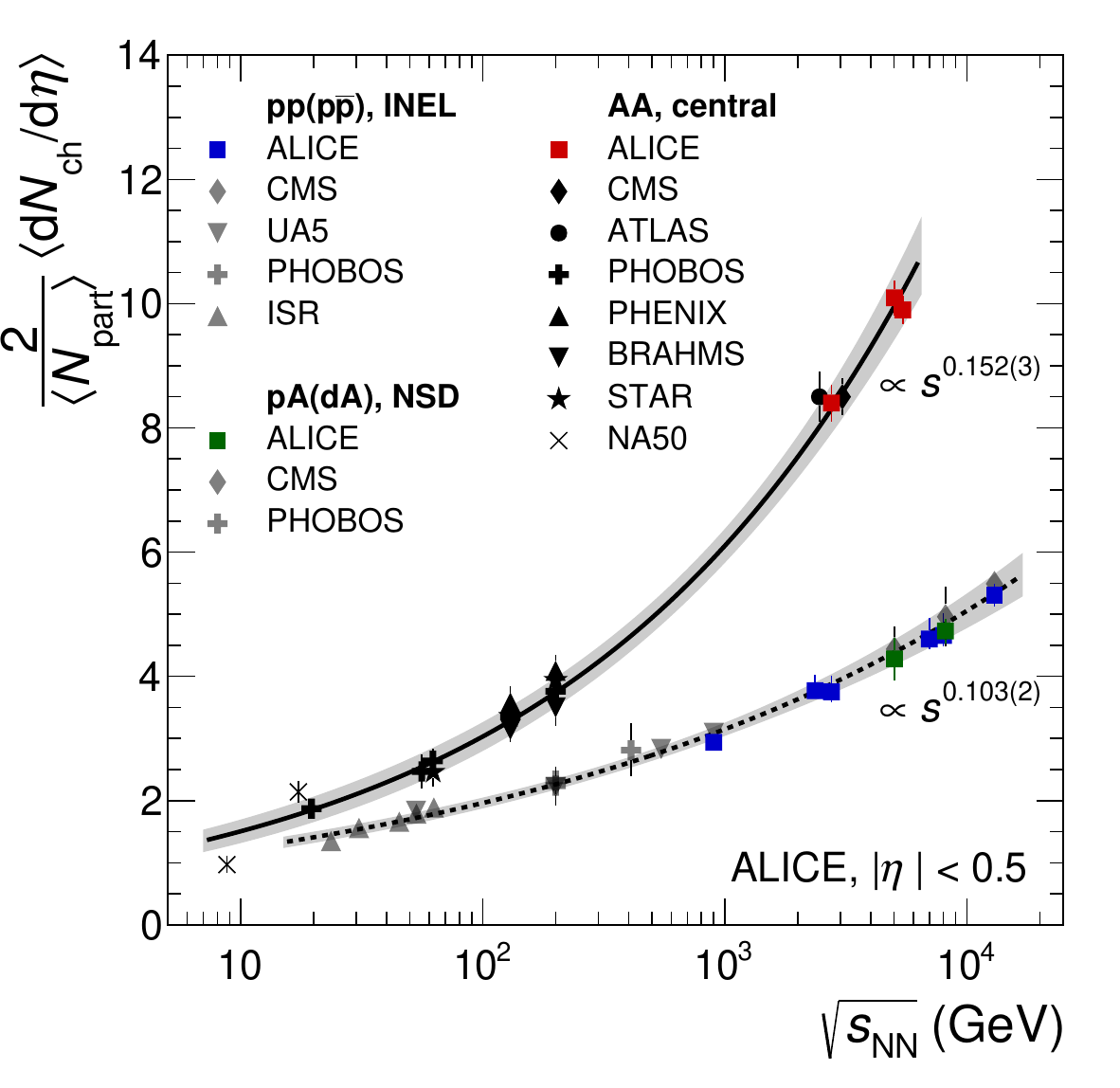}
    \end{center}
    \caption{Collision energy dependence of the charged-particle pseudorapidity density at midrapidity ($|\eta|<0.5$) normalised to the average number of participants, \dNdetape. Data from central \AA\ collisions~\cite{Acharya:2018hhy, Adam:2015ptt, Aamodt:2010cz, ATLAS:2011ag, Chatrchyan:2011pb, Abreu:2002fw, Bearden:2001xw, Bearden:2001qq, Adcox:2000sp, Alver:2010ck, Abelev:2008ab} are compared to measurements in non-single diffractive p(d)A collisions~\cite{Acharya:2018egz, Sirunyan:2017vpr, ALICE:2012xs, Back:2003hx} and inelastic (INEL) \pp\ and \ppbar~collisions~\cite{Adam:2016ddh, Khachatryan:2015jna, Adam:2015pza,Aachen-CERN-Heidelberg-Munich:1977izz,UA5:1982ygd,UA5:1986yef,UA5:1987rzq}. ALICE heavy-ion data include \PbPb\ collisions at \sqrtSnn~=~2.76 TeV and 5.02 TeV as well as \XeXe\ collisions at \sqrtSnn~=~5.44 TeV. All values of \avNpart\ used for the normalisation of the data are the results of Glauber model calculations. The lines are power law fits to the data and the bands represent the uncertainties on the extracted dependencies.}
    \label{fig:centralMulti}
\end{figure}

To compare the bulk particle production in different collision systems, the charged-particle multiplicity measured at midrapidity ($\vert \eta \vert < 0.5$) is scaled by the number of nucleon pairs participating in the collision, \avNpart$/2$.
\Figure~\ref{fig:centralMulti} shows the midrapidity charged-particle multiplicity normalised by the number of the participating nucleon pairs, \dNdetape, in \pp, \ppbar, p(d)A and in central heavy-ion collisions as a function of the centre-of-mass energy per nucleon pair, \sqrtSnn. In particular, ALICE central (0-5\%) heavy-ion data include  \PbPb\ collisions at \sqrtSnn~= 2.76 TeV and 5.02 TeV as well as \XeXe\ collisions at \sqrtSnn~=~5.44 TeV.
The dependence of \dNdetape\ on the centre-of-mass energy can be fitted with a power-law function of the form of $\alpha\times s^{\beta}$, resulting in an exponent $\beta = 0.152 \pm 0.003$ for central \AA\  collisions, under the assumption of uncorrelated uncertainties. It is a much stronger $s-$dependence than for inelastic (INEL) \pp\ and non-single-diffractive (NSD) p(d)A collisions, where a value of $\beta = 0.103 \pm 0.002$ is obtained. This indicates that heavy-ion collisions are more efficient in converting the initial beam energy into particle production at midrapidity than \pp\ or \pPb\ collisions.

\begin{figure}[htb]
    \begin{center}
    \includegraphics[width = 0.95\linewidth]{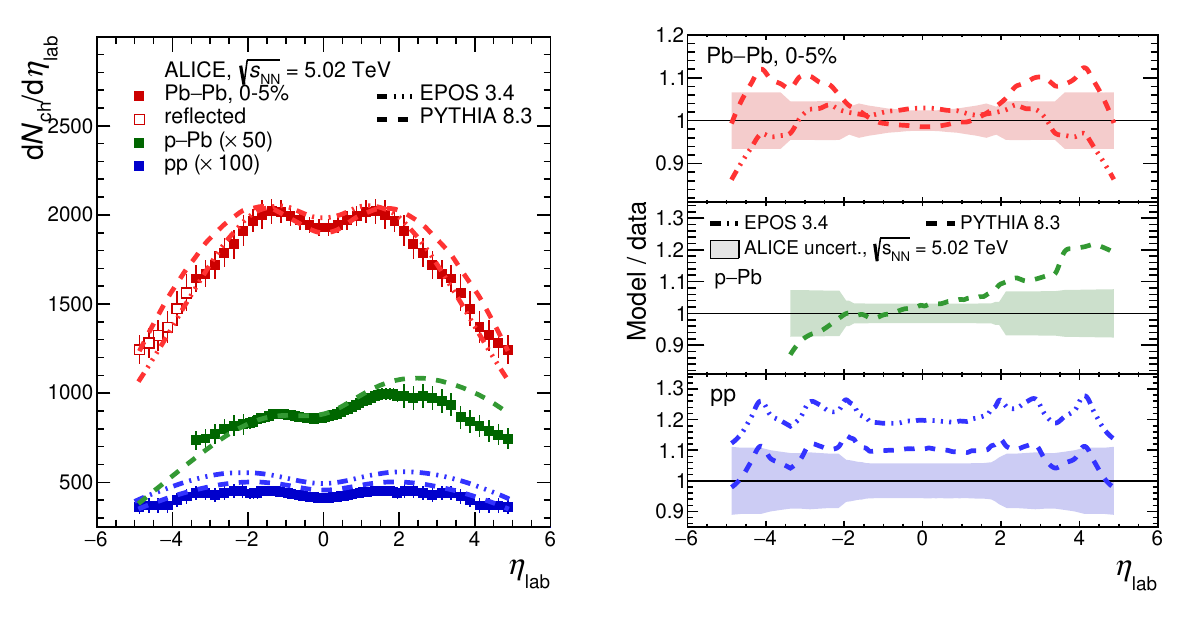}
    \end{center}
    \caption{(Left) Pseudorapidity density of charged particles in 0--5$\%$ central \PbPb\ collisions~\cite{Adam:2016ddh}, NSD \pPb\ collisions~\cite{ALICE:2012xs}, and INEL \pp\ collisions at \mbox{\sqrtSnn~= 5.02 TeV}~\cite{ALICE:2022kol} in ALICE. Dashed lines represent EPOS 3.4 model calculations~\cite{Drescher:2000ha, Werner:2010aa, Werner:2013tya}, available only for pp and \PbPb\ collisions at this energy and PYTHIA 8.3~\cite{Bierlich:2018xfw}, available for all collision systems.
    Error bars represent systematic uncertainties on data, statistical uncertainties are negligible. (Right) Ratio between EPOS 3.4 and PYTHIA 8.3 model calculations relative to \PbPb\ (top), \pPb\ (middle), and \pp\ (bottom) data. Shaded bands represent the relative systematic uncertainties on data.} 
    \label{fig:etaDistribution}
\end{figure}

Figure~\ref{fig:etaDistribution} presents the charged-particle multiplicity density \dndetalab\ as a function of pseudorapidity for central \PbPb\ collisions at \mbox{\sqrtSnn\ = 5.02 TeV}.  For symmetric collision systems, such as \PbPb\ and \pp, the laboratory reference frame coincides with the centre-of-mass one and the \dndetalab\ distributions are symmetric (hence hereafter indicated as \dndeta). When available, the data are extended into the non-measured region $-5<\eta<-3.5$ by reflecting the $3.5<\eta<5$ values around $\eta = 0$.
An asymmetry between the proton and the lead hemispheres is observed in the \pPb\ distribution, in which the number of charged particles is higher in the Pb-going side (positive $\eta_{\rm lab}$). 
The results are compared to the EPOS 3.4 model for \PbPb\ and \pp\ collisions and to PYTHIA 8.3 for all the three collision systems (see Sec.~\ref{sec:theotools} for an introduction to these models). For \PbPb~collisions, both models reproduce quantitatively  the \dndeta\ in the midrapidity region. The comparison at forward rapidities is not as good as at midrapidity. 
For \pPb\ collisions, PYTHIA 8.3 overpredicts the data in the Pb-fragmentation region, while it reproduces well the trend for midrapidity and for the p-fragmentation side. For pp collisions, EPOS 3.4 is overestimating the data by around 20\% in the full pseudorapidity range.

Particle production has been measured in small to medium-sized and large colliding systems such as \XeXe\ and \PbPb\ with unprecedentedly high precision. The \dndeta\ uncertainties range from around 3\% for central \AA\ collisions, at midrapidity, to around 10\% for peripheral results, in the forward region.

\begin{figure}[htb]
    \begin{center}
    \includegraphics[width = 0.6\linewidth]{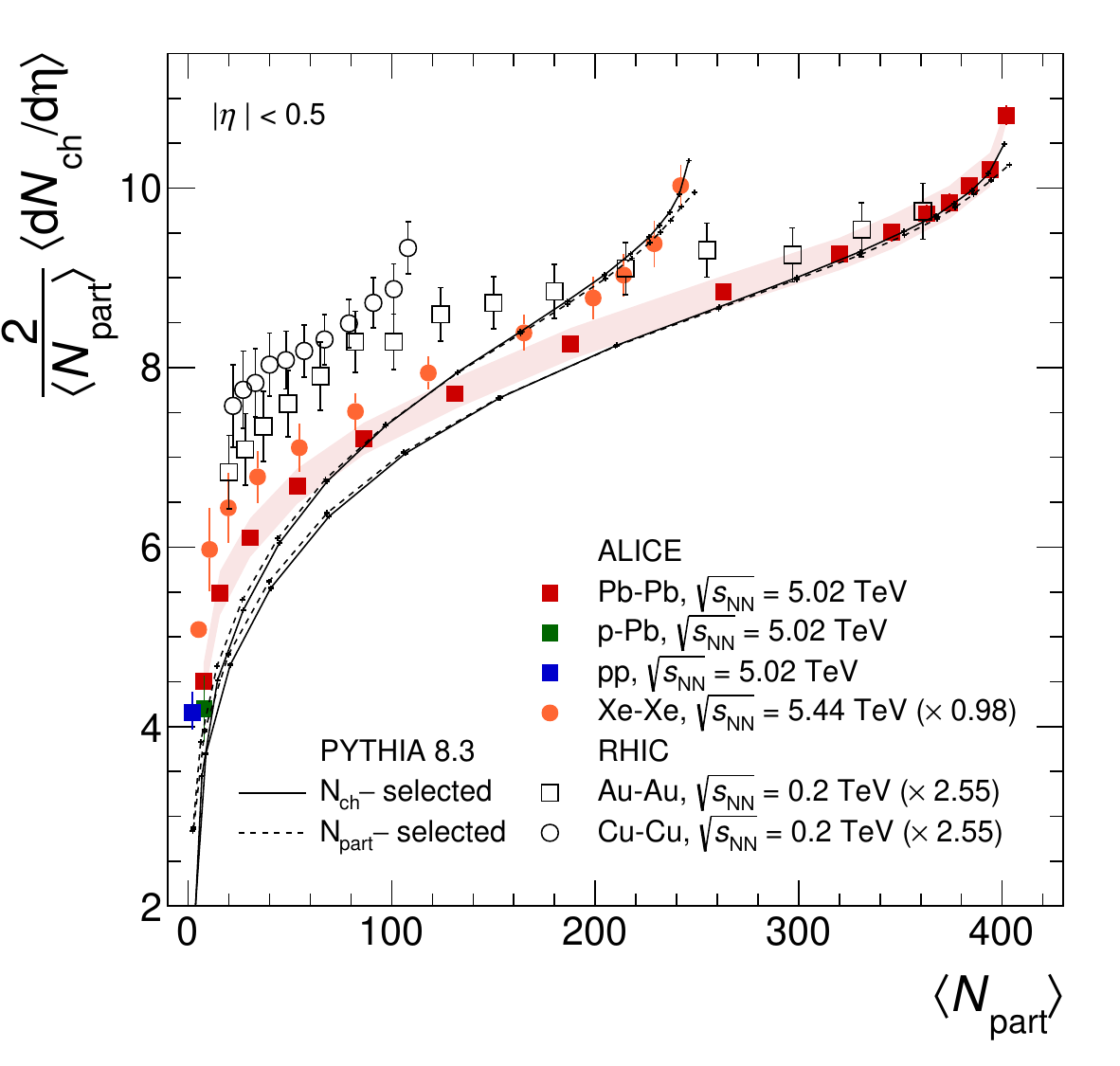}
    \end{center}
    \caption{\dNdetape\ in \PbPb\ collisions~\cite{Adam:2015ptt}, \pPb\ collisions~\cite{ALICE:2012xs} and \pp\ collisions~\cite{ALICE:2022kol} at \sqrtSnn~=~5.02 TeV, \XeXe\ collisions at \sqrtSnn~=~5.44 TeV~\cite{Acharya:2018hhy} scaled by a factor of 0.98, Au-Au and Cu-Cu collisions at \sqrtSnn~=~0.2 TeV~\cite{Alver:2010ck} scaled by 2.55 as a function of \avNpart. PYTHIA 8.3 calculations~\cite{Bierlich:2016smv} for $N_{\rm ch}-$selection are shown as black lines, dotted lines are $N_{\rm part}-$selected calculations.}
    \label{fig:npartscaling}
\end{figure}

Figure~\ref{fig:npartscaling} shows the centrality dependence of \dNdetape, with the centrality being expressed in terms of \avNpart. Data from 5.02 TeV \PbPb\ and 5.44 TeV \XeXe\ collisions are compared to lower energy results in \AuAu\ and \CuCu\ collisions by the PHOBOS experiment at RHIC. 

To disentangle energy dependence from \Npart\ dependence, the \AuAu, \CuCu\ and \XeXe\ data are scaled by factors calculated using the fit function of \Fig\ref{fig:centralMulti} for the top 5\% most central collisions.
The values of \dNdetape\ for \PbPb\ and \XeXe\ collisions decrease by a factor of two from the most central to the most peripheral collisions, where they agree with the values measured in minimum bias \pp\ and \pPb\ collisions. At the same \Npart , \PbPb\ undergoes less binary collisions than \XeXe. This causes the separation between the shapes of \XeXe\ and \PbPb\ for the \dNdetape\ as a function of \avNpart. 
Hints for a deviation between the lower energy \AuAu\ and \CuCu\ results from the LHC data for \avNpart\ $< 100$ are visible. This deviation is not significant given the large uncertainties and it could be attributed to the different Glauber model implementation used to estimate \Npart\ in ALICE~\cite{Adam:2015ptt} and PHOBOS~\cite{Alver:2010ck}. 
For the 5\% most central \XeXe\ and \PbPb\ collisions the \dNdetape\ increases steeply. A qualitatively similar feature is observed also in \CuCu\ collisions.

In \Fig ~\ref{fig:npartscaling} a comparison to PYTHIA 8.3~\cite{Bierlich:2016smv} calculations is also shown with black solid lines for the same selection in centrality as done in data ($N_{\rm ch}-$ selected). Dotted lines represent, instead, a selection performed on the $N_{\rm part}$. The calculations for the most central events exhibit a steep increase for the $N_{\rm ch}-$ selected results only, indicating that the sources of the rise are fluctuations of the charged-particle multiplicity at a fixed number of particle sources, $N_{\rm sources}$. Selection biases and auto-correlation effects are covered more in detail in Sec.~\ref{section:3.1}.
The $N_{\rm ch}/N_{\rm sources}$ fluctuation bias can be avoided by selecting centrality and calculating the $N_{\rm part}$ using the energy deposited by spectator nucleons in the Zero Degree Calorimeters. 
A methodology to derive these geometrical quantities from spectator detection near beam rapidity is under study in the Collaboration.
Measurements in light-nucleus collisions at the LHC Run~3 could allow us to constrain further the underlying particle production mechanisms, and to describe the increase with energy, see \Fig ~\ref{fig:centralMulti}, and centrality, \Fig ~\ref{fig:npartscaling}, bridging the gap between the trends observed in \pp\ and \pA\ collisions and those in the mid-sized \XeXe\ and the large \PbPb\ systems.

\subsubsection{Determination of the initial energy density}
\label{sec:TG1energydensity}

The energy density and temperature of the early collision stages are key physical quantities since they determine whether the critical conditions for the QCD phase transition are reached in the collision. In nuclear collisions, both can be, in principle, controlled by selecting events based on their centrality.
The energy density in the collision can be estimated from the total produced transverse momentum using the 'Bjorken-estimate'~\cite{Bjorken:1982qr}:

\begin{equation}
    \epsilon_\mathrm{Bj} (\tau) = \frac{1}{\ST\tau}\frac{\mathrm{d}\ET}{\mathrm{d}y}\quad,
    \label{eq:bjorkenest}
\end{equation}
where ${\mathrm{d}\ET}/{\mathrm{d}y}$ is the total produced transverse energy $\ET = \sqrt{\pt^2 + m^2}$ per unit of rapidity, and $\ST$ is the transverse size of the interaction region at proper time $\tau$. This estimate is valid for a free-streaming system that undergoes boost-invariant longitudinal expansion and no transverse expansion.

\begin{figure}[htbp!]
  \centering
  \includegraphics[width=.5\textwidth]{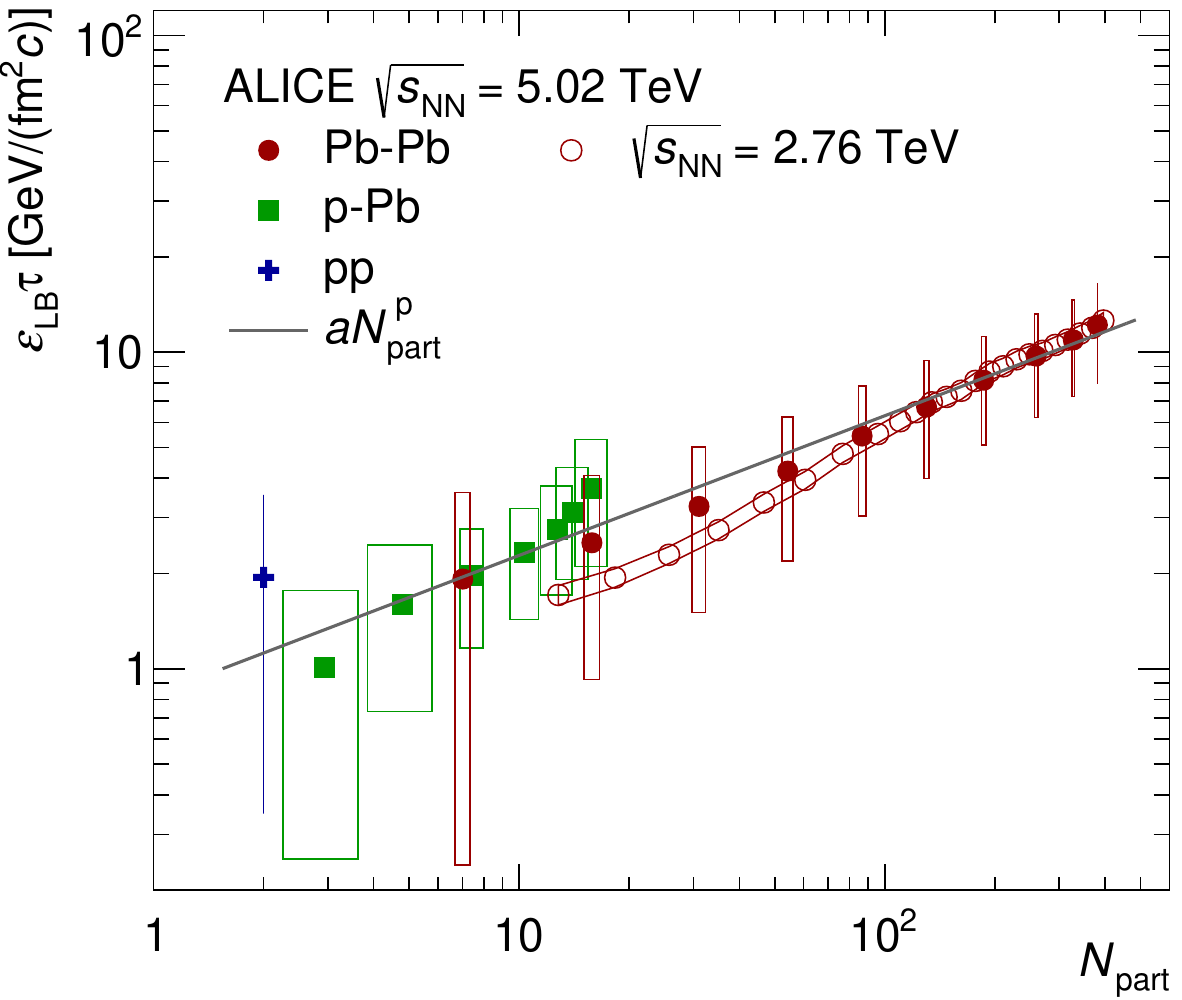}
  \caption[Lower bound estimate of the energy density in pp, \pPb, and \PbPb{} at $\sqrt{s_{\scriptscriptstyle\mathrm{NN}}}=5.02\,\TeV$]{Lower-bound estimate of the energy density times the formation time $\tau$ in pp, \pPb, and \PbPb\ collisions at $\sqrtSnn~=~5.02$ \TeV as a function of the number of participating nucleons~\cite{ALICE:2022imr}. The transverse area is calculated as the total area of overlap between participating nucleons using Eq.~\ref{eq:eps_lb}.  The open circles are from \PbPb{} collisions at $\sqrtSnn~=~2.76$~\TeV and calculated via direct measurements of $\ET$~\cite{Adam:2016thv}. A fit to a power law $a\Npart^p$ is shown with best-fit parameter values $a=0.8\pm0.2$ GeV/(fm$^2$$c$) and $p=0.44\pm0.07$, and $\chi^2/dof~=~1.2/16$.}
  \label{fig:all_ebj_2760}
\end{figure}

From the measurements of the charged-particle pseudorapidity density and assuming that the rapidity of the charged particles produced in the collisions is normally distributed~\cite{Adam:2016ddh,Abbas:2013bpa}, we extract a lower-bound estimate of the energy density $\epsilon_{\mathrm{LB}}$ times the formation time $\tau$ in the collisions~\cite{ALICE:2022imr}
\begin{equation}
    \label{eq:eps_lb}
    \epsilon_{\mathrm{LB}}\tau 
    = \frac{1}{\ST}\frac{1}{f_{\mathrm{total}}}\sqrt{1+a^2}\langle m\rangle \frac{\mathrm{d}N_{\mathrm{ch}}}{\mathrm{d}y}\quad,
\end{equation}
where $\sqrt{1+a^2}\langle m\rangle$ is the effective transverse mass, and $f_{\mathrm{total}}=0.55\pm0.01$ is the fraction of the charged out of all particles~\cite{Adam:2016thv}. Here, $a$ is the effective $\pt/m$ ratio extracted as part of the estimate. The transverse area $\ST$ is calculated using a Glauber model~\cite{Loizides:2014vua} in which the full area of participating nucleons is taken into account. The resulting $\epsilon_{\mathrm{LB}}\tau$ is shown in \Fig~\ref{fig:all_ebj_2760} for pp, \pPb, and \PbPb\ collisions at $\sqrtSnn~=~5.02$ \TeV.  
We see an increase of roughly an order of magnitude from \pp\ and most peripheral \pPb\ collisions to the most central \PbPb\ collisions. A power-law ($a\Npart^p$) fit to the data indicates a non-trivial increase in the energy density with an increasing transverse area of the initial overlap between the colliding nuclei.

\begin{figure}[htbp!]
    \begin{center}
    \includegraphics[width = 0.5\textwidth]{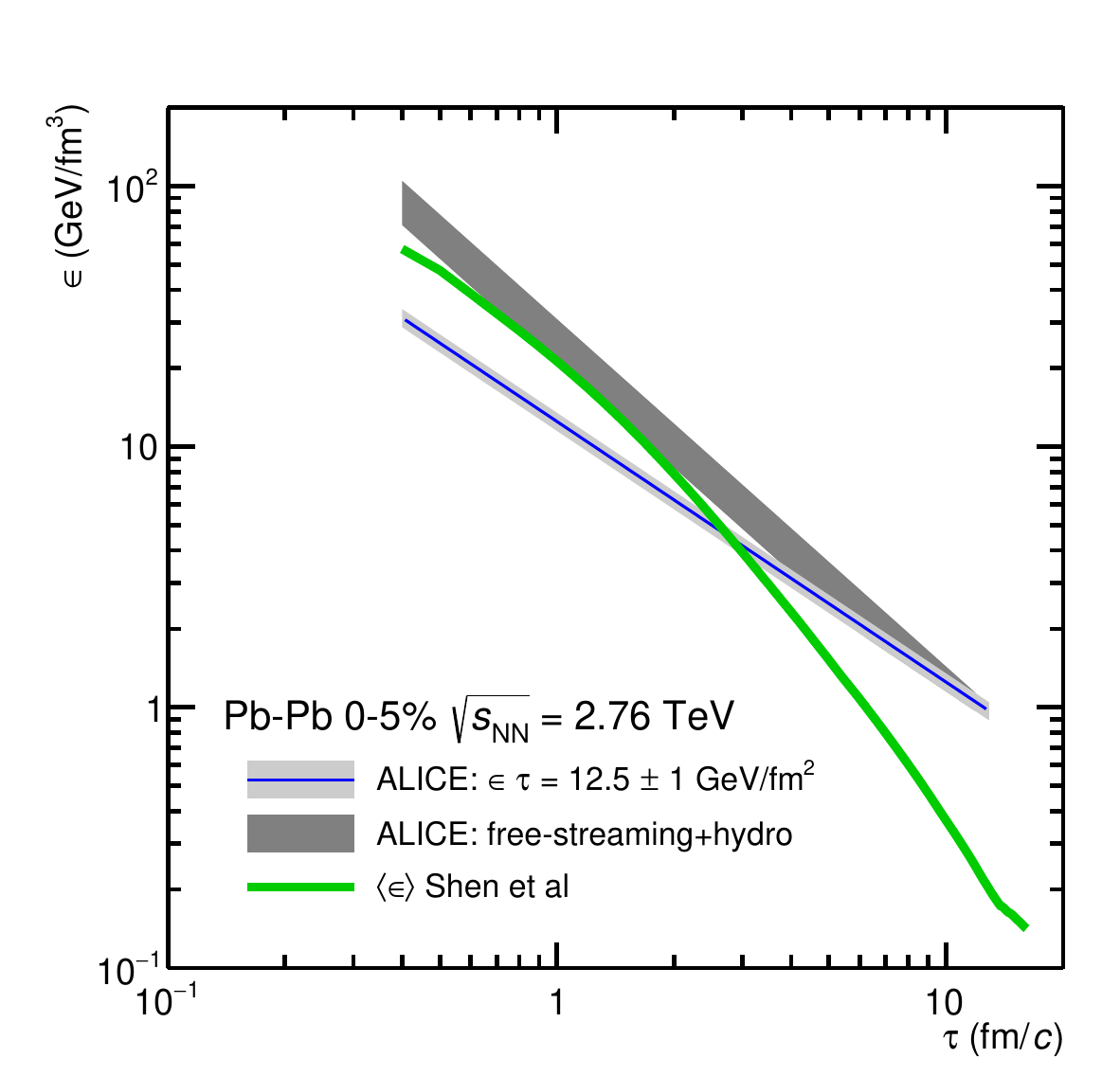}
    \end{center}
    \caption{Evolution of the average energy density as a function of the proper time of the system in hydrodynamic calculations. The green curve represents the average energy density profile from hydrodynamic model calculation from~\cite{Gale:2020dum, Gale:2020xlg}. The blue line the grey bands represent the energy density estimated from the total transverse energy, $E_{\mathrm{T}}$, measured in 0-5$\%$ central \PbPb\ collisions at $\sqrtSnn~= 2.76$~TeV~\cite{Adam:2016thv} using Eq.~\ref{eq:bjorkenest} for the free-streaming expansion and Eq.~\ref{eq:hydro_eps_evol} for hydrodynamical expansion. The hydrodynamical expansion is matched to the free-streaming behaviour at $\tau = 6$ and $12 \;\mathrm{fm}/c$.}
    \label{fig:energyDensity}
\end{figure}

In a system that follows hydrodynamic expansion, the expanding fluid performs work and the resulting time-dependence of the energy density in a system that undergoes boost-invariant longitudinal expansion (and no transverse expansion) follows a power law with $\tau^{-4/3}$ instead of $\tau^{-1}$,
\begin{equation}
    \epsilon_\mathrm{Bj,hydro}(\tau) = \epsilon_0 \left( \frac{\tau_0}{\tau} \right)^{4/3}\quad,
    \label{eq:hydro_eps_evol}
\end{equation}
where $\epsilon_0$ and $\tau_0$ are the initial energy density and the formation time, respectively.
These simple data-driven estimates are compared with the energy density from a hydrodynamical model calculation (IPGlasma + MUSIC + URQMD with the s95p-PCE equation of state~\cite{Shen:2010uy}) for hydrodynamical evolution and the UrQMD model for final state scattering) in \Fig~\ref{fig:energyDensity}. The blue line and light gray band indicate the energy density estimated using the free-streaming estimate from Eq.~\ref{eq:bjorkenest} while the dark gray band indicates the behaviour of a longitudinally expanding hydrodynamical system. Since then the system freezes out gradually, there is no single time at which the expansion switches from hydrodynamical expansion to free streaming. The band was constructed by switching at $\tau=6$ and 12 fm/$c$, which are the times at which approximately 10$\%$ and 90$\%$ of the system is frozen out according to the hydrodynamical calculation. The uncertainty from the measured $\ET$ has been added linearly. It can be seen that this simple data-driven estimate is in reasonable agreement with the full model calculation up to 2--3~fm/$c$.

\subsubsection{Temperature of the system}
\label{sec:TG1directphotons}
Temperature is a key thermodynamic property that characterises the fireball at any given time during its evolution. 
To experimentally access the temperature of the early partonic phase one can rely on sensitive probes that are produced at the early stage of the collision, such as heavy flavour \qqbar\ states (quarkonia) and electromagnetic radiation.
The study of quarkonium as a thermometer for the QGP was developed starting from the early idea~\cite{Matsui:1986dk} that the strong potential binding q and $\mathrm{\overline{q}}$ into pairs is screened by colour charges when immersed in the dense and hot coloured medium, ultimately leading to a ``melting’’ of the \qqbar\ states and offering the possibility to connect the production or suppression of quarkonia to the QGP temperature. 
The most recent developments and the ALICE results on quarkonia are discussed more in details in the Sec.~\ref{sec:Quarkonium}.
Electromagnetic radiation is emitted by a variety of sources during the evolution of the system in the form of real ``direct’’ photons or virtual photons, measurable via their internal conversion into lepton-antilepton pairs. 
The invariant mass spectrum of dileptons is studied looking for any enhancement at intermediate ($\approx$ 1-2 \GeV for dielectrons) masses relative to expectations from hadronic decays in vacuum. Such an excess could be related to thermal emission from the QGP and thus provide information on its temperature. However, as dilepton measurements at the LHC are not yet sensitive to possible thermal signals~\cite{Acharya:2018nxm}, a precise measurement of the low-mass dielectron continuum will be pursued by ALICE during the LHC Runs 3 and 4. In the rest of this section, we focus on results obtained by measuring real direct photons.

Temperature can be accessed experimentally by measuring the yield of thermal photons that are emitted by the hot plasma during its entire evolution.
As the mean free path of a photon in hot matter is much larger than the typical sizes of the created fireball, photons escape the collision zone unaffected, delivering information on the QGP conditions at early times, as well as on the development of collective flow during the evolution of the hot matter. 
Direct photons are the photons that do not originate from parton fragmentation nor hadronic decays, and are produced in electromagnetic interactions during different stages of the collision. The low-\pt~region of the spectrum (\pt~$\lesssim$ 3 \GeVc) is dominated by ``thermal'' photons and follows an approximately exponential behaviour, $d^{2}N_{\gamma_{\rm dir}}/(\pt\mathrm{d}\pt\mathrm{d}y) \propto e^{-\pt/\Teff}$, characterised by the inverse logarithmic slope 
\Teff. The latter represents the effective temperature of the fireball and can be related to the ``true'' temperature by accounting for the radial expansion of the system, which causes a blue-shift of the emitted photons~\cite{vanHees:2011vb}. 
At higher momentum (\pt~$\gtrsim$~5~\GeVc), ``prompt'' photons dominate and follow a power law spectrum. These photons are produced in the initial hard scatterings between the colliding nuclei. Other direct photon production mechanisms, like the interaction of hard scattered partons with the medium (``jet-photon conversion'')~\cite{Zakharov:2004bi}, may be important for \pt $\lesssim$ 10 \GeVc~\cite{Fries:2002kt,Turbide:2007mi}. These contributions to the spectrum need to be disentangled, such that the measured excess of thermal photons can be related to the effective temperature of the system.

\begin{figure}[htb]
    \begin{center}
     \includegraphics[width = 1.\textwidth]{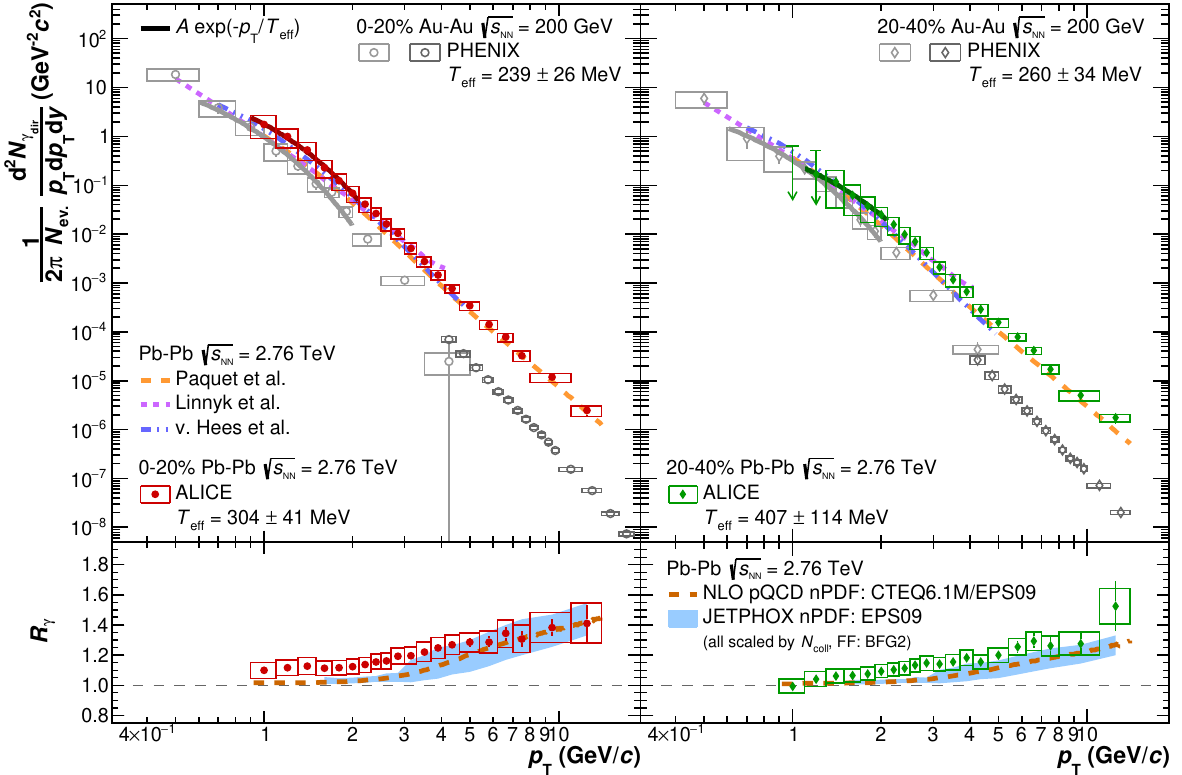}
    \end{center}
    \caption{Direct photon spectra (top) and direct photon excess \Rgamma (bottom) measured in \PbPb\ collisions at \sqrtSnn~=~2.76~TeV~\cite{Adam:2015lda} and \sqrtSnn~=~0.2~TeV~\cite{Adler:2005,Adare:2008ab} in the 0--20\% (left) and 20--40$\%$ (right) centrality classes. 
    The slope of the exponential function fitted to the ALICE data was determined without subtracting any pQCD contribution. The slope of the spectrum measured by PHENIX was determined after subtracting a pQCD contribution determined by parameterising a direct photon measurement in pp collisions.}
    \label{fig:direct-photon-spectra}
\end{figure}

The invariant yield of direct photons is determined by first measuring the direct photon excess over decay photons, $\Rgamma$, which is defined by
\begin{equation}
\Rgamma \equiv \frac{\gammainc}{\npiparam} \Big/ \frac{\gammadec}{\npiparam} = \frac{\gammainc}{\gammadec},
\end{equation}
where $\gammainc$ is the inclusive photon invariant yield, $\gammadec$ the decay photon invariant yield, and $\npiparam$ is the parametrised neutral pion invariant yield. Here the neutral pion invariant yield has been measured in the $\pi^{0}\rightarrow \gamma\gamma$ channel. The result is reported in \Fig~\ref{fig:direct-photon-spectra} (bottom) for central and semi-central \PbPb\ collisions at \sqrtSnn~=~2.76 TeV. At high-momenta, the \Rgamma is consistent with prompt photon production (pQCD) and JETPHOX calculations ~\cite{Paquet:2015lta,Aurenche_2006}. For 0--20\% collision centrality and considering all data points in $0.9~<~\pt~<~2.1~$\GeVc, the significance of the direct photon excess is about $2.6\sigma$, which indicates that there is an excess of thermal photons coming from the QGP.
From $\gammadir = (1-1/\Rgamma) \gammainc$, the invariant yield of direct photons is calculated for central and semi-central \PbPb\ collisions at \sqrtSnn~=~2.76 TeV, and is reported in \Fig~\ref{fig:direct-photon-spectra} (top) and compared to measurements in central and semi-central \AuAu\ collisions at \sqrtSnn~= 200 GeV. Harder photon spectra are measured at the LHC than those at RHIC. 
Hydrodynamical model calculations for \PbPb\ collisions~\cite{Gale:2014dfa,Linnyk:2015tha} are consistent with the data (\Fig~\ref{fig:direct-photon-spectra}). The region that is dominated by thermal direct photons is fitted by an exponential function (see solid lines in \Fig~\ref{fig:direct-photon-spectra}) to obtain the effective temperature ($T_{\mathrm{eff}}$). For central and semi-central \PbPb\ collisions, the fits lead to \Teff~=~(304~$\pm$~41)~MeV and \Teff~=~(407~$\pm$~114)~MeV, respectively, and the extracted temperatures are consistent within uncertainties. 
The large relative uncertainty on the temperature in semi-central collisions is due to the large uncertainties on the fitted photon spectrum at low-\pt.
Comparing these values to the slopes of the spectra measured by PHENIX and reported in \Fig~\ref{fig:direct-photon-spectra}, one can see an increase in the effective temperature from RHIC to the LHC.
However, obtaining the initial temperature of the fireball is only possible indirectly by invoking model calculations that incorporate the evolution of the QGP medium as well as radial flow effects that blue-shift the direct photon spectra, and has not yet been attempted.

\begin{figure}[htb]
    \begin{center}
    \includegraphics[width =.95\textwidth]{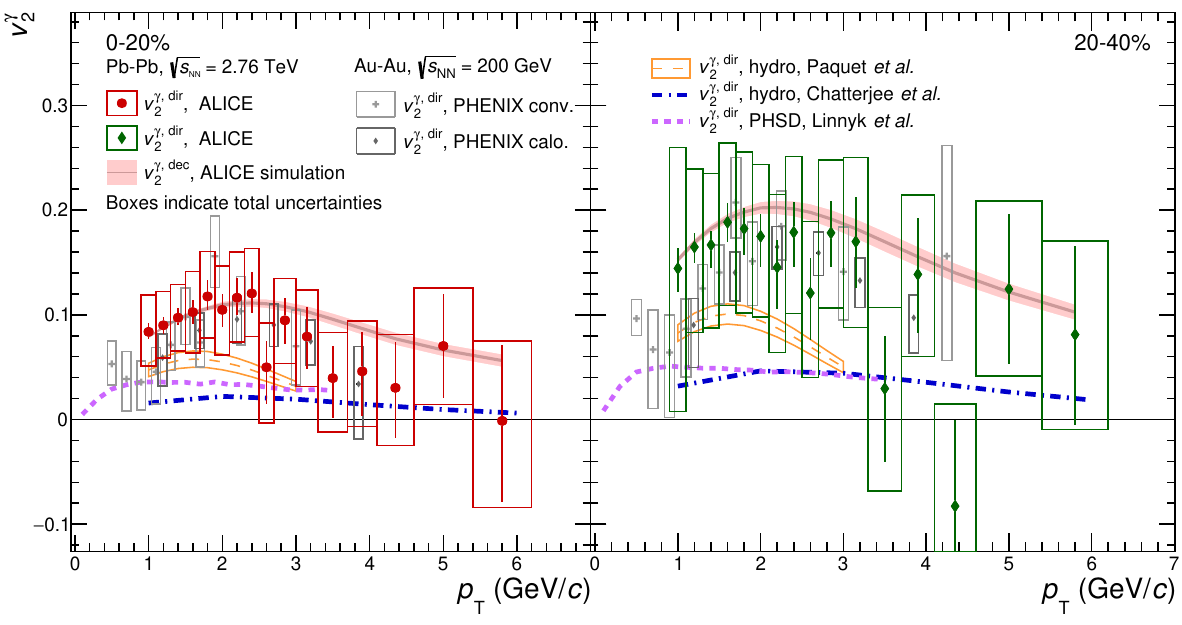}
    \end{center}
    \caption{Elliptic flow of direct photons, \vtwogammadir, in the 0--20\% (left) and 20--40\% (right) centrality classes for \PbPb\ collisions at \sqrtSnn~=~2.76~TeV~\cite{Acharya:2018bdy}. The vertical bars and the boxes indicate the statistical and the total uncertainties on the data, respectively. ALICE data are compared to model calculations~\cite{Chatterjee:2017akg,Gale:2014dfa,Linnyk:2015tha} and PHENIX data~\cite{Adare:2011zr} in the same respective centrality classes.}
    \label{fig:direct_photon_v2}
\end{figure}
As direct photons escape the medium unaffected and are predominantly produced when the system is hot, it is expected that they don't have a large flow component. Measuring otherwise poses large questions to our understanding of the space-time dynamics of the QGP.
The elliptic flow of direct photons (\vtwogammadir) was calculated by combining the elliptic flow of inclusive photons together with that of decay photons (\vtwogammadec) and the measured \Rgamma using the scalar product method (see~\cite{Acharya:2018bdy} for details). The measured \vtwogammadir, reported in \Fig~\ref{fig:direct_photon_v2}, appears to be close to the decay photon flow for both centrality classes and similar to the measurements by the PHENIX collaboration at RHIC~\cite{Adare:2011zr}. 
Hydrodynamic and transport models~\cite{Chatterjee:2017akg,Gale:2014dfa,Linnyk:2015tha} that include thermal photon production related to the local temperature of the fluid show a smaller \vtwogammadir\ than observed. 
For models to increase \vtwogammadir\, one can lower the initial temperature of the fluid, such that more photons are emitted when the system is more developed, but in turn would lead to a larger discrepancy in the prediction of the direct photon production. Especially at RHIC energies, theoretical predictions are not yet able to describe the direct photon production and the flow simultaneously. This is what is referred to as the ``direct photon puzzle''~\cite{Adam:2015lda,Acharya:2018bdy}. Due to the size of the experimental  uncertainties in \Fig ~\ref{fig:direct_photon_v2}, it is not yet confirmed if this puzzle is present at LHC energies, although the results are consistent with the measurements at RHIC. In Run 3, statistical uncertainties are expected to be reduced by a factor of ten, and the systematic uncertainties by about a factor of two. If the \Rgamma\ measured with the new datasets will more strongly reject the null hypothesis, i.e.\,$\Rgamma=1$, it will be very probable to establish if there is a direct photon puzzle at the LHC.

In summary, the results of the fits to the direct photon spectra indicate that the effective temperature of the fireball is larger than the critical temperature and increases from RHIC to LHC energies.

\subsubsection{Size and lifetime of the system}
\label{sec:TG1sizelifetime}

The size and lifetime of the system created in a collision can be inferred from femtoscopy, which measures particle momentum correlations at kinetic freeze-out.
The study of momentum correlations of particles emitted from a common source was historically referred to as ``HBT interferometry'' in the heavy-ion community, named after the original technique proposed by Hanbury-Brown and Twiss in the 1950s and 1960s to determine the size of laboratory and stellar sources by studying the interference of emitted photons~\cite{HanburyBrown:1954amm, HanburyBrown:1956bqd}.

In the case of two-particle femtoscopy~\cite{Abelev:2006gu,Lisa:2005dd,Humanic:2005ye}, if two particles are emitted from a pp or heavy-ion collision and detected, the two-particle count rate is used to form the momentum correlation function, $C(k^*)$, given by 
$C(k^*)=\EuScript{N}\frac{A(k^*)}{B(k^*)}$,
where $A(k^*)$ is the measured distribution of particle pairs from the same event,
i.e.\,the two-particle count rate, $B(k^*)$ is the reference distribution of pairs created from particles coming from different events (referred to as ``mixed events''), $\EuScript{N}$ is the normalisation factor, and $k^*$ is the magnitude of the momentum of each of the particles in their pair rest frame. Note that for identical particle pairs the invariant momentum difference is denoted as $q_{\rm inv} = 2k^*$.

In order to extract the explicit spatial information and implicit time information about the emitting 
particle source at kinetic freeze-out, the measured two-particle correlation function is, in general, fitted with a formula that includes a quantum statistics term for identical particles, and a parameterisation which incorporates strong final-state interactions between the particles (FSI), for cases where they are important~\cite{Lednicky:1981su,Lednicky:2005af}.
For example, %
for uncharged particles:

\begin{equation}
C(k^*)=1+\lambda e^{-4k^{*2}R^2}+\lambda\alpha\left[\frac{1}{2}\left|\frac{f(k^*)}{R}\right|^2+\frac{2\mathcal{R}f(k^*)}{\sqrt{\pi}R}F_1(2k^* R)-\frac{\mathcal{I}f(k^*)}{R}F_2(2k^* R)+\Delta C\right],
\label{fit1}
\end{equation}
where $f(k^*)$ is the $s$-wave scattering amplitude, %
$\alpha=0.5$ in the case of identical bosons, $R$ is the source radius parameter assuming a
spherical Gaussian source distribution, and $\lambda$ is the correlation strength. 
The term $\Delta C$ is a calculated correction factor that takes into account the deviation of the spherical
wave assumption used in the inner region of the short-range potential in the derivation of Eq.~\ref{fit1}~\cite{Abelev:2006gu}.
The second term in Eq.~\ref{fit1} describes the quantum statistics if identical bosons are considered, and the third one is the FSI term. %
In \eq~\ref{fit1}, the $R$ parameter is related to the effective size of the source. %
In order to improve the description of the particle emitting source, two-particle femtoscopy can be reformulated in three-dimensions. 
The experimental correlation function is then 
obtained in terms of the components of the pair momentum difference vector: %
$q_{\rm out}$ (along the direction of the sum of the transverse momenta of the particles) and $q_{\rm side}$ (perpendicular to the direction of the sum of the transverse momenta of the particles), and parallel to the beam direction, $q_{\rm long}$. 

Femtoscopy measures the volume of the emitting source, which is in general not equivalent to the total volume occupied by the system at freeze-out. For an expanding source, with strong flow gradients, particles with similar momenta are emitted from a region referred to as the homogeneity volume, which is smaller than the total volume~\cite{Akkelin:1995gh}.

\begin{figure}[htb]
    \begin{center}
    \includegraphics[width = 0.7\textwidth]{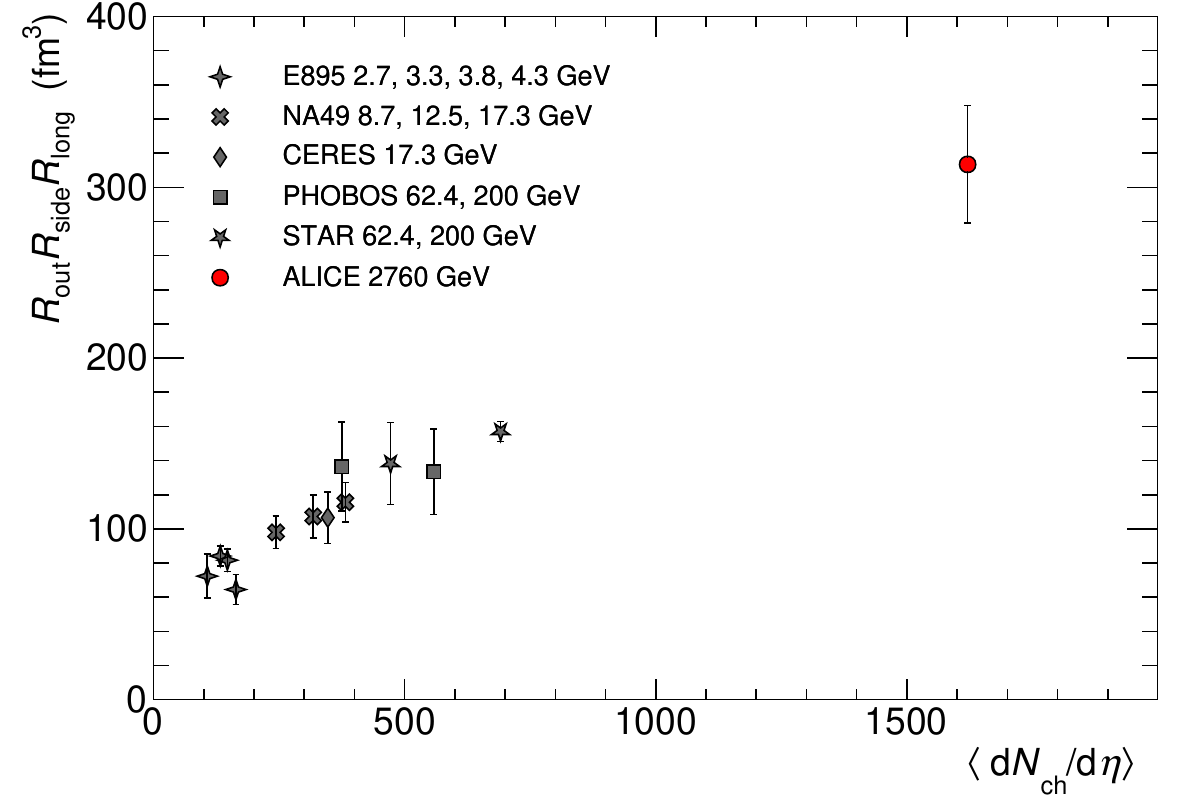}
    \includegraphics[width = 0.7\textwidth]{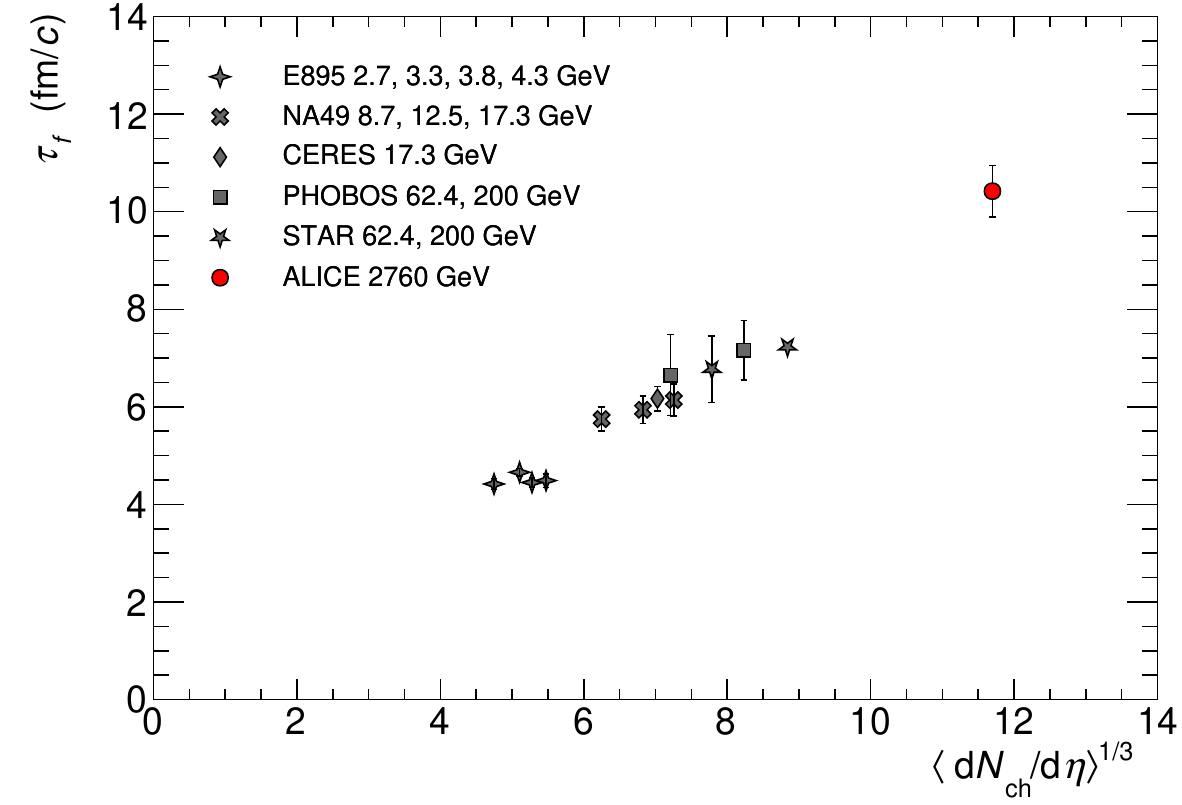}
    \end{center}
    \caption{Homogeneity volume (top) and decoupling time $\tau_f$ (bottom) measured at \sqrtSnn~=~2.76~TeV~\cite{Aamodt:2011mr,Adam:2015vna} compared to those obtained for central \AuAu\ and \PbPb\ collisions at lower energies at the AGS~\cite{Lisa:2000hw}, SPS~\cite{Alt:2007uj,Afanasiev:2002mx,Adamova:2002wi}, and RHIC~\cite{Abelev:2009tp,Back:2004ug,Back:2005hs,Back:2002wb,Abelev:2008ab}. The homogeneity region is determined as the product of the three pion femtoscopic radii at $\langle\kT\rangle~=~0.3$~\GeVc for 0--5\% central events, whereas the decoupling time $\tau_f$ is extracted from \Rlong(\kT) according to \eq~\ref{eq:rlong_mt}.}\label{fig:volume_and_time}
\end{figure}

The charged-particle pseudorapidity density dependence of the femtoscopic pion radii and the decoupling time $\tau_f$ of pions for central \PbPb\ and \AuAu\ collisions were measured at AGS~\cite{Lisa:2000hw}, SPS~\cite{Alt:2007uj,Afanasiev:2002mx,Adamova:2002wi}, RHIC~\cite{Abelev:2009tp,Back:2004ug,Back:2005hs,Back:2002wb,Abelev:2008ab} and LHC~\cite{Aamodt:2011mr,Adam:2015vna} energies. The decoupling time of the system is typically approximated with the decoupling time of pions $\tau_f$, since pions are the most abundant species ($\approx$ 80$\%$) constituting the bulk of the system.
The size of the homogeneity region, obtained as the product of the three pion radii $R_{\rm out}$, $R_{\rm side}$, and $R_{\rm long}$ measured at $\langle\kT\rangle~=~0.3$~\GeVc, has a linear dependence on the charged-particle pseudorapidity density shown in the top panel of \Fig ~\ref{fig:volume_and_time}. 
The magnitude increases three times from AGS energies to the LHC.

According to the implementation of the system evolution in hydrodynamic models, the $R_{\rm long}$ at midrapidity is proportional to the total duration of the longitudinal expansion~\cite{Herrmann:1994rr}. 
The decoupling time $\tau_f$ can be obtained from the relation between $R_{\rm long}$ and the transverse mass (\mT) that is
\begin{equation}
    R_{\rm long}^2(m_{\rm T}) = 2 \frac{\tau_f^2 \Tkin}{m_{\rm T}}\frac{K_2(m_{\rm T}/T)}{K_1(m_{\rm T}/\Tkin)},\label{eq:rlong_mt}
\end{equation}
where \Tkin\ is the kinetic freeze-out temperature, taken to be 0.12 GeV, and $K_1$ and $K_2$ are the integer order modified Bessel functions~\cite{Herrmann:1994rr}.
The decoupling time obtained from the fit to the \mT\ distribution of the pion $R_{\rm long}$ ~\cite{Aamodt:2011mr} is shown in the bottom panel of \Fig~\ref{fig:volume_and_time}, together with world data.
The $\tau_f$ increases linearly with the cube root of the charged-particle pseudorapidity density. 
The time measured at AGS is about 4--5 \fmC\ increasing gradually up to 7--8 \fmC\ at top RHIC energies, and finally it reaches 10–-11 \fmC\ in central \PbPb\ collisions at \sqrtSnn~=~2.76 TeV. 
It should be noted that corrections to \eq~\ref{eq:rlong_mt}, due to the transverse expansion and the finite pion chemical potential can increase the decoupling time by up to 25\%~\cite{Sinyukov:1996ww}. 
An additional source of uncertainty originates from the chosen value of the kinetic freeze-out temperature. 
Varying it by 0.02 GeV in both directions leads to an increase of $\tau_f$ by 13\% or a decrease by 10\%, respectively, without changing the overall observed scaling behaviour. Taking into account these uncertainties leads to the $\tau_f$ range 10--13~fm/$c$.

\begin{figure}[htb]
    \begin{center}
    \includegraphics[width = 1.0\textwidth]{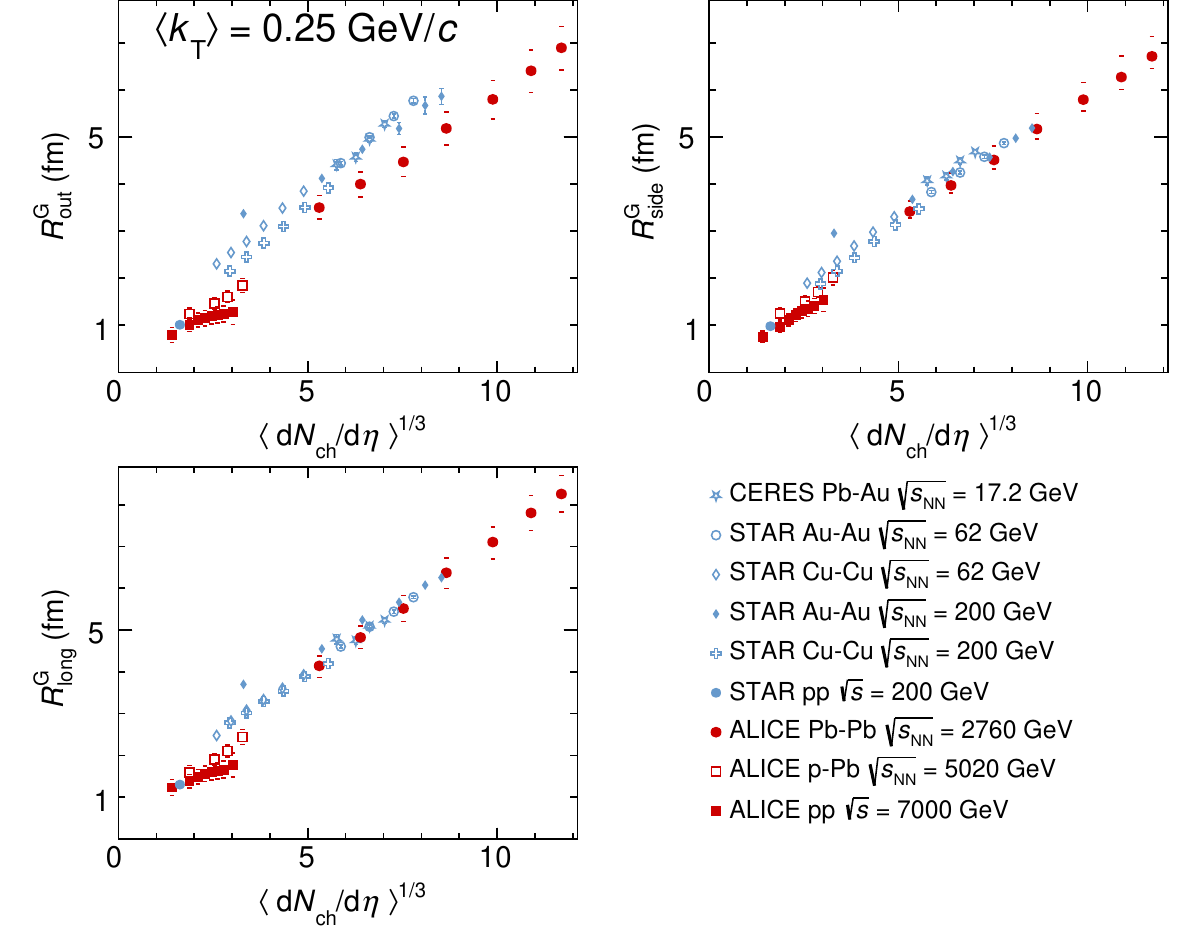}
    \end{center}
    \caption{Multiplicity dependence of pion radii in \pp, \pPb\ and \AA\ collisions for a number of collision systems and energies~\cite{Adams:2004yc,Abelev:2009tp,Aggarwal:2010aa,Adamova:2002wi,Aamodt:2011kd,Aamodt:2011mr,Adam:2015pya,Adam:2015vna}.  
    Various experiments use different \kT ranges. On the plot are the values for the range for which the average \kT is closest to the selected value of 0.25 \GeVc.}    
    \label{fig:9}
\end{figure}

The LHC centrality-dependent \PbPb\ data are compared to the world heavy-ion data, as well as to \pPb\ and \pp\ data at various energies, in Fig.~\ref{fig:9}. As argued in~\cite{Lisa:2005dd}, each of the three-dimensional radii is observed to scale roughly with the cube root of the measured charged-particle multiplicity density  (\avdndeta$^{1/3}$) across a wide range of collision energies and colliding systems with different initial geometries. 
For similar \avdndeta$^{1/3}$ in different \AA\ collision systems, differences in the magnitude of the radii, more significant in the transverse direction (\Rout), were predicted by hydrodynamic calculations~\cite{Kisiel:2008ws} and are attributed to distinct freeze-out shapes. 
The pp and \AA\ data sets exhibit significantly different scaling behaviour, although both are linear in \avdndeta$^{1/3}$. The radii in \pPb\ and \pp\ collisions agree at low multiplicities and diverge by up to 10--20$\%$ at the highest multiplicities reached in these systems. 
The application of the femtoscopic technique to the study of the source in small systems will be further discussed in Chap.~\ref{ch:NuclPhysLHC}
while the rest of this section is dedicated to discussing the radii dependence on the transverse momentum (or transverse mass).

\begin{figure}[htb]
    \begin{center}
    \includegraphics[width = 0.9\textwidth]{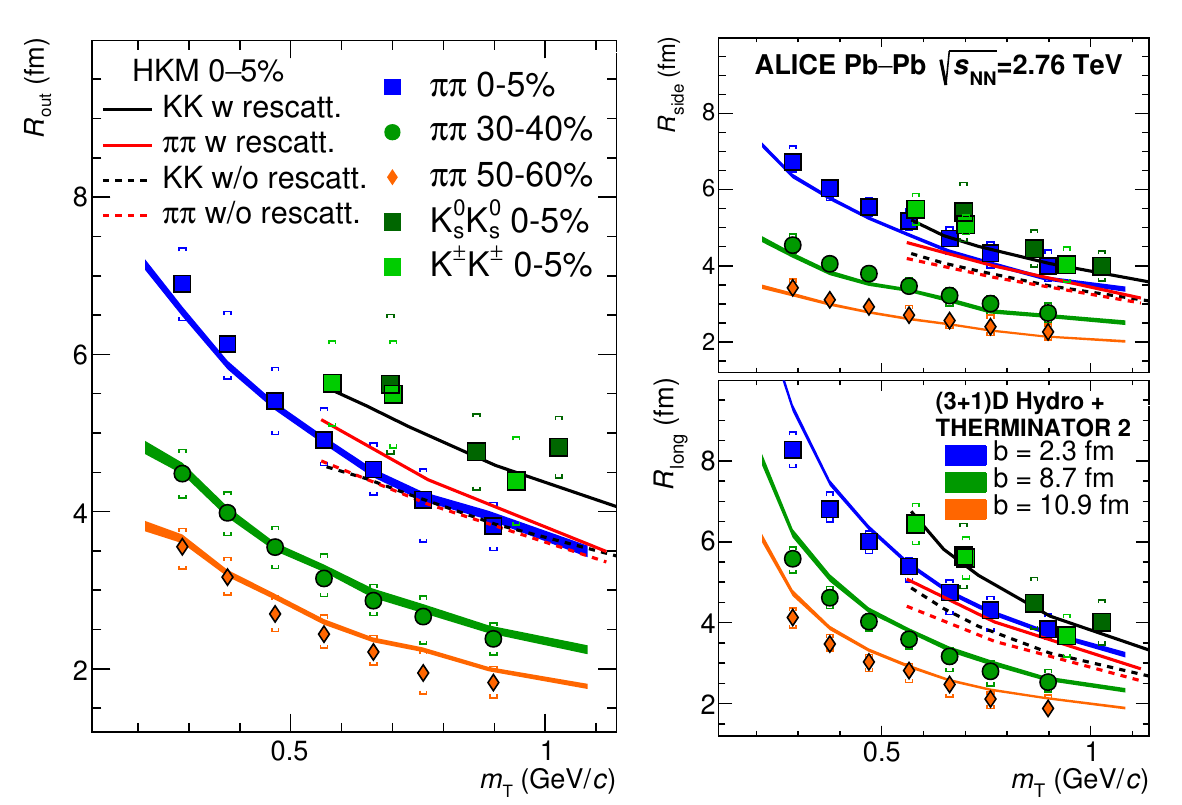}
    \end{center}
    \caption{Pair transverse mass dependence of the pion~\cite{Adam:2015vna} and kaon~\cite{Acharya:2017qtq} femtoscopic radii for different event centralities in \PbPb\ collisions at \sqrtSnn~= 2.76 \TeV. The measured $R_{\rm out}$, $R_{\rm side}$, $R_{\rm long}$ are reported in the left, right top and bottom panel, respectively. The experimental data are reported with solid symbols together with statistical and systematic uncertainties. Bands represent theoretical predictions of pion radii by a (3+1)D hydrodynamic model coupled to the Therminator code~\cite{Kisiel:2014upa} for the same centralities as in data, selected based on the impact parameter b in the calculation.
    Lines represent calculations for central collisions by the HKM model with and without rescattering~\cite{Karpenko:2012yf}.}
    \label{fig:pion_kaon_femto_rad}
\end{figure}

In a hydrodynamical picture, the collective expansion of the fireball reduces the size of the homogeneity region due to the interplay between the collective and thermal velocities of particles. Due to this, each of the femtoscopic radii is expected to decrease with \mT\ following a power-law behaviour~\cite{Lisa:2005dd}. 
This is shown in \Fig~\ref{fig:pion_kaon_femto_rad}, where the dependence of \Rout, \Rside and \Rlong\ on the pair transverse mass \mT\ is reported on the left, right top and right bottom panels, respectively, for different centralities. 

To discuss the general applicability of hydrodynamics for describing the transverse-mass dependence of the femtoscopic radii, we show that one finds similar results for different implementations of the model.
In \Fig~\ref{fig:pion_kaon_femto_rad}, data are compared to the calculations~\cite{Kisiel:2014upa} from the (3+1)D hydrodynamic model coupled to the THERMINATOR 2 statistical hadronisation code~\cite{Chojnacki:2011hb}.
The slope of the \mT~dependence of the transverse radii \Rout and \Rside, which depends on the amount of radial flow in the system, is well reproduced by the model. The calculation also reproduces well the magnitude of \Rout for all centralities, whereas it tends to underestimate the magnitude of \Rside, which lies on the lower edge of the systematic uncertainty associated with the data. The intercept of \Rside at low-\mT\ is usually associated with the overall geometrical size of the system. The model reproduces the values of \Rlong for all centralities, but it overpredicts the magnitude and the slope of the \mT\ dependence, especially at low-momentum. This is an indication that also the longitudinal dynamics is reasonably well described in the model, both in momentum and space-time sectors. Altogether, the good agreement of the model with the femtoscopy data supports the validity of the hydrodynamic approach.

The calculation from the HKM model~\cite{Shapoval:2014wya}, also based on a hydrodynamic formalism, is shown in \Fig~\ref{fig:pion_kaon_femto_rad} for the 5\% most central collisions. It differs from the previous model in the implementation of the freeze-out process. 
Two variants are compared: one in which hadronic interactions in the late hadron-gas phase are neglected (dashed lines, labelled as ``w/o rescattering'' in \Fig~\ref{fig:pion_kaon_femto_rad}), and one in which hadronic interactions are implemented using the UrQMD transport model (continuous lines, labelled as ``W/ rescattering'' in \Fig~\ref{fig:pion_kaon_femto_rad}). 
Both calculations are compatible with the pion femtoscopy data for \Rout\ within uncertainties. However, predictions of HKM with rescattering are also compatible with the measured \Rside\ and \Rlong, whereas calculations without rescattering slightly underpredict them.
Therefore, the approximate agreement of the hydrodynamics-based models with data is a universal feature of such calculations, not of a particular implementation. 
It is also to be noted that the particular choice of initial conditions and the equation of state for these models was motivated by the analysis of the RHIC femtoscopic measurements and is essential for the correct description of the data. 
The results also indicate that the details of the freeze-out process have limited influence on pion femtoscopy. Some studies suggest that femtoscopy of heavier particles might be a more sensitive probe in this case~\cite{Shapoval:2014wya}.

In addition to the pion radii for the 0-5\% most central events, \Fig~\ref{fig:pion_kaon_femto_rad} also shows the \mT\ dependence of the three radii for charged and neutral kaons in comparison with HKM predictions ~\cite{Karpenko:2012yf}. The HKM calculation without rescattering exhibits an approximate \mT-scaling but does not describe the data. The measurements are well reproduced by the full hydro-kinetic model calculations that includes the effect of the hadronic rescattering in the late stages of the system evolution, thereby showing the importance of the hadronic phase at the LHC.  

\subsubsection{Conclusions}
\paragraph{Particle production.} The multiplicity of final-state particles is closely related to the entropy produced in the collision. Central heavy-ion collisions are more effective in converting the longitudinal beam energy into particle production at midrapidity than pp collisions. The final state multiplicity spans from few tens of charged particles per unit of rapidity in peripheral heavy-ion collisions (a ``small system’’) to few thousands in central heavy-ion collisions (a ``large system’’). Centrality gives an experimental handle to control the amount of energy deposited in the collision region and thus the properties of the resulting system. %
\paragraph{Energy density and temperature.} The energy density estimate based on the free-streaming Bjorken equation, which provides a lower bound for this quantity, and with the assumption of a formation time of 1~fm/$c$~\cite{Adcox:2004mh}, results in \mbox{$\epsilon=(12.3 \pm  1.0)$ \GeV/fm$^{3}$} in 0--5$\%$ \PbPb\ collisions at \sqrtSnn~=~2.76~TeV.
This value exceeds by more than a factor of 30 the critical energy density from lattice QCD in most central \PbPb\ collisions at the LHC. The slope of the thermal direct photon spectra represents an effective temperature, which can be related to the true temperature of the system once radial flow is taken into account via full hydrodynamic model calculations. The effective temperature in central \PbPb\ collisions at the LHC is \Teff~=~(304~$\pm$~41)~MeV. The latter is larger than at RHIC and than the critical temperature from lattice QCD. The observations above provide an experimental demonstration that two necessary conditions for the formation of the quark--gluon plasma are met at the LHC. 
\paragraph{Freeze-out.} The increase in beam energy by a factor of 25 from top RHIC energy to the LHC produces a homogeneity region approximately twice larger in most central collisions. The decoupling time for midrapidity pions lies in the range 10--13~\fmC\ at the LHC, which is about 40\% larger than at RHIC. These results indicate that the fireball formed in central \PbPb\ collisions at the LHC lives longer and expands to a larger size at freeze-out compared to lower energies. The quantitative agreement between models based on relativistic hydrodynamics with statistical hadronisation and femtoscopic measurements supports this picture of the evolution of the system created in heavy-ion collisions.

\newpage

\newcommand{\sNN}{\sqrt{s_{\rm NN}}}
\newcommand{\pT}{\ensuremath{p_{\rm T}}\xspace}
\newcommand{\vodd}{v_1^{\rm odd}}
\newcommand{\veven}{v_1^{\rm even}}
\newcommand{\pspec}{{\rm p}}
\newcommand{\tspec}{{\rm t}}
\newcommand{\ptspec}{{\rm p,t}}
\newcommand{\tpspec}{{\rm t,p}}
\newcommand{\PsiPP}{\Psi_{\rm PP}^{(1)}}
\newcommand{\PsiSPp}{\Psi^\pspec_{\rm SP}}
\newcommand{\PsiSPt}{\Psi^\tspec_{\rm SP}}
\newcommand{\PsiSPtp}{\Psi^\tpspec_{\rm SP}}
\newcommand{\PsiRP}{\ensuremath{\Psi_{\rm RP}}\xspace}
\newcommand{ \bs }{{\bf s}}
\newcommand{ \la }{\langle}
\newcommand{ \ra }{\rangle}
\newcommand{\mean}[1]{\la #1 \ra}
\newcommand{\ph}{\ensuremath{P_{\rm H}}\xspace}
\newcommand{ \bomega }{{\boldsymbol{\omega}}}
\newcommand{ \grad }{{\bf \nabla}}
\newcommand{ \bv }{{\bf v}}
\newcommand{\rh}{\ensuremath{\mathrm{\rho_{00}}}~}
\newcommand{\snn}{$\sqrt{s_{\mathrm{NN}}}$~}
\newcommand{\snna}{$\sqrt{s_{\mathrm{NN}}}$}
\newcommand{\s}{$\sqrt{s}$~}
\newcommand{\gvc}{GeV/$c$~}
\newcommand{\kst}{$\rm K^{*0}$~}
\newcommand{\kstc}{$\rm K^{*\pm}$~}
\newcommand{\kstb}{$\rm \overline{K}^{*0}$~}
\newcommand{\kstba}{$\rm \overline{K}^{*0}$~}
\newcommand{\pha}{$\phi$~}
\newcommand{\pt}{$p_{\mathrm{T}}$~}
\newcommand{\pta}{$p_{\mathrm{T}}$}
\newcommand{\dndch}{$\langle\mathrm{d}N_{\mathrm{ch}}/\mathrm{d}\eta\rangle^{1/3}$~}

\newcommand{\Kzs}{\mbox{$\rm K^0_S$}~}
\newcommand{\pb}{\mbox{Pb--Pb}~}

\subsection{QGP evolution and its dynamical properties }
\label{sec:QGPevolution}

The dynamical properties of the QGP offer an essential insight into the strongly-coupled matter. They are principally characterised by measurements sensitive to anisotropic and radial flow, which contribute to the {\it collective motion} observed in heavy-ion collisions. In addition, these properties control how the initial state angular momentum manifests in the global polarisation of the produced particles. Measurements of anisotropic flow at RHIC drew immense attention~\cite{Adcox:2004mh,Arsene:2004fa,Back:2004je,Adams:2005dq} since the observed values were large enough to indicate the formation of a ``perfect liquid" in the laboratory, a claim which was further validated by the estimated values of $\eta/s$ being close to zero for the dense phase of the QGP. Around the same time, AdS/CFT calculations performed using string theoretical techniques showed that $\eta/s$ for any fluid has a lower limit of $1/4\pi$~\cite{Kovtun:2004de}. This originates from the uncertainty principle: the decrease in shear viscosity with the mean free path of the fluid constituents is limited because of the finite momentum of the constituents, thus preventing it to be zero.

Subsequent calculations have predicted the temperature dependence of $\eta/s$ and $\zeta/s$ in the frameworks of AdS/CFT (infinite coupling), QCD (strong coupling), and pQCD (weak coupling)~\cite{Arnold:2003zc, Arnold:2006fz, Meyer:2007ic, Denicol:2009am, Rougemont:2017tlu}. They generally show that the weaker the coupling, the higher the values of $\eta/s$ are and the stronger the temperature dependence is. They also show that the strong/infinite coupling approaches lead to a rise in $\zeta/s$ when the QGP is near the transition temperature to the hadronic phase. Strong and weak-coupling theories can also predict different dependencies of the $\zeta/\eta$ ratio in relation to the speed of sound~\cite{Dusling:2011fd}. These predictions can be tested in the QGP since the viscosities affect the macroscopic properties of many-body systems. Therefore, the collective motion of the QGP inferred from final state hadron measurements, coupled with comparisons to hydrodynamic calculations, offer a unique opportunity to constrain the fundamental properties of strongly interacting matter. 

Beyond what was described in Sec.~\ref{sec:expobservables}, more differential techniques characterising anisotropic and radial flow have emerged over the last ten years, which promise to place even greater constraints on QGP transport properties, and will be detailed in this section. Two key examples regarding anisotropic flow are {\it Symmetric Cumulants} and {\it Non-linear flow mode coefficients}. The former evaluates event-by-event correlations between different orders of anisotropic flow coefficients, while the latter quantifies the non-linear contributions to higher order anisotropic flow from lower orders. Measurements of two-particle correlations can provide additional constraints on the quantitative deduction of effects due to radial flow and shear viscosity. These can also reveal information about the charge diffusion as the created positive-negative charge pairs move through the QGP. An investigation into the polarisation of hyperons in the beam direction~\cite{Voloshin:2017kqp,Becattini:2017gcx} also provides a new avenue to further explore the hydrodynamic response.

\subsubsection{First ALICE results of anisotropic flow}
\label{sec:TG2flowresults}

\begin{figure}[h!t]
\centering
\hbox{\hspace{2.0cm}\includegraphics[width = 0.65\textwidth]{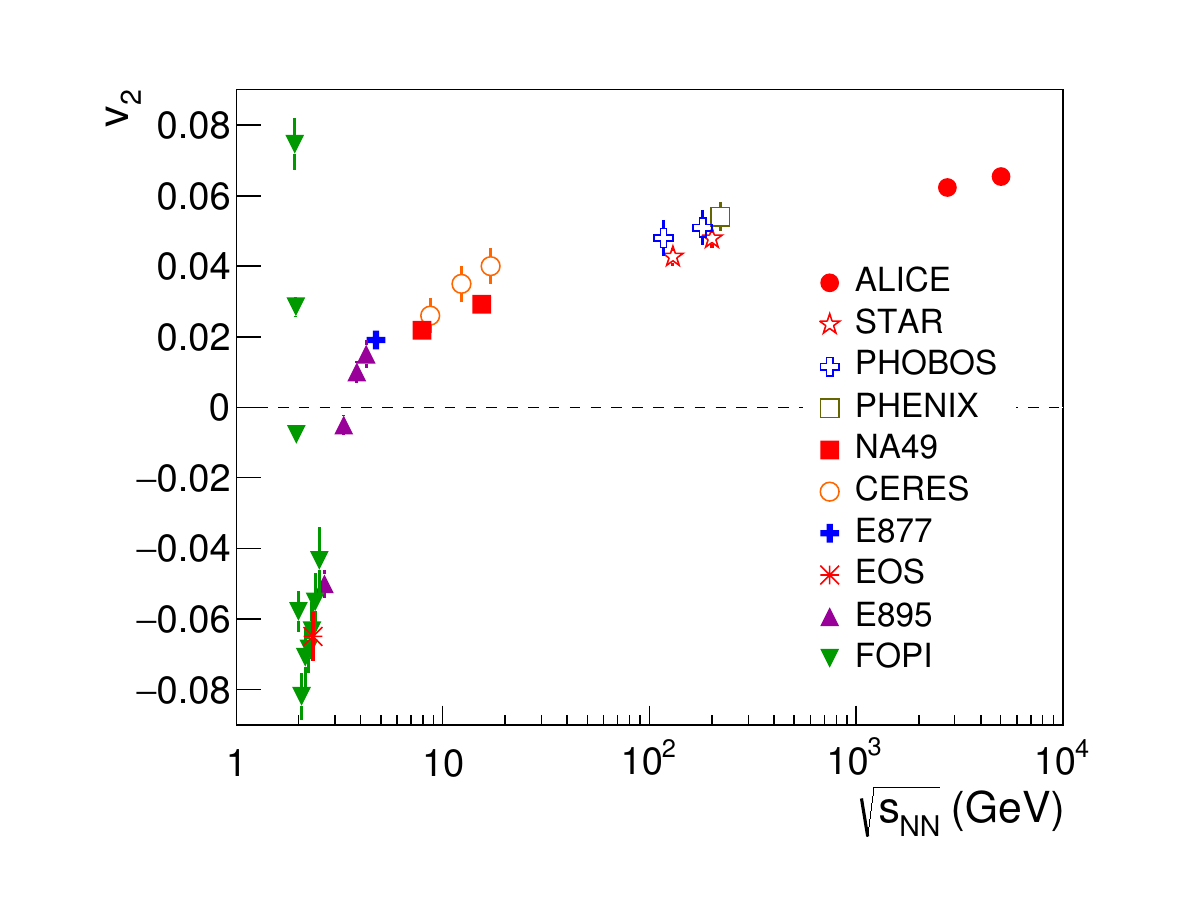}}
\hbox{\hspace{1.3cm}\includegraphics[width = 0.7\textwidth]{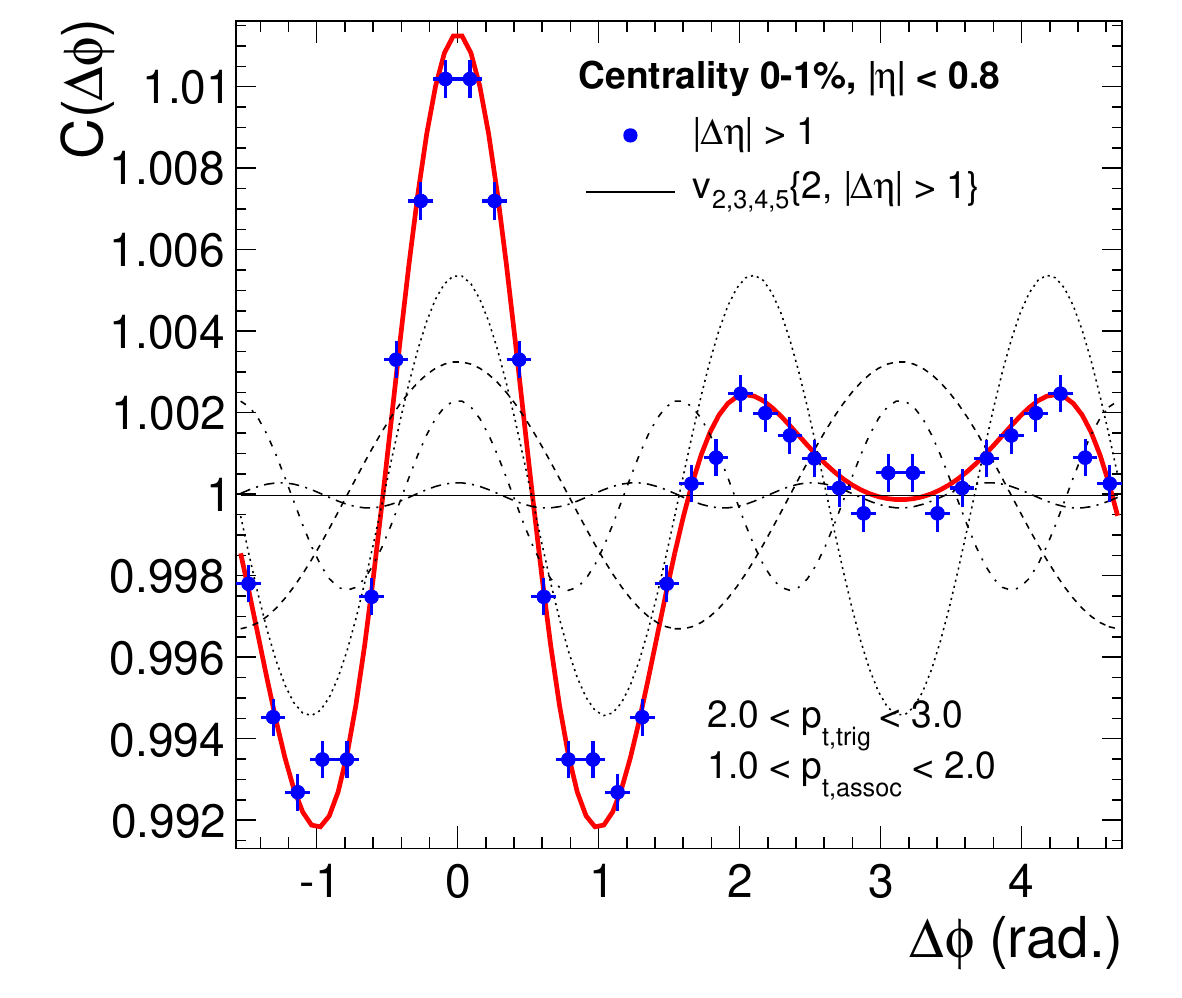}}
\caption{(Top) Energy dependence of $v_2$ for midcentral heavy-ion collisions integrated over $p_{\rm T}$~\cite{Aamodt:2010pa,FOPI:2004bfz,E895:1999ldn,CERES:2002eru,Alt:2003ab,Voloshin:2008dg,STAR:2008ftz,Adam:2016izf}. (Bottom) Decomposition of the two-particle correlation function in terms of even and odd anisotropic flow harmonics ($v_{\rm n}$) from very central Pb--Pb collisions at $\sqrt{s_{\rm{NN}}}=$ 2.76 TeV~\cite{ALICE:2011ab}.}
\label{FIGURE-1}
\end{figure}

Anisotropic flow is quantified by the $v_{\rm n}$ coefficients and symmetry plane angles $\Psi_{\rm n}$. Of particular interest are non-central heavy-ion collisions, which are characterised by a dominant ellipsoidal geometry in the initial state. Theoretical expectations for the LHC energies predicted an increase of 10--50\% for the $p_{\rm{T}}$ integrated $v_2$ when compared to the previous RHIC measurements at lower energies. As can be seen from the top panel of Fig.~\ref{FIGURE-1}, the first $v_2$ measurements at the LHC reported by ALICE in midcentral collisions show an increase of about 30\% compared to the top energy at RHIC~\cite{Aamodt:2010pa}, and this result was in agreement only with predictions from hydrodynamic models which included values of $\eta/s$ close to the AdS/CFT limit of $1/4\pi$. The further increase of $v_2$ by about 3\% in all centralities for Pb--Pb collisions at $\sqrt{s_{\rm{NN}}}=$ 5.02 TeV~\cite{Adam:2016izf}, indicated that the elliptic flow did not saturate in Pb--Pb collisions at $\sqrt{s_{\rm{NN}}}=$ 2.76 TeV, and was in qualitative agreement with most of the hydrodynamic models with similarly low values of specific shear viscosity.

Another interesting aspect of anisotropic flow is the nature of its higher order coefficients $v_3$--$v_5$. A double-peaked structure in two-particle azimuthal correlations, which was first observed at RHIC~\cite{STAR:2010mfd}, was initially interpreted as the signature of a Mach-cone response from fast partons. However, the ALICE results (bottom panel of Fig.~\ref{FIGURE-1}), in conjunction with a reanalysis of RHIC data~\cite{Alver:2010gr}, and other LHC experiments~\cite{ALICE:2011ab, Chatrchyan:2011eka, ATLAS:2012at}, demonstrated that a more natural explanation stems from the non-vanishing values of these higher flow coefficients. One of the first extractions of the two-particle correlation function at the LHC is shown in the bottom panel of Fig.~\ref{FIGURE-1}. The amplitude of the Fourier components provides a measure of $v_{n}^{2}$ for the corresponding momentum ranges. The higher harmonic contributions originate from event-by-event fluctuations in the number and distribution of nucleons in the overlap region of the colliding nuclei, and hydrodynamic models predicted that their magnitude and transverse momentum dependence are sensitive to $\eta/s$~\cite{Teaney:2010vd, Qin:2010pf}. In addition, event-by-event fluctuations violate symmetries linked to an idealistic ellipsoidal geometry, and as a consequence each symmetry plane $\Psi_{\rm n}$ is distinct. 

Measurements of spatial anisotropies of the final state using femtoscopic techniques were also performed in Pb--Pb collisions at $\sqrt{s_{\rm{NN}}}=$ 2.76 TeV~\cite{Adamova:2017opl, Acharya:2018dpu}. While the momentum anisotropies shown previously reflect the initial spatial anisotropies, the azimuthally-differential femtoscopic measurements are sensitive to spatio-temporal characteristics of the source as well as to the collective velocity fields at freeze out. These femtoscopic measurements revealed that the second order spatial anisotropy remains finite in the final state~\cite{Adamova:2017opl}, however, it is reduced compared to values expected from the initial state. The third order final state spatial anisotropy appears washed out~\cite{Acharya:2018dpu}, being significantly smaller than the typical initial state third order anisotropy. All of these observations are consistent with hydrodynamic expectations. These indicate that the finer features of the initial state anisotropy dissipate more quickly for higher eccentricity orders compared to the lower orders. The associated predictions will be compared to other data later in this section.

\subsubsection{Identified hadron spectra, radial flow, and kinetic freeze-out temperatures}
\label{sec:TG2particlespectra}

\begin{figure}[b]
    \begin{center}
    \includegraphics[width = 0.7\textwidth]{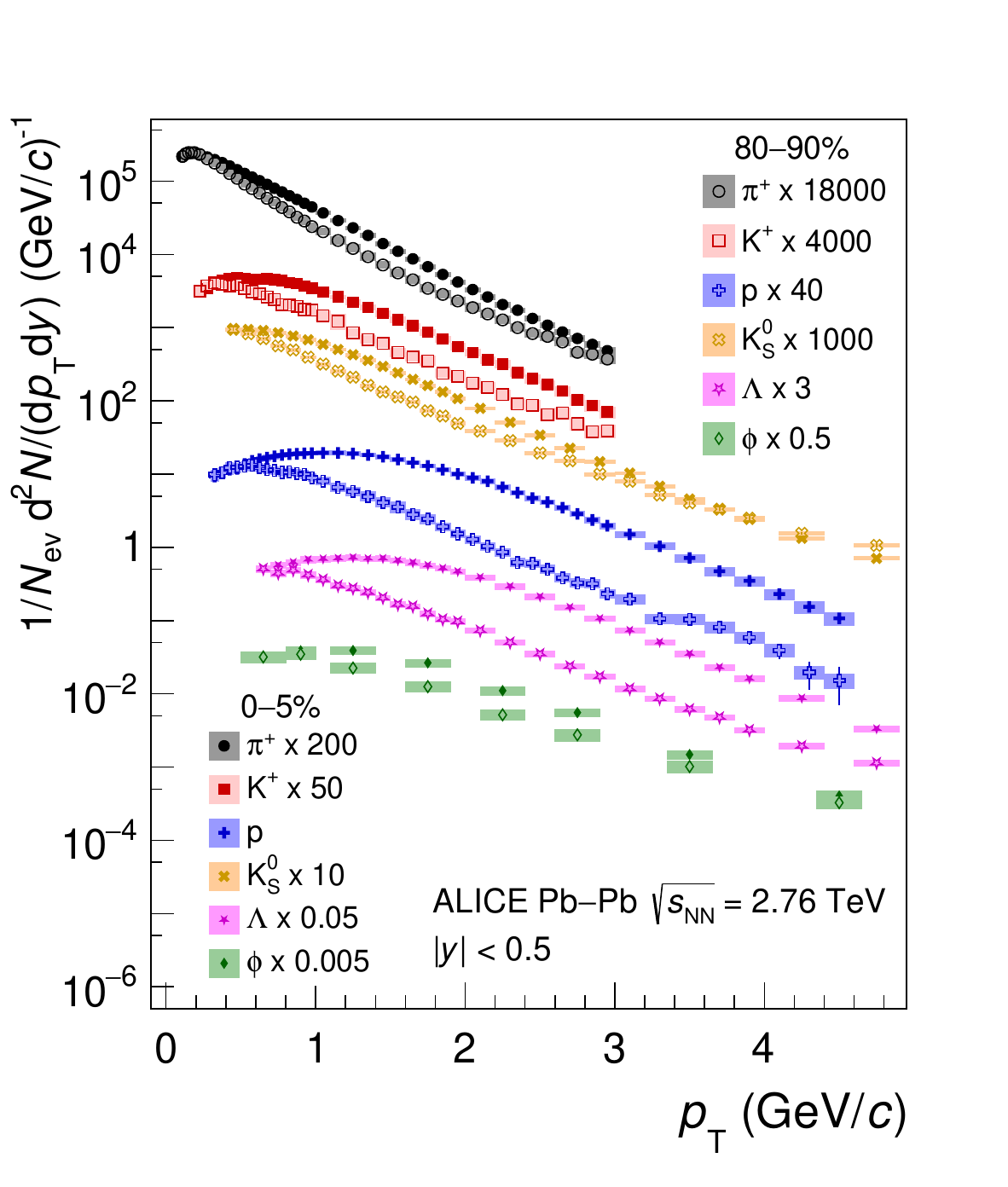}
    \end{center}
    \caption{Transverse momentum distributions of $\pi^{+}$, $\rm K^+$, p~\cite{Abelev:2013vea}, $\rm K^{0}_{S}$ and $\Lambda$~\cite{Abelev:2013xaa}, and the $\phi$ meson~\cite{Abelev:2014uua} for the 0--5\% and 80--90\% centrality intervals in Pb--Pb collisions at $\sqrt{s_{\rm{NN}}}= 2.76$ TeV. The data points are scaled by various factors for better visibility. Statistical and systematic uncertainties are shown as bars and boxes, respectively.} 
    \label{fig:spectraR1}
\end{figure}

The hydrodynamic expansion of the QGP and the late, dissipative hadronic rescattering phase are reflected in the motion of final-state hadrons. Radial flow leads to flatter transverse momentum ($p_{\mathrm{T}}$) distributions with increasing mass, particularly at low values of $p_{\mathrm{T}}$. In this region, the $p_{\mathrm{T}}$ distribution has contributions from the random thermal motion, and the collective expansion. While the former depends on the decoupling temperature, the latter is dependent on the hadron mass because all hadrons acquire an additional transverse momentum given by their mass multiplied by the common radial flow velocity. Figure~\ref{fig:spectraR1} shows the $p_{\rm T}$ spectra of various particles~\footnote{Only $\pi^+$, $\rm K^+$, and p are shown as the distributions of positively and negatively charged particles are compatible within uncertainties at all $p_{\rm T}$ at the LHC.} in Pb--Pb collisions at $\sqrt{s_{\rm{NN}}}=$ 2.76 TeV for the 0--5\% and 80--90\% centrality intervals~\cite{Abelev:2013vea, Abelev:2013xaa, Abelev:2014uua}. The spectral shapes depend on centrality with the maxima located at higher transverse momenta in central compared to peripheral collisions. Furthermore, the flattening of the spectra in the low-$p_{\rm T}$ region is mass dependent, and is more pronounced for heavier particles. This observation is in line with the expected effect of increasing radial flow with collision centrality.

\begin{figure}[t]
    \centering
    \includegraphics[width = 0.55\textwidth]{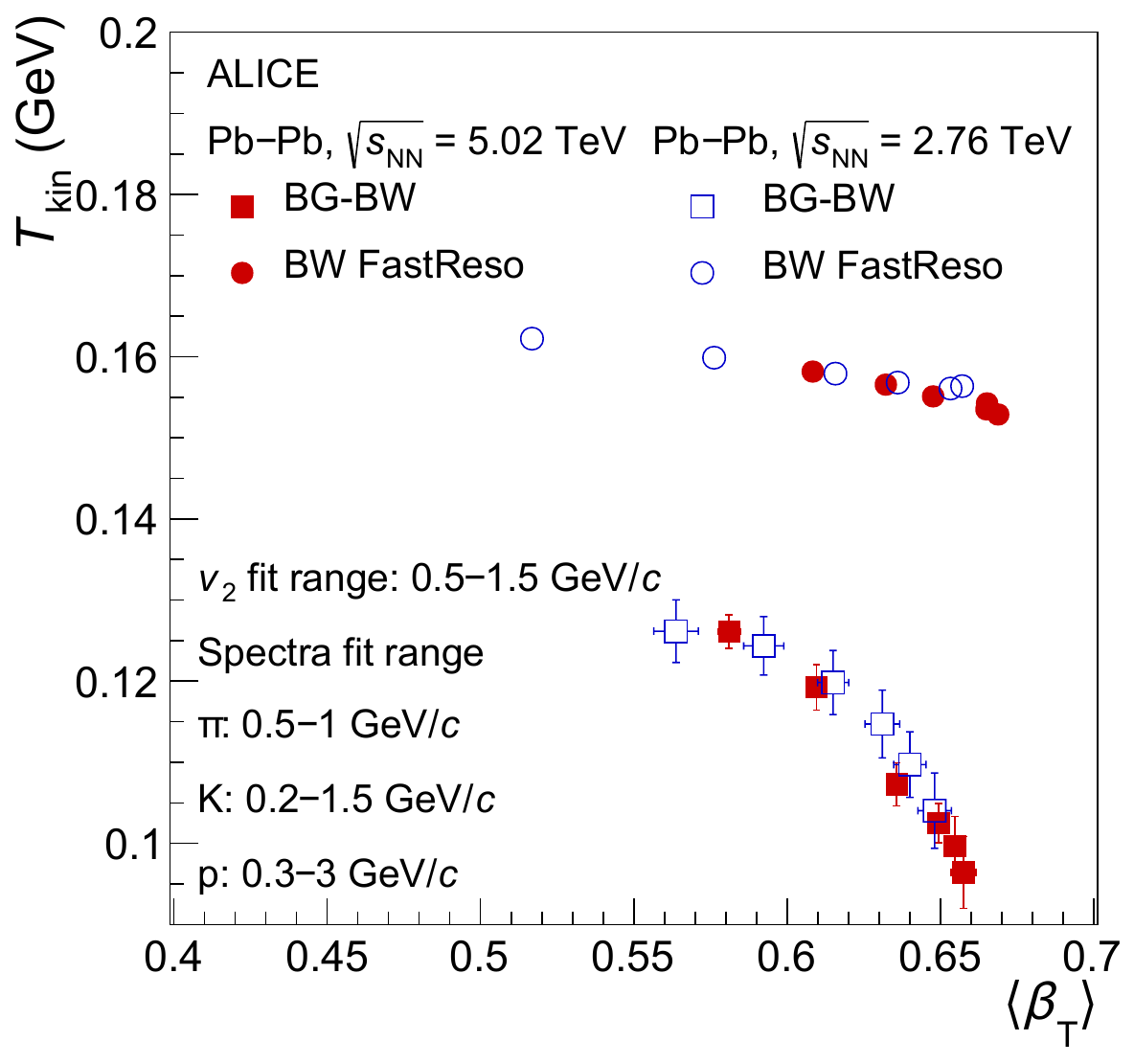}
    \caption{Blast-Wave model parameters of the kinetic freeze-out temperature $T_{\rm{kin}}$ and radial flow velocity $\beta_{\rm{T}}$. These extracted values involve simultaneous fits to $\pi$, K, and p spectra and $v_{2}$ for the two models in Pb--Pb collisions.}
    \label{fig:BWParameters}
\end{figure}

The hadron spectra can be studied in the context of the Boltzmann-Gibbs Blast-Wave (BG-BW) parameterisation~\cite{Schnedermann:1993ws}. In this parameterisation, particles are produced by a thermalised medium, which expands radially and undergoes an instantaneous kinetic freeze-out at a temperature $T_{\rm kin}$. In the simplest case, the expansion is radial and is prescribed by a common velocity field profile, $\beta(r) = \beta_{s}\left(\frac{r}{R}\right)^{n}$, where $\beta_{s}$, $R$, and $n$ are the expansion velocity on the surface of the fireball, its radius, and the exponent regulating the shape of the velocity profile, respectively. It is well known that particle spectra, especially at low-transverse momenta, are populated by the decay products of resonance decays. These decays modify the spectral shapes of hadrons, and thus the simple Boltzmann--Gibbs prescription may fall short. So far, several attempts have been made to include the products of resonance feed-down into the Blast-Wave parameterisation, but they usually treat the production of primary hadrons and short-lived hadronic resonances separately. The calculation of the latter usually resorts to either Monte Carlo generators or a semi-analytical treatment of decay integrals which are both computationally intensive. An alternative approach was proposed in Ref.~\cite{Mazeliauskas:2019ifr}, enabling a computationally-efficient fitting procedure to measured particle spectra (BW FastReso)~\cite{Mazeliauskas:2018irt}. In addition, the procedure was extended to include anisotropic flow, particularly the $v_2$ of different particle species (see Sec.~\ref{sec:TG2pidflow}). The extracted parameters from the simultaneous fits to the $p_{\rm T}$ spectra and to $v_2$ of $\pi$, K, and p in Pb--Pb collisions at $\sqrt{s_{\rm NN}}= 2.76$ and 5.02 TeV are shown in Fig.~\ref{fig:BWParameters} from both BG-BW and BW FastReso approaches. A key aspect to note is that the radial flow velocities appear rather independent of the BW parameterisation in more central collisions, which correspond to larger $\beta_{\rm T}$ values. The highest ever radial flow velocities in heavy-ion collisions are observed at the LHC (a comparison to RHIC results can be found elsewhere~\cite{Abelev:2013vea}). However, the kinetic freeze-out temperatures depend strongly on the BW parameterisation. The BG-BW approach determines the kinetic freeze-out temperature directly, whereas BW FastReso in this implementation assumes the same chemical and kinetic freeze-out temperature for hadrons. BW FastReso allows for these temperatures to be two distinct free parameters, however, they happen to be consistent for fits utilising this option (not shown). The parameters shown in Fig.~\ref{fig:BWParameters} indicate that the inclusion of resonance feed-down denoted by the filled and open circles results in an increase of the kinetic freeze-out temperature by $\sim$60 MeV. This demonstrates that this temperature is highly dependent on the inclusion of resonance feed-down.

\subsubsection{Hydrodynamic descriptions of global observables}
\label{sec:TG2hydrodescription}

\begin{figure}[b]
    \begin{center}
    \includegraphics[width = 0.95\textwidth]{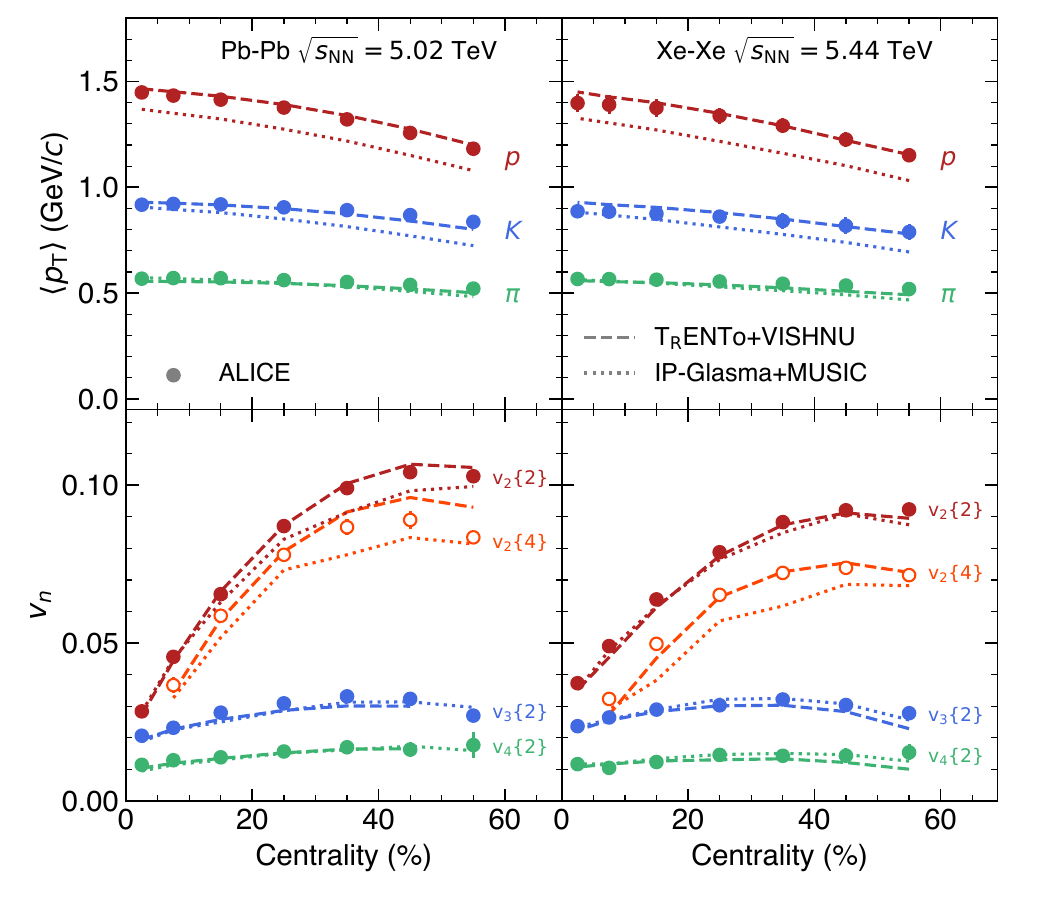}
    \end{center}
    \caption{Comparisons of ALICE measurements of identified particle $\langle p_{\mathrm{T}} \rangle$ and charged hadron $\it{v}_{n}$ coefficients in Pb--Pb collisions at $\sqrt{s_{\rm{NN}}}=$ 5.02 TeV (left) and Xe--Xe collisions at $\sqrt{s_{\rm{NN}}}=$ 5.44 TeV (right)~\cite{Adam:2016izf,Acharya:2018ihu, Acharya:2019vdf,Acharya:2021ljw} to hydrodynamic calculations ~\cite{Bernhard:2019bmu,Schenke:2020mbo}.} 
    \label{fig:HydroXePb}
\end{figure}

In Fig.~\ref{fig:HydroXePb}, hydrodynamic model calculations of the average transverse momentum of identified particles and of various anisotropic flow coefficients for charged hadrons are compared to ALICE data from heavy-ion collisions. Hydrodynamic models differ from Blast-Wave parameterisations. While the former use properties of the QGP through an equation of state, Blast-Wave models are limited and attempt to assess the final-state properties under certain assumptions. It is shown in Ref.~\cite{Ryu:2015vwa} that $v_n$ can be described by multiple combinations of $\eta/s$ and $\zeta/s$ values. However, these are constrained further by attempts to describe $\langle p_{\mathrm{T}} \rangle$ data for identified hadrons, which is why such a simultaneous comparison is important. The precision of the data is such that the uncertainty is about 2\% for most data points, while both models are able to describe the data within 10\% of the measured values. The T$_{\rm{R}}$ENTo+VISHNU calculation uses a Bayesian analysis to tune the input parameters of $\eta/s$ and $\zeta/s$ (among others) to best match the data in Pb--Pb $\sqrt{s_{\rm{NN}}}=$ 2.76 and 5.02 TeV collisions~\cite{Bernhard:2019bmu}. This tune does not include the Xe--Xe data~\cite{Acharya:2018ihu, Acharya:2019vdf,Acharya:2021ljw}, and the subsequent model calculations can be considered as predictions based on the Pb--Pb inference. The IP-Glasma+MUSIC uses a priori predictions of $\eta/s$ and $\zeta/s$ in an attempt to describe the same measurements at RHIC and the LHC~\cite{Schenke:2020mbo}. Both models require an $\eta/s$ similar to the minimum AdS/CFT value of $1/4\pi$ with little or no dependence on the temperature, and a finite $\zeta/s$ that rises with decreasing temperature as it approaches the QGP transition temperature. Those parameterisations do not depend on the collision centrality or the colliding system. Each of these approaches implicitly assumes a strongly-coupled QGP, and they appear to describe the data. It is also worth noting that the peak value of $\zeta/s$ close to the QGP transition temperature is larger in IP-Glasma+MUSIC compared to T$_{\rm{R}}$ENTo+VISHNU. This may explain why T$_{\rm{R}}$ENTo+VISHNU appears to describe $\langle p_{\mathrm{T}} \rangle$ slightly better, since lower values of $\zeta/s$ will increase $\langle p_{\mathrm{T}} \rangle$, and this pushes T$_{\rm{R}}$ENTo+VISHNU closer to the data, particularly for kaons and protons. The values of $\eta/s$ and $\zeta/s$ explored for these hydrodynamic models will be discussed in more detail in Sec.~\ref{sec:QGPsummary}, while the role of the initial-state models will be discussed in Chap.~\ref{ch:InitialState}.

\begin{figure}[t]
    \begin{center}
    \includegraphics[width = 0.55\textwidth]{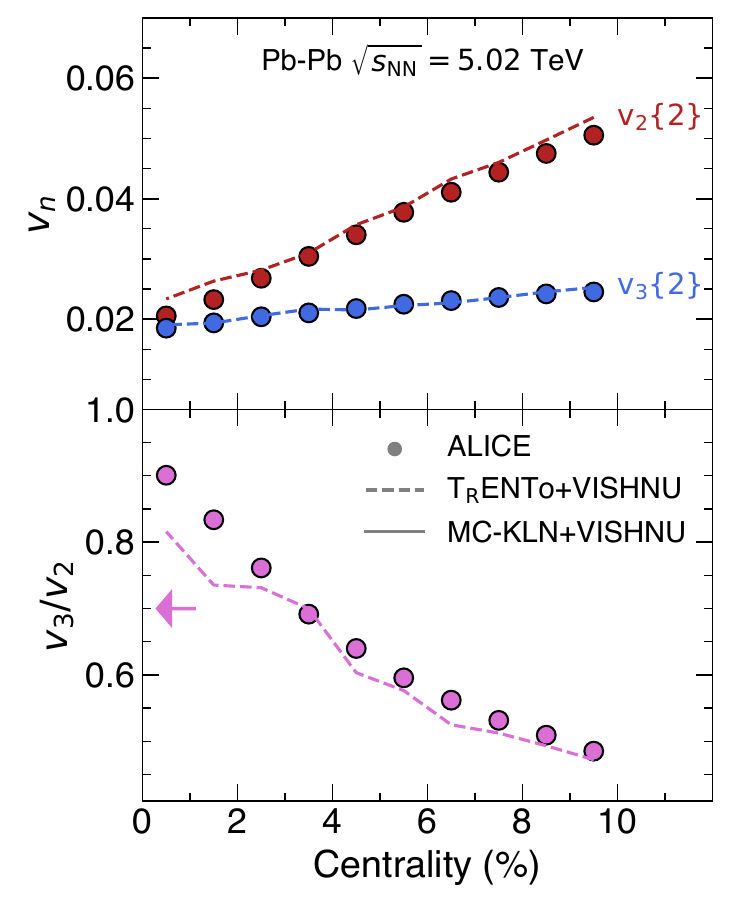}
    \end{center}
    \caption{Measurements of charged hadron $\it{v}_{n}$ coefficients in central Pb--Pb collisions at $\sqrt{s_{\rm{NN}}}=$ 5.02 TeV~\cite{Acharya:2018ihu} compared with hydrodynamic calculations~\cite{Bernhard:2019bmu,Shen:2015qta}. The arrow denotes predictions from Ref.~\cite{Shen:2015qta} for 0--0.2\% Pb--Pb collisions.} 
    \label{fig:HydroPbUC}
\end{figure}

\begin{figure}[t]
    \begin{center}
    \includegraphics[width = 0.55\textwidth]{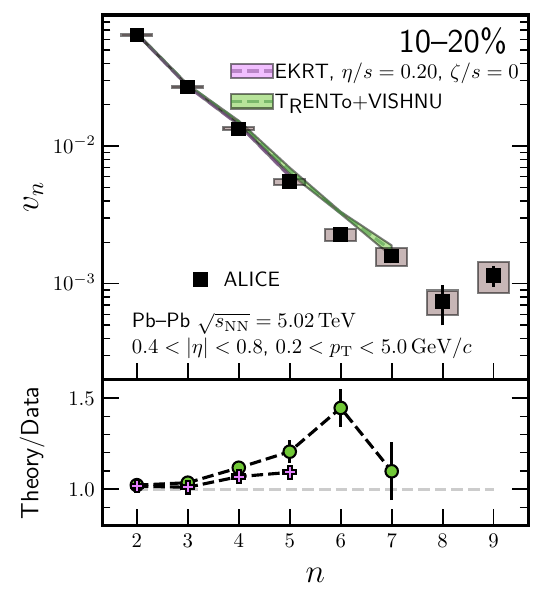}
    \end{center}
    \caption{Higher order charged hadron $v_n$ coefficients in the 10--20\% centrality interval~\cite{Acharya:2020taj}.} 
    \label{fig:vnHO}
\end{figure}

Even though the quality of these descriptions is impressive given the multitude of measurements they involve, such comparisons need to be further tested by more differential measurements. Figure~\ref{fig:HydroPbUC} shows a comparison of measurements of anisotropic flow coefficients with T$_{\rm{R}}$ENTo+VISHNU predictions for very central Pb--Pb collisions at $\sqrt{s_{\rm{NN}}}=$ 5.02 TeV. Earlier hydrodynamic calculations could not describe the $v_3/v_2$ ratio for centralities less than 1\%, a phenomenon sometimes referred to as the {\it ultra-central problem}, and the deviations compared to the data were around 30\%~\cite{Shen:2015qta}. For the ALICE measurements in the 0--1\% range, more recent hydrodynamic predictions appear to do better, and describe the ratio on the level of 10\%, which is similar to some of the other comparisons shown in Fig. ~\ref{fig:HydroXePb}. The value of $\eta/s$ in the T$_{\rm{R}}$ENTo+VISHNU approach is roughly a factor of two smaller than the earlier calculation, which leads to a larger value of $v_3/v_2$. The lower value of $\eta/s$ is related to the implementation of bulk viscosity in recent hydrodynamic calculations, as concluded to be needed for very central Pb--Pb collisions~\cite{Rose:2014fba}. These improvements might also have a contribution from advancements in the initial state modelling, which will be discussed in Chap.~\ref{ch:InitialState}. The first measurements of the anisotropic flow power spectra up to the ninth order are shown in Fig.~\ref{fig:vnHO} for the 10--20\% centrality interval in Pb--Pb collisions. As mentioned, higher order anisotropic flow coefficients are more sensitive to viscous damping, and the dampening rate depends on $\eta/s$. In addition to T$_{\rm{R}}$ENTo+VISHNU, these data are also compared with the EKRT predictions. The EKRT model~\cite{Paatelainen:2012at, Paatelainen:2013eea} describes anisotropic flow measurements equally well as T$_{\rm{R}}$ENTo+VISHNU although requiring higher shear viscosity since it does not include bulk viscosity. However, even in its most recent implementation~\cite{Auvinen:2020mpc}, it has difficulties describing average transverse momentum measurements of identified particles at the LHC, due to the assumption of $\zeta/s=0$. More broadly, all of these investigations regarding the hydrodynamic response rely on a realistic description of the initial state to extract QGP properties, and this will be addressed further in Chap.~\ref{ch:InitialState}. 

Finally, an extraction of the QGP speed of sound was performed with the ALICE data and hydrodynamic model comparisons using charged hadron $\langle p_{\mathrm{T}} \rangle$~\cite{Gardim:2019xjs} measurements. The value squared of the speed of sound obtained from Pb--Pb collisions at $\sqrt{s_{\rm{NN}}}=$ 2.76 and 5.02 TeV is $(0.24 \pm 0.04)~c^2$. This number is compatible with the Lattice QCD prediction of $0.25~c^2$ for a temperature of 222 MeV i.e.\ in the deconfinement regime~\cite{Borsanyi:2013bia}. Strong- and weak-coupling theories predict competing dependencies of the $\zeta/\eta$ ratio in relation to the speed of sound~\cite{Dusling:2011fd} even though the speed of sound is intrinsically not a dynamical physical quantity. Lattice QCD can predict its value in the QGP, which is consistent with the value obtained from ALICE data. It is also an important parameter for hydrodynamic models which require it as input for shear and bulk relaxation times. The value used in the IP-Glasma+MUSIC model~\cite{Schenke:2020mbo} is consistent with what is extracted from ALICE data. Considering the uncertainty on the value obtained from ALICE data, it is not yet possible to discern between the weak- and the strong-coupling regimes.

\subsubsection{Identified hadron anisotropic flow}
\label{sec:TG2pidflow}

\begin{figure}[b]
    \begin{center}
    \includegraphics[width = 0.99\textwidth]{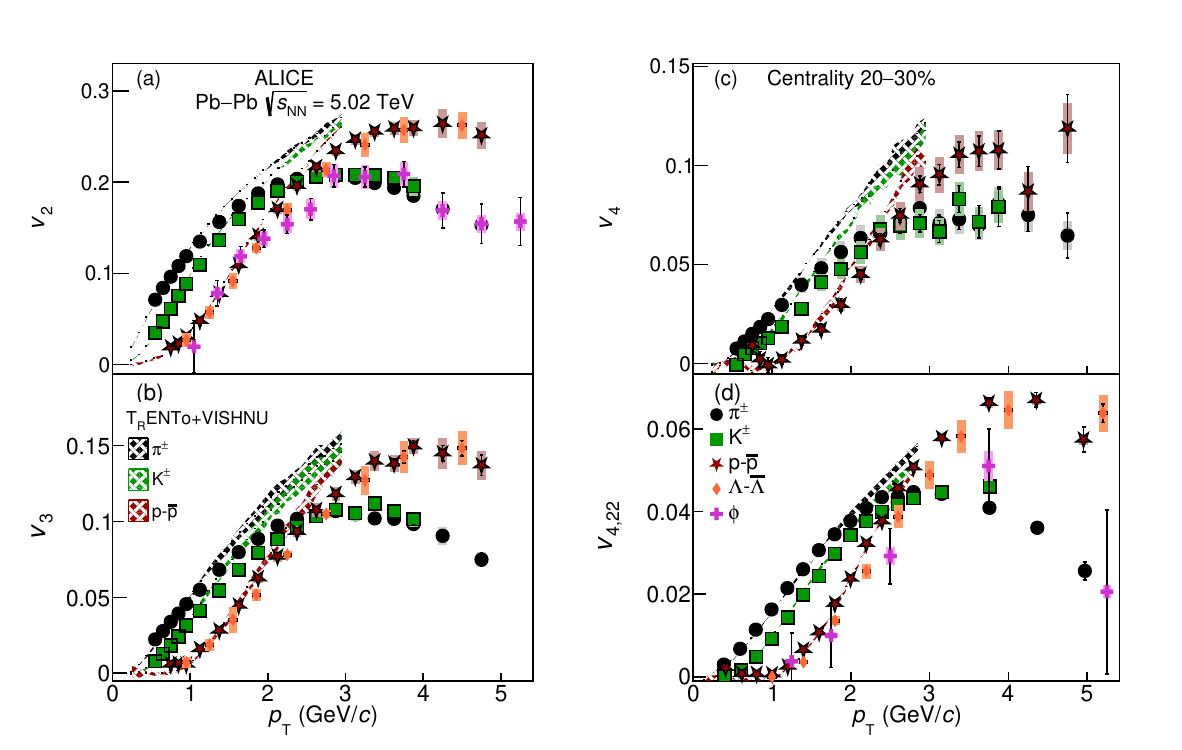}
    \end{center}
    \caption{The $p_{\mathrm{T}}$--differential $v_2$ (a), $v_3$ (b), $v_4$ (c), and $v_{4,22}$ (d) measured by ALICE in semicentral (i.e., 20--30\% centrality interval) Pb--Pb collisions at $\sqrt{s_{\mathrm{NN}}} = 5.02$~TeV~\cite{Acharya:2018zuq,Acharya:2019uia}. The data points are drawn with their statistical (error bars) and systematic (boxes) uncertainties. The curves represent estimations extracted from T$_{\rm{R}}$ENTo+VISHNU that give the best description of other anisotropic flow measurements among all models studied.} 
    \label{fig:pidVnResults}
\end{figure}
A cornerstone for investigating the strongly-coupled QGP paradigm is a set of measurements of $v_2$ for various particle species. A characteristic mass ordering at low-$p_{\rm T}$ is induced at hadronization, after the development of radial flow in the QGP phase. Radial flow leads to a depletion in the particle $p_{\mathrm{T}}$ spectrum at low values, which increases with increasing particle mass. When introduced in a system that exhibits azimuthal anisotropy, then radial flow develops an azimuthal profile, with larger velocity values in- than out-of-plane. This, in turn, leads to a depletion in the momentum spectrum that becomes larger in-plane than out-of-plane, resulting in a reduction in the differences of particle yields at a given $p_{\mathrm{T}}$ in- versus out-of-plane and consequently in a reduction of $v_2$. Due to the dependence of momentum on the mass, this effect is more pronounced for heavier particles. The net result is that at a fixed value of $p_{\mathrm{T}}$, heavier particles have smaller $v_2$ values compared to lighter ones. Therefore the focus in this section is on the low-$p_{\rm T}$ region (i.e., $p_{\rm T}<3$ GeV/$c$), while the high-$p_{\rm T}$ trends are discussed in Sec.~\ref{sec:QGPHadronization}.

Figure~\ref{fig:pidVnResults} (a), presents the transverse momentum dependence of $v_2$ as measured for various particle species in semicentral (i.e.,~20--30\% centrality interval) Pb--Pb collisions at $\sqrt{s_{\mathrm{NN}}} = 5.02$~TeV~\cite{Acharya:2018zuq}. Similar results are obtained in Pb--Pb collisions at $\sqrt{s_{\mathrm{NN}}} = 2.76$~TeV~\cite{Abelev:2014pua,Adam:2016nfo}, Xe--Xe collisions at $\sqrt{s_{\mathrm{NN}}} = 5.44$~TeV~\cite{ALICE:2021ibz}, and RHIC~\cite{Adler:2003kt, Adams:2003am}. The characteristic mass ordering that develops due to an interplay between radial flow and the anisotropic expansion of the fireball is clearly seen for $p_{\mathrm{T}} < 2.5$~GeV/$c$. 

This mass ordering develops not only during the partonic evolution of the medium but also in the late hadronic re-scattering phase. The relative contribution, though, is expected to be related to the corresponding duration time of the two phases. Hence, at the LHC, the relative impact of the highly dissipative hadronic stage is expected to be weaker than at lower, RHIC energies~\cite{Hirano:2007ei}. An excellent testing ground for the contribution of the hadronic phase is provided by studying particles that are estimated to have small hadronic cross sections, such as the $\phi$ meson and the $\Xi$ baryon, and are thus expected not to be affected by this stage~\cite{Biagi:1980ar,Bass:1999tu,Bass:2000ib,Shor:1984ui}. Figure~\ref{fig:pidVnResults} (a) illustrates that the $\phi$ meson follows the observed mass ordering (similarly for the $\Xi$ baryon, see~Ref.~\cite{Abelev:2014pua}). The current level of statistical and systematic uncertainties of these measurements, however, does not rule out the hypothesis that the mass ordering can be broken due to a significantly reduced contribution of the hadronic phase to the development of $v_2$ for the $\phi$ meson~\cite{Hirano:2007ei}. 

The interplay between radial flow and the anisotropic expansion of the system was further tested by investigating whether the mass ordering is reflected in measurements of higher order cumulants~\cite{ALICE:2022zks} or higher flow harmonics~\cite{Acharya:2018zuq,Adam:2016nfo}. Panels (b) and (c) of Fig.~\ref{fig:pidVnResults} present the transverse momentum dependence of $v_3$ and $v_4$, respectively, for the same particle species (with the exception of the $\Lambda$ baryon for $v_4$ and the $\phi$ meson for both $v_3$ and $v_4$) and the same centrality interval as before for Pb--Pb collisions at $\sqrt{s_{\mathrm{NN}}} = 5.02$~TeV~\cite{Acharya:2018zuq}. These results clearly illustrate that the effect can be seen in $v_2$, as well as $v_3$ and $v_4$ (but also for $v_5$ see Refs.~\cite{Acharya:2019uia,Adam:2016nfo}). This mass ordering develops also for the 1\% most central Pb--Pb collisions~\cite{Acharya:2018zuq,Adam:2016nfo}, a category of events referred to as ultra-central collisions, where there is no dominant second order initial state geometry. 

Hydrodynamic models have also shown that $v_2$ and to a large extent $v_3$ are linearly proportional to their corresponding initial spatial eccentricities, $\epsilon_2$ and $\epsilon_3$, for central and semicentral collisions~\cite{Niemi:2012aj}. However, the study of higher order flow coefficients (i.e., for $n > 3$) revealed that higher order eccentricities have a non-linear dependence on the lower order i.e., $\epsilon_2$ and $\epsilon_3$~\cite{Bhalerao:2014xra,Yan:2015jma}. This further supports the earlier ideas that the $v_n$ coefficients (i.e., for $n > 3$) receive contributions not only from the linear response of the system to $\epsilon_n$, but also a non-linear response proportional to the product of lower order initial spatial anisotropies~\cite{Bhalerao:2014xra,Yan:2015jma}. The non-linear contribution arises from lower order momentum anisotropies generating higher order spatial anisotropies as the system evolves, which in turn generate higher order momentum anisotropies. Figure~\ref{fig:pidVnResults} (d), presents the transverse momentum dependence of $v_{4,22}$, the non-linear flow mode of quadrangular flow, reported in~Ref.~\cite{Acharya:2019uia} for the 20--30\% centrality interval in Pb--Pb collisions at $\sqrt{s_{\mathrm{NN}}} = 5.02$~TeV. A similar mass ordering like the one reported for $v_n$ measurements develops also here for $p_{\mathrm{T}} < 2.5$~GeV/$c$. This arises from the interplay between radial flow and the initial spatial anisotropy, generated from both the geometry and the fluctuating initial energy density profile. Although the mass ordering should develop differently for $v_{4,22}$ and $v_4$ due to the dependence on $\epsilon_2^2$ of the former, it is found to be quantitatively the same with maybe a hint of differences developing for $p_{\mathrm{T}} < 0.8$ GeV/$c$ in the 0--30\% centrality interval~\cite{Acharya:2019uia}. Non-linear flow modes are discussed in more detail in Sec.~\ref{sec:TG2higherorderflowandnonflow}.

Figure~\ref{fig:pidVnResults} also present the comparison of the measurements with the same T$_{\rm{R}}$ENTo+VISHNU calculations as shown in Fig.~\ref{fig:HydroXePb}. The curves describe the trends of the data points for $v_2$, $v_3$, $v_4$, and $v_{4,22}$ within 20\% for $p_{\rm T}<2$ GeV/$c$. There are larger discrepancies at higher $p_{\rm T}$ which have two potential sources. In hydrodynamic models, the momentum distribution of hadrons at freeze-out is principally modeled under the assumption of thermal equilibrium, and that distribution is labelled as $f_{\rm thermal}$. A $\delta f$ term is introduced to account for non-equilibrium processes, thus expressing the final momentum distribution sampled as $f = f_{\rm thermal}+\delta f$. The $\delta f$ corrections grow with $p_{\mathrm{T}}$, and are highly model dependent~\cite{Noronha-Hostler:2014dqa}. Therefore, deviations at higher $p_{\mathrm{T}}$ might be due to improper $\delta f$ corrections, or sub-optimal tunes of $\eta/s$ and $\zeta/s$.

\subsubsection{Symmetric Cumulants}
\label{sec:TG2symmetriccumulants}

\begin{figure}[b]
    \begin{center}
    \includegraphics[width = 0.99\textwidth]{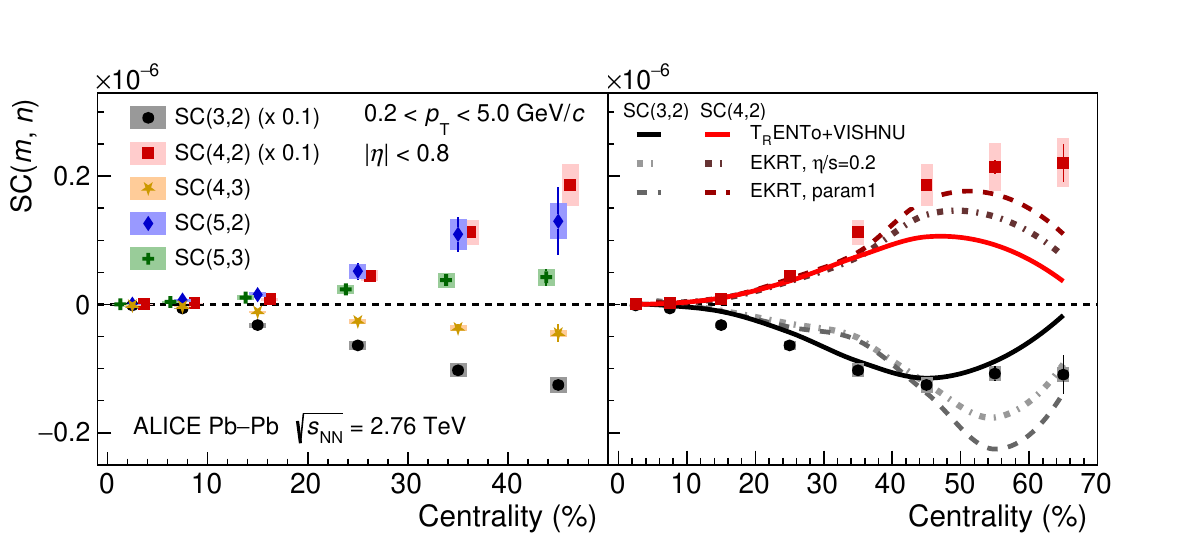}
    \end{center}
    \caption{Centrality dependence of event-by-event flow harmonic correlations measured by ALICE in Pb--Pb collisions at $\sqrt{s_{\mathrm{NN}}} = 2.76$ TeV~\cite{ALICE:2016kpq,Acharya:2017gsw} compared with various hydrodynamic calculations~\cite{Niemi:2015qia, Bernhard:2019bmu}. The SC(4,2) and SC(5,3) points are slightly shifted along the horizontal axis for better visibility in the left panel.} 
    \label{fig:SC}
\end{figure}
The studies of the individual flow amplitudes $v_n$ can be extended to correlations between event-by-event fluctuations of flow coefficients~\cite{Niemi:2012aj,Bilandzic:2013kga,Aad:2015lwa,Qian:2016pau,Zhu:2016puf}. Model calculations show that while $v_2$ and $v_3$ exhibit an approximately linear dependence on the corresponding eccentricities $\epsilon_2$ and $\epsilon_3$, respectively, the higher order $v_n$ coefficients (i.e., for $n > 3$) have also non-linear contributions from $\epsilon_2$ and $\epsilon_3$ in addition to the linear ones from $\epsilon_n$~\cite{Qin:2010pf,Qiu:2011iv,Yan:2015jma,Kolb:2003zi}. These observations lead to non-trivial correlations between different flow coefficients which result in new and independent constraints on initial conditions and $\eta/s$. In addition, they have the potential to separate the effects of $\eta/s$ from fluctuations in the initial conditions. For such novel studies, ALICE introduced \textit{Symmetric Cumulants}~(SC)~\cite{Bilandzic:2013kga,Mordasini:2019hut} and mixed harmonic cumulants~\cite{Moravcova:2020wnf}. These observables are independent of the symmetry plane angles $\Psi_n$, and are robust against systematic biases due to unwanted nonflow correlations (i.e., short-range correlations unrelated to the azimuthal asymmetry in the initial geometry, such as inter-jet correlations and resonance decays) in heavy-ion collisions.

The left panel of Fig.~\ref{fig:SC} presents the centrality dependence of correlations between $v_n$ coefficients (up to $5^{\mathrm{th}}$ order) using SC$(k,l) \equiv \langle v_k^2 v_l^2\rangle - \langle v_k^2\rangle\langle v_l^2\rangle$ in Pb--Pb collisions at $\sqrt{s_{\mathrm{NN}}} = 2.76$ TeV~\cite{ALICE:2016kpq, Acharya:2017gsw}. The correlations among different flow coefficients depend on harmonic as well as collision centrality~\cite{Acharya:2017gsw}. Positive values of SC(4,2), SC(5,2), and SC(5,3) and negative values of SC(3,2) and SC(4,3) are observed for all centralities. These indicate that event-by-event fluctuations of $v_2$ and $v_4$, $v_2$ and $v_5$, and $v_3$ and $v_5$ are correlated, while $v_2$ and $v_3$, and $v_3$ and $v_4$ are anti-correlated. Furthermore, the lower order harmonic correlations are much larger than the higher order ones. Precision measurements from Pb--Pb collisions at $\sqrt{s_{\mathrm{NN}}} = 5.02$ TeV~\cite{ALICE:2021adw} show similar trends.

The SC observables are compared with EKRT~\cite{Niemi:2015qia} and T$_{\rm{R}}$ENTo+VISHNU~\cite{Bernhard:2019bmu} predictions in the right panel of Fig.~\ref{fig:SC}. The EKRT calculations are shown for the two temperature dependent $\eta/s$ parameterisations that provide the best description of RHIC and LHC data: constant $\eta/s=0.2$ and ``param1"~\cite{Niemi:2015qia}. The ``param1" parameterisation is characterised by a moderate slope in the temperature dependence of $\eta/s$ which decreases (increases) in the hadronic (QGP) phase and the phase transition occurs around 150 MeV. The SC(3,2) and SC(4,2) are not described simultaneously in each centrality interval by the EKRT calculations, which points to a strong dependence on $\eta/s$ of both observables. The SC(3,2) is better described by the T$_{\rm{R}}$ENTo+VISHNU predictions for Pb--Pb collisions at $\sqrt{s_{\mathrm{NN}}} = 2.76$ TeV although these measurements or the subsequent ones presented later in this section are not used in the Bayesian inference. In addition, normalised symmetric cumulants were also measured in Pb--Pb collisions at $\sqrt{s_{\mathrm{NN}}} = 2.76$ TeV~\cite{ALICE:2016kpq, Acharya:2017gsw} and are reported in Chap.~\ref{ch:InitialState}. Recently, ALICE measured multiparticle cumulants with three different flow harmonics in Pb--Pb collisions at $\sqrt{s_{\mathrm{NN}}} = 2.76$ and 5.02 TeV~\cite{ALICE:2021klf, ALICE:2021adw}.

\subsubsection{Non-linear flow modes and flow vector fluctuations}
\label{sec:TG2higherorderflowandnonflow}

\begin{figure}[b]
    \begin{center}
    \includegraphics[width = 1.0\textwidth]{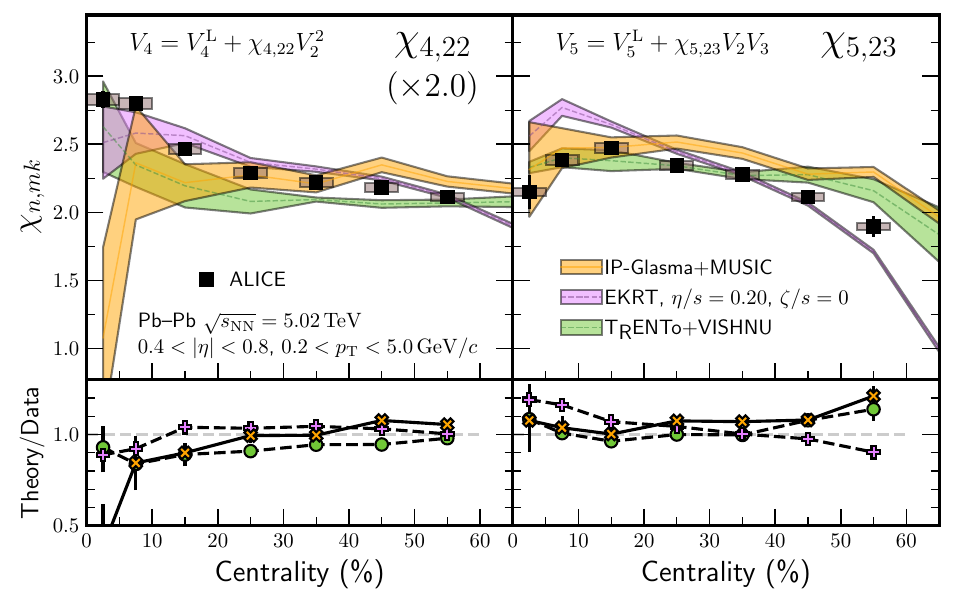}
    \end{center}
    \caption{Non-linear flow mode coefficients $\chi_{4, 22}$ (left)  and $\chi_{5, 23}$ (right) in Pb--Pb collisions at $\sqrt{s_{\mathrm{NN}}} = 5.02$ TeV compared to various hydrodynamic calculations~\cite{Acharya:2020taj}.}
    \label{fig:NL1}
\end{figure}

Higher order anisotropic flow coefficients (i.e., $n > 3$) have also contributions from the initial-state anisotropy of the lower orders as discussed in Sec.~\ref{sec:TG2pidflow}. These additional contributions can be characterised by measurements of the non-linear flow mode coefficients, $\chi_{n, mk}$ with the indices representing different harmonics, which in principle should primarily arise from the QGP evolution~\cite{Bhalerao:2014xra,Qian:2016fpi}. The lower order coefficients are more sensitive to the initial-state configuration, while the higher orders are more susceptible to the QGP transport properties at later times and therefore lower temperatures. Figure~\ref{fig:NL1} shows the centrality dependence of the non-linear flow mode coefficients $\chi_{4, 22}$ (left panel) and $\chi_{5, 23}$ (right panel) in Pb--Pb collisions at $\sqrt{s_{\mathrm{NN}}} = 5.02$~TeV~\cite{Acharya:2020taj}. A small decrease from central to peripheral collisions is found for both coefficients. In addition, the relationship of $\chi_{4, 22} \approx \chi_{5, 23}/2$ is approximately valid as predicted by hydrodynamic calculations~\cite{Yan:2015jma}. Given both the T$_{\rm{R}}$ENTo+VISHNU and IP-Glasma+MUSIC models appear equally competitive in describing the $\langle p_{\mathrm{T}} \rangle$ and $v_n$ coefficients in Pb--Pb collisions at $\sqrt{s_{\mathrm{NN}}} = 5.02$~TeV (see left panel of Fig.~\ref{fig:HydroXePb}), it is worth noting that the non-linear flow mode coefficients are also well described by both models. Furthermore, the EKRT model~\cite{Niemi:2015qia}, which as mentioned does not implement bulk viscosities, does also equally well.

Fluctuating initial conditions can also generate fluctuations of the flow vector in different transverse momentum or pseudorapidity regions. This was first tested in Ref.~\cite{Aamodt:2011by}, where two-particle correlation $V_{n\Delta}(p_{\rm T}^{\rm a}, p_{\rm T}^{\rm t})$ (one associated particle from $p_{\rm T}^{\rm a}$ and the other trigger particle from $p_{\rm T}^{\rm t}$) measurements were fitted globally, and then compared to the product $v_n(p_{\rm T}^{\rm a}) \times v_n(p_{\rm T}^{\rm t})$. If such a factorisation holds, this implies all particles flow in the same direction irrespective of transverse momentum or pseudorapidity. This was indeed observed for the case of $n \geq 2$ at low values of \pta~($\lesssim 2$ GeV/\textit{c}) ~\cite{Aamodt:2011by,Adam:2017ucq}. The degree to which factorisation is broken can also be investigated using two types of two-particle correlation measurements: i) the factorisation ratio $r_n$~\cite{Gardim:2012im}, which probes $V_{n\Delta}(p_{\rm T}^{\rm a}, p_{\rm T}^{\rm t})$ with respect to the square root of the product of $V_{n\Delta}(p_{\rm T}^{\rm a}, p_{\rm T}^{\rm a})$ and $V_{n\Delta}(p_{\rm T}^{\rm t}, p_{\rm T}^{\rm t})$ that uses both particles from either $p_{\rm T}^{\rm a}$ or $p_{\rm T}^{\rm t}$; or ii) $v_{n}\{2\}/v_{n}[2]$~\cite{Heinz:2013bua}, where $p_{\rm T}^{\rm a} \neq p_{\rm T}^{\rm t}$ for $v_{n}\{2\}$ and $p_{\rm T}^{\rm a} = p_{\rm T}^{\rm t}$ for $v_{n}[2]$. Both $v_{n}\{2\}/v_{n}[2]$ and $r_{n}$ were measured in Pb--Pb collisions at $\sqrt{s_{\rm NN}} = 2.76$ TeV~\cite{Acharya:2017ino}. Clear deviations from unity are observed for the second order flow vector in the 0--5\% centrality interval. This indicates the presence of possible $p_{\rm T}$-dependent flow vector fluctuations, which can have an influence from the hydrodynamic response. Higher order measurements ($n>2$) show no clear indication of $p_{\rm T}$-dependent fluctuations within uncertainties. The results are described fairly well by hydrodynamic calculations, including the T$_{\rm{R}}$ENTo+VISHNU approach~\cite{Zhao:2017yhj}. Recently, ALICE reported the first measurements of $p_{\rm T}$-dependent flow angle $\Psi_2$ and flow magnitude $v_2$ fluctuations, determined using new four-particle correlators, in Pb--Pb collisions at $\sqrt{s_{\rm NN}} = 5.02$ TeV~\cite{ALICE:2022smy}. Deviations from unity are observed for both flow angle and flow-magnitude fluctuations in the presented centrality intervals for $p_{\rm T} > 2$ GeV/$c$, being the largest in the most central collisions.

\subsubsection{Charge dependent and independent two-particle correlations}
\label{sec:TG2cdandcicorrelators}

Differential measurements of two-particle correlations can shed additional light on the system evolution. A number of physics processes contribute to the final shape of a differential two-particle correlation function. The two-particle correlation measurements involve the counting of the number of correlated particle pairs at relative azimuthal and longitudinal separation, $\Delta \varphi$ and $\Delta \eta$, respectively. These measurements are primarily affected by intra-jet, resonance decays, quantum-statistics correlations, and radial flow. In addition, correlations induced earlier in the evolution of the system will show longer range $\Delta \eta$ correlations compared to those induced later. The fact that particles are created in pairs due to local charge conservation introduces additional charge-dependent effects into these correlations. The balance function observable, a two-particle charge dependent correlator, which studies the distribution of balancing charges in momentum space, has been suggested to be sensitive to all these effects~\cite{Jeon:2001ue,Bass:2000az,Bozek:2004dt}. Measurements of the balance function at RHIC~\cite{Adams:2003kg} and SPS~\cite{Alt:2004gx,Alt:2007hk} energies showed a longitudinal narrowing when going from peripheral to central collisions, while other results from RHIC~\cite{Aggarwal:2010ya, Abelev:2013csa} also revealed a similar trend on the azimuthal dimension.

\begin{figure}[!t]
    \centering
    \includegraphics[width = 0.7\textwidth]{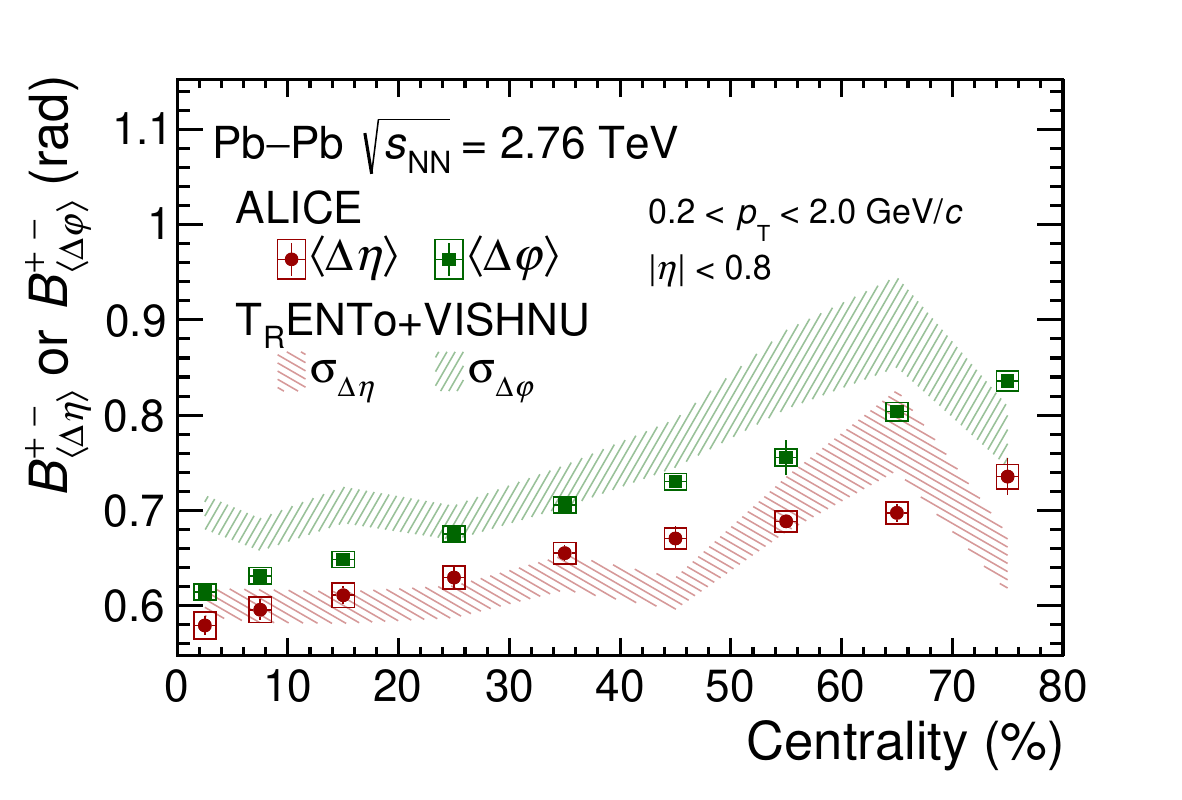}
    \includegraphics[width = 0.7\textwidth]{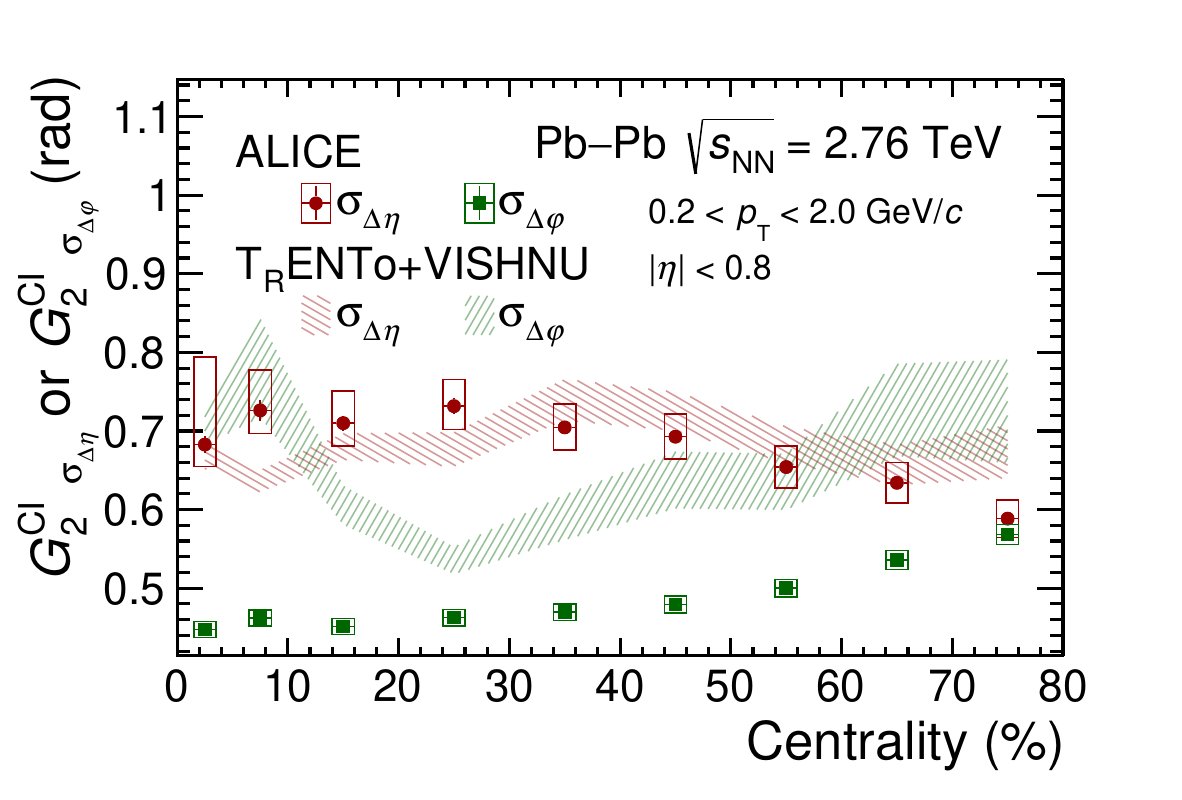}
    \caption{(Top) Evolution of longitudinal and azimuthal widths as a function of collision centrality for the balance function $B^{+-}$. (Bottom) The two-particle transverse momentum correlator $G_{2}^{\rm CI}$ as a function of centrality. 
    ALICE data~\cite{Abelev:2013csa} are compared to calculations from the T$_{\rm{R}}$ENTo+VISHNU model chain~\cite{Bernhard:2019bmu}.}
    \label{fig:BFG2}
\end{figure}

The top panel of Fig.~\ref{fig:BFG2} presents the balance function widths for Pb--Pb collisions at $\sqrt{s_{\rm NN}}=2.76$ TeV from ALICE~\cite{Abelev:2013csa}. They are compared to predictions from the T$_{\rm{R}}$ENTo+VISHNU model (initially shown in Fig.~\ref{fig:HydroXePb}). In particular, the $\langle \Delta \varphi \rangle$ measurements show some differences between the data and the predictions. Since other observables sensitive to radial flow have been well described by this model earlier in this section, these differences in $\langle \Delta \varphi \rangle$ might arise from charge-diffusion effects in the QGP, which are another QGP transport property. These effects are not implemented in this model, but were independently studied elsewhere regarding the electric diffusivity of light quarks~\cite{Pratt:2018ebf,Pratt:2019pnd}. These differences are also observed for the $\langle \Delta \eta \rangle$ measurements, where charge diffusion effects are expected to be present. The differences in both cases between the data and hydrodynamic model calculations therefore might place constraints on this specific transport property of the QGP.

The charge independent (CI) differential two-particle transverse momentum correlator $G_{2}^{\rm CI}$ was constructed to be sensitive to the diffusion of transverse momentum currents ~\cite{Gavin:2006xd,Sharma:2008qr}. Under the assumption that more central collisions generate the QGP that lasts for a longer time, the longitudinal width of $G_{2}^{\rm CI}$ can provide a measure of viscous diffusion as the effects of diffusion become more pronounced. The proposed way of extracting $\eta/s$ under this assumption contrasts with the methods shown previously, involving comparisons to results from hydrodynamic models which strongly rely on the correct modelling of the initial conditions. Measurements of $G_{2}^{\rm CI}$ by the STAR collaboration~\cite{Agakishiev:2011fs} have shown a longitudinal broadening for more central collisions, which are in agreement with the expected viscous effects inferred from a value of $\eta/s$ close to the AdS/CFT limit. 

The measurements of $G_{2}$ were performed in Pb--Pb collisions at $\sqrt{s_{\rm NN}}=2.76$ TeV for charge independent (CI) and charge dependent (CD) hadron combinations~\cite{Acharya:2019oxz}. The CI results are shown in the bottom panel of Fig.~\ref{fig:BFG2}. 
The azimuthal behaviour of $G_{2}^{\rm CI}$ shows a narrowing trend, while the longitudinal evolution conversely demonstrates a clear broadening towards central collisions. Extracting the expected widths of the correlator for the most central collisions for different values of $\eta/s$ and interpreting the longitudinal broadening of $G_{2}^{\rm CI}$ as originating exclusively from viscous effects, suggests that the measured longitudinal width is compatible with a value of $\eta/s$ close to the AdS/CFT limit. These features are not reproduced by the HIJING and AMPT models (not shown). The bottom panel of Fig.~\ref{fig:BFG2} also shows that the longitudinal width is compatible with the T$_{\rm{R}}$ENTo+VISHNU model, which uses an $\eta/s$ close to the AdS/CFT limit, and therefore should incorporate the aforementioned viscous effects. On the other hand, the same model cannot reproduce the azimuthal width, which requires further investigation.

\subsubsection{Polarisation of hyperons and vector mesons}
\label{sec:TG2globalpolarization}

The spin-orbital angular momentum interaction is one of the most well-known effects in nuclear, atomic, and condensed matter physics. In non-central heavy-ion collisions, a large initial angular momentum ({\it O}(10$^{7}$) $\hslash$) is expected to be created perpendicular to the reaction plane. Due to any possible spin-orbit coupling, particles produced in such a collision can become globally polarised~\cite{Liang:2004ph,Voloshin:2004ha,Liang:2004xn,Gao:2007bc}. Global polarisation measurements of produced particles provide important information about the initial conditions and dynamics of the QGP, as well as the hadronisation processes~\cite{Becattini:2015ska,Li:2017slc,Xie:2017upb}. Considering the importance of initial conditions, it is imperative that these are determined using other measurements. The three-dimensional spatial profile of the initial conditions and its effect on the space--time evolution of a heavy-ion collision can be further probed by the directed flow, $v_1$, of hadrons containing light (u, d, and s) and heavy (c) quarks. The charge dependence of $v_1$, as well as the different magnitude of global polarisation of particles and anti-particles, is sensitive to the effects of the strong magnetic fields caused by colliding nuclei and will be discussed in Sec.~\ref{sec:NovelQCD}. 

The collision symmetry requires that the directed flow is an anti-symmetric function of pseudorapidity, \mbox{$\vodd(\eta) = - \vodd(-\eta)$}. Due to event-by-event fluctuations in the initial energy density of the collision, the participant plane angle defined by the dipole asymmetry of the initial energy density~\cite{Teaney:2010vd,Luzum:2010fb} and that of projectile and target spectators, are different from the geometrical reaction plane angle $\PsiRP$. As a consequence, the directed flow can develop a rapidity-symmetric component, $\veven(\eta) = \veven(-\eta)$, which does not vanish at midrapidity~\cite{Teaney:2010vd,Luzum:2010fb,Gardim:2011qn,Retinskaya:2012ky}. Results for rapidity-even $v_1$ of charged particles at midrapidity in Pb--Pb collisions at $\sNN=$ 2.76~TeV~\cite{Abelev:2013cva} show strong evidence of dipole-like initial density fluctuations in the overlap zone of the nuclei. Similar trends in the rapidity-even $v_1$ and the estimate from two-particle correlation measurements at midrapidity~\cite{Agakishiev:2010ur,Aamodt:2011by,ATLAS:2012at,Chatrchyan:2012wg} indicate a weak correlation between fluctuating participant and spectator symmetry planes. The observed negative slope of the rapidity-odd $v_1$, with approximately a 3 times smaller magnitude than found at the highest RHIC energy~\cite{Abelev:2008jga}, suggests a smaller longitudinal tilt of the initial system and disfavors the strong fireball rotation predicted for the LHC energies~\cite{Adamczyk:2016eux, Adil:2005bb, Adil:2005qn}.
\begin{figure}[!t]
\center
\hbox{\hspace{2.0cm} \includegraphics[width=0.85\textwidth]{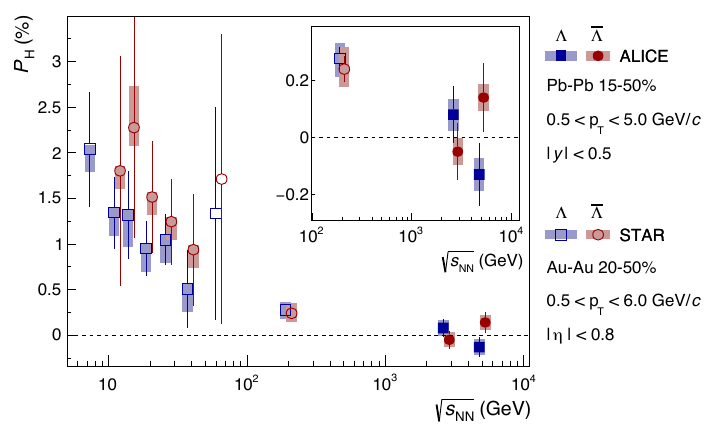}}
\hbox{\hspace{1.9cm}\includegraphics[width=0.66\textwidth]{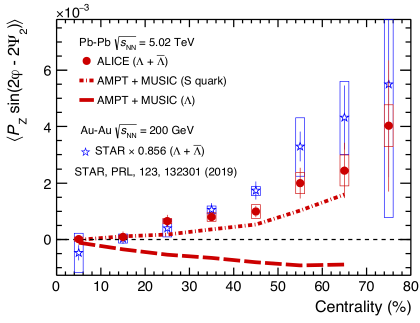}}
\caption{
(Top) The global hyperon polarisation (\ph) as a function of collision energy compared with the STAR data at lower energies~\cite{STAR:2017ckg,Adam:2018ivw}. The insert shows zoomed-in comparison with the data at the top RHIC energy. The systematic uncertainties are shown as shaded boxes. Points are slightly shifted along the horizontal axis for better visibility. (Bottom) Centrality dependence of the hyperon polarisation component along the beam direction $P_z$ measured in Pb--Pb collisions at $\sNN=$ 5.02~TeV compared with STAR data and two hydrodynamic calculations~\cite{ALICE:2021pzu}.}
\label{fig:polarization}
\end{figure}

Most of the recent calculations of the global polarisation assume complete thermal equilibrium and the validity of the hydrodynamic description of the system~\cite{Karpenko:2016jyx,Becattini:2016gvu,Florkowski:2017ruc,Li:2017slc,Baznat:2017jfj}. They relate polarisation of a charged particle to the thermal vorticity of the system at the hadronisation time. The global polarisation is determined by the average vorticity component along the orbital momentum of the system and is perpendicular to the collision reaction plane. Both the magnitude and the direction of the vorticity can strongly vary within the system~\cite{Voloshin:2017kqp}. In particular, a significant component along the beam direction can be acquired due to the transverse anisotropic flow~\cite{Voloshin:2017kqp,Becattini:2017gcx}. The vorticity of the system can be linked with $v_1$ through the asymmetries in the initial velocity fields. Hydrodynamic simulations show that the orbital angular momentum stored in the system and the directed flow of charged particles are almost directly proportional to each other~\cite{Becattini:2015ska}. This allows for an empirical estimate of the collision energy dependence of the global polarisation~\cite{Voloshin:2017kqp}. The STAR results for the directed flow~\cite{Adamczyk:2014ipa,Shanmuganathan:2015qxb} and the hyperon global polarisation~\cite{STAR:2017ckg,Adam:2018ivw} show that the slopes of $v_1$ at midrapidity (${\rm d}v_1/{\rm d}\eta$) for charged hadrons (pions) and the hyperon polarisation are indeed strongly correlated.

The global polarisation of the $\Lambda$ and $\overline\Lambda$ hyperons was measured in Pb--Pb collisions at $\sNN=2.76$ and 5.02~TeV~\cite{Acharya:2019ryw}. The results are reported differentially as a function of collision centrality and transverse momentum of the hyperon for the 5--50\% centrality interval, and the dependence on collision energy is shown in the top panel of Fig.~\ref{fig:polarization}. The average hyperon global polarisation for Pb--Pb collisions at $\sNN=2.76$ and 5.02~TeV is found to be consistent with zero, $\mean{\ph}(\%)\approx -0.01\pm 0.05~\mbox{(stat.)} \pm 0.03~\mbox{(syst.)}$ in the 15--50\% centrality interval. The results are compatible with expectations based on an extrapolation from measurements at lower collision energies at RHIC~\cite{Abelev:2013cva}, hydrodynamic model calculations, and empirical estimates based on collision energy dependence of directed flow~\cite{Voloshin:2017kqp}. All these calculations predict the global polarisation values at the LHC energies of the order of $0.01\%$. 

In addition to the global hyperon polarisation, a polarisation of $\Lambda$ and $\overline{\Lambda}$ along the beam direction was predicted in non-central heavy-ion collisions due to the strong elliptic flow~\cite{Voloshin:2017kqp, Becattini:2017gcx}. The bottom panel of Fig.~\ref{fig:polarization} presents the polarisation of hyperons in the $z$-direction of Pb--Pb collisions at $\sNN=5.02$ TeV~\cite{ALICE:2021pzu}. The polarisation is non zero, thus exhibits a clear second harmonic sine modulation as expected due to elliptic flow, and is of similar magnitude as the one measured at RHIC~\cite{Adam:2019srw}. Comparisons with two different hydrodynamic calculations~\cite{Fu:2021pok} are also shown. The strange quark scheme assumes that the hyperon polarisation is inherited from the strange quark prior to hadronisation, while in the other approach the polarisation is calculated for the $\Lambda$ and $\overline{\Lambda}$ at the freeze-out using the hyperon mass for the mass of the spin carrier. The data favour the strange quark scheme, both quantitatively and qualitatively, which perhaps demonstrates that these measurements are sensitive to the $z$-polarisation of strange quarks in the QGP phase.

Another consequence of quarks being polarised due to the spin-orbital angular momentum interactions leads to a preferential alignment of the intrinsic angular momentum (spin) of vector mesons. These are formed by hadronisation of quarks, along the direction of global angular momentum~\cite{Voloshin:2004ha,Liang:2004ph,Liang:2004xn,Liang:2007ma}. The spin alignment of a vector meson is quantified by measuring the spin density matrix element \rh which is the probability of finding a vector meson in spin state 0 out of 3 possible spin states (-1, 0, 1). In the absence of spin alignment all 3 spin states are equally probable which makes $\rho_{00}$ = 1/3. 
\begin{figure}[!t]
    \begin{center}
    \includegraphics[width = 1.0\textwidth]{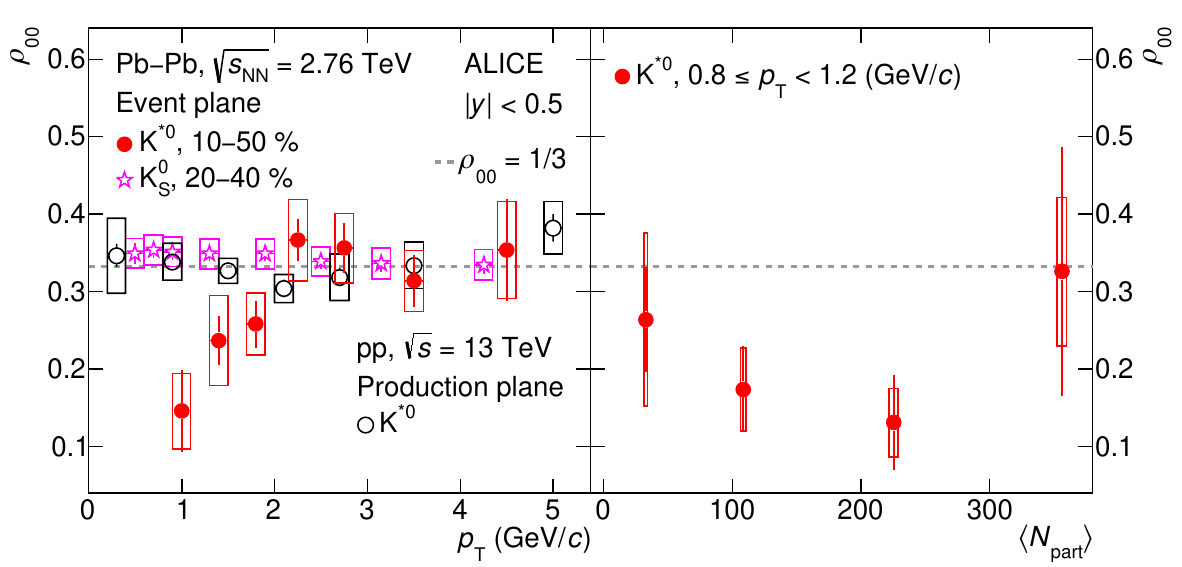}
    \end{center}
    \caption{(Left) \rh as a function of \pt for \kst in Pb--Pb and pp collisions, and for \Kzs in Pb--Pb collisions. (Right) Centrality dependence of \rh for \kst mesons in Pb--Pb collisions. From Ref.~\cite{Acharya:2019vpe}.} 
    \label{fig:SOA}
\end{figure}
The present measurements of the \rh values for \kst mesons are shown in Fig.~\ref{fig:SOA}. They are found to deviate from 1/3 at low-\pt in midcentral Pb--Pb collisions, whereas at high-\pt the \rh values are consistent with 1/3. Various systematic tests such as \rh measurements for \Kzs (meson with spin zero) in Pb--Pb and \rh measurements for \kst mesons in pp collisions (see the left panel of Fig.~\ref{fig:SOA}) yield \rh = 1/3, indicating no spin alignment as expected. The observed \pt dependence of \rh is qualitatively consistent with the expectation from the hadronisation of polarised quarks via a recombination mechanism in the presence of initial angular momentum. At low-\pta, the maximum deviation (at a 3$\sigma$ level) of \rh from 1/3 occurs in midcentral collisions, whereas in central and peripheral collisions, the measurements are consistent with 1/3 (see the right panel of Fig.~\ref{fig:SOA}). The centrality dependence of \rh is consistent with the impact-parameter dependence of initial angular momentum~\cite{Becattini:2007sr}. The expected order of magnitude of \rh can be estimated from the previously shown polarisation measurements of $\Lambda$ hyperons. In the quark recombination model~\cite{Liang:2004ph,Liang:2004xn,Liang:2007ma}, the $\Lambda$ polarisation ($P_{\Lambda}$) is equivalent to strange quark polarisation. Considering the same quark polarisation ($P_{q}$) for a light and strange quark the \rh of vector mesons are related to quark polarisation as $\rh = \frac{1-P^{2}_{q}}{3+P^{2}_{q}}$. In a non-relativistic thermal approach~\cite{Becattini:2016gvu}, the vorticity ($\omega$) and temperature ($T$) are related to $P_{\Lambda}$ and \rh as $P_{\Lambda} = \frac{\omega}{4T}$ and $\rh \simeq\frac{1}{3}\left(1-\frac{\left( \omega/T\right)^{2}}{3}\right)$. Substituting the inputs from the $\Lambda$ hyperon measurements shown previously, the estimated \rh value from these models are found to be close to 1/3. Therefore, the large effect observed for the \rh of vector mesons requires further theoretical input.

Recently, spin alignment studies were extended to the charmonium sector, by performing measurements on the angular distribution of decay muons of the J/$\psi$, in the rapidity interval $2.5<y<4$~\cite{ALICE:2022sli}. Results were obtained for Pb--Pb collisions at $\sqrt{s_{\rm NN}}=5.02$ TeV, for three transverse momentum intervals in $2<p_{\rm T}<10$ GeV/$c$ and as a function of centrality, in $2<p_{\rm T}<6$ GeV/$c$. The $\lambda_\theta\propto 1+\cos^2\theta$ quantity was studied, with $\theta$ being the polar angle emission of the positive decay muon, in the J/$\psi$ rest frame. The $\theta$ angle was measured with respect to an axis perpendicular to the event plane. 
It can be shown that the finite spin-alignment condition $\rho_{00}\ne 1/3$ is equivalent to having $\lambda_\theta\ne 0$. As a function of centrality $\lambda_\theta$ deviates from zero with a 3.5$\sigma$ significance in the 40--60\% centrality interval. For the measurement as a function of $p_{\rm T}$, the significance reaches 3.9$\sigma$ for $2<p_{\rm T}<4$ GeV/$c$ (30--50\% centrality), where $\lambda_\theta = 0.23\pm 0.05~\mbox{(stat.)} \pm 0.03~\mbox{(syst.)}$.
Having $\lambda_\theta>0$ implies a partially transverse polarisation for the J/$\psi$. For a decay involving spin-1 particles, such a positive value also implies $\rho_{00}< 1/3$, i.e. an effect similar to that observed for the K$^{*0}$ and ascribed to vorticity effects. However, the J/$\psi$ production process in nuclear collisions at the LHC is different from that of K$^{*0}$ (and $\phi$), with the charm-anticharm pair being created very early in the collision process and the bound ${\rm c\overline c}$ state being produced both in the QGP phase and at hadronisation. Due to the earlier production, charmonia may, in addition to vorticity, be more sensitive than strange hadrons to the large magnetic field present in the first instants of the Pb--Pb collision. Quantitative theory estimates, still not available, will be necessary to elucidate this possibility. 

\subsubsection{Conclusions}

\paragraph{Global collectivity and kinetic freeze-out temperatures.} The highest ever values of anisotropic and radial flow in heavy-ion collisions are achieved at the LHC. The radial-flow velocities derived from $p_{\rm T}$ spectra are up to about 70\% of the speed of light (in the lab frame), while light hadron $v_{2}$ measurements, which determine the magnitude of elliptic flow, are 30\% higher than at the top RHIC energy. The kinetic freeze-out temperature extracted from Blast-Wave parameterisations depends strongly on whether resonances are included for feed-down. The differences are on the order of 60 MeV for central collisions, yielding the range $90 < T_{\rm kin} < 150$~MeV. The extracted radial flow velocities are much less sensitive to such an inclusion.

\paragraph{Emergent features of the dynamical evolution of the QGP.} New observables measured (and in some cases developed) by ALICE over the last decade such as finite higher-order anisotropies $v_{n\geq3}$, correlations between different-order anisotropies, and a translation of the angular momentum of the QGP to the polarisation of its outgoing hadrons, reveal an extremely rich pattern of the dynamical evolution of the QGP. Higher-order anisotropies were measured up to the ninth order, and show an exponential decrease, which can be understood in the context of a finite shear viscosity over entropy ratio $\eta/s$. It was also demonstrated how higher-order anisotropies can be understood from lower contributions, with such a decomposition playing an important role in accessing the properties of the QGP just before hadronisation. The observations of correlations between different-order anisotropies and hadron polarisation offer unique constraints on the initial state, on the QGP evolution, and on the hadronisation mechanisms.

\paragraph{Hydrodynamic description.} Hydrodynamic calculations, invoking the QGP equation of state, describe a wide variety of these results. Such results include measurements sensitive to anisotropic and radial flow, the energy dependence of $\Lambda$ global polarisation values, and $\Lambda$ longitudinal polarisation measurements at the LHC. The assumed dependence of $\eta/s$ and $\zeta/s$ on temperature used in these calculations for such a description will be further discussed in Sec.~\ref{sec:QGPsummary}. In a number of cases, the original assertion assumes the modeling of the initial state is sufficiently accurate and will be addressed in Chap.~\ref{ch:InitialState}. On the other hand, the same hydrodynamic calculations cannot describe balance-function widths, which might be explained by charge diffusion effects not implemented in these calculations. The differences with respect to the data can therefore provide a crude measure of charge diffusion, a key QGP transport parameter, notwithstanding other correlation mechanisms not present in hydrodynamic models (e.g.\,mini-jets).
 
\paragraph{Global polarisation and spin alignment.} Measurements of the hyperon global polarisation at the LHC are consistent with zero, and compatible with expectations based on an extrapolation from measurements at lower collision energies at RHIC, hydrodynamic model calculations, and empirical estimates based on collision energy dependence of directed flow. On the other hand, the \kst spin alignment measurements show a potential sensitivity to the global angular momentum of the system, not seen in the polarisation measurements. These differences remain a challenge to interpret, and are in need of further theoretical attention. The significant transverse polarisation of the J/$\psi$ with respect to the event plane may provide insight on the relative contributions of the global angular momentum and of the initial magnetic field.

\newpage

\newcommand {\stat}     {({\it stat.})~}
\newcommand {\syst}     {({\it syst.})~}

\newcommand{\Fig}       {\textsc{f}ig.~}
\newcommand{\fig}       {\Fig}
\newcommand{\Figs}      {\textsc{f}igs.~}
\newcommand{\figs}      {\Figs}
\newcommand{\Figure}    {\textsc{f}igure~}
\newcommand{\Figures}   {\textsc{f}igures~}
\newcommand{\Tab}       {\textsc{t}ab.~}
\newcommand{\tab}       {\Tab}
\newcommand{\Table}     {\textsc{t}able~}
\newcommand{\Tables}    {\textsc{t}ables~}
\newcommand{\chap}      {\textsc{c}hap.~}
\newcommand{\parag}     {\textsc{p}ar.~}
\newcommand{\appdx}     {\textsc{a}pp.~}
\newcommand{\eq}        {\textsc{e}q.~}
\newcommand{\opt}        {\textsc{o}pt.~}

\newcommand{\tune}      {\emph{tune}}
\newcommand{\tunes}     {\emph{tunes}\xspace}

\newcommand{\orderOf}[1]{\ensuremath{\mathcal{O}}(#1)}
\newcommand{\gsim}{\,{\buildrel > \over {_\sim}}\,}

\newcommand{\CheckGr}   {\textcolor{Green}{\normalsize \CheckmarkBold}} 
\newcommand{\SurpriseGr}{\textcolor{Green}{\textbf{!}{\scriptsize !}}}
\newcommand{\Wtf}       {\textcolor{IndianRed}{\textbf{?}!}}
\newcommand{\NB}        {\textcolor{IndianRed}{\HandRight}}
\newcommand{\Caution}   {\textcolor{IndianRed}{\scriptsize \textlhdbend}}
\newcommand{\NoWay}     {\textcolor{IndianRed}{\normalsize \XSolidBrush}}
\newcommand{\UnderWork} {\textcolor{Blue}{\Large \dsagricultural}}
\newcommand{\ToDo}      {\textcolor{Gray}{\scriptsize \textsc{ToDo}}}

\newcommand {\pT}           {\ensuremath{p_{\rm T}}}
\newcommand {\pTlw}         {\ensuremath{p_{\rm T}^{lw} }}
\newcommand {\pTIdx}[1]     {\ensuremath{p_{\text{\rm T,#1}}}}
\newcommand {\pTExp}[1]     {\ensuremath{p_{\text{\rm T}}^{#1}}}
\newcommand {\pTIdxExp}[2]  {\ensuremath{p_{\text{\rm T,#1}}^{#2}}}
\newcommand {\pTof}[1]      {\ensuremath{p_{\text{\rm T}} \text{(#1)}}}
\newcommand {\pTchJet}      {\ensuremath{p_{\text{\rm T,jet}}^{\text{ch}}}}
\newcommand {\pTchEmJet}    {\ensuremath{p_{\text{\rm T,jet}}^{\text{ch+em}}}}
\newcommand {\meanpT}       {\ensuremath{\langle p_{\rm T} \kern-0.1em\rangle}}
\newcommand {\mean}[1]      {\ensuremath{\langle #1 \kern-0.1em\rangle}} 
\newcommand {\sqrtSnn}      {\ensuremath{\sqrt{s_{\rm NN}}}}
\newcommand {\sqrtS}        {\ensuremath{\sqrt{s}}}
\newcommand {\vTwo}         {\ensuremath{v_{\text{2}}}}
\newcommand {\vThree}       {\ensuremath{v_{\text{3}}}}
\newcommand {\vFour}        {\ensuremath{v_{\text{4}}}}
\newcommand {\vFive}        {\ensuremath{v_{\text{5}}}}
\newcommand {\vSix}         {\ensuremath{v_{\text{6}}}}
\newcommand {\vN}           {\ensuremath{v_{\text{n}}}}
\newcommand {\eT}           {\ensuremath{E_{\text{\textsc{t}}}}}
\newcommand {\mT}           {\ensuremath{m_{\text{\textsc{t}}}}}
\newcommand {\mTmZero}      {\ensuremath{m_{\text{\textsc{t}}} - m_0}}
\newcommand {\minv}         {\mbox{$m_{\ee}$}}
\newcommand {\rap}          {\mbox{$y$}}
\newcommand {\rapLab}       {\mbox{$y_{\text{lab}}$}}
\newcommand {\rapCms}       {\mbox{$y_{\text{\textsc{cms} }}$}}
\newcommand {\absrap}       {\mbox{$\left | y \right | $}}
\newcommand {\rapPart}[1]   {\mbox{$\left | y\text{(#1)} \right | $}}
\newcommand {\rapXi}        {\mbox{$\left | y(\rmXi) \right | $}}
\newcommand {\rapJpsi}      {\mbox{$y_{\tiny \rmJpsi}$}}
\newcommand {\abspseudorap} {\mbox{$\left | \eta \right | $}}
\newcommand {\pseudorap}    {\mbox{$\eta$}}
\newcommand {\pseudorapLab} {\mbox{$\eta_{\,\text{lab}}$}}
\newcommand {\pseudorapCms} {\mbox{$\eta_{\text{\textsc{cms} }}$}}
\newcommand {\cTau}         {\ensuremath{c.\tau}}
\newcommand {\sigee}        {$\sigma_E$/$E$}
\newcommand {\AxEff}        {\ensuremath{\mathscr{A}.\varepsilon}}

\newcommand{\dd}{\mathop{}\!\mathrm{d}}
\newcommand {\oneOverpipT}  {\ensuremath{1/2\pi\pT}}
\newcommand {\crossSec}[1]  {\mbox{$\sigma_{\scriptsize \rm #1}$}}
\newcommand {\visCrossSec}[1]  {\mbox{$\sigma_{\scriptsize \rm #1}^{visible}$}}
\newcommand {\dsigmady}     {\ensuremath{\mathrm{d}\sigma/\mathrm{d}y}}
\newcommand {\dsigmadpt}    {\ensuremath{\mathrm{d}^{2}\sigma/\mathrm{d}\pT}}
\newcommand {\dsigmadptdy}  {\ensuremath{\mathrm{d}^{2}\sigma/\mathrm{d}\pT\mathrm{d}y}}
\newcommand {\dsigmadptdeta}{\ensuremath{\mathrm{d}^{2}\sigma/\mathrm{d}\pT\mathrm{d}\eta}}
\newcommand {\dsigmaXdy}[1] {\ensuremath{\mathrm{d}\sigma\text{(#1)}/\mathrm{d}y}}
\newcommand {\dsigmaXdpt}[1]{\ensuremath{\mathrm{d}\sigma\text{(#1)}/\mathrm{d}\pT}}
\newcommand {\dsigmaXdptdy}[1]  {\ensuremath{\mathrm{d}^{2}\sigma\text{(#1)}/\mathrm{d}\pT\mathrm{d}y}}
\newcommand {\dNdy}         {\ensuremath{\mathrm{d}N/\mathrm{d}y}}
\newcommand {\dNXdy}[1]     {\ensuremath{\mathrm{d}N\text{(#1)}/\mathrm{d}y}}
\newcommand {\dNJpsidy}     {\dNXdy{\rmJpsi}}
\newcommand {\dNdpt}        {\ensuremath{\mathrm{d}N/\mathrm{d}\pT }}
\newcommand {\dNXdptdy}[1]  {\ensuremath{\mathrm{d}^{2}N\text{(#1)}/\mathrm{d}\pT\mathrm{d}y}}
\newcommand {\dNdptdy}      {\ensuremath{\mathrm{d}^{2}N/\mathrm{d}\pT\mathrm{d}y }}
\newcommand {\dNdptdeta}    {\ensuremath{\mathrm{d}^{2}N/\mathrm{d}\pT\mathrm{d}\eta }}
\newcommand {\dNdptdphidy}  {\ensuremath{\mathrm{d}^{3}N/\mathrm{d}\pT\mathrm{d}\varphi\mathrm{d}y}}
\newcommand {\fracdsigmadptdy}  {\ensuremath{ \frac{\mathrm{d}^{2}\sigma}{\mathrm{d}\pT\mathrm{d}y}}}
\newcommand {\fracdNdptdy}  {\ensuremath{ \frac{\mathrm{d}^{2}N}{\mathrm{d}\pT\mathrm{d}y } }}
\newcommand {\fracdNdy}     {\ensuremath{ \frac{\dN}{\dy}}}
\newcommand {\fracdNdyBold} {\ensuremath{ \frac{\bm{\dN}}{\bm{\dy}}}}

\newcommand {\dNdmtdy}      {\ensuremath{\mathrm{d}^{2}N/\mathrm{d}\mT\mathrm{d}y }}
\newcommand {\dN}           {\ensuremath{\mathrm{d}N }}
\newcommand {\Npp}          {\ensuremath{N_{\textsc{\pp}}}}
\newcommand {\dNsquared}    {\ensuremath{\mathrm{d}^{2}N }}
\newcommand {\dsquared}     {\ensuremath{\mathrm{d}^{2} }}
\newcommand {\dpT}          {\ensuremath{\mathrm{d}\pT }}
\newcommand {\dy}           {\ensuremath{\mathrm{d}y}}
\newcommand {\dNdyBold}     {\ensuremath{\bm{\dN/\dy}}}
\newcommand {\dNchdy}       {\ensuremath{\mathrm{d}N_\text{ch}/\mathrm{d}y }}
\newcommand {\dNchdeta}     {\ensuremath{\mathrm{d}N_\text{ch}/\mathrm{d}\eta }}
\newcommand {\dNchdptdeta}  {\ensuremath{\mathrm{d}^{2}N_\text{ch}/\mathrm{d}\pT\mathrm{d}\eta }}
\newcommand {\RAA}          {\ensuremath{R_\text{AA}}}
\newcommand {\RpA}          {\ensuremath{R_\text{pA}}}
\newcommand {\RpPb}         {\ensuremath{R_\text{pPb}}}
\newcommand {\RPbp}         {\ensuremath{R_\text{Pbp}}}
\newcommand {\RAuAu}        {\ensuremath{R_\text{AuAu}}}
\newcommand {\RPbPb}        {\ensuremath{R_\text{PbPb}}}
\newcommand {\Rcp}          {\ensuremath{R_\text{CP}}}
\newcommand {\hPMVzsCorrel} {\ensuremath{(\text{\hPM-V0})}}
\newcommand {\hVzsCorrel}   {\ensuremath{(\text{h-V0})}}
\newcommand {\mPDG}[1]      {\ensuremath{m_{\text{pdg}}(#1)}}

\newcommand {\Nevt}         {\ensuremath{N_\text{evt}}}
\newcommand {\NevtINEL}     {\ensuremath{N_\text{evt}(\textsc{inel})}}
\newcommand {\NevtNSD}      {\ensuremath{N_\text{evt}(\textsc{nsd})}}
\newcommand {\INEL}         {\ensuremath{\textsc{inel}}}
\newcommand {\INELZero}     {\ensuremath{\textsc{inel}>0}}
\newcommand {\NSD}          {\ensuremath{\textsc{nsd}}}
\newcommand {\dEdx}         {\ensuremath{\textup{d}E/\textup{d}x }}

\newcommand {\bsTe}      {\ensuremath{\bm{T_e}}}
\newcommand {\bsTb}      {\ensuremath{\bm{T_b}}}
\newcommand {\bsTt}      {\ensuremath{\bm{T_T}}}
\newcommand {\bsCt}      {\ensuremath{\bm{C_T}}}
\newcommand {\bsNt}      {\ensuremath{\bm{n}}}
\newcommand {\bsQt}      {\ensuremath{\bm{q}}}
\newcommand {\bsVt}      {\ensuremath{\bm{V}}}
\newcommand {\bsgVt}     {\ensuremath{\bm{g.V}}}
\newcommand {\bsNb}      {\ensuremath{\bm{n}}}
\newcommand {\bsFp}      {\ensuremath{\bm{f_P}}}
\newcommand {\bsCp}      {\ensuremath{\bm{C_P}}}
\newcommand {\TpseudoCritic}    {\ensuremath{T_\text{pc}}}
\newcommand {\Tchem}     {\ensuremath{T_\text{chem}}}
\newcommand {\Tkin}     {\ensuremath{T_\text{kin}}}

\newcommand {\Lint}         {\ensuremath{L_{\text{int}}}}

\newcommand {\rphi}         {\mbox{\ensuremath{(r,\varphi)}}}
\newcommand {\alphaS}       {\ensuremath{ \alpha_s}}
\newcommand {\chLeptonAsymm}{\ensuremath{ A_{\ell\ell} }}

\newcommand {\MeanNpart}    {\mbox{\ensuremath{\langle\kern-0.05em N_{part} \kern-0.05em \rangle}}}
\newcommand {\MeanNcoll}    {\mbox{\ensuremath{\langle\kern-0.05em N_{coll} \kern-0.05em \rangle}}}
\newcommand {\sigmaBarlow}  {\ensuremath{\sigma_{Barlow}}}
\newcommand {\sigmaStat}    {\ensuremath{\sigma_{stat}}}
\newcommand {\sigmaSyst}    {\ensuremath{\sigma_{syst}}}
\newcommand {\sigmaTot}     {\ensuremath{\sqrt{\sigmaStat^2 + \sigmaSyst^2}}}

\newcommand {\xXzero}     {\ensuremath{\textsc{x}/X_0}}

\newcommand {\eX}    {\ensuremath{\vec{e}_{\textsc{x}}}}
\newcommand {\eY}    {\ensuremath{\vec{e}_{\textsc{y}}}}
\newcommand {\eZ}    {\ensuremath{\vec{e}_{\textsc{z}}}}

\newcommand {\ePhi}  {\ensuremath{\vec{e}_{\varphi}}}
\newcommand {\eR}    {\ensuremath{\vec{e}_{r}}}

\newcommand{\Sherpa}        {\textsc{Sherpa}\xspace}
\newcommand{\Herwig}        {\textsc{Herwig7}\xspace}
\newcommand{\Epos}          {\textsc{Epos}\xspace}
\newcommand{\Pythia}        {\textsc{Pythia}\xspace}
\newcommand{\Pythiaeight}   {\Pythia~8\xspace}
\newcommand{\Rivet}         {\textsc{Rivet}\xspace}
\newcommand{\HepMC}         {\textsc{HepMc}\xspace}

\newcommand{\Root}          {\textsc{Root}\xspace}
\newcommand{\GeantFour}     {\textsc{Geant4}\xspace}
\newcommand{\Fluka}         {\textsc{Fluka}\xspace}

\newcommand {\pp}        {\ensuremath{\mbox{\text {p\kern-0.05em p}}}\xspace}
\newcommand {\rmpp}      {\ensuremath{\text{p\kern-0.05em p} }}
\newcommand {\ppbar}     {\mbox{$\text{p}\overline{\text{p}}$}}
\newcommand {\PbPb}      {\ensuremath{\mbox{\text{Pb--Pb}} }}
\newcommand {\rmPbPb}    {\ensuremath{\text{PbPb} }}
\newcommand {\AuAu}      {\ensuremath{\mbox{\text{Au--Au}} }}
\newcommand {\ArAr}      {\ensuremath{\mbox{\text{Ar--Ar}} }}
\newcommand {\CuCu}      {\ensuremath{\mbox{\text{Cu--Cu}} }}
\newcommand {\UU}        {\ensuremath{\mbox{\text{U--U}}   }}
\newcommand {\XeXe}      {\ensuremath{\mbox{\text{Xe--Xe}} }}
\renewcommand {\AA}      {\ensuremath{\text{A--A}          }} %
\newcommand {\rmAA}      {\ensuremath{\text{AA}            }} %
\newcommand {\pA}        {\ensuremath{\mbox{\text{p--A}}   }}
\newcommand {\rmpA}      {\ensuremath{\text{pA}            }} %
\newcommand {\dA}        {\ensuremath{\mbox{\text{d--A}}   }}
\newcommand {\pPb}       {\ensuremath{\mbox{\text{p--Pb}}  }}
\newcommand {\Pbp}       {\ensuremath{\mbox{\text{Pb--p}}  }}
\newcommand {\dAu}       {\ensuremath{\mbox{\text{d--Au}}  }}
\newcommand {\pAu}       {\ensuremath{\mbox{\text{p--Au}}  }}
\newcommand {\ePMproton}             {\ensuremath{\mbox{$\text{e}^{\pm}\text{p}$}}}
\newcommand {\EplusEminus}      {\ee}
\newcommand {\ee}               {\mbox{$\text{e}^+\text{e}^-$}}
\newcommand {\Eplus}            {\mbox{$\text{e}^+$}}
\newcommand {\Eminus}           {\mbox{$\text{e}^-$}}
\newcommand {\MuPlusMuMinus}    {\mbox{$\mu^+\mu^-$}}
\newcommand {\MuPlus}           {\mbox{$\mu^+$}}
\newcommand {\MuMinus}          {\mbox{$\mu^-$}}

\newcommand {\massStyle}[1] {\mbox{\ensuremath{\text{#1}\kern-0.1em /\kern-0.12em c^2}}}
\newcommand {\mass}     {\massStyle{MeV}}
\newcommand {\kmass}    {\massStyle{keV}}
\newcommand {\mmass}    {\massStyle{MeV}}
\newcommand {\gmass}    {\massStyle{GeV}}

\newcommand {\unitStyle}[1] {\mbox{\ensuremath{\text{#1}}}}
\newcommand {\tev}      {\unitStyle{TeV}}
\newcommand {\gev}      {\unitStyle{GeV}}
\newcommand {\mev}      {\unitStyle{MeV}}
\newcommand {\kev}      {\unitStyle{keV}}
\newcommand {\J}        {\unitStyle{J}}

\newcommand {\momStyle}[1] {\mbox{\ensuremath{\text{#1}\kern-0.1em /\kern-0.12em c}}}
\newcommand {\mmom}     {\momStyle{MeV}}
\newcommand {\gmom}     {\momStyle{GeV}}

\newcommand {\fsec}       {\unitStyle{fs}}
\newcommand {\psec}       {\unitStyle{ps}}
\newcommand {\nsec}       {\unitStyle{ns}}
\newcommand {\musec}      {\mbox{$\mu\unitStyle{s}$}}
\newcommand {\millisec}   {\unitStyle{ms}}
\newcommand {\second}     {\unitStyle{s}}

\newcommand {\MHz}     {\unitStyle{MHz}}
\newcommand {\kHz}     {\unitStyle{kHz}}

\newcommand {\fmC}      {\mbox{$\unitStyle{fm}/\kern-0.12em c$}}

\newcommand {\fm}       {\unitStyle{fm}}
\newcommand {\nm}       {\unitStyle{nm}}
\newcommand {\mum}      {\mbox{$\mu\unitStyle{m}$}}
\newcommand {\mm}       {\unitStyle{mm}}
\newcommand {\cm}       {\unitStyle{cm}}
\newcommand {\m}        {\unitStyle{m}}

\newcommand {\cmq}      {\mbox{$\cm^2$}}
\newcommand {\mmq}      {\mbox{$\mm^2$}}
\newcommand {\mumq}     {\mbox{$\mum^2$}}

\newcommand {\fmCube}   {\mbox{\unitStyle{fm}$^3$}}

\newcommand {\mug}      {\mbox{$\mu\unitStyle{g}$}}
\newcommand {\mg}       {\unitStyle{mg}}
\newcommand {\gram}     {\unitStyle{g}}
\newcommand {\kg}       {\unitStyle{kg}}

\newcommand {\dens}     {\mbox{$\unitStyle{g}/\unitStyle{cm}^{3}$}}

\newcommand {\dg}       {\mbox{$\kern+0.1em ^\circ$}}

\newcommand {\lumi}     {\mbox{$\cm^{-2}\second^{-1}$}}

\newcommand {\barn}     {\unitStyle{b}}
\newcommand {\fb}       {\unitStyle{fb}}
\newcommand {\pb}       {\unitStyle{pb}}
\newcommand {\nb}       {\unitStyle{nb}}
\newcommand {\mub}      {\mbox{$\mu\unitStyle{b}$}}
\newcommand {\mb}       {\unitStyle{mb}}
\newcommand {\kb}       {\unitStyle{kb}}

\newcommand {\invmub}   {\mbox{$\mub^{-1}$}}
\newcommand {\invnb}    {\mbox{$\nb^{-1}$}}
\newcommand {\invpb}    {\mbox{$\pb^{-1}$}}
\newcommand {\invfb}    {\mbox{$\fb^{-1}$}}

\newcommand{\hPM}           {\ensuremath{h^{\pm}}}
\newcommand{\ePlusMinus}    {\mbox{$\mathrm {e^{\pm}}$}}
\newcommand{\muPlusMinus}   {\mbox{$\mathrm {\mu^{\pm}}$}}

\newcommand{\piZero}        {\mbox{$\mathrm {\pi^0}$}}
\newcommand{\piMinus}       {\mbox{$\mathrm {\pi^-}$}}
\newcommand{\piPlus}        {\mbox{$\mathrm {\pi^+}$}}
\newcommand{\piPlusMinus}   {\mbox{$\mathrm {\pi^{\pm}}$}}
\newcommand{\Pion}          {\mbox{$\mathrm \pi$}}
\newcommand{\rmPiPlusMinus} {\piPlusMinus}
\newcommand{\rmPiPM}        {\piPlusMinus}
\newcommand{\rmPiMinus}     {\piMinus}
\newcommand{\rmPiPlus}      {\piPlus}

\newcommand{\Kzs}        {\mbox{$\mathrm {K^0_S}$}}
\newcommand{\rmKzero}    {\Kzs}
\newcommand{\Kzl}        {\mbox{$\mathrm {K^0_L}$}}
\newcommand{\rmKzeroL}   {\Kzl}
\newcommand{\Kminus}     {\mbox{$\mathrm {K^-}$}}
\newcommand{\rmKminus}   {\Kminus}
\newcommand{\Kplus}      {\mbox{$\mathrm {K^+}$}}
\newcommand{\rmKplus}    {\Kplus}
\newcommand{\Kplusmin}   {\mbox{$\mathrm {K^{\pm}}$}}
\newcommand{\rmKpm}      {\Kplusmin}
\newcommand{\rmRhoZero}  {\mbox{$\mathrm{\rho(770)}^0$}}
\newcommand{\rmKstar}    {\mbox{$\mathrm{K}^*\mathrm{(892)}^0$}}
\newcommand{\rmPhiMes}   {\mbox{$\mathrm {\phi(1020)}$}}
\newcommand{\rmPhi}      {\mbox{$\mathrm {\phi}$}}

\newcommand{\proton}    {\mbox{$\mathrm {p}$}}
\newcommand{\pbar}      {\mbox{$\mathrm {\overline{p}}$}}
\newcommand{\pOrPbar}   {\mbox{$\mathrm {p^{\pm}}$}}
\newcommand{\rmPpm}     {\pOrPbar}

\newcommand{\rmLambdaZ}         {\mbox{$\mathrm {\Lambda}$}}
\newcommand{\rmAlambdaZ}        {\mbox{$\mathrm {\overline{\Lambda}}$}}
\newcommand{\rmLambda}          {\mbox{$\mathrm {\Lambda}$}}
\newcommand{\rmAlambda}         {\mbox{$\mathrm {\overline{\Lambda}}$}}
\newcommand{\rmLambdas}         {\mbox{$\mathrm {\Lambda \kern-0.2em + \kern-0.2em \overline{\Lambda}}$}}
\newcommand{\ratioLamOverKzs}   {\rmLambda/\rmKzero}

\newcommand{\rmSigma}       {\mbox{$\mathrm {\Sigma}$}}
\newcommand{\rmSigmaM}      {\mbox{$\mathrm{\Sigma}^{-}$}}
\newcommand{\rmSigmaP}      {\mbox{$\mathrm{\Sigma}^{+}$}}
\newcommand{\rmSigmaZero}   {\mbox{$\mathrm{\Sigma}^{0}$}}
\newcommand{\rmSigmaMres}   {\mbox{$\mathrm{\Sigma(1385)}^{-}$}}
\newcommand{\rmSigmaPres}   {\mbox{$\mathrm{\Sigma(1385)}^{+}$}}

\newcommand{\rmLambdaStar}  {\mbox{$\mathrm {\Lambda(1520)}$}}

\newcommand{\rmXi}      {\mbox{$\mathrm{\Xi}$}}
\newcommand{\rmXiM}     {\mbox{$\mathrm{\Xi}^{-}$}}
\newcommand{\rmXiPM}    {\mbox{$\mathrm {\Xi^{\pm}}$}}

\newcommand{\rmAxiP}    {\mbox{$\mathrm {\overline{\Xi}^{+}}$}}
\newcommand{\rmXis}     {\mbox{$\mathrm {\Xi^{-} \kern-0.3em + \kern-0.1em \overline{\Xi}^{+}}$}}
\newcommand{\rmXiZero}  {\mbox{$\mathrm {\Xi^{0}}$}}
\newcommand{\rmXiZ}     {\rmXiZero}
\newcommand{\rmXiZres}  {\mbox{$\mathrm {\Xi (1530)^{0}}$}}
\newcommand{\rmAxiZres} {\mbox{$\mathrm {\overline{\Xi} (1530)^{0}}$}}
\newcommand{\rmXiMres}  {\mbox{$\mathrm {\Xi (1530)^{-}}$}}
\newcommand{\rmAxiPres} {\mbox{$\mathrm {\overline{\Xi}(1530)^{+}}$}}

\newcommand{\rmOmega}   {\mbox{$\mathrm {\Omega}$}}
\newcommand{\rmOmegaM}  {\mbox{$\mathrm {\Omega^{-}}$}}
\newcommand{\rmAomegaP} {\mbox{$\mathrm {\overline{\Omega}^{+}}$}}
\newcommand{\rmOmegas}  {\mbox{$\mathrm {\Omega^{-} \kern-0.3em +  \kern-0.1em \overline{\Omega}^{+}}$}}
\newcommand{\rmOmegaPM} {\mbox{$\mathrm {\Omega^{\pm}}$}}

\newcommand{\rmDeuton}   {\mbox{$\mathrm {d}$}}
\newcommand{\rmDeutonPM} {\mbox{$\mathrm {d}^{\pm}$}}
\newcommand{\rmTriton}   {\mbox{$\mathrm {t}$}}
\newcommand{\rmTritonPM} {\mbox{$\mathrm {t}^{\pm}$}}
\newcommand{\rmHeThree}  {\mbox{$\mathrm {^3He}$}}
\newcommand{\rmHeThreePM}{\mbox{$\mathrm {^3He^{2\pm}}$}}
\newcommand{\rmHeFour}   {\mbox{$\mathrm {^4He}$}}
\newcommand{\rmHeFourPM} {\mbox{$\mathrm {^4He^{2\pm}}$}}

\newcommand{\rmHypertriton}  {\mbox{$^{3}_{\Lambda}\mathrm{H}$}}

\newcommand{\rmJpsi}    {\mbox{$\mathrm{J\kern-0.05em /\kern-0.05em\psi}$}}
\newcommand{\rmPsiTwoS} {\mbox{$\mathrm {\psi(2S)}$}}
\newcommand{\rmChicZero}{\mbox{$\mathrm {\chi_{c_0}}$}}
\newcommand{\rmChicOne} {\mbox{$\mathrm {\chi_{c_1}}$}}
\newcommand{\rmChicTwo} {\mbox{$\mathrm {\chi_{c_2}}$}}
\newcommand{\rmChicJ}   {\mbox{$\mathrm {\chi_{c_J}}$}}

\newcommand{\rmLambdaC}         {\mbox{$\mathrm {\Lambda_{c}^{+}}$}}
\newcommand{\rmXiCplus}         {\mbox{$\mathrm {\Xi_{c}^{+}}$}}
\newcommand{\rmXiCzero}         {\mbox{$\mathrm {\Xi_{c}^{0}}$}}
\newcommand{\rmOmegaCzero}      {\mbox{$\mathrm {\Omega_{c}^{0}}$}}
\newcommand{\rmXiCCtwoPlus}     {\mbox{$\mathrm {\Xi_{cc}^{2+}}$}}
\newcommand{\rmOmegaCCtwoPlus}  {\mbox{$\mathrm {\Omega_{ccc}^{2+}}$}}

\newcommand{\rmDzero}   {\mbox{$\mathrm {D}^{0}$}}
\newcommand{\rmDzeroBar}{\mbox{$\mathrm {\overline{D}}^{0}$}}
\newcommand{\rmDplus}   {\mbox{$\mathrm {D}^{+}$}}
\newcommand{\rmDminus}  {\mbox{$\mathrm {D}^{+}$}}
\newcommand{\rmDpm}     {\mbox{$\mathrm {D}^{\pm}$}}
\newcommand{\rmDstar}   {\mbox{$\mathrm{D}^*\mathrm{(2010)}^+$}}
\newcommand{\rmDs}      {\mbox{$\mathrm {D^{+}_{s}}$}}
\newcommand{\rmBs}      {\mbox{$\mathrm {B^{0}_{s}}$}}

\newcommand{\rmBzero}       {\mbox{$\mathrm {B^{0}}$}}
\newcommand{\rmBplus}       {\mbox{$\mathrm {B^{+}}$}}
\newcommand{\rmBminus}      {\mbox{$\mathrm {B^{-}}$}}
\newcommand{\rmBplusMinus}  {\mbox{$\mathrm {B^{\pm}}$}}
\newcommand{\rmBzeroS}      {\mbox{$\mathrm {B^{0}_s}$}}

\newcommand{\rmUpsOneS}         {\mbox{$\mathrm {\Upsilon(1S)}$}}
\newcommand{\rmUpsTwoS}         {\mbox{$\mathrm {\Upsilon(2S)}$}}
\newcommand{\rmUpsThreeS}       {\mbox{$\mathrm {\Upsilon(3S)}$}}
\newcommand{\rmUpsTwoThreeS}    {\mbox{$\mathrm {\Upsilon(2S,3S)}$}}
\newcommand{\rmUpsnS}           {\mbox{$\mathrm {\Upsilon(nS)}$}}
\newcommand{\rmUpsOneTwoThreeS} {\mbox{$\mathrm {\Upsilon(1S,2S,3S)}$}}

\newcommand{\rmPhoton}      {\mbox{$\mathrm {\gamma}$}}
\newcommand{\rmWplus}       {\mbox{$\mathrm {W^{+}}$}}
\newcommand{\rmWminus}      {\mbox{$\mathrm {W^{-}}$}}
\newcommand{\rmWPlusMinus}  {\mbox{$\mathrm {W^{\pm}}$}}
\newcommand{\rmZzero}       {\mbox{$\mathrm {Z}$}}

\newcommand{\qqbar}             {\mbox{$q\overline{q}$}}
\newcommand{\ccbar}             {\mbox{$c\overline{c}$}}
\newcommand{\bbbar}             {\mbox{$b\overline{b}$}}
\newcommand{\DDbar}             {\mbox{$\mathrm {D\overline{D}}$}}

\newcommand{\trento}            {\mbox{T$_{\mathrm{R}}$ENTo}}

\subsection{Hadronisation of the QGP}
\label{sec:QGPHadronization}

\label{sec:IntroTG05-QGPHadronisation}

While the QGP formed from a heavy-ion collision, with a finite energy,
pursues its explosive expansion, 
the energy density of the medium reaches the pseudo-critical level of the transition 
($\approx 0.5-1~\gev/\fmCube$ according to lattice-QCD calculations~\cite{Bazavov:2009zn, Bazavov:2018mes}).
The \emph{hadronisation} of the plasma, i.e.~the process of hadron formation from the previous partonic phase, starts and makes the transition from the deconfined medium to hadronic matter; 
such a process leads to a system which is still collective but has different degrees of freedom: a \emph{hadron gas}. 
The hadronisation, in which the confinement phenomenon of QCD sets in, involves quark and gluon processes characterised by small momentum transfers and hence large values of the strong coupling constant, such that a perturbative approach is not applicable.
Thus, one must resort to phenomenological models, such as the statistical hadronisation models and the recombination models presented in this section. These models describe successfully several properties of the final-state hadrons produced in heavy-ion collisions, suggesting that they can capture some key features of the process of hadron formation~\cite{Becattini:2009fv}.
After the hadronisation, which at LHC energies occurs 7--10~\fmC\ (see Sec.~\ref{sec:TG1sizelifetime}) after the initial collision, 
the created hadrons can still interact via inelastic processes,
implying that the overall chemical composition can evolve further.
Such interactions live on, until the temperature of the \emph{chemical freeze-out} is reached,
i.e.~until the moment at which the hadronic species and their respective populations become settled.
If the most vigorous inelastic interactions have markedly ceased by then, 
the mildest inelastic ones can continue, affecting at most the resonance population.
In parallel, though, significant elastic collisions between hadrons still occur,
altering the momentum distributions of any given species.
Such momentum transfers persist down to the \emph{kinetic freeze-out}, 
occurring at a typical time of $\gsim 10$ \fmC~\cite{Aamodt:2011mr}.
After this ultimate freeze-out, the system vanishes into free-streaming hadrons that propagate towards the detection apparatus.

The late stages of the collision that are presented here come along with several questions.
To illustrate the current spurs, a few typical questions can be underlined: 
~\emph{i)} what is the level of chemical equilibration and/or kinetic thermalisation achieved by the hadron populations, be it for the light-flavour or, conversely, for the heavy-flavour species?
,~\emph{ii)} to what extent is the temperature for the chemical freeze-out, \Tchem, a unique temperature, shared by any hadron species?
,~\emph{iii)} what is the level of interplay among the quark flavours at the hadronisation stage?
,~\emph{iv)} what is the level of interplay between the momentum domains connecting the soft and hard sectors?
,~\emph{v)} similarly to \Tchem, can we assume a univocal value of \Tkin{}, the temperature of a sudden kinetic freeze-out?
and~\emph{vi)} how long does the hadronic phase last?

In concrete terms, the properties of both aspects, the hadronisation itself as well as the resulting medium, 
can be studied in high-energy nuclear collisions utilising a set of observables 
essentially related to the production of the different hadron species.
To that end, the total multiplicity per rapidity unit \dNdy\ together with the differential distributions 
(in momentum, \dNdptdy\ and/or in azimuth, \dNdptdphidy) can be analysed. This will be discussed further in the following subsections.

\subsubsection{Hadron gas composition and QCD thermodynamics}
\label{sec:SHM}

The measurement of the multiplicity per rapidity unit \dNdy\ of different hadron species provides access to the chemical composition of the hadron gas at the stage where the aforementioned inelastic collisions cease and the abundances of the different hadron species are frozen (except for resonance decays).
QCD thermodynamics in the confined phase, below the pseudo-critical temperature, can be well approximated as an \emph{ideal} hadron-resonance gas (HRG) of all known hadrons and resonances~\cite{Karsch:2003vd,HotQCD:2012fhj}.
The relative abundances of particle species measured in heavy-ion collisions over a broad centre-of-mass energy range were observed to closely follow the equilibrium populations of a HRG~\cite{Andronic:2017pug}.
This is very suggestive of a thermal origin of particle production, which implies that the properties of the fireball are governed by statistical QCD, and allows for the calculation of the chemical composition of the system within the framework of the statistical hadronisation approach. This approach is based on the partition function of the HRG and on the assumption that 
the hadro-chemical composition of the fireball is frozen out on a hypersurface defined by a uniform temperature~\cite{Cleymans:1992zc,BraunMunzinger:1999qy,Becattini:2000jw,BraunMunzinger:2001ip,Andronic:2017pug}.
This %
allows for the determination of thermodynamic parameters characterising the chemical freeze-out, which are relevant for the QCD phase diagram.
In particular, such freeze-out can be characterised by only three macroscopic parameters: the temperature \Tchem\ and the volume $V$ of the fireball, together with the baryochemical potential $\mu_{\rm B}$, which guarantees the conservation on average of the baryon number in the grand-canonical formulation of the partition function (other chemical potentials are fixed by the conservation laws). Notably, the statistical hadronisation model with a canonical formulation, incorporating local quantum number conservation, can also describe the measured yields of hadron species in pp and e$^+$e$^-$ collisions~\cite{Becattini:1995if,Becattini:1996gy,Andronic:2008ev}, albeit with a worse $\chi^2/$ndf compared to heavy-ion reactions. In small collision systems, a fireball in thermal and chemical equilibrium is not formed; for instance, in e$^+$e$^-$  annihilation, quark--antiquark pairs are produced and subsequently fragment into jets of hadrons. The observation of statistical features in these cases suggests that the underlying hadronisation process (i.e.\ the way an excited system populates the hadronic states) follows for a good part statistical laws, which are governed by entropy maximization and determined by phase space dominance. Moreover, the temperatures resulting from statistical model fits to pp and e$^+$e$^-$ data (of about 175--180 MeV~\cite{Becattini:1995if,Becattini:1996gy}) exceed the pseudocritical temperature from lattice QCD calculations, indicating that this case is different from the hadronisation of a QGP at the crossover boundary.

\begin{figure}[tb!]
    \begin{center}
        \includegraphics[width=0.9\textwidth, angle=0, clip=true, trim=0cm 0 0 0]{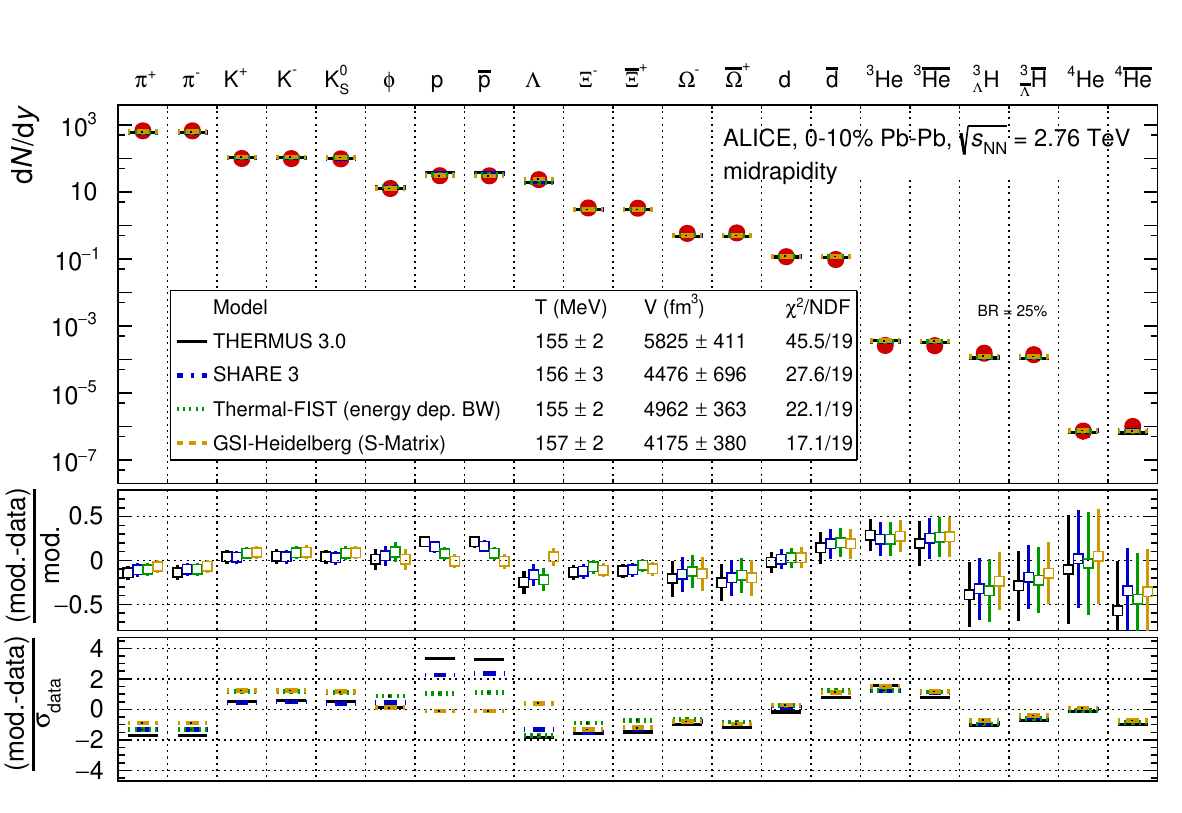}
    \caption{Measured multiplicity per unit of rapidity of different hadron species and light nuclei~\cite{Abelev:2013vea,Abelev:2013xaa,ABELEV:2013zaa,Abelev:2014uua,Adam:2015vda,ALICE:2015oer,Acharya:2017bso} compared to SHM fits from 
    THERMUS~\cite{Wheaton:2011rw, Cleymans:2006xj, Wheaton:2011rw}, 
    SHARE~\cite{Torrieri:2004zz, Torrieri:2006xi, Petran:2013dva, Petran:2013lja},
    Thermal-FIST~\cite{Vovchenko:2018fmh,Vovchenko:2019pjl}, and
    GSI-Heidelberg~\cite{BraunMunzinger:2003zd,Andronic:2005yp,Andronic:2018qqt}. 
    Differences between the model calculations and the measured yields are shown in the bottom panels. The hypertriton yield is obtained using a theoretical estimation of BR~=~25\% for the branching ratio of the ${}^3_\Lambda{\rm H} \rightarrow {}^3{\rm He}\,\pi^{-}$ decay channel~\cite{Kamada:1997rv}.}
    \label{fig:yieldsLF-SHM}
    \end{center}
\end{figure}

\paragraph{\textbf{\textit {Light-flavour hadron yields and chemical freeze-out properties.}}} 
The multiplicities of hadron species containing only light (u, d and s) quarks measured at midrapidity in central (0--10\%) \PbPb\ collisions at \sqrtSnn{}~=~2.76~\tev{} are shown in Fig.~\ref{fig:yieldsLF-SHM} for non-strange and strange mesons (\piPlusMinus{}~\cite{Abelev:2013vea}, \Kplusmin{}~\cite{Abelev:2013vea}, \Kzs{}~\cite{Abelev:2013xaa}, \rmPhiMes{}~\cite{Abelev:2014uua}) and baryons (\proton~\cite{Abelev:2013vea}, \rmLambda{}~\cite{Abelev:2013xaa}) including multi-strange hyperons (\rmXiPM{}~\cite{ABELEV:2013zaa}, \rmOmegaPM~\cite{ABELEV:2013zaa}), as well as light nuclei (d, $^3$He~\cite{Adam:2015vda} and $^4$He~\cite{Acharya:2017bso})\footnote{The yields of antinuclei $\overline{\mathrm{d}}$ and $^3\overline{\mathrm{He}}$ have been derived based on published data~\cite{Adam:2015vda}.} and ${}^{3}_{\Lambda}$H hypernuclei~\cite{ALICE:2015oer} and their antiparticles.
The particle yields at midrapidity, integrated over the full \pT{} range, were obtained from their measured \pT-differential distributions, which were fitted individually with a blast-wave function~\cite{Schnedermann:1993ws}, used for the  extrapolation to zero \pT{} (except for $^4$He and ${}^{3}_{\Lambda}$H for which the blast-wave parameters from the combined fit to d and $^3$He were used).
The yields were then calculated by adding to the integral of the data in the measured \pT{} region, the integral of the fit function outside that region.
The fraction of extrapolated yield is small for most hadron species (ranging from 4\% for protons to 20\% for \rmXiPM{}) and it is larger for ${}^{3}_{\Lambda}$H (about 28\%) and \rmOmegaPM{} (about 50\%).
Particle and antiparticle measurements are reported separately wherever possible to underline that matter and antimatter are observed to be produced in equal amounts within experimental uncertainties at LHC energies~\cite{Aamodt:2010dx}.

The measured multiplicities are well described by statistical hadronisation models (SHM),
as shown in Fig.~\ref{fig:yieldsLF-SHM}, where the data are compared to the results of four different implementations of such models, namely 
THERMUS~\cite{Wheaton:2011rw, Cleymans:2006xj, Wheaton:2011rw}, 
SHARE~\cite{Torrieri:2004zz, Torrieri:2006xi, Petran:2013dva, Petran:2013lja},
Thermal-FIST~\cite{Vovchenko:2018fmh,Vovchenko:2019pjl}, and
GSI-Heidelberg~\cite{BraunMunzinger:2003zd,Andronic:2005yp,Andronic:2018qqt}.
The SHM describes the yields of all the measured species over nine orders of magnitude in abundance values.
Among the parameters, the value for the baryochemical potential has been fixed to $\mu_{\rm B} \approx 0$ in most models, given the almost equal abundances of particles and antiparticles at LHC energies; as a consequence, the corresponding fits are performed using the average of particle and antiparticle yields. An exception is the GSI-Heidelberg model, where $\mu_{\rm B}$ is a free parameter, determined to be zero with an uncertainty of about 4~\mev.
For a comparison on equal basis, the $\chi^2$/NDF value was recalculated for the other models considering both particles and antiparticles.
The chemical freeze-out temperature obtained from the fit is $\Tchem \approx 156$~\mev, with uncertainties of 2--3~\mev{} and very small differences ($\pm$1~\mev) among the different implementations of the SHM. 
It should be noted that \Tchem{} is very close to the pseudo-critical temperature \TpseudoCritic~=~156--158~\mev{} for the transition from the QGP to the hadron gas obtained from lattice-QCD calculations for $\mu_{\rm B}=0$~\cite{Bazavov:2018mes,Borsanyi:2020fev}.
As pointed out in~\cite{Braun-Munzinger:2003htr}, this can be related to the rapid drop of the particle density and, consequently, of the multi-particle scattering rates with decreasing temperature. In particular, multi-hadron scatterings are expected to be substantial only in the vicinity of \TpseudoCritic, while for temperatures lower than \Tchem{} only two-particle interactions and decays play a relevant role, but they are too slow to maintain the system in equilibrium catching-up with the decreasing temperature.
Therefore, the chemical freeze-out hallmarks a moment near the hadronisation of the QGP itself.
The last parameter estimated from the SHM fits is the volume of the fireball for one unit of rapidity at the chemical freeze-out, which is of about 4500 fm$^3$ in SHARE, Thermal-FIST, and GSI-Heidelberg results. 
A significantly larger volume is obtained from the fits with THERMUS, which comprises an excluded-volume (Van-der-Waals like) correction~\cite{Rischke:1991ke} to account for the short-range repulsive interactions between hadrons, resulting in a lower particle density in the fireball.
As discussed in more detail in Sec.~\ref{sec:QGPsummary}, the volume of one unit of rapidity at chemical freeze-out extracted from the SHM fit can not be directly compared to the homogeneity region estimated from the femtoscopy measurements described in Sec.~\ref{sec:TG1sizelifetime}, since the latter does not represent the source volume at a precise instant during the fireball evolution.
A SHM fit to the yields of pions, kaons, and protons measured in \PbPb\ collisions at \sqrtSnn{}~=~5.02~\tev{}~\cite{Acharya:2019yoi} gives a larger volume and a value of \Tchem{} compatible with the one observed at \sqrtSnn{}~=~2.76~\tev{}, as expected from the trend established from the statistical hadronisation analysis of the measured hadron yields at lower collision energies, which shows that \Tchem{} increases with increasing energy for \sqrtSnn{}~$<$~20~\gev{} and saturates at a value of 155--160~\mev{} for higher \sqrtSnn{}~\cite{Andronic:2017pug}.

Experimental information on the QCD phase boundary, complementary to the one extracted from the analysis of hadron yields within the SHM, can be obtained from measurements of event-by-event fluctuations of conserved quantum charges and their probability distributions (see e.g.~\cite{Braun-Munzinger:2014lba}).
The second-order fluctuations can be directly connected to susceptibilities, defined as the second-order derivatives of thermodynamic pressure, which can be calculated on the lattice, thus providing a direct link between measurements and predictions from lattice-QCD calculations.
Second-order fluctuations and correlations of conserved charges, which are established at the pseudo-critical temperature \TpseudoCritic{}, can be extracted from data notably if the number of particles $N_{\rm q}$ and antiparticles $N_{\overline{\rm q}}$ are uncorrelated and both distributed according to Poisson statistics.
In this case, the probability $P(N_{\rm q}-N_{\overline{\rm q}})$ becomes a Skellam distribution, which is completely determined by the average values $\langle N_{\rm q} \rangle$  and $\langle N_{\overline{\rm q}} \rangle$.
So, under the assumption of Skellam distribution (which is supported by event-by-event measurements of net-proton fluctuations from the STAR Collaboration at RHIC~\cite{Aggarwal:2010wy,Adamczyk:2013dal} and by ALICE at the LHC~\cite{Acharya:2019izy}), inclusive measurements of particle yields can provide access to susceptibilities, which can be directly compared with lattice-QCD predictions at \TpseudoCritic{}~\cite{HotQCD:2012fhj,Bazavov:2014xya}. 
The fluctuations of conserved charges (e.g.~in net baryon and net strangeness numbers) extracted from the measured yields of different hadron species 
(\Kplusmin, \Kzs, \rmKstar, \rmPhiMes, \proton+\pbar, \rmLambdas, \rmXiPM, \rmOmegaPM) in \PbPb\ collisions at \sqrtSnn{}~=~2.76~\tev{} were calculated in~\cite{Braun-Munzinger:2014lba}. They were found to be consistent with the susceptibilities from lattice-QCD calculations in the interval of pseudo-critical temperature $150~<~\TpseudoCritic~\leq~163$~\mev, comparable to the chemical freeze-out temperature obtained from the SHM fit to the hadron yields. 
This indicates that the hadrons produced in central \PbPb{} collisions at the LHC originate from the hadronisation of the QGP.

\paragraph{\textbf{\textit {Production of protons, nuclei, and strange particles.}}} 
A significant deviation between the measured yields and the SHM calculations can be seen  in Fig.~\ref{fig:yieldsLF-SHM} for protons in the THERMUS and SHARE models, which are based on the ideal HRG implementation consisting of non-interacting hadrons and resonances. 
Such calculations predict a 25\% higher proton yield as compared to the measured one, resulting in a data-to-model difference of about 2--3.5 standard deviations.
This tension between the data and the equilibrium value of the p/$\pi$ ratio has been dubbed the ``proton-yield anomaly'' and different explanations were proposed for it in the past years.
It was argued that it was possibly connected to the annihilation of baryons and antibaryons in the hadronic phase after the chemical freeze-out~\cite{Steinheimer:2012rd,Becattini:2012xb}.
Using simulations with the microscopic transport model for hadrons, UrQMD~\cite{Bleicher:1999xi, Bass:1998ca}, correction factors were determined for the effects of interactions before the kinetic freeze-out, resulting in an improved description of the data, but with a value of the hadrochemical equilibrium temperature of about 165~\mev{}~\cite{Becattini:2014hla}.
The effect of baryon-antibaryon annihilation was also studied in the framework of a hydrodynamic calculation forking to UrQMD for the hadronic phase, showing that the annihilations help in achieving a better description of the data~\cite{Sinyukov:2017nht}.
Another proposed explanation was based on different particle eigenvolumes for different hadron species, giving rise to a species-dependent excluded-volume correction, which provides an improved description of the data~\cite{Alba:2016hwx}.
In the Thermal-FIST results shown in Fig.~\ref{fig:yieldsLF-SHM}, an energy-dependent Breit-Wigner (eBW) scheme has been implemented to model the influence of finite resonance widths in the HRG, which leads to a suppression of the proton yields, mainly due to a reduced feed-down from $\Delta$ resonances~\cite{Vovchenko:2018fmh}. With this approach, a significantly improved agreement between the SHM and the data for the proton yield is obtained at $\Tchem = 155$~\mev. The eBW scheme, however, induces a larger discrepancy with the data for the yield of \rmLambda{} hyperons. According to Ref.~\cite{Vovchenko:2018fmh}, this discrepancy may be mitigated by including in the SHM calculations additional, yet undiscovered, strange baryonic resonances not reported in the PDG lists.
Along the line of improving the description of the interactions in the HRG, the S-matrix formulation of statistical mechanics offers a natural implementation of interactions, with proper treatment of resonance widths and of non-resonant contributions. 
In particular, pion--nucleon interactions can be incorporated into the model via the empirical scattering phase shifts based on \rmPiPM--N scattering data.
As pointed out in~\cite{Lo:2017lym}, the S-matrix approach allows one to obtain an improved matching with the lattice-QCD results for the correlation of the net baryon number with the electric charge, which is not well reproduced by an ideal HRG model.
At a temperature of 156~\mev, the treatment of $\pi$--N interactions via the S-matrix approach leads to a reduction of the proton yield by about 17\% relative to the ideal HRG result~\cite{Andronic:2018qqt}.
The corresponding reduction of the pion yield is only about 1\%.
The implementation of the S-matrix correction in the GSI-Heidelberg SHM fits, shown in Fig.~\ref{fig:yieldsLF-SHM}, provides here the best match to the measured proton yields, together with a good description of all other measured particles.

An alternative approach was proposed considering that, in addition to the overestimation of the measured proton yield, 
the various SHM calculations %
tend to underestimate the data for strange baryon production, as it can be seen in Fig.~\ref{fig:yieldsLF-SHM} for the \rmLambda{}, \rmXiPM{} and \rmOmegaPM{} yields, which differ by about 1.5--2 standard deviations from the
predictions. %
This approach advocates different hadronisation temperatures for up and down versus strange quarks, based on studies of flavour-specific fluctuations and correlations of conserved charges in lattice-QCD simulations with physical quark masses. These simulations indicate that strange quarks experience deconfinement at slightly larger temperatures compared to light quarks, implying that strange hadrons may be formed in the QGP slightly above \TpseudoCritic{}~\cite{Ratti:2011au}.
As a consequence, a univocal temperature cannot be assigned to the QCD crossover transition and a flavour hierarchy in the hadronisation is anticipated, with the characteristic temperature being higher by 10--15~\mev\ for the strange quarks compared to up and down quarks~\cite{Bellwied:2013cta}.
These studies motivated the usage of two different temperatures for strange and non-strange hadrons in the SHM fit~\cite{Chatterjee:2016cog, Flor:2020fdw}. 
This provided an improved description of the data at RHIC and LHC energies compared to the single-temperature description, with the difference between the two temperatures decreasing with decreasing $\sqrtSnn$, i.e.~increasing $\mu_{\rm B}$.  
However, it should be pointed out that the data-based S-matrix correction to treat the interactions in the HRG was not included in the fits with two freeze-out temperatures. 
In this context, a recent study of the effect of S-matrix corrections employing a coupled-channel model for K, \rmLambda{} and \rmSigma\ interactions predicts an enhancement in the \rmAlambda+\rmSigmaZero\ yields relative to the HRG baseline, thus allowing for a more accurate description of the measured particle yields within the THERMUS SHM framework~\cite{Cleymans:2020fsc}.

The production of light nuclei is quantitatively well described by the statistical hadronisation model.
However, it is debatable whether the light nuclei should be included in the thermal model fit. The production mechanism for such loosely bound states (\orderOf{\mev}) is presently a subject of intense investigation. Models have been proposed which predict that nuclei are formed at the late stages of the fireball expansion via coalescence of nucleons that are close to each other in phase space~\cite{Scheibl:1998tk} (see Chap.~\ref{ch:NuclPhysLHC} for a detailed discussion).
It should be noted that if the yields of light nuclei are excluded from the fit, the temperature and volume extracted from the SHM do not change significantly.
Therefore, the inclusion of nuclei in the fit does not alter the considerations that were made above on the characteristics of the hadron gas at the chemical freeze-out.
Conversely, the yields of nuclei are very sensitive to the temperature: a small variation in temperature leads to large variations of the yields of nuclei.
In particular, a SHM fit with the GSI-Heidelberg framework including only the yields of light nuclei gives $\Tchem^{\rm nuclei}=159\pm5$~\mev, which is consistent with the value of \Tchem{} extracted from all particles~\cite{Andronic:2017pug}.
These observations indicate that the same thermal parameters governing light-hadron yields also determine the production of light composite objects which are loosely bound states.
However, this poses a question about how the relatively loosely-bound light nuclei formed at the QGP hadronisation can survive in the hostile environment of the hadronic phase that follows the chemical freeze-out.
This was dubbed the ``snowball in hell'' problem, as the binding energies of light nuclei (e.g.~2.2~\mev\ for the deuteron) are much lower than the temperature of the hadron gas at the chemical freeze-out.
A possible explanation assumes that the reactions involving  break-up and formation of light nuclei proceed in relative chemical equilibrium after the chemical freeze-out, so that nuclei are disintegrated and regenerated at similar rates during the hadronic phase, thus preserving the relative abundances of nuclei and nucleons at \Tchem{}~\cite{Oliinychenko:2018ugs,Xu:2018jff,Vovchenko:2019aoz}.
A different proposed possibility is that, at the QGP hadronisation, compact and colourless bound states of quarks are produced and evolve into the final state nuclei only after a formation time of some \fmC~\cite{Andronic:2017pug}; such compact objects can survive a short-lived hadronic phase after hadronisation.

It is important to note that the measured yields of strange and multi-strange particles at midrapidity in \PbPb\ collisions are described (within 2 standard deviations) by the full-equilibrium SHM model, corresponding to the \emph{grand-canonical} ensemble.
The relative abundance of strange hadrons in heavy-ion collisions is larger compared to the suppressed strangeness yield observed in collisions between elementary particles, as discussed in detail in Sec.~\ref{section:3.2}.
Strangeness production in \pp\ collisions is consistent with a \emph{canonical} formulation of the statistical model, in which a correlation volume parameter is introduced to account for the locality of strangeness conservation (a strangeness undersaturation parameter is also often employed for lower energies), see~\cite{Cleymans:2020fsc} for a recent analysis.
In the framework of the SHM, the lifting of strangeness suppression from pp to A--A collisions can therefore be understood in terms of the transition from the canonical to the grand-canonical limit~\cite{Hamieh:2000tk}.
An enhanced strangeness production in heavy-ion collisions, as compared to pp collisions, was one of the earliest proposed signals for the QGP formation~\cite{Rafelski:1982pu,Koch:1986ud,Koch:1986hf,Rafelski:1991rh}.
In particular, in a deconfined state, the abundances of light partons, including the strange quark, are expected to quickly reach their equilibrium values due to the low energy threshold to produce ${\rm s}\overline{\rm s}$ pairs.
It was also shown that the strangeness content of the QGP in equilibrium is similar to that of a chemically equilibrated hadron gas at the same entropy-to-baryon ratio, although in both cases the production level of strange hadrons is higher than that observed in pp collisions~\cite{Redlich:1985zg,McLerran:1986aq,Lee:1987mj}.
However, it should be considered that the time needed to achieve the equilibrium for multi-strange baryons via two-body hadronic collisions is estimated to be much longer than the lifetime of the fireball produced in heavy-ion collisions~\cite{Koch:1986ud}.
The equilibration times for \rmLambda, \rmXi, and \rmOmega\ baryons via hadronic processes are instead expected to be substantially smaller ($\lesssim$~1~\fmC) if multi-hadron scatterings play a relevant role~\cite{Braun-Munzinger:2003htr}, which is the case when the particle densities are high. Such conditions are realised only for temperatures near to \TpseudoCritic. This confirms the conclusions in~\cite{Koch:1986ud,Koch:1986hf} that strange hadron abundances are established very close in time to the transition from the QGP to hadronic matter and are not appreciably changed by hadronic interactions at temperatures below \TpseudoCritic.
Hence, the observed validity of a grand-canonical description of strange-hadron production in \AA\ collisions can be seen as a natural consequence of the formation of the QGP in the early phases of the collision evolution~\cite{BraunMunzinger:2003zd}.

\paragraph{\textbf{\textit {Heavy-flavour hadron yields.}}} 
An additional question of the statistical nature of hadron production is whether the yields of hadrons containing heavy quarks can be described with the statistical hadronisation approach discussed above for light-flavour particles.
The masses of charm and beauty quarks are substantially larger than the temperature scale of the QCD medium attained at LHC energies, and therefore thermal production of these quarks in the QGP is strongly suppressed.
Thus, the charm and beauty content of the fireball is determined by the initial production of c and b quarks in hard-scattering processes, i.e.~very far from a chemical equilibrium population for a temperature of the order of \TpseudoCritic{}.
The heavy c and b quarks produced in initial hard scatterings traverse the QGP and interact with its constituents, exchanging energy and momentum with the medium. These interactions can lead the charm quarks to (partially) thermalise in the QGP~\cite{Batsouli:2002qf, Moore:2004tg,Prino:2016cni,Rapp:2018qla,Cao:2018ews} such that, at the phase boundary, they can be distributed into hadrons with thermal weights as discussed above for the light quarks. 
In the statistical hadronisation approach, the absence
of \emph{chemical} equilibrium for heavy quarks is accounted
for by a fugacity factor $g_{\rm c}$, which ensures that all initially produced charm quarks are distributed into hadrons at the phase boundary~\cite{Andronic:2021erx}.
The value of $g_{\rm c}$ is determined from a balance equation containing the charm quark production cross section as input.
With this approach, the yield of hadrons containing charm quarks can thus be calculated using as input values: \emph{i}) the thermal parameters obtained from the analysis of light-flavour hadrons, and \emph{ii}) the charm-quark production cross section, which is an external input to the model. %

\begin{figure}[!tb]
    \begin{center}
        \includegraphics[width=0.95\textwidth, angle=0, clip=true, trim=0cm 0 0 0]{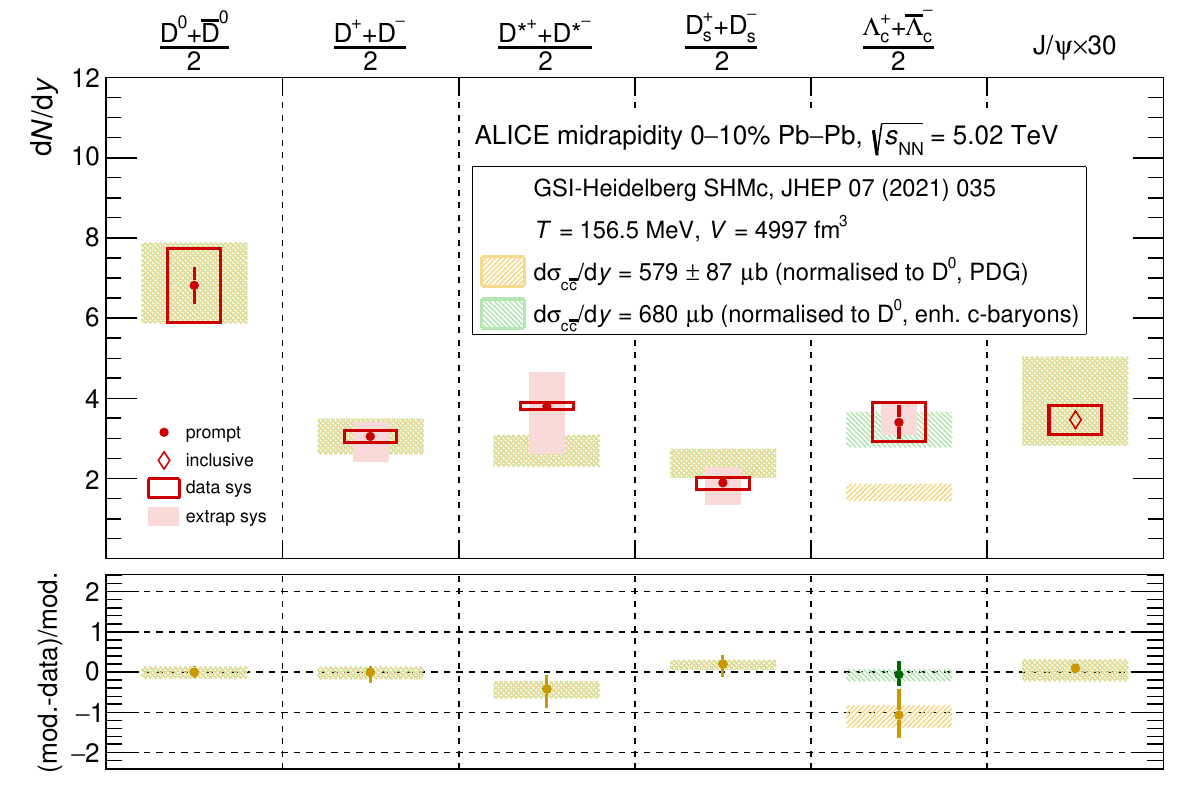}
    \caption{\pT-integrated yields per unit of rapidity measured at midrapidity for different charm-hadron species in the 10\% most central \PbPb\ collisions at $\sqrtSnn=5.02~\tev$~\cite{ALICE:2021rxa,ALICE:2021kfc,ALICE:2021bib,ALICE:Jpsieefwref} compared to SHM predictions~\cite{Andronic:2021erx}. Systematic uncertainties from data (open boxes) and from the \pT{} extrapolation (shaded boxes) are shown separately. The coloured band in the SHM predictions represents the uncertainty on the charm-quark production cross section.}
    \label{fig:yieldsHF-SHM}
    \end{center}
\end{figure}

The yields of charm hadrons (non-strange~\cite{ALICE:2021rxa} and strange D mesons~\cite{ALICE:2021kfc}, \rmLambdaC{} baryons~\cite{ALICE:2021bib} and \rmJpsi{} mesons~\cite{ALICE:Jpsieefwref}) measured in central (0--10\%) \PbPb\ collisions at \sqrtSnn{}~=~5.02~\tev{} are shown in Fig.~\ref{fig:yieldsHF-SHM}. The yields per unit of rapidity of \rmDzero{} and \rmJpsi{} mesons were obtained by integrating the \pT-differential measurements performed down to $\pT=0$. Instead, the \pT-integrated yields of \rmDplus{}, \rmDstar{}, \rmDs{} mesons, and \rmLambdaC{} baryons were extrapolated exploiting the measured \pT-differential production ratios relative to the \rmDzero{} mesons, using different assumptions on their shapes in the \pT{} range  where the measurements were not performed (see~\cite{ALICE:2021rxa, ALICE:2021kfc, ALICE:2021bib} for details). The fraction of extrapolated yield is different for the different hadron species, and it is about 19\% for the \rmLambdaC{} (which is measured down to \pT~=~1~\gmom{}), about 70\% for \rmDplus{} and \rmDs{} (measured down to \pT~=~2~\gmom{}), and about 85\% for \rmDstar{} mesons, which are measured for \pT~$>$~3~\gmom{}.

The data are compared to SHM calculations~\cite{Andronic:2021erx} where: \emph{i}) thermal parameters come from the fit to light-flavour hadrons at \sqrtSnn{}~=~2.76~\tev{} extrapolated to \sqrtSnn{}~=~5.02~\tev{} (same \Tchem{} and larger volume $V$ following the increase of \dNchdeta\ with increasing \sqrtSnn{}), and \emph{ii}) the charm production cross section is determined from the measured ${\rm D^0}$ yield in the 10\% most central Pb--Pb collisions~\cite{ALICE:2021rxa}, leading to $\mathrm{d}\sigma_{\rm c\overline{c}}/\mathrm{d}y=0.579~\mb$. 
The solid band in the SHM predictions in Fig.~\ref{fig:yieldsHF-SHM} represents the uncertainty related to the determination of the charm production cross section at midrapidity. For the measured data points, the empty boxes represent the systematic uncertainty associated to the measurement in the visible \pT{} range, while the shaded boxes stand for those associated to the \pT-extrapolation procedure.
The measured yields of open-charm mesons are compatible with the SHM calculations within the uncertainties, which is to some extent expected since the charm-quark production cross section used in the model is tuned to reproduce the measured \rmDzero{}-meson yield. A larger difference is observed for the production yield of \rmLambdaC{} baryons, which is underestimated by the SHM predictions. A possible solution for this difference within the framework of the statistical hadronisation model was proposed in~\cite{He:2019tik}, where an augmented set of excited charm baryon states beyond those listed in the PDG~\cite{Zyla:2020zbs} is considered, as discussed in Sec.~\ref{sec:HFpTdiffratios}. In fact, as presented in~\cite{Andronic:2021erx}, if a charm-quark production cross section of $\mathrm{d}\sigma_{\rm c\overline{c}}/\mathrm{d}y=0.68~\mb$ is considered instead of $\mathrm{d}\sigma_{\rm c\overline{c}}/\mathrm{d}y=0.579~\mb$ and the augmented set of charm baryon states is used, the SHM provides a satisfactory description of the measured \rmLambdaC{} baryon yield. %
Particularly interesting for the statistical hadronisation approach is the \rmJpsi{} meson. In fact, an enhanced production of hidden-charm states has been observed in \PbPb\ collisions at LHC energies compared to the predicted suppression with respect to pp collisions due to the colour-charge screening in the QGP~\cite{Matsui:1986dk} (see Sec.~\ref{sec:charmonia}).
The SHM provides a good description of the \rmJpsi{} production yield within uncertainties, 
supporting the statistical (re)combination of charm and anticharm quark pairs at the phase boundary~\cite{BraunMunzinger:2000px}. %

Recalling that charm quarks need to be (to some extent) thermalised in order to be distributed into hadrons according to thermal weights, one can conclude that the fair description of the measured open charm and charmonium yields within the SHM suggests that the interactions with the medium constituents bring charm quarks in (or close to) kinetic equilibrium with the QGP. This is further supported by the measurements of charm-hadron \RAA{} and $v_2$ at low-\pT{} reported in Sec.~\ref{sec:Transport}.

\subsubsection{Particle momentum and angular distributions: connection to microscopic hadronisation mechanisms}
\label{sec:MicroscopicHadronization}

\label{sec:PtDiffRatioAndv2}
Measurements of the production of different hadron species as a function of \pT{} can provide insight into the transition, at the pseudo-critical temperature \TpseudoCritic, from the partonic degrees of freedom of the QGP to the hadron gas phase.
The process of hadron formation from the QGP medium is expected to be different from other cases of hadronisation, such as the (in-vacuum) fragmentation of hard-scattered partons in elementary collisions, in which no bulk of thermalised partons is present~\cite{Fries:2008hs}.
For processes with a momentum transfer much larger than $\Lambda_{\rm QCD}$, the QCD factorisation theorems allow one to separate short and long distance dynamics and to express the cross section for inclusive hadron production as a convolution of two terms: the hard-scattering cross section at the partonic level, which can be calculated as a perturbative series in powers of the strong coupling constant \alphaS{}, 
and the fragmentation functions, which account for the non-perturbative evolution of a hard-scattered parton into the given hadron in the final state.
Fragmentation functions are not calculable in QCD, but they can be parameterised after experimental measurements, mostly from \ee{} collisions.
In microscopic models (i.e.~models involving explicitly quarks and gluons, which are the underlying fundamental degrees of freedom in QCD), the (in-vacuum) fragmentation of the hard-scattered parton occurs via the creation of ${\rm q \overline{q}}$ pairs through string breaking or gluon radiation and splitting, thus forming colour singlet states that evolve into hadrons.
In the case of the hadronisation of the QGP created in heavy-ion collisions, a bulk of deconfined partons is present when the transition temperature \TpseudoCritic{} is reached. So, there is no need for the creation of additional partons through string breaking, and the quarks of the bulk that are close to each other in phase space can simply combine into hadrons~\cite{Fries:2003vb,Hwa:2002tu,Greco:2003xt,Molnar:2003ff}. 
Models of hadronisation via recombination (also known as coalescence) were originally developed~\cite{Das:1977cp} to describe the relative abundances of hadrons measured at very forward rapidity in $\pi$-nucleus collisions at the SPS~\cite{Adamovich:1993kc} and at FNAL~\cite{Aitala:1996hf}. These models are based on the concept that the presence of a reservoir of partons (the valence quarks of the projectile in the considered case, concerning a kinematic region at very forward rapidity) can induce significant modifications in the hadronisation process.
In the case of the hadronisation of the  QGP, the bulk of deconfined partons constitutes this reservoir.

Recombination models are built on the assumption that quarks, at the hadronisation of the QGP, can be treated as effective degrees of freedom having a dynamical mass approaching the constituent mass~\cite{Bowman:2004xi,Diakonov:1985eg}, while gluons disappear as dynamical degrees of freedom and are converted into $\mathrm{q}\overline{\mathrm{q}}$ pairs~\cite{Greco:2003xt,Fries:2003kq}.
With the additional assumption of dominance of the lowest Fock states in the hadron wave function, effective constituent quarks (and antiquarks) are combined into mesons and baryons taking into account only the valence structure of the hadron. Hence, the probability of emitting a hadron from the QGP is proportional to the probability of finding its valence quarks in the bulk medium.
As discussed in~\cite{Muller:2005pv}, the hadron spectra produced via recombination of quarks from a thermal medium are not significantly affected by the inclusion of higher Fock states to take into account the complexity of the internal structure of hadrons.
The recombination models have provided a natural explanation for some unexpected measurements of baryon and meson production yields and elliptic flow in the intermediate-\pT{} region (1.5 $<$ \pT{}{} $<$ 5~\gmom) at RHIC~\cite{Arsene:2004fa, Adcox:2004mh, Back:2004je, Adams:2005dq}.
Notably, also global polarisation measurements are expected to be sensitive to the hadronisation mechanism and, as discussed in Sec.~\ref{sec:TG2globalpolarization}, the results on \rmKstar{} polarisation as a function of \pT{} are qualitatively consistent with the expectation of hadronisation of polarised quarks via recombination~\cite{Acharya:2019vpe}.

In the hadronisation of the fireball, the recombination process competes with the hadronisation of energetic (high-\pT) partons that escape from the QGP and hadronise in the vacuum via fragmentation.
Final hadron spectra are a mixture of hadrons from recombination (with a momentum essentially given by the sum of the momenta of the valence quarks) and from string fragmentation (having a momentum lower than that of the parent parton).
As discussed in~\cite{Fries:2003vb}, the recombination process is expected to dominate over fragmentation at low- and intermediate-\pT{} (up to few \gmom), while at higher momenta ($p_{\rm T} > 8-10$~GeV$/c$), where the particle spectra exhibit a power-law trend, a transition occurs to a regime dominated by fragmentation of jets. 
This transition is predicted to take place at higher values of \pT{} for baryons as compared to mesons. This naturally explains the baryon versus meson grouping of \vTwo{} and the baryon-to-meson enhancement at intermediate-\pT{} that will be discussed later in this paragraph. 
At low-\pT{}, the hadronisation is expected to occur completely via recombination.
In this region, the hadrons formed by combination of thermalised quarks are found to have momentum spectra reproducing the thermal equilibrium limit, and this naturally leads to a good agreement with spectra from hydrodynamic calculations~\cite{Ravagli:2007xx}.
Also note that the dominance of recombination at low- and intermediate-\pT{} is completely consistent with what was discussed in Sec.~\ref{sec:SHM} about the thermal origin of particle production, based on the agreement between SHM calculations and the measured integrated yields (see Fig.~\ref{fig:yieldsLF-SHM} and Fig.~\ref{fig:yieldsHF-SHM}). 
As pointed out in~\cite{Fries:2003kq}, the recombination mechanism connects a thermal partonic phase with the observed thermal hadronic phase, and therefore it can be seen as a microscopic manifestation of statistical hadron production.
In this thermal-limit regime, the hadron \pT{} spectra and \vTwo{} are the result of a system which evolved through equilibrium stages, thus not preserving the memory of the previous stages and of the underlying microscopic dynamics that drove the system evolution.
Instead, the intermediate-\pT{} region, which is located above the thermal-limit regime but below the high-momentum region where fragmentation dominates, offers a window where non-equilibrium effects in the hadronic observables can provide sensitivity to the hadronisation mechanism at a microscopic level.\\

\begin{figure}[tb!]
    \begin{center}
        \includegraphics[width=0.48\textwidth, angle=0, clip=true, trim=0cm 0 0 0]{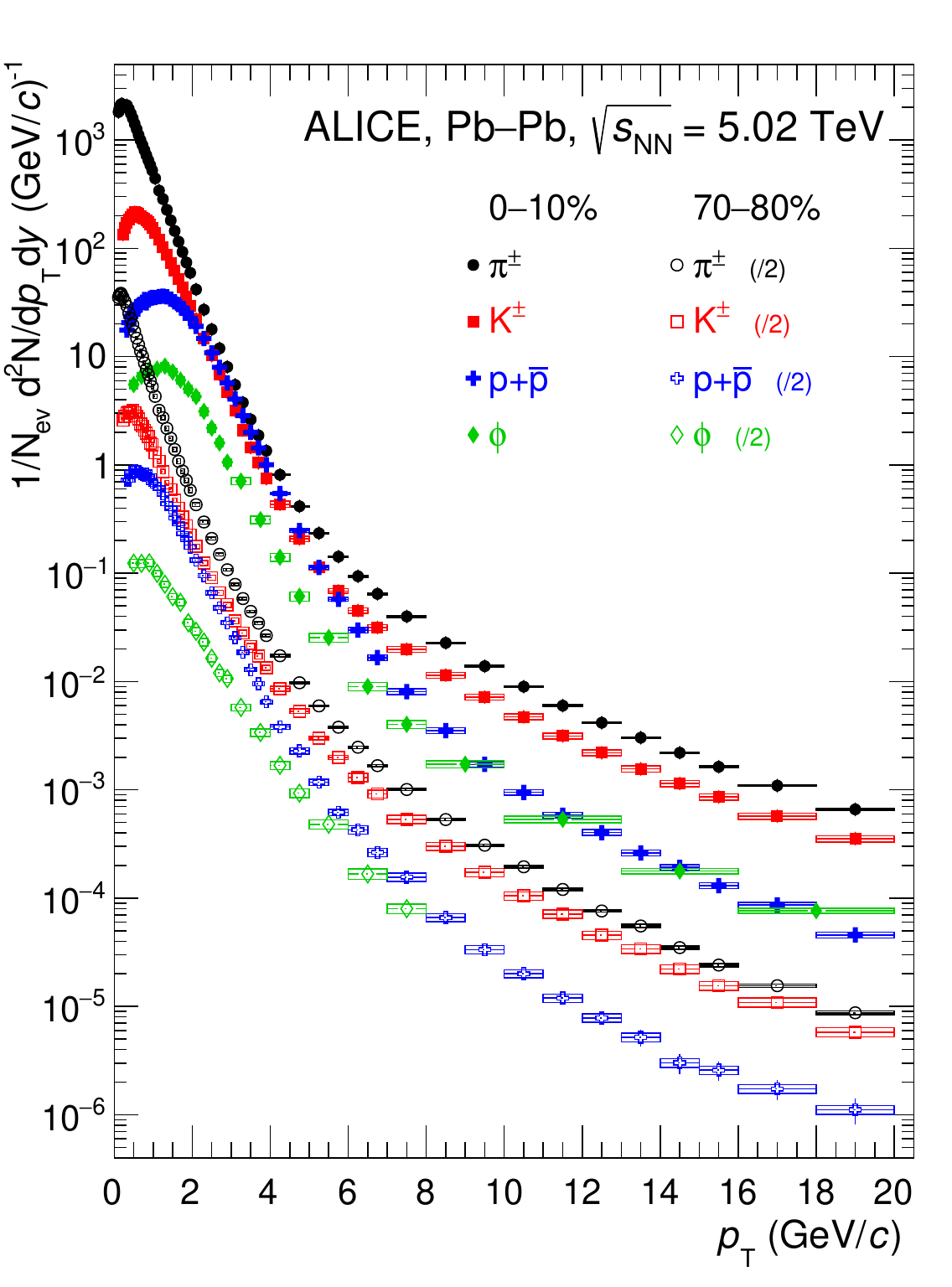}
        \includegraphics[width=0.48\textwidth, angle=0, clip=true, trim=0cm 0 0 0]{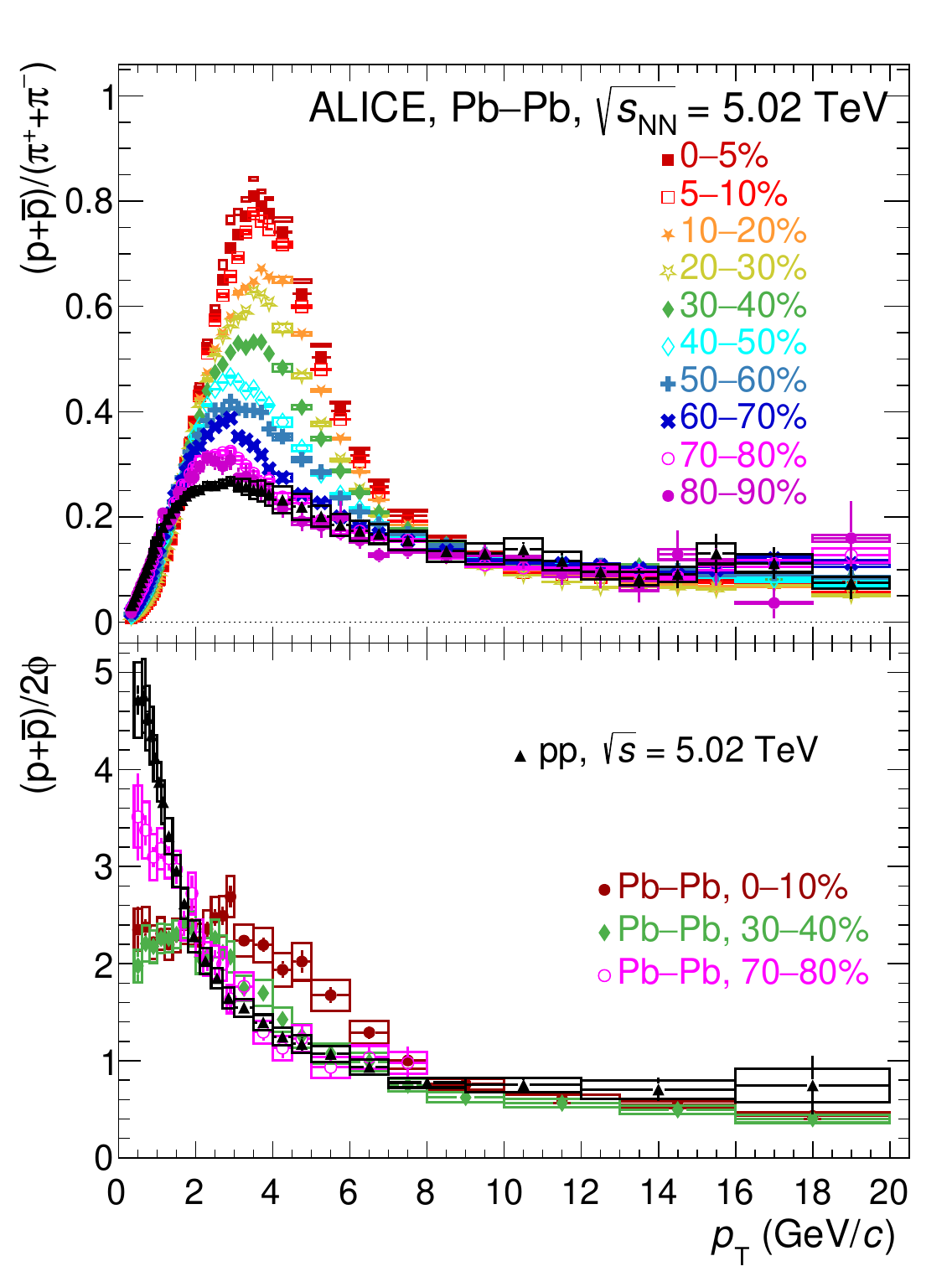}

    \caption{(Left) Transverse-momentum spectra of \rmPiPM, \rmKpm, p+\pbar~\cite{Acharya:2019yoi}
    and \rmPhiMes~\cite{ALICE:2019xyr} in 0--10\% (filled markers) and \mbox{70--80\%} (empty markers) central \PbPb\ collisions at \sqrtSnn{} = 5.02~\tev. (Right) Proton-to-pion and proton-to-phi \pT-differential ratios in inelastic pp collisions at \sqrtS{} = 5.02~\tev\ and in several centrality intervals in \PbPb\ collisions at \sqrtSnn{} = 5.02~\tev. Systematic uncertainties are shown as boxes while statistical uncertainties as vertical bars. In the (p+\pbar)/(\piPlus+\piMinus) ratio, only the systematic uncertainties uncorrelated across centrality intervals are shown in the error boxes, in order to better highlight the evolution of the ratio with the collision centrality.}
    \label{fig:spectraANDratios}
    \end{center}
\end{figure}

\paragraph{\textbf{\textit {Light-flavour particle spectra and ratios.}}} 
The transverse-momentum spectra of \rmPiPM, \rmKpm, \proton+\pbar, and \rmPhiMes{} in \mbox{0--10\%} and 70--80\% central \PbPb\ collisions at \sqrtSnn{} = 5.02 ~\tev~\cite{ALICE:2019xyr,Acharya:2019yoi} are shown in the left panel of Fig.~\ref{fig:spectraANDratios}. As commented in Sec.~\ref{sec:TG2particlespectra}, the spectra harden with increasing centrality and multiplicity (radial flow signature, more evident for heavier particles): the maximum of the proton spectrum shifts from $\sim$0.6~\gmom\ to $\sim$1.3 \gmom\ going from peripheral to central collisions.   
The right panel of Fig.~\ref{fig:spectraANDratios} shows the \pT-differential (p+\pbar)/(\piPlus+\piMinus) and (p+\pbar)/(2\rmPhi) ratios (referred to as p/\Pion{} and p/(2\rmPhi) in the following\footnote{At LHC, the particle yield ratios are quoted for the sum of particles and antiparticles since both populations of a same species are produced in almost equal amount. See for example discussion in Sec.~\ref{sec:SHM} or in~\cite{Abelev:2013vea}}) in inelastic pp and centrality-dependent \PbPb\ collisions at \sqrtSnn~=~5.02~\tev. 
The p/\Pion{} ratios in \PbPb\ collisions exhibit a bump structure in 1.5 $\lesssim$ \pT{} $\lesssim$ 8.0~\gmom, gradually increasing with increasing centrality, reaching a maximum of about 0.82 at \pT{}~$\approx$~3.5~\gmom\ in the most central collisions. This can be explained as an interplay between the collective motion of the system (heavier particles are boosted to higher momenta) and the recombination mechanism that is expected to be the dominant one for the hadronisation in this momentum region. 
It is interesting to note that the fraction of yield in the peak is about 1.5-2\% for pions and 7-21\% for protons, depending on centrality, and therefore the \pT-integrated p/\Pion{} ratio shows only a mild dependence on centrality. In particular, the \pT-integrated ratio decreases by about 20\% from the most peripheral to the most central collisions~\cite{Acharya:2019yoi}, consistent with SHM calculations including either antibaryon-baryon annihilations in the hadronic phase~\cite{Steinheimer:2012rd,Becattini:2012xb} or the S-matrix approach to account for non-resonant interactions in the HRG~\cite{Andronic:2018qqt}. 
This suggests that baryon enhancement at intermediate-\pT\ is mainly due to the redistribution of baryons and mesons over the momentum range.
Further insight can be obtained from the ratios of the yields of p and \rmPhi{}, which have similar masses but different quark content.
At low-\pT\ (\pT~$\lesssim$~3.5~\gmom), the \proton/(2\rmPhi)\ ratio is rather independent of \pT\ in central \PbPb\ collisions. Such behaviour can be expected from hydrodynamic-based models~\cite{Shen:2011eg, Minissale:2015zwa} in which the particle mass is the main variable in the determination of the spectral shapes (even though interactions in the hadronic phase could modify this picture because they are expected to affect differently protons and \rmPhi{} mesons).
In contrast to this and as a confirmation of the radial flow signature, the 70--80\% \PbPb\ ratio is similar to that in pp collisions showing a decreasing trend with \pT.
At higher momenta, \pT~$\gtrsim$~8~\gmom, all the \PbPb\ and pp baryon-to-meson ratios shown in Fig.~\ref{fig:spectraANDratios} are consistent within the uncertainties suggesting the dominance of vacuum-like fragmentation processes in this region~\cite{Abelev:2014laa}.

\begin{sloppypar}
The \pT-differential p/\Pion{} and p/(2\rmPhi){} yield ratios were also measured in Xe--Xe collisions at \sqrtSnn~=~5.44~\tev~\cite{Acharya:2021ljw} and compared to the results in \PbPb\ collisions at \sqrtSnn~=~5.02~\tev~\cite{Acharya:2019yoi}. The baryon-to-meson ratios in the two collision systems are compatible between each other within the uncertainties, if events with similar charged-particle multiplicity densities are selected. As concluded in~\cite{Acharya:2021ljw}, this indicates that the radial flow and the hadronisation dynamics, which determine the shape of the \pT-spectra of different hadron species, are mainly driven by the density of produced particles, in contrast to the elliptic flow which depends also on the eccentricity of the collision region. 
\end{sloppypar}

\begin{figure}[tb!]
    \begin{center}
        \includegraphics[width=0.48\textwidth, angle=0, clip=true, trim=0cm 0 0 0]{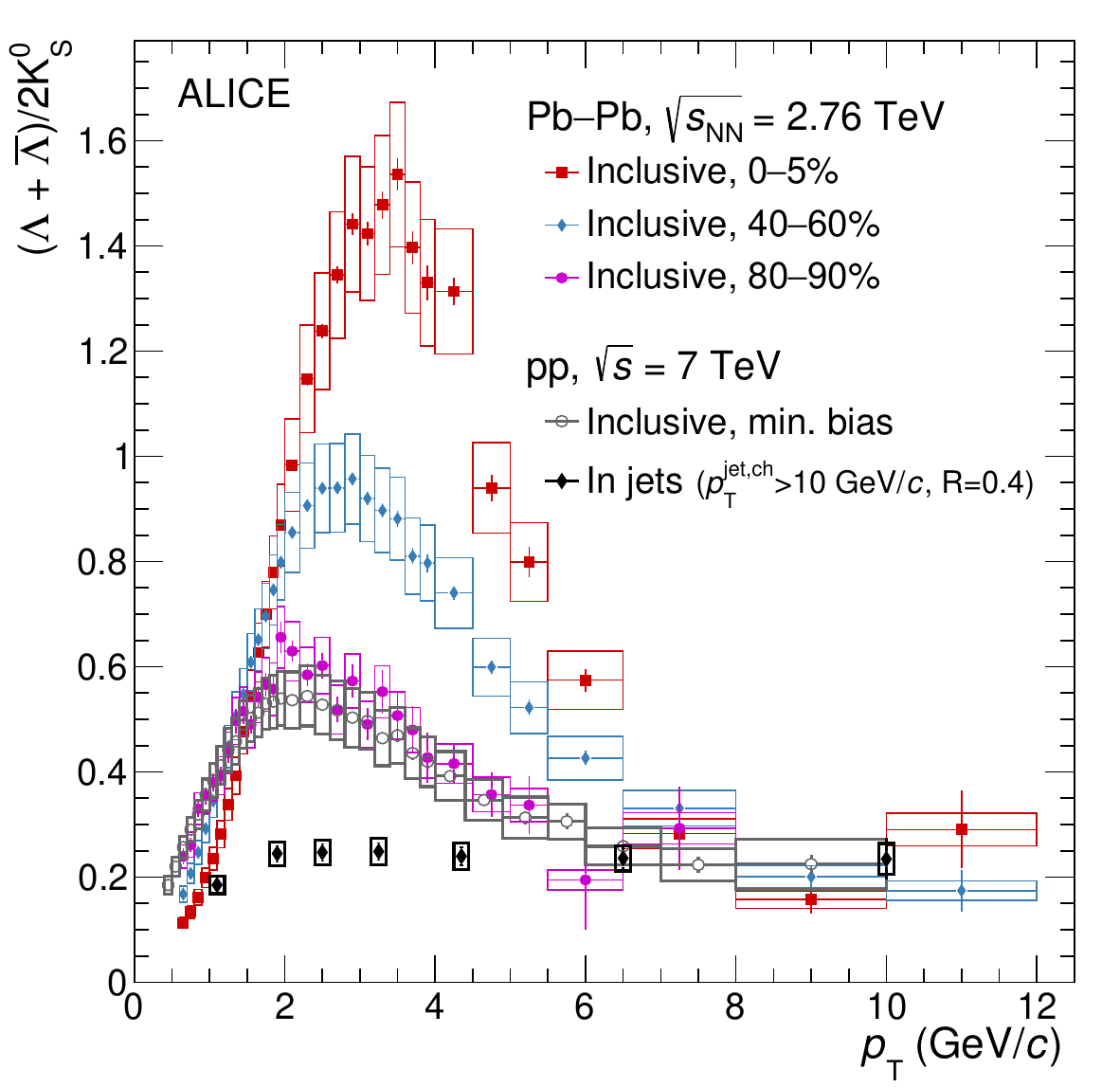}
        \includegraphics[width=0.48\textwidth, angle=0, clip=true, trim=0cm 0 0 0]{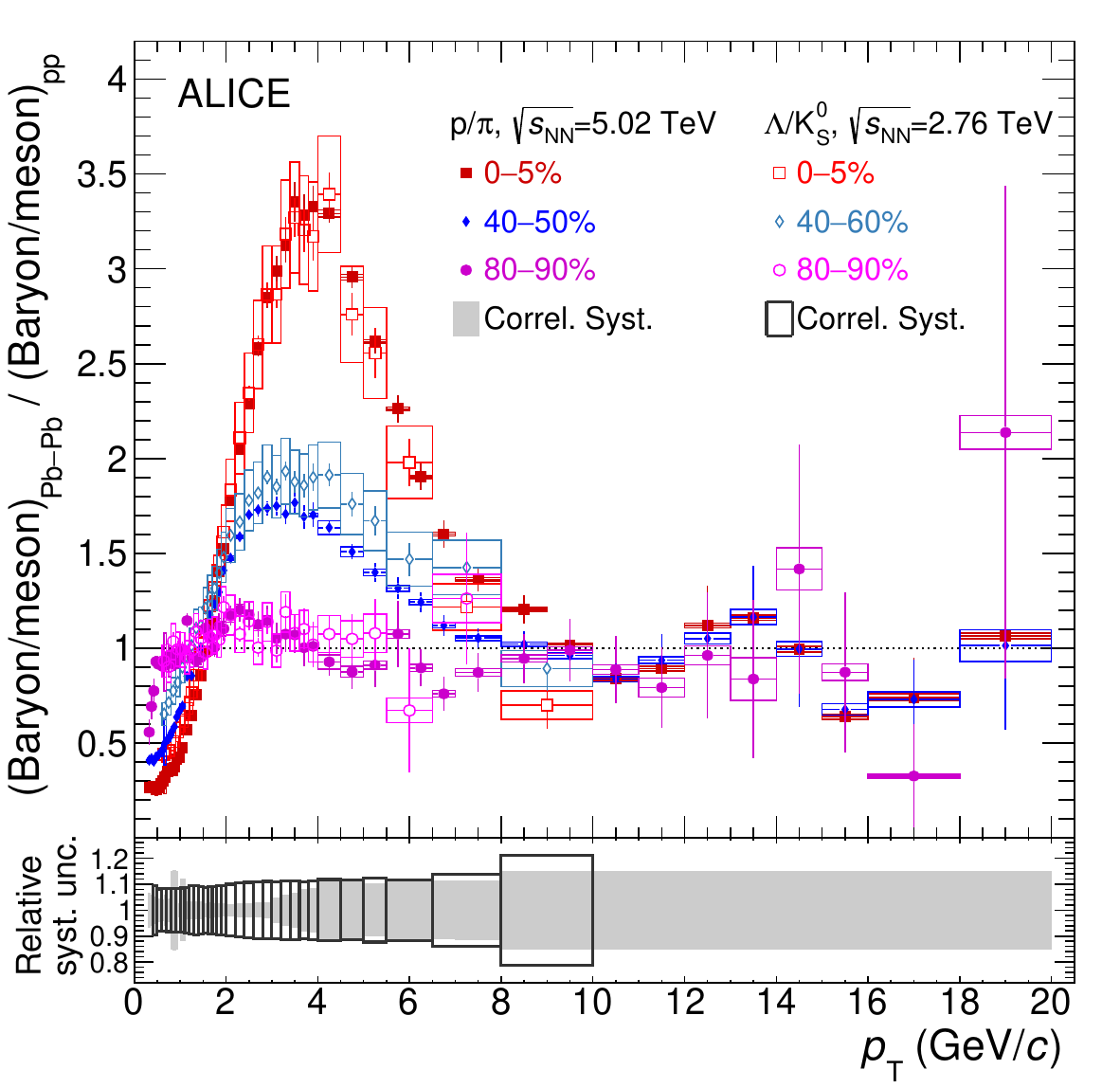}

    \caption{(Left) \pT-differential \ratioLamOverKzs~\cite{Abelev:2013xaa} ratio in \PbPb\ collisions at \sqrtSnn~=~2.76~\tev\ compared to pp collisions at \sqrtS~=~7~\tev\ for the inclusive production~\cite{Abelev:2013xaa} and for the production in jets~\cite{Acharya:2021oaa}. 
    (Right) \pT-differential double ratios (baryon/meson ratios in \PbPb\ divided by the \pp\ ones at a same colliding energy) for \proton/\Pion{} at \sqrtSnn~=~5.02~\tev~\cite{Acharya:2019yoi} and \ratioLamOverKzs{} at \sqrtSnn~=~2.76~\tev~\cite{Abelev:2013xaa}. Results are shown for several centrality intervals. Systematic uncertainties are shown as boxes while statistical uncertainties as vertical bars. The relative systematic uncertainty on the pp results in the double ratios, which is correlated among the different centrality intervals of the \PbPb\ measurements is displayed separately in the lower panel.}
    \label{fig:baryontomeson}
    \end{center}
\end{figure}

The baryon-to-meson ratios in the strange-hadron sector show similar features as those discussed above for the p/\Pion{} ratios. 
The measured $(\Lambda+\overline{\Lambda})/(2\Kzs)$ ratios (referred to as \ratioLamOverKzs\ in the following) as a function of \pT{} in central (0--5\%), semi-central (40--60\%), and peripheral (80--90\%) \PbPb\ collisions at \sqrtSnn{} = 2.76~\tev\ are reported in the left panel of Fig.~\ref{fig:baryontomeson} and compared to the ratio measured in minimum-bias pp collisions at \sqrtS{} = 7~\tev~\cite{Abelev:2013xaa}.
The ratio measured in the most peripheral \PbPb\ collisions is compatible with the pp result. 
For more central collisions, the peak at low/intermediate-\pT{} becomes more pronounced and its position shifts towards higher momenta, in qualitative agreement with the effect of increased radial flow.
The ratios of the \pT-integrated \rmLambda{} and \Kzs{} yields (after extrapolation to $\pT=0$) are independent of centrality within the experimental uncertainties~\cite{Abelev:2013xaa}, suggesting, as mentioned before, that the baryon enhancement at intermediate-\pT{} is predominantly due to a redistribution of baryons and mesons over \pT\ rather than to an enhanced production of baryons.
In addition, the \ratioLamOverKzs\ ratio measured in pp collisions at \sqrtS{} = 7~\tev\ within jets with \pTchJet\ $>$ 10 \gmom\ reconstructed with a resolution parameter $R$ = 0.4~\cite{Acharya:2021oaa} is shown for comparison. 
The ratio in jets is almost independent of the transverse momentum of the particles produced in the jet and it does not show a maximum at intermediate-\pT. 
For \pT~$\lesssim$~5~\gmom, it is lower than the ratio measured in inclusive (minimum bias) pp collisions as discussed in Sec.~\ref{section:3.2}.
For \pT~$>$~7--8~\gmom, the \ratioLamOverKzs\ ratio in jets is consistent with the inclusive one in pp collisions, as well as with the values measured in \PbPb\ collisions independently of the centrality.
This confirms that the high-\pT hadrons stem from the fragmentation of hard-scattered partons, which escape from the QGP, and indicates that the relative fragmentation into baryons and mesons at high-\pT{} is essentially vacuum-like and is unmodified -- or at least, not significantly -- by the medium formed in the heavy-ion collisions.

A further illustration of the previously outlined physics message concerning particle yield ratios can be observed in the right panel of Fig.~\ref{fig:baryontomeson} which allows for a direct comparison of the \pT-differential baryon-to-meson ratios for non-strange and strange hadrons (p/\Pion{} and \ratioLamOverKzs)  thanks to a double ratio \PbPb/pp calculated at \sqrtSnn~=~5.02~\tev\ and 2.76~\tev, respectively~\cite{Acharya:2019yoi, Abelev:2013xaa}. The two double ratios are compatible between each other within uncertainties in all centrality intervals and they are consistent with unity for \pT~$>$~8--10~\gmom, where fragmentation dominates. For \pT~$\lesssim$~1.5~\gmom, a hierarchy of the baryon-to-meson ratios is observed as a function of centrality: the ratios are strongly reduced in central collisions compared to peripheral collisions, thus compensating the enhancement at intermediate-\pT. 

\begin{figure}[tb!] 
    \begin{center}
        \includegraphics[width=0.68\textwidth, angle=0, clip=true, trim=0cm 0 0 0]{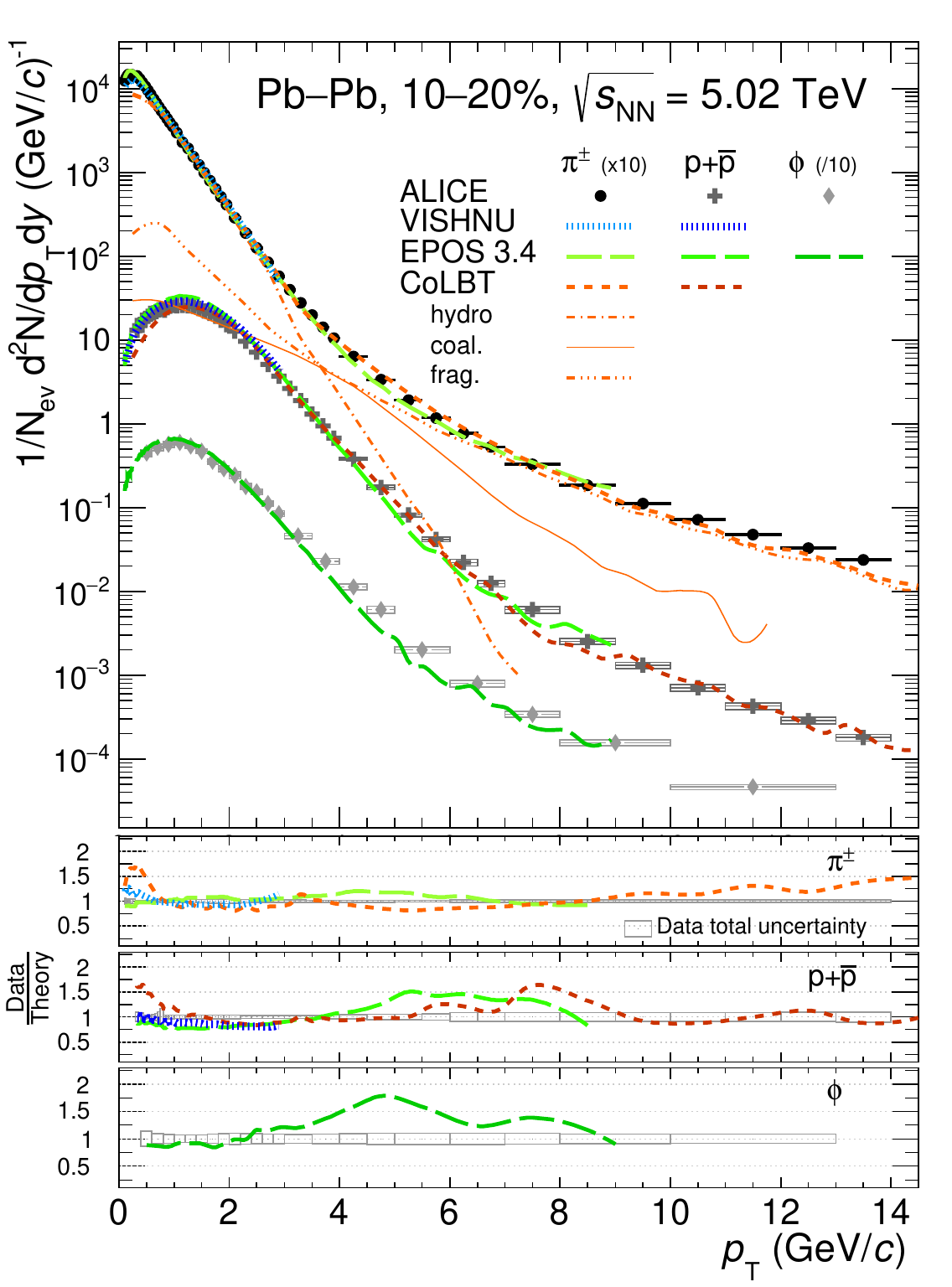}

 \caption{Comparison of \rmPiPM, p+\pbar~\cite{Acharya:2019yoi}, and \rmPhiMes~\cite{ALICE:2019xyr}
 transverse-momentum spectra in 10--20\% central \PbPb\ collisions at \sqrtSnn~=~5.02~\tev\ to VISHNU~\cite{Shen:2014vra, Zhao:2017yhj}, EPOS v3.4~\cite{Werner:2013tya}, and CoLBT~\cite{Zhao:2021vmu} models. The systematic uncertainties are represented as boxes while the statistical uncertainties are shown as vertical bars.}
    \label{fig:comptomodelch2_4}
    \end{center}
\end{figure}

\begin{figure}[tb!] 
    \begin{center}
        \includegraphics[width= 0.9\textwidth, angle=0, clip=true, trim=0cm 0 0 0]{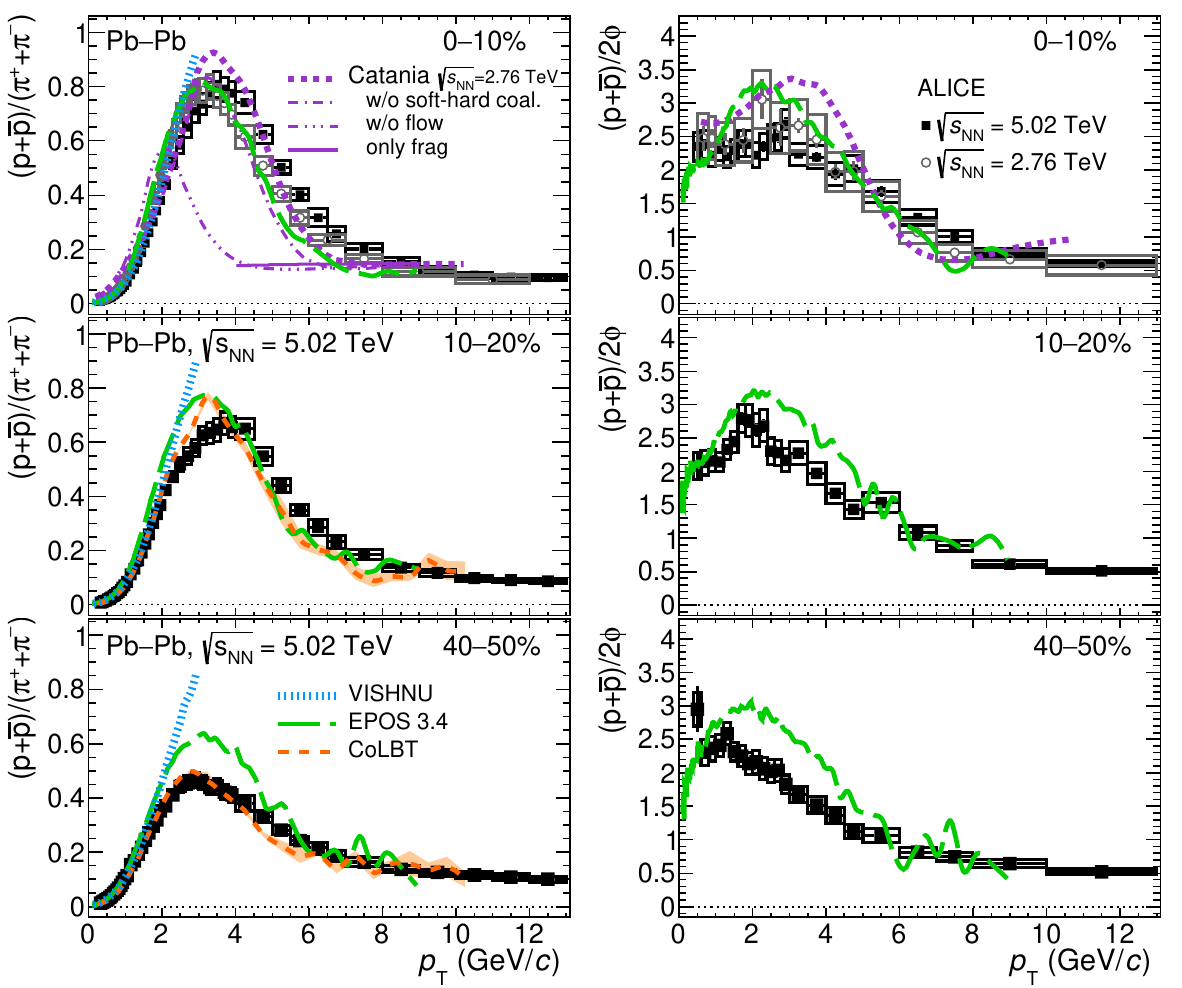}

 \caption{Comparison of \proton/\Pion{} (left) and (\proton+\pbar)/(2\rmPhi) (right) in \mbox{0--10\%}, \mbox{10--20\%}, and 40--50\% \PbPb\ collisions to the VISHNU~\cite{Shen:2014vra, Zhao:2017yhj}, EPOS v3.4~\cite{Werner:2013tya}, Catania~\cite{Minissale:2015zwa}, and CoLBT~\cite{Zhao:2021vmu} models as illustrated in the legend. More details on the different Catania model components can be found in the text. The systematic uncertainties are represented as boxes while the statistical uncertainties are shown as vertical bars.}
    \label{fig:comptomodelchratio2_4}
    \end{center}
\end{figure}

\paragraph{\textbf{\textit {Comparison to models.}}} 
The results related to \rmPiPM, \proton+\pbar, and \rmPhiMes{} spectra and yield ratios in different centrality classes (0--10\%, 10--20\%, and 40--50\%) are compared with several models in Figs.~\ref{fig:comptomodelch2_4} and~\ref{fig:comptomodelchratio2_4}. These models aim at bridging the gap between the partonic phase with its hydrodynamic motion and the ultimate hadronic stages. The models considered in this section are: VISHNU~\cite{Shen:2014vra, Zhao:2017yhj}, EPOS v3.4 (simply called EPOS3 in the following)~\cite{Werner:2013tya}, Catania coalescence plus independent-fragmentation model~\cite{Minissale:2015zwa}, and CoLBT~\cite{Zhao:2021vmu}. In the following, the description of the main features of each model is provided:
\begin{itemize}
    \item The VISHNU model, introduced in Sec.~\ref{sec:TG2hydrodescription}, is a viscous hydrodynamic calculation of the expansion of the fireball, handing over to a microscopic hadron cascade model (UrQMD)~\cite{Bleicher:1999xi, Bass:1998ca} for the late phases of the expansion of the hadron resonance gas. The initial energy density profile is modelled with \trento{}~\cite{Moreland:2014oya}. To convert the hydrodynamic output into particles (propagated then with UrQMD) first an isothermal freeze-out surface is found and then the hadron spectra are calculated with the Cooper-Frye formula~\cite{Song:2010aq}. Note that this model addresses only the low-\pT{} part of the spectra ($<$ 3 \gmom).
    
    \item The EPOS3 model is a general-purpose event generator based on the Gribov-Regge theory of multiple scattering~\cite{Drescher:2000ha, Werner:2005jf}, perturbative QCD, and string fragmentation~\cite{Liu:2001hz}. 
    The dense region in the system, the so-called \emph{core}, is treated as the QGP~\cite{Werner:2007bf} and modelled with a hydrodynamic evolution (since EPOS2~\cite{Werner:2010aa}) incorporating viscosity (EPOS3~\cite{Werner:2013tya}), followed by statistical hadronisation which conserves energy, momentum, and flavours thanks to a microcanonical formulation~\cite{Werner:2007bf}. Particle production from the low-density regions of the system, called \emph{corona}, is treated as in proton--proton collisions. EPOS3 implements further a saturation in the initial state as predicted in the Colour Glass Condensate model~\cite{Gelis:2012ri} and a hadronic cascade for the late stages. The model also accounts for jets and their interactions with the hydrodynamically expanding bulk matter. This is important for particle production at intermediate-\pT{}~\cite{Werner:2012xh} and it is reminiscent of the recombination mechanism~\cite{Fries:2003vb, Greco:2003xt}.
    
    \item The Catania model~\cite{Minissale:2015zwa} describes the QGP hadronisation via the competing processes of parton fragmentation and quark recombination. The latter is implemented by employing the instantaneous coalescence approach described in~\cite{Greco:2003xt, Greco:2007nu}, which is based on the Wigner-function formalism~\cite{Dover:1991zn} to calculate the spectrum of hadrons from that of quarks. The approach is based on a Monte Carlo implementation that allows one to include a 3D geometry as well as the radial flow correlation in the partonic spectra and the effect of the main resonance decays. At low-\pT, the partonic spectra are obtained with a blast-wave approach assigning the mass of \emph{constituent} quarks (330~\mev/$c^2$ for u and d quarks and 450~\mev/$c^2$ for s quarks). 
    The contribution of mini-jets is also considered, which is important to model the intermediate momentum region where soft--hard recombination processes (involving a quark from a mini-jet and one from the bulk) play a role in the determination of the spectral shapes until independent fragmentation takes over at high-\pT. This is achieved by using a parton \pT{} spectrum calculated at next-to-leading order in a pQCD scheme. In A--A collisions, the modification due to the jet quenching mechanism is included (see Eq. 3 in~\cite{Minissale:2015zwa}).
    Note that this approach does not guarantee the hadronisation of the full bulk of particles of the QGP medium. Moreover, the underlying kinematics of the instantaneous projection of massive quarks into hadrons, which does not include interactions with the surrounding medium, makes it impossible to conserve the four-momentum in the instantaneous coalescence process. Therefore, the model has to be considered fully applicable only at \pT~$>$~1.5~\gmom\ even if the description of the spectra at low-\pT{} is reasonable. 
    
    \item The CoLBT model~\cite{Zhao:2021vmu} is based on a state-of-the-art coupled linear Boltzmann transport model (CoLBT-hydro)~\cite{Chen:2017zte} to simulate the concurrent evolution of the bulk medium and parton showers. The hadronisation is modelled via the Hydro-Coal-Frag hybrid approach~\cite{Zhao:2020wcd} that includes: \emph{i}) particle formation from the hydrodynamical fields at the freeze-out hypersurface at low-\pT, \emph{ii}) instantaneous coalescence of effective constituent quarks at intermediate-\pT, and \emph{iii}) fragmentation at high-\pT. The interplay between hadron formation at the hydrodynamical freeze-out and parton dynamics is regulated with a \pT\ threshold for the effective constituent quarks. Above this threshold, viscous corrections to the equilibrium distributions are large and recombination and fragmentation become the relevant hadronisation mechanisms. The recombination includes thermal--thermal, thermal--shower, and shower--shower coalescence. Shower partons that do not coalesce are hadronised via string fragmentation. Finally, UrQMD is used to model the hadronic phase until the kinetic freeze-out.

\end{itemize}

Figure~\ref{fig:comptomodelch2_4} shows the transverse-momentum spectra of \rmPiPM, \proton+\pbar, and \rmPhiMes{} in \PbPb\ collisions at \sqrtSnn{}~=~5.02~\tev\ for the 10--20\% centrality class compared with the VISHNU, EPOS3, and CoLBT models described above. For the CoLBT model, the curves obtained with the separate contributions of hydrodynamics, coalescence and fragmentation are also shown. 
The VISHNU calculations reproduce the pion and proton spectra fairly well in the low-\pT{} region (\mbox{\pT~$<3$~\gmom}), where a hydrodynamic description of the fireball evolution is expected to be valid. It should be noted that the measured spectral shapes of protons could not be reproduced by viscous hydrodynamic calculations lacking an explicit description of the hadronic phase, and therefore the inclusion of UrQMD to model the hadron transport has been crucial to obtain a good description of the measured spectra~\cite{Abelev:2012wca}. 
EPOS3 includes both soft and hard physics processes and, as a consequence, it is anticipated to give a good description of data from low- to high-\pT; this is observed here up to 4~\gmom\ or 8~\gmom\ depending on the considered particle species.
The CoLBT model provides a good description of the measured pion and proton spectra. The hydrodynamics contribution is the dominant one for \pT~$<$~2--3~\gmom, while fragmentation dominates for \pT~$>$~6--8~\gmom. At intermediate-\pT\ the coalescence contribution is needed to reproduce the data.
As noted in~\cite{Zhao:2021vmu}, in the CoLBT model the transition of the hadron production mechanism from the hydrodynamics domain to the region dominated by coalescence and fragmentation occurs at higher \pT\ for more central collisions. This is due to the stronger radial flow in central collisions that pushes the hydrodynamics regime up to higher momenta.

Despite some difficulties in the description of the individual transverse-momentum spectra~\cite{Acharya:2019yoi}, the majority of the models considered here are able to describe adequately (in their \pT\ range of applicability) the baryon-to-meson ratios \proton/\Pion{} shown in the left panels of Fig.~\ref{fig:comptomodelchratio2_4}.
For the 0--10\% centrality class, the measured \proton/\Pion{} ratios are shown for two different centre-of-mass energies, namely \sqrtSnn{}~=~5.02 and 2.76~\tev.
The trend and the magnitude of the yield ratios are similar at the two collision energies, notably with a slight shift of the peak towards higher \pT\ at the higher \sqrtSnn{}, consistent with the larger radial flow.
The VISHNU model is able to reproduce the low-\pT{} rise of the p/\Pion{} ratio, but it can quantitatively reproduce the data only for $\pT{}<1.0$--1.5~\gmom\ since, above this range, the contributions from jet showers and mini-jets to the pion spectra start to be important~\cite{McDonald:2016vlt}. 
For \sqrtSnn{}~=~2.76~\tev, the Catania model with the contributions of radial flow and soft--hard coalescence qualitatively describes the peak in the p/\Pion{} ratio in the $2<\pT{}<4$~\gmom\ region, where recombination processes are dominant. On the other hand, as expected, the Catania calculations without flow cannot reproduce the peak, while the removal of the soft--hard coalescence worsens the agreement with data in 6~$<$~\pT{}~$<$~8~\gmom. The Catania model with only the fragmentation component can reproduce the data only for \pT{}~$>$~8~\gmom. 
The EPOS3 event generator is able to describe the shape of the \proton/\Pion{} ratio, despite the peak position being shifted to lower \pT{} indicating lower radial-flow effects in the model than that measured in data. On the other hand, in the 40--50\% centrality class, where radial flow is smaller than in central collisions, EPOS3 accurately describes the peak position but overestimates the magnitude of the \proton/\Pion{} ratio at intermediate-\pT{}.
The CoLBT model describes well the measured yield ratios both in the 10--20\% and 40--50\% centrality classes, with a steep increase at low-\pT\ due to the mass ordering induced by the radial flow, and a decrease for \pT{}~$>$~3~\gmom, which results from the interplay among hydrodynamic expansion, quark recombination, and parton fragmentation.

Considering the \proton/(2\rmPhi) ratio shown in the right panels of Fig.~\ref{fig:comptomodelchratio2_4}, both EPOS3 and Catania can reproduce the measured ratio at very low-\pT{} and at high-\pT, but fail in describing the intermediate-\pT{} region, mostly because the spectral shapes are underestimated for protons. The same considerations hold true in 40--50\% \PbPb\ collisions concerning the EPOS3 model.
It should be noted that, as pointed out in~\cite{Minissale:2015zwa}, similar slopes of proton and \rmPhi{} meson spectra (i.e.~a flat p/(2\rmPhi{}) ratio as a function of \pT{}) are expected for $\pT < 2$~\gmom\ in case of coalescence (as in the case of hydrodynamics) because the difference between a combination of three quarks with 330~\mev/$c^2$ mass and that of two quarks with 450~\mev/$c^2$ mass flowing with the same collective velocity is marginal.

\begin{figure}[htb!]
    \begin{center}
        \includegraphics[width=0.96\textwidth, angle=0, clip=true, trim=0cm 0 0 0]{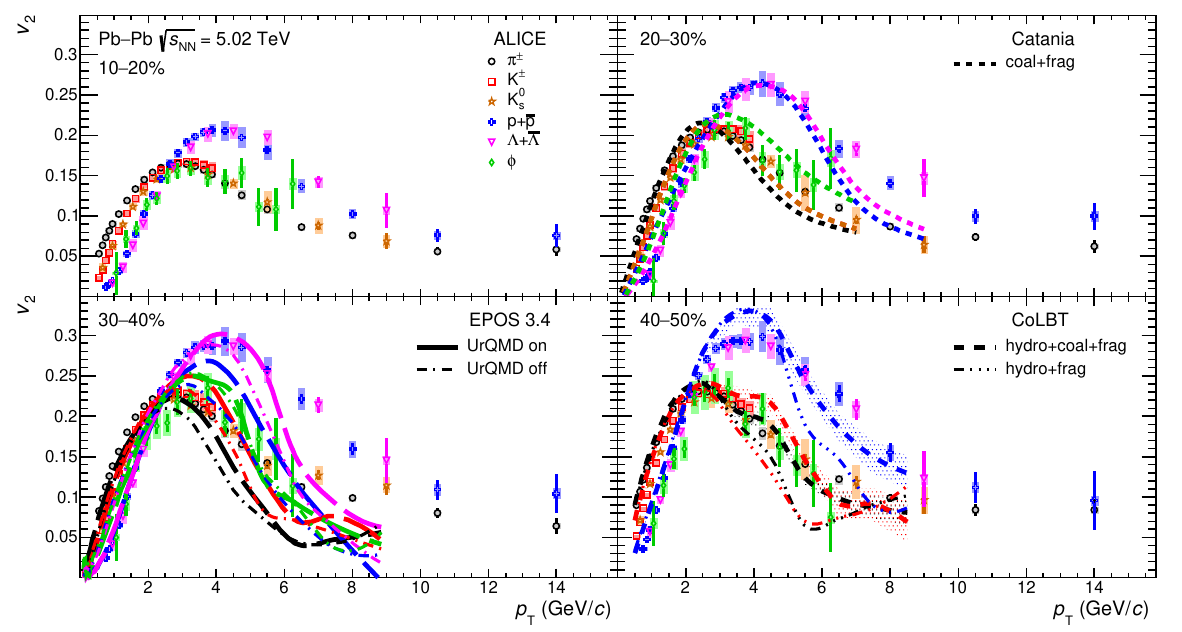}

    \caption{\pT-differential \vTwo{} for several particle species in \PbPb{} collisions at \sqrtSnn~=~5.02~\tev\ for different centrality classes~\cite{Acharya:2018zuq} compared to the Catania~\cite{Minissale:2015zwa}, EPOS 3.4~\cite{Werner:2013tya}, and CoLBT~\cite{Zhao:2021vmu} models. %
    }
    \label{fig:v2PbPb}
    \end{center}
\end{figure}

\paragraph{\textbf{\textit {Anisotropic flow coefficients.}}} 
A further benchmark for the recombination/fragmentation scenario is provided by the comparison of the measured elliptic flow coefficient \vTwo{} of identified particles with theoretical models. Figure~\ref{fig:v2PbPb} shows the \vTwo{} as a function of \pT\ for \rmPiPM, \rmKpm, \Kzs{}, \proton+\pbar, \rmLambda+\rmAlambda, and \rmPhi{} in \PbPb{} collisions at \sqrtSnn~=~5.02~\tev\ for different centrality classes~\cite{Acharya:2018zuq}. The main features of the \pT\ dependence of \vTwo{} are the same in all the centrality intervals displayed in the figure.
For \pT~$<$~2-3~\gmom, the strong radial flow results in a mass-ordered \vTwo{} as reproduced by the VISHNU hydrodynamic model with the initial energy density profile from \trento{} (see Sec.~\ref{sec:TG2pidflow}). At intermediate momenta ($3 \lesssim \pT < 8-10$~\gmom), a grouping of the baryons versus mesons is observed, as expected in the case of hadron formation via recombination. 
The \rmPhi{} meson plays a special role because its mass is similar to that of the proton. This is reflected in the trend of its \vTwo{} which follows the one of the proton at low-\pT{}; on the other hand, at intermediate-\pT{}, the \vTwo{} of the \rmPhi{} becomes compatible with meson \vTwo{} indicating that the elliptic flow is driven by quark content rather than by mass in this momentum range.
At higher \pT{}  (\pT{}$~>~$8--10 \gmom), where \vTwo{} originates mainly from the path-length dependence of the in-medium energy loss of partons produced in hard-scattering processes and fragmentation takes over from recombination as the dominant hadronisation mechanism, the \vTwo{} values of the different hadron species become compatible within uncertainties.

As shown in Sec.~\ref{sec:TG2higherorderflowandnonflow}, the same features discussed above for the \pT{}-differential elliptic flow \vTwo, namely the mass ordering at low-\pT{} followed by baryon versus meson grouping at intermediate-\pT{}, are observed also for the higher-order flow harmonics $v_3$ and $v_4$, as well as for the non-linear flow modes $v_{4,22}$ and $v_{5,32}$~\cite{Acharya:2019uia}. In particular, the observation of the same particle type grouping for $v_4$ and $v_{4,22}$ is in line with the expectation that quark recombination is the dominant particle production mechanism in the intermediate-\pT{} region and affects both flow modes similarly.

The top right and the bottom panels of Fig.~\ref{fig:v2PbPb} show the comparison of \vTwo{} to the Catania, EPOS3, and CoLBT models %
for the 20--30\%, 30--40\% and 40--50\% centrality classes, respectively. For EPOS3, the predictions from a simulation without the hadronic cascade stage (modelled with UrQMD) are also shown and they indicate that the interactions in the hadronic phase are crucial to obtain a good description of the measured \vTwo{}. EPOS3 is able to reproduce the mass-ordering observed up to \pT{}~=~2--3~\gmom. It cannot quantitatively reproduce the \vTwo{} of pions and protons at intermediate-\pT{} but a good agreement with the data is observed for the other hadrons up to \pT~=~5--6~\gmom. At higher momenta, EPOS3 always underestimates the data, indicating that a larger amount of energy loss is needed in the model.  
On the other hand, the Catania and CoLBT models, which both include recombination and fragmentation, describe well the measured \vTwo{} of the different hadron species in a wide \pT\ interval, extending from the hydrodynamic realm at low momentum to the fragmentation region at high-\pT.
Conversely, the results of CoLBT simulations without the coalescence component, shown in the bottom-right panel of Fig.~\ref{fig:v2PbPb}, underestimate the measurements for \pT~$>$~4~\gmom, indicating that in the hybrid hadronisation approach implemented in this model quark recombination is crucial to describe the data.

The recombination of flowing quarks provides a natural explanation for the observed \pT{} dependence of \vTwo{} and of the baryon versus meson grouping at intermediate-\pT~\cite{Fries:2003vb, Voloshin:2002wa}.
In the recombination picture, mesons (M) and baryons (B) at a transverse momentum \pT{} reflect the properties of partons with an average transverse momentum of \pT/2 (for mesons) or \pT/3 (for baryons). At low-\pT{}, their elliptic flow, to a first approximation, should be determined by that of their constituent quarks (q): $v_{2,\rm{M}}(\pT)=2v_{2,\rm{q}}(\pT/2)$ for mesons and $v_{2,\rm{B}}(\pT)=3v_{2,\rm{q}}(\pT/3)$ for baryons. As a consequence, \vTwo{} attains its climax value at higher momenta for baryons than for mesons. 
These simple formulae for \vTwo{} in the recombination picture hold true under the assumptions of coalescence of quarks with the same velocity (i.e.~neglecting the width of the wave functions) and of completely direct production (i.e.~neglecting the contributions from resonance decays). 
This motivated the picture of constituent-quark-number scaling of \vTwo{} as a function of the transverse kinetic
energy, which was observed to hold with good approximation (within $\pm$~10\% at intermediate-\pT) in the data at RHIC energies~\cite{Adcox:2004mh, Back:2004je, Adams:2005dq}.
At the LHC, up to $\pm20\%$ deviations from the exact scaling were observed~\cite{Acharya:2018zuq}. 
After the precise measurements obtained at the LHC for heavy-flavour hadrons, the studies of the scaling of the flow coefficients with the number of constituent quarks could be extended to charm mesons and charmonia. Such studies (see e.g.\ those reported in Sec.~\ref{sec:2.5flow}) can shed light on the role of the hadronisation mechanism in the hierarchy observed in the \vTwo{} of pions, D, and \rmJpsi{} mesons at low- and intermediate-\pT{}.

It should be noted that the constituent-quark number scaling is a prediction based on na\"{i}ve coalescence calculations with several simplifying assumptions. 
Deviations from this simple scaling and a consequent break-up of the quark-number scaling are expected to appear when some of these assumptions are relaxed by performing a full calculation with the Wigner function formalism and by accounting for multiple extra-considerations, such as:
the effects of the resonance decays~\cite{Greco:2004ex}, 
quark momentum distributions in hadrons~\cite{Greco:2004ex}, 
higher Fock states in the hadron wave function~\cite{Muller:2005pv}, 
the interactions in the hadronic phase,
the space-momentum correlations (e.g.~due to radial flow), 
the high phase-space density of quarks~\cite{Singha:2016aim}, 
the different \vTwo{} of strange quarks compared to up and down quarks, 
and the contribution of hadrons produced by fragmentation.

\paragraph{\textbf{\textit {Heavy-flavour production.}}}
\label{sec:HFpTdiffratios}
In order to extend the investigations of light-flavour hadron production, the measurement of hadrons containing heavy quarks can provide additional insights in the hadronisation mechanisms of the QGP. In fact, as already mentioned in Sec.~\ref{sec:SHM}, charm and beauty quarks are created in hard-scattering processes characterised by shorter timescales compared to the QGP formation time. Subsequently, they experience the full system evolution, undergoing multiple elastic (collisional) and inelastic (gluon radiation) interactions with the medium constituents, which can lead to their, at least partial, thermalisation in the QGP, as supported by the comparison of the measured yields and \vTwo\ of charm hadrons with model calculations~\cite{Batsouli:2002qf, Moore:2004tg,Prino:2016cni,Rapp:2018qla,Cao:2018ews}. As for the light-flavour partons, the hadronisation of heavy quarks takes place in the medium when the phase boundary is reached and can happen via two competing mechanisms, namely fragmentation and recombination with quarks from the medium. The former is expected to be the dominant mechanism at high-\pT{} ($\pT>6-8$~\gmom{}), while the latter at low- and intermediate-\pT{}. In particular, the hadronisation via recombination is expected to be most probable for heavy and light quarks close in momentum and space.
Since heavy quarks are produced essentially in the early stage of the collisions, and not in soft processes at later stages (such as thermal production in the QGP or string-breaking processes in the hadronisation as described in fragmentation models), they are bound to be especially sensitive to recombination effects. In particular, recombination is expected to affect the momentum distributions and the abundances of different heavy-flavour hadron species compared to those measured in \pp\ collisions~\cite{He:2012df}. If heavy quarks hadronise via recombination, the production of baryons relative to that of mesons is expected to be enhanced at intermediate-\pT{}. In addition, the yield of charm and beauty hadrons with strange-quark content (e.g.~\rmDs{} and \rmBs{} mesons) relative to non-strange hadrons is expected to be larger in heavy-ion collisions compared to \pp\ collisions, because of the larger production of strange quarks~\cite{Kuznetsova:2006bh,Andronic:2007zu}.  
In the case of hadronisation via recombination with flowing light quarks from the medium, open heavy-flavour hadrons are also expected to acquire some of the collective behaviour of the expanding system in addition to that inherited via their (partial) thermalisation, thus leaving a signature in the \RAA{} and \vTwo{}, as discussed in Sec.~\ref{sec:Transport}. A charm quark can also combine, during the medium evolution or at the phase boundary, with a $\overline{\rm c}$ antiquark originating from a different hard scattering, giving rise to charmonium bound states (e.g.~\rmJpsi{} mesons). Thus, recombination constitutes an additional quarkonium production mechanism in the QGP, which could counterbalance the predicted suppression of the initially produced quarkonia due to the colour-charge screening in the QGP (see Sec.~\ref{sec:charmonia}).

\begin{figure}[!tb]
    \begin{center}
        \includegraphics[width=0.48\textwidth, angle=0, clip=true, trim=0cm 0 0 0]{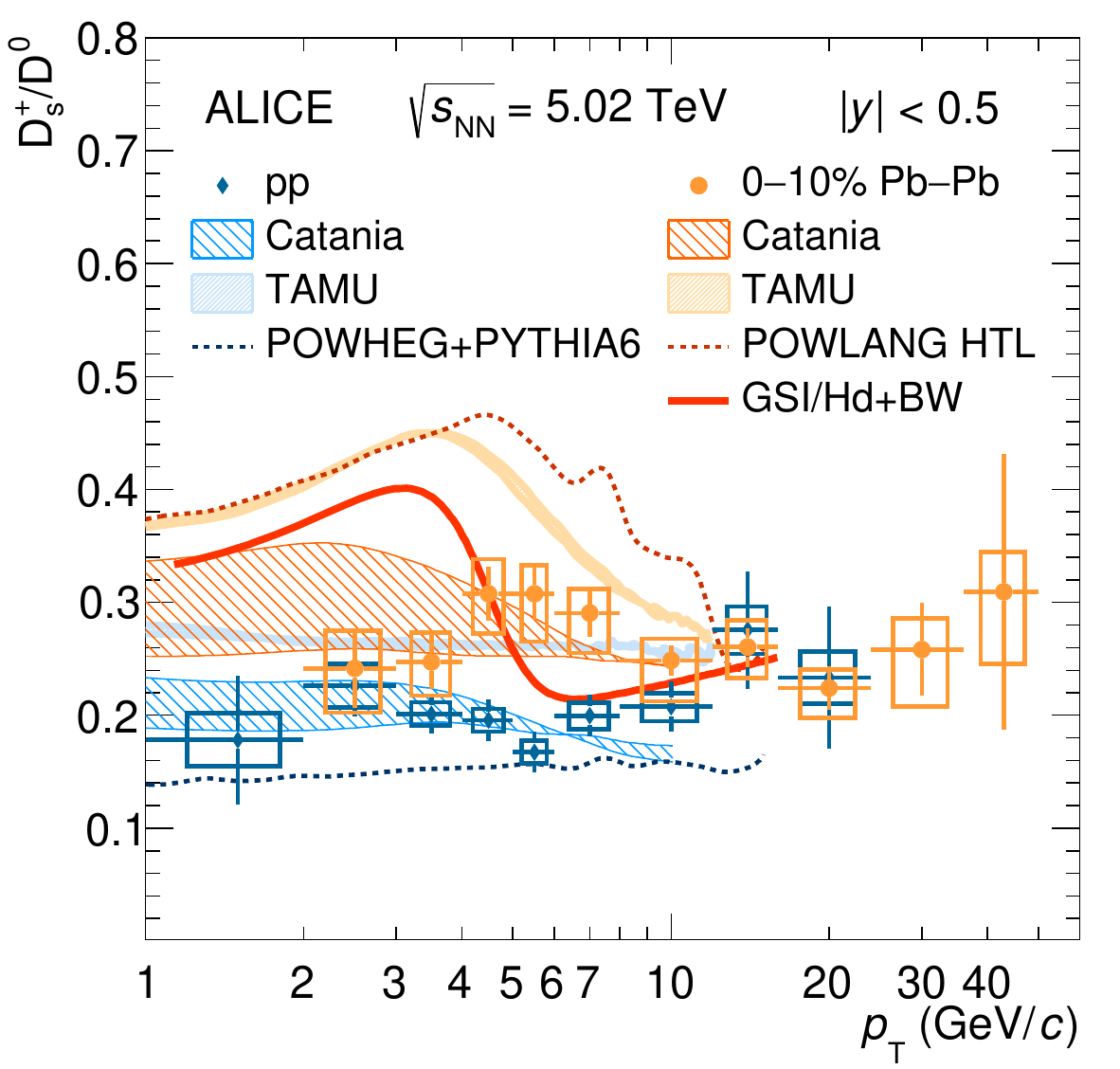}
        \includegraphics[width=0.48\textwidth, angle=0, clip=true, trim=0cm 0 0 0]{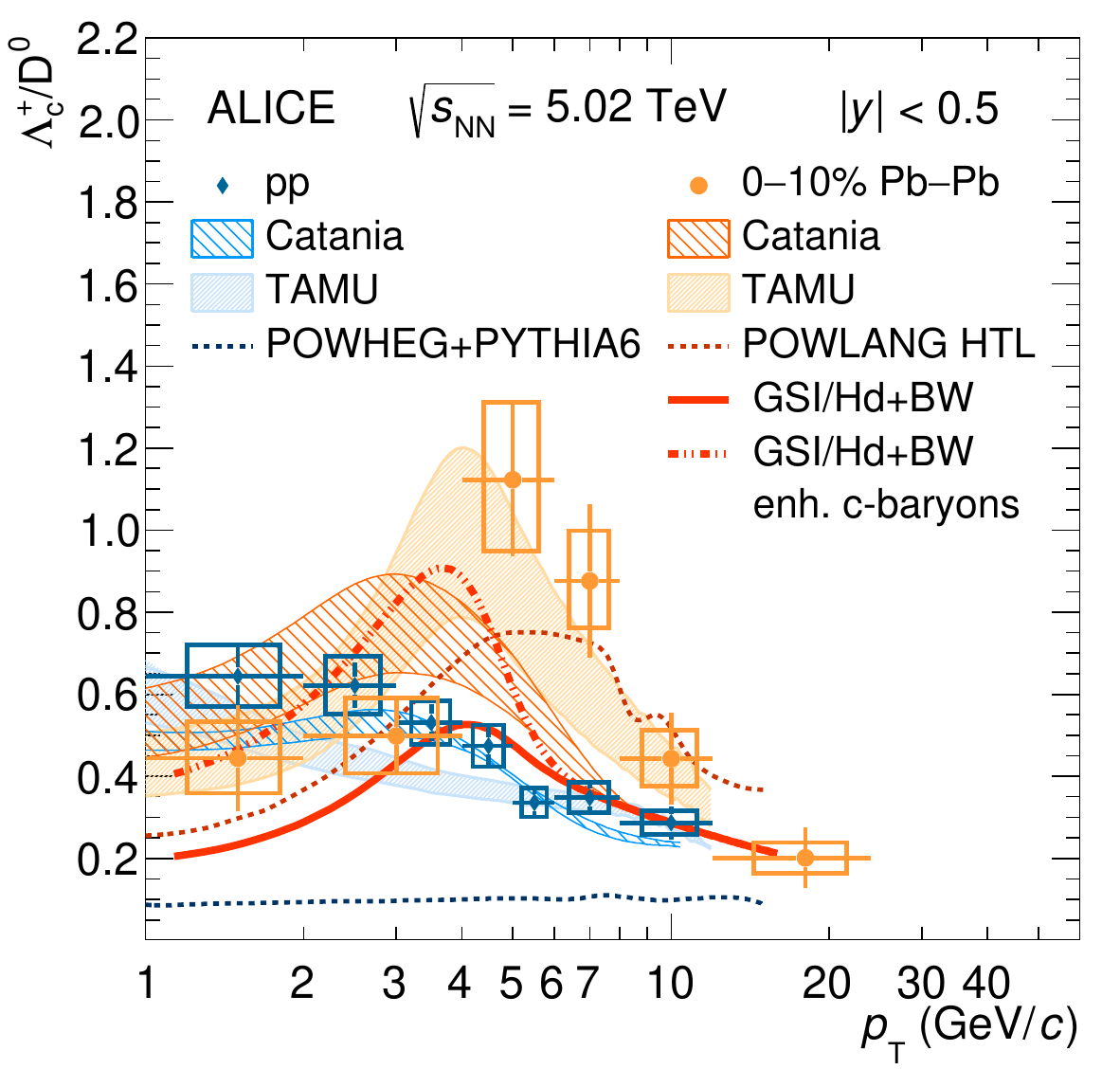}

    \caption{ \rmDs/\rmDzero~\cite{ALICE:2019nxm,ALICE:2021mgk,ALICE:2021kfc} (left) and \rmLambdaC/\rmDzero~\cite{Acharya:2020uqi,Acharya:2020lrg,ALICE:2021bib} (right) yield ratios as a function of \pT{} in pp collisions and in the 10\% most central \PbPb{} collisions at \sqrtSnn~=~5.02~\tev\ compared to different model calculations, namely Catania~\cite{Plumari:2017ntm, Minissale:2020bif}, TAMU~\cite{He:2019vgs}, POWLANG~\cite{Beraudo:2022dpz}, and the GSI-Heidelberg statistical hadronisation model~\cite{Andronic:2021erx}.}
    \label{fig:HF_ratios}
    \end{center}
\end{figure}

\begin{sloppypar}
Figure~\ref{fig:HF_ratios} shows the \pT-differential yield ratios \rmDs/\rmDzero{} (left panel) and \rmLambdaC/\rmDzero{} (right panel) measured in minimum-bias pp collisions~\cite{Acharya:2020lrg,Acharya:2020uqi,ALICE:2019nxm,ALICE:2021mgk} and in the 10\% most central \PbPb\ collisions at \sqrtSnn~=~5.02~\tev{}~\cite{ALICE:2021bib,ALICE:2021kfc}. The ratio between the \pT-differential yields of strange and non-strange D mesons in pp collisions does not show any marked \pT{} dependence within the experimental uncertainties and is compatible with previous measurements at \EplusEminus{} colliders~\cite{Gladilin:2014tba}. A hint of a larger \rmDs/\rmDzero\ ratio (at $2.3\sigma$ level of the combined statistical and systematic uncertainties) is observed for $\pT<8~\gmom$ in central \PbPb\ collisions compared to \pp\ collisions, which can be explained in the case of charm-quark hadronisation via recombination within a strangeness-rich medium. At higher \pT{}, the \rmDs/\rmDzero{} ratio converges to the values measured in \pp\ collisions.
\end{sloppypar}

Unlike the ratio between charm meson yields, the \rmLambdaC/\rmDzero{} ratio measured in minimum-bias pp collisions exhibits a \pT{} dependence, which is around  0.4--0.6 at low-\pT{} and decreases to 0.2--0.3 at high-\pT, similarly to what was observed in the light-flavour sector for the baryon-over-meson ratios p/\Pion{} and \ratioLamOverKzs~\cite{Acharya:2020uqi}. In addition, the measured \rmLambdaC/\rmDzero{} ratio is significantly higher than the measurements done at \EplusEminus{} and ep  colliders~\cite{Gladilin:2014tba,Lisovyi:2015uqa}. This enhancement, observed for different charm-baryon species in pp collisions~\cite{ALICE:2021rzj,ALICE:2021psx,ALICE:2021bli}, as discussed in Sec.~\ref{sec:open-heavy-flavor},
suggests that the fragmentation of charm quarks into hadrons is not universal across collision systems, and further mechanisms that enhance the baryon productions have to be taken into account with respect to those considered for the \EplusEminus{} and \ePMproton\  predictions.   %
A hint of enhancement in the baryon-over-meson ratio is observed for $4<\pT<8~\gmom$ in the 0--10\% most central \PbPb{} collisions with respect to pp collisions. This possible enhancement at intermediate-\pT{} is consistent both with the radial expansion of the system that imparts the same velocity to all particle species, boosting heavier particles to higher momenta, and the hadronisation of the charm quarks via the recombination mechanism that is expected to be dominant in this momentum region.

In Fig.~\ref{fig:HF_ratios}, the \pT-differential \rmDs/\rmDzero{} and \rmLambdaC/\rmDzero{} yield ratios are compared to  theoretical predictions obtained with four different models. 
The Catania model~\cite{Plumari:2017ntm, Minissale:2020bif} assumes that a colour-deconfined state of matter is formed in both pp and \PbPb{} collisions 
and implements the heavy-quark transport via the Boltzmann equation (see Sec.~\ref{sec:PartonInteractions} for more details). 
As for the light-flavour particles, the hadronisation in the Catania model can occur via instantaneous coalescence, implemented through the Wigner formalism, in addition to the fragmentation. 
The coalescence mechanism is predicted to be the dominant one at low- and intermediate-\pT{}, 
while at high-\pT{} charm quarks are expected to mainly hadronise via fragmentation. 
The Catania model reproduces within the uncertainties the measured \rmDs/\rmDzero{} ratios both in pp and \PbPb{} collisions. They also describe the \rmLambdaC/\rmDzero{} ratios in pp collisions while, in central \PbPb{} collisions, they slightly overestimate the measurements at low-\pT{} and underestimate them at intermediate-\pT{}. A similar description of the heavy-quark transport and hadronisation in \PbPb{} collisions is provided by the TAMU model~\cite{He:2019vgs}. In this case, the charm-quark transport in a hydrodynamically expanding medium is described by the Langevin equation and the hadronisation via recombination is implemented with a Resonance Recombination Model (RRM)~\cite{Ravagli:2007xx}. Within this formalism, the hadronisation proceeds via formation of resonant states when approaching the pseudo-critical temperature, whose rates are governed by the Boltzmann equation. In this case, the recombination process is not instantaneous and it is governed by a time scale which is the inverse of the width of the resonant states formed and is therefore different for different resonant states. The resonant states include heavy-flavour baryon states that have not been measured so far, but they are predicted by the Relativistic Quark Model (RQM)~\cite{Ebert:2011kk} and consistent with lattice-QCD predictions. It is also important to notice that, differently from the Wigner formalism, the RRM guarantees 4-momentum conservation for all \pT{} values. In pp collisions, in contrast to the \PbPb{} case, the hadronisation in the TAMU model is not described with the RRM: the abundances of the different charm-hadron species are instead determined with a statistical hadronisation approach~\cite{He:2019tik}, which takes into account an augmented set of charm-baryon excited states based on guidance from the RQM and is able to reproduce the measured \rmLambdaC/\rmDzero{} ratio. The TAMU predictions reproduce the magnitude and shape of the \pT-differential \rmLambdaC/\rmDzero{}, while they overestimate the \rmDs/\rmDzero{} ratios both in pp and \PbPb{} collisions.
The POWLANG model~\cite{Beraudo:2022dpz} implements a Langevin-based transport of heavy quarks in the QGP followed by in-medium hadronisation. At the hadronisation stage, occurring around the QCD pseudo-critical temperature, charm quarks are recombined with light thermal quark or di-quark states from the medium into colour-singlet clusters. Low-mass clusters are then decayed into a charm hadron and a light particle, while heavy clusters are fragmented according to the Lund model~\cite{Andersson:1983ia,Andersson:1997xwk}. In order to provide a benchmark relevant for pp collisions, the POWLANG model is compared to the results obtained with the POWHEG-BOX event generator~\cite{Alioli:2010xd} matched to PYTHIA~6~\cite{Sjostrand:2006za} for the parton shower and for the hadronisation, the latter being implemented only via fragmentation. In central Pb--Pb collisions the model overestimates the \rmDs/\rmDzero{} ratios, while it better describes the \rmLambdaC/\rmDzero{} measurements. 
Finally, the predictions for the \rmDs/\rmDzero{} and \rmLambdaC/\rmDzero{} ratios in \PbPb{} collisions obtained from an approach based on the statistical hadronisation model are reported in Fig.~\ref{fig:HF_ratios} as solid lines  for the calculations using the list of charm-baryon states from the PDG~\cite{Zyla:2020zbs}, and as a dash-dotted line for the \rmLambdaC/\rmDzero{} ratio obtained considering an augmented set of excited charm baryon states along with a larger charm production cross section.
Such calculations use the GSI-Heidelberg SHM described in Sec.~\ref{sec:SHM} for the hadron yields, while the \pT{} spectra of charm hadrons are modelled with a core-corona approach~\cite{Andronic:2021erx}. The core contribution, especially important at low-\pT{}, is parameterised with a blast-wave function under the assumption of local thermal equilibrium of charm quarks in the fireball formed in the collision. Hence, the charm hadrons at hadronisation inherit the thermal motion of the charm quarks superimposed with the collective velocity of the hydrodynamically expanding QGP. The corona contribution is instead parameterised from measurements in pp collisions. Overall, this approach provides a fair description of the gross features of the data, even though it predicts a \pT\ dependence of the \rmDs/\rmDzero{} ratio that is more pronounced than the measured one and it underestimates the measured \rmLambdaC/\rmDzero{} ratio at intermediate-\pT. In particular, the calculations with an augmented set of charm baryons predict a larger \rmLambdaC/\rmDzero{} ratio than the ones using the PDG list for $\pT<6$~\gmom, but both of them undershoot the data points in $4<\pT<8$~\gmom.

The $p_{\rm T}$-integrated $\Lambda_{\rm c}/{\rm D}^0$ ratios in Pb--Pb collisions in the two considered centrality classes were calculated by extrapolating the measured $\Lambda_{\rm c}$ $p_{\rm T}$-differential yields down to $p_{\rm T}=0$ as discussed in Ref.~\cite{ALICE:2021bib}. These ratios are compatible with the ones measured in pp and p--Pb collisions~\cite{Acharya:2020lrg,Acharya:2020uqi} within one standard deviation of the combined uncertainties. This behaviour is similar to the one discussed above for the \ratioLamOverKzs\ baryon-to-meson ratio in the strangeness sector and suggests that, also in the charm sector, the modifications of the hadronisation in presence of the QGP mainly result in a redistribution of the $\Lambda_{\rm c}$ and D$^0$ yields over $p_{\rm T}$ rather than an overall increase in baryon production. However, considering the relatively large uncertainty of the $p_{\rm T}$-integrated $\Lambda_{\rm c}$ yield in Pb--Pb collisions, the current data do not allow for a discrimination among different charm-baryon formation scenarios.

Insight into the hadronisation of beauty quarks was obtained from recent measurements of the \pT{}-differential yield ratios of non-prompt \rmDs{} and non-prompt \rmDzero{} mesons~\cite{ALICE:2021mgk,ALICE:2022xrg}, i.e.\ D mesons originating from the decay of beauty hadrons. This study is complementary to the measurements of the ${\rm B_s^0}/{\rm B^+}$ ratios reported by the CMS Collaboration~\cite{CMS:2021mzx} for $p_{\rm T}>7$ GeV/$c$, which show a hint of enhancement of beauty-strange meson production in Pb--Pb collisions. The measurement of non-prompt D mesons provides sensitivity 
to the production yields of different B-meson species because in pp collisions about 50\% of non-prompt \rmDs{} mesons are produced in \rmBs{}
decays, while most of the non-prompt \rmDzero{} mesons originate from non-strange B-meson decays~\cite{ALICE:2021mgk}.
The measurements reported in Ref.~\cite{ALICE:2022xrg} show a hint of a larger non-prompt \rmDs/\rmDzero{} yield ratio in central \PbPb{} collisions relative to pp interactions, with a significance of $1.7\sigma$.
This is consistent with an enhanced production of beauty-strange mesons in heavy-ion collisions, as expected in a scenario in which beauty quarks hadronise via recombination with quarks from the strangeness-rich QGP.
The TAMU model~\cite{He:2014cla} can describe qualitatively the data points, capturing the enhancement of non-prompt \rmDs{} mesons and its trend with \pT{}.

The models that implement the heavy-quark hadronisation via recombination and fragmentation in the QGP provide a good description of the main features of heavy-flavour hadron production measurements down to low transverse momentum, supplying important additional insights into the hadronisation mechanisms.

\subsubsection{The hadronic phase}
\label{sec:HadrPhase}

The hadron-gas phase lasts approximately 5--10~\fmC\ from the chemical freeze-out, right after the hadronisation of the QGP, to the kinetic freeze-out, when all interactions cease and hadrons stream freely. 
The resonances with lifetimes of the same timescale are likely to decay before the kinetic freeze-out. They are therefore good probes of the dynamics of the hadronic phase~\cite{Torrieri:2001ue}. In fact, the decay products of resonances are subject to elastic interactions in the hadron gas, which modify their momenta and prevent the reconstruction of the resonance signal by means of an invariant mass analysis. As a consequence, the measured resonance yield is suppressed with respect to the amount produced at the chemical freeze-out. In this context, it is also worth recalling that the line shapes of the invariant mass peaks of the reconstructed resonances in Pb--Pb collisions do not show any significant deviation, within the experimental uncertainties, with respect to the measurements in pp collisions and to the expectations based on the resonance mass and width from the PDG. 
The suppression effect due to \emph{rescattering} can be compensated by \emph{regeneration} processes, by which hadrons from the medium interact and form a resonance that will decay after the kinetic freeze-out. Regeneration could even be the dominant effect: in this case, the measured resonance yield would be enhanced. Resonance decays and regeneration reactions are not in balance, though, since chemical equilibrium is lost at the chemical freeze-out. A partial chemical equilibrium (PCE), however, has been proposed~\cite{Motornenko:2019jha}, for which the decays and the regeneration of the short-lived resonances obey the law of mass action, i.e.~the abundances of the different resonances stay in equilibrium with the particles that are formed in the decays of these resonances. 
In this approach, the yields of resonances are related to the temperature at the kinetic freeze-out (see also~\cite{Rapp:2003ar}).

\begin{figure}[tb!]
    \begin{center}
       \includegraphics[width=0.48\textwidth, angle=0, clip=true, trim=0cm 0 0 0]{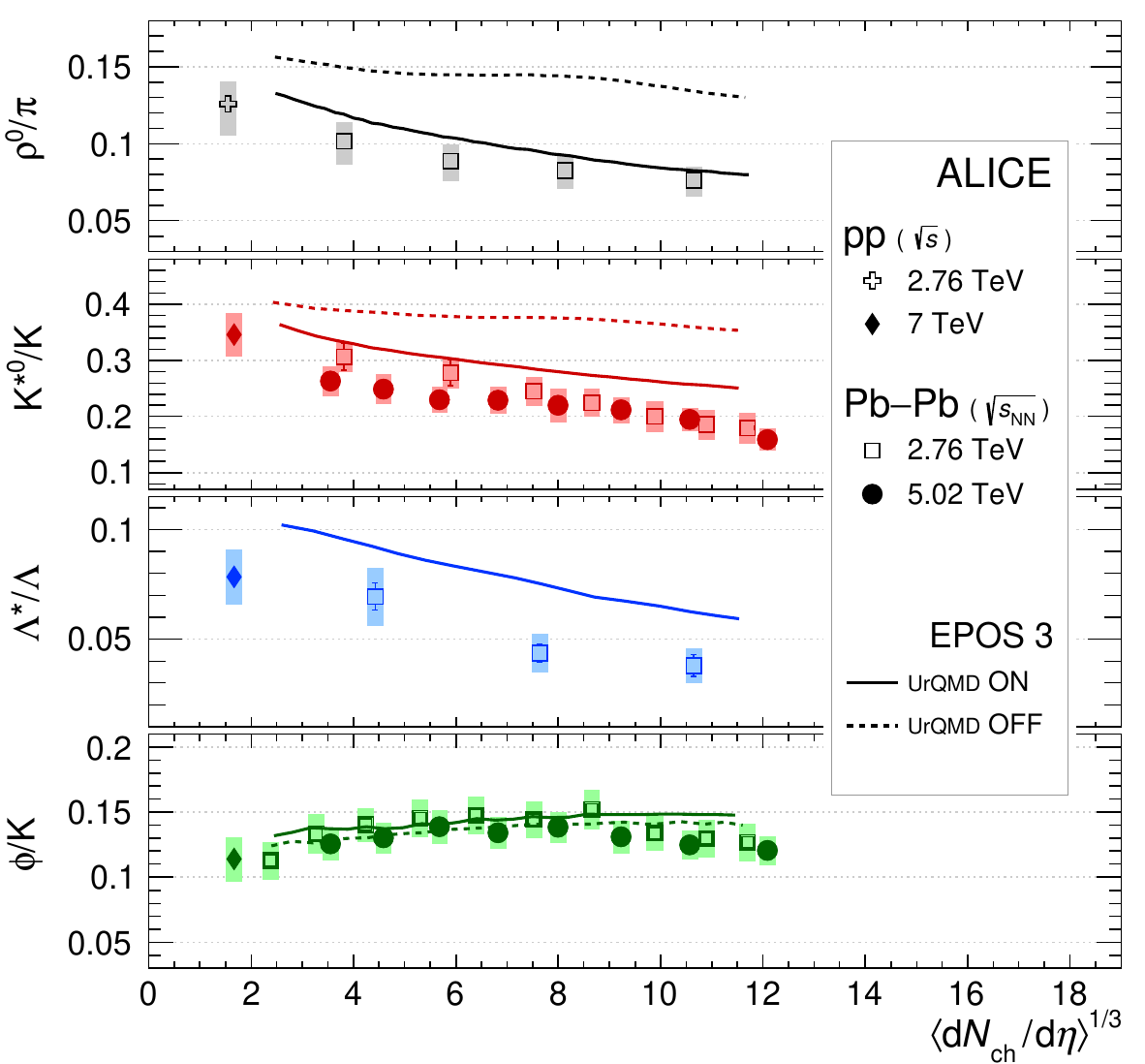}
       \includegraphics[width=0.48\textwidth, angle=0, clip=true, trim=0cm 0 0 0]{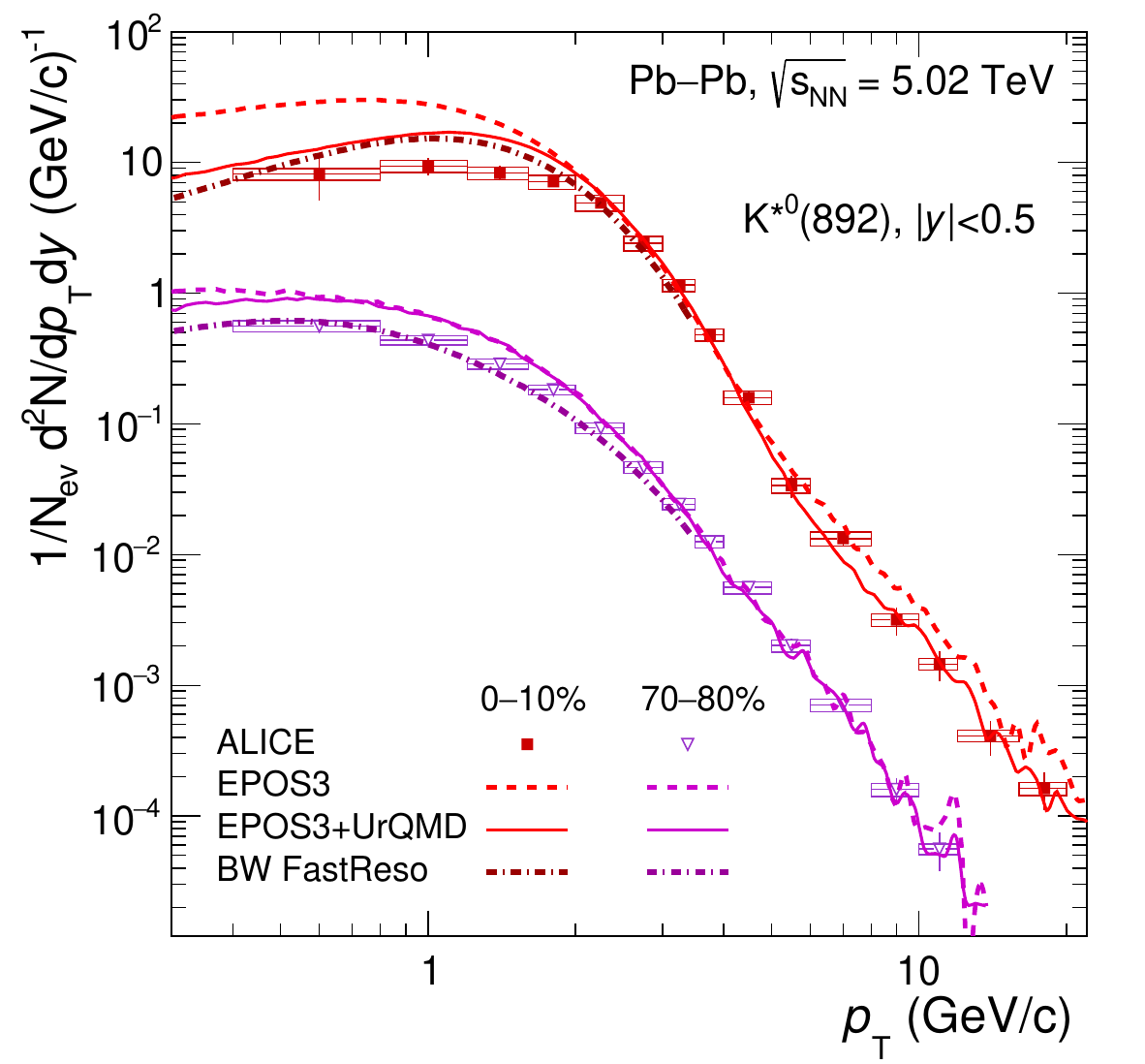}
        \caption{(Left)  
        Ratios of midrapidity yields between resonances and ground-state hadrons of similar valence-quark content, as a function of $\langle$\dNchdeta$\rangle^{1/3}$. 
        These are obtained for inelastic \pp{} collisions (\sqrtS{} = 2.76~\tev\ and 7~\tev~\cite{ALICE:2017jyt, Acharya:2018orn, ALICE:2019etb, ALICE:2020nkc}) and for \PbPb\ collisions (\sqrtSnn{}~=~2.76~\tev~\cite{ABELEV:2013zaa, ALICE:2018ewo, ALICE:2018qdv} and 5.02~\tev~\cite{Abelev:2013vea,Acharya:2019yoi, ALICE:2019xyr}) ranging from peripheral to most central overlaps. Error bars and boxes represent the statistical and the total systematic uncertainty, respectively. Model predictions are from EPOS3 with and without UrQMD~\cite{Knospe:2015nva}. 
        (Right) Transverse-momentum distributions of \rmKstar{} at \sqrtSnn{} =  5.02~\tev~\cite{ALICE:2019xyr} for central and peripheral collisions compared with predictions from EPOS3 (with and without UrQMD) and from a blast-wave model with a dedicated treatment of resonances (BW FastReso)~\cite{Mazeliauskas:2019ifr}. 
        }
        \label{fig:resonances_yield_ratio}
    \end{center}
\end{figure}

In a simple scenario of a sudden and instantaneous kinetic freeze-out common to all particle species, 
the amount of suppressed (or enhanced) yield that is measured depends on 
\emph{i}) the lifetime of the resonance, 
\emph{ii}) the cross sections for rescattering and regeneration processes,
and \emph{iii}) the time span (duration) of the hadronic phase. 
The latter is expected to be longer for central collisions, where larger system volumes are created, than for peripheral collisions. 
The best figure of merit to quantify the net effect is the yield ratios of resonances to long-lived hadrons with the same quark composition;
by measuring such ratios with resonances of increasing proper lifetime, like
\rmRhoZero{} (\mbox{$\tau^{\rm{rf}}    = 1.3~\fmC$}), 
\rmKstar{} (\mbox{$\tau^{\rm{rf}}      = 4.16~\fmC$}),
\rmLambdaStar{} (\mbox{$\tau^{\rm{rf}} = 12.6~\fmC$})
and \rmPhiMes{} (\mbox{$\tau^{\rm{rf}} = 46.3~\fmC$})~\cite{Zyla:2020zbs}, 
one can make use of different sensitivities to the hadronic phase and its interactions.
For simplicity, these resonances will be denoted as $\rho$, K$^*$, $\Lambda^*$, and $\rmPhi$ in the following.
The results of this investigation are shown in the left panel of Fig.~\ref{fig:resonances_yield_ratio}, where the ratios of \pT-integrated\footnotemark\ yields $\rho/\pi$~\cite{ALICE:2018qdv}, K$^*/$K~\cite{ALICE:2017ban}, $\Lambda^*/\Lambda$~\cite{ALICE:2019smg,ALICE:2018ewo} and $\rmPhi/K$~\cite{ALICE:2017ban} are presented as function of the cubic root of the charged-particle multiplicity at $\absrap \approx 0$, $\langle$\dNchdeta$\rangle^{1/3}$, which represents the radial extent of the system (see Fig.~\ref{fig:volume_and_time} in Sec.~\ref{sec:TG1sizelifetime}). Results are shown for \PbPb\ collisions at two different colliding energies, \sqrtSnn~=~2.76~\tev\ and 5.02~\tev, and for pp collisions at \sqrtS~=~2.76~\tev\ and 7~\tev. Note that the \rmPhi\ meson is a hidden-strangeness hadron but, as far as hadro-chemistry is concerned, it behaves like the kaon that has strangeness $S=1$. 

\footnotetext{~Like for the ground states (\Pion, K, \rmLambda) used in the present ratio, the fraction of extrapolated yield to zero \pT, in \pp\ as well as in \PbPb, remains small to moderate for the various resonances (typically 5\% for \rmKstar, 6\% to 16\% for \rmLambdaStar, 14\% for \rmPhiMes{} and 20--30\% for \rmRhoZero).}

If no final-state effects such as rescattering and regeneration intervened during the hadron-gas phase to modify the resonance yields, the ratios in Fig.~\ref{fig:resonances_yield_ratio} would be constant as a function of $\langle$\dNchdeta$\rangle^{1/3}$ due to the similar production mechanisms for particles with the same strangeness and quark composition.
What is instead observed in Fig.~\ref{fig:resonances_yield_ratio} is that the yields of $\rho$, K$^*$, and $\Lambda^*$ are progressively more suppressed with respect to the reference long-lived hadrons when going from peripheral to central collisions, whereas the \rmPhi/K ratio is almost flat. For the $\rho/\pi$, K$^*/$K, and $\Lambda^*/\Lambda$ ratios, the difference between the values measured in pp and in the most central collisions is larger than $3\sigma$ of the combined statistical and systematic uncertainties.
The fact that the yield ratios involving $\rho$, K$^*$, and $\Lambda^*$ resonances decrease with increasing $\langle$\dNchdeta$\rangle^{1/3}$ suggests that,
in the hadronic phase, the rescattering of the decay products is dominant over regeneration 
and that it is more effective in central collisions, where the hadronic phase lasts longer.
The fact that the \rmPhi/K ratio remains approximately constant indicates instead that
\rmPhi\ mesons are not affected by final state effects in the hadronic phase. 
This is likely due to the \rmPhi\-meson lifetime, one order of magnitude larger than that of the K$^*$ resonance, 
which causes \rmPhi\ mesons to decay after the kinetic freeze-out. Hence, the \rmPhi\ decay products do not rescatter and no regeneration of \rmPhi\ mesons occurs (since the regeneration of a resonance is likely to originate from rescatterings of resonance decay products). This is consistent with the PCE picture of Ref.~\cite{Motornenko:2019jha}, where the yield of the long-lived \rmPhi\ mesons relative to pions is expected to have a very mild dependence on the kinetic freeze-out temperature, i.e.\ to be minimally modified during the hadronic phase.

The yield ratios in Fig.~\ref{fig:resonances_yield_ratio} are also compared to EPOS3 simulations with and without a hadronic cascade phase modelled by UrQMD~\cite{Knospe:2015nva}. The ratios obtained from EPOS3 without UrQMD are independent of multiplicity as expected for all resonances, including a flat behaviour for \rmPhi/K, that further supports the similarity between \rmPhi\ and kaon as far as hadro-chemistry is concerned. When the hadronic cascade phase (UrQMD) is switched on, the model reproduces, at least qualitatively, the observed decreasing trend of the $\rho/\pi$, K$^*/$K and $\Lambda^*/\Lambda$  ratios with increasing multiplicity. The \rmPhi/K ratio remains flat also with UrQMD, confirming the hypothesis of no final-state effects acting on the \rmPhi\ meson.

The hadronic rescattering effect is expected to be momentum dependent, with greater strength for low-\pT\;resonances (\pT~$<$~2~\gmom)~\cite{Abelev:2014uua}. 
This can be studied by comparing the measured resonance \pT\;spectra with the expectations from different models.
The \pT\;spectra of K$^*$ resonances are shown in the right panel of Fig.~\ref{fig:resonances_yield_ratio} 
for central (0--10\%) and peripheral (70--80\%) \PbPb\ collisions at \sqrtSnn{} = 5.02~\tev. 
They are compared with predictions from the EPOS3 event generator (with and without UrQMD) and from an improved blast-wave model (BW FastReso)~\cite{Mazeliauskas:2019ifr} hinging on the FastReso computation for resonance production~\cite{Mazeliauskas:2018irt} (described in Sec.~\ref{sec:TG2flowresults}). 
In the BW FastReso calculations, the common radial velocity and temperature at the kinetic freeze-out are evaluated from a simultaneous blast-wave fit to the measured \pT\;spectra of pions, kaons, and protons and used to predict the shape of resonance spectra. The resulting distributions are then normalised so that their integrals are equal to the measured yield of charged kaons in \PbPb\ collisions multiplied by the K$^*/$K ratios given by a SHM fit to ALICE data~\cite{Andronic:2017pug}. 
For peripheral collisions, model predictions are in fairly good agreement with data over the full measured range. 
For central collisions, agreement is found only for \pT~$>$~2~\gmom, whereas at low-\pT, as expected, the measured yields are reduced due to the rescattering effects. 
EPOS3 without UrQMD fails to describe the data at low-\pT, for both central and peripheral collisions, 
whereas good agreement is obtained at high-\pT. 
When switching UrQMD on, a suppression is indeed observed at low-\pT, with increasing strength for decreasing \pT. 

\begin{figure}[tb!]
    \begin{center}
        \includegraphics[width=0.48\textwidth, angle=0, clip=true, trim=0cm 0 0 0]{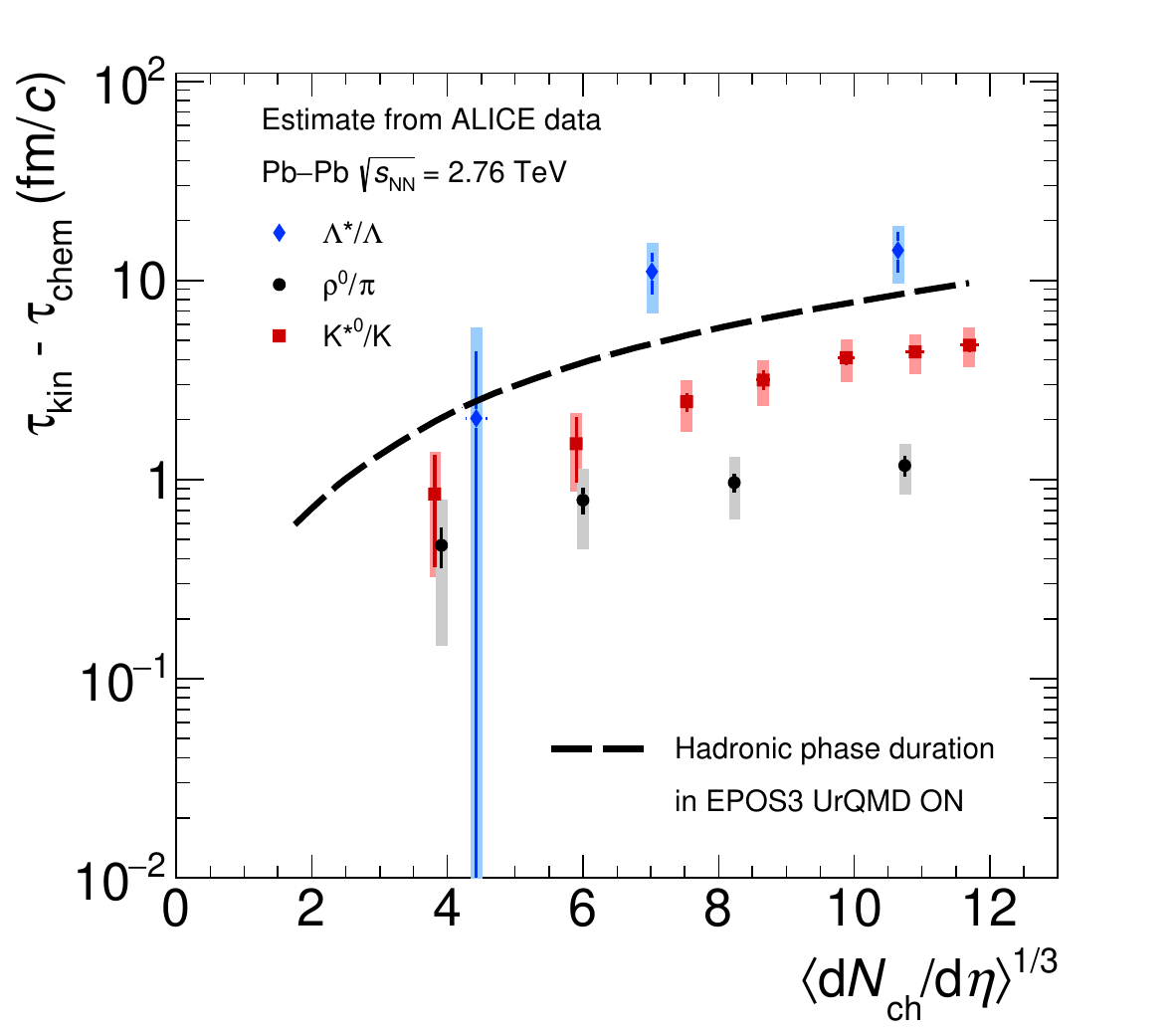}
       \includegraphics[width=0.48\textwidth, angle=0, clip=true, trim=0cm 0 0 0]{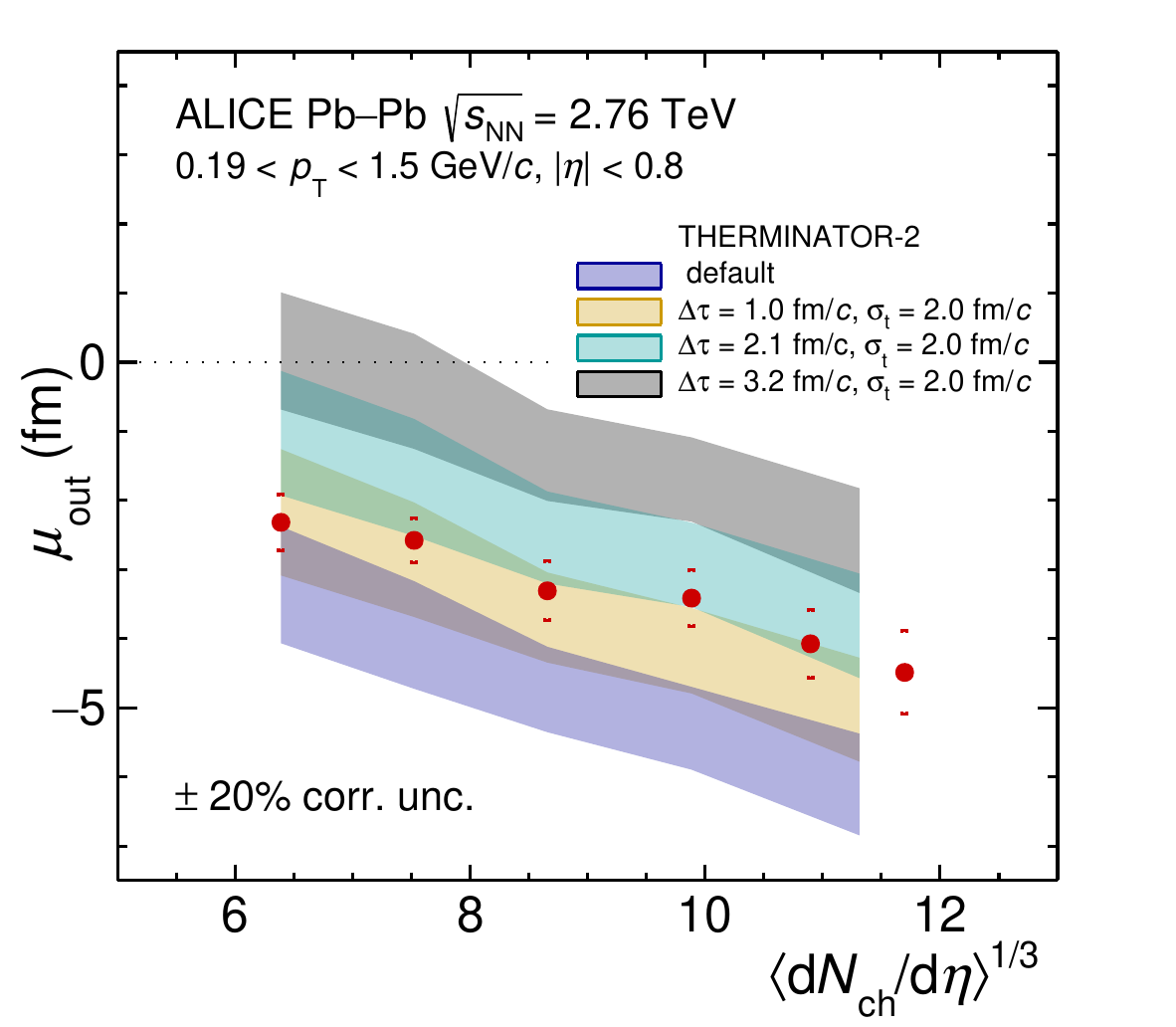}
        \caption{(Left) Hadronic phase duration (time span between chemical and kinetic freeze-out) as a function of $\langle$\dNchdeta$\rangle^{1/3}$ calculated from the $\rho/\pi$, K$^*/$K and $\Lambda^*/\Lambda$ ratios in \PbPb\ collisions at \sqrtSnn{} = 2.76~\tev~\cite{ABELEV:2013zaa, ALICE:2018ewo, ALICE:2018qdv}. The $\rho/\pi$ results are shifted horizontally by 0.1 units for visibility. (Right) Pion–kaon emission asymmetry for \PbPb\ collisions at \sqrtSnn{} = 2.76~\tev\ as a function of $\langle$\dNchdeta$\rangle^{1/3}$~\cite{Acharya:2020fmt}. The shaded areas show predictions from the THERMINATOR 2 model with default and selected values of additional delay $\Delta\tau$ for kaons~\cite{Kisiel:2018wie}.}
        \label{fig:lifetime}
    \end{center}
\end{figure}

Rescattering in the hadronic phase is seen as the most probable cause of the measured suppression. 
This suggests to further use the $\rho/\pi$, K$^*/$K, and $\Lambda^*/\Lambda$ ratios 
to estimate the time span between chemical and kinetic freeze-out ($\tau_{\rm kin}-\tau_{\rm chem}$).
This estimation is performed by means of an exponential decay law~\cite{ALICE:2019xyr}, 
$r_{\rm{kin}} = r_{\rm{chem}} \times e^{-(\tau_{\rm kin}-\tau_{\rm chem})/\tau_{\rm{res}}}$, under the assumptions of negligible regeneration, and sudden chemical and kinetic freeze-outs occurring at the same instant for all particle species.
The ratios $r_{\rm{kin}}$ and $r_{\rm{chem}}$ are any of the $\rho/\pi$, K$^*/$K or $\Lambda^*/\Lambda$ measured yield ratios at the kinetic and chemical freeze-out, respectively. 
In the calculations, the ratio in pp collisions is taken as a substitute for the ratio $r_{\rm{chem}}$ at the chemical freeze-out, while $r_{\rm{kin}}$ is the measured ratio in \PbPb\ collisions. 
The lifetime of the resonance is dilated by the Lorentz factor, when going from the resonance rest frame to the laboratory reference frame;
such a factor is associated with \meanpT\ of the individual \pT-spectrum of each resonance, so that $\tau_{\rm{res}} = \tau^{\rm{rf}}_{\rm{res}}. \gamma^{\meanpT}_{\rm{res}}$.
Results for the estimated duration of the hadronic phase are shown in the left panel of Fig.~\ref{fig:lifetime} as a function of $\langle$\dNchdeta$\rangle^{1/3}$. 
For all particles an increasing trend is observed, which mirrors the decreasing behaviour of the yield ratios for increasing system size. 
Unlike the na\"{i}ve expectation of a hadronic-phase duration common to all resonances for a given system size, different values are obtained from the $\rho$, K$^*$, and $\Lambda^*$ yields.
The estimated times differ by up to an order of magnitude, being larger for longer-lived resonances. 
The predictions for the time span between chemical and kinetic freeze-out obtained from EPOS3 simulations with UrQMD~\cite{Knospe:2015nva} are also shown for reference in the left panel of  Fig.~\ref{fig:lifetime}.
The hadronic-phase duration from EPOS3+UrQMD has a similar magnitude (of few \fmC) and trend with multiplicity as compared to the estimates obtained from the measured resonance yields, being closer to the estimate calculated from the \rmLambdaStar{}, which is the longest lived among the considered resonances.

The reason why the time durations obtained from $\rho/\pi$ and K$^*/$K ratios are shorter than the time calculated from the $\Lambda^{*}/\Lambda$ ratio might be due to regeneration effects, neglected in the simple exponential decay model and affecting more the $\rho$ and K$^*$ resonances than the $\Lambda^*$. If regeneration plays a role, the hadronic-phase duration estimated from the yield ratios can be interpreted as the lapse of time between the on-average delayed resonance production due to regeneration and the kinetic freeze-out, therefore representing a lower limit for the duration of the hadronic phase. In order to estimate the actual duration of the hadronic phase, the delay due to the regeneration has to be added to the time span estimated from the ratios. 

Sensitivity to the delay in the production of K$^*$ due to regeneration can be obtained from femtoscopy measurements (see Sec.~\ref{sec:TG1sizelifetime} and also~\cite{Acharya:2017qtq}). In particular, it is expected from hydrokinetic model (HKM) simulations~\cite{Shapoval:2014wya,Sinyukov:2002if} that the interferometry radii are little affected by the decay of K$^*$ resonances, but they are sensitive to regeneration, which shifts the emission time of pions and kaons. Since pions from K$^*$ decays have a negligible effect on the pion interferometry measurements due to the large amount of primary pions, a longer emission time for kaons than for pions is expected due to regeneration through the K$^*$ resonance. Regeneration can also explain~\cite{Sinyukov:2015kga} the \mT-scaling violation seen in Fig.~\ref{fig:pion_kaon_femto_rad} in Sec.~\ref{sec:TG1sizelifetime}, where the radii of kaons are systematically higher than those of pions in the same centrality class. The difference in the emission time of pions and kaons is quantified by the pion--kaon emission asymmetry, presented in the right panel of Fig.~\ref{fig:lifetime} for Pb–Pb collisions at \sqrtSnn{} = 2.76~\tev\ as a function of $\langle$\dNchdeta$\rangle^{1/3}$~\cite{Acharya:2020fmt}. The non-zero emission asymmetry observed in the figure is the consequence of several effects, including the collective expansion of the system, the presence of short-lived resonances decaying into the considered particles, and the radial flow of these resonances. All these features are only qualitatively reproduced by a (3+1) viscous hydrodynamic model~\cite{Bozek:2011ua}, coupled to the statistical hadronisation, resonance decay, and propagation code THERMINATOR~2~\cite{Kisiel:2018wie}. 
In order to reproduce quantitatively the measured emission asymmetry, an emission delay $\Delta\tau$ of 1.0–-2.1~\fmC\ for the kaons has to be specifically included in the model. This $\Delta\tau$ can be interpreted as the delay due to the decay of K$^*$ resonances produced by regeneration in the late stages of the hadronic phase.

The interpretation of the femtoscopy results suggests that regeneration plays a relevant role until the late stages of the hadron gas evolution.
This challenges the assumption of the simple formula used to estimate the duration of the hadronic phase from the resonance yields, in which regeneration was neglected, and confirms that regeneration provides a plausible explanation for the different values estimated from different resonances.
A further consideration is that the findings from the resonance-yield and femtoscopy measurements may disfavour a scenario of a sudden kinetic freeze-out of all particle species at the same time. A more complex scenario could be figured out in which the kinetic freeze-out spans over a range of time, possibly different for different particle species. This is supported by hydrodynamic calculations coupled with UrQMD to model the hadron cascade~\cite{Zhu:2015dfa} and by considering that the decoupling of particles from an expanding system is a continuous process, which takes place over a finite range of temperatures~\cite{Motornenko:2019jha}.

\subsubsection{Conclusions}

    \paragraph{Equilibration of light-flavour hadrons.} High-precision measurements of the yields of several species of hadrons composed of light (up, down, and strange) quarks were carried out, spanning nine orders of magnitude in production rate from pions to light nuclei and hypernuclei. 
    The abundances are well described by statistical hadronisation models, confirming the thermal nature of particle production in heavy-ion collisions already observed at lower collision energies. 
    The thermodynamic (macroscopic) parameters characterising the chemical freeze-out (\Tchem{} and $\mu_{\rm B}$) follow the trend with \sqrtSnn{} established at SPS and RHIC. In particular,  values of $\Tchem \approx 156$~\mev and $\mu_{\rm B} \approx 0$ are extracted from the SHM analysis of the hadron abundances measured in \PbPb\ collisions at \sqrtSnn{}~=~2.76~\tev{}.
    The high precision of the data revealed small tensions between the measured data points and the SHM predictions for protons, and to a lesser extent for strange baryons. 
    This motivated several phenomenological studies, which pointed out the importance of including the interactions among hadrons (relevant for temperatures close to \TpseudoCritic) in the modelling of the hadron-resonance gas.
    
    \paragraph{Sensitivity of heavy quarks to thermalisation.} Heavy quarks (charm and beauty) are not produced in chemical equilibrium with the quark--gluon plasma. 
    Nevertheless, the yields of charm hadrons can be described by the SHM using as input the production cross section of charm quarks in the initial hard-scattering processes, which determines the charm content of the fireball, and accounting for charm conservation in the fireball evolution.
    This supports a scenario in which charm quarks are to some extent thermalised in the QGP.
    
    \paragraph{Mechanisms of light-flavour hadron production in different $\mathbf{\emph{p}_{\rm T}}$ domains.} In the light-flavour sector, the measurements in a wide momentum interval of the \pT-differential spectra of mesons and baryons, the baryon-to-meson ratios, as well as the anisotropic flow harmonics of different hadron species, allowed for a significant step forward in the understanding of the mechanisms of hadron formation at a microscopic level. 
    At high-\pT{} ($> 8$--$10$~\gmom{}), the measurements in Pb--Pb collisions show that the \vTwo{} coefficient is the same for all particle species and that particle yield ratios are consistent with those measured in pp collisions and in the fragmentation of jets. 
    This confirms that, in this region, hadron formation is dominated by fragmentation. 
    At low-\pT{} ($< 2$~\gmom{}), the measured spectra and flow coefficients are described by hydrodynamic calculations and are thus understood in a picture of a system which evolves in equilibrium. 
    The intermediate-\pT{} region, where equilibrium does not hold, provides a window into the scenario in which the momentum and the angular distributions of hadrons can preserve a memory of their evolution, providing in particular sensitivity to the mechanism of hadron formation at the microscopic level. 
    The features of the data in this \pT{} region are \emph{qualitatively}, and to a good extent also \emph{quantitatively}, captured by models that include a hadronisation via recombination of effective constituent quarks at the phase boundary. The recombination involves soft partons emerging from the collective expansion of the QGP as well as partons which are produced in hard-scattering process and are quenched while traversing the hot and dense plasma.
    
    \paragraph{Charm-hadron production in different $\mathbf{\emph{p}_{\rm T}}$ domains.} Measurements of the \pT{}-differential yields of several charm hadron species in \PbPb\ collisions show hints of an enhancement in the production charm-strange mesons (\rmDs{}) and charm baryons (\rmLambdaC) at low and intermediate momentum relative to \rmDzero{} mesons with respect to pp collisions. This effect is naturally expected for the hadrons produced via quark recombination, and the measured particle yield ratios are described by models implementing heavy-quark hadronisation via both recombination and fragmentation. This is consistent with the conclusions drawn from measurements of \rmJpsi{} production and angular distributions, which demonstrate the important role of the recombination mechanism to generate hidden-charm states during the QGP expansion or at the phase boundary.
    
    \paragraph{Resonance production altered by the hadronic phase.} Measurements of the yield of strongly-decaying resonances as a function of the collision centrality show a suppression of short-lived resonances ($c\tau \sim 1$ to a few fm/$c$), increasing from peripheral to central collisions. This can be interpreted as a result of the rescattering of the resonance decay products in the hadronic phase, which dominates over regeneration. This allowed for a simple quantitative estimation of the duration of the hadronic phase, spanning from the chemical to the kinetic freeze-out, under the assumption that regeneration can be neglected. The hadronic phase duration estimated with this simple approach ranges from $\approx 1$ to $\approx 10$ fm/$c$, increasing from peripheral to central collisions. However, the times calculated from the yields of different resonances differ by up to an order of magnitude, and are larger for longer-lived resonances. These results, along with the ones obtained from kaon--pion femtoscopy, indicate that the regeneration contribution is likely to play a relevant role. Moreover, these results challenge the simple scenario of a sudden kinetic freeze-out of all particle species at the same time and may indicate that the decoupling of particles from the expanding hadron gas is a continuous process, which takes place over a range of times and temperatures that is different for different hadron species.

\footnotesize

\normalsize

\newpage

\newcommand{\CommentBlock}[1]{}

\newcommand{\redtext}[1]{\textcolor{red}{#1}}
\newcommand{\bluetext}[1]{\textcolor{blue}{#1}}
\newcommand{\greentext}[1]{\textcolor{green}{#1}}

\newcommand{\dAu}{d--Au}
\newcommand{\AuAu}{Au--Au}
\newcommand{\PbPb}{Pb--Pb}
\newcommand{\pPb}{p--Pb}
\newcommand{\pbarp}{bar{p}p}
\newcommand{\pp}{pp}
\newcommand{\pA}{p--A}
\newcommand{\aaa}{\ensuremath{A+A}}

\newcommand{\sqrtsNN}{\ensuremath{\sqrt{s_{\rm NN}}}}
\newcommand{\sqrts}{\ensuremath{\sqrt{s}}}

\newcommand{\intlumi}{\ensuremath{L_{\mathrm{int}}}}
\newcommand{\invmub}{\ensuremath{\mu{b}^{-1}}}
\newcommand{\invnb}{{nb}$^{-1}$}
\newcommand{\gev}{\ensuremath{\mathrm{GeV/}c}}

\newcommand{\alphas}{\ensuremath{\alpha_{s}}}

\newcommand{\Nbin}{\ensuremath{N_{bin}}}
\newcommand{\Ncoll}{\ensuremath{N_{coll}}}
\newcommand{\Npart}{\ensuremath{N_{part}}}
\newcommand{\Nbinavg}{\ensuremath{\left<N_{bin}\right>}}

\renewcommand{\Nbinavg}{\ensuremath{\left<N_{bin}\right>}}
\newcommand{\TAA}{\ensuremath{\left<T_{\mathrm{AA}}\right>}}
\newcommand{\TpPb}{\ensuremath{\left<T_{\mathrm{pPb}}\right>}}
\newcommand{\TdAu}{\ensuremath{\left<T_{\mathrm{dAu}}\right>}}
\newcommand{\TpA}{\ensuremath{\left<T_{\mathrm{p}A}\right>}}

\newcommand{\rr}{\ensuremath{R}}
\newcommand{\kT}{\ensuremath{k_\mathrm{T}}}
\newcommand{\antikT}{anti-\ensuremath{k_\mathrm{T}}}

\newcommand{\Ajet}{\ensuremath{A_\mathrm{jet}}}
\newcommand{\rhoA}{\ensuremath{\rho\times\Ajet}}

\newcommand{\pT}{\ensuremath{p_\mathrm{T}}}
\newcommand{\meanpT}{\ensuremath{\left<p_\mathrm{T}\right>}}

\newcommand{\yLAB}{\ensuremath{y_\mathrm{LAB}}}
\newcommand{\yNN}{\ensuremath{y_\mathrm{NN}}}
\newcommand{\ystar}{\ensuremath{y^*}}

\newcommand{\pTjet}{\ensuremath{p_{\mathrm{T,jet}}}}
\newcommand{\pTjetch}{\ensuremath{p_\mathrm{T,jet}^\mathrm{ch}}}
\newcommand{\pTjetchemb}{\ensuremath{p_\mathrm{T,jet}^\mathrm{ch,emb}}}
\newcommand{\pTtrig}{\ensuremath{p_{\mathrm{T,trig}}}}
\newcommand{\pTassoc}{\ensuremath{p_{\mathrm{T,assoc}}}}
\newcommand{\pTtrk}{\ensuremath{p_{\mathrm{T,trk}}}}
\newcommand{\pTemb}{\ensuremath{p_{\mathrm{T,emb}}}}

\newcommand{\pTjetpart}{\ensuremath{p_{\mathrm{T,jet}^\mathrm{part}}}}
\newcommand{\pTjetdet}{\ensuremath{p_{\mathrm{T,jet}^\mathrm{det}}}}

\newcommand{\pThadron}{\ensuremath{p_{\mathrm{T,hadron}}}}
\newcommand{\ET}{\ensuremath{E_{\mathrm T}}}

\newcommand{\dNdpT}{\ensuremath{\frac{\rm{d}N}{\mathrm{d} \pT}}}

\newcommand{\pizero}{\ensuremath{\pi^0}}
\newcommand{\kzerol}{\ensuremath{\mathrm{K}^{0}_\mathrm{L}}}
\newcommand{\kzeros}{\ensuremath{\mathrm{K}^{0}_\mathrm{S}}}
\newcommand{\Dzero}{\ensuremath{\mathrm{D}^{0}}}

\newcommand{\JPsi}{\ensuremath{\mathrm{J}/\psi}}

\newcommand{\zleading}{\ensuremath{z_\mathrm{leading}}}

\newcommand{\DCAxy}{\ensuremath{DCA_{\mathrm xy}}}
\newcommand{\sDCAxy}{\ensuremath{sDCA_{\mathrm xy}}}

\newcommand{\qhat}{\ensuremath{\hat{q}}}
\newcommand{\vtwo}{\ensuremath{v_2}}

\newcommand{\sigdab}{\ensuremath{\sigma_{\delta,ab}}}

\newcommand{\RAA}{\ensuremath{R_\mathrm{AA}}}
\newcommand{\RCP}{\ensuremath{R_\mathrm{CP}}}
\newcommand{\ICP}{\ensuremath{I_\mathrm{CP}}}
\newcommand{\IAA}{\ensuremath{I_{\rm{AA}}}}

\newcommand{\RpPb}{\ensuremath{R_\mathrm{pPb}}}
\newcommand{\QpPb}{\ensuremath{Q_\mathrm{pPb}}}

\newcommand{\Drecoil}{\ensuremath{\Delta_\mathrm{recoil}}}
\newcommand{\dphi}{\ensuremath{\Delta\varphi}}

\newcommand{\tg}{\ensuremath{\theta_{\mathrm{g}}}}
\newcommand{\rg}{\ensuremath{R_{\mathrm{g}}}}
\newcommand{\zg}{\ensuremath{z_{\mathrm{g}}}}
\newcommand{\pTsub}{\ensuremath{p_{\mathrm{T,sub-leading}}}}
\newcommand{\pTlead}{\ensuremath{p_{\mathrm{T,leading}}}}

\newcommand{\vzero}{\ensuremath{\mathrm{V0}}}
\newcommand{\vzeroA}{\ensuremath{\mathrm{V0A}}}
\newcommand{\vzeroC}{\ensuremath{\mathrm{V0C}}}
\newcommand{\vzeroM}{\ensuremath{\mathrm{V0M}}}
\newcommand{\vzeroAscaled}{\ensuremath{\mathrm{V0A}/\left<\mathrm{V0A}\right>}}
\newcommand{\vzeroCscaled}{\ensuremath{\mathrm{V0C}/\left<\mathrm{V0C}\right>}}
\newcommand{\vzeroMscaled}{\ensuremath{\mathrm{V0M}/\left<\mathrm{V0M}\right>}}

\subsection{Partonic interactions in matter}
\label{sec:PartonInteractions}

This section discusses the application of processes generated in high momentum-transfer (high-$Q^2$) interactions as hard probes of the QGP. Hard probes are created far out of equilibrium with the QGP, and their in-medium interactions can be used to measure effects due to dynamical processes such as energy transport and equilibration. 
Because of their high-momentum they are short-wavelength phenomena, and their in-medium interactions are therefore sensitive to the microscopic structure of the QGP and its quasi-particle nature. It is important to note, however, that high-$Q^2$ refers only to the process of generation of such probes but not to their in-medium interactions, which in general span a broad range in $Q^2$. 

The focus of this section is on the production and propagation in the QGP of high-momentum jets, where the high-$Q^2$ scale requires high-\pT, and heavy quarks, i.e. charm and beauty quarks, where the high-$Q^2$ scale is imposed by the quark masses. The essential characteristics of high-$Q^2$ processes as probes of the QGP are as follows:

\begin{enumerate}

\item \label{enum:pQCD} Their production cross sections in \pp\ collisions are calculable with controlled and improvable accuracy using the tools of pQCD;

\item  \label{enum:early} Based on considerations of the uncertainty principle and QCD factorisation, high-$Q^2$ processes are expected to occur at the earliest stage of a heavy-ion collision, prior to equilibration of the QGP, and therefore probe its hottest and densest phase; 
 
\item  \label{enum:model} The interaction between a hard probe and the QGP can be calculated theoretically starting from the pQCD formulation for elementary collisions or from transport theory, giving a firm conceptual basis to such modeling approaches.

\end{enumerate}

\paragraph{\textbf{\textit {Open heavy flavour in the QGP.}}}
\label{sec:QGPHF}

Because the masses of charm and beauty quarks are much larger than both $\Lambda_{\rm QCD}$ and the medium temperature, their production even at low \pT\ is governed predominantly by hard scattering processes early in the evolution of the system and additional thermal production is negligible. Heavy-flavour measurements probe the medium over a wide range of wavelengths, depending upon \pT. Theoretically, at low \pT\ the interaction is governed largely by  elastic (collisional) processes, whereas inelastic (radiative) processes predominate at high-\pT\ (see~\cite{Prino:2016cni,Dong:2019byy} for recent topical reviews).

At long wavelengths, i.e. quark momenta less than a few GeV/$c$, heavy quarks exchange energy and momentum via multiple soft and incoherent collisions within the hydrodynamically expanding medium, picking up collective flow and approaching thermalisation. The relaxation time $\tau_{\rm Q}$ of heavy quarks is expected to be close to the lifetime of the QGP, with $\tau_{\rm Q}$ being significantly longer for beauty than for charm quarks due to their larger mass. Heavy-flavour observables thereby retain a memory of quark-medium interactions in an early phase of the QGP. The long-wavelength physics of the interaction of heavy quarks with the medium is treated theoretically in a diffusion approach based on Fokker-Planck or Langevin dynamics. The typical momentum exchange is small in individual scattering processes and, therefore, the dynamical behaviour can be described in terms of Brownian motion in the QGP medium~\cite{Svetitsky:1987gq,Rapp:2009my}. In the intermediate-wavelength region, corresponding to heavy-flavour hadron transverse momenta up to about 10~GeV/$c$, measurements probe the mechanism of hadronisation from a deconfined medium (see Sec.~\ref{sec:QGPHadronization}). At short wavelengths, i.e. heavy-flavour hadron $\pT>$ 10~GeV/$c$, heavy-quark interactions with the QGP probe the physics of jet quenching, notably the quark-mass dependence of medium induced energy loss, which is expected to be influenced by the dead-cone effect~\cite{Dokshitzer:2001zm,Armesto:2003jh}.

The theoretical description of heavy-flavour propagation through the medium is commonly realised in transport models, which start from the production of heavy quarks in hard scattering processes, followed by transport in a hydrodynamically expanding medium, hadronisation, and interactions during the hadronic phase. 

The Brownian motion of heavy particles in a medium of light constituents is characterised by a diffusion process. The coupling between the medium and the heavy particle in this regime can be expressed by the dependence of the average squared displacement of the particle, $\langle \vec{r}^{\ 2} \rangle$, on the time $t$

\begin{equation}
    \langle \vec{r}^{\ 2} \rangle = 6\,D_{\rm s}\,t
\end{equation}

\noindent
where the spatial diffusion coefficient $D_{\rm s}$ encodes the transport properties of the medium.

The description of heavy-quark transport in the QGP starts from the Boltzmann equation, which describes the space--time evolution of the heavy-quark phase space distribution in terms of external forces and a collision integral, encoding the interaction of the heavy quark with the medium partons. For moderate medium temperatures, the momentum transfer between the heat bath and a heavy quark is small in individual interactions, motivating the picture of soft and incoherent scatterings. In that case, the Boltzmann equation can be approximated by the Fokker-Planck equation~\cite{Prino:2016cni}. This formulation incorporates momentum-dependent transport coefficients $A(\vec{p})$ and $B(\vec{p})$ characterising momentum friction (drag) and diffusion, which are denoted $\gamma$ and $D_{\rm p}$ in the non-relativistic, momentum-independent limit. In this limit one also obtains the dissipation-fluctuation theorem, often called Einstein relation, 

\begin{equation}
    D_{\rm p} = m_{\rm Q}\,\gamma\,T\,,
\end{equation}

\noindent
which highlights the role of momentum frictions and diffusion by connecting the temperature $T$ of the heat bath with the momentum distribution of heavy quarks with mass $m_{\rm Q}$.

Alternatively, heavy-quark transport in the QGP is implemented as a Langevin process with drag and diffusion terms. To ensure that the heavy-quark momentum distribution asymptotically approaches the equilibrium limit, the Einstein relation is usually enforced in such calculations, i.e. only the drag coefficient is treated as an independent parameter, with the momentum diffusion coefficient obtained via the equation above. In turn, the drag and spatial diffusion coefficients are related via

\begin{equation}
    D_{\rm s} \propto\frac{T}{m_{\rm Q}\;\gamma}.
\end{equation}

\noindent
Since $\gamma$ is approximately proportional to $1/m_{\rm Q}$, $D_{\rm s}$ is almost independent of the quark mass and is therefore a parameter that characterises the transport properties of the medium. Furthermore, $D_{\rm s}$ is proportional to the relaxation time $\tau_{\rm Q}$ of heavy quarks in the medium, i.e. $\tau_{\rm Q} = (m_{\rm Q}/T) D_{\rm s}$. If $D_{\rm s}$ is sufficiently small, the relaxation time is smaller than the expansion rate of the medium, meaning that a heavy quark is likely to remain in the same fluid cell throughout the expansion of the medium. Consequently, heavy quarks pick up the large collective flow of the medium. If, on the contrary, $D_{\rm s}$ and the relaxation time are large, both flow and yield suppression of heavy-flavour probes are small. Therefore, this QGP transport property can be determined by comparing measurements of the heavy-flavour yield suppression and flow coefficients with corresponding results from model calculations. The product of $D_{\rm s}$ and the thermal wavelength of the medium, $\lambda_{\rm th} = 1/(2 \pi T)$, is a dimensionless quantity that is proportional to the specific shear viscosity $\eta/s$; such investigations thereby also constrain $\eta/s$~\cite{Moore:2004tg, Rapp:2009my}. 

ALICE measures open heavy-flavour production in a variety of channels. Heavy-flavour hadrons are fully reconstructed from their hadronic decays~\cite{Abelev:2012vra,ALICE:2012ab,Abelev:2014hha,Abelev:2012tca,Acharya:2017kfy,Acharya:2017lwf,Acharya:2017qps,Acharya:2018hre,Acharya:2018ckj,ALICE:2021mgk}. In addition, heavy-flavour production is measured using semi-leptonic decays into electrons~\cite{Abelev:2012xe,Adam:2015qda,ALICE:2020hdw,Acharya:2019mom} and muons~\cite{Abelev:2012qh,Acharya:2017hdv,ALICE:2020sjb}, including partial reconstruction of semi-electronic weak decays of heavy-flavour baryons~\cite{Acharya:2017kfy,ALICE:2021psx,ALICE:2021bli}. 
In addition, beauty production is measured via non-prompt J/$\psi$ mesons from the decay mode B $\rightarrow$ J/$\psi + X$~\cite{Abelev:2012gx,Adam:2015rba,Acharya:2018yud} and non-prompt D mesons from the decay mode B $\rightarrow$ D $+ X$~\cite{ALICE:2021mgk,ALICE:2022tji}. For many of these measurements the excellent spatial resolution of the ITS~\cite{Abelev:2014ffa} is instrumental since it allows to resolve the decay vertex of charm and beauty hadrons from the primary collision vertex.

\paragraph{\textbf{\textit {Jets in the QGP.}}}
\label{sec:QGPJets}

A scattered parton with \pT\ greater than a few \gev\ propagates and evolves independently of other products of the same high-$Q^2$ interaction, starting at time earlier than 1\,fm/$c$. The QGP has an extended size, with a lifetime in the order of 10\,fm/$c$ (see Sec.~\ref{sec:TG1sizelifetime}), so that the jet shower propagates through the expanding and cooling QGP. During this process the jet shower itself evolves and its colour-charged constituents interact with the colour-charged constituents of the QGP, resulting in modification of the shower. %
Such modifications, called ``jet quenching,'' are observable experimentally and calculable theoretically; comparisons of jet quenching data and calculations provide unique, penetrating probes of QGP structure and dynamics. See Refs.~\cite{Majumder:2010qh,Cunqueiro:2021wls,Harris:2023tti} for recent reviews of jet quenching.

Experimentally, jet quenching is manifest in several ways:

\begin{enumerate}

\item Medium-induced energy transport to large angles to the hard parton or jet direction, commonly called ``energy loss'', observed through inclusive yield suppression;

\item Medium-induced modification of the distribution of jet constituents, observed through the radial energy profile, jet substructure, and fragmentation functions;

\item Jet centroid deflection due to soft multiple scattering or scattering from quasi-particles in the QGP, observed as medium-induced acoplanarity in coincidence measurements;

\item Response of the QGP medium to deposited energy, observed as energy and momentum flow over a large range in phase space relative to the jet axis.
\end{enumerate}

\noindent
This many-pronged approach to quantifying jet quenching is a valuable opportunity: it must provide a consistent picture of jet quenching, thereby constraining significantly our understanding of its underlying processes. 

Theoretically, interactions of high-energy partons with the QGP are broadly categorised as elastic interactions between the propagating parton and the QGP constituents ($2 \rightarrow 2$ scatterings), and inelastic scattering, or medium-induced gluon radiation events ($2 \rightarrow 3$ processes). These medium-induced processes are interleaved with the spontaneous splittings of the parton shower process that also occur in \pp\ and other elementary collisions without a QCD medium (vacuum emissions).

At high-\pT{}, radiative processes are the dominant energy loss mechanism. This has been explored for instance in the BDMPS-Z framework~\cite{Baier:1996kr,Zakharov:1997uu}, which shows that interference between scattering and emission processes generate a characteristic dependence of the energy loss on the square of the in-medium path length. Various calculational approaches have been developed: multiple soft scattering approximation (BDMPS)~\cite{Baier:1996kr,Salgado:2003gb}; opacity expansion (GLV)~\cite{Gyulassy:2000fs}; thermal-field theory (AMY)~\cite{Arnold:2002ja}; higher twist (HT)~\cite{Wang:2001ifa,Qin:2009gw}; and Soft-Collinear Effective Theory with Glauber gluons (SCETg)~\cite{Idilbi:2008vm,Ovanesyan:2011kn,Kang:2014xsa,Kang:2016ehg}. 
While these different formalisms give numerically different results, their relationship is largely understood. In some cases explicit correspondence between the calculations has been established~\cite{Armesto:2011ht,Majumder:2010qh}, for example by evaluating the all-order opacity expansion~\cite{CaronHuot:2010bp,Feal:2019xfl}.

Each such formulation has its own natural parameter characterising the strength of the jet-medium interaction, the Debye screening mass, or the effective value of the strong coupling. Different formalisms can be compared via the transport coefficient \qhat, which is the mean squared transverse momentum transfer $\mathbf{q}_\perp$ per unit path length. 
In models, \qhat\ is calculated as the ratio of $\langle\mathbf{q}_\perp^2\rangle$ and mean free path $\lambda$, or by averaging the scattering rate $\mathrm{d}\Gamma/\mathrm{d}\mathbf{q}_{\perp}\mathrm{d}z = \rho \,\mathrm{d}\sigma/\mathrm{d}\mathbf{q}_{\perp}$,
\begin{equation}
\hat{q} = \frac{\langle q_\perp^2 \rangle}{\lambda} = \int \mathrm{d}^2\mathbf{q}_\perp\, q^2_\perp \frac{\mathrm{d}\Gamma}{\mathrm{d}\mathbf{q}_{\perp}\mathrm{d}z}.
\label{eq:qhat_def}
\end{equation}

Theoretical calculations indicate that the following characteristics are expected for medium-induced radiative energy loss:

\begin{itemize}

\item {\bf Path length dependence:} in the BDMPS formalism, for short path lengths the medium-induced energy loss varies linearly with \qhat\ and quadratically with path length, which arises from the LPM effect~\cite{Baier:1996kr}, $\langle \Delta E \rangle \propto \alpha_{\rm S} \, \hat{q} \, L^2$. Similar dependencies are found for other formalisms~\cite{Armesto:2011ht,CaronHuot:2010bp}.

\item {\bf Colour-charge dependence:} the energy loss is proportional to the colour factor of the propagating parton. Energy loss is expected to be 9/4 times larger for gluons than for quarks.
    
\item {\bf Quark mass dependence:} the phase space for gluon emission is limited to the region where the gluon does not outrun the charge source, implying a minimum emission angle for gluons (dead-cone effect, see Sec.~\ref{sec:pp-deadcone}). Consequently, charm and bottom quarks are expected to lose less energy than light quarks at moderate momentum, i.e. at a momentum less than a few times the quark mass.

\end{itemize}

While analytical calculations are required to connect measurements to the fundamental underlying theory, such calculations do not provide a straightforward way to implement kinematic limits and they have to be worked out specifically for each observable. 
Monte Carlo event generators such as JEWEL~\cite{Zapp:2008gi}, JETSCAPE~\cite{JETSCAPE:2020mzn} and the Hybrid Model~\cite{Casalderrey-Solana:2016jvj,HybridModelResolution} provide a complementary set of tools, generating full events to which any analysis algorithm can be applied.

Further detailed discussion of theoretical approaches to jet quenching and their implementation in numerical modeling can be found in Ref.~\cite{Cunqueiro:2021wls} and references therein. Extraction of the in-medium jet transport coefficient \qhat\ from the comparison of models and data can be found in Refs.~\cite{Burke:2013yra,Xu:2014ica,Armesto:2009zi,JETSCAPE:2021ehl}.

ALICE has reported jet measurements using both charged-particle tracking and cluster energy from the highly granular Electromagnetic Calorimeter (EMCal)~\cite{Abelev:2013fn,Adam:2015ewa,Acharya:2019jyg}, as well as jet measurements of charged-particle jets~\cite{Abelev:2013kqa,Adam:2015hoa,Adam:2016jfp,Acharya:2019tku,Adam:2015doa,Acharya:2017goa,Acharya:2017okq,Acharya:2018uvf,Acharya:2019djg}. For coincidence measurements between a trigger hadron and recoil jets, ALICE employs a semi-inclusive approach. In these measurements the large uncorrelated jet background at low-\pTjet\ and large jet radius \rr\ is suppressed by using a new observable, \Drecoil, which is the difference of two such trigger-normalised distributions in exclusive intervals in hadron trigger $p_\mathrm{T,trig}$ ($\mathrm{TT_{sig}}$ and $\mathrm{TT_{ref}}$)~\cite{Adam:2015doa}:

\begin{align}
\Drecoil = 
\frac{1}{N^\mathrm{AA}_\mathrm{trig}} & \frac{\mathrm{d^3}N^\mathrm{AA}_\mathrm{{jet}}}{\mathrm{d}p^{\mathrm{ch}}_\mathrm{T,jet} \mathrm{d}\Delta\varphi \mathrm{d}\Delta\eta} \bigg|_{p_\mathrm{T,trig} \in \mathrm{TT_{Sig}}}  \nonumber \\
& - c_\mathrm{ref} \, \frac{1}{N^\mathrm{AA}_\mathrm{trig}} \frac{\mathrm{d^3}N^\mathrm{AA}_\mathrm{{jet}}}{\mathrm{d}p^{\mathrm{ch}}_\mathrm{T,jet} \mathrm{d}\Delta\varphi \mathrm{d}\Delta\eta} \bigg|_{p_\mathrm{T,trig} \in \mathrm{TT_{Ref}}},
\label{eq:DeltaRecoil}
\end{align}

\noindent
where $N^\mathrm{AA}_\mathrm{trig}$ is the number of trigger hadrons, $p^{\mathrm{ch}}_\mathrm{T,jet}$ is the recoil jet \pT,  $\Delta\varphi$ is the azimuthal separation between trigger hadron and recoil jet axis, and $c_\mathrm{ref}$ is a constant determined from data to account for finite area effects~\cite{Adam:2015doa}. This data-driven suppression of uncorrelated jet yield enables measurements in central \PbPb\ collisions of recoil jets with large $R$ and low-\pTjet, without fragmentation bias.

\subsubsection{Heavy-quark transport and diffusion measurements}
\label{sec:Transport}

In this section ALICE measurements of heavy-quark diffusion in the QGP and their comparison to theoretical calculations are presented. Theoretical concepts and experimental techniques are sketched in Sec.~\ref{sec:QGPHF}.

\begin{figure}[tb]
\centering
\includegraphics[width=1.\textwidth]{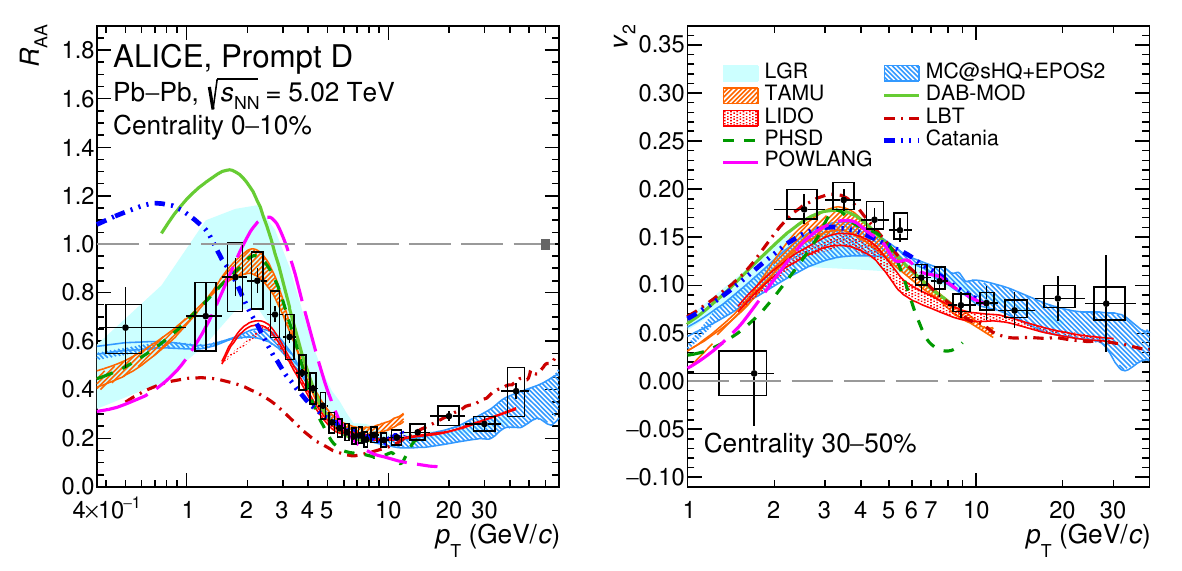}
\caption{D-meson production in \PbPb\ collisions at \sqrtsNN~=~5.02~TeV~\cite{ALICE:2021rxa,ALICE:2020iug}. (Left) \RAA\ in 0--10\% collisions; (right) $v_2$ in 30--50\% collisions.  Model calculations implement charm-quark transport in a hydrodynamically expanding QGP (see text for references).}
\label{Fig:HFTransport}
\end{figure}

Figure~\ref{Fig:HFTransport} shows the nuclear modification factor \RAA\ 
and the elliptic-flow parameter \vtwo\ 
as a function of \pT\ for prompt D mesons in central (0--10\%) and semi-central (30--50\%) \PbPb\ collisions at \sqrtsNN~=~5.02 TeV~\cite{ALICE:2021rxa,ALICE:2020iug}. These measurements are compatible with previous results by the ALICE and CMS Collaborations at the same \sqrtsNN~\cite{Acharya:2018hre,Acharya:2017qps, Acharya:2018bxo,Sirunyan:2017plt,Sirunyan:2017xss}.

At low- and intermediate-\pT, the \pT-differential yields and azimuthal distributions of hadrons carrying heavy quarks are expected to be governed by collisional interactions within the quark--gluon plasma. In order to test this expectation the measurements are compared in Fig.~\ref{Fig:HFTransport} with various model calculations based on charm-quark transport in a hydrodynamically expanding QGP (TAMU~\cite{He:2019vgs}, LIDO~\cite{Ke:2018jem},
POWLANG~\cite{Beraudo:2014boa,Beraudo:2018tpr,Beraudo:2021ont},
PHSD~\cite{Song:2015sfa}, MC@sHQ~\cite{Nahrgang:2013xaa}, Catania~\cite{Scardina:2017ipo,Plumari:2019hzp}, LBT~\cite{Cao:2016gvr,Cao:2017hhk}, LGR~\cite{Li:2019lex}, and DAB-MOD~\cite{Katz:2019fkc}). All models are qualitatively in agreement with the data, although tension is observed in particular at low-\pT. 

\begin{figure}[tb]
\centering
\includegraphics[width=0.99\textwidth]{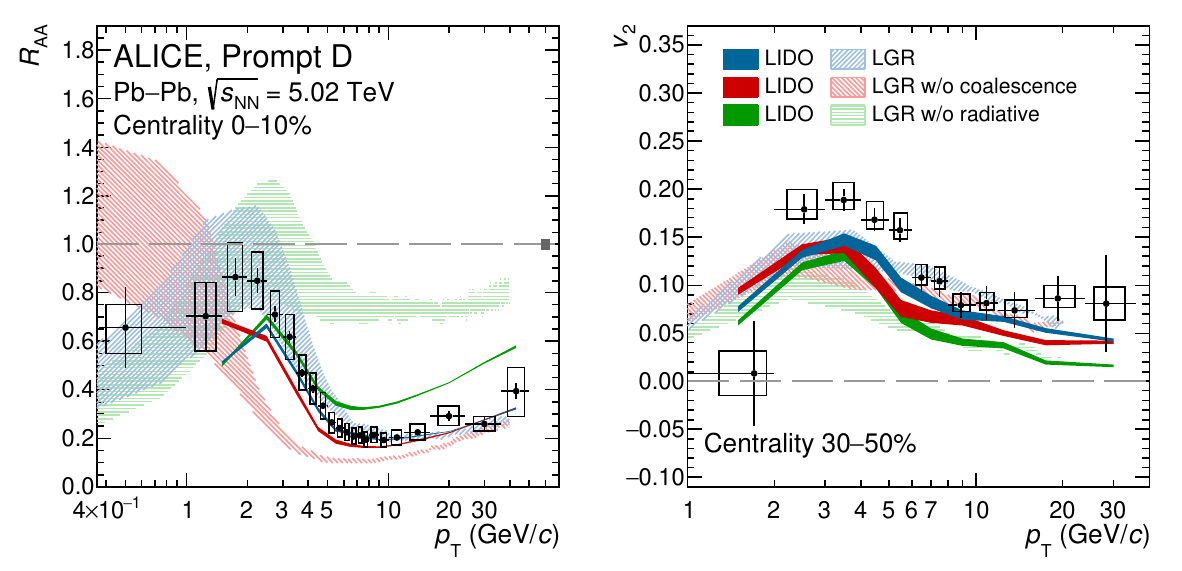}
\caption{Data from Fig.~\ref{Fig:HFTransport}
compared with calculations from the LIDO~\cite{Ke:2018jem} and LGR~\cite{Li:2019lex} models using several variations of charm-quark interactions and hadronisation; see text for details.}
\label{Fig:HFModelIngredients}
\end{figure}

The models differ in their implementation of the interaction between charm quarks with the medium (Boltzmann vs. Langevin transport, inclusion of collisional and/or radiative interactions), nuclear PDFs and shadowing, bulk evolution of the medium (ideal or viscous hydrodynamics, Boltzmann quasiparticles or parton transport, off-shell parton transport), charm hadronisation (fragmentation and/or recombination or in-medium string formation), and hadronic scattering. In particular, in the TAMU, POWLANG, PHSD, and Catania calculations the interactions between the charm quarks and the medium constituents are modelled with collisional processes only, while the MC@sHQ, LBT, LIDO, DAB-MOD, and LGR models consider radiative processes as well. All the models include the hadronisation of the charm quark via coalescence, in addition to the fragmentation mechanism. With the exception of DAB-MOD, all the models employ nuclear PDFs in the calculation of the initial $p_{\rm T}$ distributions of charm quarks in order to include initial-state effects. Initial-state event-by-event fluctuations are included in the POWLANG, LIDO, PHSD, MC@sHQ, LBT, and DAB-MOD models.
In order to disentangle the effects of elastic and radiative interactions of the heavy quarks and of hadronisation, calculations of the LIDO and LGR models are compared for configurations including all physics processes with those excluding radiative processes or hadronisation via coalescence. Figure~\ref{Fig:HFModelIngredients} shows the comparison of these model variants with the data from Fig.~\ref{Fig:HFTransport}.

The predictions for prompt D-meson \RAA\ and $v_2$ in the range $\pT < 4$--$5~\gev$ are similar for LGR and LIDO, with and without radiative processes. They describe reasonably well the \RAA\ and slightly underestimate the $v_2$. This indicates that elastic collisions are the predominant interaction process in that \pT{} range. Similar conclusions are drawn from models that only contain collisional processes (TAMU, PHSD, Catania, POWLANG).

In this \pT\ range, which is dominated by elastic collisions, charm quarks may acquire radial and anisotropic azimuthal flow, due to multiple interactions within the QGP medium. Momentum exchanges between charm quarks and the heat bath in such interactions typically have a magnitude similar to the medium temperature, $q^2 \approx T^2$. This value is smaller than the thermal momentum, due to the large charm-quark mass ($m_\mathrm{charm}$). Heavy quarks therefore experience Brownian motion in the QGP, with a thermal relaxation time, $\tau_\mathrm{charm} = \tau_\mathrm{th} \, m_\mathrm{charm}/T$, that is significantly longer than the thermalisation time of the bulk medium ($\tau_\mathrm{th}$). In other words, heavy-flavour hadrons probe the equilibration process. At high $p_{\rm T}$, the LGR model underestimates the energy loss if radiative processes are not included. The role of radiative processes in describing heavy quark production and flow at higher \pT\ is discussed in Sec.~\ref{sec:ElossHadrons}. 

Hadronisation effects must also be taken into account for modelling D-meson \RAA\ and $\vtwo$. Jet fragmentation is expected to dominate D-meson production at very high- \pT, while at intermediate-\pT\ ($\sim$few \gev), recombination effects are expected to play a significant role (see Sec.~\ref{sec:QGPHadronization}).
In the TAMU model, hadronisation at intermediate-\pT\ is implemented as a process in which heavy quarks interact with light quarks in the QGP, leading to formation and dissociation of broad bound hadronic states whose rates are governed by the Boltzmann equation~\cite{He:2019vgs}. 
In the POWLANG model, a heavy quark propagates stochastically in the fireball, which simultaneously expands and cools. Once the heavy quark is in a fluid cell whose temperature is below the decoupling temperature $T_\mathrm{dec}$ of the medium, it is forced to hadronise with a light quark extracted from the thermal momentum distribution corresponding to the temperature $T_\mathrm{dec}$~\cite{Beraudo:2014boa}.

In the recombination process, the transverse momentum of the generated D meson is typically larger than that of the initiating charm quark. Hadronisation via recombination therefore corresponds to a hardening of the D-meson \pT\ distribution relative to jet fragmentation in vacuum, generating a characteristic peak in \RAA\ at $\pT\approx 2~\gev$. This peak results from the interplay between coalescence, whose probability relative to fragmentation decreases with increasing \pT , the shape of the \pT-differential spectra of light and heavy quarks at hadronisation, and the radial flow. A similar effect occurs for azimuthally asymmetric flow of heavy-flavour hadrons formed via recombination: the hadron inherits the flow of both the heavy and the light quarks. Further discussion of coalescence effects in charm-quark hadronisation via comparison of the abundances of different heavy-flavour hadron species in pp and heavy--ion collisions is reported in Sec.~\ref{sec:QGPHadronization}.

As discussed above, the heavy-quark spatial diffusion coefficient $D_{\rm s}$ does not depend significantly on the heavy-quark mass and is  therefore a property of the QGP medium. The coefficient $D_{\rm s}$, which is calculable within the framework of lattice QCD~\cite{Banerjee:2011ra,Ding:2012sp,Kaczmarek:2014jga}, can be determined by comparison of model calculations with data, via a comprehensive treatment of charm-quark transport in the QGP and hadronisation (see~\cite{Rapp:2018qla} for a recent review). 
 
 Experimental data from the LHC and theoretical efforts have enabled significant progress in the determination of $D_{\rm s}$. Recent studies utilised ALICE measurements of elliptic and triangular flow of D mesons~\cite{ALICE:2020iug} as well as \RAA\ and $v_2$ measurements of prompt D mesons~\cite{ALICE:2021rxa}. 
 Models that agree with the data at the level $\chi^2/{\rm ndf} < 2$ yield a value $1.5 < 2 \pi D_{\rm s}(T) T < 4.5$. The corresponding charm quark relaxation time $\tau_{\rm charm} = (m_{\rm charm}/T) D_{\rm s}(T)$ is in the range $3 < \tau_{\rm charm} < 9\; {\rm fm}/c$ with $T = T_\mathrm{c}$ and $m_{\rm charm}$ = 1.5~GeV/$c^2$. A similar range for $D_{\rm s}$ was obtained for \AuAu\ collisions at $\sqrtsNN = 200~\mathrm{GeV}$, based on measurements by the STAR Collaboration of \vtwo\ and \RAA\ of \Dzero\ mesons~\cite{Adamczyk:2017xur}. These values are similar in magnitude to the estimated lifetime of the QGP at the LHC of about 10~fm/$c$~\cite{Aamodt:2011mr}, indicating that the charm quark may thermalise completely in the medium. The extracted value for $D_{\rm s}$ is likewise consistent with calculations based on lattice QCD, which yield values in the range $2 < 2 \pi D_{\rm s}(T) T < 6$~\cite{Banerjee:2011ra,Ding:2012sp,Kaczmarek:2014jga}, and up to an order of magnitude smaller than values predicted by pQCD calculations at leading order~\cite{Svetitsky:1987gq,Moore:2004tg,vanHees:2004gq}. This provides significant evidence in the heavy-quark sector in favor of the formation of a strongly coupled QGP being produced in heavy-ion collisions at the LHC (see Sec.~\ref{sec:QGPsummary} for more details).

Future high-precision measurements of heavy-flavour hadron production with the upgraded ALICE detector in Run 3 down to very low transverse momenta are expected to set more stringent constraints on the production of charm quarks and their interactions in the QGP~\cite{Citron:2018lsq}. 

\subsubsection{Jet quenching measurements}
\label{sec:JetsMatter}

In this section ALICE measurements of jet quenching are presented, whose physics motivation and theoretical tools are discussed above.%

\paragraph{\textbf{\textit {2.4.2.1 Energy loss}}}
\label{sec:Eloss}

\paragraph{\textbf{\textit{High-p$_{\rm\bf T}$ inclusive hadron production and radiative energy loss.}}}
\label{sec:ElossHadrons}

\begin{figure}[htb!]
\centering
\includegraphics[width=0.75\textwidth]{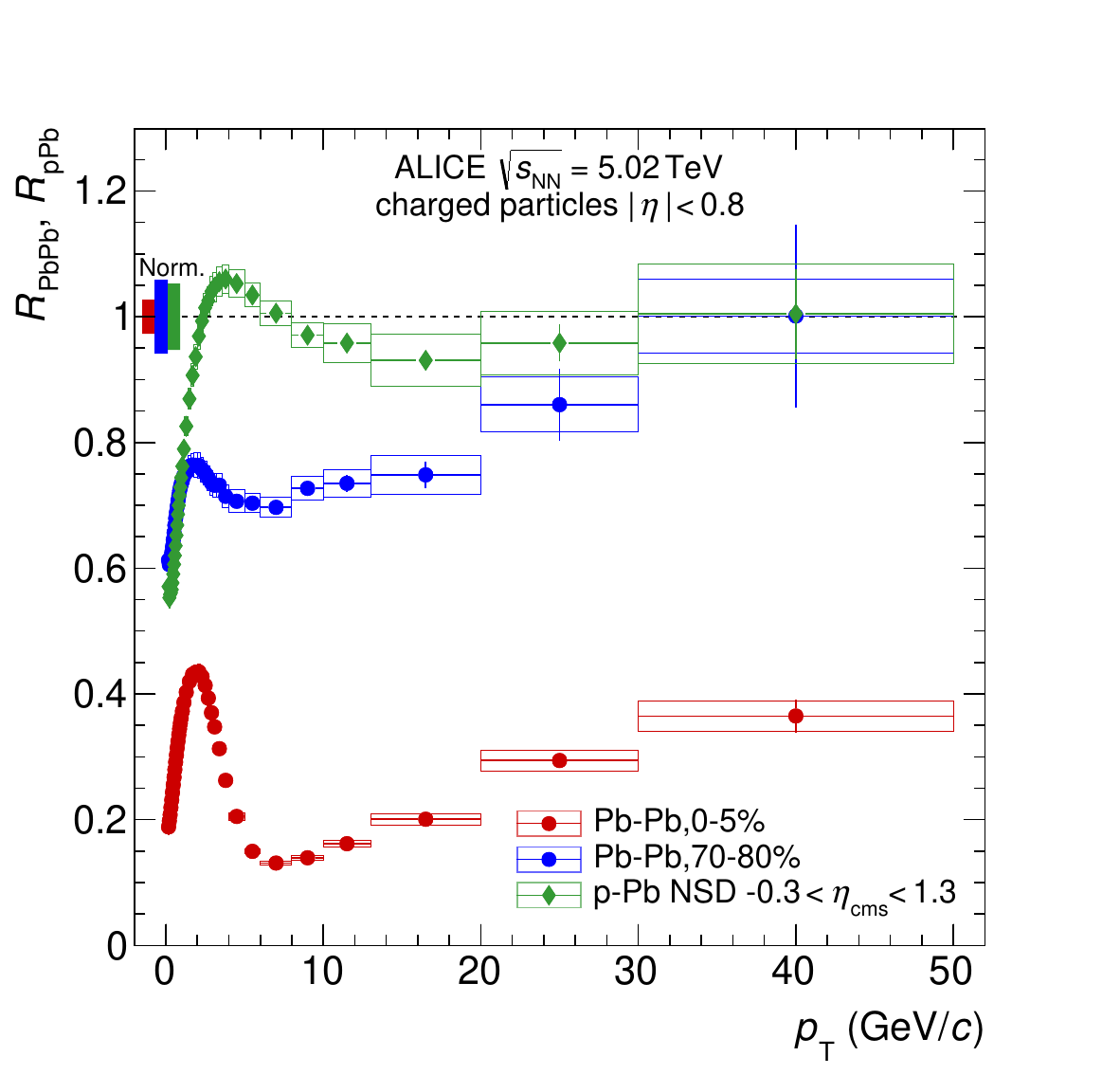}
\includegraphics[width=0.7\textwidth]{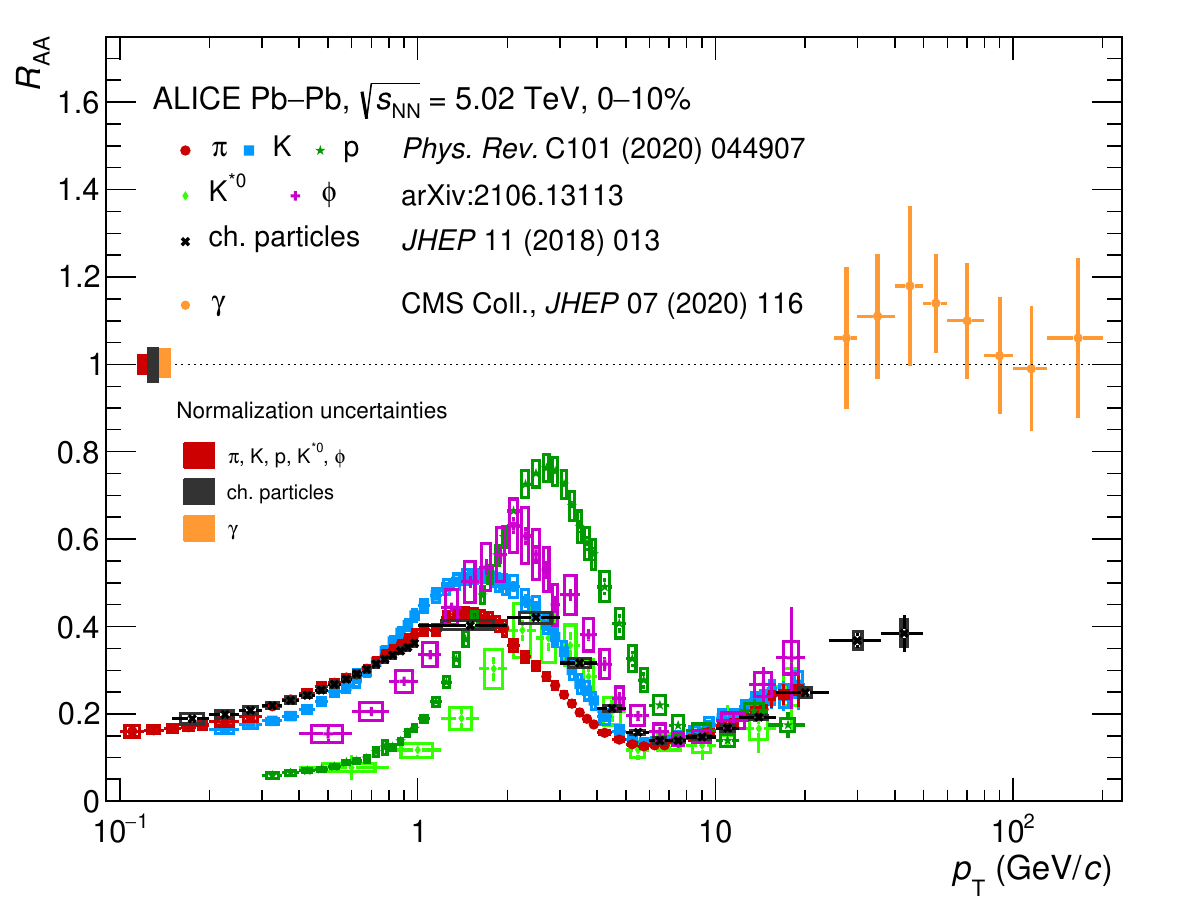}
\caption{Nuclear modification factor \RAA\ for charged hadrons~\cite{Acharya:2018qsh} in central and peripheral \PbPb{} collisions and NSD \pPb{} collisions (top), and for various particle species in central \PbPb{} collisions~\cite{Acharya:2019yoi,ALICE:2017ban,ALICE:2021ptz} (bottom). Isolated photons measured by CMS are also included for comparison~\cite{Sirunyan:2020ycu}.}
\label{Fig:RAAhadrons}
\end{figure}

Radiative energy loss is expected to be the dominant mechanism responsible for the suppression of inclusive particle production at high-\pT.
Figure~\ref{Fig:RAAhadrons}, top panel, shows the nuclear modification factor \RAA\ as a function of transverse momentum for charged particles at $\sqrtsNN~=~5.02$ TeV for central and peripheral \PbPb\ collisions, and \pPb\ collisions. In \pPb{} collisions, particle production approximately follows $N_{\rm coll}$ scaling, with \RpPb{} within 10\% of unity in the region dominated by hard processes, $\pT>2~$~\gev. 
In  central \PbPb{} collisions a peak is seen at intermediate transverse momentum, with yield suppression of a factor five observed at high-\pT. The peak at 2--3 \gev\ for charged hadrons is partly due to collective radial flow and enhancement in the baryon-to-meson yield ratio, as discussed in the next paragraph. The value of \RAA\ has a minimum value of 0.18 at $\pT \approx 6$~\gev, and then increases for $\pT>10$~\gev. The increase suggests that energy loss as observed by measuring high-\pT{} hadrons is not proportional to the parent parton energy, but closer to constant~\cite{Spousta:2015fca}. 
In peripheral \PbPb{} collisions (the class 70--80\% is reported in the figure) \RAA\ is about 30\% below unity at $\pT \approx 6$~\gev\ and increases to unity at high- \pT. The interpretation of this measurement will be discussed at the end of the section.

Figure~\ref{Fig:RAAhadrons}, bottom panel, shows \RAA\ for identified light-flavour mesons and protons (ALICE~\cite{ALICE:2017ban}), and isolated photons (CMS~\cite{Sirunyan:2020ycu}). The peak in the yields in the range $2<\pT<6$~\gev\ is most pronounced for protons, with maximum value ordered by mass. This ordering corresponds to the expected effect of radial flow (see Sec.~\ref{sec:QGPevolution}), but inclusive yields in this region may also be affected by initial–state nuclear effects, in particular shadowing. 
Figure~\ref{Fig:RAAhadrons} shows that all hadron species exhibit the same suppression for $p_{\rm T} > 8$ GeV/$c$. This observation, combined with identified particle measurements in small systems~\cite{Acharya:2021oaa}, indicate that hadron formation in jets occurs outside of the QGP in this kinematic range.

In contrast to charged hadrons, which are jet fragments, electroweak particles do not carry colour charge and do not interact in the QGP. Indeed, as shown in Fig.~\ref{Fig:RAAhadrons}, bottom panel, the inclusive production yield of isolated photons at high-$\pT$ is compatible with that expected from binary-scaled pp collisions within uncertainties ($\RAA{}$~$\approx$~1). Their production rate is however affected by the initial state of the collision, notably nuclear modification of PDFs (see Chap.~\ref{ch:InitialState}). Similar trends are also reported for Z bosons measured at midrapidity~\cite{Aad:2019lan,Chatrchyan:2014csa}, while at forward rapidity the $\RAA{}$ deviates from unity due to shadowing effects~\cite{Acharya:2020puh}. On the other hand, the $\RAA{}$ of $\rm W^+$ and $\rm W^-$ also differs from unity due to isospin effects and, depending on the rapidity coverage, shadowing effects~\cite{ALICE:2022cxs,Aad:2019sfe} (see also Chap.~\ref{ch:InitialState}).

\begin{figure}[htb]
\centering
\includegraphics[width=0.85\textwidth]{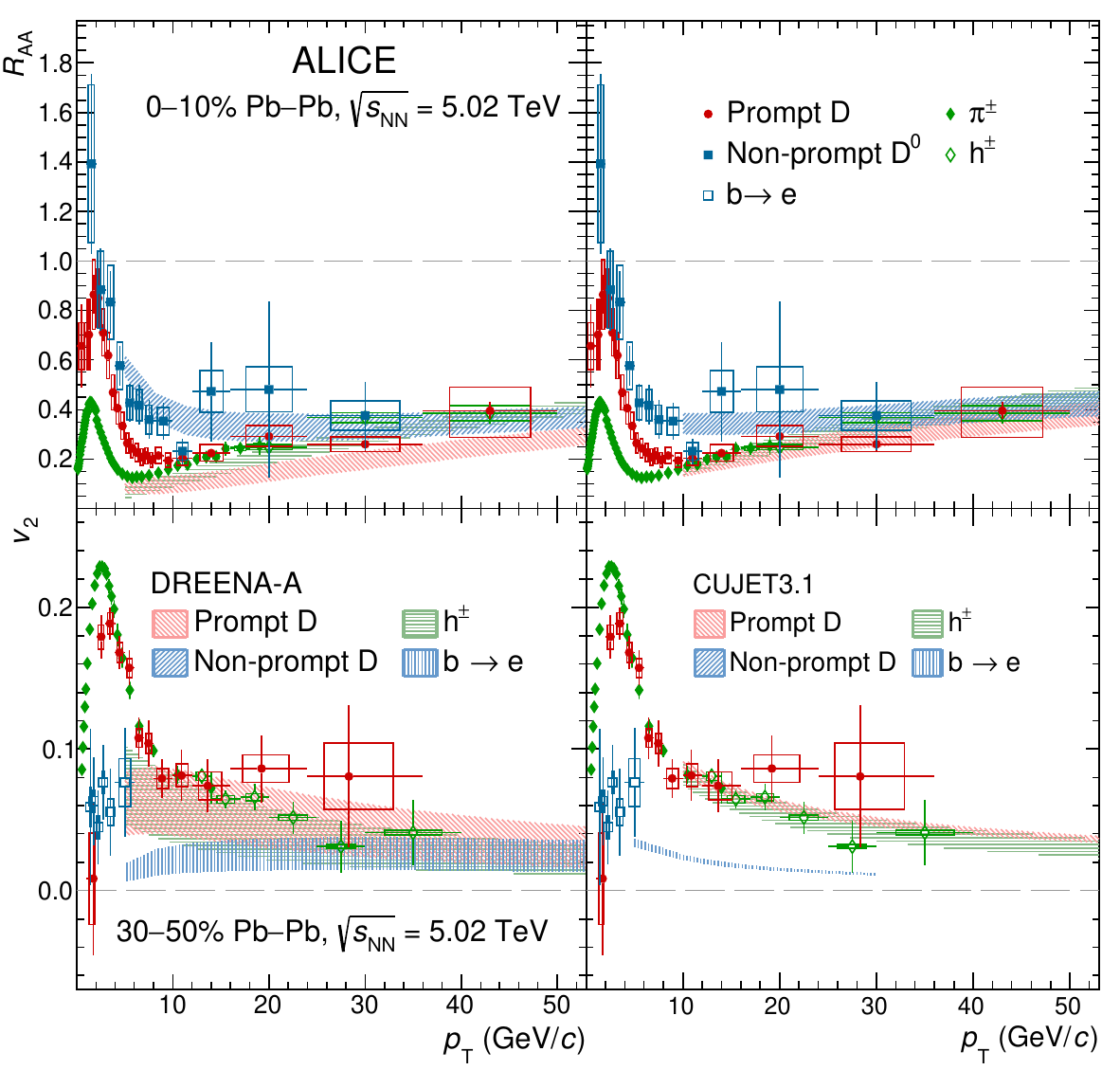}
\caption{Comparisons of the \pT-differential \RAA\ (upper panels) and $v_2$ (lower panels) measured for charged pions~\cite{Acharya:2019yoi,Acharya:2018zuq} and hadrons, prompt D mesons~\cite{ALICE:2021rxa,ALICE:2020iug}, non-prompt D$^0$ mesons (\RAA\ only~\cite{ALICE:2022tji}), and electrons from beauty-hadron decays ($v_2$ only~\cite{ALICE:2020hdw}) in the 10\% most central (\RAA\,) and 30--50\% semi-central ($v_2$) Pb--Pb collisions at \sqrtsNN~=~5.02 TeV with calculations using the DREENA-A~\cite{Stojku:2020tuk} (left panels) and CUJET3.1~\cite{Shi:2019nyp} (right panels) models. A global normalisation uncertainty of 3\% on \RAA\ is not shown in the figure.}
\label{Fig:Eloss_models}
\end{figure}

For a more comprehensive understanding of the mechanisms responsible for parton energy loss in the QGP, jet quenching effects are compared for inclusive charged particles, which are mostly produced by gluon jets, and for prompt and non-prompt open charm mesons, which originate from charm  and beauty quarks. In addition, model comparisons are made to explore the dependence of energy loss on colour charge (gluon vs. quark) and quark mass. Both the normalised yield \RAA\ and the azimuthal anisotropy $v_2$ are compared to model calculations, to clarify the role of effects such as in-medium path length and density fluctuations. 

Figure~\ref{Fig:Eloss_models} shows the \pT\,-dependence of \RAA\, (upper panels) and $v_2$ (lower panels) in the 0--10\% and 30--50\% \PbPb\ collisions
at \sqrtsNN~=~5.02 TeV, for charged hadrons and pions, electrons from beauty-hadron decays, and for prompt and non-prompt D mesons. 
It can be seen in the figure that at high-\pT{}, the production yields of light and heavy-flavour hadrons are suppressed by a similar amount. At $\pT<8$~GeV/$c$, a number of different effects comes into play, as discussed above and in Sec.~\ref{sec:QGPHF}. 

In the high-\pT\, region, \RAA\ increases slowly with \pT\, and the observed suppression is similar for all species, although the relative contributions of elastic and inelastic energy loss mechanisms, as well as the difference between gluon and quark quenching, have a different impact on the suppression pattern of the different hadron species. 

To further explore path-length and fluctuation effects on energy loss, model calculations are compared with \pT\,-differential $v_2$ measurements in semi-central collisions. Figure~\ref{Fig:Eloss_models} shows results from two different energy loss model calculations that provide results for \RAA{} and $v_2$ of both light and heavy flavours: CUJET\,3.1~\cite{Shi:2019nyp,Xu:2014tda,Xu:2015bbz} and DREENA-A~\cite{Stojku:2020tuk}. Both models include radiative energy loss using the opacity expansion framework and collisional energy loss, coupled to a full hydrodynamic description of the medium density profile as a function of time.

For radiative energy loss in a homogeneous medium, a quadratic dependence of the energy loss on the path length is expected for short path lengths due to formation time effects~\cite{Baier:1996kr,Gyulassy:2000fs,Salgado:2003gb,Zapp:2011ya} (Sec.~\ref{sec:QGPJets}), while collisional energy loss depends linearly on the path length in the medium. 
In the first studies with RHIC data, it was found that the calculated azimuthal anisotropy at high-\pT{} was smaller than the observed $v_2$, even with quadratic path-length dependence of the energy loss~\cite{Xu:2014ica,Andres:2016iys}. 
This triggered detailed investigations of the models which showed that the predicted $v_2$ is highly sensitive to details of the medium density evolution~\cite{Renk:2010qx} and the effects of density fluctuations~\cite{Renk:2011qi,Betz:2011tu,Noronha-Hostler:2016eow}.
It was recently realised that the pre-equilibrium phase may also play an important role. Specifically, in most energy loss models, it is assumed that the QGP density builds up during the very early times and energy loss is small or negligible during this time, and it was found that the predicted azimuthal asymmetry depends strongly on the starting time of the evolution~\cite{Andres:2019eus}. This dependence has also been explored within the DREENA model~\cite{Stojku:2020tuk}.

With these considerations in mind, the predictions of CUJET3.1 and DREENA-A are compared to the measurements in the high-\pT{} region, where the approximations used in the models are valid. In this regime, the models describe the measurements shown in Fig.~\ref{Fig:Eloss_models} reasonably well, though tension is observed between the data and the DREENA-A calculation, which predicts larger suppression in central collisions.
The observed \RAA{} and $v_2$ are in agreement with the expected behaviour, including the expected mass dependence due to the dead-cone effect~\cite{Dokshitzer:2001zm,Armesto:2003jh,Djordjevic:2003zk}. Furthermore, the predicted $v_2$ of beauty hadrons (measured via non-prompt $\rm D^0$ and leptons) is much smaller than that of charm and light-flavour hadrons. 
Also at low-\pT\ the $v_2$ of electrons from beauty-hadron decays is smaller than the one of electrons originating from charm-hadron decays, consistent with the expected longer beauty quark relaxation time in the QGP.

Several other groups are exploring different approximations to energy loss calculations, such as the multiple-soft scattering approach~\cite{Armesto:2009zi,Andres:2016iys}, hard thermal loop calculations~\cite{Arnold:2002ja,Schenke:2009gb}, higher-twist calculations~\cite{Wang:2001ifa, Qin:2009gw}, and soft collinear effective theory approaches~\cite{Kang:2016ofv}. 
Work is currently ongoing to extend the formalism beyond the eikonal limit~\cite{Apolinario:2014csa,Apolinario:2017jme}. 
These calculations lead to similar conclusions as those presented here.

\begin{figure}[ptb]
\centering
\includegraphics[width=0.49\textwidth]{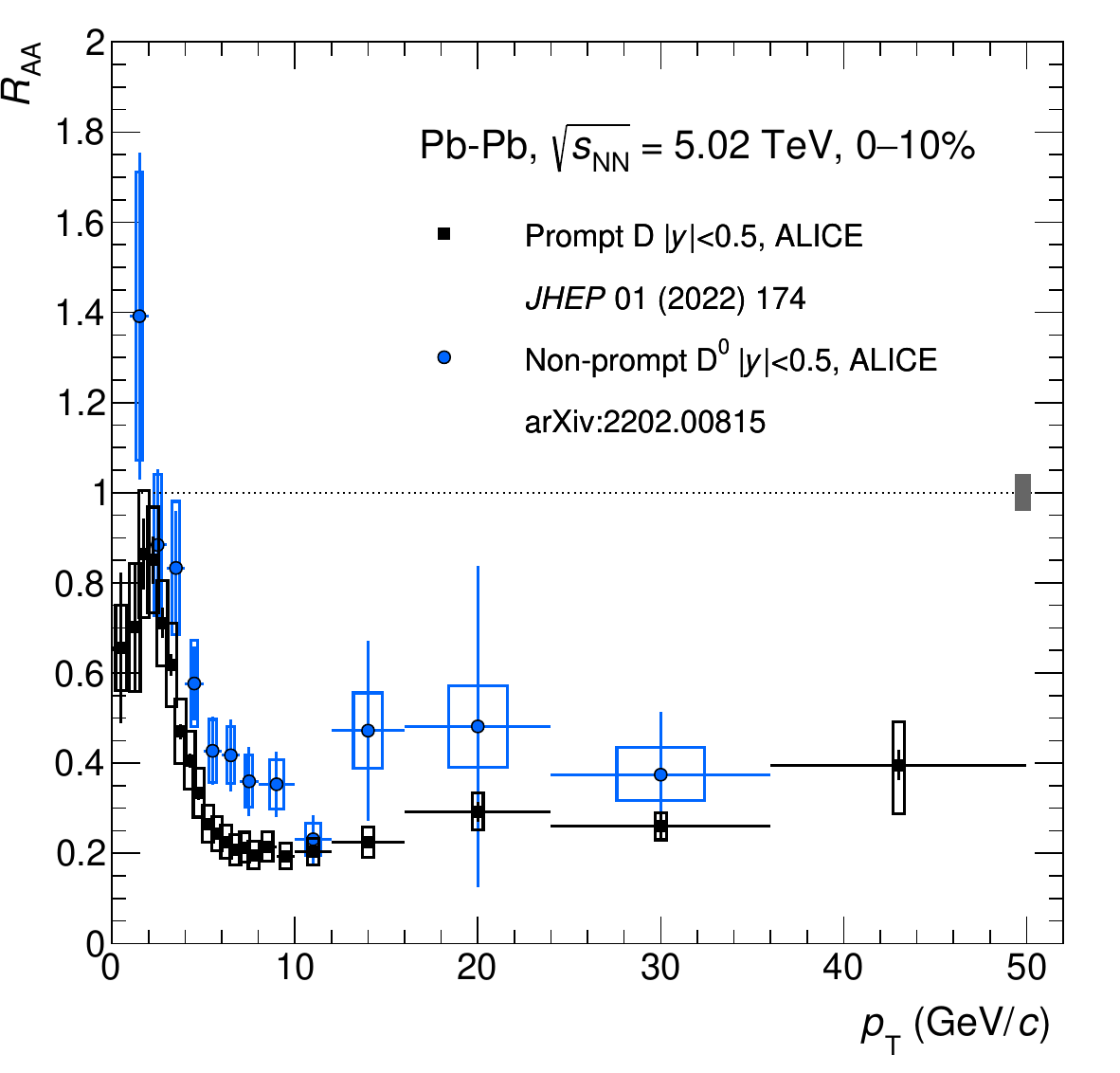}
\includegraphics[width=0.48\textwidth]{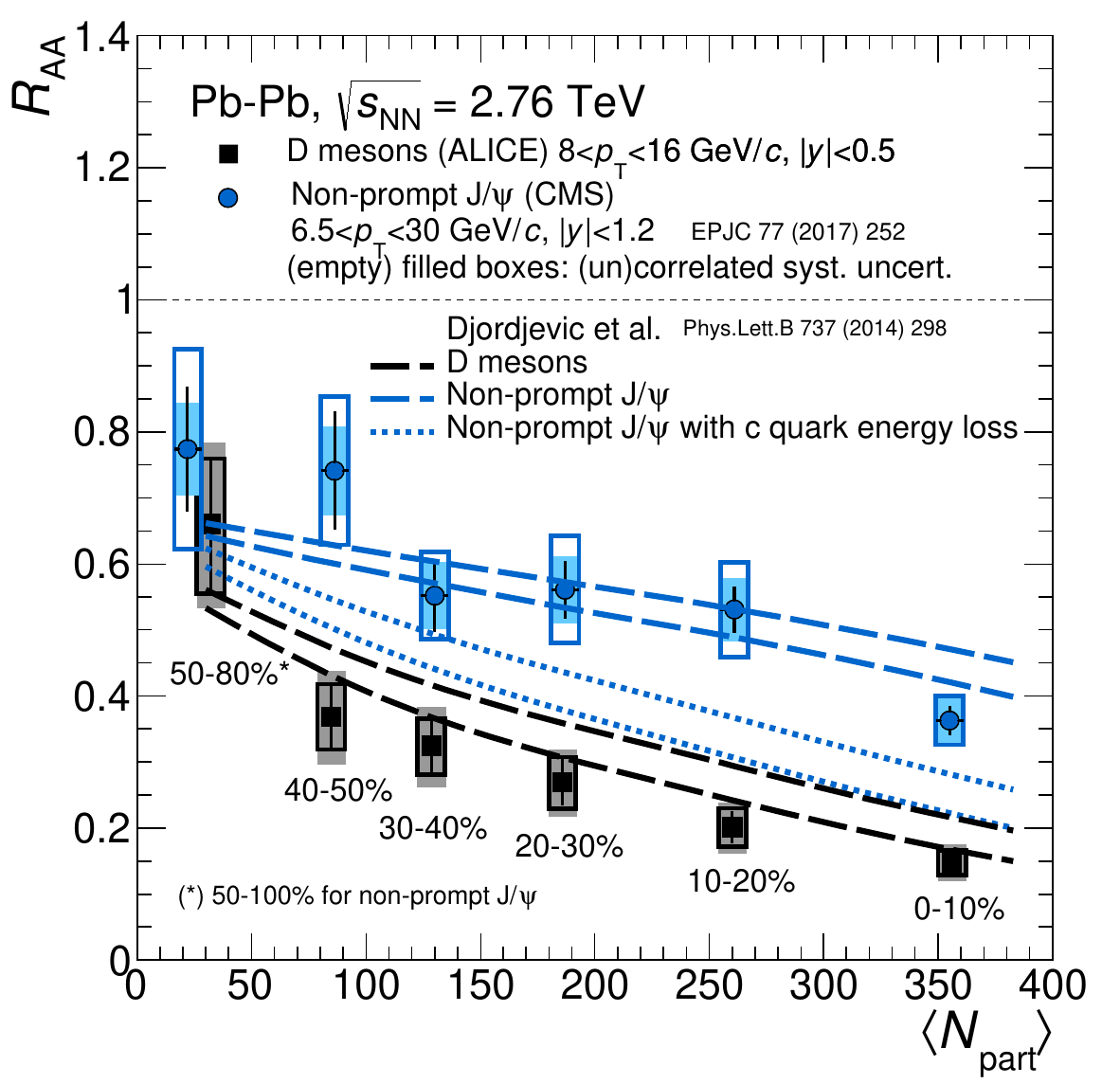}
\caption{(Left) \RAA\ for prompt and non-prompt D mesons as a function of \pT\, in the 10\% most central \PbPb\ collisions at \sqrtsNN~=~5.02~TeV~\cite{ALICE:2021rxa,ALICE:2022tji}. (Right) \RAA\ for prompt D mesons~\cite{Adam:2015nna} and non-prompt J/$\psi$ mesons (CMS~\cite{Khachatryan:2016ypw}) as a function of the collision centrality for Pb--Pb collisions at \sqrtsNN~=~2.76 TeV, compared to a model calculation with mass-dependent energy loss~\cite{Djordjevic:2014tka}.}
\label{Fig:RAA_DPrompt_JpsiNonPrompt}
\end{figure}

\begin{figure}[ptb]
\centering
\includegraphics[width=.65\textwidth]{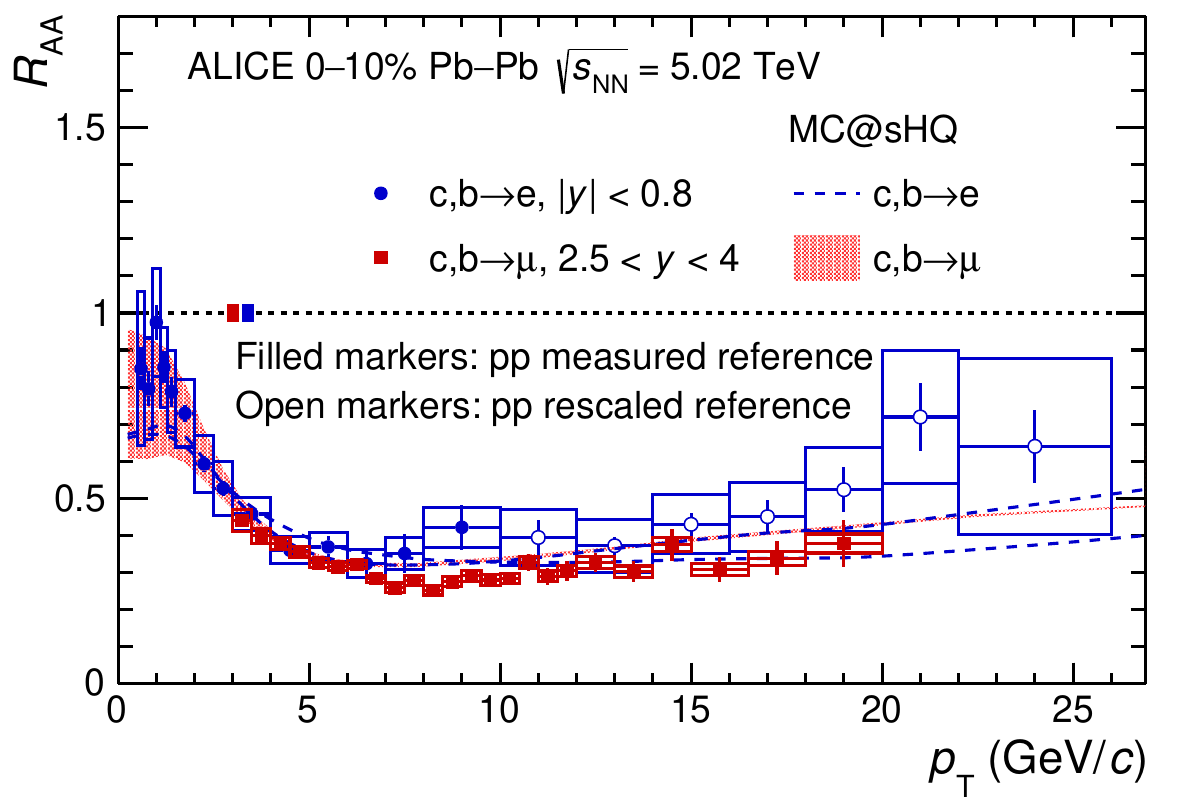}
\hbox{\hspace{0.1cm}\includegraphics[width=.75\textwidth]{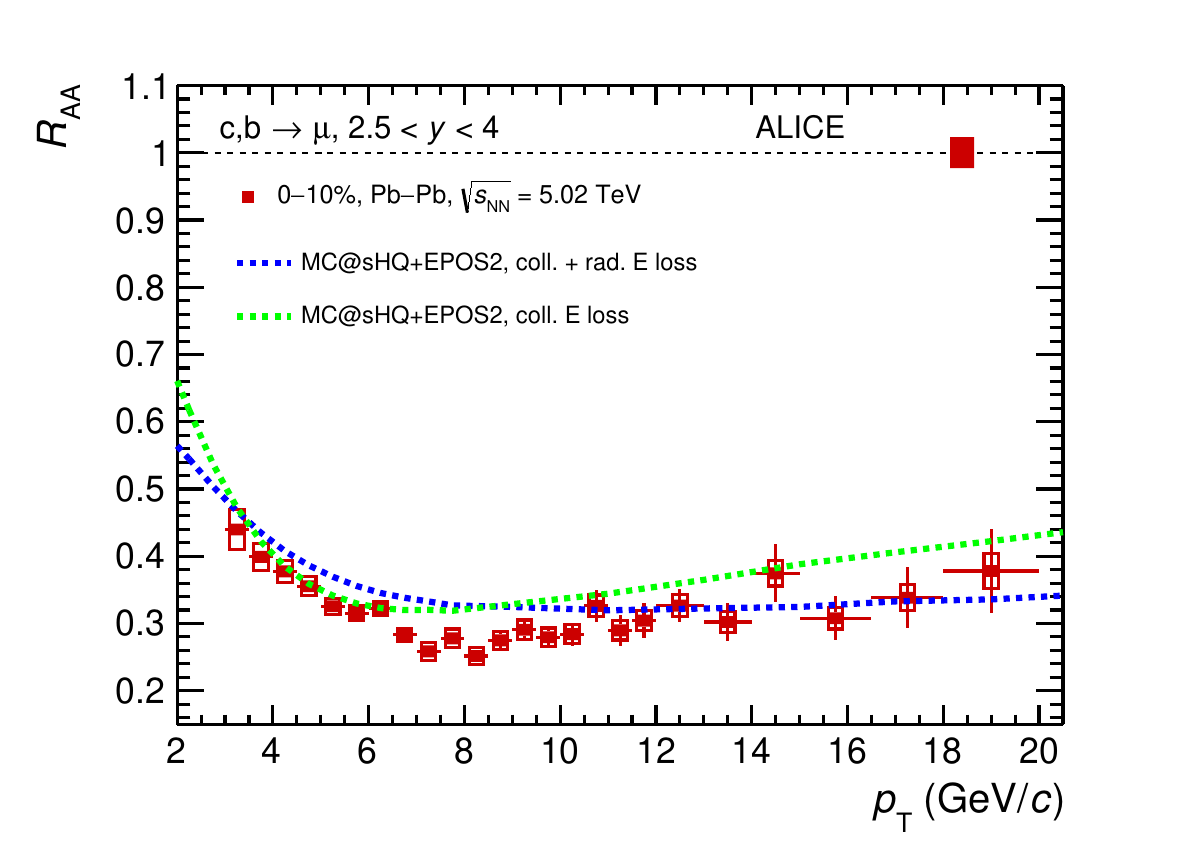}}
\caption{(Top) Measurement of the \pT-differential \RAA\ for muons and electrons from heavy-flavour hadron decays at forward ($2.5 < y < 4$) and midrapidity ($|y| < 0.8$), respectively, in $0-10\%$ Pb--Pb collisions at $\sqrt{s_{\rm NN}} = 5.02$ TeV~\cite{ALICE:2020sjb,Acharya:2019mom}, with comparison to MC@sHQ+EPOS2 calculations~\cite{Nahrgang:2013xaa,Nahrgang:2013saa}. Statistical (vertical bars) and systematic (open boxes) uncertainties are shown. The filled boxes centered at \RAA\ = 1 represent the normalisation uncertainty. (Bottom) \pT-differential \RAA\ of muons from heavy-flavour hadron decays is compared with MC@sHQ+EPOS2 calculations with pure collisional energy loss (dashed lines) and a combination of collisional and radiative energy loss.}
\label{Fig:HFleptons}
\end{figure}

The quark-mass dependence of parton energy loss can be tested by measurements of prompt D mesons, that originate from charm quarks, and non-prompt  D or J/$\psi$ mesons from decays of B mesons, that originate from beauty quarks. Figure~\ref{Fig:RAA_DPrompt_JpsiNonPrompt}, left panel, compares \RAA\ of prompt and non-prompt D mesons~\cite{ALICE:2021rxa,ALICE:2022tji},
indicating that the latter have a larger \RAA\ at intermediate-\pT\ of 5--10~GeV/$c$.
The right panel of the same figure shows the centrality dependence of high-\pT\ \RAA\ for prompt D mesons and non-prompt J/$\psi$ mesons in Pb--Pb collisions at $\sqrt{s_{\rm NN}} = 2.76$ TeV~\cite{Khachatryan:2016ypw}. 
The kinematic ranges ($8< \pT < 16$\,\gev\ for D, $6.5< \pT < 30$\,\gev\ for J/$\psi$) were chosen such that the average momenta of the B mesons that decay into J/$\psi$ are similar to the D-meson momenta; the difference in rapidity interval width is expected to have negligible effect. 
The different magnitudes of suppression indicate larger in-medium energy loss of the lighter charm quarks compared to beauty quarks. The figure also shows results of calculations which include the dead-cone effect~\cite{Djordjevic:2014tka}, which are in reasonable agreement with the data. 
The dotted lines show the calculation for non-prompt J/$\psi$ mesons with the same initial momentum distribution and fragmentation function for beauty quarks but using the charm instead of the beauty-quark mass for the calculation of the energy loss. This results in larger suppression, demonstrating the mass dependence of radiative energy loss directly. 
As similar study is reported in Ref.~\cite{ALICE:2022tji}, showing that models can achieve a larger \RAA\ for non-prompt $\rm D^0$ mesons with respect to prompt ones at $\pT$ of 5--10~GeV/$c$ only if they include mass-dependent energy loss.
It can thus be concluded that this observed \RAA\, hierarchy is indeed consistent with the expected mass dependence of radiative energy loss.

ALICE has also studied the rapidity dependence of in-medium heavy-quark energy loss using muons from heavy-flavour hadron decays in the rapidity interval $2.5 < y < 4$. Figure~\ref{Fig:HFleptons}, top panel, shows the \pT\,-differential \RAA\ 
of muons ($2.5 < y <4$)~\cite{ALICE:2020sjb} and electrons ($\vert y \vert < 0.8$)~\cite{Acharya:2019mom} from heavy-flavour hadron decays in 0-10\%  central Pb--Pb collisions at $\sqrt{s_{\rm NN}}$ = 5.02 TeV. The magnitude of suppression is similar at forward and midrapidity, within uncertainties. This shows that heavy quarks experience in-medium energy loss over a wide rapidity interval. The similar magnitude of suppression does not necessarily correspond to similar energy loss, however, since the shape of \pT-spectrum also influences the suppression~\cite{Prado:2019ste}. The calculation based on the MC@sHQ+EPOS2 transport model~\cite{Nahrgang:2013xaa,Nahrgang:2013saa} agrees with data at both central and forward rapidity.
Figure~\ref{Fig:HFleptons}, bottom panel, compares the measured \RAA\ of forward muons from heavy-flavour hadron decay to calculations based on two configurations of the MC@sHQ+EPOS2: (i) energy loss from both medium-induced gluon radiation and collisional processes, or (ii) collisional energy loss only. Both calculations agree with the data, though the configuration with both collisional and radiative energy loss is closer to the data for $\pT > 10$~\gev{}, indicating the importance of radiative processes in the high-\pT\ region. 

\paragraph{\textbf{\textit {Interplay of inclusive jets and hadrons.}}}

As discussed in Sec.~\ref{sec:QGPJets}, high-\pT\ hadrons and reconstructed jets explore different aspects of jet quenching: hadrons are sensitive principally to energy loss in the hardest branch of the jet shower, while jets, which subtend an area approximately $\pi{R^2}$ for jet resolution parameter $R$, are sensitive more broadly to modification of the shower. A comprehensive understanding of jet quenching requires measurements of both high-\pT\ hadrons and reconstructed jets, with the latter spanning significant range in \pTjet\ and $R$. In this section, measurements of inclusive jets after a statistical subtraction of the underlying event are discussed and compared to the inclusive hadron measurements presented in Sec.~\ref{sec:ElossHadrons}.

\begin{figure}[ptb]
\centering
\includegraphics[width=0.65\textwidth]{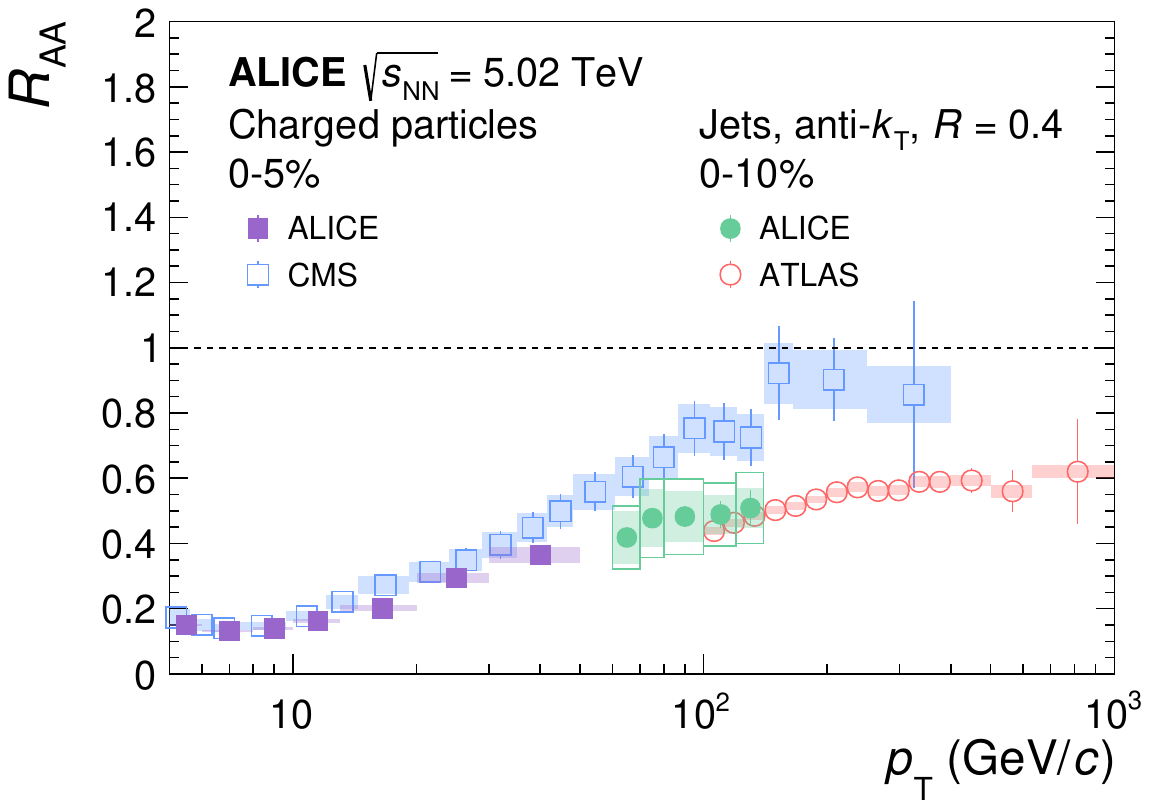}
\caption{Measurement of \RAA\ for charged hadrons (ALICE~\cite{Acharya:2018qsh} and CMS~\cite{Khachatryan:2016odn}) and jets (ALICE~\cite{Acharya:2019jyg} and ATLAS~\cite{Aaboud:2018twu}) in central \PbPb\ collisions. A global normalisation uncertainty of 3\% on \RAA\ is not shown in the figure.}
\label{Fig:RAAhadronsjets}
\end{figure}

Figure~\ref{Fig:RAAhadronsjets} shows \RAA\ in central \PbPb\ collisions at \sqrtsNN~=~5.02 TeV for inclusive jets with $R=$ 0.4, together with charged hadrons. Jet \RAA\ exhibits larger suppression than hadrons at the same \pT, with the ALICE jet spectrum extending down to $\pT = 60$ GeV/$c$. At higher \pT\, the ALICE~\cite{Acharya:2019jyg} and ATLAS~\cite{Aaboud:2018twu} jet data are consistent, and show slowly increasing \RAA\ with increasing \pT.

In general, a reconstructed jet will catch a fraction of medium-induced radiation, and the inclusive \pT-spectrum of jets is significantly harder than that of hadrons.
Both factors suggest that \RAA\ for inclusive jets will be larger than that for inclusive hadrons. It is to be noted that, while the opposite is observed in the \pT\ region where the measurements of hadrons and jets overlap, such direct comparison is not meaningful
since hadrons and jets at any given \pT\ originate from different parton energies. For a proper interpretation of this observation, the inclusive hadron population must be mapped to that of jets, taking into account the bias imposed by selecting high-\pT\ hadrons.  

\begin{figure}[htb]
\centering
\includegraphics[width=0.49\textwidth]{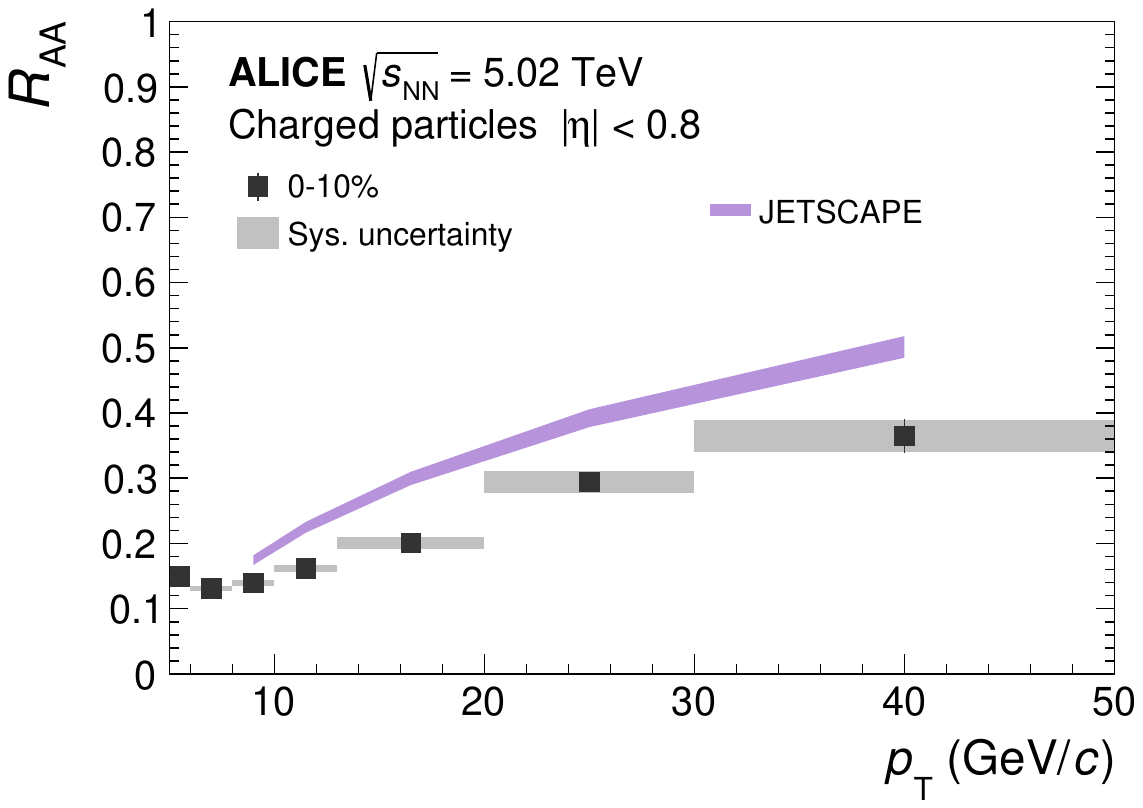}
\includegraphics[width=0.49\textwidth]{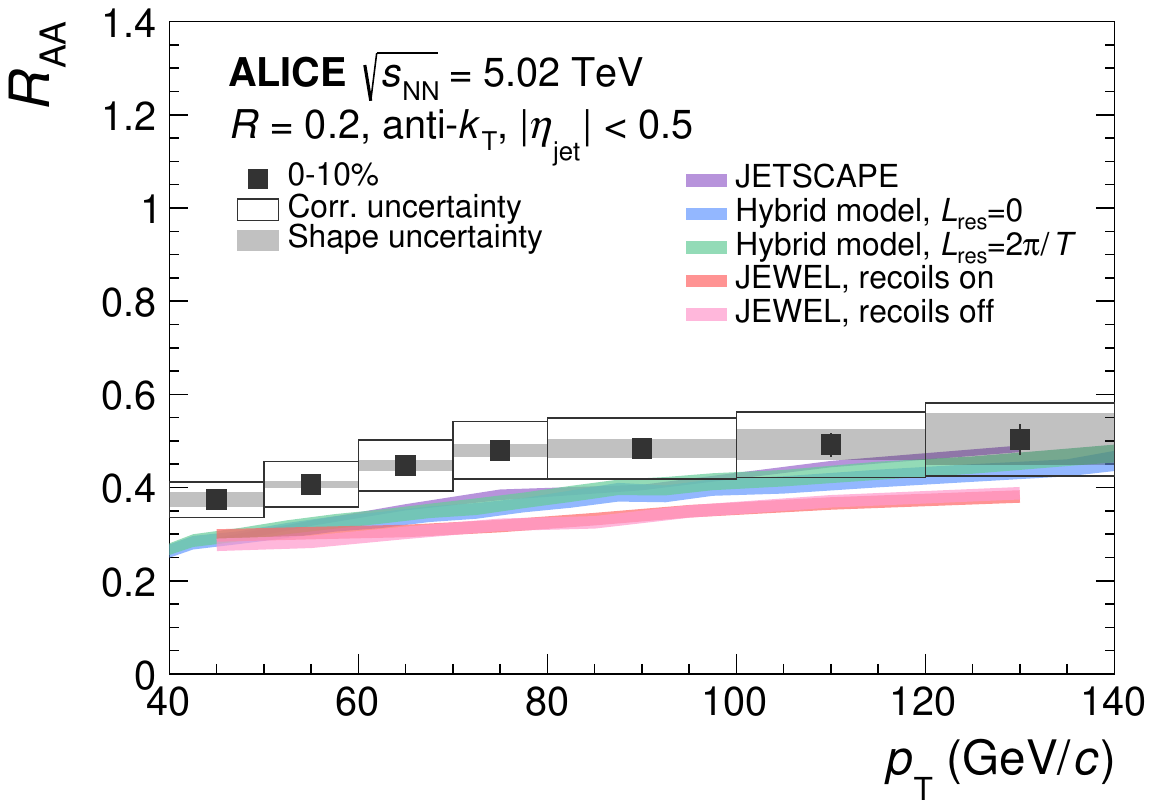}
\caption{ALICE measurements of \RAA\ for charged hadrons~\cite{Acharya:2018qsh} and jets~\cite{Acharya:2019jyg} in central \PbPb\ collisions compared to calculations from JETSCAPE, JEWEL, and the hybrid model. A global normalisation uncertainty of 3\% on \RAA\ is not shown in the figure.}
\label{Fig:RAAhadronsjetsWithModels}
\end{figure}

Model calculations indicate that high-\pT\ hadrons are more likely to originate from narrow, hard fragmenting jets, which on average undergo fewer interactions with the QGP. These fewer interactions result in less energy loss~\cite{Milhano:2015mng,Chesler:2015nqz,Rajagopal:2016uip}. %
Detailed model comparisons provide the opportunity to elucidate the relationship between hadron and jet $\RAA{}$, and the impact of $\qhat{}$ on those observables. Figure~\ref{Fig:RAAhadronsjetsWithModels} presents comparisons of the measured $\RAA{}$ of charged hadrons (left) and reconstructed $R =$ 0.2 jets (right) with JETSCAPE~\cite{JETSCAPE:2021ehl}, JEWEL~\cite{Zapp:2008gi,Zapp:2011ek,Zapp:2012ak} and the hybrid model~\cite{Sirunyan:2018gct,Casalderrey-Solana:2016jvj}. Jets with $R =$ 0.2 are utilised for this comparison due to their higher precision and larger $\pT{}$ range.

JETSCAPE employs detailed hydrodynamic modelling of the QGP, together with a two-stage jet quenching calculation based on MATTER and LBT which use the Higher Twist formalism for virtuality-dependent jet-medium interactions. The LBT model also incorporates medium response to the deposited energy via the Boltzmann transport approach. Transition between MATTER (high virtuality) and LBT (low virtuality) occurs at a fixed parton virtuality $Q_{\rm switch}$, which is determined from data~\cite{JETSCAPE:2021ehl}. 
As shown in Fig.~\ref{Fig:RAAhadronsjetsWithModels}, the JETSCAPE calculations predict slightly larger \RAA\ for charged hadrons over the entire measured $\pT{}$ interval and broadly describes the jet measurements. 
Implementation of the $Q$-dependence of $\qhat{}$ was found to be critical to simultaneously describe both measurements. Without such a dependence, the energy loss at high-$\pT{}$ increases in the calculation, leading to an underprediction of the \RAA.

JEWEL overestimates the suppression for jet \RAA{}. This conclusion holds regardless of the inclusion of medium recoil, although the model with recoil tends to get closer to the data. The hybrid model describes the trends of the jet \RAA{}, but exhibits some tension. 

\paragraph{\textbf{\textit {Measurement of jet energy loss.}}}
\label{sec:elossmeas}

As discussed above, yield suppression measurements are only indirectly related to jet energy loss, since yield suppression depends on both the population-averaged energy loss and the shape of the spectrum. However, these effects can be disentangled for jet yield suppression measurements since reconstructed jets account for all correlated hadronic energy within the jet area (colloquially, within the ``jet cone''), and jet yield suppression therefore must arise from energy transport out of the jet cone, i.e. jet energy loss. Several recent papers have carried out such a phenomenological extraction of energy loss, by converting yield suppression to the equivalent \pT-shift of the spectrum~\cite{Adam:2015doa,Acharya:2017okq,Adamczyk:2017yhe,Adam:2020wen}.

\begin{figure}[htb]
\centering
\includegraphics[width=0.49\textwidth]{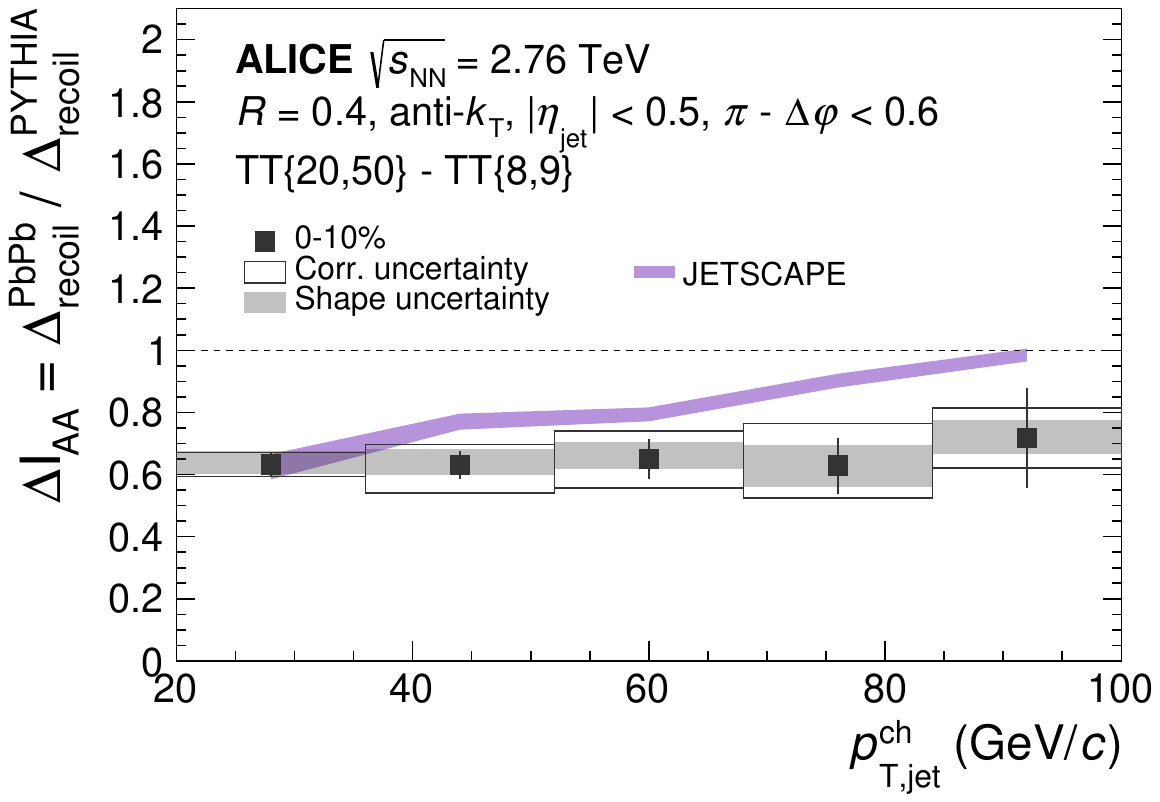}
\includegraphics[width=0.49\textwidth]{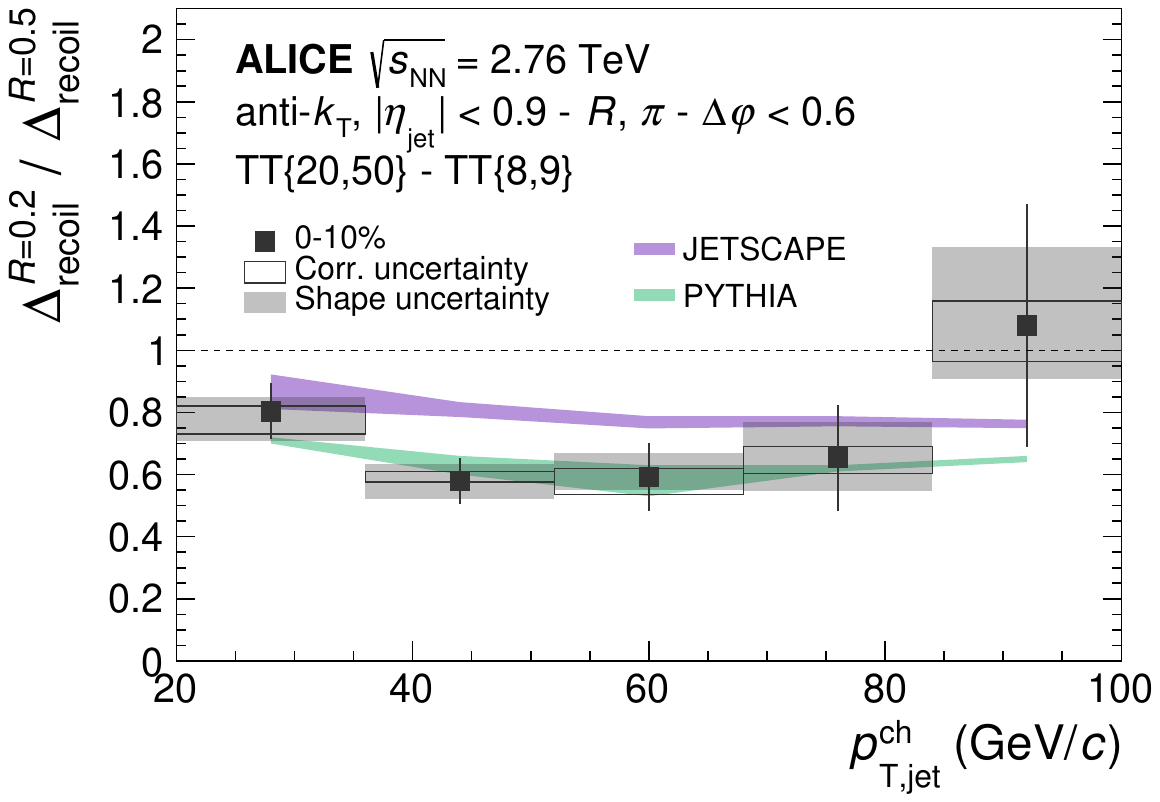}
\caption{Semi-inclusive hadron+jet coincidence measurement for \PbPb\ collisions at \sqrtsNN~=~2.76 TeV~\cite{Adam:2015doa} compared to JETSCAPE expectations. The observable $\Delta_\mathrm{recoil}$ is defined in Eq.~\ref{eq:DeltaRecoil}. The notation “${\rm TT}\{a,b\}$” refers to a trigger hadron within the interval $a<p_\mathrm{T}<b$, and $\Delta\varphi$ refers to the azimuthal separation between trigger hadron and recoil jet. (Left) Ratio of \Drecoil\ distribution to that for \pp\ collisions calculated with the PYTHIA6 event generator. (Right) Ratio of \Drecoil\ distributions for recoil jets with $R=0.2$ and 0.5, measured for central \PbPb\ and calculated with PYTHIA for \pp\ collisions.} %
\label{fig:hJetyield}
\end{figure}

Figure~\ref{fig:hJetyield}, left panel, shows the ratio of \pT-distributions of the \Drecoil\ observable (Eq.~\ref{eq:DeltaRecoil}), for a semi-inclusive hadron+jet measurement in central \PbPb\ collisions at \sqrtsNN~=~2.76 TeV compared to that for \pp\ collisions calculated by PYTHIA~\cite{Adam:2015doa}. A \pT-independent yield suppression of 0.6 is observed for central \PbPb\ collisions. Both \Drecoil\ spectra  are well approximated over the measured \pT range by an exponential function $\sim{e^{-p_{\mathrm{T}}/b}}$ with  $b\sim16$~\gev, so that a single number for the \pT\ shift, $(8\pm2)$~\gev, characterises the measured suppression~\cite{Adam:2015doa}. JETSCAPE calculations reproduce the data at low-\pT\, but show a rising trend with increasing \pT\ that is  not seen in the data.

A similar analysis was carried out for high multiplicity \pPb\ collisions at \sqrtsNN~=~5.02 TeV~\cite{Acharya:2017okq}) (see Sec.~\ref{Sect:JetQuenchSmallSystems}), giving a limit of 0.4 \gev\ at 90\% CL for medium-induced energy transport to angles greater than $R=0.4$ for such collisions. A similar \pT-shift analysis was likewise carried out for jet measurements in central \AuAu\ collisions at \sqrtsNN~=~200 GeV, giving shift values of $(3.3\pm0.3\pm0.7)$~\gev\ for an inclusive jet population~\cite{Adam:2020wen} and $(5.0\pm0.5\pm1.2)$~\gev\ for a semi-inclusive recoil jet population~\cite{Adamczyk:2017yhe}, both with $R=0.4$. These values suggest that energy loss due to jet quenching is larger in central nucleus--nucleus collisions at the LHC than at RHIC, though with limited significance. Future measurements with improved systematic uncertainties and extending over a broader \pT\ range, together with theoretical calculations, will provide more insight into this comparison.

Figure~\ref{fig:hJetyield}, right panel, shows the ratio of \Drecoil\ distributions for recoil jets with $R=0.2$ and 0.5, separately for \PbPb\ and \pp\ collisions. The value of this ratio reflects the transverse jet shape 
and can be modified by medium-induced energy transport. However, these ratios for \PbPb\ collisions and for \pp\ collisions as calculated using PYTHIA are consistent within uncertainties, suggesting that the medium-induced lost energy due to quenching is predominantly transported to angles greater than 0.5 rad.  

\paragraph{\textbf{\textit {Jet quenching in peripheral \PbPb\ collisions.}}}
\label{sec:jetquenchperiphPbPb}

Jet energy loss in a static medium is expected to vary parametrically as $L^2$, where $L$ is the in-medium path length~\cite{Baier:1996kr,Armesto:2011ht,CaronHuot:2010bp}. Detailed modeling is required to account for additional effects due to the spatial distribution and orientation of propagating partons, and the dynamical expansion of the QGP. A valuable tool to test and constrain such calculations is the measurement of jet quenching observables with variation in system size. 
Jet quenching measurements in small collision systems (\pp, \pPb) are discussed in Sec.~\ref{Sect:JetQuenchSmallSystems}; to date, no significant signals of jet quenching have been observed in such systems. In contrast, inclusive yield measurements in peripheral \PbPb\ collisions exhibit values of \RAA\ less than unity, as shown in Fig.~\ref{Fig:RAAhadrons} (top) for the 70--80\% centrality class. This measurement
suggests significant jet quenching effects even in the most peripheral  (large impact parameter) \PbPb\ collisions (see also Ref.~\cite{ALICE:2018ekf}). However, most of the suppression observed in this centrality class and in more peripheral collisions can be ascribed to biases caused by the event selection and collision geometry. In essence, for peripheral collisions selected using charged-particle multiplicity, the nucleon--nucleon impact parameter distribution is biased towards larger values, leading to a lower yield of hard processes. A PYTHIA-based prediction of this effect~\cite{Loizides:2017sqq} describes a measurement of \RAA\ in fine peripheral centrality classes carried out by ALICE~\cite{ALICE:2018ekf}.
This could explain the unintuitive observation that \RAA\ is below unity in peripheral Pb--Pb, but equal to unity in minimum-bias p--Pb collisions despite similar charged-particle multiplicities.

\subsubsection*{\textbf{\textit {2.4.2.2 Jet substructure modification}}}
\label{sec:Substr}

Modifications to the internal structure of jets can be studied with jet-substructure observables -- defined by
first clustering a jet, and then constructing an observable
as a function of the properties of the constituents of that jet~\cite{Acharya:2019djg, Acharya:2018uvf, Aaboud:2018hpb, Chatrchyan:2013kwa, PhysRevLett.120.142302, Sirunyan:2018gct, Sirunyan:2020qvi, Tripathee:2017ybi}.
Jet substructure observables 
can be constructed to be sensitive to specific regions
of jet radiation phase space in a way that is theoretically
calculable from first principles~\cite{Larkoski:2017jix, Dasgupta:2013ihk, Larkoski:2015lea, Dreyer:2018nbf, Kang:2019prh}, and can target limited
regions of phase space to explore specific jet quenching mechanisms that cannot be resolved using 
jet \pT{} measurements alone
\cite{Mehtar-Tani:2016aco, Casalderrey-Solana:2012evi, Andrews:2018jcm, Ringer:2019rfk, Caucal_2018, Sirunyan:2018gct, HybridModelResolution, Casalderrey-Solana:2019ubu, JEWEL2017, KunnawalkamElayavalli:2017hxo}. Questions addressed by jet substructure measurements in nucleus--nucleus collisions include the strength of the jet-medium coupling, the rate of medium-induced emissions,
and constraints on medium properties such as
coherence scales and the nature of the QGP degrees of freedom.

There is a close connection between substructure modifications due to quenching 
and in-medium scattering and deflection measurements: 
deflection of jet constituents from in-medium interactions can be seen with jet substructure observables, and modification of a branch of a jet that has split must generate acoplanarity of the reconstructed jet centroid.

The fact that jet substructure observables depend on the detailed distribution of
jet constituents brings an experimental challenge, however, since
the underlying event can dramatically distort the reconstructed observables~\cite{Acharya:2018uvf, Acharya:2017goa} (see Sec.~\ref{sec:QGPJets}).
Therefore, precise measurements of jet substructure observables utilise background subtraction
techniques that remove constituents of the jet event-by-event, such as constituent subtraction~\cite{Berta:2014eza, Berta:2019hnj} and subtraction in moment space~\cite{Cacciari:2012mu}. Each observable must be verified individually to ensure that it is 
robust to the bias introduced by such procedures. Jet substructure observables which tag a specific substructure object, 
such as groomed jet observables, face an additional challenge in that 
the underlying event can cause the object to be incorrectly identified~\cite{Mulligan:2020tim}. This underscores the importance of selecting observables that can both be calculated theoretically and measured experimentally.

ALICE jet substructure measurements are carried out using charged-particle jets, to take advantage of the precise spatial resolution down to small angular scales within the jet core (Sec.~\ref{sec:QGPJets}). These measurements are unfolded for detector and background effects in two dimensions (\pT{} and the substructure observable), enabling direct
comparison to theoretical jet quenching calculations.

\paragraph{\textbf{\textit {Groomed jet substructure.}}}

Jet substructure can be used as a starting point
to understand jet quenching by constructing observables 
that isolate the perturbative part of the substructure.
In this approach, jet grooming algorithms such as Soft Drop 
(SD)~\cite{Larkoski:2014wba, Dasgupta:2013ihk, Larkoski:2015lea} are applied to remove soft, wide-angle radiation and identify a single hard ``splitting''.
Monte Carlo event generators suggest that groomed jet splittings are correlated to parton shower splittings in vacuum~\cite{Larkoski:2015lea, Apolinario:2020uvt, Dokshitzer:1991wu}. In heavy-ion collisions this is complicated by both medium and background effects, but recent theoretical studies show that this correlation of the time structure and splittings remains~\cite{Apolinario:2020uvt}, and groomed jet splittings may provide a handle on the space--time evolution of jet propagation through the QGP~\cite{Andrews:2018jcm}.

The SD grooming algorithm  identifies a single splitting by re-clustering the constituents of a jet. 
The splitting is selected from the history of the re-clustering with a grooming condition, $z > z_\mathrm{cut} \theta^\beta$, where $\beta$ and $z_\mathrm{cut}$ are tunable parameters and $z$ is defined as the fraction of transverse momentum carried by the sub-leading prong,

\begin{equation} \label{eq:2}
z \equiv \frac{\pTsub} {\pTlead + \pTsub},
\end{equation}

\noindent
and the angle $\theta$ is defined as the angular
distance between the two branches of the identified splitting, 

\begin{equation} \label{eq:parton1}
\tg \equiv \frac{\rg}{R} \equiv \frac{\sqrt{\Delta y ^2 + \Delta \varphi ^2}}{R},
\end{equation}

\noindent
where $y$ is the rapidity, $\varphi$ is the azimuthal angle, and $R$ is the jet resolution parameter. The groomed splitting can be characterised by two kinematic observables: 
the groomed momentum fraction, \zg{},
and the groomed jet radius, \tg{}, with $z$ and $\theta$ for the groomed splitting as defined in Eq.~\ref{eq:2} and~\ref{eq:parton1}. 

Several different physics mechanisms have been conjectured for
modifications of the \zg{} distribution due to jet quenching, such as
medium-induced radiation being hard enough to pass the grooming condition
and thereby enhancing the number of
asymmetric splittings. Additionally, \zg{} may be sensitive to effects due to the medium response, breaking of colour coherence, and modification to the DGLAP splitting function in the QGP~\cite{Chien:2016led,Caucal:2019uvr,Chang:2019nrx,Casalderrey-Solana:2019ubu}.

\begin{figure}[htb]
\centering
\includegraphics[width=0.49\textwidth]{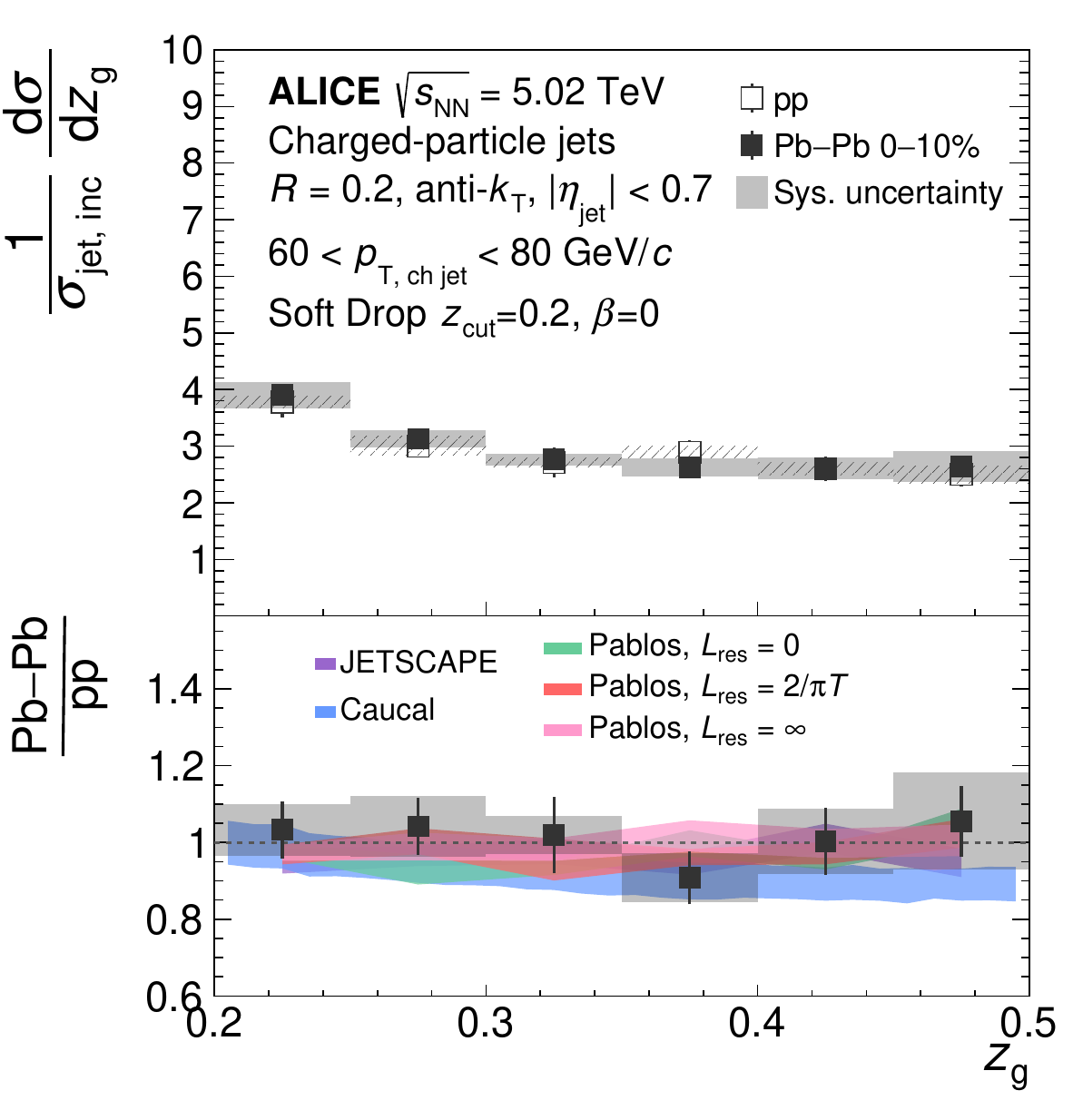}
\includegraphics[width=0.49\textwidth]{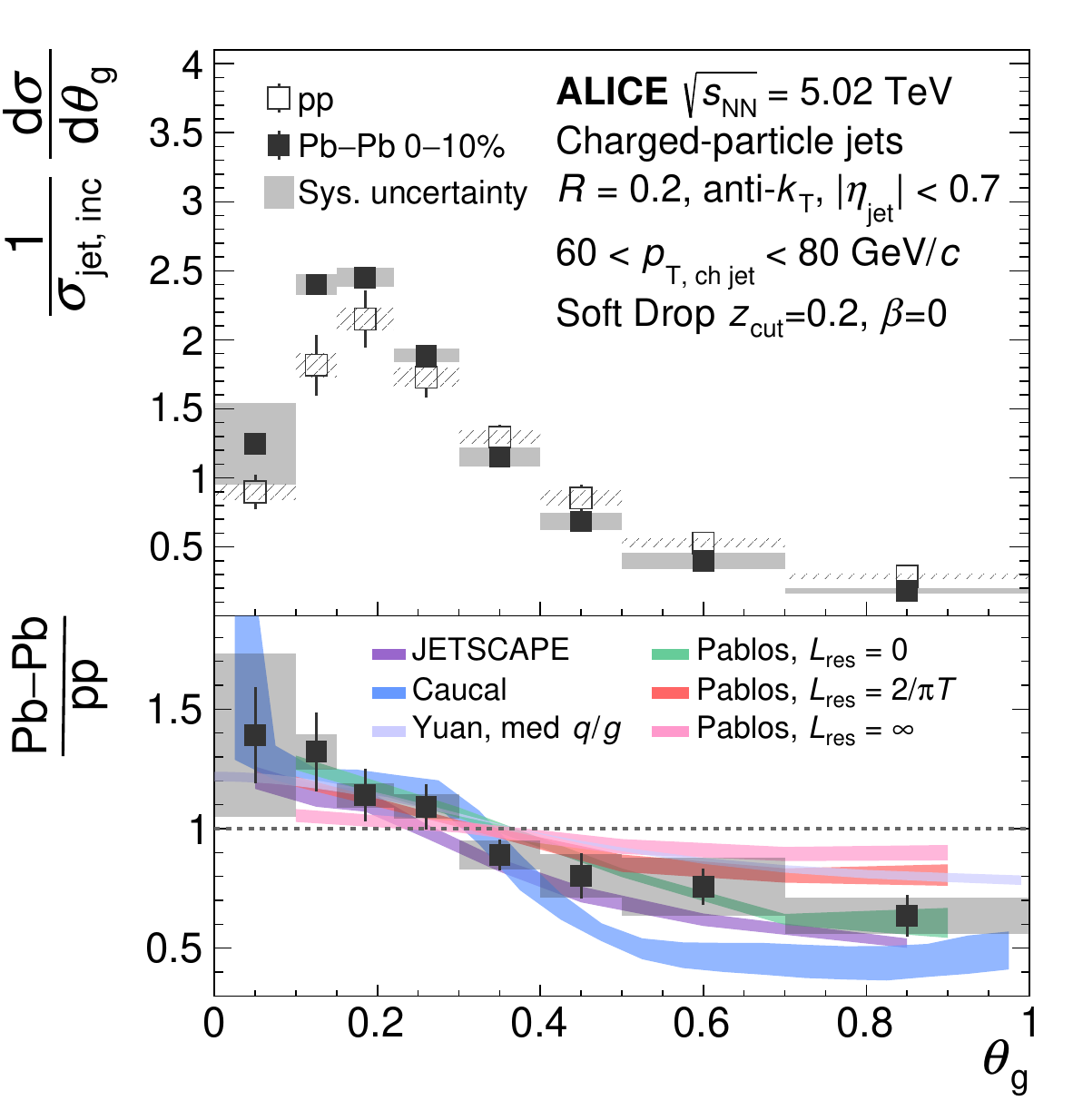}
\caption{Jet $z_{\rm g}$ (left) and $\theta_{\rm g}$ (right) in 0--10\% centrality for $R=0.2$ charged-particle jets~\cite{ALICE:2021obz}. %
The ratio of the distributions in Pb--Pb and pp collisions is shown in the bottom panels and is compared to various jet quenching calculations.}
\label{Fig:JetSubstr}
\end{figure}

Initial heavy-ion jet substructure measurements by CMS~\cite{PhysRevLett.120.142302} and ALICE~\cite{Acharya:2019djg} indicated a suppression of symmetric splittings relative to asymmetric splittings in \PbPb{} collisions compared to \pp{} collisions. However, further analysis revealed that this effect could arise from a cut on the sub-jet separation distance, and that there were additional background contributions from mistagged splittings arising from the underlying event~\cite{Mulligan:2020tim}. More recently, ALICE has addressed these mistagging effects by using stronger grooming conditions and smaller $R$, and carried out measurements in more peripheral collisions. Figure~\ref{Fig:JetSubstr}, left panel, shows the measurement of \zg{}, which exhibits no significant modification of the $\zg$ distribution in \PbPb{} collisions compared to \pp{} collisions. This is consistent within uncertainties with a variety of jet quenching models, also shown.

\begin{figure}[htb]
\centering
\includegraphics[width=0.49\textwidth]{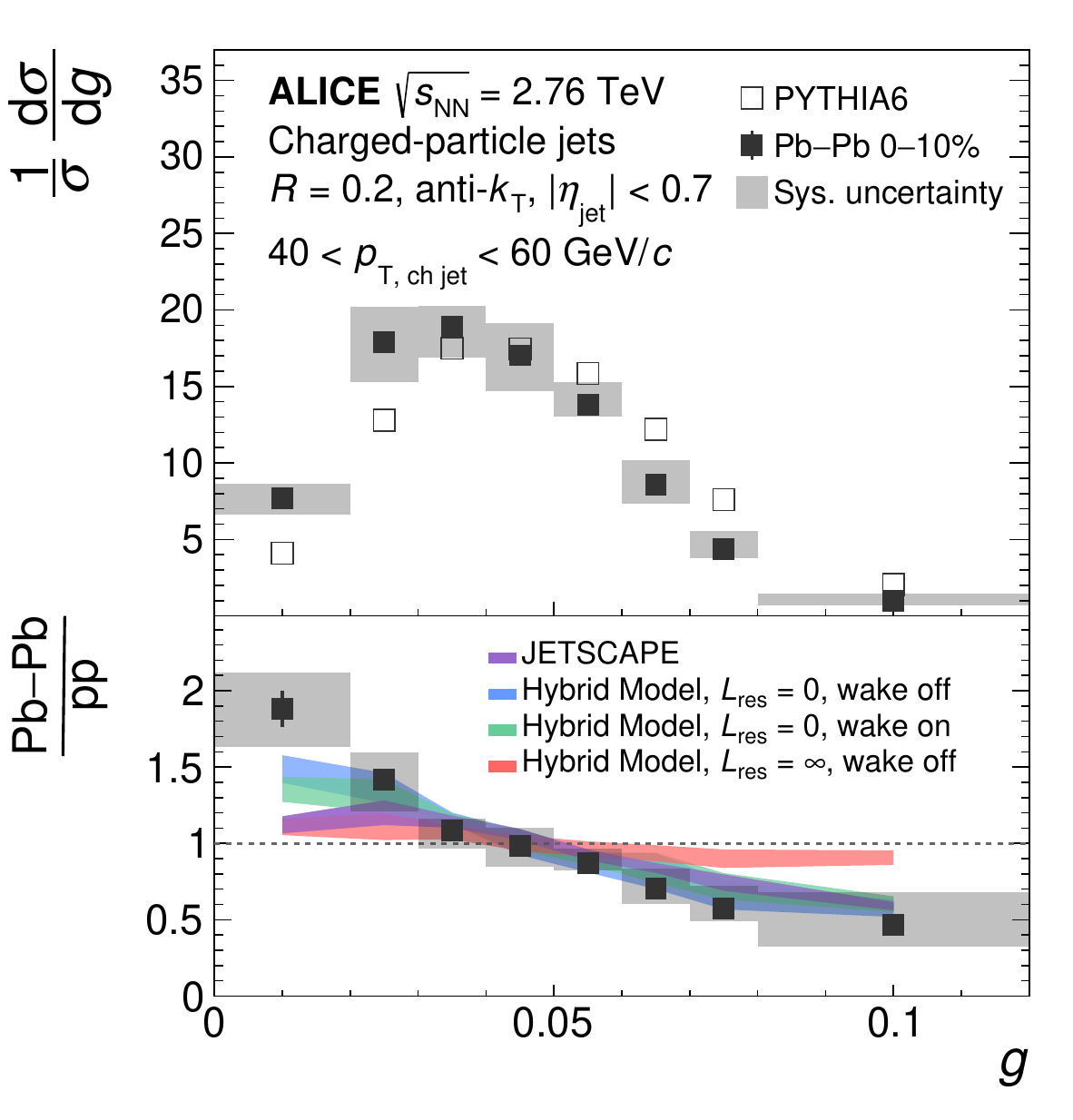}
\raisebox{0.46\height}{\includegraphics[width=0.49\textwidth]{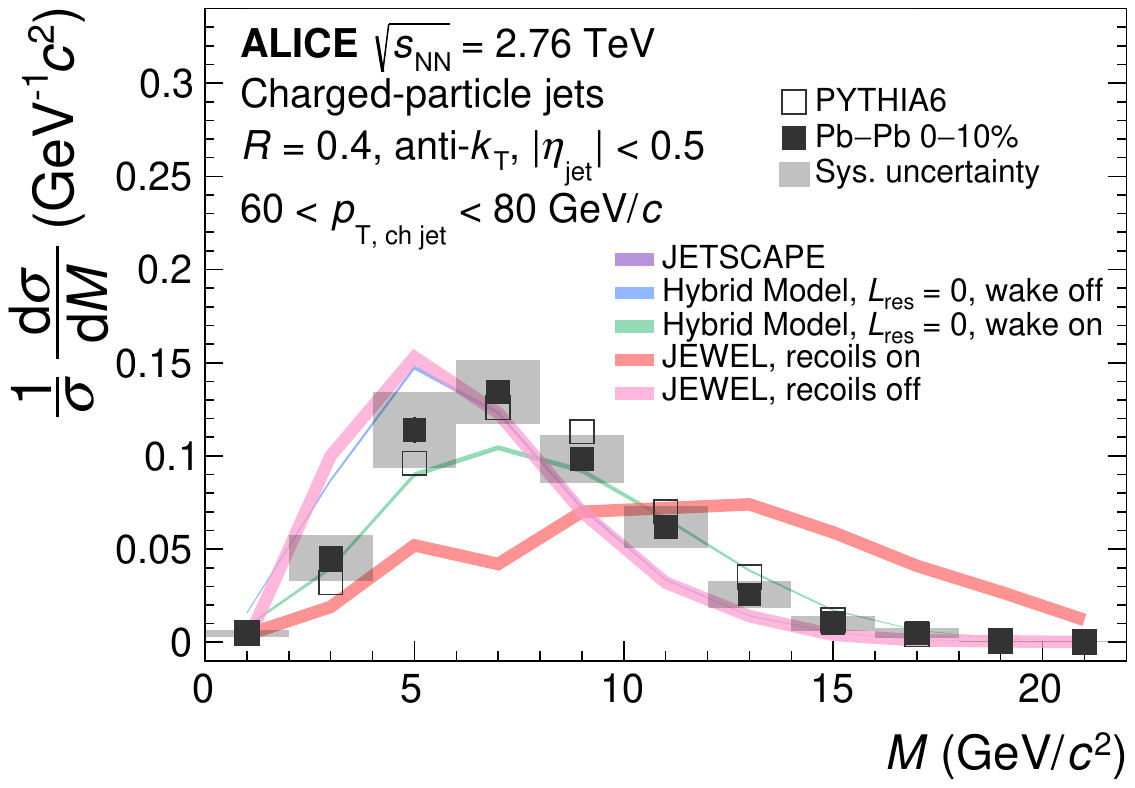}}
\caption{(Left) Distribution of jet angularity for $R=0.2$ charged-particle jets in 0--10\% central \PbPb\ collisions  compared to that for \pp\ collisions calculated with the PYTHIA6 event generator~\cite{Acharya:2018uvf}. (Right) Jet mass for $R=0.4$ charged-particle jets in 0--10\% central compared to PYTHIA calculations and various jet quenching models~\cite{Acharya:2017goa}.
Systematic uncertainties have not been assigned to the PYTHIA calculations in both panels.}
\label{Fig:Jetshape}
\end{figure}

These analysis techniques also enable measurements of the angular distribution \tg{}, which has been predicted to be sensitive to the quark-gluon fraction, splitting formation time, and colour coherence~\cite{Ringer:2019rfk, Casalderrey-Solana:2012evi}.
In contrast to the \zg{} distribution, Fig.~\ref{Fig:JetSubstr} (right) shows a significant narrowing of 
the \tg{} (\rg) distribution in central \PbPb{} collisions compared to \pp{} 
collisions, which may arise from modification of the angular scale of jets in the quark–gluon plasma.
This narrowing is consistent with models implementing (transverse) incoherent interaction
of the jet shower constituents with the medium, but also with
medium-modified quark-to-gluon fractions and fully coherent energy loss;
further measurements will be needed to characterise the mechanism underlying the narrowing.
Taken together, these measurements suggest that the hard substructure
of jets is consistent with (i) little-to-no modification of the 
momentum splitting, and (ii) stronger suppression of jets with wide
fragmentation patterns. This indicates that the medium has a significant resolving power for splittings with a particular dependence on the angular scale, preserving narrow jets or filtering out wider jets.

\paragraph{\textbf{\textit {Jet shapes.}}}

Jet substructure observables can also be used to probe soft non-perturbative physics by studying the distribution of radiation inside the jet, without selecting on the hard structure. Such ungroomed observables can probe the interplay between the modification of the jet structure and the response of the medium to the jet propagating through it. 

Jet shape is measured both by jet-by-jet functions of the jet constituent momentum, such as jet mass and width, and by inclusive and semi-inclusive measurements of intra- and inter-jet distributions, such as the ratio of jet yields measured with different $R$. The jet-by-jet jet shapes in particular are sensitive to the underlying soft physics. The first radial moment~\cite{Larkoski:2014pca}, or the angularity (girth or width), probes the radial energy profile of the jet. The jet mass, which is related to the second radial moment~\cite{Larkoski:2014pca}, captures the virtuality of the original parton that produced the jet and increases with increasing contribution of large-angle, typically soft particles~\cite{Majumder:2014gda}. These observables are complementary to measurements of the jet fragmentation using the longitudinal momentum fraction of the jet by ATLAS~\cite{Aaboud:2018hpb} and CMS~\cite{Chatrchyan:2014ava} that suggest that the jet energy in heavy-ion collisions is transferred to soft particles and wider angles inside the jet. 
ATLAS and CMS have also investigated the radial dependence of charged particles within a jet, demonstrating enhancement of softer particles at larger angles~\cite{Aad:2019igg, Chatrchyan:2013kwa} and enhanced yield of high-\pT{} charged particles in the jet core~\cite{Aad:2019igg}.

ALICE has measured jet angularity for charged-particle jets with $R=0.2$ in central \PbPb\ collisions at \sqrtsNN~=~2.76 TeV~\cite{Acharya:2018uvf}. The angularity is defined as 

\begin{equation}
    g = \sum_{i\in{\rm jet}} \frac{p_{\rm{T}, i}}{p_{\rm T, jet}} \Delta{R}_{\rm{jet}, i}\,\,,
\end{equation}

\noindent
where $p_{\rm{T}, i}$ is the transverse momentum of the i$^{\rm th}$ constituent and $\Delta{R}_{\rm{jet}, i}$ is the distance in ($\eta$, $\varphi$) space between the i$^{\rm th}$ constituent and the jet axis. Smaller values of $g$ correspond to
jets with more collinear fragmentation. Figure~\ref{Fig:Jetshape}, left panel, shows the jet-angularity distribution measured in central \PbPb\ collisions at \sqrtsNN~=~2.76 TeV compared to that for \pp\ collisions calculated with the PYTHIA6 event generator (Perugia 2011 tune)~\cite{Acharya:2018uvf}. The width is observed to be reduced in \PbPb\ collisions, which is consistent with jets in heavy-ion collisions being narrower or having harder fragments; this may arise for example from greater suppression of gluon jets relative to quark jets. 
This is qualitatively consistent with the observations in Fig.~\ref{Fig:JetSubstr}, right panel, showing narrowing of the groomed core of jets. This measurement is also compared to calculations based on the Hybrid Model~\cite{HybridModelResolution} and JETSCAPE~\cite{JETSCAPE:2020mzn} and favors an incoherent energy loss picture, similar to the measurement of the groomed jet radius. Comparison is also made to Hybrid Model calculations with and without medium recoil (referred to as a wake), with no difference observed between these alternative model components. This effect may arise because soft particles from the wake are generated at large angles to the jet centroid and therefore make little contribution to small radius jets.

ALICE has measured the jet mass for charged-particle jets with $R=0.4$ in central Pb--Pb collisions at $\sqrt{s_{\rm NN}} = 2.76$\,TeV~\cite{Acharya:2017goa}. The jet mass is defined as 
 
\begin{equation}
    M = \sqrt{E^{2} - p_{\rm T}^{2} - p_{\rm z}^{2}}\,\,,
\end{equation}

\noindent
where $E$ is the energy of the jet, $p_{\rm T}$ is the transverse and $p_{\rm z}$ is the longitudinal momentum of the jet. Figure~\ref{Fig:Jetshape}, right panel, shows the jet-mass distribution measured in central \PbPb\ collisions at \sqrtsNN~=~2.76 TeV, compared to model calculations~\cite{Acharya:2017goa}. The calculations include PYTHIA~\cite{Skands:2010ak}, which represents pp collisions without jet quenching,and  models with medium-induced jet energy loss: Hybrid Model~\cite{HybridModelResolution};  JEWEL~\cite{Zapp:2011ek, Zapp:2012ak}; and JETSCAPE~\cite{JETSCAPE:2021ehl}. The JEWEL and Hybrid Models are run both with and without medium recoil. 

The jet mass is observed to be larger with recoils present, which is consistent with the picture that the recoils add large-angle soft particles to the jet. The jet mass is however overestimated by both the Hybrid model and JEWEL with medium recoil, and is underestimated without the medium recoil, except in the case of the Hybrid model with coherent energy loss and no wake. Rather, it is consistent within systematic uncertainties with the PYTHIA simulation, suggesting no significant modification of the jet mass in heavy-ion collisions. Alternatively, this could be due to partial cancellation of in-medium effects; specifically, momentum broadening could lead to energy loss outside the jet and a smaller mass, and medium response could add soft particles inside the jet and increase the mass~\cite{KunnawalkamElayavalli:2017hxo}. %
Note that, in contrast, the jet angularity exhibits strong modification in Pb--Pb compared to \pp\ collisions, though with smaller jet resolution parameter ($R=0.2$ instead of $R=0.4$), which is expected to be less sensitive to the medium response. 
 
Additionally, ALICE has measured the N-subjettiness jet shape distribution~\cite{Acharya:2021ibn}, which quantifies the degree to which a jet corresponds to an N-pronged substructure~\cite{Thaler:2010tr}. Specifically, measurement of the 2-subjettiness to 1-subjettiness ratio ($\tau_{2}/\tau_{1}$) suggests a relative reduction in the rate of 2-prong jets for central \PbPb\ collisions. However, the absence of strong modification of $\tau_{2}/\tau_{1}$ suggests that medium induced radiation
is not sufficiently hard to produce an independent hard prong, similar
to the case observed for \zg{}.
The $\tau_{2}/\tau_{1}$ observable is only weakly correlated with other substructure observables discussed in this section, with a slightly stronger correlation with \zg{} than the other observables~\cite{Sirunyan:2018asm}, and thus adds independent information on in-medium jet substructure modification. 
 
 Jet substructure observables contain some degree of sensitivity to the same
 physics mechanisms as other jet observables, including ensemble-based measurements using inclusive or semi-inclusive jet distributions at different jet $R$~\cite{Aad:2012vca, Abelev:2013kqa,Adam:2015doa}. Figure~\ref{fig:hJetyield}, right panel, shows one such ratio, for semi-inclusive yields of jets recoiling from a high-\pT{} hadron for jet $R = 0.2$ and 0.5~\cite{Adam:2015doa}. No significant modification of the intra-jet energy profile for different jet $R$ is observed, indicating no significant in-medium transfer of energy to large angles.
In addition, the theoretical descriptions of jet substructure and (semi-)inclusive jet observables have 
 significant differences for jets in vacuum;
for instance, the jet shape (profile) requires significantly
larger soft power corrections than the inclusive jet cross section as a function
of $R$~\cite{Cal:2019hjc}.

\subsubsection*{\textbf{\textit {2.4.2.3 Jet multiple scattering and deflection}}}
\label{sec:Acoplanarity}

Medium-induced modification of the jet shower is studied via intrajet shapes and substructure, which are sensitive to the redistribution of jet momentum and constituents to wider angles. Such effects will likewise generate a change in the jet direction as a whole, which is explored in this section via the semi-inclusive azimuthal angular distributions of jets recoiling from a hadron trigger~\cite{Adam:2015doa,Adamczyk:2017yhe}.

In vacuum, the width of the azimuthal distribution of recoil jets in this observable arises largely from soft radiation (Sudakov radiation~\cite{Chen:2016vem}), while for jets in-medium, modification of the vacuum angular distribution arises from (inelastic) gluon emission and elastic scattering off the medium constituents. Measurements which disentangle vacuum and medium-induced azimuthal decorrelation effects provide direct sensitivity to  the transport coefficient $\hat{q}$. Since Sudakov radiation dominates at high jet \pT\ with respect to medium effects~\cite{Chen:2016vem}, measurement of the azimuthal decorrelation at low values of recoil \pTjet\ is desirable; however, measurements at low-\pTjet\ in heavy-ion collisions are challenging because of the large uncorrelated background, and new experimental approaches are needed.
The expected difference in the parametric dependence of jet energy loss and momentum broadening on the medium path length $L$~\cite{Gyulassy:2018qhr,Gyulassy:2020jlb} also motivates the simultaneous measurement of observables sensitive to both energy loss and momentum broadening, to discriminate between weakly- and strongly-coupled scenarios. 

Measurement of the rate of jet scattering to large angles with respect to the trigger axis may also provide evidence of weakly-coupled degrees of freedom within the strongly-coupled QGP (``quasi-particles''), analogous to the Rutherford scattering experiment that revealed the existence of the atomic nucleus~\cite{DEramo:2012uzl,DEramo:2018eoy}. The deflection of an energetic quark projectile in the QGP is expected to be Gaussian if the QGP is strongly coupled at all scales. However, QCD is asymptotically free and thus weakly-coupled quark and gluon quasi-particle degrees of freedom are expected to emerge when the QGP is probed at sufficiently short distances. The scattering off point-like quasi-particles will lead to a power-law tail in the momentum transfer ($1/k_{\rm T}^{4}$), the so-called Moliere scattering~\cite{DEramo:2012uzl,DEramo:2018eoy}. 
An excess of large-angle deflections observed in \PbPb\ relative to \pp\ collisions would be a direct observation of such quasi-particles in the QGP.

The large-angle scattering signal is expected to be small, however, requiring high experimental sensitivity to observe it. The \Drecoil\ observable (Eq.~\ref{eq:DeltaRecoil}) has several features crucial for such a high-sensitivity search: (i) absolute normalisation, rather than normalisation to the total number of pairs, meaning that each data-point is independently measured; (ii) fully data-driven correction for background; and (iii) full correction for the contribution of multiple partonic interactions, which are uncorrelated with the trigger and, therefore, generate an azimuthally uniform background~\cite{Adam:2015doa}.

\begin{figure}[htb]
\centering
\includegraphics[width=0.70\textwidth]{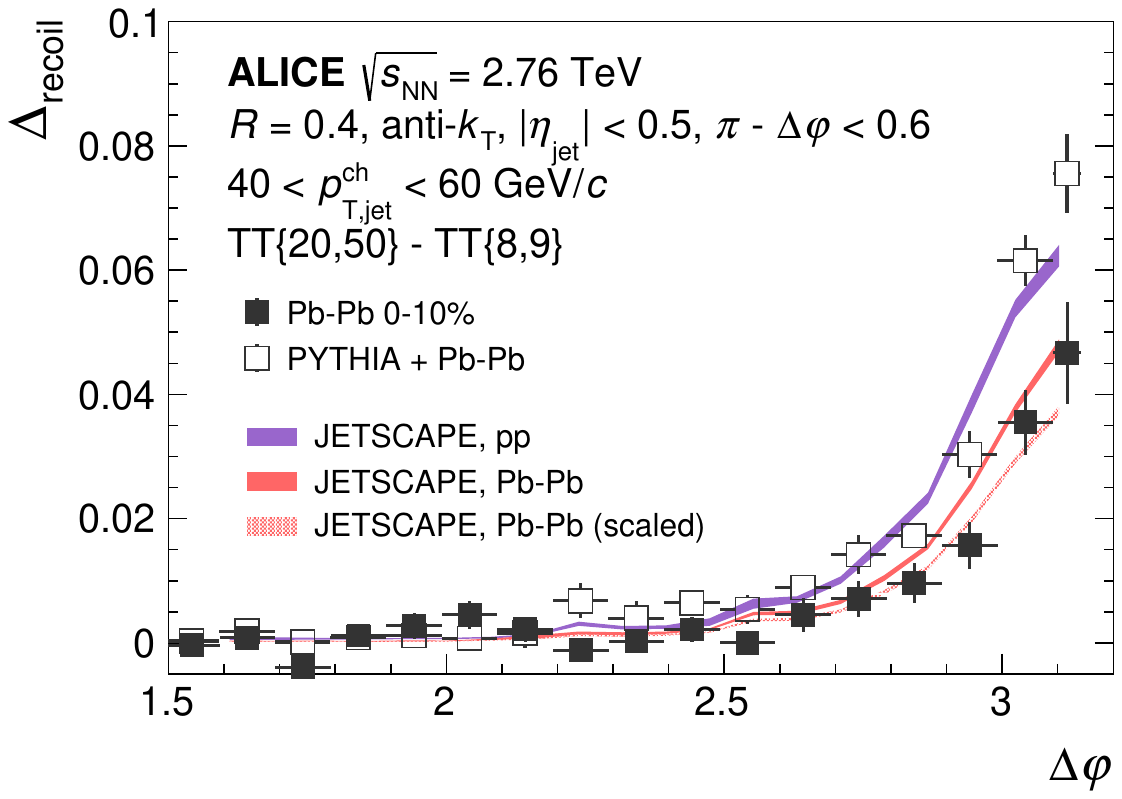}
\caption{ALICE measurement of h+jet acoplanarity in 0-10\% \PbPb~collisions at \sqrtsNN~=~2.76 TeV~\cite{Adam:2015doa} compared to an embedded PYTHIA+\PbPb~reference and JETSCAPE calculations, in pp and \PbPb~collisions for charged-particle recoil jets with $R=0.4$ and $40<\pTjet<60$~\gev. See text for discussion of JETSCAPE calculations.
}
\label{fig:hJetaco}
\end{figure}

Figure~\ref{fig:hJetaco} shows the first ALICE measurement of the azimuthal decorrelation in 0--10\% \PbPb~collisions at \sqrtsNN~=~2.76 TeV~\cite{Adam:2015doa}. The horizontal axis represents the azimuthal angle \dphi\ between a high-\pT\ charged hadron and the recoiling jets. The vertical axis corresponds to the \Drecoil\ observable (Eq.~\ref{eq:DeltaRecoil}) measured differentially in \dphi. The widths of the distributions in \PbPb\ data and in the vacuum reference, which in this case is a PYTHIA calculation for \pp\ collisions embedded into \PbPb\ data, are statistically compatible. The absolute yield of the \PbPb\ distribution is seen to be smaller than that of the \pp\ reference, indicating that the recoiling jet population is suppressed in \PbPb\ collisions
(see also Fig.~\ref{fig:hJetyield}).

For the selected kinematic cuts and jet resolution $R$, the strong energy loss is therefore not accompanied by a medium-induced acoplanarity within the statistical and systematic limits of our measurement. The change of recoil jet yield at large angles with respect to the PYTHIA reference was also studied, and within the current experimental uncertainties no evidence of large-angle scattering was found.

The figure also shows the result of JETSCAPE calculations which have been smeared to account for instrumental effects and background fluctuations for comparison, since these data have not been unfolded for such effects. The JETSCAPE calculation matches the distribution measured in \pp\ collisions well, and slightly overshoots the distribution measured in \PbPb\ collisions in the region near $\dphi\sim\pi$. In order to compare shapes directly as a function of \dphi, which could reveal quasi-particle effects, the JETSCAPE \PbPb\ distribution is also shown scaled to the same integral as the measured \PbPb\ distribution, which factors out the effect of an overall yield suppression. We note that the JETSCAPE calculation of jet quenching is based on the LBT model at low virtuality, which includes elastic scattering in the QGP. JETSCAPE therefore includes a specific implementation of Moliere scattering. The scaled JETSCAPE distribution reproduces the data well, likewise exhibiting no evidence of in-medium acoplanarity broadening or quasi-particle scattering.

The uncertainties in this measurement are dominated by the statistical uncertainty, so that future measurements with higher-statistics data sets will improve its precision. 
On the theoretical side, the parametrically different dependence in some models on in-medium path length of energy loss $\propto{L^2}$ and acoplanarity due to multiple scattering $\propto{L}$ suggest that simultaneous consideration of yield suppression and acoplanarity in data-to-model comparisons 
may provide significant discrimination of weakly- and strongly-coupled QGP~\cite{Gyulassy:2020jlb}. Extensions of the analysis to jets recoiling from different triggers (e.g. high-\pT\ photons/EW bosons) will provide complementary ways to study in-medium broadening.

\subsubsection{Conclusions}
\label{Sect:Summary}

\paragraph{Heavy-quark diffusion.} Differential studies of the yield suppression and flow of D mesons highlight the predominance of elastic charm quark-QGP interactions for \pT\ less than a few \gev. Heavy quarks in this range experience Brownian motion in the QGP, and therefore provide incisive probes of the equilibration process. This interaction is characterised by a spatial diffusion constant $D_{\rm s}$. Its value, $1.5 < 2 \pi D_{\rm s}(T) T < 4.5$,  is smaller than that determined from a pQCD calculation for a gas of weakly-coupled quarks and gluons, providing clear evidence in the heavy-flavour sector for a strongly-coupled QGP. Recent lattice QCD calculations~\cite{Altenkort:2023oms} give even slightly smaller values for $D_{\rm s}$.

\paragraph{Jet quenching: energy loss.} The magnitude of hadron suppression for $\pT>8$~\gev\ in central \PbPb\ collisions is the same for a wide range of light-flavour hadrons, providing definitive evidence that jet quenching occurs at the partonic level. This is likewise strong evidence for the existence of a dense, extended QGP phase. Suppression measurements in the heavy-flavour sector at intermediate-\pT\ indicate that beauty quarks lose less energy than charm quarks. These measurements are described by models that include mass-dependent elastic energy loss and a reduction of gluon radiation off heavy quarks. The latter mechanism is known as the QCD dead-cone effect and has been directly measured by ALICE for radiation in vacuum using pp collisions. Measurements of reconstructed jets provide unique insight into the jet--medium interaction. Since jets subtend finite area, such modifications reflect the spatial distribution of medium-induced energy loss and the medium response. Inclusive jet suppression has been observed up to a \pT\ of several hundred \gev. Jet energy loss has been measured for the semi-inclusive distribution of jets recoiling from a hadron trigger to be ($8\pm2$)~\gev\ for central \PbPb\ collisions. This value is larger than that determined from similar analyses at RHIC~\cite{Sahoo:2020kwh}, though the comparison currently has limited significance. 

\paragraph{Jet quenching: substructure modification.} The hard substructure of jets, measured using groomed observables, shows a strong angular modification -- jets with wide substructure are suppressed -- but weak momentum modification ($z_{\rm g}$). 
Measurements of soft jet substructure, which are carried out using ungroomed observables, exhibit effects that are consistent with the suppression of wide-angle radiation due to quenching. Soft jet substructure also probes the dynamics of soft radiation in the jet, in particular the contribution of medium response, that cannot be calculated using purely perturbative techniques.
Such groomed and ungroomed approaches are therefore complementary, and they show that the medium resolves hard substructure and promotes narrower spatial structures. Theoretical calculations suggest that this may be due to colour decoherence, i.e. the gradual loss of the coherence of radiation emitted by a colour antenna while propagating through a dense medium, or to modification of the quark-to-gluon fraction. The space--time structure of jets also plays a role, since wider splittings may occur earlier and their products therefore traverse longer paths in the QGP.

\paragraph{Jet quenching: acoplanarity.} The dijet acoplanarity, which has direct sensitivity to the transport coefficient \qhat\ and large-angle scattering off quasiparticles, has been measured via hadron+jet correlations. No angular modifications have been observed within the currently statistics-dominated uncertainties. These measurements will be extended with larger data samples and kinematic coverage and this approach remains a promising tool to explore the nature of the QGP constituents.

\newpage

\newcommand{\pp}           {pp\xspace}
\newcommand{\ppbar}        {\mbox{$\mathrm {p\overline{p}}$}\xspace}
\newcommand{\XeXe}         {\mbox{Xe--Xe}\xspace}
\newcommand{\PbPb}         {\mbox{Pb--Pb}\xspace}
\newcommand{\pA}           {\mbox{pA}\xspace}
\newcommand{\pPb}          {\mbox{p--Pb}\xspace}
\newcommand{\Pbp}          {\mbox{Pb--p}\xspace}
\newcommand{\AuAu}         {\mbox{Au--Au}\xspace}
\newcommand{\dAu}          {\mbox{d--Au}\xspace}

\newcommand{\s}            {\ensuremath{\sqrt{s}}\xspace}
\newcommand{\snn}          {\ensuremath{\sqrt{s_{\mathrm{NN}}}}\xspace}
\newcommand{\pt}           {\ensuremath{p_{\rm T}}\xspace}
\newcommand{\meanpt}       {$\langle p_{\mathrm{T}}\rangle$\xspace}
\newcommand{\minv}         {\ensuremath{m_{\rm inv.}}\xspace}
\newcommand{\ycms}         {\ensuremath{y_{\rm CMS}}\xspace}
\newcommand{\ylab}         {\ensuremath{y_{\rm lab}}\xspace}
\newcommand{\etarange}[1]  {\mbox{$\left | \eta \right |~<~#1$}}
\newcommand{\yrange}[1]    {\mbox{$\left | y \right |~<~#1$}}
\newcommand{\dndy}         {\ensuremath{\mathrm{d}N_\mathrm{ch}/\mathrm{d}y}\xspace}
\newcommand{\dndeta}       {\ensuremath{\mathrm{d}N_\mathrm{ch}/\mathrm{d}\eta}\xspace}
\newcommand{\avdndeta}     {\ensuremath{\langle\dndeta\rangle}\xspace}
\newcommand{\dNdy}         {\ensuremath{\mathrm{d}N_\mathrm{ch}/\mathrm{d}y}\xspace}
\newcommand{\Npart}        {\ensuremath{N_\mathrm{part}}\xspace}
\newcommand{\Ncoll}        {\ensuremath{N_\mathrm{coll}}\xspace}
\newcommand{\dEdx}         {\ensuremath{\textrm{d}E/\textrm{d}x}\xspace}
\newcommand{\RpPb}         {\ensuremath{R_{\rm pPb}}\xspace}
\newcommand{\mpt}          {\ensuremath{\langle p_{\rm T}\rangle}\xspace}
\newcommand{\mptsquared}   {\ensuremath{\langle p^2_{\rm T}\rangle}\xspace}
\newcommand{\RAA}          {\ensuremath{R_{\rm AA}}\xspace}
\newcommand{\RpA}          {\ensuremath{R_{\rm pA}}\xspace}

\newcommand{\nineH}        {$\sqrt{s}~=~0.9$~Te\kern-.1emV\xspace}
\newcommand{\seven}        {$\sqrt{s}~=~7$~Te\kern-.1emV\xspace}
\newcommand{\twoH}         {$\sqrt{s}~=~0.2$~Te\kern-.1emV\xspace}
\newcommand{\twosevensix}  {$\sqrt{s}~=~2.76$~Te\kern-.1emV\xspace}
\newcommand{\five}         {$\sqrt{s}~=~5.02$~Te\kern-.1emV\xspace}
\newcommand{\thirteen}     {$\sqrt{s}~=~13$~Te\kern-.1emV\xspace}
\newcommand{\twosevensixnn}{$\sqrt{s_{\mathrm{NN}}}~=~2.76$~Te\kern-.1emV\xspace}
\newcommand{\fivenn}       {$\sqrt{s_{\mathrm{NN}}}~=~5.02$~Te\kern-.1emV\xspace}
\newcommand{\eightnn}       {$\sqrt{s_{\mathrm{NN}}}~=~8.16$~Te\kern-.1emV\xspace}
\newcommand{\LT}           {L{\'e}vy-Tsallis\xspace}
\newcommand{\GeVc}         {Ge\kern-.1emV/$c$\xspace}
\newcommand{\MeVc}         {Me\kern-.1emV/$c$\xspace}
\newcommand{\TeV}          {Te\kern-.1emV\xspace}
\newcommand{\GeV}          {Ge\kern-.1emV\xspace}
\newcommand{\MeV}          {Me\kern-.1emV\xspace}
\newcommand{\GeVmass}      {Ge\kern-.2emV/$c^2$\xspace}
\newcommand{\MeVmass}      {Me\kern-.2emV/$c^2$\xspace}
\newcommand{\lumi}         {\ensuremath{\mathcal{L}}\xspace}

\newcommand{\ITS}          {\rm{ITS}\xspace}
\newcommand{\TOF}          {\rm{TOF}\xspace}
\newcommand{\ZDC}          {\rm{ZDC}\xspace}
\newcommand{\ZDCs}         {\rm{ZDCs}\xspace}
\newcommand{\ZNA}          {\rm{ZNA}\xspace}
\newcommand{\ZNC}          {\rm{ZNC}\xspace}
\newcommand{\SPD}          {\rm{SPD}\xspace}
\newcommand{\SDD}          {\rm{SDD}\xspace}
\newcommand{\SSD}          {\rm{SSD}\xspace}
\newcommand{\TPC}          {\rm{TPC}\xspace}
\newcommand{\TRD}          {\rm{TRD}\xspace}
\newcommand{\VZERO}        {\rm{V0}\xspace}
\newcommand{\VZEROA}       {\rm{V0A}\xspace}
\newcommand{\VZEROC}       {\rm{V0C}\xspace}
\newcommand{\Vdecay} 	   {\ensuremath{V^{0}}\xspace}

\newcommand{\jpsi}         {\ensuremath{\text{J}/\psi}\xspace}
\newcommand{\psiprime}     {\ensuremath{\psi(2\text{S})}\xspace}
\newcommand{\ccbar}        {\ensuremath{\text{c}\overline{\text{c}}}\xspace}
\newcommand{\bbbar}        {\ensuremath{\text{b}\overline{\text{b}}}\xspace}
\newcommand{\ee}           {\ensuremath{\text{e}^{+}\text{e}^{-}}\xspace} 
\newcommand{\mumu}           {\ensuremath{\mu^{+}\mu^{-}}\xspace} 
\newcommand{\pip}          {\ensuremath{\pi^{+}}\xspace}
\newcommand{\pim}          {\ensuremath{\pi^{-}}\xspace}
\newcommand{\kap}          {\ensuremath{\rm{K}^{+}}\xspace}
\newcommand{\kam}          {\ensuremath{\rm{K}^{-}}\xspace}
\newcommand{\pbar}         {\ensuremath{\rm\overline{p}}\xspace}
\newcommand{\kzero}        {\ensuremath{{\rm K}^{0}_{\rm{S}}}\xspace}
\newcommand{\lmb}          {\ensuremath{\Lambda}\xspace}
\newcommand{\almb}         {\ensuremath{\overline{\Lambda}}\xspace}
\newcommand{\Om}           {\ensuremath{\Omega^-}\xspace}
\newcommand{\Mo}           {\ensuremath{\overline{\Omega}^+}\xspace}
\newcommand{\X}            {\ensuremath{\Xi^-}\xspace}
\newcommand{\Ix}           {\ensuremath{\overline{\Xi}^+}\xspace}
\newcommand{\Xis}          {\ensuremath{\Xi^{\pm}}\xspace}
\newcommand{\Oms}          {\ensuremath{\Omega^{\pm}}\xspace}
\newcommand{\degree}       {\ensuremath{^{\rm o}}\xspace}
\newcommand{\upsone}         {\ensuremath{\Upsilon(1\text{S})}\xspace}
\newcommand{\upstwo}         {\ensuremath{\Upsilon(2\text{S})}\xspace}
\newcommand{\upsthree}         {\ensuremath{\Upsilon(3\text{S})}\xspace}
\newcommand{\rmJpsi}    {\mbox{$\mathrm{J\kern-0.05em /\kern-0.05em\psi}$}}
\newcommand{\velip}         {\ensuremath{v_{\rm 2}}\xspace}

\newcommand{\note}[1]{{\color{red}\textbf{#1}}}

\subsection{Deconfinement and modification of the QCD force}
\label{sec:Quarkonium}

Heavy quarkonia, the bound states of a \ccbar (charmonium) or \bbbar pair (bottomonium), have been the subject of intense studies  since their discovery in the '70s. The investigation, in the frame of QCD, of their production processes, of the rich spectroscopy of the various states, and of their decay modes is a lively field until today, and although great progress has been  accomplished, a complete understanding of their properties is still to be reached (for a general review see~\cite{Brambilla:2010cs}).

Quarkonium states also represent a very important tool for the study of the QGP and of its properties (for a recent review see~\cite{Rothkopf:2019ipj}). It was early realised that the binding of the heavy-quark pair can be affected to various extents when quarkonia are immersed in a deconfined medium. The high density of free colour charges in the QGP leads to a screening of the QCD force and ultimately to the dissolution of the quarkonium~\cite{Matsui:1986dk}. This simple but profound idea has led to a wealth of theoretical and experimental studies that have revealed new and somewhat unexpected effects.
Early studies focused on establishing a direct connection between the suppression of the quarkonium states and the temperature of the deconfined phase~\cite{Digal:2001ue}. The rich spectroscopic structure of quarkonia, with binding energies varying from a few MeV (\psiprime) to more than 1 GeV (\upsone), may lead to a ``sequential suppression'' with increasing temperature, with the more strongly bound states surviving up to a dissociation temperature $T_{\rm diss}\sim 2 T_{\rm pc}$ and the weakly bound states melting at temperatures close to $T_{\rm pc}$. In nuclear collisions, the temperature of the QGP can in principle be varied by selecting the centrality of the collision or its energy. If the ``melting temperature'' of each state could be precisely singled out by lattice QCD studies~\cite{Asakawa:2000tr,Burnier:2013nla,Aarts:2014cda,Kim:2018yhk,Larsen:2019zqv}, quarkonium would represent an ideal thermometer for the medium. 

The above considerations are valid in a static picture of %
the medium in which the quarkonium states are assumed to be immersed.
When moving to a study of the dynamics of the bound states
and their interaction with an evolving medium, several effects lead to a more complex description. In particular, the formation of the quarkonium states is a multi-stage process~\cite{Bodwin:1994jh} (production of the q$\overline {\rm q}$ pair and formation of the bound state) that spans over a time covering a significant fraction of the collision history. In addition, the quarkonium potential, as calculated at $T>0$ in effective field theories, has also an imaginary part, corresponding to the collisional damping of the states, which leads to a loss of correlation in the pair and consequently to an in-medium modification of the spectral functions~\cite{Laine:2006ns}. Furthermore, in a system with a high multiplicity of heavy quarks, the effects related to the combination of uncorrelated pairs originated from different hard scattering processes, or the recombination of previously destroyed ones, can lead to a significant increase of the quarkonium yields~\cite{Thews:2000rj,BraunMunzinger:2000px}, counterbalancing the suppression. If a partial or full kinetic equilibration of the deconfined heavy quarks in the medium~\cite{Zhou:2014kka} takes place, collective flow effects can be inherited by quarkonia produced in the (re)combination process~\footnote{The terms (re)combination and (re)generation are both used to describe this process}. Finally, also in the hadronic stage of the collisions the yields might be altered, due to quarkonium-meson break-up effects which could play a role in particular for weakly bound states~\cite{Capella:2007jv,Maiani:2004qj}. 
The theoretical treatment of quarkonium production in the hot QCD medium is multi-faceted, comprising statistical hadronisation, transport models, hydrodynamics and the recently emerging approach through quantum dynamics~\cite{Rothkopf:2019ipj,Akamatsu:2020ypb}.

At the LHC start-up, quarkonium production in nuclear collisions, and in particular \jpsi studies, had already been pursued for many years in the frame of QGP-related studies. Early experiments at SPS energies resulted in high precision results on the ratio between the \jpsi and the Drell-Yan yields, with the latter used as a reference process because of its electromagnetic nature. In particular, the NA50 collaboration had shown a suppression effect of  $\sim 30$\% for central \PbPb collisions at \snn= 17.3 \GeV~\cite{Abreu:2000ni,Alessandro:2004ap}, with respect to the expected size of cold nuclear matter effects, evaluated via p--A studies at the same centre-of-mass energy by the NA60 experiment~\cite{Arnaldi:2010ky}. The size of the suppression  approximately corresponds to the feed-down to \jpsi from $\chi_{\rm c}$ and \psiprime states~\cite{Lansberg:2019adr}, leading to the interpretation of the measured effect as due to the melting of these weakly bound states in the medium~\cite{Digal:2001ue}. Indeed a hierarchy between the \jpsi and \psiprime, with the latter experiencing stronger suppression, was  observed~\cite{Abreu:2000ni,Alessandro:2004ap,Alessandro:2006ju}. The exact nature of this medium, either a deconfined phase or a dense hadron gas, sparked a considerable controversy, due to competing theoretical approaches being able to fairly reproduce the data~\cite{Kharzeev:1996yx,Blaizot:1996nq,Capella:2000zp}.

The \jpsi studies received a second considerable boost with the availability of RHIC data for Au--Au collisions at \snn= 200 \GeV by PHENIX~\cite{Adare:2006ns,Adare:2014hje}, with important contributions also from STAR~\cite{Adamczyk:2012ey,Adamczyk:2013tvk,Adam:2019rbk}. At RHIC, the charmonium suppression was evaluated via the study of the nuclear modification factor, as the Drell-Yan contribution becomes increasingly negligible with respect to competing sources such as semileptonic heavy-quark decays. The RHIC measurements featured a similar suppression level to that observed at SPS energies and a significantly larger suppression in the forward ($1.2<|y|<2.2)$ than in the central rapidity region ($|y|<0.35$). %
Detailed studies were performed to interpret the RHIC results, with a possible explanation assuming that the $\it direct$ suppression of the tightly bound J/$\psi$ is partly compensated by a significant (re)generation effect~\cite{Capella:2007jv,Zhao:2007hh,Thews:2005vj,Sharma:2012dy,Yan:2006ve}.
However, the rather complex energy dependent interplay between hot and cold matter effects~\cite{Adare:2012wf,Adare:2007gn,Acharya:2019zjt} does not allow for a significant conclusion on the magnitude of either the \jpsi suppression or (re)generation.    

In this situation, the studies at LHC energies are of paramount importance to settle these ambiguities in the interpretation of the results. The very large charm-quark multiplicity ($\sim 100$ \ccbar  pairs in a central \PbPb collision at $\sqrt{s_{\rm NN}}=5.02$~TeV, a $>10$ factor larger with respect to central Au--Au collisions at top RHIC energy) provides a test for the (re)generation mechanism which, if unambiguously demonstrated, would implicitly provide a strong evidence for the existence of a deconfined phase of nuclear matter, implying  that coloured partons can roam freely over distances much larger than the hadronic scale. 

Another fundamental area where LHC is expected to provide a decisive step forward is the study of the bottomonium sector, since the increase of the production cross sections with collision energy make such a measurement feasible with good enough statistical precision. 
The binding energies of the bottomonium vector states range from $\sim 1.1$ down to about 0.2 \GeV and consequently $\Upsilon({\rm 1S,2S,3S})$ represent an excellent test for the sequential suppression mechanism~\cite{Satz:2005hx}. The (re)generation component, contrary to charmonia, is expected to be small, due to the much lower multiplicity of b${\rm{\overline b}}$ pairs (5--10 in central Pb--Pb collisions) compared to  c${\rm{\overline c}}$.  
In addition, bottomonia are very promising from a theoretical point of view, thanks to a much more pronounced separation of scales between the heavy-quark mass, the typical bottomonium size and the radial or orbital
angular-momentum excitation energy ($M \gg Mv \gg Mv^2$, where M is the heavy-quark mass and $v$ the relative velocity of the pair in the bound state). That makes them good candidates for a direct application of effective field theory approaches~\cite{Rothkopf:2019ipj}. 

Finally, at lower energies, competing non-QGP effects like quarkonium break-up by the nucleons of the colliding nuclei played a significant role in the description of the results~\cite{Alessandro:2003pi}. At the LHC, due to the extremely short crossing time of the nuclei, the only sizeable effect not related to the medium is represented by nuclear modification of gluon densities (shadowing~\cite{Kovarik:2015cma,Eskola:2016oht,AbdulKhalek:2019mzd} or by CGC-related effects~\cite{Fujii:2013gxa,Ma:2015sia,Ducloue:2015gfa}), or coherent energy loss~\cite{Arleo:2012hn}. These effects can be investigated in the study of \pPb collisions. 

In the following sections we review the main results obtained by ALICE for various charmonium and bottomonium states, comparing them to experimental results at lower energies and to various theory calculations.

\subsubsection{Study of the charmonium ground state: evidence for the (re)generation and demonstration of deconfinement at LHC energies}
\label{sec:charmonia}

As outlined in the previous section, results at LHC energies were expected to clarify, in the charmonium sector, the presence of a (re)generation mechanism related to (re)combination of deconfined charm quarks. In ALICE, charmonia are measured at both central ($|y|<0.9$) and forward rapidity ($2.5<y<4)$, and down to zero \pt. This phase space coverage is ideal for such studies because the amount of (re)generation is expected to depend on rapidity and \pt, with a stronger effect around $y=0$ and at small \pt, due to the higher charm-quark multiplicity in these kinematic regions. It should be remarked that the results shown in these sections refer to inclusive production, i.e., the contribution of decays of hadrons containing a b quark was not subtracted. Such a contribution has a small effect on the nuclear modification factor at moderate \pt ($<10$\% for $p_{\rm T}<5 $ GeV/$c$)~\cite{Adam:2015isa} and increases at higher \pt~\cite{Abelev:2012rv}.

\begin{figure}[ht]
\begin{center}
\includegraphics[width=0.69\linewidth]{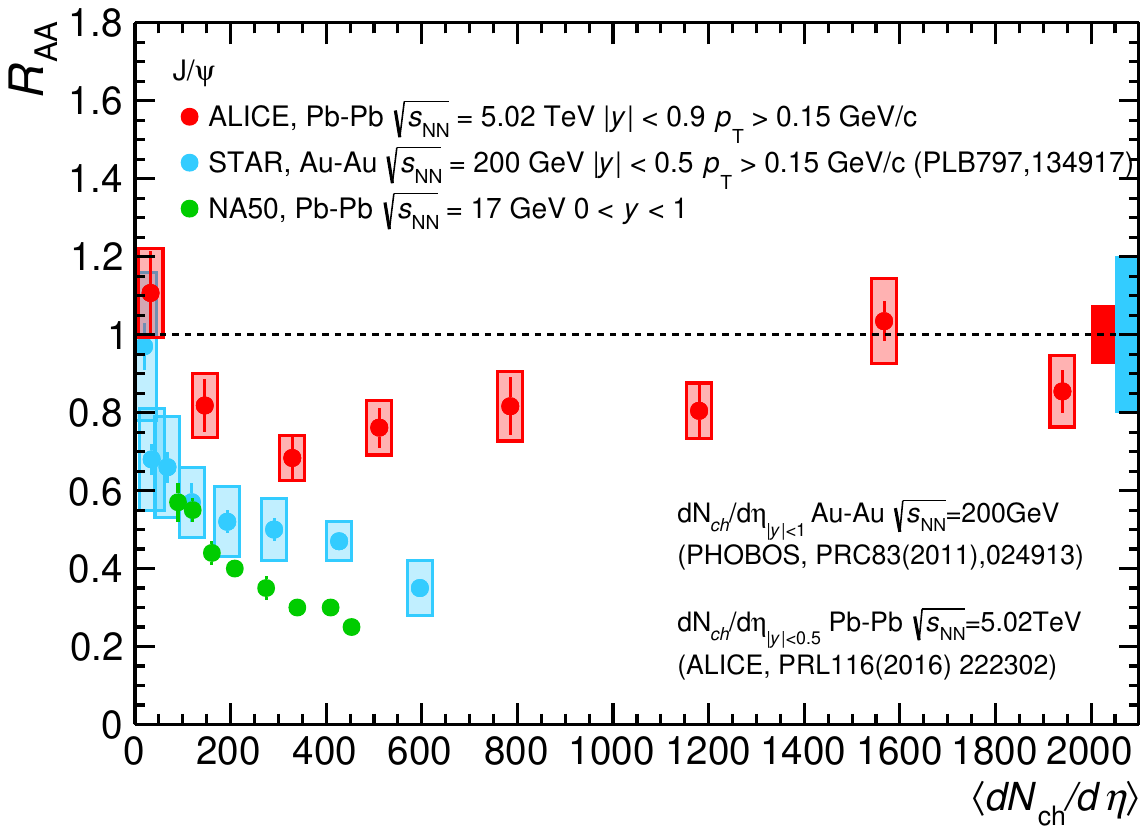}
\includegraphics[width=0.69\linewidth]{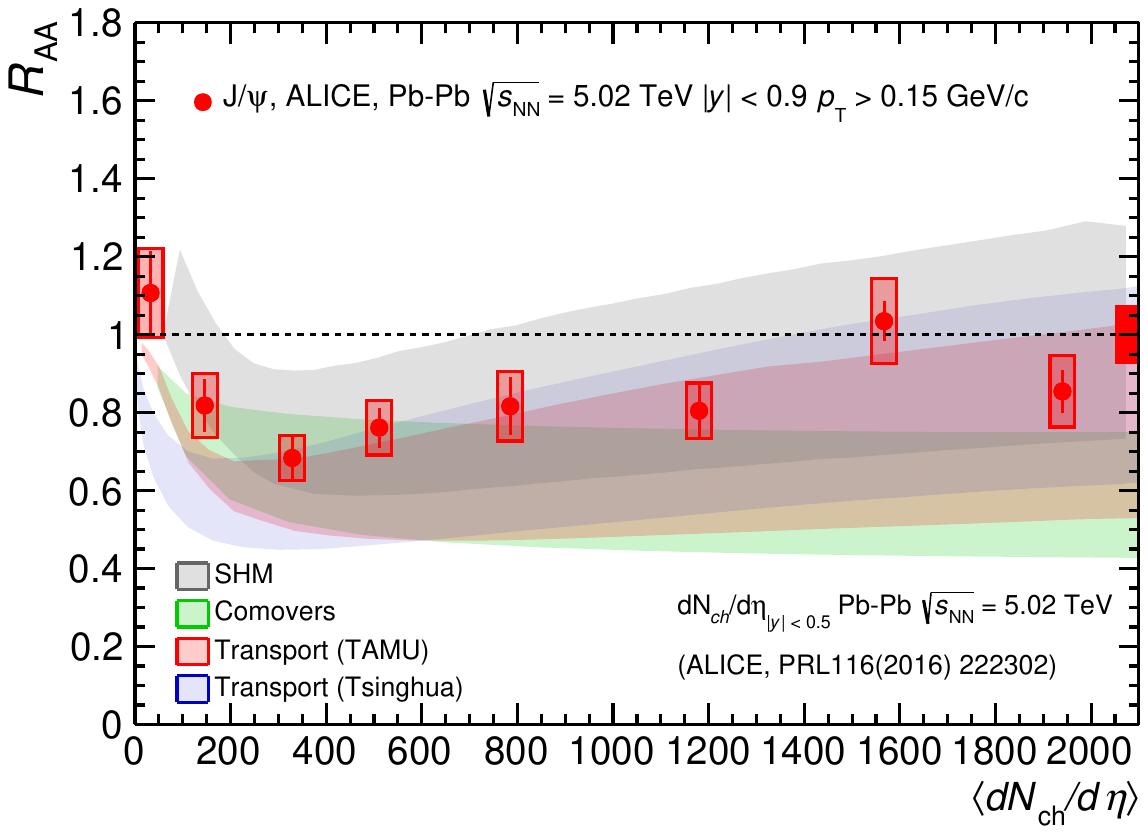}  
\caption{The nuclear modification factor as a function of the charged hadron multiplicity. For ALICE and STAR results, the selection $p_{\rm T}>0.15$ GeV/$c$ minimises the contribution of photoproduced J/$\psi$ ~\cite{Adam:2015gba}. (Top) Comparison between SPS (NA50)~\cite{Alessandro:2004ap}, RHIC (STAR)~\cite{Adam:2019rbk} and LHC (ALICE)~\cite{ALICE:Jpsieefwref} results; (bottom) the ALICE results compared with theoretical calculations (SHM~\cite{Andronic:2019wva}, Comovers~\cite{Ferreiro:2012rq}, TM-TAMU~\cite{Du:2015wha}, TM-Tsinghua~\cite{Zhou:2014kka}). 
}
\label{fig:RAAJpsivsdndeta} 
\end{center}
\end{figure}

One of the main results on \jpsi production is shown in the top panel of Fig.~\ref{fig:RAAJpsivsdndeta}, where the nuclear modification factor for inclusive \jpsi measured by ALICE in \PbPb collisions, at \fivenn at central rapidity in the \ee decay channel, is shown as a function of the charged hadron pseudorapidity density \avdndeta at midrapidity~\cite{ALICE:Jpsieefwref}.
The latter quantity is directly related, for a certain collision system at a given energy, to the centrality of the interaction and, for different collision systems, is roughly proportional to the initial energy density~\cite{Bjorken:1982qr}.
The ALICE result is compared with the \RAA values measured at RHIC by STAR (Au--Au at \snn= 0.2 \TeV) in $|y|<0.5$~\cite{Adam:2019rbk} and those obtained from NA50 results at the SPS in \PbPb at \snn= 0.017 \TeV for $0<y<1$~\cite{Alessandro:2004ap}, using a pp reference cross section extrapolated from NA60 \pA measurements ranging from p--Be to p--U~\cite{Arnaldi:2010ky}. The selection $p_{\rm T}>0.15 $ GeV/$c$ is meant to remove a non-negligible contribution due to \jpsi photoproduction in peripheral hadronic Pb--Pb collisions~\cite{Adam:2015gba}. The prominent feature of this set of results is the strong decrease of the \jpsi suppression moving from low- to high-energy experiments and, at the LHC, the disappearance of suppression effects when going towards central collisions. Both effects provide a strong indication of the presence of  (re)generation effects on the \jpsi. Similar observations were first carried out by ALICE at forward rapidity~\cite{Abelev:2012rv}. As discussed later in Sec.~\ref{sec:2.5nonQGP}, the contribution of non-QGP effects, in particular of nuclear shadowing, does not alter this conclusion.

A more quantitative assessment of the physics mechanisms at play requires a comparison of ALICE results with models. In the bottom panel of Fig.~\ref{fig:RAAJpsivsdndeta}, calculations of \RAA carried out in the frame of transport models (TM), of a comover model  and of a statistical hadronisation model (SHM) are presented. The transport models (TM-TAMU~\cite{Du:2015wha},TM-Tsinghua~\cite{Zhou:2014kka}) rely on the solutions of macroscopic transport equations. Dissociation/(re)generation thermal rates for quarkonium states in the QGP are calculated taking into account a lattice-QCD inspired evaluation of the dependence of their spectral properties on the evolving thermodynamical properties of the medium. The two models discussed here mainly differ in the choice of the rate equation and of the open charm cross section. In the comover model~\cite{Ferreiro:2012rq}, the
scattering of the nascent charmonium states with ``comovers'' (partonic or hadronic) produced in the same kinematic region is at the origin of the suppression, with the comover density being tuned on the measured hadron yields. The dissociation cross section, extracted from results at lower energy, is assumed to be energy independent and the (re)generation effects are implemented through a gain and loss differential transport equation. Finally, in the SHM~\cite{Andronic:2019wva}, charmonium yields are assumed to be determined at chemical freeze-out according to their statistical weights, introducing a charm fugacity factor related to charm conservation and obtained from the charm production cross section. The comparison between data and models shows a fair agreement with most of the approaches, with TM-TAMU and SHM giving the best reproduction of the \avdndeta dependence. The uncertainties of the models are clearly large when compared to the experimental data, mainly as a direct consequence of the uncertainty on the total charm cross section for \PbPb collisions, which is a fundamental input of the calculations. Getting a precise evaluation of this quantity is not easy, due to the problem of measuring open charm mesons and baryons down to zero \pt in nuclear collisions. Recent measurements by ALICE~\cite{ALICE:2021dhb} have shown strong modifications of the fragmentation fractions of the c quark to the various final states from \ee to pp collisions and allowed a fairly precise ($\sim 10$\% uncertainty) estimate of the total charm cross section for the latter collision system (see also Sec.~\ref{sec:SHM} and~\ref{sec:open-heavy-flavor} for more details on the measurements).
However, the extrapolation from pp collisions is not trivial as it involves an estimate of the non-negligible shadowing effects on the initial state~\cite{Adam:2015iga}. 

\begin{figure}[ht]
\begin{center}
\includegraphics[width=0.65\linewidth]{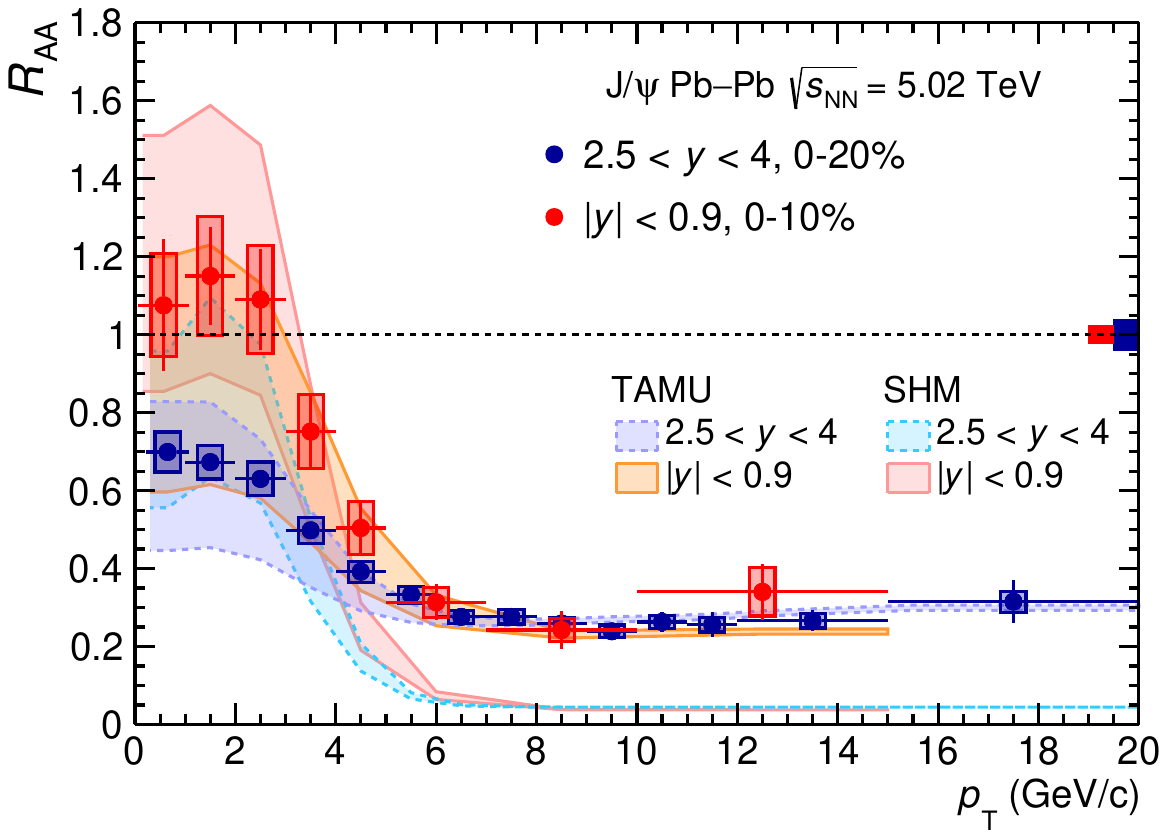}
\hbox{\hspace{0.4cm}\includegraphics[width=0.76\linewidth]{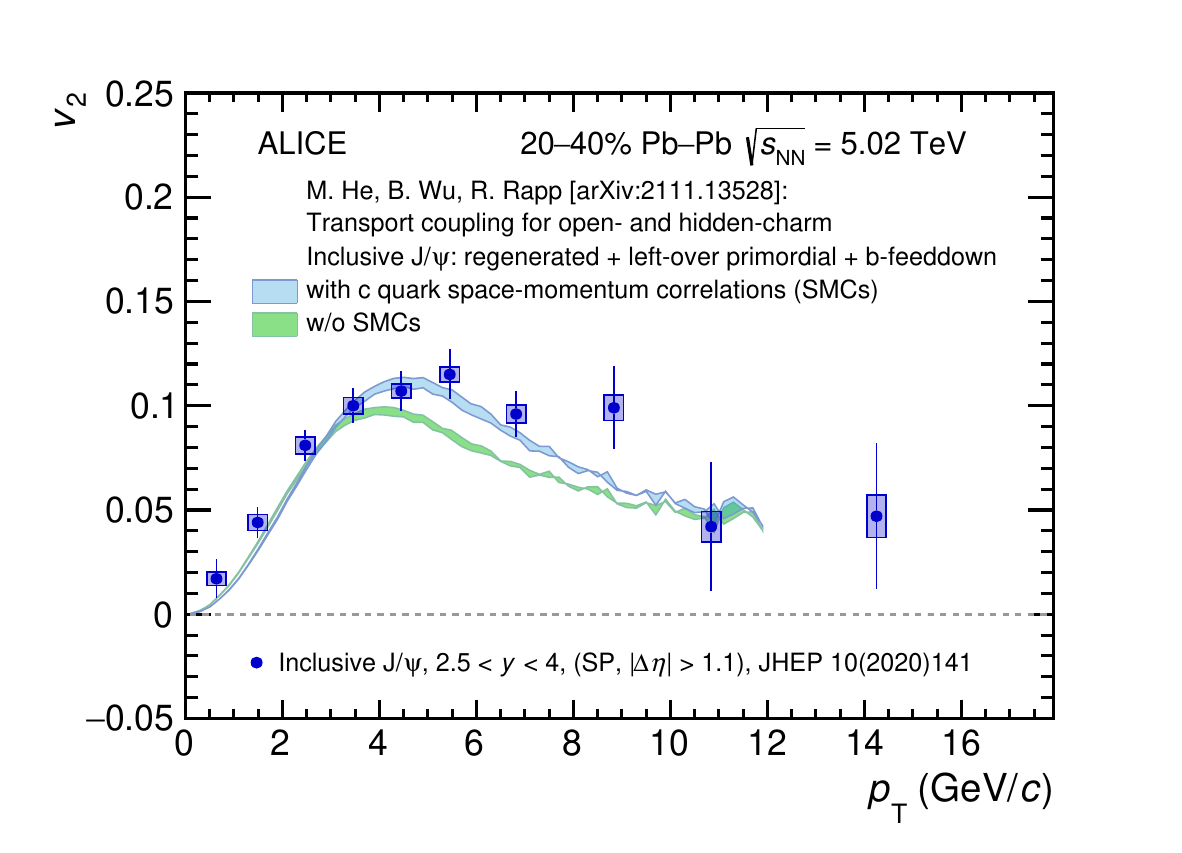}} 
\caption{(Top) The nuclear modification factor as a function of the transverse momentum, for the central ($|y|<0.9$) and forward ($2.5<y<4$) rapidity regions~\cite{ALICE:Jpsieefwref}, compared to TM-TAMU and SHM model calculations.
(Bottom) The $p_{\rm T}$ dependence of the inclusive J/$\psi$ $v_{\rm 2}$, compared to TM-TAMU model calculations~\cite{Acharya:2020jil,He:2021zej}. }
\label{fig:RAAv2Jpsivspt} 
\end{center}
\end{figure} 

Other valuable insights on the mechanisms at play can be obtained by considering the \pt dependence of the \RAA and \velip of the  \jpsi. In the top panel of Fig.~\ref{fig:RAAv2Jpsivspt}  results on the nuclear modification factor are presented for the two rapidity ranges accessed by ALICE, $|y|<0.9$ (\ee decay) and $2.5<y<4$ (\mumu decay), for central \PbPb events~\cite{ALICE:Jpsieefwref}.
The significant rise of $R_{\rm AA}$ at low-\pt, where the bulk of charm production occurs, with this effect being stronger at midrapidity, 
strongly hints at the presence of (re)generation as a dominant mechanism in this transverse momentum range. Remarkably, at higher \pt ($>8-10$ GeV/$c$) there is no rapidity dependence for the strong suppression, and the \RAA values become compatible with those 
measured for charged hadrons (see Fig.~\ref{Fig:RAAhadrons} in Sec.~\ref{sec:JetsMatter}). This observation may suggest parton energy loss, as in the case of \jpsi production from the splitting of a hard gluon, as a significant effect in this kinematic range~\cite{Spousta:2016agr,Arleo:2017ntr}, even if a quantitative description inspired to NRQCD underestimate the measured $R_{\rm AA}$~\cite{Makris:2019ttx} . 
In the bottom panel of Fig.~\ref{fig:RAAv2Jpsivspt} the \pt dependence of \velip for \PbPb collisions in the centrality interval 20--40\% is shown~\cite{Acharya:2020jil}. The measurement is performed using the scalar product method with a pseudorapidity gap $|\Delta\eta|>1.1$ between the \jpsi and the event flow vector, to suppress non-flow effects. A large \velip reaching $\sim 0.1$ at $p_{\rm T}\sim 5$ \GeVc is observed, which could be explained if a large fraction of the detected \jpsi originates from the (re)combination of charm quarks, which acquire their anisotropy by taking part in the collective expansion of the system. At the same time, the large \jpsi \velip values, according to Ref.~\cite{Du:2015wha}, favour a late \jpsi formation time, since the charm-quark anisotropy requires time to build up.

The results of Fig.~\ref{fig:RAAv2Jpsivspt} are compared with theoretical calculations from TM-TAMU and SHM (only \RAA for the latter). For \RAA, the models  reproduce the different size of the rise at low-\pt. At large \pt only TM-TAMU is in agreement with data, due to the absence in the SHM of additional production mechanisms such as \jpsi production from gluon fragmentation in jet, which are expected to contribute in that range. In the high-\pt range the slight rise of \RAA in TM-TAMU is mainly due to the longer resonance formation time, that is Lorentz-dilated, coupled to the expected lower suppression rate for the pre-resonant state~\cite{Zhao:2007hh,Sharma:2012dy}.  For \velip, TM-TAMU is able to reproduce the data over the explored \pt range. In the low- and intermediate-\pt region, where the \jpsi is produced mostly via (re)combination, space-momentum correlations of the diffusing charm and anti-charm quarks in the hydrodynamically expanding fireball represent an important ingredient of the model~\cite{He:2021zej}.
At high-\pt, where the \velip is likely determined by path-length dependent effects~\cite{Noronha-Hostler:2016eow}, the calculations describe the data within the large statistical uncertainties. It is worth noting that no energy loss effects are implemented in the TM-TAMU calculations, so the positive \velip in the high-\pt region can only be originated in this model from path-length dependent dissociation.

\subsubsection{Study of the bottomonium ground state: strong suppression and small (re)generation effects}
\label{sec:bottgrstate}

\begin{figure}[ht!]
\begin{center}
\includegraphics[width=0.7\linewidth]{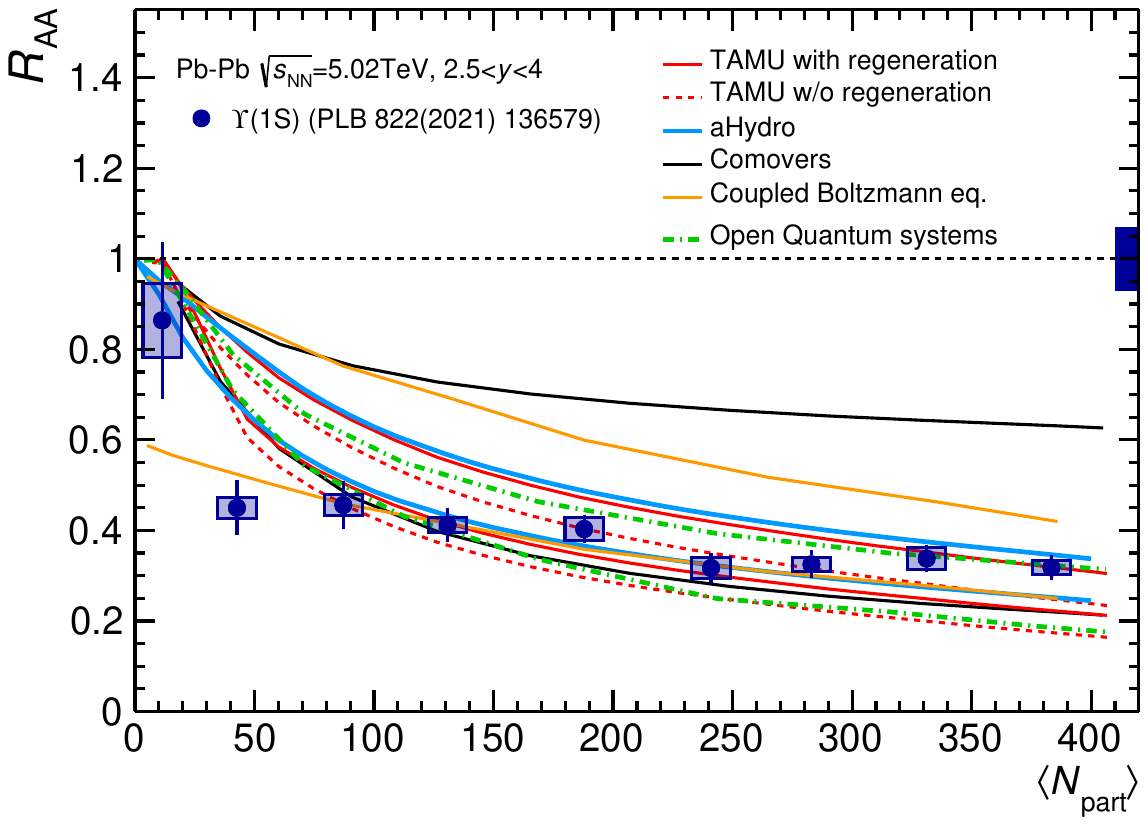}
\caption{The nuclear modification factor for inclusive \upsone as a function of \Npart~\cite{Acharya:2020kls}, compared to model calculations~\cite{Du:2017qkv,Krouppa:2016jcl,Ferreiro:2018wbd,Yao:2020xzw,Brambilla:2021wkt}. }
\label{fig:RAAUpsilonvscent} 
\end{center}
\end{figure}

A remarkable observation at the LHC was that the $\Upsilon$ states are strongly suppressed, in a hierarchy apparently dictated by their binding energies~\cite{Digal:2001ue}. 
First observed by the CMS Collaboration at midrapidity~\cite{Chatrchyan:2012lxa}, this hierarchy of suppression was confirmed by the ALICE Collaboration at forward rapidity~\cite{Abelev:2014nua,Acharya:2020kls}.
In Fig.~\ref{fig:RAAUpsilonvscent} the $\langle N_{\rm part}\rangle$ dependence of the \upsone \RAA measurements performed by ALICE at $\sqrt{s_{\rm NN}}=5.02$ TeV is shown and indicates a quick onset of the suppression of the \upsone already in semi-peripheral collisions. 
The results are compared with various models. For each of them two lines are shown, corresponding to upper and lower uncertainty limits. All the models describe the data quite well, although the theory uncertainties are rather large.
First, the same comover model~\cite{Ferreiro:2018wbd} introduced in Sec.~\ref{sec:charmonia} is applied to the bottomonia, but, in contrast to the charmonium case, no (re)generation component is introduced.
The model uncertainties are related to the used nPDF parameterisation and to the comover--$\Upsilon$ dissociation cross section. 
Second, the transport model introduced above for the \jpsi (TM-TAMU~\cite{Du:2017qkv}) is also used for the \upsone. 
The model results are shown with and without a (small) (re)generation component. The former is favoured by the measured \RAA in central collisions. 
Third, a calculation of another transport model based on the framework of coupled Boltzmann equations~\cite{Yao:2020xzw} is also shown. In this approach the (re)generation is dominated by real-time (re)combination of correlated heavy-quark pairs, and the uncertainties correspond to those on the EPPS16 nPDF parametrisation~\cite{Eskola:2016oht}.
Fourth, the model labelled ``aHydro'' in the figures implements a thermal modification of a complex heavy-quark potential inside an anisotropic plasma~\cite{Krouppa:2016jcl}. The survival probability of bottomonia is evaluated based on the local energy density, integrating
a rate equation over the proper time of each state.
The uncertainties represent the envelope of calculations performed with values of the viscosity-to-entropy density ratio $\eta/s$ ranging from 1/(4$\pi$) to 3/(4$\pi$).
No modifications of nuclear PDFs or any (re)generation phenomenon are included.
Finally, the ``Open Quantum Systems'' model results are obtained using potential Non-Relativistic QCD (pNRQCD) and the  formalism of open quantum systems to treat the interaction of the \upsone state with the QGP~\cite{Brambilla:2020qwo,Brambilla:2021wkt}. The uncertainties shown here correspond to variations of the non-perturbative transport coefficient $\kappa$ (heavy-quark momentum diffusion coefficient), keeping the value of its dispersive counterpart $\gamma$ fixed at its central value.

In pp collisions, up to 30--50\% of the measured \upsone 
yield results from the feed-down from other states~\cite{Aaij:2014caa,Andronic:2015wma,Lansberg:2019adr}. Consequently, a significant amount of \upsone suppression may arise from the very strong suppression of the excited states (see Secs.~\ref{sec:2.5nonQGP} and~\ref{sec:qqexcstates}). The models mentioned above follow similar approaches for the evaluation of the feed-down contribution, based on PDG values~\cite{ParticleDataGroup:2022pth} for the branching ratios of the excited states, and information on the production cross sections of those states from pp data taken at LHC energy.

\begin{figure}[ht!]
\begin{center}
\includegraphics[width=0.49\linewidth]{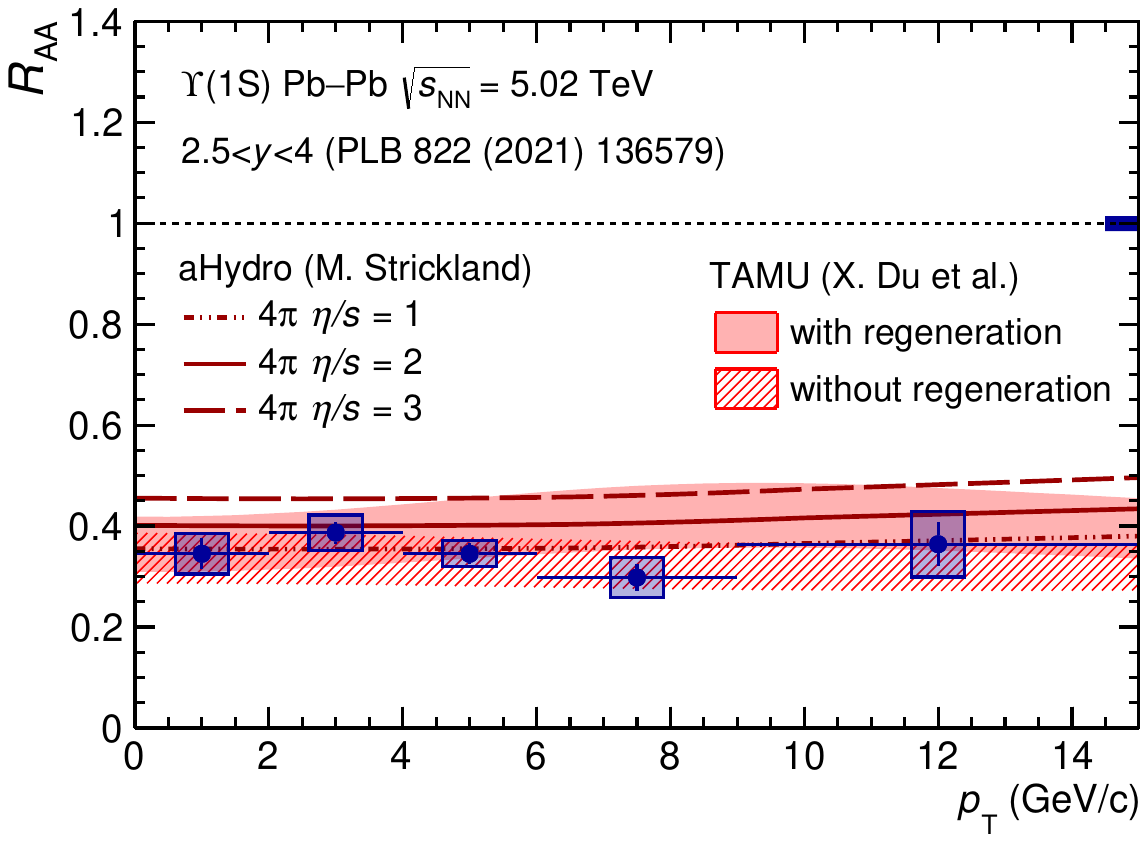}
\includegraphics[width=0.49\linewidth]{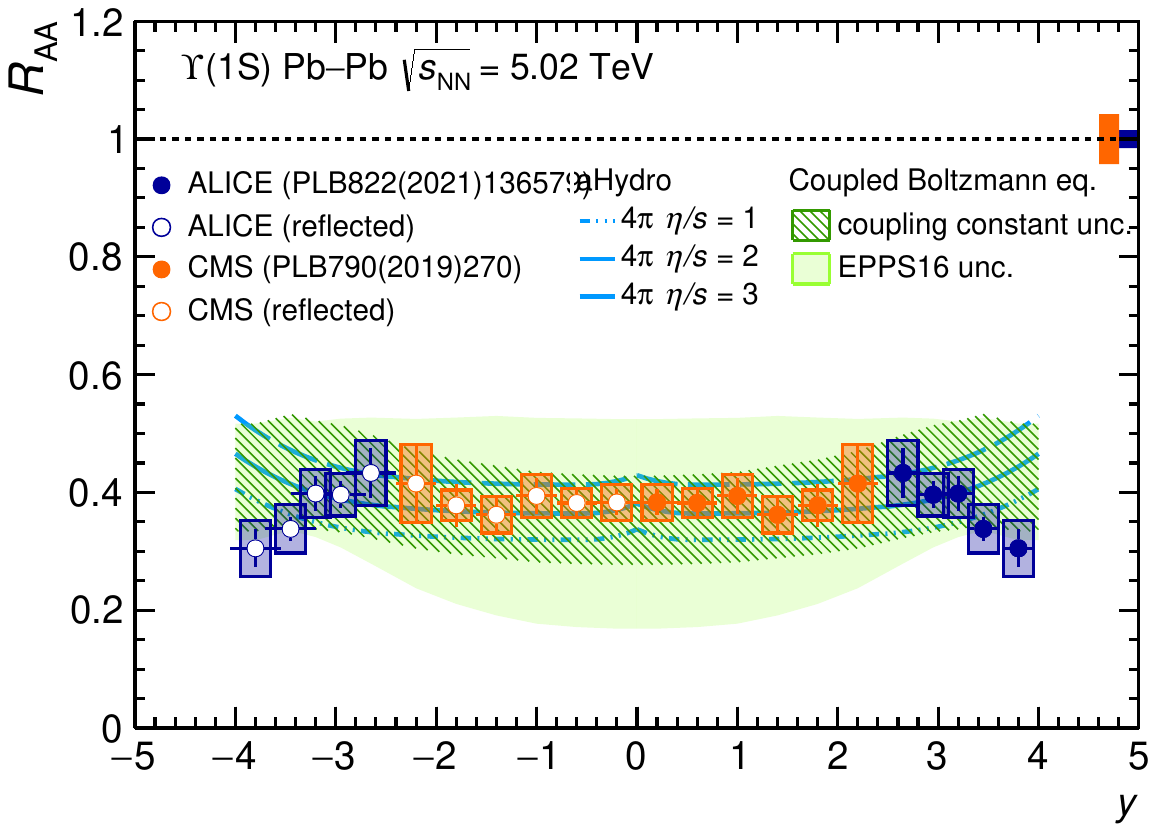}
\caption{The transverse momentum (left, data from ALICE~\cite{Acharya:2020kls}) and rapidity (right, data from ALICE~\cite{Acharya:2020kls} and CMS~\cite{Sirunyan:2018nsz}) dependence of the inclusive \upsone nuclear modification factor, compared to model calculations~\cite{Du:2017qkv,Krouppa:2016jcl,Yao:2020xzw} }
\label{fig:RAAUpsilonvspty} 
\end{center}
\end{figure} 

The \pt and rapidity dependence of the \RAA for \upsone are shown in Fig.~\ref{fig:RAAUpsilonvspty}~\cite{Acharya:2020kls}.
The data exhibit a very weak (if any) \pt dependence and a weak rapidity dependence, apparent only for the most forward rapidities $3<y<4$ covered by ALICE.
The \pt dependence is reproduced by both TM-TAMU and aHydro  models~\cite{Du:2017qkv,Krouppa:2017jlg}. 
It is worth noting that in the TM-TAMU model~\cite{Du:2017qkv}, the only one to implement (re)generation, its (small) effect becomes visible at \pt of the order of the \upsone mass, where the uncertainties are still rather large.
More interesting is the rapidity dependence of the \upsone \RAA. 
Complemented by the CMS results~\cite{Sirunyan:2018nsz}, the measurement spans four units of rapidity. 
While both the hydrodynamic~\cite{Krouppa:2017jlg} and the transport model~\cite{Yao:2020xzw} describe the \upsone \RAA at midrapidity rather well, they predict an opposite trend at forward rapidity than suggested by the data. In the models, the rapidity dependence of the suppression is determined by the rapidity dependence of the energy density in the QGP phase, which is expected to decrease slightly towards forward $y$. The opposite trend suggested by the data may find a natural explanation in case a significant (re)generation component is present, even if the size of this effect, as visible in the left panel of Fig.~\ref{fig:RAAUpsilonvspty}, is predicted to be relatively small in TM-TAMU  calculations. Another effect that would, in principle, lead to the observed behaviour for $R_{\rm AA}$ is the presence of significant energy loss effects that may lead to a shrinking of the rapidity distribution in Pb--Pb collisions. Models that implement coherent energy loss effects in cold nuclear matter~\cite{Arleo:2014oha} suggest a decrease of $R_{\rm AA}$ at forward rapidity, but the effect is much smaller than observed.

The bottomonium data were recently used for a first explicit extraction of the heavy-quark potential in the hot QCD medium~\cite{Du:2019tjf}, found to be characterised by significant remnants of the long-range confining force in the QGP.
The bottomonium family is of utmost relevance for the next advances in the theoretical treatment, based on a quantum evolution in the deconfined medium, where first quantitative results are just becoming available~\cite{Brambilla:2020qwo,Brambilla:2021wkt}.

\subsubsection{Study of non-QGP effects on ground state suppression}
\label{sec:2.5nonQGP}

The \jpsi and \upsone production yields are, to various extents,  modified in \PbPb collisions. As detailed in Sec.~\ref{sec:charmonia} and~\ref{sec:bottgrstate}, the \jpsi nuclear modification factor presents a suppression of the charmonium ground state significantly increasing with \pt (see Fig.~\ref{fig:RAAv2Jpsivspt} top), while for the \upsone the \pt dependence is weaker (see Fig.~\ref{fig:RAAUpsilonvspty} left). 

\begin{figure}[ht!]
\begin{center}
\includegraphics[width=0.69\linewidth]{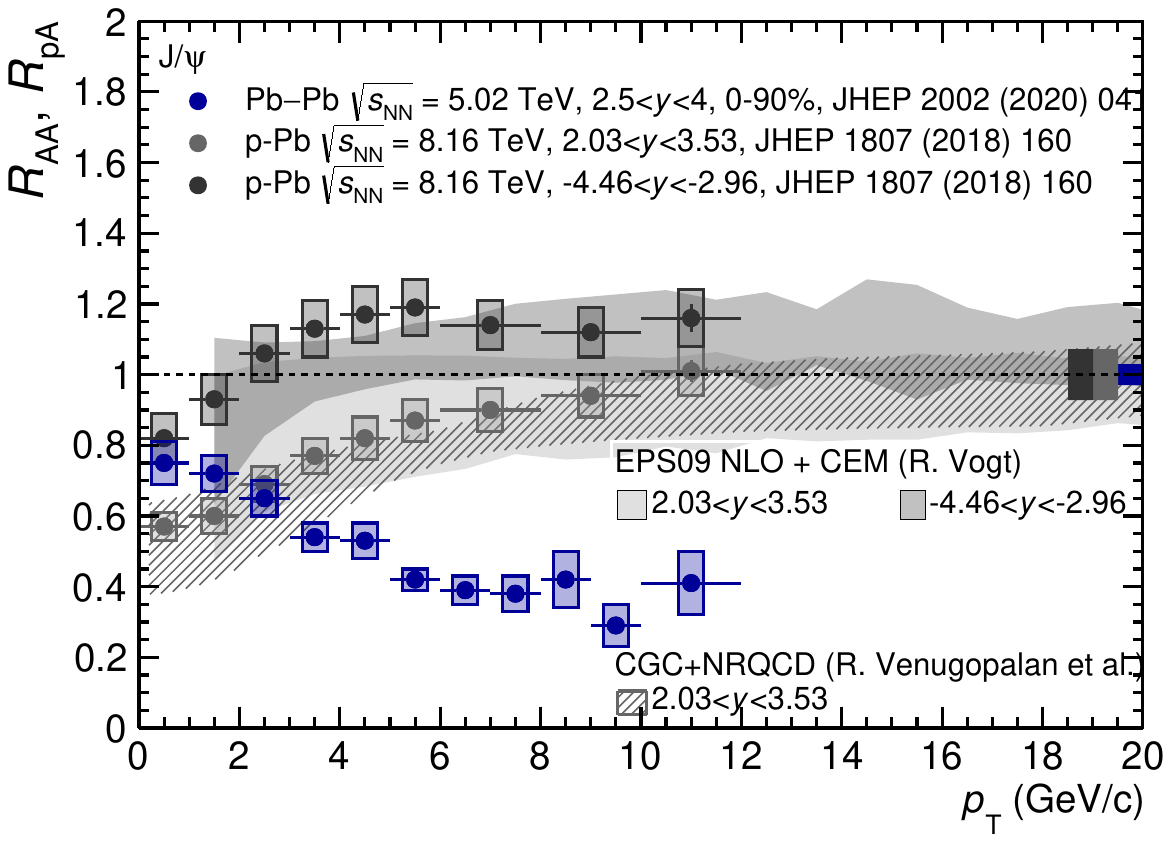}
\includegraphics[width=0.69\linewidth]{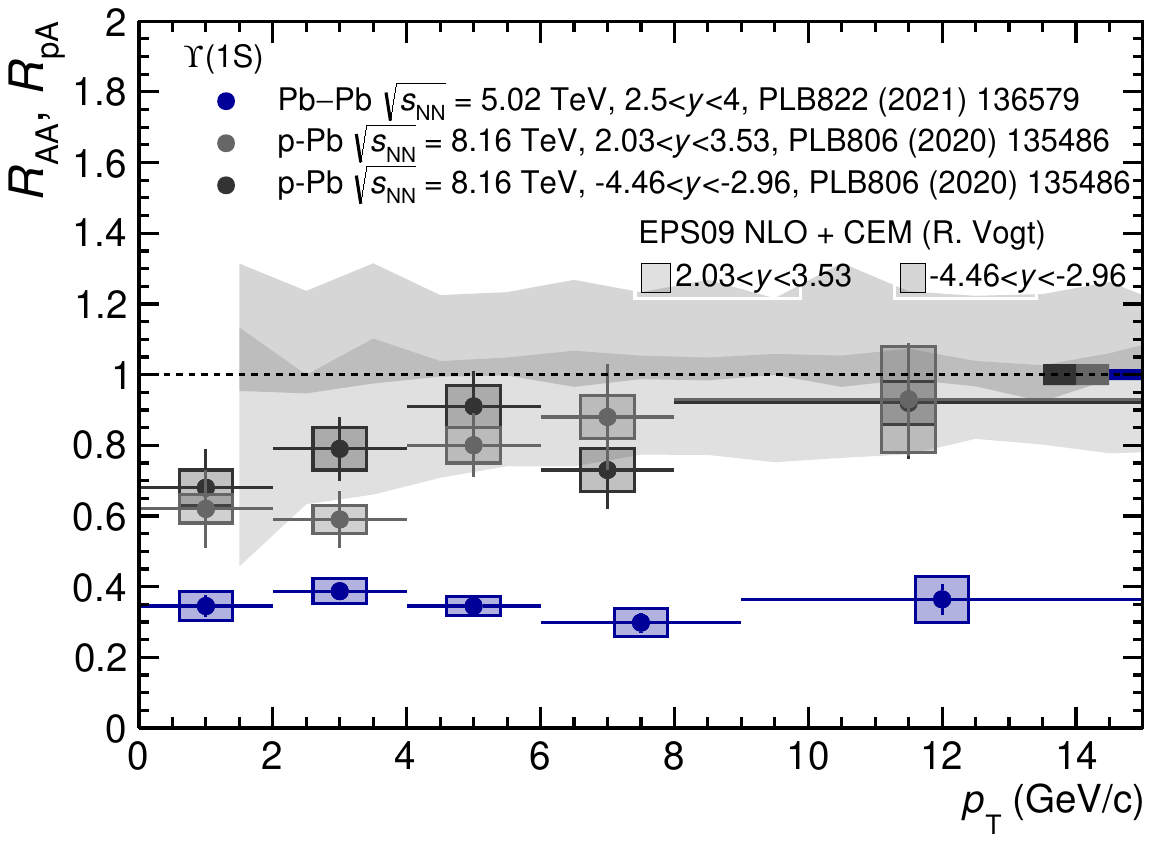}  
\caption{\jpsi (left) and \upsone (right) \RAA as a function of \pt in centrality-integrated \PbPb collisions at \fivenn in $2.5 < y_{\mathrm{cms}} < 4$ (blue circles)~\cite{Acharya:2020jps,Acharya:2020kls}. Results are compared to the corresponding \RpA values obtained in \pPb collisions at \eightnn in $2.03 < y_{\mathrm{cms}} < 3.53$ (light gray circles) and $-4.46 < y_{\mathrm{cms}} < -2.96$ (dark gray circles)~\cite{Acharya:2018kxc,Acharya:2019lqc}. The gray bands correspond to model calculations based on a pure nuclear shadowing scenario using EPS09LO as set of nuclear parton distribution functions (filled bands)~\cite{Albacete:2017qng}. For the \jpsi case, a theory calculation based on a CGC approach coupled with a NRQCD production mechanism is also shown (dashed band)~\cite{Ma:2017rsu}. }
\label{fig:fig5} 
\end{center}
\end{figure} 

Even if  QGP effects are significant at LHC energies,  other non-QGP mechanisms have influence on the production of quarkoniua. Commonly defined as ``cold nuclear matter'' (CNM) effects, in contrast to the hot-matter effects related to the presence of the QGP, they can be relevant in collisions involving light and heavy nuclei. Therefore, the quantification of their size is important to correctly define the net impact of QGP effects on the production of quarkonia. Experimentally, CNM effects are studied in minimum-bias \pA collisions, where hot-matter mechanisms are not expected to dominate.

ALICE has studied quarkonium production in \pPb collisions at $\sqrt{s_{\mathrm{NN}}}~=~5.02$~TeV\ and $8.16$~TeV~\cite{
Abelev:2013yxa,
Adam:2015jsa,
Adam:2015iga,
Adamova:2017uhu,
Acharya:2018yud,
Acharya:2018kxc,
Abelev:2014oea,
Acharya:2019lqc}. The \pt dependence of the nuclear modification factor  of the \jpsi and \upsone, at the highest energy, is shown in Fig.~\ref{fig:fig5} at  forward (p-going) and backward (Pb-going) rapidity. For both \jpsi and \upsone a significant \pt dependence, with a stronger suppression in the low-\pt region, can be seen. Furthermore, for the \jpsi the effect is more important in the p-going ($2.03 < y_{\mathrm{cms}} < 3.53$) than in the Pb-going direction  ($-4.46<y_{\mathrm{cms}} < 3.53$), while for the \upsone no significantly different behaviour, within the %
statistical and systematic uncertainties, is observed in the two accessible rapidity domains. The comparison of the \jpsi and \upsone \RpA with theoretical models that implement only nuclear modification of the gluon PDFs~\cite{Albacete:2017qng} confirms the negligible role of QGP-induced mechanisms in pA collisions, and it underlines the importance of CNM effects related to gluon saturation and/or nuclear shadowing. 
 
Significantly different nuclear modification factors are measured in \pPb and \PbPb collisions, with stronger suppressions observed in \PbPb, except in the low-\pt region. At high-\pt, the \jpsi suppression measured in \PbPb collisions can certainly not be due to the 
CNM effects alone and, even in the limits of this qualitative comparison, this is a clear confirmation of the role played by mechanisms related to the hot matter formation. At low-\pt, $R_{\rm AA}$ and $R_{\rm pPb}$ become quite similar. This does not imply that $R_{\rm AA}$ is dominated by CNM effects, but rather that (re)combination, which was shown to be strong in this kinematic region (see left panel of Fig.~\ref{fig:RAAv2Jpsivspt}) completely balances suppression effects, leading to an approximate recovery of the binary collision scaling once CNM effects are taken into account.
We also remark that the comparison of  \pPb and \PbPb results at two different energies is meaningful when considering the shadowing effects, since the variation in the Bjorken-$x$ ranges due to the different $\sqrt{s_{\rm NN}}$ is compensated by the different $y_{\rm cms}$ coverage (about half a unit rapidity shift).

For the \upsone, the difference in the \pt dependence of \RAA and \RpA, in particular at high-\pt, gives an indication that hot-matter final state effects are significant. At low-\pt, the $R_{\rm AA}$ and $R_{\rm pPb}$ values become more similar. Since (re)combination effects are negligible for bottomonium, this observation implies that a significant fraction of the observed \upsone suppression in \PbPb collisions might be ascribed to CNM effects.

\begin{figure}[ht!]
\begin{center}
\includegraphics[width=0.6\linewidth]{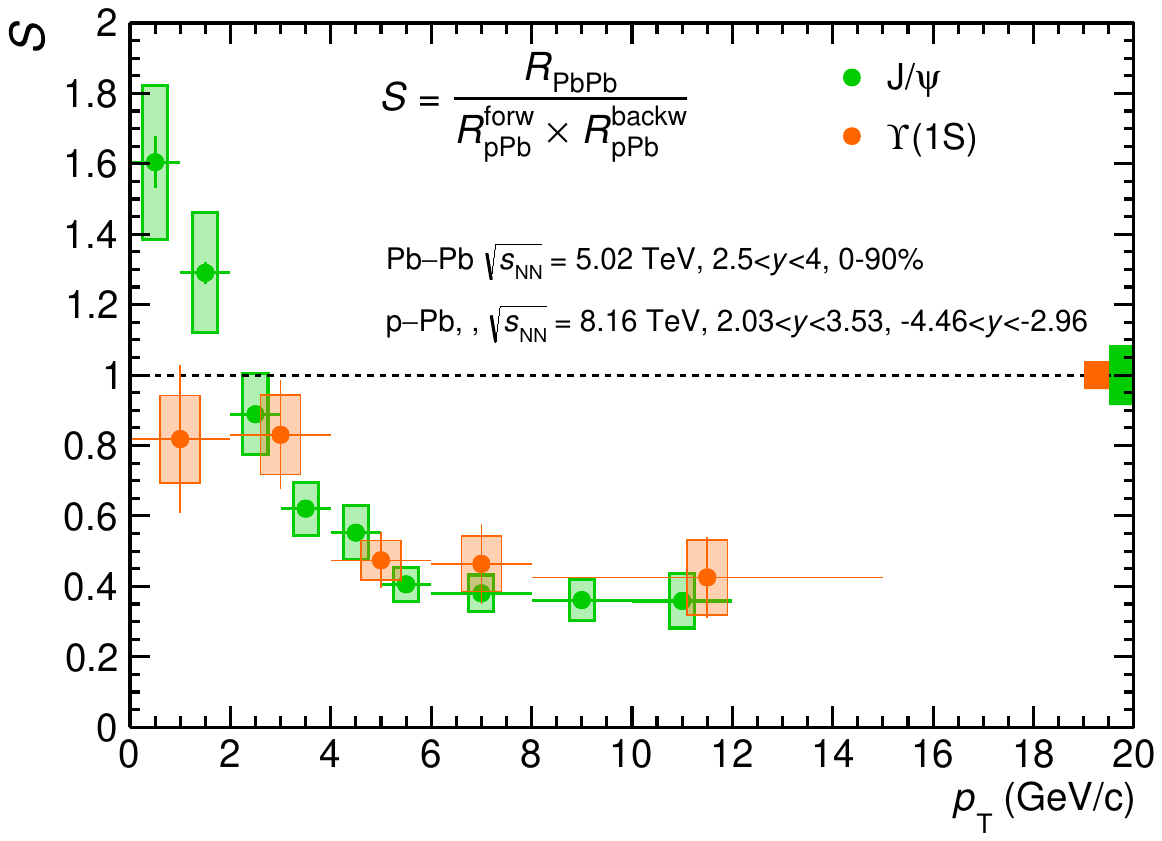}
\caption{$S_{J/\psi}$ (green symbols) and $S_{\Upsilon(1S)}$ (orange symbols) as a function of \pt. See text for further details on the definition of this quantity.}
\label{fig:figS} 
\end{center}
\end{figure} 

A more quantitative estimate of the magnitude of CNM effects in nucleus--nucleus collisions can be performed assuming that such effects are dominated by nuclear shadowing. 
Under this hypothesis, the magnitude of shadowing effects in \PbPb collisions can be obtained as the product $R_{\rm{pPb}}^{\rm forw} \times R_{\rm{pPb}}^{\rm backw}$~\cite{Adam:2015iga,Abelev:2013yxa}. By normalising the measured $R_{\rm AA}$ to this product, the quantity $S=R_{\rm AA}/(R_{\rm{pPb}}^{\rm forw} \times R_{\rm{pPb}}^{\rm backw})$ can be computed.
 This quantity, shown in Fig.~\ref{fig:figS} for both \jpsi and \upsone, can be considered as the Pb--Pb nuclear modification factor, corrected for shadowing effects. For the \jpsi, values larger than unity are obtained at low-\pt. This indicates a net \jpsi enhancement, clearly related to  (re)generation effects~\cite{Adam:2015iga}. At high-\pt, values as low as $S_{\jpsi}\sim 0.3$ are measured, corresponding to a strong suppression.
 
Also $S_{\upsone}$ shows a significant \pt dependence. At high-\pt its value is similar to $S_{\jpsi}$, while at low-\pt it becomes compatible with unity. Taking into account that for Pb--Pb collisions  the contribution from the (re)combination of b quarks should be negligible, this result could imply that once shadowing effects are divided out, only a modest suppression effect survives for the strongly bound \upsone. This tentative conclusion can be further quantified by taking  into account the feed-down contribution from the excited bottomonium states. A thorough investigation of these effects can be found in Sec.~2.1 of Ref.~\cite{Lansberg:2019adr}. Indeed, both $\Upsilon(2S)$ and $\Upsilon(3S)$, as well as various $\chi_{\rm b}$ states have a non-negligible branching ratio towards the 1$^{--}$ ground state. The feed-down fractions were measured in pp collisions by ATLAS, CMS and LHCb, with the resulting contribution from all these states being $\sim 30$\% at relatively low-\pt, increasing to $\sim 50$\% at high-\pt~\cite{Lansberg:2019adr}. A precise evaluation of the direct \upsone $R_{\rm AA}$ would require the knowledge of the corresponding quantity for the excited states, which is available for the $\Upsilon(2S)$ and $\Upsilon(3S)$~\cite{Acharya:2020kls,Sirunyan:2018nsz}, but not for the various $\chi_{\rm b}$. Assuming  $R_{\rm AA}(\chi_{\rm b}(1P))\sim R_{\rm AA}(\Upsilon(2S))$ and $R_{\rm AA}(\chi_{\rm b}(2P))\sim R_{\rm AA}(\Upsilon(3S))$, as it may be suggested by the similar binding energies, one may evaluate the direct $\Upsilon(1S)$ nuclear modification factor by subtracting out the contribution of the excited states as $R_{\rm AA}(\Upsilon(1S))^{\rm dir}=(R_{\rm AA}(\Upsilon(1S))^{\rm incl}-R_{\rm AA}(\Upsilon(2S) \, F_{\Upsilon(2S)\rightarrow\Upsilon(1S)}-...)/F_{\Upsilon(1S)^{\rm dir}}$. A numerical evaluation for the lowest $p_{\rm T}$ interval ($p_{\rm T}<2$ GeV/$c$) in Fig.~\ref{fig:fig5} (bottom) gives $R_{\rm AA}(\Upsilon(1S))^{\rm dir}\sim 0.45\pm 0.08$, to be compared with the product of the forward and backward $R_{\rm pPb}$ which yields $0.42\pm 0.11$. This approximate estimate suggests that shadowing and feed-down effects could be responsible for most of the suppression observed at low-$p_{\rm T}$ for inclusive $\Upsilon(1S)$ production. At larger \pt CNM effects become weaker but the feed-down fractions increase. The corresponding results, affected by larger uncertainties, are $R_{\rm AA}(\Upsilon(1S))^{\rm dir}\sim 0.68\pm 0.22$ while $R_{\rm pPb}^{\rm forw}\times R_{\rm pPb}^{\rm backw}\sim 0.86\pm 0.26$.

\subsubsection{Excited quarkonium states}
\label{sec:qqexcstates}

Both \ccbar and \bbbar states have complex spectroscopies. The various states have very different binding energies, ranging from about 50~MeV for the \psiprime to about 650~MeV for the \jpsi in the charmonium sector and from about 200~MeV for the loosely bound \upsthree up to about 1~GeV for the tightly-bound \upsone in the bottomonium sector. Hence the comparison of their behaviours in a hot and dense medium can shed light on various properties of the created system. In particular, by linking the sequential disappearance of these resonances to the melting temperature, as evaluated by lattice QCD studies~\cite{Kim:2018yhk}, one might also have a tool to access the temperature reached in the collision. In addition,
also the details of the production process itself in nuclear collisions, with the contribution of direct and regenerated states, can be different for the various quarkonium resonances. Thus, it is expected that a precise measurement of both ground and excited states will allow distinguishing between statistical hadronisation~\cite{Andronic:2007bi}, which assumes that all quarkonium states are produced according to thermal weights determined at the chemical freezeout, and microscopic transport models~\cite{Du:2015wha}. In the latter, the production of quarkonia is continuous through the lifetime of the fireball and can thus occur out of chemical equilibrium.

\begin{figure}[ht!]
\begin{center}
\includegraphics[width=0.69\linewidth]{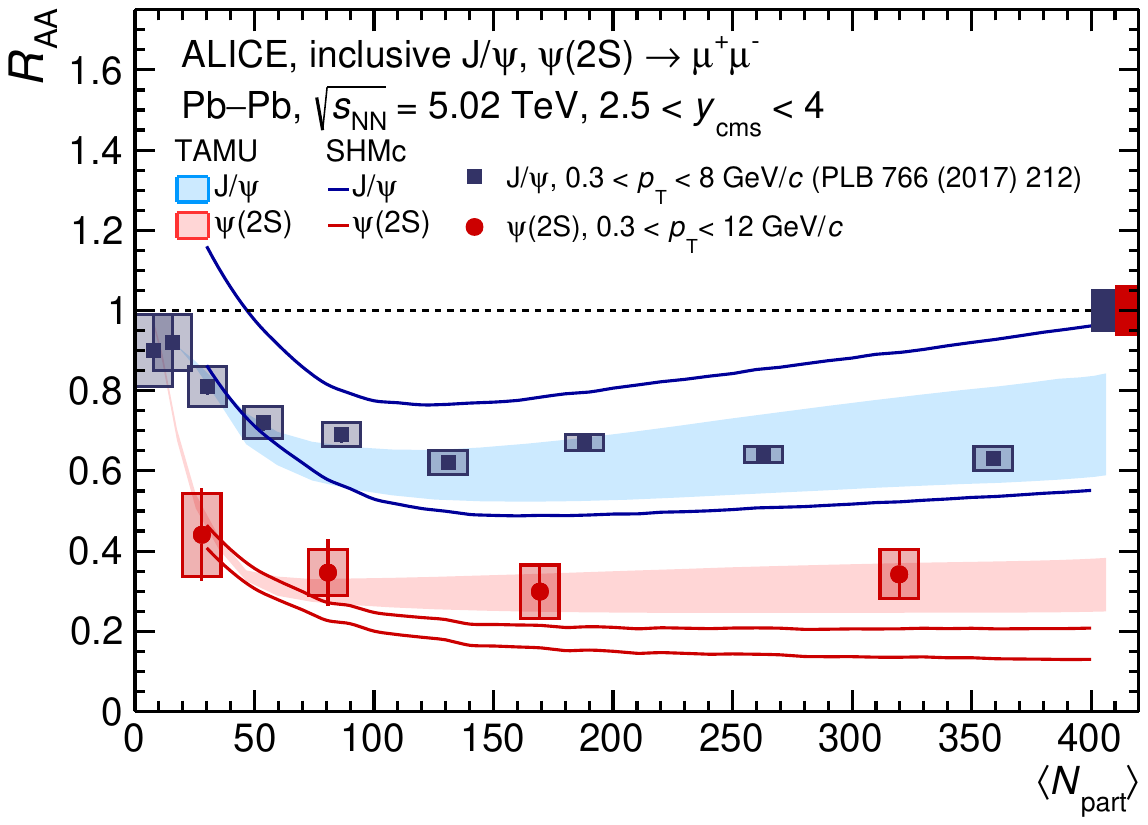}
\includegraphics[width=0.69\linewidth]{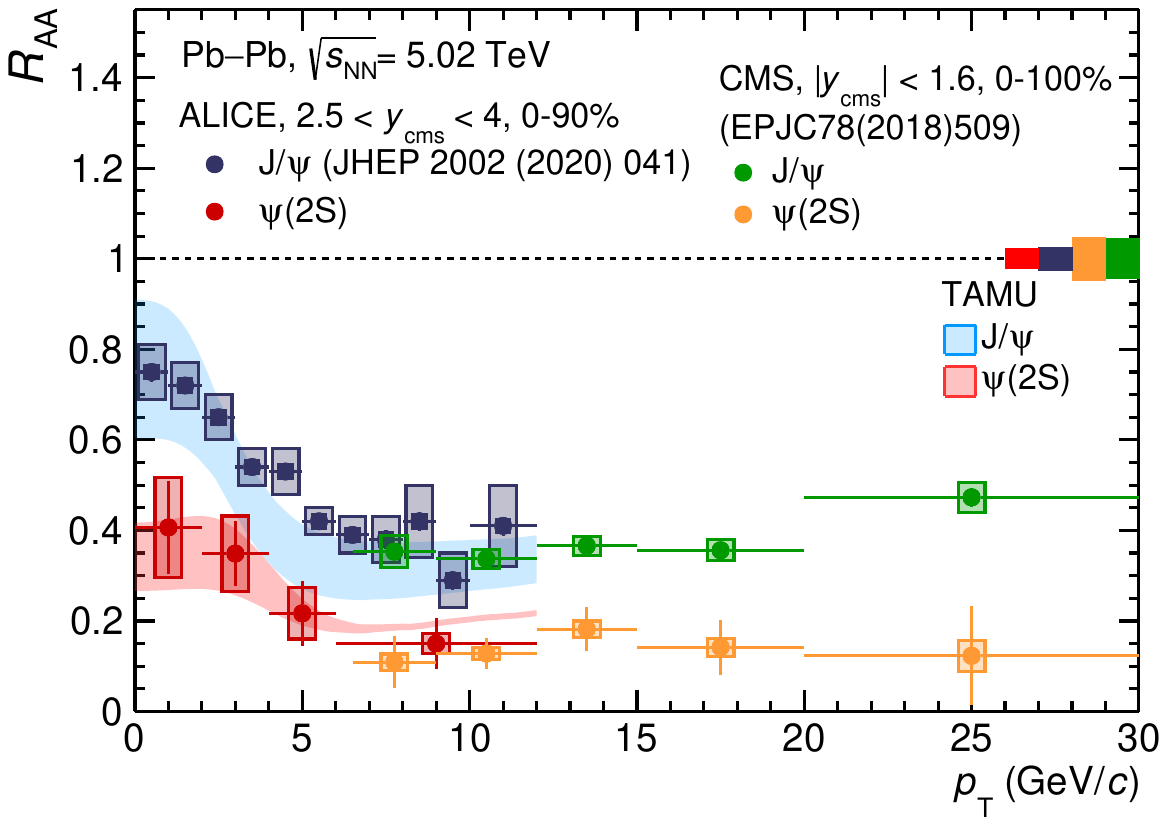}
\caption{(Top) The nuclear modification factor for inclusive $\psi(2S)$ and J/$\psi$, measured in Pb--Pb collisions at $\sqrt{s_{\rm NN}}=5.02$ TeV as a function of $N_{\rm part}$~\cite{ALICE:2022jeh}, in the range $2.5<y<4$. Predictions from the TM-TAMU~\cite{Du:2015wha} and SHM~\cite{Andronic:2019wva} models are also shown (Bottom) The $p_{\rm T}$ dependence of the $\psi(2S)$ $R_{\rm AA}$, measured by ALICE in $2.5<y<4$ for 0--90\% Pb--Pb collisions~\cite{ALICE:2022jeh}, compared with corresponding high-$p_{\rm T}$ results in $|y|<1.6$ and 0--100\% centrality from CMS~\cite{Sirunyan:2017isk}. Model comparisons are also shown.}  %
\label{fig:RAApsi2svscent} 
\end{center}
\end{figure}

ALICE has measured both the \jpsi and the \psiprime at forward rapidity (2.5$<y<$4) and down to zero transverse momentum~\cite{ALICE:2022jeh}. These results extend the measurements of the CMS experiment, which has studied the \jpsi and the \psiprime  in the kinematic ranges $|y|<1.6$, $6.5<p_{\rm T}<30$ \GeVc and  $1.6<|y|<2.4$, $3<p_{\rm T}<30$ \GeVc~\cite{Sirunyan:2016znt}.
As visible in Fig.~\ref{fig:RAApsi2svscent}, where the nuclear modification factors of the \jpsi and \psiprime measured in the rapidity range $2.5<y<4$ are shown, the 
\psiprime is significantly more suppressed than the \jpsi over all the centrality and \pt ranges explored by ALICE. 
Furthermore, a rise of $R_{\rm AA}$ at low-\pt can be observed also for the \psiprime, suggesting that (re)combination effects can be sizeable also for this state.
The calculations of the TM-TAMU model~\cite{Du:2015wha} are in good agreement with the experimental results as a function of both $N_{\rm part}$ and \pt. The SHM approach~\cite{Andronic:2019wva} tends to overestimate the suppression effects for the \psiprime in central events, while a good agreement, as discussed previously, is found for the \jpsi.
In the region $6.5<p_{\rm T}<12$ GeV/$c$ the results agree with those of CMS~\cite{Sirunyan:2017isk}, obtained at midrapidity, which also explore a higher-$p_{\rm T}$ range, where the \psiprime $R_{\rm AA}$ remains roughly constant.

\begin{figure}[ht!]
\begin{center}
\includegraphics[width=0.78\linewidth]{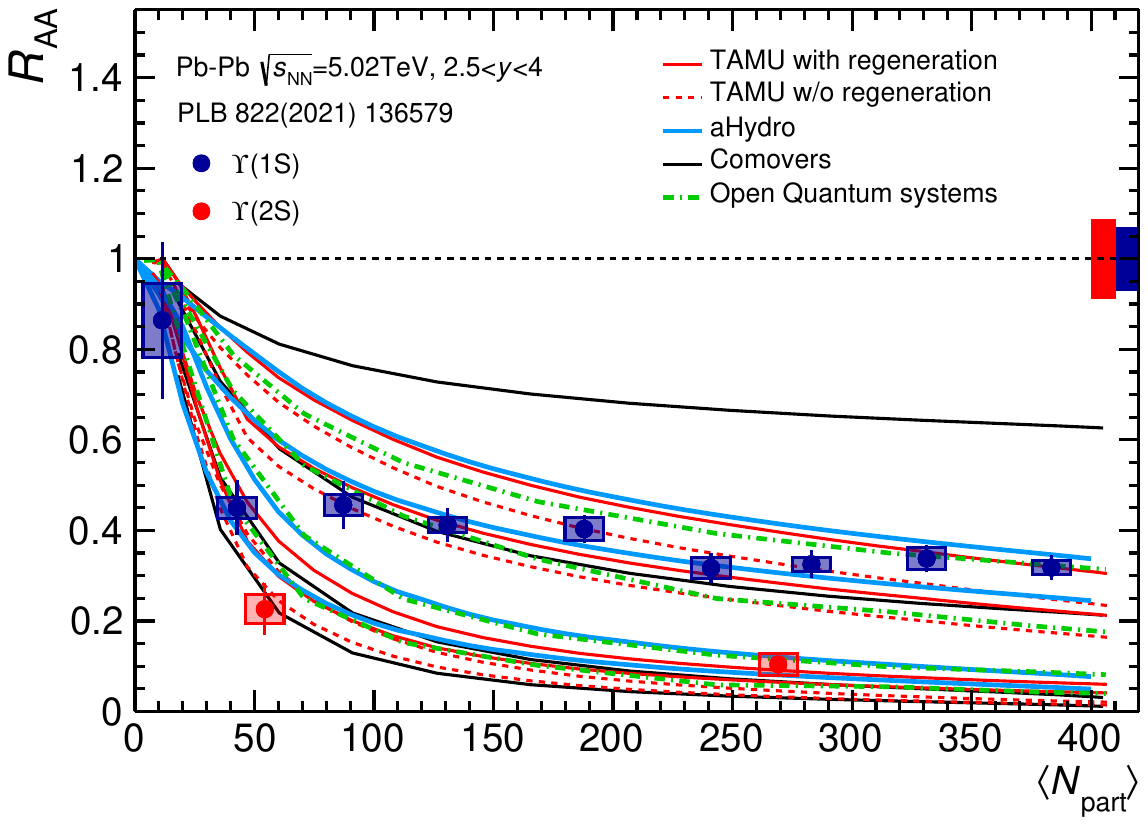}
\caption{The nuclear modification factor for inclusive $\Upsilon(2S)$ and $\Upsilon(1S)$ as a function of $N_{\rm part}$~\cite{Acharya:2020kls}, compared to model calculations~\cite{Du:2017qkv,Krouppa:2016jcl,Ferreiro:2018wbd}.}
\label{fig:RAAupsilon2s2svscent} 
\end{center}
\end{figure}

A similar comparison between the \RAA centrality dependence for ground and excited states has been carried out by ALICE also in the bottomonium sector in \PbPb collisions at \fivenn~\cite{Acharya:2020kls}. Bottomonium states are reconstructed at forward rapidity (2.5$<y<$4). As for charmonium states, also the \upsone and \upstwo show a strong suppression in their production, with a more significant reduction of their yields in the most central collisions. Even if the measurement of the \upstwo yields in \PbPb is still statistically limited, allowing only a small number of centrality intervals to be defined, a stronger suppression of the excited state, with respect to the \upsone, is visible in Fig.~\ref{fig:RAAupsilon2s2svscent}. 

It can be noted that the size of the measured suppression of the \upsone and \upstwo production in \PbPb with respect to \pp collisions is well described by several theory models, whose predictions are shown in Fig.~\ref{fig:RAAupsilon2s2svscent}, and it is similar to the one  measured by the CMS experiment~\cite{Acharya:2018mni}, in a complementary rapidity region $|y|<2.4$~\cite{Sirunyan:2018nsz}. Some of the models shown in Fig.~\ref{fig:RAAupsilon2s2svscent} give explicit predictions for the temperatures needed to describe the relative suppression of the two states.
In particular, in the aHydro model~\cite{Krouppa:2016jcl}, the hydrodynamic fireball evolution starts at $\tau_0=0.3$~fm/$c$, with an initial system temperature and shear viscosity to entropy ratio tuned simultaneously to describe the measured charged particle density and anisotropic flow. The most probable values for the above quantities in this model are $T=641$~MeV and $\eta/s = 1/4\pi$, in agreement with other hydrodynamic calculations which favour the hypothesis of a nearly perfect QGP fluid with a shear viscosity close to the lowest bound conjectured by the AdS/CFT correspondence~\cite{Kovtun:2004de}. Other approaches, as TM-TAMU~\cite{Du:2017qkv}, start the space--time evolution of the medium at a slightly lower temperature, $T\sim 540$ MeV. 
In any case, more quantitative constraints on the initial QGP temperature and its shear viscosity require a better understanding of all the differential experimental results, especially the rapidity dependence which, as previously shown in Sec.~\ref{sec:bottgrstate} for the \upsone, does not seem to be well reproduced in this model.

\subsubsection{Discussion of further flow results for quarkonia}
\label{sec:2.5flow}

A detailed study of the \jpsi anisotropic flow, reported in Refs.~\cite{Acharya:2020jil,Acharya:2017tgv} offers additional insights into the dynamics of the production process. 
The centrality dependence of the elliptic flow coefficients allows us the study of the sensitivity of the \jpsi production mechanisms to the system size and initial spatial anisotropy. The \velip of light flavoured hadrons was shown to be proportional to the initial state spatial ellipticity~\cite{Niemi:2012aj}, as expected from the evolution of a nearly ideal hydrodynamic system, with the proportionality factor being dependent on the medium properties, like the equation of state and viscosity. Since the \jpsi \velip should be determined by the flow of charm quarks in the low- and intermediate-\pt regions where (re)combination processes dominate its production, its centrality dependence provides information on the degree of charm-quark thermalisation/equilibration in the medium as well as on the charmonium formation time. 

\begin{figure}[ht!]
\begin{center}
\includegraphics[width=0.95\linewidth]{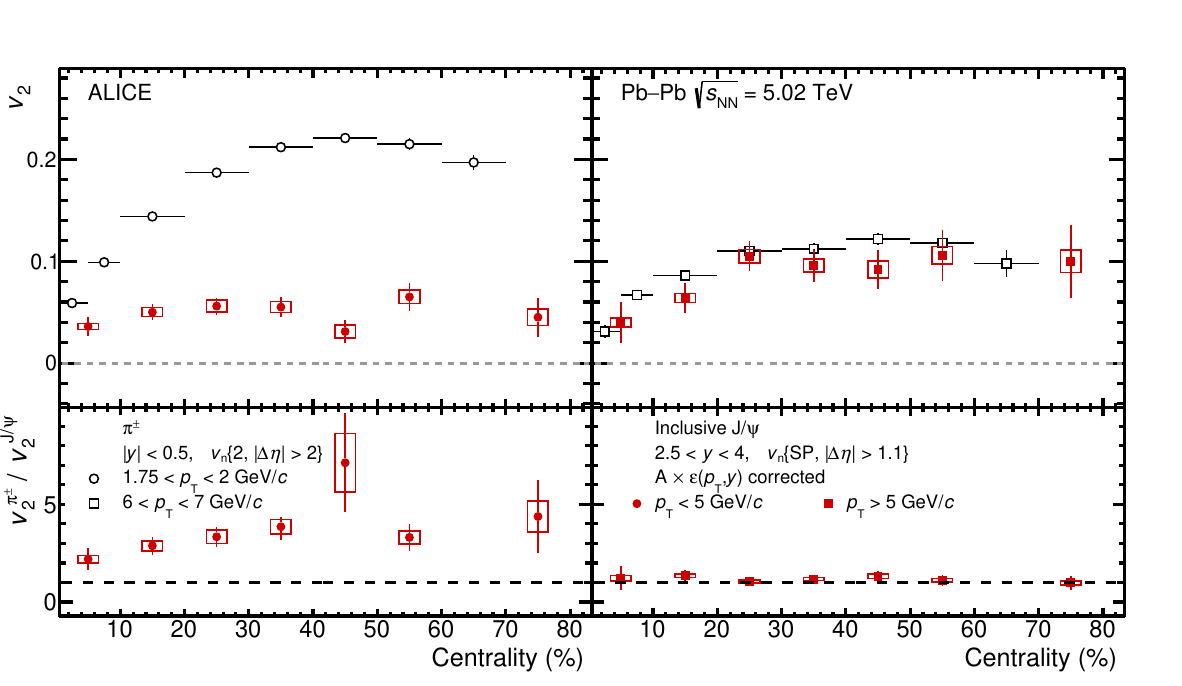}
\caption{The centrality dependence of inclusive J/$\psi$ $v_{\rm 2}$ for two transverse momentum regions~\cite{Acharya:2020jil}, compared with the corresponding results for charged pions~~\cite{Acharya:2018zuq}.}
\label{fig:v2jpsipions} 
\end{center}
\end{figure} 

Figure~\ref{fig:v2jpsipions} presents the centrality dependence of the \pt-integrated \jpsi $v_2$ in \PbPb collisions at \snn = 5.02~TeV, measured at forward rapidity in the ranges $\pt<5$~\GeVc (left) and $5<\pt<20$~\GeVc (right)~\cite{Acharya:2020jil}. These results are compared to the \velip of the charged pions measured at midrapidity~\cite{Acharya:2018zuq} and at a \pt corresponding to the \jpsi average \pt in the above mentioned intervals. The $v_2$ for low-\pt pions grows fast from the most central towards peripheral collisions, reaching a maximum in the 40--50\% centrality interval. For the \jpsi, however, a milder centrality dependence is observed, illustrated quantitatively by the $v_2^{\pi} / v_2^{\jpsi}$ ratio shown in the bottom panel, which grows linearly from central to peripheral collisions. The significance of the trend is larger than 2\,$\sigma$. Such a trend for the \jpsi may be interpreted as the consequence of an increase towards more central collisions of the fraction of regenerated \jpsi and/or the degree of charm-quark thermalisation, with both favouring the development of anisotropic flow in more central collisions. An incomplete charm thermalisation, arguably affecting more the peripheral collisions, would lead, as the measurements suggest, to a centrality dependent conversion factor of the initial state spatial anisotropy into final state momentum anisotropy.  At high-\pt, the $v_2$ shows a similar centrality dependence for \jpsi and charged pions, suggesting a possible common origin.

\begin{figure}[ht!]
\begin{center}
\includegraphics[width=0.65\linewidth]{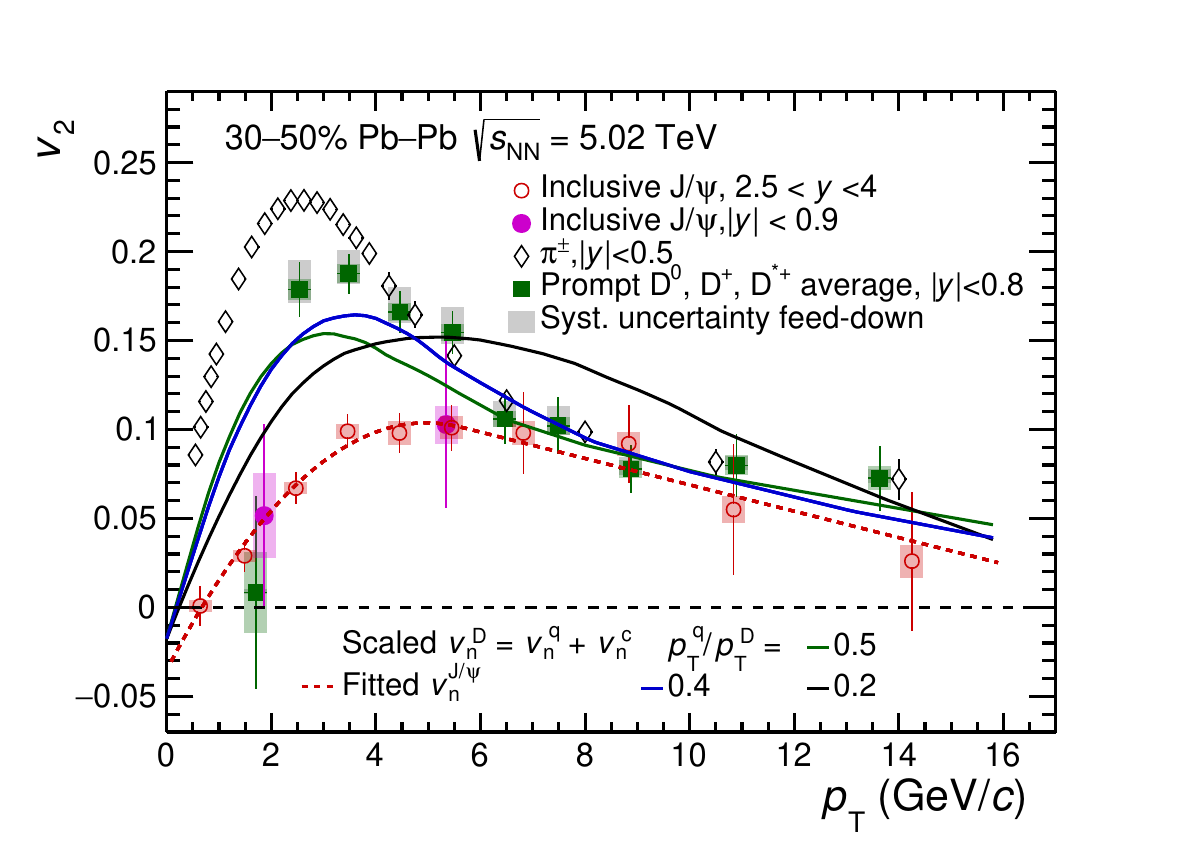}
\caption{The $p_{\rm T}$ dependence of inclusive J/$\psi$ $v_{\rm 2}$~\cite{Acharya:2020jil} compared with the corresponding results for prompt D mesons~\cite{ALICE:2020iug} and charged pions~\cite{Acharya:2018zuq}. 
The expected $v_{\rm 2}$ for open charm in the constituent quark approach, and for various values of the ratio $p_{\rm T}^{\rm q}/p_{\rm T}^{\rm D}$ are also shown.} %
\label{fig:NCQscaling} 
\end{center}
\end{figure} 

An extension to the picture of the number of constituent quark (NCQ) scaling of the anisotropic flow, first proposed at RHIC~\cite{Adams:2003am,Afanasiev:2007tv}, which assumes that the charm-quark flow magnitude is different than that of the lighter quark species, can be tested as suggested in Refs.\cite{Lin:2003jy,Acharya:2020jil}. Within the NCQ ansatz, the \velip of the charm ($\velip^c$) and of light quarks ($\velip^q$) can be inferred as described in Ref.~\cite{Acharya:2020jil}, based on the measured \velip of \jpsi and charged pions, respectively. The \velip of D mesons can be obtained as a superposition of the charm and light-quark flow
\begin{equation}
    \velip^D(\pt^D) = \velip^{\rm q}(\pt^{\rm q}) + \velip^{\rm c}(\pt^{\rm c}),
\label{eq:ncqDmeson}
\end{equation}
where q and c superscripts denote the light and charm quarks, respectively. Unlike for light mesons and charmonia, where the momentum fractions carried by the constituent quarks are assumed equal from symmetry reasons, for D mesons different assumptions need to be made. In coalescence motivated models of hadronisation, constituent quarks have equal velocities and thus carry \pt fractions proportional to their effective masses. For D mesons, a simplistic approach considering the mass of the light quark being one third of the proton mass leads to a light-quark \pt fraction of about 0.2 of the D-meson \pt~\cite{Lin:2003jy}. Figure~\ref{fig:NCQscaling} shows the \pt dependence of \velip in the 30--50\% centrality range measured by ALICE at midrapidity for charged pions~\cite{Acharya:2018zuq} and D-mesons~\cite{ALICE:2020iug} (average of prompt D$^0$, D$^+$ and D$^{*+}$ mesons), and at both central and forward rapidity for \jpsi mesons~\cite{Acharya:2020jil}. The red dashed curve is a fit of the \jpsi \velip~\cite{Acharya:2020jil} used to obtain a parameterisation of the charm-quark \velip, while the black solid curve corresponds to the D-meson \velip obtained using Eq.~\ref{eq:ncqDmeson} and assuming $\pt^q = 0.2 \pt^D$. This calculation underestimates the measured D-meson flow at low-\pt and overestimates it at high-\pt, which disfavours the constituent quark coalescence inspired assumption. This approach however largely ignores the contribution of gluons to the D meson kinematics. In this naive study, an attempt to compensate for this missing contribution was made by roughly tuning the momentum share carried by the light quark. Surprisingly, the assumption of equal \pt sharing between the light and charm quarks provides a much better description of the data, with the best fit suggesting a value of $\pt^q = 0.4 \pt^D$. A very good compatibility with the measurements in different centrality ranges was observed for both \velip and $v_3$~\cite{Acharya:2020jil}, which suggests that the flow of charmonia and open charm mesons can be described assuming a common underlying charm-quark flow different from that of the light quarks.

\begin{figure}[ht!]
\begin{center}
\includegraphics[width=0.64\linewidth]{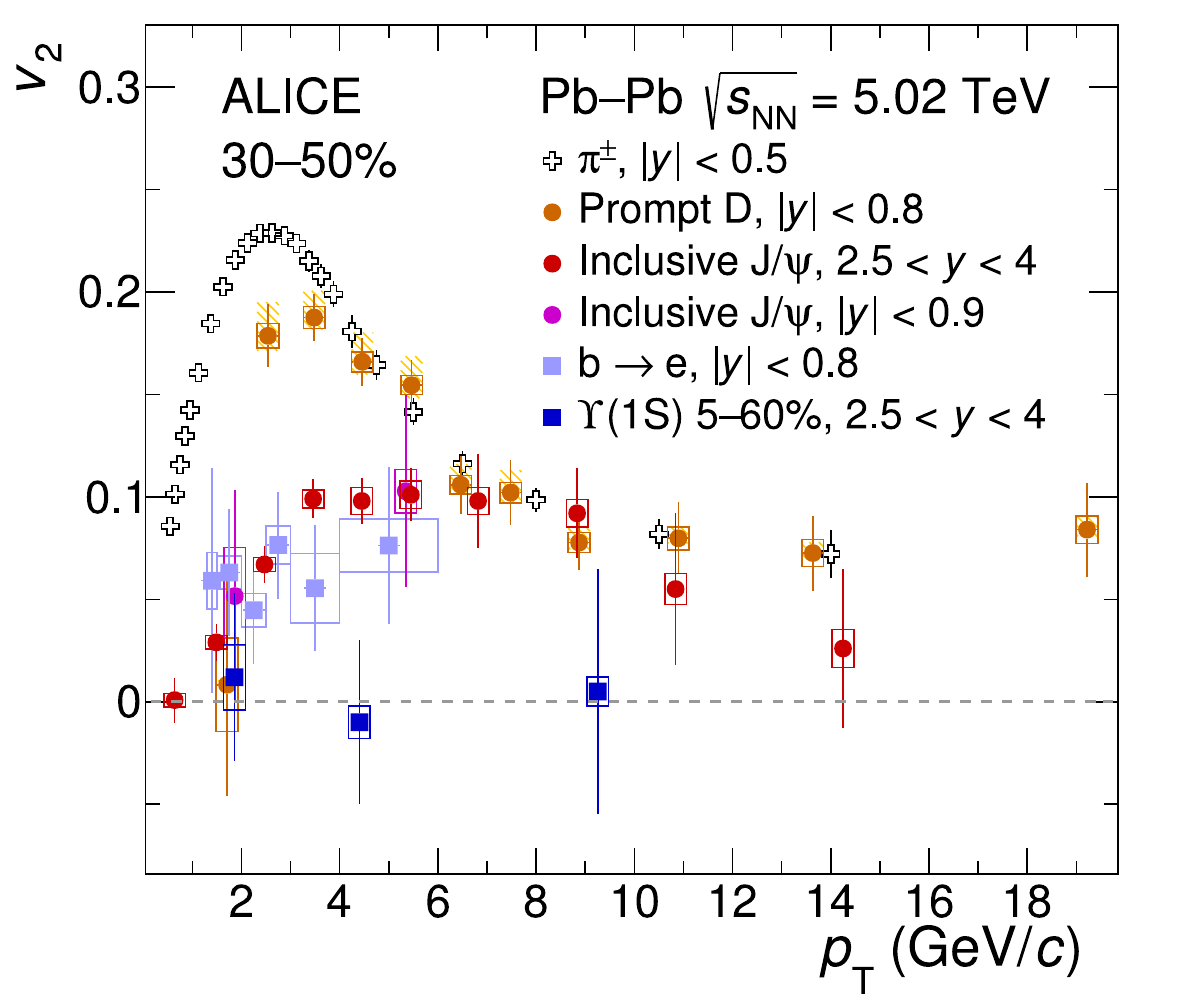}
\caption{The $p_{\rm T}$ dependence of $v_{\rm 2}$ for the centrality range 30--50\% in Pb--Pb collisions at $\sqrt{s_{\rm NN}}=5.02$ TeV. Results are shown for open and hidden charm (D-mesons and J/$\psi$), for open and hidden beauty (electrons from semi-leptonic beauty-hadron decays and $\Upsilon(1S)$), and for pions~\cite{Acharya:2020jil,ALICE:2020iug,ALICE:2020hdw,ALICE:2019pox}.}
\label{fig:LightAndHeavyV2} 
\end{center}
\end{figure} 

Finally, Fig.~\ref{fig:LightAndHeavyV2} shows a comparison of the \pt dependent \velip measurements for charged pions, prompt D mesons, inclusive \jpsi, electrons from beauty-hadron decays and \upsone~\cite{ALICE:2020iug,ALICE:2020hdw,ALICE:2019pox}. A rather clear quark flavour hierarchy is observed in the low-\pt range, for both open and hidden heavy flavour hadron species, with the beauty hadrons exhibiting the least amount of flow. Both open and hidden charm hadrons show a significant amount of anisotropic flow, suggesting that charm quarks are at least partly thermalised in the QGP medium. In addition, since the low-\pt \jpsi are not expected to develop a significant collective anisotropic flow on their own, their relatively large observed flow supports the scenario of \jpsi formation via (re)combination during the late stages of the collision, when the charm-quark flow is fully developed.
The elliptic flow of the electrons from beauty hadron decays, measured up to the electron $\pt=6$~\GeVc, is a convolution of the \velip of beauty hadrons  mostly from the \pt range $\pt<15$~\GeVc and their decay kinematics. Although the \pt dependence of the open beauty hadrons \velip cannot be easily inferred, this measurement clearly shows that open beauty hadrons exhibit flow, even if significantly smaller with respect to open charm hadrons. This observation constitutes also a hint that beauty quarks may participate to some extent to the collective flow of the plasma, 
although it is difficult to disentangle the contribution of the b and of light quarks in producing the observed b-hadron flow. In the case of hidden beauty hadrons, namely inclusive \upsone, the \velip measurements using Run 2 data are compatible with zero elliptic flow and are estimated to be 2.6$\,\sigma$ lower than the \jpsi \velip. Within the large uncertainties of these measurements, such an observation is compatible with the expectations of a negligible contribution from (re)generation in the beauty sector. 

\subsubsection{Conclusions}

    \paragraph{Charmonium production.} In Pb--Pb collisions at LHC energies the J/$\psi$ production and the observed anisotropy in its azimuthal distributions  is dominated, at low-$p_{\rm T}$ and for central events, by a (re)generation effect due to the (re)combination of the charm quarks abundantly produced in the collision and at least partially thermalised. This observation constitutes a proof of deconfinement, as it implies that coloured partons can move freely over distances much larger than the hadronic scale. Results on the weakly bound $\psi(2S)$ state show that this resonance exhibits a stronger suppression compared to J/$\psi$ in all the explored $p_{\rm T}$ interval. Hints for the presence of (re)combination effects were also detected. 
    
   \paragraph{Bottomonium production.} The bottomonium state \upsone is suppressed by more than 50\% in central Pb--Pb collisions, with a nearly flat transverse momentum dependence of the $R_{\rm AA}$ that indicates that (re)generation effects play a small role, as expected because of the much lower multiplicity of b with respect to c quarks. The less strongly bound \upstwo is found to be more strongly suppressed (up to about 90\% in central Pb--Pb collisions), an observation consistent with a sequential suppression of the bottomonium states.
    
    \paragraph{Initial state effects.} The influence of gluon shadowing/saturation is not negligible, with a 30\% suppression measured at low-$p_{\rm T}$ for \jpsi and \upsone. Also, feed-down from higher mass charmonium and bottomonium resonances contributes about 20\% and 30\% at low-$p_{\rm T}$, respectively. When considering these effects in a semi-quantitative way, it turns out that: $i)$ they could be responsible for a significant fraction of the measured inclusive \upsone suppression, $ii)$ that production of low-\pt direct \jpsi is larger than expected from binary scaled pp collisions. 

\newpage

\subsection{Early electromagnetic fields and novel QCD phenomena}
\label{sec:NovelQCD}

\definecolor{dgreen}{cmyk}{1.,0.,1.,0.25}        %
\definecolor{dblue}{rgb}{0.,0.,1.}       
\definecolor{orange}{cmyk}{0.,0.453,1.,0.}    %
\newcommand{\blue}[1]{\textcolor[named]{Blue}{#1}} %
\newcommand{\red}[1]{\textcolor[named]{red}{#1}}
\newcommand{\magenta}[1]{\textcolor[named]{magenta}{#1}}
\newcommand{\green}[1]{\textcolor[named]{Green}{#1}}
\newcommand{\orange}[1]{\textcolor{orange}{#1}}
\newcommand{\dblue}[1]{\textcolor{dblue}{#1}}
\newcommand{\dgreen}[1]{\textcolor{dgreen}{#1}}
\def \new {\dblue}
\def \old {\orange}
\def \ask#1{\magenta{\bf [#1]}}
\def \note#1{\dgreen{\bf [#1]}}

Heavy-ion collisions provide a unique opportunity to study intriguing, novel QCD phenomena that are not
directly accessible elsewhere, including local parity (P) and charge conjugation parity symmetry (CP) violating
effects in the strong
interaction~\cite{Lee:1973iz,Lee:1974ma,Morley:1983wr,Kharzeev:1998kz,
  Kharzeev:1999cz,Kharzeev:2015kna,Kharzeev:2007tn,Kharzeev:2007jp,
  Fukushima:2008xe}. Even though many potential explanations have been
proposed~\cite{Peccei:1977hh,Weinberg:1977ma,Wilczek:1977pj,Kim:1979if,
  Shifman:1979if,Cheng:1987gp,Mannel:2007zz}, it is still unclear why 
P and CP invariances are 
respected in the strong interaction. This is known as the strong-CP 
problem~\cite{Shifman:1979if,Banerjee:2000qw} which still continues to be 
one of the remaining puzzles of the Standard Model.

The uniqueness of heavy-ion collisions as a tool for the study of the strong P and CP violation effects stems from 
 the presence of the strongest magnetic
field in nature, of the order of
$10^{19}$~gauss~\cite{Skokov:2009qp,Bzdak:2011yy,Deng:2012pc}, 
produced by the colliding positively charged nucleons, mostly protons, that do not participate in the interaction
(so-called spectators) in non-central collisions. 
This field develops perpendicular to the reaction plane, the plane defined by the impact parameter and the beam axis. The magnetic field, that is
rapidly decaying~\cite{Toneev:2011aa,Shi:2017cpu} with a rate that
depends on the electric conductivity of the medium, besides leading to observable effects of the strong P and CP violation, could have direct
implications on the kinematics of final state particles. 
The latter is used for detecting and estimating the strength of the electromagnetic fields.
While so far there exists no unambiguous experimental evidence for the existence of such strong electromagnetic fields, ALICE measurements reveal several important hints, which are discussed below.

The results from ALICE searches for the effects of the early stage electromagnetic fields will be discussed first. This section will be followed by the description of our studies for the discovery of novel QCD phenomena, associated with local P and CP violating effects in the strong interaction, referred to in the literature as the Chiral Magnetic Effect (CME) and the Chiral Magnetic Wave (CMW).

\subsubsection{Searches for the early stage electromagnetic fields}

\textit{\textbf{Charge-dependent directed flow measurements}}

It is proposed in Refs.~\cite{Gursoy:2014aka,Gursoy:2018yai} that the strong initial state electromagnetic fields significantly affect the evolution of the produced QGP, thus leaving distinct imprints in the final state particle distributions. Charged particles
propagating in the electromagnetic field experience a Lorentz force. Strong
electric fields can also be generated due to the decay of the
magnetic field with time through Faraday's law. In the
centre-of-mass of the collision system the two effects (i.e., the Lorentz force and the
force due to the induced electric field) have opposite sign; both effects
are also antisymmetric in rapidity (i.e., rapidity-odd) and could be detected experimentally. 
This would, consequently, lead to the first experimental constrains on the value of the electric conductivity of the QGP.

A way to probe the electromagnetic fields and their rapidity-odd induced effects is the measurement of the
charge-dependent anisotropic flow $v_n$.  In particular, the charge
dependence of the directed flow, $v_1(\eta)$, of the produced
particles relative to the spectator plane is directly
sensitive to the presence of the electromagnetic fields.  The
spectator plane is defined by the deflection direction of the
collision spectators, measured with the Zero Degree Calorimeter (ZDC),
and is strongly correlated with the direction of the magnetic
field. As the production times for light (mainly produced through
gluon-splitting processes throughout the entire evolution of the
system) and heavy quarks (predominantly produced in early stage hard
scattering processes) are vastly different, the directed flow of
hadrons containing a c-quark is expected to be more sensitive to the
early times of the system evolution, of the order of 0.1~fm$/c$.

\begin{figure}[t!b]
    \begin{center}
    \includegraphics[width = 0.8\textwidth]{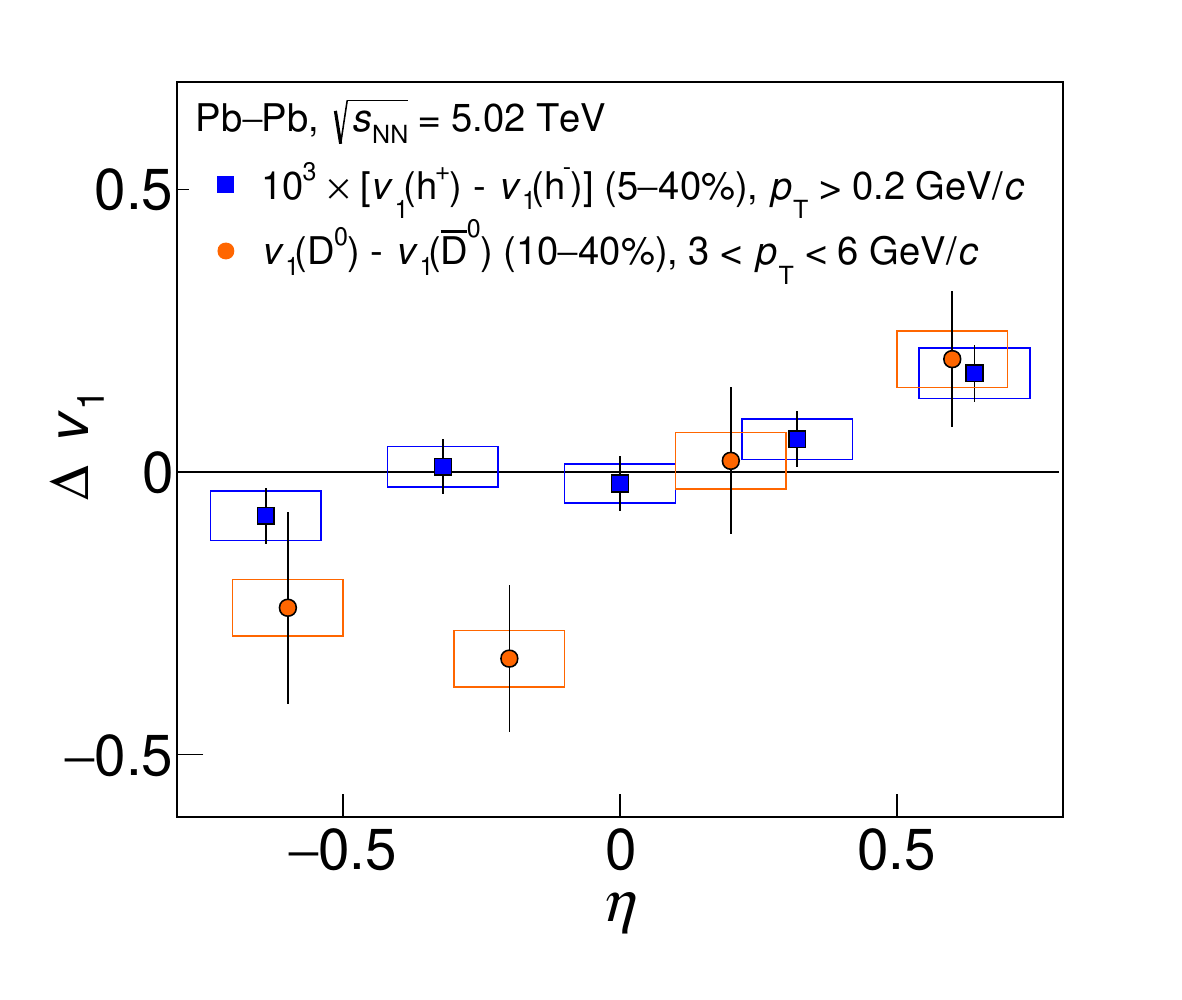}
    \end{center}
    \caption{The pseudorapidity dependence of $\Delta v_1$ for charged
      particles (blue markers) and D mesons (orange markers)
      in mid-central Pb--Pb collisions at $\sqrt{s_{\mathrm{NN}}} =
      5.02$~TeV~\cite{Acharya:2019ijj}. The data points for charged
      particles are scaled by a factor $10^3$. Statistical and systematic uncertainties are represented by the error bars and the boxes around each data point, respectively.}
    \label{fig:deltaV1}
\end{figure}

Figure~\ref{fig:deltaV1} presents the differences of the
charge-dependent $v_1$, denoted as $\Delta v_1$, in mid-central Pb--Pb
collisions at $\sqrt{s_{\mathrm{NN}}} = 5.02$~TeV as a function of
pseudorapidity for charged particles and
D$^0$ mesons~\cite{Acharya:2019ijj}. The magnitude of the effect is smaller for charged particles than for D$^0$ mesons, as expected from the fact that charm
quarks are more sensitive probes of the magnetic field at earlier
times, and is thus scaled up by a factor $10^3$ for visibility. The
rapidity slope $\mathrm{d}\Delta v_1/\mathrm{d}\eta$, extracted with a
linear fit function, yields $[1.68 \pm 0.49~\mathrm{(stat.)}~\pm 0.41~\mathrm{(syst.)}]\times 10^{-4}$ for charged hadrons with
$p_{\mathrm{T}} > 0.2$~GeV$/c$ for the 5--40\% centrality interval and $[4.9 \pm 1.7~\mathrm{(stat.)}~\pm 0.6~\mathrm{(syst.)}]\times 10^{-1}$ for D mesons with $3 <
p_{\mathrm{T}} < 6$~GeV$/c$ in a centrality interval of 10--40\%, resulting in a significance of
2.6$\sigma$ and 2.7$\sigma$ for having a positive value,
respectively. This apparent large slope could point to a stronger
effect of the Lorentz force relative to the contribution from the
induced electric field and the initial tilt of the source. This
measurement constitutes the first experimental hint of the existence
of the initial state electromagnetic fields at the LHC. The data samples that
will be collected in the upcoming heavy-ion runs at the LHC will give
us the potential to make a decisive step in this direction.

\textit{\textbf{Hyperon polarisation}}

An independent way of probing and constraining the value of the
magnetic field is provided by the measurements of the hyperon global
polarisation (see also Sec.~\ref{sec:TG2globalpolarization}).

Fluid vorticity, the main mechanism for the global polarisation
effect, leads to the same polarisation between particles and
antiparticles. The magnetic field, which is oriented in the same
direction as the average vorticity, on the other hand, couples to the
magnetic moment ($\mu$) of a particle $q$ and leads to a particle 
polarisation $P^{B}_{q} = \mu_q B/2T = Q_q B/(2m_q T)$, opposite for
particles and antiparticles with charge $Q_q$. In the previous
formula, $T$ is the temperature of the system at the time of the
particle formation and $B$ is the value of the magnetic field. This
opens up the possibility to probe the magnetic field with a different,
independent observable.

\begin{figure}[t!b]
    \begin{center}
    \includegraphics[width = 0.8\textwidth]{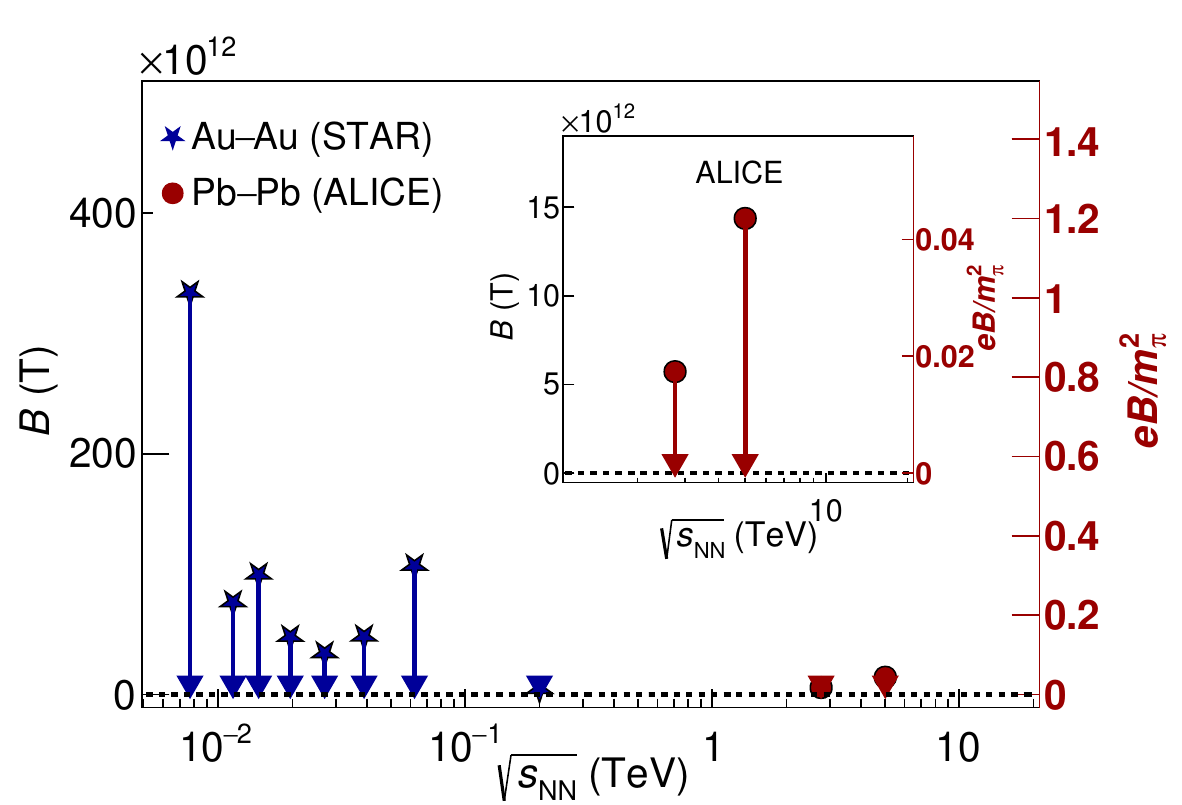}
    \end{center}
    \caption{The $\sqrt{s_{\mathrm{NN}}}$ dependence of the upper limit for the magnitude
      of the magnetic field at freeze-out, estimated from the global polarisation
      results of $\Lambda$ and $\overline{\Lambda}$. The data points
      are extracted from the results reported in
      Refs.~\cite{STAR:2017ckg,Acharya:2019ryw}.}
    \label{fig:bField}
\end{figure}

The first observation of the finite global polarisation of
$\Lambda$ hyperons in Au--Au collisions at RHIC~\cite{STAR:2017ckg}
revealed that the perfect liquid created in such collisions is also
the most vortical one. The ALICE measurement of global polarisation at
the LHC~\cite{Acharya:2019ryw} although presently by large
statistical and systematic uncertainties 
indicates a decrease with increasing
collision energy, consistent with expectations based on the correlation between
vorticity and the slope of directed flow as a function of
centre-of-mass energy~\cite{Voloshin:2017kqp}. Although no significant
difference was found in the polarisation between $\Lambda$ and
$\overline{\Lambda}$, one can use the relationships $P_{\Lambda}
\approx 0.5\omega/T + |\mu_{\Lambda}|B/T$ and $P_{\overline{\Lambda}}
\approx 0.5\omega/T - |\mu_{\Lambda}|B/T$, where $\omega$ is the
vorticity of the QGP, to constrain the value of the magnetic field at
freeze-out by evaluating $(P_{\Lambda} -
P_{\overline{\Lambda}})$. Figure~\ref{fig:bField} presents the
centre-of-mass energy dependence of such an estimate of the value of
the magnetic field. At LHC energies, one obtains an upper limit of 
$eB/m_{\pi}^2$ of 0.017 (corresponding to $5.7 \times 10^{12}$~T) and
0.044 (or $14.4 \times 10^{12}$~T) at a 95\% confidence level for Pb--Pb 
collisions at $\sqrt{s_{\mathrm{NN}}} = 2.76$~and 5.02~TeV, respectively.

\subsubsection{Searches for chiral anomalies}

The ground state of the QCD vacuum consists of a superposition of
topologically distinct states~\cite{Belavin:1975fg}. These states,
characterised by a topological charge, are separated by a potential
barrier and can be connected through tunneling transitions, called
instantons~\cite{Belavin:1975fg,tHooft:1976rip,tHooft:1976snw,tHooft:1986ooh,Gross:1980br,Rapp:1999qa}.
However, due to the height of the potential barrier, which is of the
order of the QCD scale over the strong coupling constant ($\Lambda_{\mathrm{QCD}}/\alpha_\mathrm{s}$), instanton
transitions are suppressed~\cite{tHooft:1976rip,tHooft:1976snw}. At
high temperatures, like the ones reached when the QGP is created in
collisions between two heavy ions, the transitions between two states
with different topological quantum numbers can take place by jumping
over the barrier, and are known as
sphalerons~\cite{Manton:1983nd,Klinkhamer:1984di,
  Kuzmin:1985mm,Arnold:1987mh,Arnold:1987zg}. In electroweak theory,
transitions between vacuum states with different topological charges
are associated with baryon and lepton number violation~\cite{Rubakov:1996vz}. In QCD, they
are connected to the non-conservation of chirality. The existence of
these transitions is very well motivated by theoretical studies and would lead to the QGP that possessing domains with non-zero net chirality, with a sign that changes
from event to event i.e., leading to an event-by-event local P and 
CP violation.

\textit{\textbf{ Chiral magnetic effect studies}}

The possibility of observing parity violation in the strong interaction 
in relativistic heavy-ion collisions was first pointed out 
in Refs.~\cite{Lee:1973iz,Lee:1974ma,Morley:1983wr} and was further
discussed in Refs.~\cite{Kharzeev:1998kz,Kharzeev:1999cz,Kharzeev:2015kna,
  Kharzeev:2007tn,Kharzeev:2007jp,Fukushima:2008xe}. 
A significant development happened in 2004, when it was noticed that in the presence of a chiraly restored medium,
a system of quarks with non-zero net-chirality immersed in a
strong magnetic field leads to an excess of positively-charged
particles moving along the magnetic field direction and an excess of
negative particles moving in the opposite direction. This introduces a
net electromagnetic current and creates an electric dipole
moment of QCD matter. This phenomenon is called the Chiral Magnetic Effect
(CME)~\cite{Fukushima:2008xe}, and its existence was recently reported
in semi-metals like zirconium pentatelluride
($\rm ZrTe_5$)~\cite{Li:2014bha}.

The resulting charge separation can be identified by studying the
$P$-odd sine terms in the Fourier decomposition of the particle
azimuthal distribution~\cite{Voloshin:1994mz,Voloshin:2004vk}
according to
\begin{eqnarray}
\frac{{\rm d}N}{{\rm d}\varphi_{\alpha}} \propto 1 + 2\sum_n \left[v_{\it n,\alpha}
  \cos(n \Delta \varphi_{\alpha}) + \mathit a_{\it n,\alpha} \sin(n \Delta 
\varphi_{\alpha})\right],
\label{Eq:Fourier}
\end{eqnarray}
where $\Delta \varphi_{\alpha}=\varphi_\alpha - \Psi_{\rm RP}$ is the
azimuthal angle $\varphi_\alpha$ of the particle of type $\alpha$
(either positively or negatively charged) relative to the reaction
plane angle $\Psi_{\rm RP}$. The coefficients $v_{n,\alpha}$ are the
$n$-th order Fourier harmonics that characterise the anisotropies in
momentum space. The leading order P-odd coefficient $a_{1,\alpha}$
reflects the magnitude of the effects from local parity violation,
while higher order terms ($a_{\it n,\alpha}$ for $n > 1$) describe the
specific shape in azimuth. The chiral imbalance that leads to the
creation of the CME fluctuates from event to event, and the event
average $\langle a_{1,\alpha} \rangle$ is consistent with zero,
i.e. compatible with the observation of global parity conservation in strong interactions. Consequently, the effect can be detected only by
correlations studies.

In Ref.~\cite{Voloshin:2004vk}, it was suggested that a suitable way
to probe the CME is via two-particle correlations measured relative to
the second-order symmetry plane of the form $\gamma_{1,1} = \langle
\cos(\varphi_{\alpha} + \varphi_{\beta} - 2\Psi_{2}) \rangle$, where
the brackets indicate an average over all events, and $\alpha$ and
$\beta$ denote particles with the same or opposite electric
charge. The advantage of using this observable is that it probes
correlations between two leading order P-odd coefficients
$a_{1,\alpha}$ and $a_{1,\beta}$ which do not average to zero over all
events, while suppressing correlation that do not depend on the symmetry 
plane orientation. In order to independently evaluate the
contributions from correlations in- and out-of-plane, one measures at
the same time a two-particle correlator of the form $\delta_1 =
\langle \cos(\varphi_{\alpha} - \varphi_{\beta}) \rangle$. Both
correlators can be generalised according to:
\begin{eqnarray}
\label{eq:moments}
\gamma_{m,n} = \langle \cos(m\varphi_{\alpha} +
n\varphi_{\beta} - (m + n)\Psi_{|m+n|})\rangle , 
\label{Eq:Generalised3ParticleCorrelator}
\end{eqnarray}
\begin{eqnarray}
\delta_m = \langle \cos[m(\varphi_{\alpha} -
\varphi_{\beta})] \rangle ,
\label{Eq:Generalised2ParticleCorrelator}
\end{eqnarray}
\noindent where $\mathrm{m}$ and $\mathrm{n}$ are integers.

The observable of Eq.~\ref{Eq:Generalised3ParticleCorrelator} is
constructed as the difference between correlations in- and
out-of-plane and is thus expected to suppress background effects
approximately by a factor of $v_2$. On the other hand, the correlator
of Eq.~\ref{Eq:Generalised2ParticleCorrelator}, owing to its
construction, is affected (if not dominated) by background
contributions. This background was recently identified as stemming from an interplay between local charge conservation embedded in an environment that exhibits azimuthal anisotropy~\cite{Schlichting:2010qia}. Its isolation and the subsequent quantification of the CME contribution to these observables has been the main focus of this line of research at both RHIC and LHC.

\begin{figure}[t!b]
    \begin{center}
     \includegraphics[width = 1\textwidth]{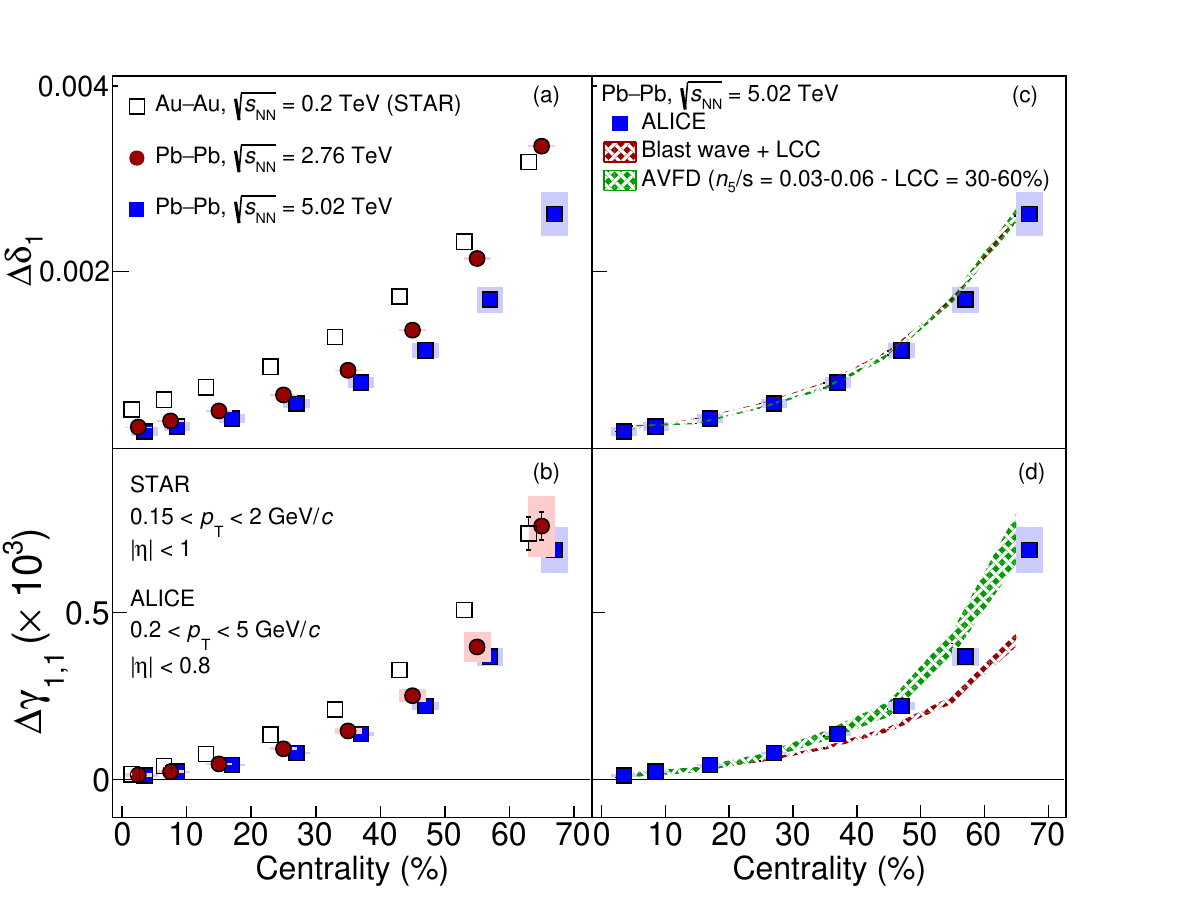}
    \end{center}
    \caption{The centrality dependence of $\Delta \delta_1$ (a) and
      $\Delta \gamma_{1,1}$ (b), measured at
      RHIC~\cite{Abelev:2009ac,Abelev:2009ad} and LHC
      energies~\cite{Abelev:2012pa,ALICE:2020siw}. Statistical and systematic uncertainties are represented by the error bars and the boxes around each data point, respectively. Comparison of the
      centrality dependence of $\Delta \delta_1$ (c) and $\Delta
      \gamma_{1,1}$ (d), measured at the highest LHC
      energy~\cite{ALICE:2020siw} and estimated with a blast wave~\cite{Retiere:2003kf} and the AVFD~\cite{Shi:2017cpu,Jiang:2016wve} 
      models (see text for details).}
    \label{fig:cmeResults}
\end{figure}

Figure~\ref{fig:cmeResults} presents the centrality dependence of
$\Delta \delta_1$ and $\Delta \gamma_{1,1}$ i.e., the difference
of $\delta_1$ and $\gamma_{1,1}$ between opposite- and same-sign pairs
in panels (a) and (b), respectively. The data points that correspond to
results from the STAR Collaboration in Au--Au collisions at
$\sqrt{s_{\mathrm{NN}}} = 200$~GeV~\cite{Abelev:2009ac,Abelev:2009ad}
and the ones from Pb--Pb collisions at $\sqrt{s_{\mathrm{NN}}} =
2.76$~and 5.02~TeV~\cite{Abelev:2012pa,ALICE:2020siw} have a
characteristic and significant centrality dependence.
The correlator $\Delta \delta_1$ is related to the balance function
also studied at the LHC~\cite{Abelev:2013csa,Adam:2015gda}. The
results of $\Delta \delta_1$ are qualitatively consistent with the ones in
Refs.~\cite{Abelev:2013csa,Adam:2015gda}, i.e. oppositely
charged particles are more tightly correlated in central events
resulting in a narrowing of the balance function width in $\Delta
\varphi$ which results in the observed centrality evolution of $\Delta
\delta_1$. There is a clear evolution with centre-of-mass energy for
$\Delta \delta_1$, consistent with what is reported in
Ref.~\cite{Abelev:2013csa}. This energy dependence, however, is less
pronounced for $\Delta \gamma_{1,1}$, between the two
LHC energies. This lack of a significant energy dependence for $\Delta \gamma_{1,1}$ initially came as a surprise considering the
differences in centre-of-mass energy, which between RHIC and LHC is
one order of magnitude, and that the magnitude of the magnetic field and the way it
evolves is, in principle, different between the two energies. In
addition, the particle density is almost three times larger at the LHC
compared to RHIC. This might lead to a significant dilution of the
CME signal, if any, at this higher energy. Finally, the acceptance
used in the two experiments is slightly different, with ALICE using a
smaller range in $\eta$. This, in turn, could lead to a more
pronounced contribution from both any potential CME signal and
background. At the same time, however, there is no significant energy
dependence in the effects that constitute the background to these
measurements. In particular, preliminary studies indicate that the
correlations between balancing charges, as reflected in the width of
the balance function, do not exhibit any significant dependence on
collision energy. Furthermore, the values of $v_2$ measured in Pb--Pb
collisions at $\sqrt{s_\mathrm{{\rm NN}}} = 5.02$~TeV are around 2\%
higher than the values at $\sqrt{s_\mathrm{{\rm NN}}} =
2.76$~TeV~\cite{Acharya:2018lmh}.

Figure~\ref{fig:cmeResults}(c) and (d) present the comparison between
the experimental data points of both $\Delta\delta_1$ and $\Delta
\gamma_{1,1}$ in Pb--Pb collisions at $\sqrt{s_\mathrm{{\rm NN}}} =
5.02$~TeV with expectations from a blast wave (BW) model based on a
parameterisation from Ref.~\cite{Retiere:2003kf}.  Local charge
conservation (LCC) is additionally incorporated by generating
ensembles of particles with zero net charge produced at the same
spatial location, uniformly distributed over the anisotropic particle radiating source.  The input
parameters of the model are tuned to describe the $p_{\rm T}$
spectra~\cite{Acharya:2019yoi} and the $p_{\rm T}$-differential
$v_{2}$ values~\cite{Acharya:2018zuq} for charged pions, kaons, and
protons (antiprotons) measured in Pb--Pb collisions at
$\sqrt{s_\mathrm{{\rm NN}}} = 5.02$~TeV (see
Ref.~\cite{ALICE:2020siw} for details). In addition, the number of
sources that emit balancing pairs is tuned separately for each
centrality interval to reproduce the centrality dependence of $\Delta
\delta_1$, as presented in the panel (c) of
Fig.~\ref{fig:cmeResults}. The tuned model is then used to extract the
expectation for the centrality dependence of $\Delta \gamma_{1,1}$,
shown in the panel (d) of Fig.~\ref{fig:cmeResults}. The BW model
underestimates the measured data points by as much as
$\approx$~40$\%$, with the disagreement increasing progressively for
more peripheral events. A similar study performed in Xe--Xe collisions at $\sqrt{s_\mathrm{{\rm NN}}} =
5.44$~TeV~\cite{Collaboration:2022flo}, revealed that the data points can be described quantitatively by the tuned BW model.

The same figure presents the results from the Anomalous Viscous Fluid
Dynamics (AVFD) model~\cite{Shi:2017cpu,Jiang:2016wve}. This model
simulates the evolution of the chiral fermion currents in the QGP on
top of the hydrodynamic evolution of a heavy-ion collision as
prescribed by VISHNU~\cite{Shen:2014vra} which couples 2+1 dimensional
viscous hydrodynamics (VISH2+1) to a hadron cascade model
(UrQMD)~\cite{Bass:1998ca}. The model allows one to tune the value of the
axial current density to entropy ratio ($n_5/\mathrm{s}$)
that controls the imbalance between left- and right-handed quarks. In
addition, it includes local charge conservation effects by
emitting particle pairs of opposite charges from the same fluid
element. The procedure that was followed relied on finding the proper value of both
$n_5/\mathrm{s}$ and the percentage of balancing charged
particles (LCC) to reproduce simultaneously the centrality dependence
of $\Delta \delta_1$ and $\Delta \gamma_{1,1}$. This is illustrated in
panels (c) and (d) of Fig.~\ref{fig:cmeResults}, where the AVFD curves
are represented by the green bands~\cite{Christakoglou:2021nhe}. The 
model is able to describe the measurements with a set of input parameters 
that ranges between 0.03 and 0.06 for $n_5/\mathrm{s}$ throughout 
all centrality intervals while the percentage of charged particles emitted 
as part of an oppositely charged pair relative to the total multiplicity 
of an event varies from 30$\%$ up to 60$\%$ with an increasing trend for
more central events.  In addition, a background only scenario, i.e.,
$n_5/\mathrm{s} = 0$ for all centralities, is not compatible
with the measurements according to this model~\cite{Christakoglou:2021nhe}. 
Finally, a similar study performed in Xe--Xe collisions at 
$\sqrt{s_\mathrm{{\rm NN}}} = 5.44$~TeV~\cite{Collaboration:2022flo} has 
yielded systematically lower values of $n_5/\mathrm{s}$, and a 
background-only scenario being, at this time, compatible with the data~\cite{Christakoglou:2021nhe}.

\begin{figure}[t!b]
    \begin{center}
    \includegraphics[width = 0.83\textwidth]{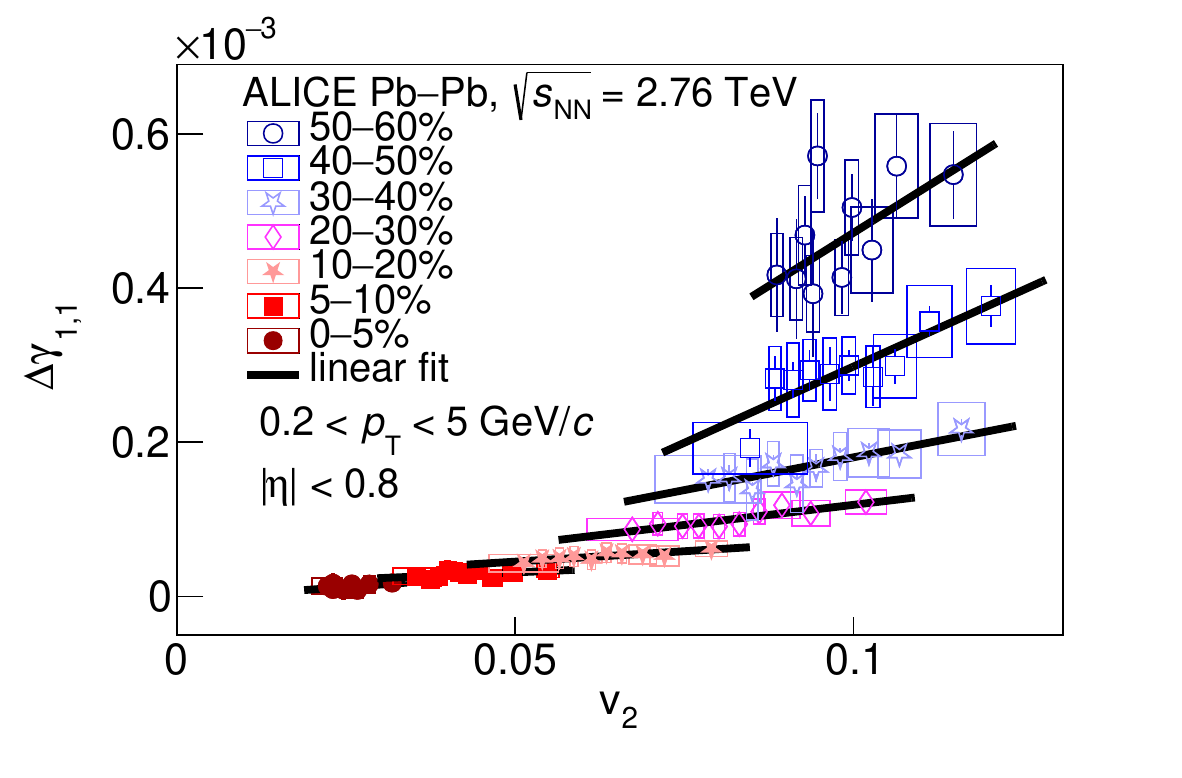}
    \includegraphics[width = 0.83\textwidth]{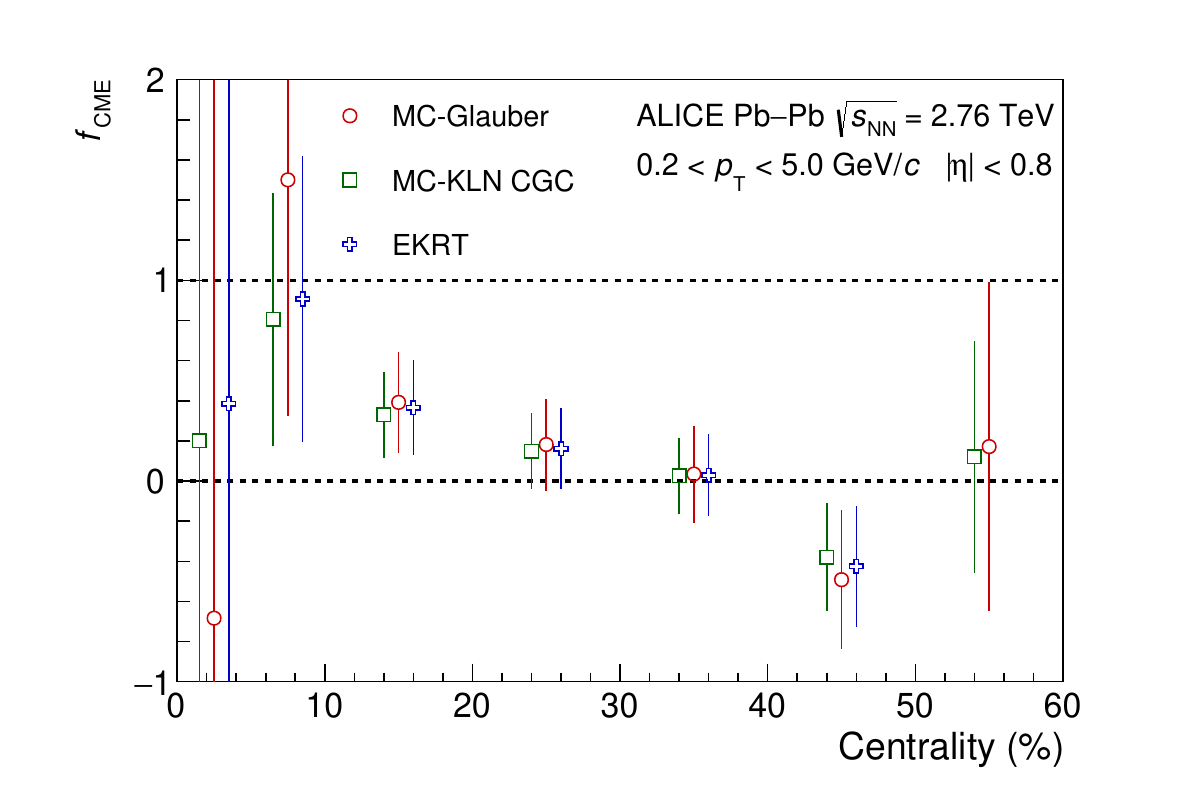}
    \end{center}
    \caption{(Top) The dependence of $\Delta \gamma_{1,1}$ on $v_2$ for shape selected events together with a linear fit (dashed lines) for various centrality classes. (Bottom) The centrality dependence of the CME
      fraction, $f_{\mathrm{CME}}$, extracted from the slope parameter
      of fits to data and MC-Glauber, MC-KLN CGC, and EKRT~models,
      respectively. The dashed lines indicate the physical parameter
      space of the CME fraction, while the points are slightly shifted
      along the horizontal axis for better visibility. Figures from Ref.~\cite{Acharya:2017fau}. 
        }
    \label{fig:cmeLimitsESE}
\end{figure}

\begin{figure}[t!b]
    \begin{center}
    \includegraphics[width = 0.8\textwidth]{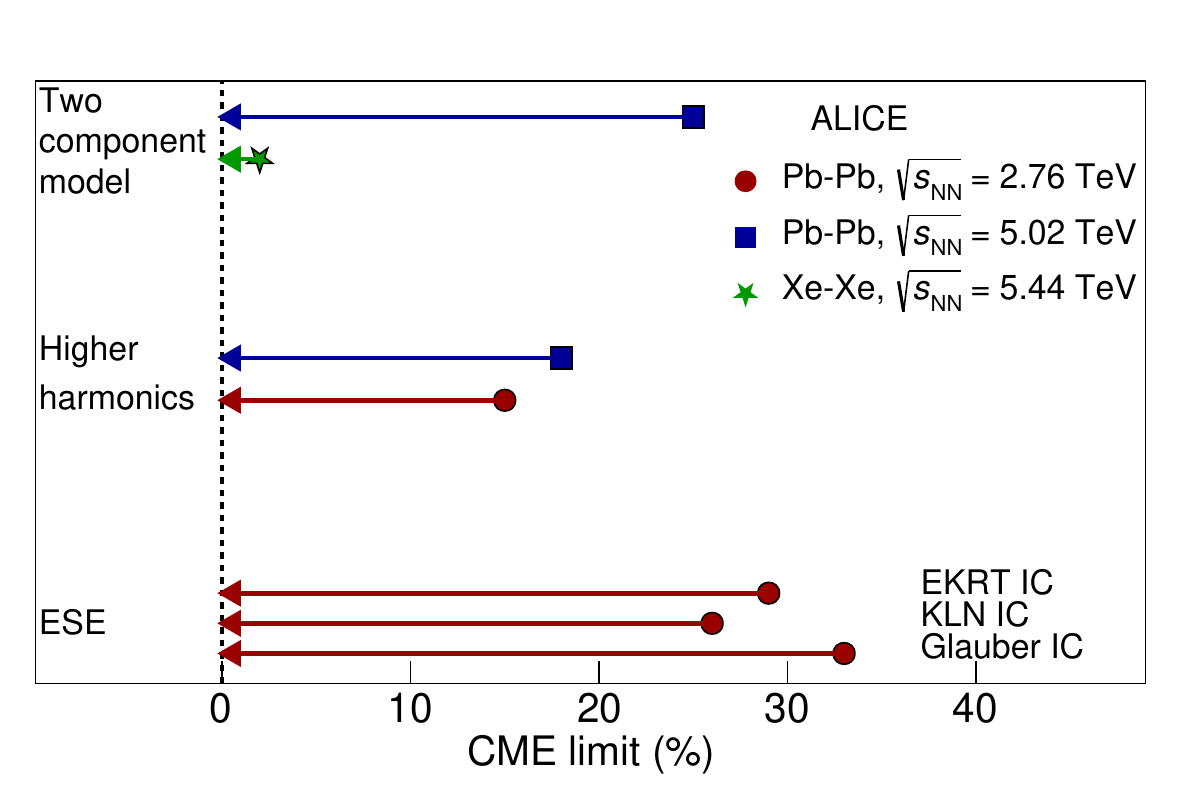}
    \end{center}
    \caption{Summary of the results for the CME limit obtained from different
      analyses performed at various LHC energies and colliding
      systems integrated over centralities (see text for details). 
        }
    \label{fig:cmeLimits}
\end{figure}

Data-driven methods could offer a more robust approach to constrain
and quantify the dominating background while in parallel isolating the
signal that comes from the CME. The first measurement in this direction
was presented by the ALICE Collaboration in Ref.~\cite{Acharya:2017fau},
using an innovative method proposed and developed
in~\cite{Schukraft:2012ah}. This method, called Event Shape
Engineering (ESE), utilises the strong fluctuations of initial
geometry (i.e.,~the position of participating nucleons) even at a fixed
impact parameter and allows one to select events with different initial
system shapes, e.g., central Pb--Pb collisions with large initial
anisotropy.
This allows one to select events where the main component of
the background, the value of $v_2$, can be varied. The top panel of Fig.~\ref{fig:cmeLimitsESE} illustrates the linear dependence of $\Delta \gamma_{1,1}$ on $v_2$ for different shaped events and centrality ranges, indicative of the dominance of background contributions. This becomes more evident seeing all data points converging on a single line when $\Delta \gamma_{1,1}$ is scaled with the relevant particle density for each centrality interval to account for the dilution of these correlations stemming from the increased multiplicity~\cite{Acharya:2017fau}. The bottom panel of Fig.~\ref{fig:cmeLimitsESE} presents the CME fraction,
$f_{\mathrm{CME}}$ which is extracted from the slope parameters of
fits to data and MC-Glauber, MC-KLN CGC, and EKRT~models, for
various centrality intervals of Pb--Pb collisions at
$\sqrt{s_\mathrm{{\rm NN}}} = 2.76$~TeV~\cite{Acharya:2017fau}. All data points corresponding to different initial state models (MC-Glauber, MC-KLN CGC, and EKRT), are consistent with zero within the
uncertainties. This study, by combining the data points in the 10--50\% centrality interval, set an upper limit of 26-33$\%$ at $95\%$
confidence level for the CME signal contribution to $\Delta
\gamma_{1,1}$.

Figure~\ref{fig:cmeLimits} presents a summary of
the upper limits for the CME signal 
contribution to $\Delta \gamma_{1,1}$ obtained from different analyses
performed at various LHC energies and colliding systems integrated over centralities. 
These include the data points from studies using the ESE
technique~\cite{Acharya:2017fau} for the three initial state models
used, analysis of the correlators including higher harmonics at the two different LHC energies~\cite{ALICE:2020siw}, 
and from the comparison of the correlations in Xe--Xe and Pb--Pb   collisions in the two-component model~\cite{Collaboration:2022flo}. The higher
harmonics results are based on measuring charge-dependent correlations 
relative to the third order symmetry plane of the form $\gamma_{1,2} = \langle\cos(\varphi_{\alpha} + 2\varphi_{\beta} - 3\Psi_3)\rangle$. Considering 
that the third order symmetry plane
($\Psi_3$) is very weakly correlated with $\Psi_2$~\cite{Aad:2014fla},
the charge separation effect relative to the third harmonic symmetry
plane is expected to be negligible and the relevant correlations are
expected to reflect mainly, if not solely, background effects. In this
background-only scenario, one could approximate $\Delta \gamma_{1,1}$
and $\Delta \gamma_{1,2}$ according to $\Delta \gamma_{1,1} \propto
\kappa_2 \Delta \delta_1 v_2$ and $\Delta \gamma_{1,2} \propto
\kappa_3 \Delta \delta_1 v_3$, with $\kappa_2$ and $\kappa_3$ being
proportionality constants~\cite{ALICE:2020siw}. The method relies heavily on the
assumption that $\kappa_2 \approx \kappa_3$ if both correlators are
studied in the same experimental setup and for the same kinematic
region and provides an upper limit of 15--18$\%$ at 95$\%$ confidence
level for the 0--40$\%$ centrality interval. Finally, the analysis of
Xe--Xe collisions at $\sqrt{s_\mathrm{{\rm NN}}} = 5.44$~TeV~\cite{Collaboration:2022flo}
revealed similar results for $\Delta \gamma_{1,1}$ as the ones in
Pb--Pb collisions at $\sqrt{s_\mathrm{{\rm NN}}} = 5.02$~TeV at the
same multiplicity. The CME fraction was estimated in this analysis 
using a two-component model, similar to the one proposed in 
Ref.~\cite{Deng:2016knn}. The first component is associated to the 
CME contribution in the measurement of $\Delta \gamma_{1,1}$ which 
was considered to be correlated with the value of the magnetic field, 
modelled with a similar prescription as the one followed in
Ref.~\cite{Acharya:2017fau}. On the other hand, the background
contribution to $\Delta \gamma_{1,1}$ was associated to the values of
$v_2$ in the two systems~\cite{Acharya:2018ihu} and was scaled by the 
corresponding charged particle multiplicity density~\cite{Acharya:2018hhy} 
to account for dilution effects. The study resulted in an upper limit 
of around 2\% and 25$\%$ at 95$\%$ confidence level for the 0--70$\%$ centrality 
interval in Xe--Xe and Pb--Pb collisions, respectively.

Searches for the Chiral Magnetic Effect have been also pursued by
the CMS Collaboration at the LHC and the STAR Collaboration at RHIC. In
particular the CMS Collaboration also used the Event Shape
Engineering technique, obtaining somewhat lower limits on the CME
contribution~\cite{Sirunyan:2017quh}. Considering that the
CMS Collaboration did not account for a possible signal dependence
on the elliptic flow value, their upper limits are too low and accounting for this effect might lead to results fully compatible
with the ALICE measurements. The CMS
Collaboration~\cite{Khachatryan:2016got,Sirunyan:2017quh} and the
STAR Collaboration~\cite{STAR:2020crk} also attempted to put limits
on the CME signal by comparing the measurements in Pb--Pb and Au--Au
collisions with corresponding measurements in p--Pb and p(d)--Au
collisions, where the CME signal contribution to  $\Delta \gamma_{1,1}$ correlator is expected to be
negligible due to a very weak correlation between the direction of the magnetic field and the elliptic flow plane. 
Unfortunately, the measurements of the $\gamma_{1,1}$ correlator in small systems are also plagued by the reaction plane independent background (see
Refs.~\cite{Abelev:2009ac,Abelev:2009ad}), 
without reliable estimate of which there is no possibility to
make a definite conclusion from such measurements. 
The STAR Collaboration has recently published results~\cite{STAR:2021mii} from the analysis of the isobar samples collected in 2018. 
The two isobar nuclei used, $^{96}_{44}\mathrm{Zr}$ and $^{96}_{40}\mathrm{Ru}$, were expected to result into similar flow-driven background contribution due to the same mass number. The difference, however, in the atomic number between the two nuclei would be reflected in a smaller magnitude of the magnetic field in the case of $^{96}_{40}\mathrm{Ru}$ compared to $^{96}_{44}\mathrm{Zr}$, which would translate into a difference in the CME contribution between these two systems. Comparing the outcome of this blind analysis to a set of predefined CME-signal
criteria they concluded that the results are consistent with no CME signal.  Further work is needed to account for the differences of the background in the two systems, reflected by the corresponding differences in both the multiplicity and flow harmonics at the same centrality, in order to estimate the final limits.

\textit{\textbf{Chiral magnetic wave studies}}

The presence of a net positive electric charge can induce a positive
axial current along the direction of the magnetic field i.e., leading 
to flow of chirality. This is caused by the Chiral Separation Effect
(CSE)~\cite{Son:2009tf}. The coupling between the CME and the CSE
leads to a wave propagation of the electric charge, resulting in an
electric charge quadrupole moment of the system, the Chiral Magnetic
Wave (CMW)~\cite{Kharzeev:2010gd,Burnier:2011bf,Burnier:2012ae}.

The azimuthal distribution of charged particles due to the presence of
the CMW can be written as
\begin{equation}
\frac{{\rm d}N^{\pm}}{{\rm d}\varphi} = N^{\pm}[1 + (2v_2 \mp
  rA)\cos\Big(2(\varphi - \Psi_2)\Big)],
    \label{Eq:CMW}
\end{equation}
where $A = (N^+ - N^-)/(N^+ + N^-)$ is the charge asymmetry, and $r$
is the parameter that encodes the strength of the electric quadrupole
due to the CMW. Therefore, one can probe the value of $r$ by measuring
the $v_2$ values for different charges as a function of the charge
asymmetry. Experimentally, the latter is measured in a specific
kinematic region and should be corrected for detector efficiency,
leading to the introduction of additional systematic
uncertainties. Instead, in Ref.~\cite{Voloshin:2014gja}, the authors
suggested to measure the covariance of $v_n$ and $A$ that is a
  robust observable and does not depend on detector inefficiencies.
 The first results for
  this observable measured in Pb--Pb
collisions at $\sqrt{s_\mathrm{{\rm NN}}} = 2.76$~TeV~were reported in
Ref.~\cite{Adam:2015vje}. The correlator for the second harmonic,
$\langle v_2\{2\}A \rangle - \langle A \rangle \langle v_2\{2\}
\rangle$, exhibits a significant centrality dependence compatible with
expectations from a CMW signal which becomes more pronounced for more
peripheral events where the magnetic field strength increases. On the
other hand, local charge conservation effects could also
contribute to the development of the measured signal~\cite{Voloshin:2014gja}.

\begin{figure}[t!b]
    \begin{center}
    \includegraphics[width = 0.49\textwidth]{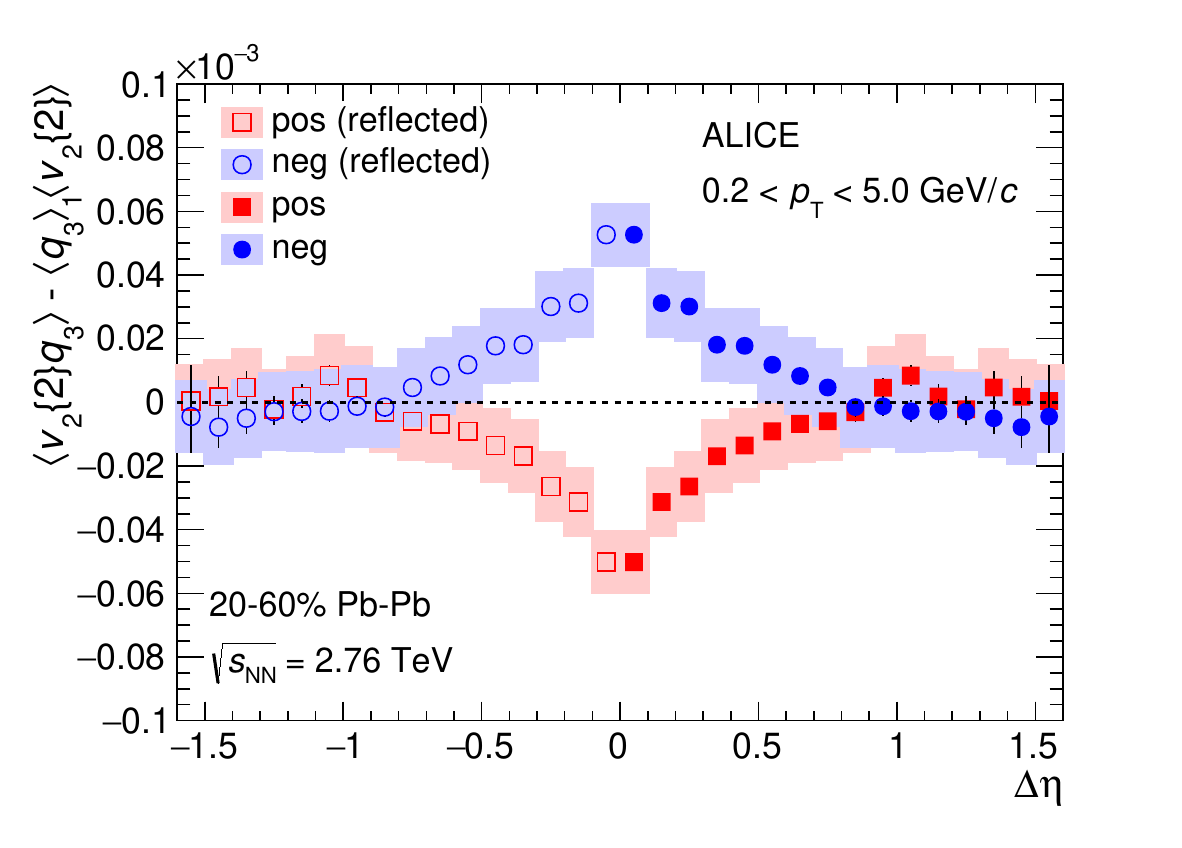}
    \includegraphics[width = 0.49\textwidth]{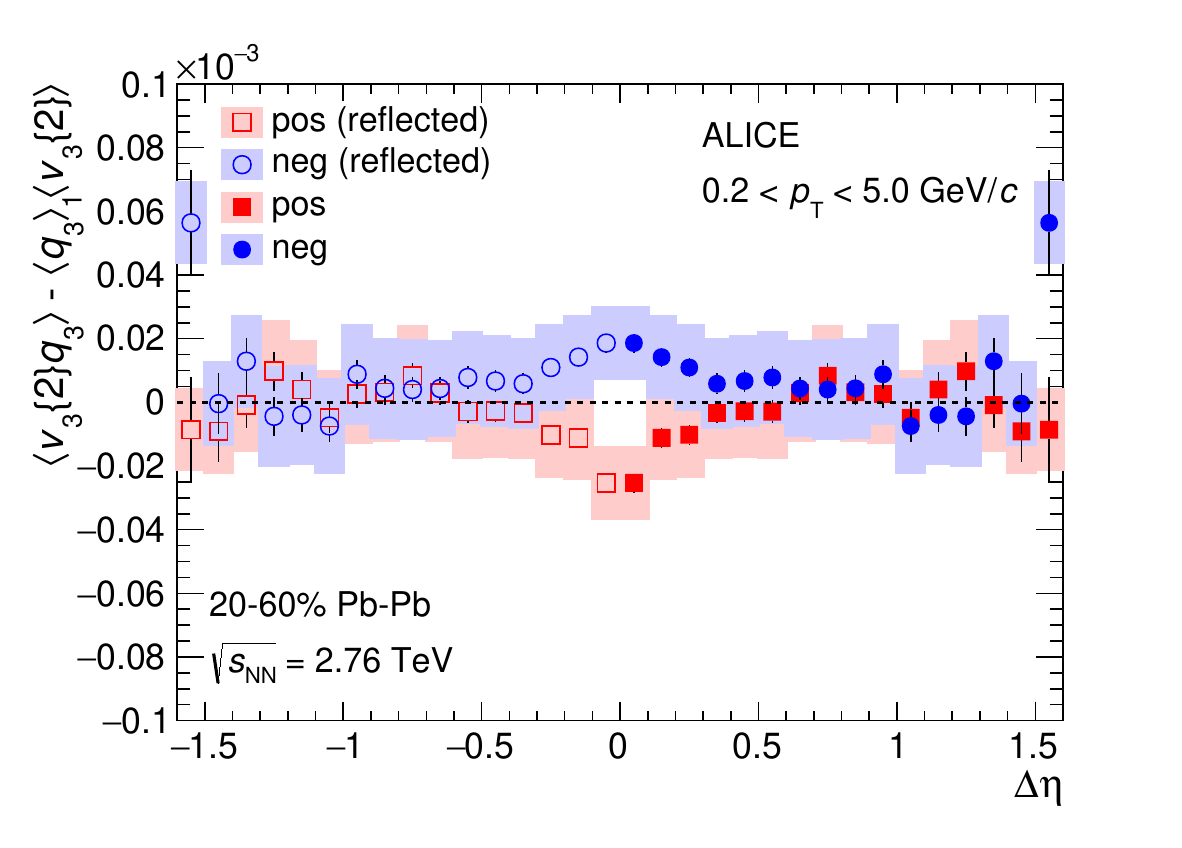}
    \end{center}
    \caption{The pseudorapidity dependence of the results for the
      second (left) and third harmonic (right) for positive (red
      squares) and negative (blue circles) particles measured in
      the 20--60$\%$ centrality interval of Pb--Pb collisions at $\sqrt{s_\mathrm{{\rm NN}}} =
      2.76$~TeV~\cite{Adam:2015vje}.}
    \label{fig:cmwResults}
\end{figure}

One direction that could help in identifying whether the CMW
contributes significantly to the measured correlations is to perform
this study differentially, e.g. as a function of the pseudorapidity
gap between the particle for which elliptic flow is measured and
another charged particle. Such measurements are expected to
differentiate contributions from the CMW signal and the
background, especially those accompanied by higher harmonics measurements. The expectation from higher harmonics results
is that the CMW contribution will be significantly suppressed due to
the symmetry of the CMW effect~\cite{Adam:2015vje}. The left panel of Fig.~\ref{fig:cmwResults}~presents the results for the second harmonic as a function of $\Delta
\eta$. These correlations exhibit a peak with a ``typical hadronic width'' of
about 0.5–1 units of rapidity which qualitatively agrees with possible
background contributions from local charge conservation combined with
strong radial and elliptic flow~\cite{Voloshin:2014gja}. In parallel, noticeable correlations in the
third harmonic, presented in the right panel of Fig.~\ref{fig:cmwResults}, indicate a significant contribution from LCC, but the
final conclusion requires a detailed model comparison. Future
$p_{\mathrm{T}}$--differential measurements as well as studies employing the ESE technique would also be very
helpful to quantify the background and provide a CMW limit. Note that the contribution of the LCC to this correlator and
the $\gamma$ correlator used for CME studies are strongly correlated,
which can be further used for quantitative characterisation of the
background in both measurements.

\subsubsection{Conclusions}
 \paragraph{Electromagnetic fields in heavy-ion collisions.} Measurements of directed flow for oppositely charged particles as well as for $\mathrm{D}^0$ and $\overline{\mathrm{D}}^0$ provide the first indication that the early stage electromagnetic fields can affect the motion of the final state particles. The differences in the measured global polarisation of $\Lambda$ and $\overline{\Lambda}$ provide an upper limit for 
    the magnitude of the magnetic field at freeze-out of $5.7 \times 10^{12}$~T~and~$14.4 \times 10^{12}$~T at a 95\% confidence level in Pb--Pb collisions at $\sqrt{s_{\mathrm{NN}}}= 2.76$~and 5.02~TeV, respectively.
  
 \paragraph{Chiral Magentic Effect.} Direct studies for the existence of the CME in heavy-ion collisions revealed that background effects are dominating at 
    the LHC. The contribution from the CME to the measurement 
    of charge dependent correlations relative to the second order symmetry plane ($\gamma_{1,1}$) is constrained to an upper limit of 15--33$\%$ 
    in Pb--Pb and 2\% in Xe--Xe collisions at LHC energies at 95$\%$ confidence level. The combination of measurements and model studies in Pb--Pb and Xe--Xe collisions indicates that there is a bigger potential to discover the CME in large
collision systems (e.g. Pb--Pb) than in their small collision counterparts, mainly due to the significant differences they exhibit in the value of the magnetic field.
  
 \paragraph{Chiral Magnetic Wave.} The combination of the second and third harmonic results for CMW studies indicates a significant background contribution from local charge conservation.

\newpage

\subsection{Quantitative characterisation of the QGP at the LHC}
\label{sec:QGPsummary}

\newcommand{\pp}           {pp\xspace}
\newcommand{\ppbar}        {\mbox{$\mathrm {p\overline{p}}$}\xspace}
\newcommand{\XeXe}         {\mbox{Xe--Xe}\xspace}
\newcommand{\PbPb}         {\mbox{Pb--Pb}\xspace}
\newcommand{\pA}           {\mbox{pA}\xspace}
\newcommand {\pPb}       {\ensuremath{\mbox{\text{p--Pb}}  }}
\newcommand{\AuAu}         {\mbox{Au--Au}\xspace}
\newcommand{\dAu}          {\mbox{d--Au}\xspace}

\newcommand{\s}            {\ensuremath{\sqrt{s}}\xspace}
\newcommand{\snn}          {\ensuremath{\sqrt{s_{\mathrm{NN}}}}\xspace}
\newcommand{\pt}           {\ensuremath{p_{\rm T}}\xspace}
\newcommand{\meanpt}       {$\langle p_{\mathrm{T}}\rangle$\xspace}
\newcommand{\ycms}         {\ensuremath{y_{\rm CMS}}\xspace}
\newcommand{\ylab}         {\ensuremath{y_{\rm lab}}\xspace}
\newcommand{\etarange}[1]  {\mbox{$\left | \eta \right |~<~#1$}}
\newcommand{\yrange}[1]    {\mbox{$\left | y \right |~<~#1$}}
\newcommand{\dndy}         {\ensuremath{\mathrm{d}N_\mathrm{ch}/\mathrm{d}y}\xspace}
\newcommand{\dndeta}       {\ensuremath{\mathrm{d}N_\mathrm{ch}/\mathrm{d}\eta}\xspace}
\newcommand{\avdndeta}     {\ensuremath{\langle\dndeta\rangle}\xspace}
\newcommand{\avNpart}      {\ensuremath{\langle N_\mathrm{part} \rangle}}
\newcommand{\avNcoll}      {\ensuremath{\langle N_\mathrm{coll} \rangle}}
\newcommand{\dNdetape}     {\ensuremath{\frac{2}{\avNpart}\avdndeta}}
\newcommand{\dndetalab}    {\ensuremath{\mathrm{d}N_\mathrm{ch}/\mathrm{d}\eta_\mathrm{lab}}}
\newcommand{\dNdy}         {\ensuremath{\mathrm{d}N_\mathrm{ch}/\mathrm{d}y}\xspace}
\newcommand{\Npart}        {\ensuremath{N_\mathrm{part}}\xspace}
\newcommand{\Ncoll}        {\ensuremath{N_\mathrm{coll}}\xspace}
\newcommand{\dEdx}         {\ensuremath{\textrm{d}E/\textrm{d}x}\xspace}
\newcommand{\RpPb}         {\ensuremath{R_{\rm pPb}}\xspace}

\newcommand{\Rgamma}        {\ensuremath{\mathrm{R}_{\gamma}}\xspace}
\newcommand{\vtwogammadir} {\ensuremath{v_{2}^{\gamma, dir}}\xspace}
\newcommand{\vtwogammadec} {\ensuremath{v_{2}^{\gamma, dec}}\xspace}
\newcommand{\vtwogammainc} {\ensuremath{v_{2}^{\gamma, inc}}\xspace}

\newcommand {\sqrtSnn}      {\ensuremath{\sqrt{s_\text{\textsc{nn}}}}}
\newcommand{\nineH}        {$\sqrt{s}~=~0.9$~Te\kern-.1emV\xspace}
\newcommand{\seven}        {$\sqrt{s}~=~7$~Te\kern-.1emV\xspace}
\newcommand{\twoH}         {$\sqrt{s}~=~0.2$~Te\kern-.1emV\xspace}
\newcommand{\twosevensix}  {$\sqrt{s}~=~2.76$~Te\kern-.1emV\xspace}
\newcommand{\five}         {$\sqrt{s}~=~5.02$~Te\kern-.1emV\xspace}
\newcommand{\twosevensixnn}{$\sqrt{s_{\mathrm{NN}}}~=~2.76$~Te\kern-.1emV\xspace}
\newcommand{\fivenn}       {$\sqrt{s_{\mathrm{NN}}}~=~5.02$~Te\kern-.1emV\xspace}
\newcommand{\LT}           {L{\'e}vy-Tsallis\xspace}
\newcommand{\GeVc}         {Ge\kern-.1emV/$c$\xspace}
\newcommand{\MeVc}         {Me\kern-.1emV/$c$\xspace}
\newcommand{\TeV}          {Te\kern-.1emV\xspace}
\newcommand{\GeV}          {Ge\kern-.1emV\xspace}
\newcommand{\MeV}          {Me\kern-.1emV\xspace}
\newcommand{\GeVmass}      {Ge\kern-.2emV/$c^2$\xspace}
\newcommand{\MeVmass}      {Me\kern-.2emV/$c^2$\xspace}
\newcommand{\lumi}         {\ensuremath{\mathcal{L}}\xspace}

\newcommand{\ITS}          {\rm{ITS}\xspace}
\newcommand{\TOF}          {\rm{TOF}\xspace}
\newcommand{\ZDC}          {\rm{ZDC}\xspace}
\newcommand{\ZDCs}         {\rm{ZDCs}\xspace}
\newcommand{\ZNA}          {\rm{ZNA}\xspace}
\newcommand{\ZNC}          {\rm{ZNC}\xspace}
\newcommand{\SPD}          {\rm{SPD}\xspace}
\newcommand{\SDD}          {\rm{SDD}\xspace}
\newcommand{\SSD}          {\rm{SSD}\xspace}
\newcommand{\TPC}          {\rm{TPC}\xspace}
\newcommand{\TRD}          {\rm{TRD}\xspace}
\newcommand{\VZERO}        {\rm{V0}\xspace}
\newcommand{\VZEROA}       {\rm{V0A}\xspace}
\newcommand{\VZEROC}       {\rm{V0C}\xspace}
\newcommand{\Vdecay} 	   {\ensuremath{V^{0}}\xspace}

\newcommand{\ee}           {\ensuremath{e^{+}e^{-}}} 
\newcommand{\pip}          {\ensuremath{\pi^{+}}\xspace}
\newcommand{\pim}          {\ensuremath{\pi^{-}}\xspace}
\newcommand{\kap}          {\ensuremath{\rm{K}^{+}}\xspace}
\newcommand{\kam}          {\ensuremath{\rm{K}^{-}}\xspace}
\newcommand{\pbar}         {\ensuremath{\rm\overline{p}}\xspace}
\newcommand{\kzero}        {\ensuremath{{\rm K}^{0}_{\rm{S}}}\xspace}
\newcommand{\lmb}          {\ensuremath{\Lambda}\xspace}
\newcommand{\almb}         {\ensuremath{\overline{\Lambda}}\xspace}
\newcommand{\Om}           {\ensuremath{\Omega^-}\xspace}
\newcommand{\Mo}           {\ensuremath{\overline{\Omega}^+}\xspace}
\newcommand{\X}            {\ensuremath{\Xi^-}\xspace}
\newcommand{\Ix}           {\ensuremath{\overline{\Xi}^+}\xspace}
\newcommand{\Xis}          {\ensuremath{\Xi^{\pm}}\xspace}
\newcommand{\Oms}          {\ensuremath{\Omega^{\pm}}\xspace}
\newcommand{\degree}       {\ensuremath{^{\rm o}}\xspace}

\subsubsection{Macroscopic properties}

Macroscopic properties pertain to the global features of the system, such as the temperature, volume and system lifetime. The system formed in heavy-ion collisions undergoes a very rapid radial expansion with velocity up to  $v\sim 2/3\,c$ on a timescale of $10^{-23}$~s. Therefore, each derived macroscopic property will either correspond to some average over this evolution, or to a particular stage. On the other hand, quantities subject to conservation laws such as the total energy or the entropy (assuming dissipative effects are small) will remain constant. In this section, we will review some of these properties that are determined either directly from ALICE measurements, or from theoretical models that are validated by measurements. We will then compare the results to predictions from various theoretical descriptions of the QGP and hadronic stages.

\paragraph{Temperature evolution of the system.}

\begin{figure}
    \centering
    \includegraphics[width=0.7\textwidth]{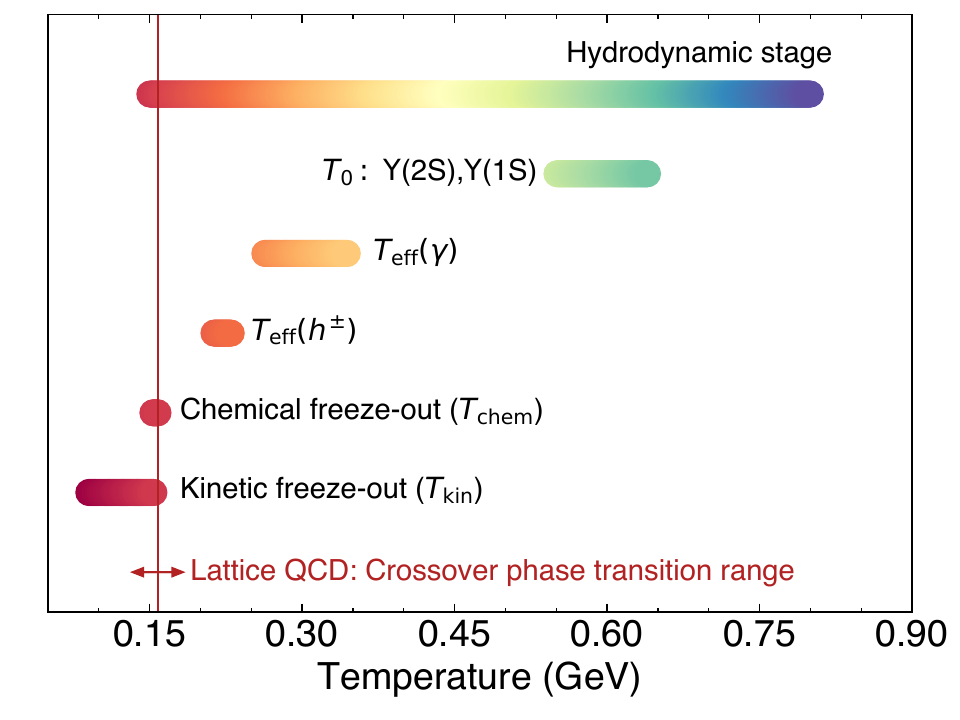}
    \caption{The temperatures ranges probed by central heavy-ion collisions at the LHC derived from ALICE measurements shown throughout the chapter. The region to the right of the red line represents the QGP phase, while the left is the hadronic phase. A lattice QCD calculation for the pseudocritical temperature $T_{\mathrm{pc}}$ and crossover temperature range where the QGP and hadron gas coexist (denoted by the arrows) is also shown~\cite{Borsanyi:2020fev}.}
    \label{fig:tempHI}
\end{figure}

The deconfined system created in heavy-ion collisions will cool and transition to a hadronic phase as time progresses, with various experimental probes being sensitive to the temperature evolution. In Fig.~\ref{fig:tempHI}, we show the temperature ranges probed using ALICE measurements from heavy-ion collisions at the LHC, which have varying dependencies on the uncertainties from measurements and theoretical modelling. As shown previously in this chapter, hydrodynamic model frameworks successfully describe many low-\pt observables such as the anisotropic flow coefficients and average \pt of identified particles. These models can be used to determine QGP temperature throughout the system evolution, and therefore, using constraints from data, they can provide an indirect assessment of the entire temperature range of the QGP created at the LHC. This range is shown in Fig.~\ref{fig:tempHI} for the IP-Glasma+MUSIC framework~\cite{Schenke:2020mbo}. It indicates that at the earliest times, which correspond to the highest temperatures of the QGP, the temperatures obtained in Pb--Pb collisions are $\sim 0.8$~GeV, that is roughly five times higher than the pseudocritical QGP transition temperature predicted by lattice QCD, $T_{\rm pc} \approx 0.155$~GeV~\cite{Bazavov:2018mes,Borsanyi:2020fev}. The hydrodynamic range is dependent on the choice of initial-state modelling and the pre-equilibrium descriptions, as well as on the time the hydrodynamic stage occurs, and the lattice-QCD equation of state used. The T$_{\rm{R}}$ENTo+VISHNU chain~\cite{Bernhard:2019bmu} (not shown in Fig.\ref{fig:tempHI}) has a smaller temperature range, with an upper limit of $T=0.35$~GeV. Among other things, this is related to the time the hydrodynamic evolution starts. This occurs at a time $\tau_0\sim1$~fm/$c$ for T$_{\rm{R}}$ENTo+VISHNU, and at $\tau_0\sim 0.3$~fm/$c$ for IP-Glasma+MUSIC. Other models have also been used to predict the temperatures needed to describe the relative bottomonia $\mathrm{\Upsilon}$(2S) and $\mathrm{\Upsilon}$(1S) yields (see Sec.~\ref{sec:Quarkonium}). The respective model calculations, one of which uses hydrodynamics~\cite{Krouppa:2016jcl} and the other is a non-perturbative transport model~\cite{Du:2017qkv}, use starting temperatures ($T_0$) of 0.55--0.64~GeV. This range is shown in Fig.~\ref{fig:tempHI}, and can be used as an additional input  to hydrodynamic descriptions of low-\pt light-flavour observables to constrain the temperature range probed by heavy-ion collisions. 

Effective temperatures ($T_{\rm{eff}}$) derived from prompt-photon and charged-hadron spectra are  shown in Fig.~\ref{fig:tempHI} as well. The photon exponential inverse slope is obtained directly from data (see Sec.~\ref{sec:MacroscopicProperties}), and due to effects related to the emission time and blue shift from the expanding source, provides an effective temperature averaged over the whole system lifetime. The effective charged-hadron temperature was obtained using a hydrodynamic model~\cite{Gardim:2019xjs} to describe ALICE spectra measurements. It can also be interpreted as an effective temperature averaged over the entire space--time evolution of the medium. It is lower than the photon effective temperature, which might be a consequence of different sensitivities to each stage of the medium evolution. The chemical freeze-out temperature ($T_{\rm chem}$) range obtained from a statistical model fit to particle yields measured by ALICE (see Sec.~\ref{sec:QGPHadronization}) is shown in Fig.~\ref{fig:tempHI}. This temperature is close to the deconfinement crossover temperature computed from lattice QCD. Finally, the kinetic freeze-out temperature ($T_{\rm kin}$) range derived from Blast-Wave fits to spectra and identified particle $v_{n}$ measurements are shown. It is found to be rather highly model dependent (see Sec.~\ref{sec:QGPevolution}). The kinetic freeze-out can either occur around the time of the chemical freeze-out or at later times that correspond to lower temperatures. We will investigate this difference by reviewing additional measurements in the next subsection.

\paragraph{Space--time extent.}

The space--time extent of the system created in heavy-ion collisions refers to the properties of the system towards the end of its evolution. It can be characterised by a hypersurface. Such a surface encodes the four-dimensional properties of the corresponding freeze-out space--time coordinates of the system at the freeze-out temperature. It depends on the position, time, and velocity of the element that freezes out. An example of hypersurface profiles from a hydrodynamic model chain can be found in Ref.~\cite{Karpenko:2012yf}. The freeze-out time on this surface will be longer at freeze-out positions closer to the centre of the originally created QGP, compared with positions at the outskirts. The reason for this is that at the centre, the temperatures are the highest, therefore the time associated with freeze-out is longer. As the velocity of a medium element increases, the freeze-out positions relative to the collision centre decrease, which means higher velocity elements decouple from the system at earlier times. This is directly observed from measurements of the femtoscopic radii in Sec.~\ref{sec:MacroscopicProperties}, which decrease as the femtoscopic particle pair velocity (characterised by $k_{\rm T}$) increases.

\begin{table}[h]
\centering
\caption{System volumes in central Pb--Pb at \twosevensixnn collisions for the chemical and kinetic freeze-out temperatures.}
\begin{tabular}{l|l|l}
\multicolumn{3}{c}{{\bf Volume:} $\mathbf{\mathrm{d}\it{V}/\mathrm{d}y}$ at $y=0$}  \\ \hline
Static at $T_{\rm chem}=156$ MeV & Dynamic at $T_{\mathrm{kin}}=156$ MeV & Dynamic at $T_{\mathrm{kin}}=100$ MeV\\ \hline
$4730\pm500$ fm$^3$    & $6210\pm360$ fm$^3$ & $7960\pm460$ fm$^3$  \\
\end{tabular}
\label{foVol}
\end{table}

Table~\ref{foVol} shows a summary of the obtained volumes that corresponds to one unit of rapidity at midrapidity for central Pb--Pb collisions, at about the freeze-out temperatures shown in this section. The volume associated with $T_{\rm chem}$ is determined from statistical models of the identified particle yields described in Sec.~\ref{sec:QGPHadronization}, which corresponds to a static volume~\footnote{The four model values shown in Sec.~\ref{sec:QGPHadronization} are averaged with the uncertainty determined from the variance.}. As noted in the previous subsection, the corresponding $T_{\rm chem}$ temperature is consistent with the deconfinement temperature. This implies two things: the QGP is born in thermal equilibrium, and the hadronic inelastic cross sections are too small for the system to maintain chemical equilibrium after the QGP hadronises and the temperature decreases (otherwise $T_{\rm chem} < T_{\mathrm{pc}}$). This volume can therefore be interpreted as the volume when the QGP hadronises. The volumes associated with $T_{\mathrm{kin}}$ are obtained using measurements of femtoscopic radii shown in Sec.~\ref{sec:MacroscopicProperties} at $k_{\rm T} = 0.25$ GeV/$c$ and for the two temperatures shown. They are determined using $V_{\rm fo} = (2\pi)^{3/2}R_{\mathrm{side}}^2R_{\mathrm{long}}$. From the temperature $T_{\mathrm{kin}}$, one can determine the mean thermal velocity and the corresponding rapidity interval~\cite{CERES:2002rfr}. In addition, as these pairs are moving, the volume they characterise is reduced compared with the static case. It has been estimated that the transverse expansion leads to a 20--25\% reduction in $R_{\rm side}$~\cite{Chapman:1995nz,Csorgo:1995bi,Scheibl:1998tk}. For the case where $T_{\rm chem}=T_{\mathrm{kin}}$, which corresponds to the largest values of $T_{\mathrm{kin}}$ from the Blast-Wave fits shown in ~Sec.~\ref{sec:QGPevolution}, the dynamic volume is larger than the static volume. Given the reverse is expected if $T_{\rm chem}=T_{\mathrm{kin}}$, this then implies $T_{\mathrm{kin}} < T_{\rm chem}$. Therefore, based on the available phenomenological models, the hypothesis of $T_{\rm chem}=T_{\mathrm{kin}}$, is unlikely, implying the existence of a prolonged time period when the system is still coupled via elastic scatterings. For $T_{\mathrm{kin}}=100$ MeV, which corresponds to the smallest values of $T_{\mathrm{kin}}$ from the Blast-Wave fits, these results imply at least a $\sim$70\% increase in the kinetic freeze-out volume compared with chemical freeze-out. 

\begin{table}[h]
\centering
\caption{Estimates of the kinetic freeze-out time from pion decoupling and of the duration of the hadronic stage implied from resonances and delay in particle emissions from femtoscopic measurements in central Pb--Pb collisions at \twosevensixnn.}
\begin{tabular}{l|l|l|l|l}
\textbf{Kinetic freeze-out time}  & \multicolumn{4}{c}{\textbf{Lifetimes of the hadronic stage}} \\ \hline
$\pi$ decoupling & $\rm K^{0*}/K^{\pm}$        & $\rho^{0}/\pi^{\pm}$    & $\Lambda^{*}/\Lambda$  & $\rm K-\pi$ emission delay      \\ 
\hline
10--13~fm/$c$ & $\sim$ 3 fm/$c$ & $\sim$ 1 fm/$c$ & $\sim$ 10 fm/$c$ & $1.4\pm0.6$ fm/$c$          \\ 
\end{tabular}
\label{foTime}
\end{table}

An estimate of the lifetime of the system is given by the decoupling (or kinetic freeze-out) time of pions, the most abundant species produced. This decoupling time is in the range 10--13~fm/$c$ in central collisions, based on the measurements of $R_{\mathrm{long}}$ discussed in Sec.~\ref{sec:MacroscopicProperties}, and on the assumption of a hydrodynamic expansion. The hadronic stage of the collision can be further investigated by resonance and femtoscopic measurements, which are summarised in Tab.~\ref{foTime} and discussed in more detail in Sec.~\ref{sec:QGPHadronization} and~\ref{sec:MacroscopicProperties}, respectively. In both cases, the data can be explained by the modelling of the hadronic cascade~\cite{Bleicher:1999xi, Chojnacki:2011hb} preceded by a hydrodynamic evolution of the system. The lifetimes derived from hadron resonances offer an estimate of the duration of the hadronic stage, assuming negligible regeneration. While all the times are above zero, there are considerable differences. This could reflect a varying competition between the re-scattering and regenerating processes for the various resonances - if re-scattering dominates, one would infer shorter lifetimes, however if regeneration contributes significantly, this would imply longer lifetimes due to the delay in the formation of regenerated resonances. The delay in the emission of $\pi$ and K hadrons was measured with femtoscopic techniques, and was associated with $\rm K^{0*}$ formation and decay via comparison to calculations from a hadronic afterburner, as discussed in Sec.~\ref{sec:MacroscopicProperties}. We finally note that this observation also requires a reconciliation with the results from the FastReso Blast-Wave model shown in Sec.~\ref{sec:QGPevolution}, which implies instead the possibility of $T_{\mathrm{chem}}=T_{\mathrm{kin}}$, namely the absence of such a hadronic stage. While the exact timescale of the hadronic stage remains an open question, evidence based mainly on femtoscopic and resonance measurements suggest indeed that the fireball continues to behave like a coupled system for an extended time after the chemical freeze-out.

\subsubsection{Microscopic properties}

The transport properties of the QGP provide information on how it responds to microscopic excitations, and such excitations can arise from either soft or hard processes. Before their review, it is useful to provide a distinction between the weak and strong coupling regimes that these transport properties probe:

\begin{itemize}
    \item {\bf Weakly-coupled system.} In this case, the QGP can be considered as a system of well defined quasi-particles (quarks and gluons), whose interactions involve few body processes e.g.\,$2\rightarrow$ 2 scattering or $2\rightarrow$ 3 collision-induced gluon radiation, which sometimes are modeled as scatterings off discrete scattering centres.  It can be described by perturbative QCD at couplings of $\alpha_{\rm S} \lesssim 0.3$, and is considered a gas.
    \item {\bf Strongly-coupled system.} The coupling between QGP constituents is large and therefore dominated by higher-order processes. Such a coupling induces strong correlations between the neighboring constituents, therefore the quasi-particle description is no longer valid, and QGP matter can be treated as a liquid.
\end{itemize}

It is also important to note that these regimes represent idealised limits - an intermediately coupled system where perturbative and non-perturbative processes compete is of course plausible. A special case of a strongly-coupled system is at infinite coupling. While physically unrealistic, it provides a limiting case for strongly-coupled systems, as will become apparent later in this section. One of the defining features of QCD is the scale dependence e.g.\,asymptotic freedom. The processes associated with a transport property that involve large momentum transfers resolve medium structures at short length scales and will by definition ``see'' a weakly-coupled QGP, with the opposite being true for processes involving small momentum transfers. In addition, as the temperature of the QGP increases, all transport properties are expected to probe an increasingly-weakly-coupled QGP. Therefore, exploring the temperature dependence of transport properties is critical in establishing whether the QGP remains strongly coupled at the temperatures accessible at the LHC, or whether a weakly-coupled QGP emerges at the highest temperatures possible in the laboratory.

\paragraph{Viscosity of the QGP.}

The success of hydrodynamic model chains in describing a wide variety of soft observables, implies that the system is strongly coupled at momentum scales corresponding to the QGP temperature, as hydrodynamics describes the evolution of the QGP in terms of a liquid. The question then becomes quantitative -- how strong is the coupling -- and values of the shear ($\eta/s$) and bulk ($\zeta/s$) viscosities per entropy density play a central role in addressing this question. In the strong-coupling picture, both transport properties $\eta/s$ and $\zeta/s$ are proportional to the shear and bulk relaxation times. As the initial state is highly non-uniform, this leads to large non-equilibrium shear and bulk excitations at early times. The shear and bulk relaxation times determine how quickly the system can respond to achieve equilibrium -- if the coupling is large, these times will be small, which corresponds to small values of $\eta/s$ and $\zeta/s$, and vice versa. 

\begin{figure}
    \centering
    \includegraphics[width=0.73\textwidth]{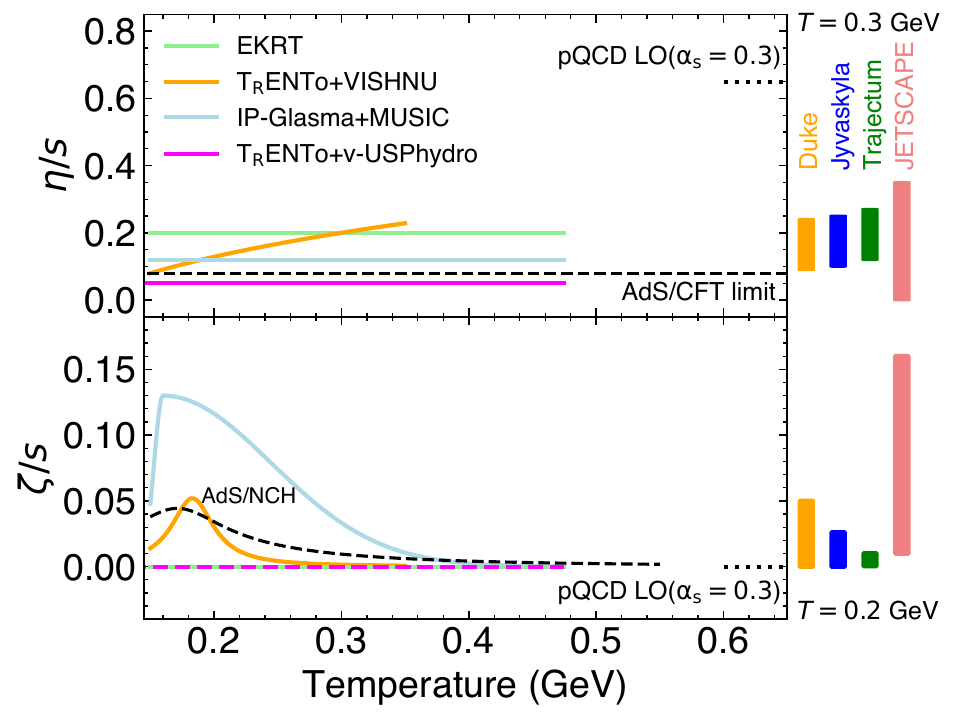}
    \caption{The temperature dependence of the shear (top panel) and bulk (bottom) viscosities over entropy density in the QGP phase constrained by the ALICE measurements shown in Secs.~\ref{sec:MacroscopicProperties},~\ref{sec:QGPevolution}, and~\ref{sec:Quarkonium} from various hydrodynamic models described in the text. Limits from pQCD~\cite{Arnold:2006fz}, AdS/CFT~\cite{Kovtun:2004de}, and AdS/Non-Conformal Holographic~\cite{Finazzo:2014cna} approaches are also shown. The ranges on the right of the plot represent 90\% posterior intervals from the Bayesian analyses.}
    \label{fig:eta_zetas}
\end{figure}

Measurements of anisotropic flow and hadron spectra by ALICE have allowed for an exploration of the temperature dependence of $\eta/s$ and $\zeta/s$. Fig.~\ref{fig:eta_zetas} compares these dependencies for four hydrodynamic model chain calculations constrained by ALICE data. Regarding $\eta/s$, many other fluids exhibit a temperature dependence for $\eta/s$, with a minimum occurring at the phase transition temperature between the gas and liquid phases. Whether such a temperature dependence exists for the QGP phase is an open question for the temperatures probed by the LHC, as demonstrated in Fig.~\ref{fig:eta_zetas}. The models  assume either an independence (IP-Glasma+MUSIC and EKRT~\cite{Niemi:2015qia}) or a weak dependence (T$_{\rm{R}}$ENTo+VISHNU). All provide fair descriptions of the anisotropic flow measurements. The corresponding shear relaxation time range is $\tau_{\pi}=$ 0.15--0.40~fm/$c$ at $T=0.5$~GeV, which implies that spatial anisotropies from the initial state are rather quickly smoothed out in the QGP phase. Two coupling limits are shown: an AdS/CFT limit which is calculated for infinite coupling, and a pQCD limit, which is determined in leading order for $\alpha_{\rm S}=0.3$. In the weak-coupling picture, $\eta/s$ is inversely related to the coupling strength -- therefore the $\alpha_{\rm S}$ chosen corresponds to the strongest possible coupling (therefore the lowest $\eta/s$) in that scheme. It is clear that the extracted $\eta/s$ values are closer to the infinite-coupling limit. Nonetheless, it should be pointed out that next-to-leading-order corrections for the pQCD determination of $\eta/s$ are very large~\cite{Ghiglieri:2018dib}. We refrain from showing them here, as these corrections largely cancel when ratios of transport parameters are explored, which will be the subject of discussion later.
 The values of $\eta/s$ from the QGP are roughly four times smaller than for Helium after it transitions to a superfluid~\cite{lemmon_etal:1998}.

The constraints on $\zeta/s$ from ALICE measurements are shown in the bottom panel of Fig.~\ref{fig:eta_zetas}. Predictions from an infinitely-coupled AdS Non-Conformal Holographic approach (AdS/NCH) are also shown for comparison. As the applicability of Conformal Symmetry regarding the strong potential assumed in the AdS/CFT scheme at temperatures close to the $T_{\rm pc}$ is expected to break down, an alternative approach is needed to determine $\zeta/s$. The breaking of conformal symmetry leads to $\zeta/s$ rising near $T_{\rm pc}$ (it is zero otherwise). This approach was also used to reevaluate the $\eta/s$ in the limit of infinite coupling at all temperatures, and was found to also give $1/4\pi$, which suggests this limit is universal. Its prediction that $\zeta/s$ should depend strongly on the temperature in this region is utilised for the T$_{\rm{R}}$ENTo+VISHNU and IP-Glasma+MUSIC models. The T$_{\rm{R}}$ENTo+VISHNU model provides the best description of ALICE identified-particle mean-\pt measurements (as shown in Sec.~\ref{sec:QGPevolution}), to which $\zeta/s$ is sensitive. This suggests that the ranges from both IP-Glasma+MUSIC and EKRT ($\zeta/s =0$) provide a conservative estimate of the uncertainty for this parameter. The high-temperature pQCD limit for $\zeta/s$ is close to 0, which appears to apply for all the models shown at temperatures above 0.4 GeV. This then implies that bulk excitations in the initial state are washed out in the QGP phase even more quickly than the shear excitations e.g $\tau_{\pi} < 0.1$~fm/$c$ for IP-Glasma+MUSIC at $T=0.4$ GeV. We also note that the validity of the initial state models used in each hydrodynamic model chain will be investigated in Chap.~\ref{ch:InitialState}, using multiplicity and anisotropic-flow measurements that are mainly sensitive to the features of the initial state. 

In addition, the posterior distributions for $\eta/s$ and $\zeta/s$ have been evaluated using Bayesian parameter estimation techniques on ALICE data. They have been carried out by the Duke~\cite{Bernhard:2019bmu}\footnote{The maximum a posterior values were tested with ALICE measurements for the T$_{\rm{R}}$ENTo+VISHNU throughout this chapter, and are shown in Fig.~\ref{fig:eta_zetas}}, JETSCAPE~\cite{JETSCAPE:2020mzn}, Trajectum~\cite{Nijs:2020ors}, and Jyv\"{a}skyl\"{a}~\cite{Parkkila:2021tqq} groups. These are shown on the right of Fig.~\ref{fig:eta_zetas}, at $T=0.3$~GeV for $\eta/s$ and $T=0.2$~GeV for $\zeta/s$. The size of these posterior ranges are influenced by the prior ranges and data-sets included. For example, the JETSCAPE prior ranges were larger than those by the Duke and Trajectum groups, and yielded a larger upper limit for $\eta/s$. The Jyv\"{a}skyl\"{a} group used an even larger prior range, but included measurements of Symmetric Cumulants, which, as mentioned in Sec.~\ref{sec:QGPevolution}, highly constrain the temperature dependence of $\eta/s$ and reduce the upper limit. The $\zeta/s$ ranges differ more than the $\eta/s$ ranges. The JETSCAPE group also found that the duration of the pre-equilibrium phase has a strong impact on the extracted viscosity transport parameters. This is clearly demonstrated by the v-USPhydro chain~\cite{Noronha-Hostler:2013gga}; the hydrodynamic evolution starts without any pre-equilibrium phase, and therefore requires low values of $\eta/s=0.05$ and $\zeta/s=0$ in order to develop enough flow to describe the ALICE data~\cite{Giacalone:2017dud}. Those values are also shown in Fig.~\ref{fig:eta_zetas}.

\paragraph{Charm spatial diffusion coefficient $D_{\rm s}$.}

Heavy quark transport models attempt to describe how charm and beauty quarks interact with the QGP, given that heavy quarks are produced early and therefore initially out of equilibrium with the system. As described in the Introduction and Sec.~\ref{sec:PartonInteractions}, the key quantity for investigating this processes is the charm relaxation time $\tau_{\rm charm}$, which can be evaluated in the low-momentum limit ($p=0$). It is proportional to the charm diffusion coefficient $D_{\rm s}$. In the strong coupling picture, this time will be small, and the charm quark will equilibrate with the QGP and participate in the collective motion, with the reverse being true in the weak coupling picture. In both limits, the dimensionless quantity $2\pi T\,D_{\rm s}$ is proportional to $\eta/s$, and the proportionality factor depends on whether the system is strongly or weakly coupled as follows~\cite{Liu:2016ysz}:

\begin{align*}
    2\pi T\,D_{\rm s} &= 4\pi \times \eta/s = 1 \text{\quad Infinite Coupling at all $T$ and Strong Coupling at $T_{\rm pc}$} \\
    &\approx  1.5\times 4\pi \times \eta/s \text{\quad Strong Coupling at $T=400$ MeV}& \\
    &\approx  2.5\times 4\pi \times \eta/s \text{\quad Weak Coupling at all $T$}&
\end{align*}

\begin{figure}
    \centering
    \includegraphics[width=0.7\textwidth]{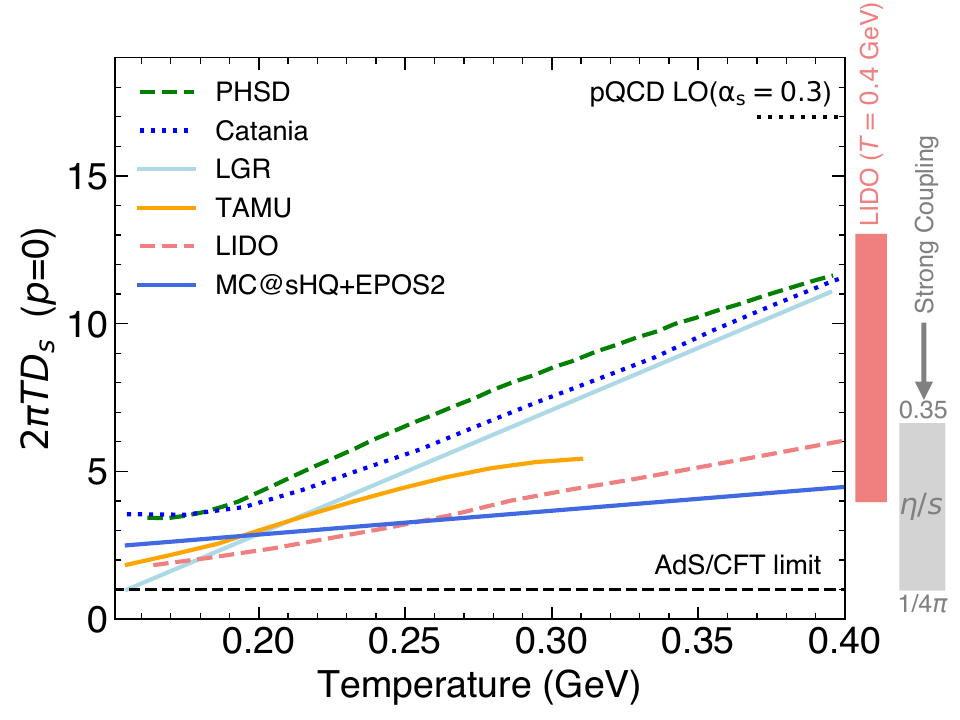}
    \caption{The temperature dependence of the spatial charm diffusion coefficient $D_{\rm s}$ in the QGP phase constrained by the ALICE measurements shown in Sec.~\ref{sec:PartonInteractions} from various transport models described in the text. The range on the right of the plot represents a 90\% posterior interval from the Bayesian analysis at $T=0.4$~GeV. Limits from pQCD~\cite{Moore:2004tg}, and AdS/CFT~\cite{Kovtun:2004de}, are also shown.}
    \label{fig:Dsubs}
\end{figure}

The determination of the heavy-quark spatial diffusion coefficient through model descriptions of experimental measurements is a major goal of recent dedicated initiatives~\cite{Rapp:2018qla,Cao:2018ews}.
In Fig.~\ref{fig:Dsubs}, we show the temperature dependence of $2\pi T\,D_{\rm s}$ from models shown in Sec.~\ref{sec:PartonInteractions}. The model calculations describe the D-meson $v_2$ and $R_{\rm AA}$ measurements. As discussed in Sec.~\ref{sec:PartonInteractions}, the models differ regarding the treatment of the charm-medium interactions. Some assume a quasi-particle description of QGP constituents e.g.\,PHSD~\cite{Song:2015sfa}, while others use a hydrodynamic treatment e.g.\,TAMU~\cite{He:2019vgs}. The Catania model~\cite{Scardina:2017ipo,Plumari:2019hzp} uses Boltzmann equations for the treatment of charm transport for the predictions shown, while LGR~\cite{Li:2019lex} is Langevin-based, and some e.g.\,MC@sHG+EPOS2~\cite{Nahrgang:2013xaa} include radiative interactions (all include collisional). Other features such as the hadronisation scheme used play an important role. Figure~\ref{fig:Dsubs} also shows a leading-order pQCD prediction. As for $\eta/s$, $2\pi T\,D_{\rm s}$ is inversely related to the coupling strength in the weakly-coupled limit. The pQCD calculation is consistent with the weak coupling limit regarding $\eta/s$, given the relations shown at the start of this subsection. All of the curves are below the pQCD charm diffusion limit, especially around the deconfinement temperature. Nonetheless, as with $\eta/s$, next-to-leading-order expansions can be large~\cite{Rapp:2018qla}, and there are other leading-order calculations~\cite{vanHees:2004gq} that give higher values at high couplings i.e.\,$2\pi T\,D_{\rm s}\sim 30$ at $\alpha_{\rm S}=0.4$. An alternative approach~\cite{Liu:2016ysz} (not shown), that attempts to include higher order corrections using the non-perturbative equations with a weak potential (T-matrix), yields predictions at least two times higher than any of curves shown. Therefore, a number of examples of weak-coupling approaches are inconsistent with the $2\pi T\,D_{\rm s}$ values constrained by our data.

On the right of Fig.~\ref{fig:Dsubs}, the strong-coupling limit is shown, based on the range of $\eta/s$ values constrained in the soft sector, observed in the previous subsection. The infinitely-coupled AdS/CFT limit is also shown. All of the models are close to the strong-coupling limit around $T_{\rm pc}$. At higher temperatures, the curves start to deviate, with LIDO~\cite{Ke:2018tsh} and MC@sHQ+EPOS2 being the closest to this limit, while the other curves lie significantly above. This observation is consistent with two scenarios: the low-momentum charm quarks either strongly couple with the QGP at all temperatures, or they couple strongly with the QGP at lower temperatures only. The aforementioned model differences play a critical role in addressing these scenarios. The LIDO and MC@sHQ+EPOS2 models include radiative interactions; these are less effective in driving the charm quark to participate in the collective motion of the medium, and therefore require smaller values of $D_{\rm s}$ (therefore smaller relaxation times) to describe the data. On the other hand, LGR also includes radiative interactions, however unlike MC@sHQ+EPOS2, which uses the Boltzmann equations, it is Langevin-based. The difference between both approaches has been investigated within the Catania model, which led to the finding that the Langevin equations lead to higher values of $D_{\rm s}$. In addition, using the LIDO model, a Bayesian posterior estimation was used to determine the $2\pi T\,D_{\rm s}$ within that scheme~\cite{Ke:2018tsh}. The posterior range  at 90\% is shown on the right of Fig.~\ref{fig:Dsubs} at $T=0.4$ GeV, and is within the range of all of curves at the corresponding temperature. It is also important to point out that the hadronisation mechanisms of charm quarks are one of the main sources of uncertainty in the estimated values of $D_{\rm s}$; all models shown in Fig.~\ref{fig:Dsubs} include a contribution of charm hadronisation from coalescence, but the present data only mildly constrain the hadronisation dynamics~\cite{Citron:2018lsq}. Providing more constraining hadronisation measurements is a major goal of future heavy-flavour programmes.

Finally, the charm relaxation times will be discussed. In Sec.~\ref{sec:PartonInteractions}, these were reported to be in the range of $\tau_{\rm{charm}}= 3$--9~fm/$c$ at the pseudo-critical temperature of $T_{\rm pc} \approx 0.155$~GeV. At higher temperatures (around the start of the QGP phase), the models in Fig.~\ref{fig:Dsubs} give $\tau_{\rm{charm}}= 1.3$--2.4~fm/$c$ at $T=0.4$~GeV. These times are significantly larger than the shear and bulk relaxation times, which is expected due to the large mass of the charm quark. Nonetheless, despite the uncertainties regarding when low-momentum charm quarks strongly couple to the QGP, all models imply these quarks will readily participate in the collective motion of the QGP after their production in the initial stages.

\paragraph{Jet transport coefficient $\hat{q}$.}

High-energy partons that propagate through the QGP interact with the colour charges in the plasma, leading to changes in both the magnitude and direction of their momenta via elastic interactions, as well as the stimulated radiation of gluons. The resulting modifications of the parton shower and the emerging jets are referred to as jet quenching, and are discussed in Sec.~\ref{sec:PartonInteractions}. The dependence of the gluon-radiation probability and energy distribution on the QGP properties can be encoded in the transport coefficient $\hat{q}$, which is the average of the square of the transverse momentum exchanged with the QGP per unit mean free path (see Eq.~\ref{eq:qhat_def}). The total amount of radiative energy loss ($\Delta E$) of a parton passing through the medium is approximately proportional to $\hat{q}$, see also Sec.~\ref{sec:QGPJets}. 

The dependence of $\hat{q}$ on the temperature of the QGP and the energy of the propagating parton has been explored in both the weak-coupling regime, using Hard Thermal Loop field theory~\cite{Arnold:2008vd,Majumder:2007zh,Muller:2021wri} and the strong-coupling regime, using the AdS/CFT correspondence~\cite{Liu:2006he}. In both regimes, $\hat{q}$ depends linearly on the temperature cubed ($T^{3}$) and the jet--medium coupling strength. As a result, the dimensionless ratio $\hat{q}/T^3$ increases with coupling strength and does not saturate to a limiting value at large coupling, unlike $\eta/s$ and $2\pi T\,D_{\rm s}$. Specific relations between $\eta/s$ and $\hat{q}/T^{3}$ have been explored at leading and next-to-leading order for weak- and strong-coupling scenarios~\cite{Majumder:2007zh,Muller:2021wri}. Given the soft nature of the interactions in the QGP at the temperatures that are experimentally explored, higher-order contributions, large logarithms and non-perturbative effects may have significant quantitative impact on the energy loss. A full mapping of these effects is actively pursued by the theory community. For example higher-order contributions to $\hat{q}$ have been explored~\cite{Caron-Huot:2008zna}, and large logarithms and effects of overlapping formation time are being explored in more detail~\cite{Arnold:2020uzm,Arnold:2021mow}. In addition, approaches are being developed to take non-perturbative effects into account by calculating $\hat{q}$ using lattice QCD~\cite{Laine:2013apa,Panero:2013pla,Kumar:2020wvb}.

\begin{figure}
    \centering
    \includegraphics[width=0.7\textwidth]{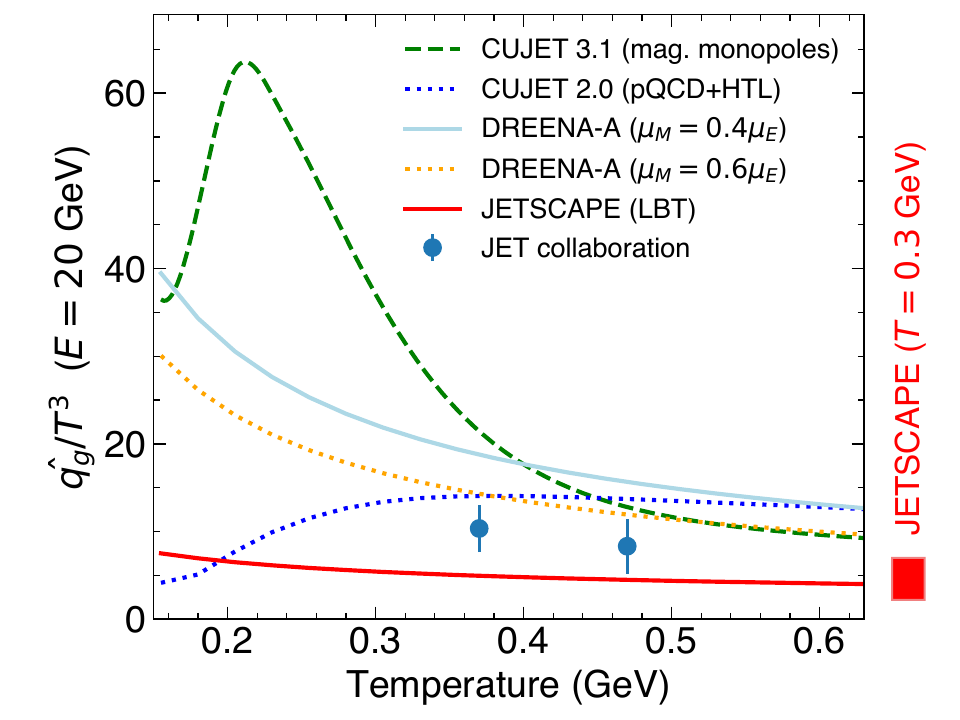}
    \caption{The temperature dependence of $\hat{q}/T^{3}$ for gluons at an energy of 20~GeV in the QGP phase from various models that describe the ALICE data (see text for details). The range on the right of the plot represents a 90\% posterior interval from the Bayesian analysis at $T=0.3$~GeV.}
    \label{fig:qhat}
\end{figure}

Figure~\ref{fig:qhat} gives an overview of the temperature dependence of $\hat{q}/T^{3}$ for gluons with an energy of $E=20$~GeV, that are used in the radiative+collisional energy loss models, which are compared to the ALICE data in Sec.~\ref{sec:PartonInteractions}. Most of these models describe the $R_{\rm AA}$ and $v_n$ for light and charmed hadrons at high-\pt{}. Only CUJET 2.0 underestimates $v_2$ at high-\pt{} systematically. In these model calculations, $\hat{q}(T,E)$ is used as input, together with the time-dependent medium density profiles from hydrodynamical calculations. The observables $R_{\rm AA}$ and $v_2$ are calculated by integrating over underlying variables, such as the location of the hard scattering in space, the initial parton momentum, the parton trajectories, as well as intrinsic fluctuations in the energy loss and the fragment distributions. While the different curves in Fig.~\ref{fig:qhat} are relatively close at high temperature, a large spread in $\hat{q}/T^3$ is seen at temperatures closer to the phase transition temperature, indicating that the transport coefficient is not well constrained by current data in that range, possibly because the value of $\hat{q}$ is smaller and the contribution to the total energy loss is small.
More systematic comparisons of different measurements with the model calculations using advanced statistical techniques such as Gaussian emulators and Bayesian parameter estimation should allow to constrain the temperature dependence of $\hat{q}$ in more detail and are being pursued, for example by the JETSCAPE collaboration~\cite{JETSCAPE:2021ehl}. The result of a first determination of $\hat{q}$ by the JETSCAPE collaboration using $R_{\rm AA}$ measurements at the LHC and RHIC via the LBT model is shown in Fig.~\ref{fig:qhat}, with the 90\% posterior range at a temperature $T=0.3$~GeV shown on the right. The curve itself is shown for median parameter values in that analysis. The earlier results from the JET collaboration~\cite{Burke:2013yra} are also shown as data points at the average temperatures from RHIC (left) and the LHC (right).

Most of the models for which $\hat{q}/T^3$ is shown in Fig.~\ref{fig:qhat} are based on weak-coupling approaches, except for CUJET 3.1~\cite{Shi:2019nyp} that incorporates the effect of colour-magnetic monopoles. These monopoles can be interpreted as a strong-coupling mechanism, that only affects $\hat{q}/T^3$ in the temperature range near the phase transition. While the data are consistent with CUJET 3.1, they also agree with the DREENA-A model~\cite{Stojku:2020tuk}, which uses a purely weak-coupling approach including chromo-magnetic fields, and the JETSCAPE tune based on the MATTER and/or LBT models~\cite{JETSCAPE:2021ehl}. The good description of the data by these weak-coupling models and the agreement between the $\hat{q}$ values at high temperature, suggest that strong-coupling scenarios which imply a much larger value of $\hat{q}$ at high-$T$ would lead to a lower $R_{\rm AA}$ (more suppression) and/or higher $v_2$ than seen in the measurements. 

To summarise, it can be concluded that there are some indications that the QGP is weakly coupled when probed with high-energy jets at the highest temperatures accessible at the LHC, as all of the models that describe ALICE data are dominated by weakly-coupled interactions in that limit. However, it is an open question whether the same applies around the deconfinement temperature. To further constrain the behaviour of the medium as seen by hard probes in this regime, additional exploration of the temperature dependence of the medium properties is needed, both with theoretical methods and with additional measurements and model comparisons, for example as a function of collision centrality and/or collision energy. At the same time, measurements of the modification of jet structure provide further insight into the microscopic dynamics of parton energy loss. 

\newpage

\setcounter{section}{2}
\section{High-density QCD effects in proton--proton and proton--nucleus collisions}
\label{ch:SmallSystems}

Our understanding of the conditions necessary for QGP formation was overthrown in 2010 with the arrival of the first LHC data. Until then, small systems such as proton--proton (pp) and proton--lead (p--Pb) collisions were considered as plain references, systems in which the conditions to form the QGP could not be reached. At LHC energies, in a small fraction of these collisions, the number of produced particles is of the same order as that in peripheral nucleus--nucleus collisions. Surprisingly, emblematic signatures of QGP formation, such as a double ridge extended in pseudorapidity~\cite{Aad:2015gqa, Abelev:2012ola, CMS:2010ifv, Aad:2014lta} and strangeness enhancement~\cite{ALICE:2017jyt}, were observed. This opens two main questions: could the QGP be formed in small systems? What are the mechanisms involved in the initial stages of the collision that might create a sufficiently-high energy density for the potential phase transition of nuclear matter?

These fundamental issues have received great attention and their investigation can be considered among the main novel aspects of the LHC physics programme, with the related discussions extending outside the community involved in QGP studies.
In this chapter, we will describe and discuss the main ALICE contributions to this exciting endeavour, also in relation to complementary results obtained by the other LHC experiments.

Section~\ref{section:3.1} addresses the event classification, a fundamental prerequisite that allows direct comparison of the studied phenomena independently of the considered system, be it (high-multiplicity) pp collisions, p--Pb or Pb--Pb. In Sec.~\ref{section:3.2} a broad discussion of the results on particle production across collision systems, as a function of charged-hadron multiplicity, is reported. Special emphasis is given to the role of strange particles, but the discussion is extended also to results on charmed hadrons relevant in this context. Quantities such as the baryon-to-meson ratios, known to be sensitive to the radial boost in the hot partonic system, are also considered in the presentation of the ALICE results. One of the main avenues for the investigation of collective effects is azimuthal anisotropy studies; they are discussed extensively in Sec.~\ref{section:3.3}, in terms of the multiplicity dependence of various flow orders ($n$) calculated with $m$-particle cumulants, $v_{n}\{m\}$, for different collision systems. Elliptic flow measurements for particles, ranging from inclusive to strange hadrons to J/$\psi$, and showing indications of a mass-ordering effect, are then presented. The last two sections of this chapter report studies of the production of hard probes in p--Pb collisions and the search for effects that may be related to the formation of the QGP. Section~\ref{sec:quarkoniumpA} addresses the quarkonium production and, in particular, the investigation of possible final-state suppression of charmonium- and bottomonium-excited states, which has been established for the $\psi$(2S) loosely-bound state. Finally, in Sec.~\ref{Sect:JetQuenchSmallSystems}, measurements of observables related to jet quenching are reported, via yield measurements of inclusive hadrons or jets and via coincidence of charged-particle jets recoiling from a high-transverse-momentum trigger hadron.

\subsection{Event classification for small collision systems}
\label{section:3.1}

A big success in the exploration of large systems reported in the previous chapter has been the development of the Glauber picture (Sec.~\ref{sec:TG1centrality}), which creates a map between the experimentally-measured multiplicity associated with centrality and key physics quantities such as $N_{\rm part}$, $N_{\rm coll}$ and the initial anisotropies. This approach captures essential features that have simplified the interpretation of several physics measurements in heavy-ion collisions. This simple mapping requires that nuclear geometry dominates over subnucleonic structure and that fluctuations average out. In small systems, neither of these requirements are fulfilled and, therefore, one becomes very sensitive to the modelling of nucleon--nucleon collisions.
Charged-particle multiplicity appears to be the simplest, even if imperfect, event classifier. It is a well defined measurement and allows for a direct comparison among all systems without any model dependence. Multiplicity was largely utilised by the ALICE Collaboration for event classification in small systems and we will discuss our current understanding and the associated biases.

In the case of pp collisions, it is generally accepted in models that initial scatterings give rise to some number of subnucleonic interactions that produce the observed hadrons~\footnote{These interactions can be between partons, dense colour fields in the CGC, or similar.}. These subnucleonic interactions vary from hard partonic interactions, giving rise to high-$p_{\rm T}$ jets that produce many particles and to colour exchanges involving soft interactions, which produce mainly longitudinal colour fields that decay into a few hadrons. The number of initial interactions, their hardness, and stochastic hadronisation effects mean that a specific event multiplicity can result from several different combinations of processes. Furthermore, combinations may vary as a function of charged-particle multiplicity leading to non-trivial modifications of measured observables as a function of multiplicity, independently of final state effects.

As the QGP is a state of matter dominated by soft interactions, the event-classification scheme should be sensitive to the number of initial interactions. 
The dynamics involved in these interactions can be experimentally quantified by comparing the shape of the $p_{\rm T}$ spectra of charged particles, in particular at high-$p_{\rm T}$, when performing a certain event selection and when measuring instead all inelastic pp and p--Pb collisions. In the following we focus first on pp collisions before moving on to p--Pb collisions. 

\begin{figure}[ht]
\centering
\includegraphics[width=0.6\textwidth]{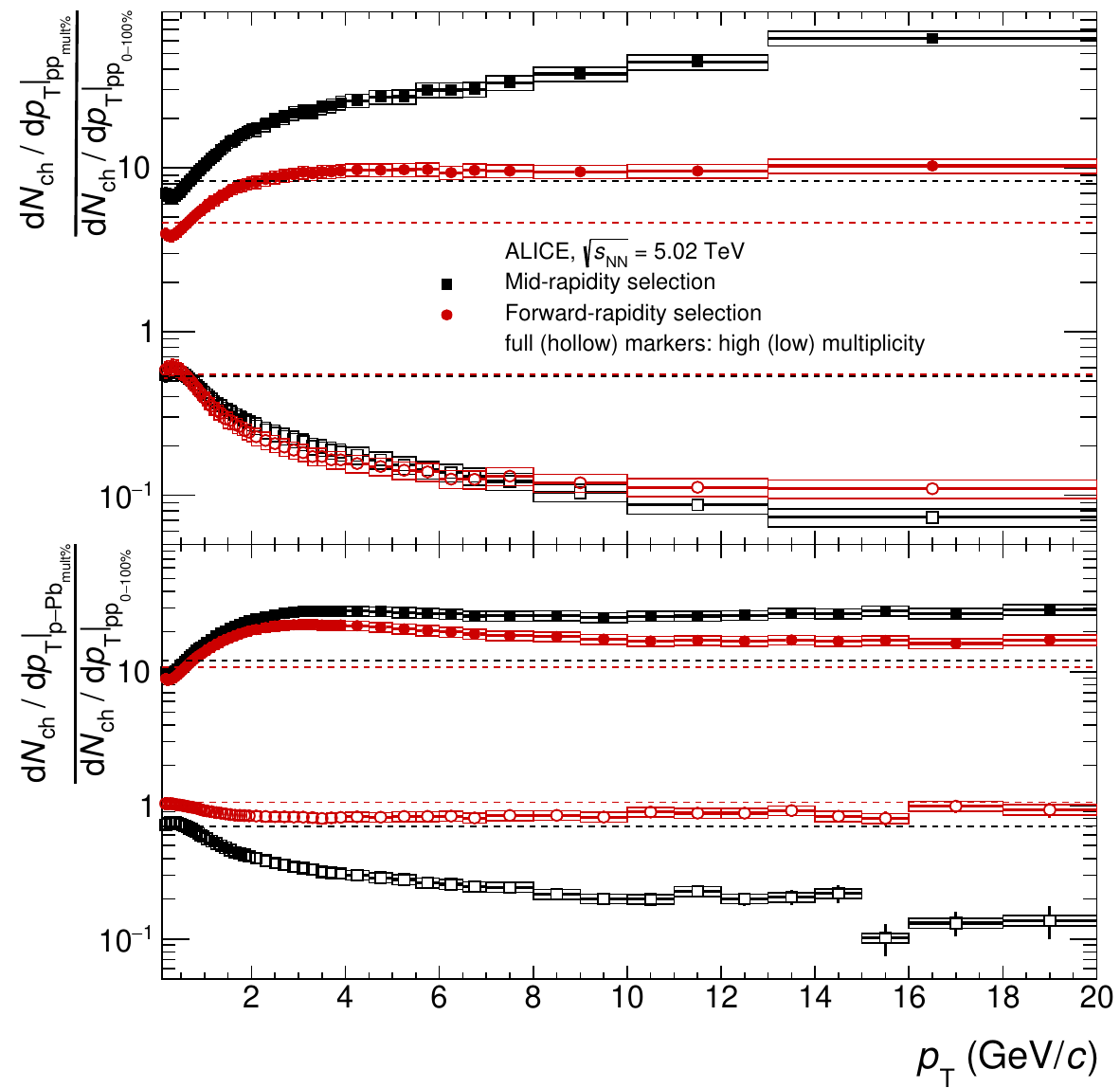}
\caption{Ratios of multiplicity-selected transverse-momentum distributions of charged particles for pp collisions (top)~\cite{Acharya:2019mzb} and p--Pb collisions (bottom)~\cite{Adam:2014qja} to inelastic (INEL) pp collisions, shown in logarithmic scale. For pp collisions, the forward-rapidity event selection is done using the V0M multiplicity estimator ($-3.7 <\eta< -1.7$, $2.8 <\eta< 5.1$), while the midrapidity event selection is done using the SPD tracklets in $|\eta|<0.8$. For p--Pb collisions, the forward-rapidity event selection is done using the A-side of the V0 multiplicity estimator ($2.8 <\eta< 5.1$), corresponding to the Pb-going side, while the midrapidity event selection is done using the first layer of the SPD. The horizontal dashed lines denote the 
integrated charged-particle multiplicity density ratios as a reference. For this figure, the pp high-multiplicity event classes correspond to 0--1\% (V0M-based, forward-rapidity selection) and 0.009--0.088\% (SPD-based, midrapidity selection) and the low-multiplicity pp event classes correspond to 70--100\% (V0M) and 49.5--100.0\%, where percentiles are defined with respect to the total pp INEL$>$0 cross section. For all event selections, the p--Pb high-multiplicity and low-multiplicity event classes 
correspond to 0--5\% and 70--100\%, respectively, with percentiles defined with respect to the total visible V0A cross section.}
\label{fig3.1a} 
\end{figure}

Figure~\ref{fig3.1a} (top) shows the ratio of the multiplicity-dependent $p_{\rm T}$ spectra of charged particles in the midrapidity region to the minimum bias (MB) spectrum. The spectra are obtained by slicing according to the multiplicity at mid (black) and forward (red) rapidity. In this selection, the midrapidity multiplicity estimator refers to particle trajectories within $|\eta|<0.8$ and forward rapidity 
refers to the multiplicity in the acceptance of the V0C and V0A scintillators, which is of $-3.7<\eta<-1.7$ and $2.8<\eta<5.1$, respectively. 
The multiplicity classes are expressed in terms of percentiles of the total cross section of inelastic pp collision with at least one charged particle in $|\eta|<1$ (INEL$>$0).
The high-multiplicity ratio for the midrapidity estimator is significantly higher than for the forward estimator. One expects a trivial difference as the $p_{\rm T}$ spectra are being measured at midrapidity in the same kinematic region where the midrapidity multiplicity selection is done. However, the slope of the $p_{\rm T}$ spectra at high-$p_{\rm T}$ indicates that the midrapidity estimator selects harder and harder subnucleonic interactions as the multiplicity increases. The ratios obtained with the forward estimator do not show a change in slope at high-$p_{\rm T}$. Still, the hard high-$p_{\rm T}$ production is more enhanced than soft low-$p_{\rm T}$ production in high-multiplicity collisions and vice versa in low-multiplicity collisions. This implies that the scaling with multiplicity of soft and hard processes is fundamentally different in pp compared to nucleus--nucleus collisions.

\begin{figure}[htbp!]
\centering
\includegraphics[width=0.6\textwidth]{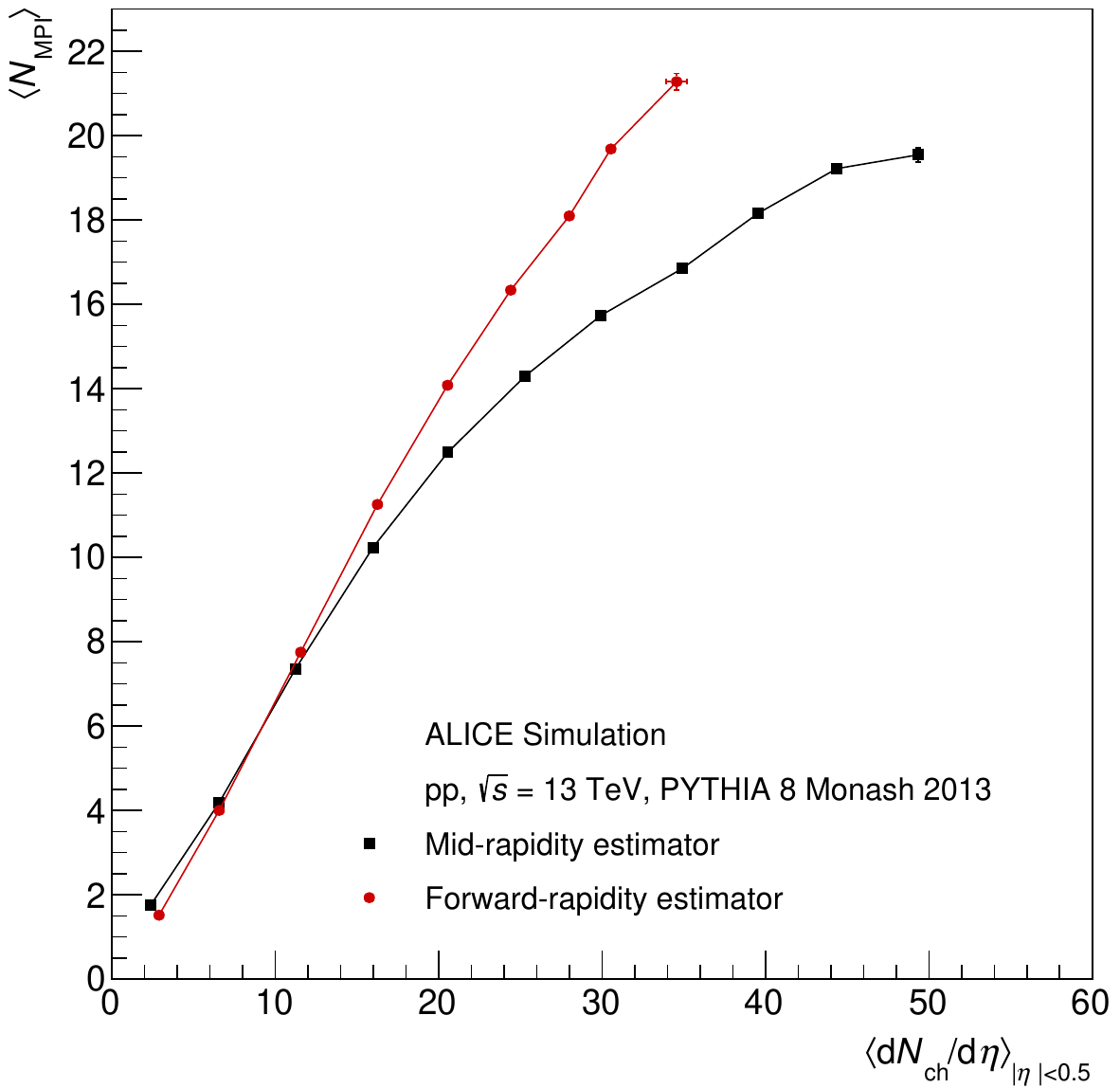}
\caption{Mean number of Multiple Parton Interactions, $\langle N_{\rm{MPI}}\rangle$, as a function of mean charged particle multiplicity at midrapidity, $\langle {\rm d}N_{\rm{ch}}/\rm{d}\eta \rangle_{|\eta|<0.5}$, for different event selections, calculated with the PYTHIA 8.2 model (Monash 2013 tune)~\cite{Sjostrand:2014zea}.}
\label{fig3.1b} 
\end{figure}

To gain insight into the origin of the difference between the spectra ratios for the midrapidity and forward estimator, it is interesting to use an event generator in which the initial number of Multiple Parton Interactions (MPIs) utilised in the simulation can be directly related to the charged-particle multiplicity. Figure~\ref{fig3.1b} reports a study with PYTHIA 8 (Monash 2013 tune) which shows that high-multiplicity events are preferably related to large number of MPIs, but are also influenced by other effects, such as the fragmentation of partons into a large number of final state particles.
The PYTHIA 8 simulations show that the forward multiplicity estimator has the strongest correlation between the number of MPIs and the multiplicity. For this reason, the forward multiplicity slicing is used for multiplicity selection in the rest of this section unless specifically noted otherwise. As the multiplicity selection is done on charged particles, a second advantage of the forward selection is that it does not create an imbalance between charged and neutral particles at midrapidity, as described in detail in~\cite{Acharya:2018orn}.

As discussed in Sec.~\ref{sec:TG1centrality}, and contrary to the MPI description of
pp collisions, nucleus--nucleus collisions are traditionally conceived as multiple interactions of individual nucleons. The p--Pb collisions act as a bridge between the two limiting descriptions where both views would need to be reconciled. 

In the same way as done for the pp collisions, p--Pb collisions are classified by making use of the V0A detector signal from the pseudorapidity range of 2.8~$< \eta <$~5.1, corresponding to the Pb-going side (direction of the fragmentation of the Pb nucleus), where the larger signal amplitudes associated to the nucleus fragmentation region will stabilise short-range multiplicity fluctuations.  
Figure~\ref{fig3.1a} (bottom) shows the ratio of the $p_{\rm T}$ spectra obtained in p--Pb, by slicing in multiplicity at mid (black) and forward (red) rapidity, to the MB spectrum obtained in pp collisions at the same $\sqrt{s}$. The midrapidity-estimator ratio has a higher multiplicity dependence than the forward-estimator ratio, although the effect is reduced with respect to the situation in pp collisions, shown in the upper panel. Focusing on the low-multiplicity sample, the ratio is below one in particular for the midrapidity estimator. This poses a fundamental challenge to the Glauber model as one cannot have $N_{\rm part}$ and $N_{\rm coll}$ smaller than that of a single pp collision. It is also clear that for the midrapidity estimator, there is no easy fix for this as the scaling for soft and hard processes are different. This indicates that one has to be careful when applying the Glauber model to interpret the results obtained with multiplicity slicing in small collision systems where the effect of particle-production fluctuations for individual nucleon--nucleon collisions, as shown in Fig.~\ref{fig3.1a} (top), can play a big role. 
Also in peripheral Pb--Pb collisions, the centrality selection based on charged-particle multiplicity, leads to a bias in the determination of the collision geometry parameters~\cite{Loizides:2017sqq}, which has to be taken into account when interpreting measurements that use these parameters, like the nuclear modification factor (see Sec.~\ref{sec:ElossHadrons}).

\begin{figure}[htbp!]
\centering
\includegraphics[width=0.45\textwidth]{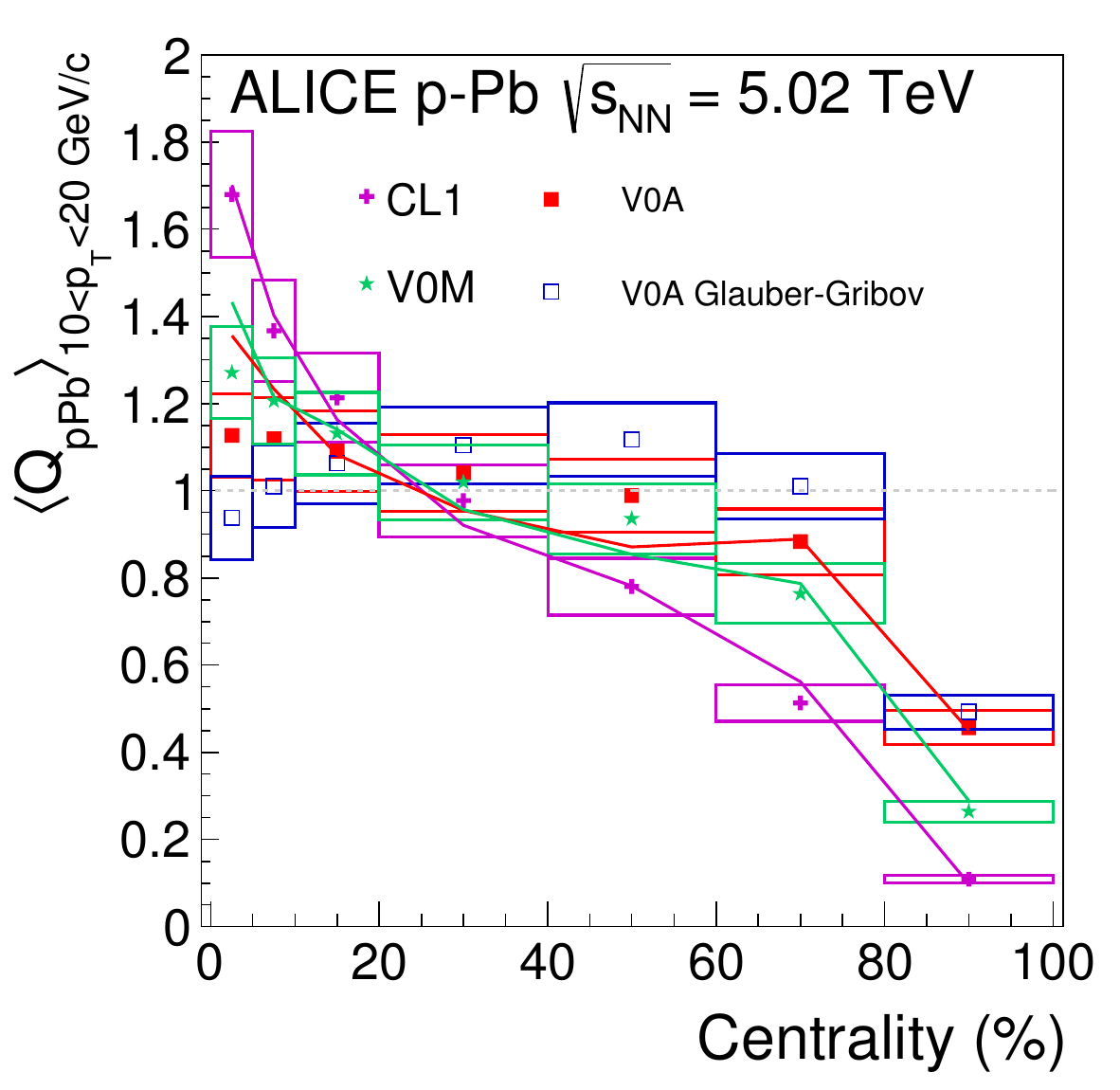}
\includegraphics[width=0.45\textwidth]{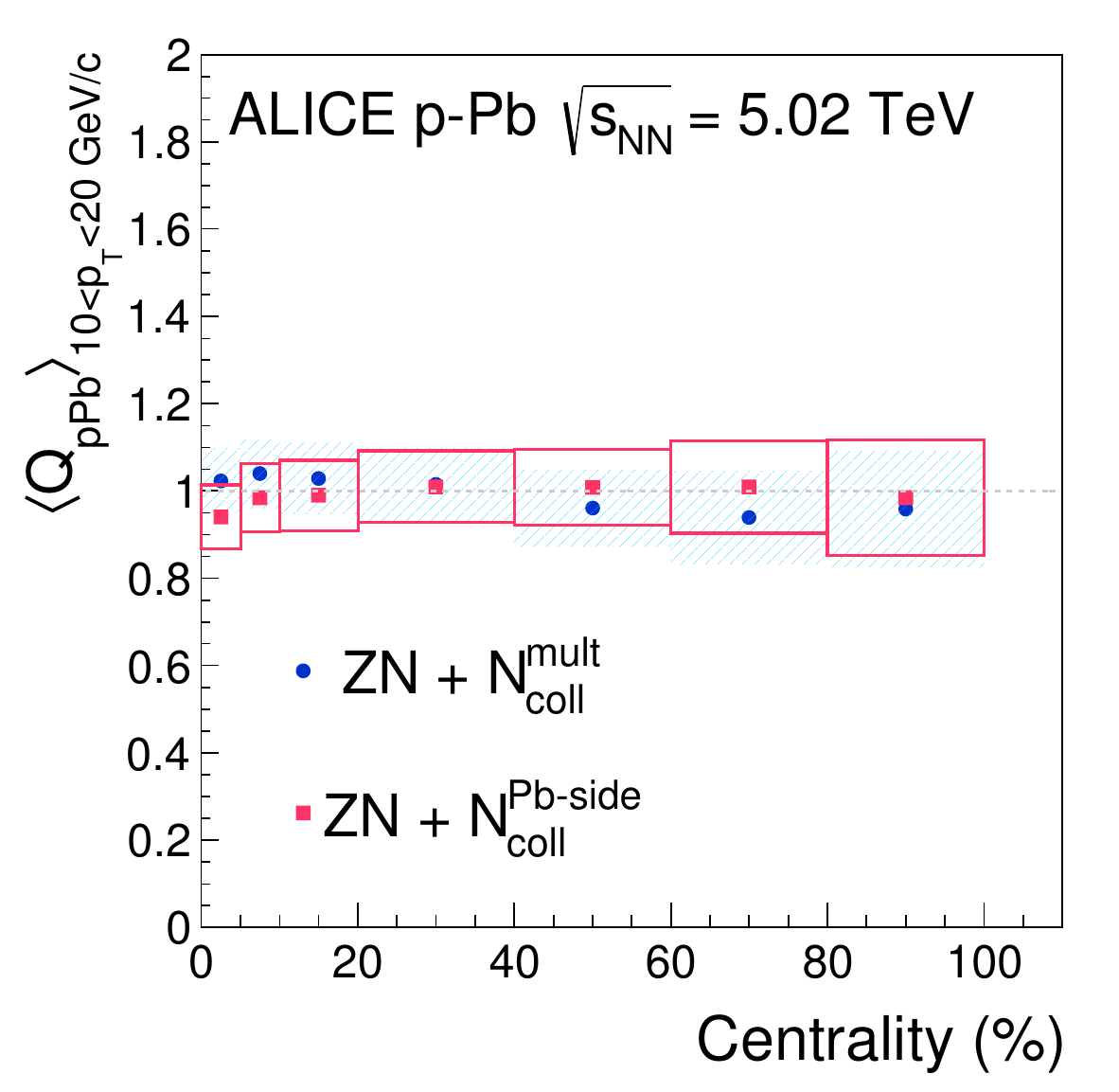}
\caption{(Left) $Q_{\rm{pPb}}$ at high transverse momenta calculated using various multiplicity selections (points) and predicted with the Q-PYTHIA model (curves)~\cite{Adam:2014qja}. 
(Right) $Q_{\rm{pPb}}$ calculated with the hybrid method. Spectra are calculated in event classes based on the ZNA energy (neutron ZDC on the A side, Pb-going) and $N_{\rm{coll}}$ is 
obtained assuming $N_{\rm{part}}$ scaling at midrapidity. Figure from~\cite{Adam:2014qja}.}
\label{fig3.1bis} 
\end{figure}

To circumvent these limitations, the ALICE Collaboration has developed an event-selection method for p--Pb collisions based on the energy deposited from the Pb nuclei in the Zero Degree Calorimeters (ZDCs)~\cite{Adam:2014qja}. Such an approach has the ability to classify event activity with an even larger rapidity gap between actual measurement (acceptance of the ALICE central barrel or muon spectrometer) and event selection and a very low sensitivity to fluctuations in particle production introduced by subnucleonic interactions. This method, coupled to an assumption 
that the total charged-particle multiplicity, dominated by low-$p_{\rm T}$ particles, scales with $N_{\rm{part}}$ - a combination henceforth known as the `hybrid' method - has been shown to 
recover the expected scaling of particle yields at high momenta. This can be observed when calculating the $Q_{\rm{pPb}}$, a quantity analogous to the nuclear modification factor $R_{\rm{AA}}$, but calculated in proton--nucleus collisions. While the $Q_{\rm{pPb}}$ at high momenta deviates from unity in data using typical multiplicity selections, such behaviour is well described by Q-PYTHIA, a model that can be used to do a baseline check including multiplicity fluctuations. When performing ZDC-based event selections and utilising the hybrid method~\cite{Adam:2014qja}, on the other hand, the high-$p_{\rm{T}}$ $Q_{\rm{pPb}}$ values are observed to be compatible with unity. These results are shown in Fig.~\ref{fig3.1bis} and the selection based on the ZN signals (neutron ZDCs) will serve as a baseline for the p--Pb studies performed in Sec.~\ref{Sect:JetQuenchSmallSystems}. 

Multiplicity was extensively used as an event classifier during the LHC Run 2, although it introduces some difficulties when comparing data and theory. In fact, multiplicity cannot be directly related to the initial energy density of the collision, and the same multiplicity in small and large collision systems can result from completely different initial energy densities. %

While the event classification studies performed during Runs 1 and 2 were
highly successful, more advanced selection strategies are being developed
which will allow for a better determination of the phase space 
in which several phenomena appear in small collision systems. Notably, 
the transverse activity estimator $R_{\rm{T}}$~\cite{Acharya:2019nqn} and spherocity selections~\cite{Acharya:2019mzb}
may shed further light on the transition between low and high multiplicities, with some measurements already underway and many more to appear in the near future. In the next sections, we will discuss the results based on 
the charged-particle multiplicity event classification scheme, since 
most of these are published at this time.

\subsection{Dynamics and hadrochemistry of particle production}
\label{section:3.2}

\begin{figure}[htbp!]
\begin{center}
\includegraphics[width=0.9\linewidth]{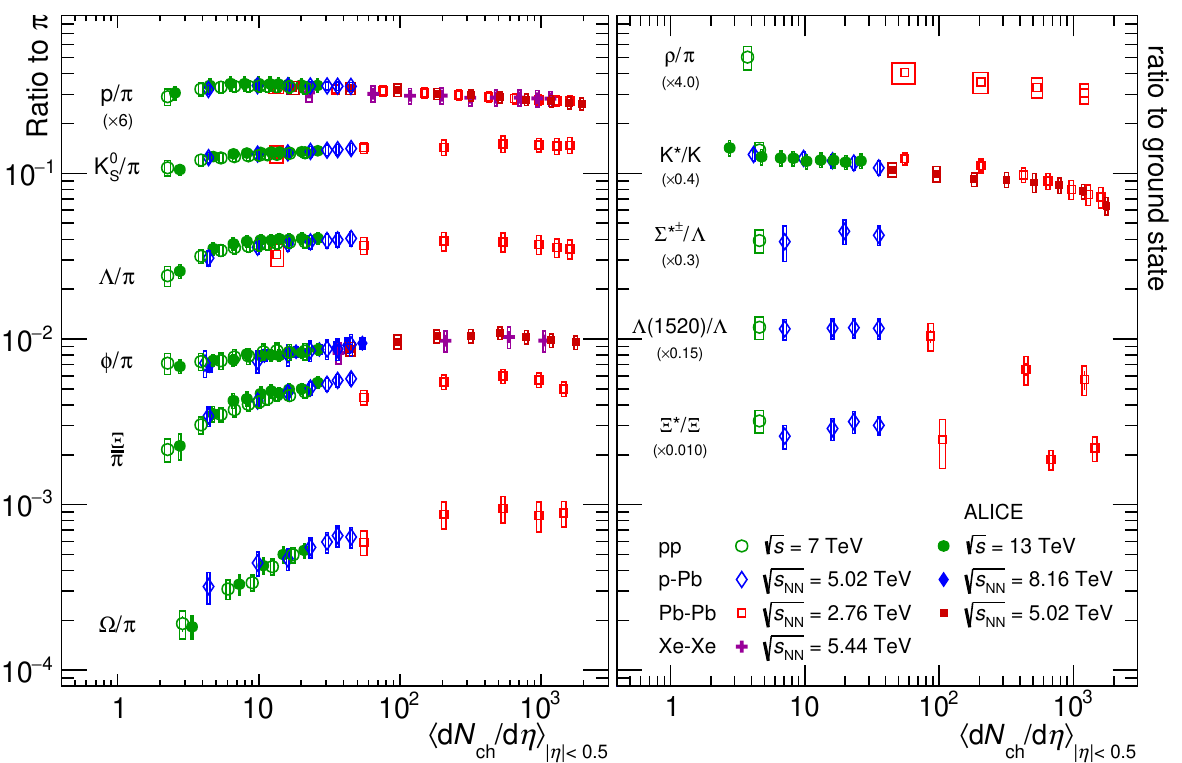}
\caption{(Left) $p_{\rm T}$-integrated yield ratios to pions ($\pi^+$ + $\pi^-$) and $p_{\rm T}$-integrated yield ratios between resonance and corresponding ground state as a function of $\langle {\rm d}N_{\rm{ch}}/\rm{d}\eta\rangle$ measured in $|\eta| < 0.5$ in pp collisions at $\sqrt{s} = 7$ and 13 TeV~\cite{ALICE:2017jyt, Acharya:2018orn, ALICE:2019etb, ALICE:2020nkc}, p--Pb collisions at $\sqrt{s_{\rm{NN}}} = 5.02$ TeV~\cite{Abelev:2013haa,Adam:2015vsf, ALICE:2019smg} and at $\sqrt{s_{\rm{NN}}} = 8.16$ TeV~\cite{ALICE:2021uyz}, Xe--Xe collisions at $\sqrt{s_{\rm{NN}}} = 5.44$~TeV~\cite{Acharya:2021ljw} and Pb--Pb collisions at $\sqrt{s_{\rm{NN}}} = 2.76$ TeV~\cite{ABELEV:2013zaa, ALICE:2018ewo, ALICE:2018qdv} and at $\sqrt{s_{\rm{NN}}} = 5.02$~TeV~\cite{Abelev:2013vea,Acharya:2019yoi, ALICE:2019xyr}. %
(Right) Ratios involving $\rho$ mesons, $\rm K^*$, 
$\Sigma^{*\pm}$, $\Lambda(1520)$ and $\Xi^{*}$. All yields are obtained at midrapidity, i.e.\,$|y|<0.5$. The
error bars show the statistical uncertainty, whereas the empty and dark-shaded boxes show the total systematic
uncertainty and the contribution uncorrelated across multiplicity bins, respectively.}
\label{fig3.2a} 
\end{center}
\end{figure}

\begin{figure}[ht]
\begin{center}
\includegraphics[width=0.9\linewidth]{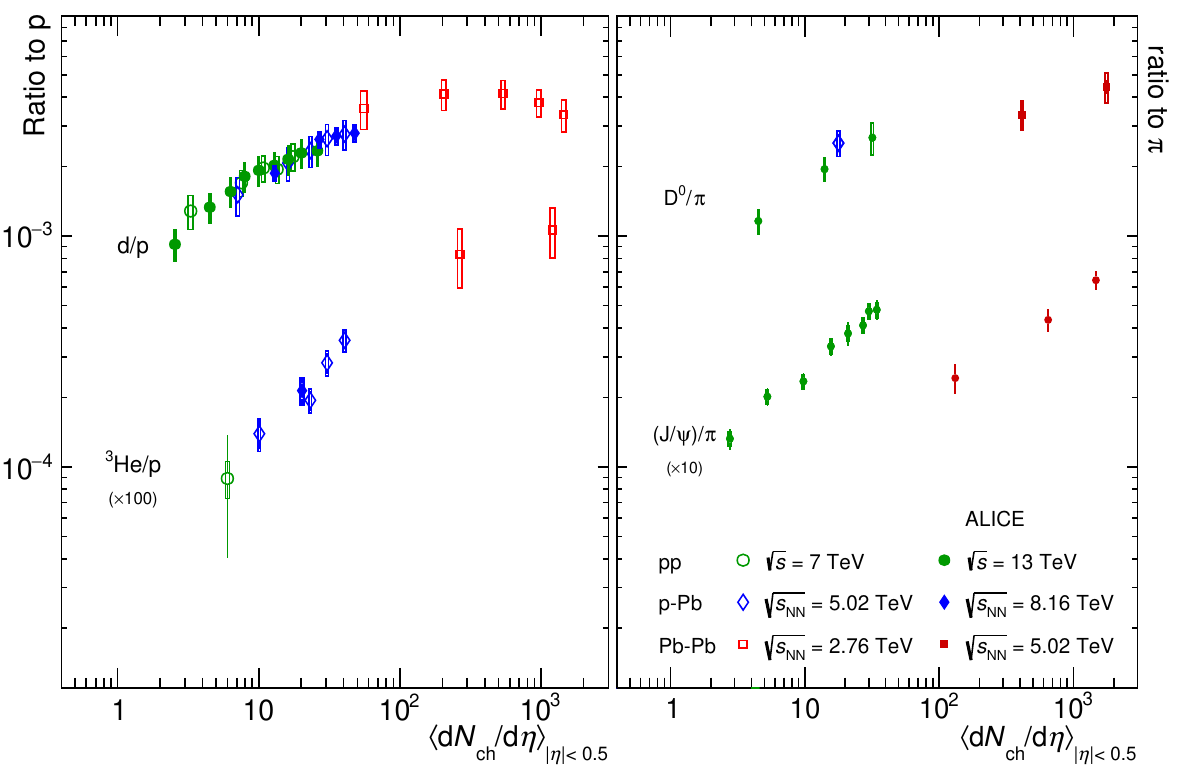}
\caption{$p_\mathrm{T}$-integrated yield ratios to protons or charged pions (particles + antiparticles) as a function of $\langle {\rm d}N_{\rm{ch}}/\rm{d}\eta\rangle$ measured in $|\eta| < 0.5$ in pp collisions~\cite{ALICE:2020nkc, ALICE:2020msa, ALICE:2021npz}, p--Pb collisions~\cite{Abelev:2013haa, ALICE:2019bnp, ALICE:2019fhe} and Pb--Pb collisions~\cite{Acharya:2019yoi, ALICE:2021rxa, ALICE:2019nrq, Adam:2015vda} for nuclei (left) and for $\rm{D}^{0}$ and $\rm{J}/\psi$ mesons (right). 
All yields are obtained at midrapidity, i.e.\,$|y|<0.5$. The error bars show the statistical uncertainty, whereas the empty and dark-shaded boxes show the total systematic
uncertainty and the contribution uncorrelated across multiplicity bins, respectively.}
\label{fig3.2b} 
\end{center}
\end{figure}

The ALICE Collaboration studied the production of a large variety of particle species across a broad range of collision energies ($2.76 < \sqrt{s_{\rm NN}} < 13$~TeV) for various hadronic systems (pp, p--Pb, Pb--Pb and Xe--Xe) as a function of the charged-particle multiplicity of the collision at midrapidity in event classes defined by selections on the signal in the forward detectors V0A and V0C ($2.8 < \eta < 5.1$ and $-3.7 < \eta < -1.7$). These results are shown in Figs.~\ref{fig3.2a} and~\ref{fig3.2b}.

\begin{figure}[ht]
\begin{center}
\begin{minipage}{0.5\textwidth}
\centering
\includegraphics[width=0.99\textwidth]{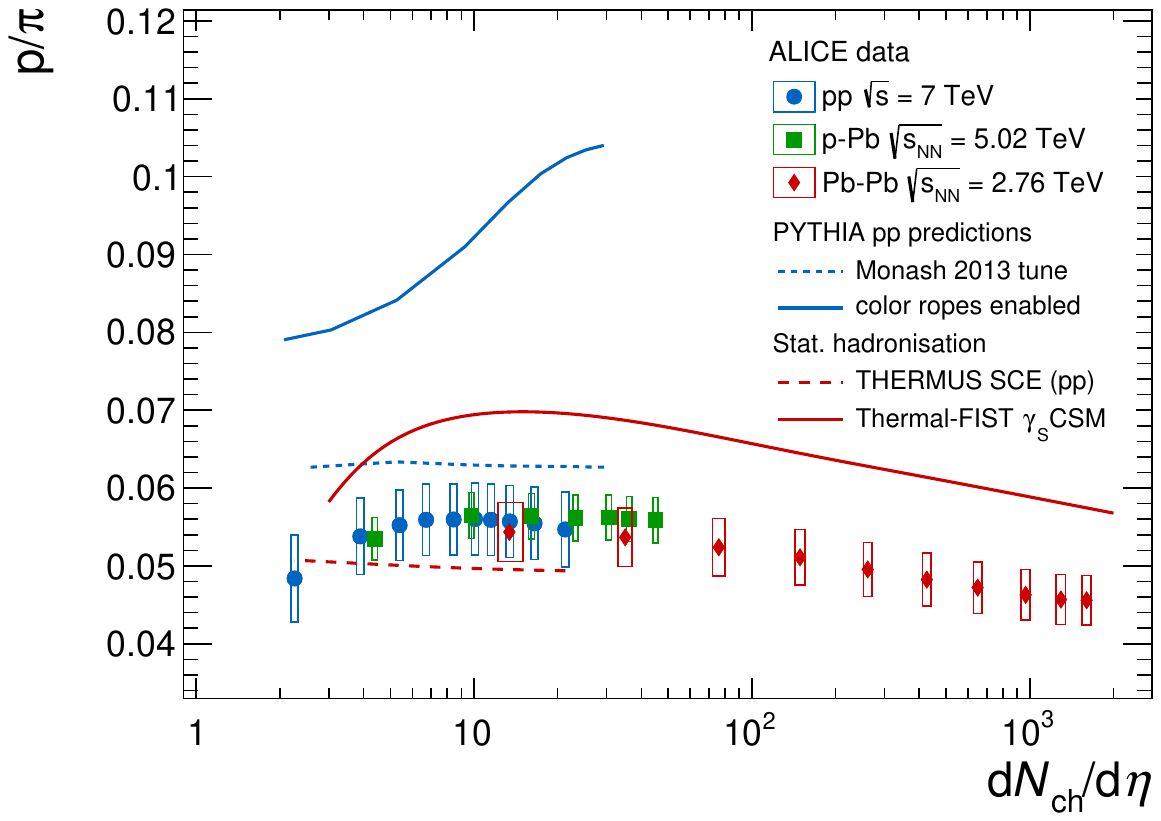}
\end{minipage}\hfill
\begin{minipage}{0.5\textwidth}
\centering
\includegraphics[width=0.99\textwidth]{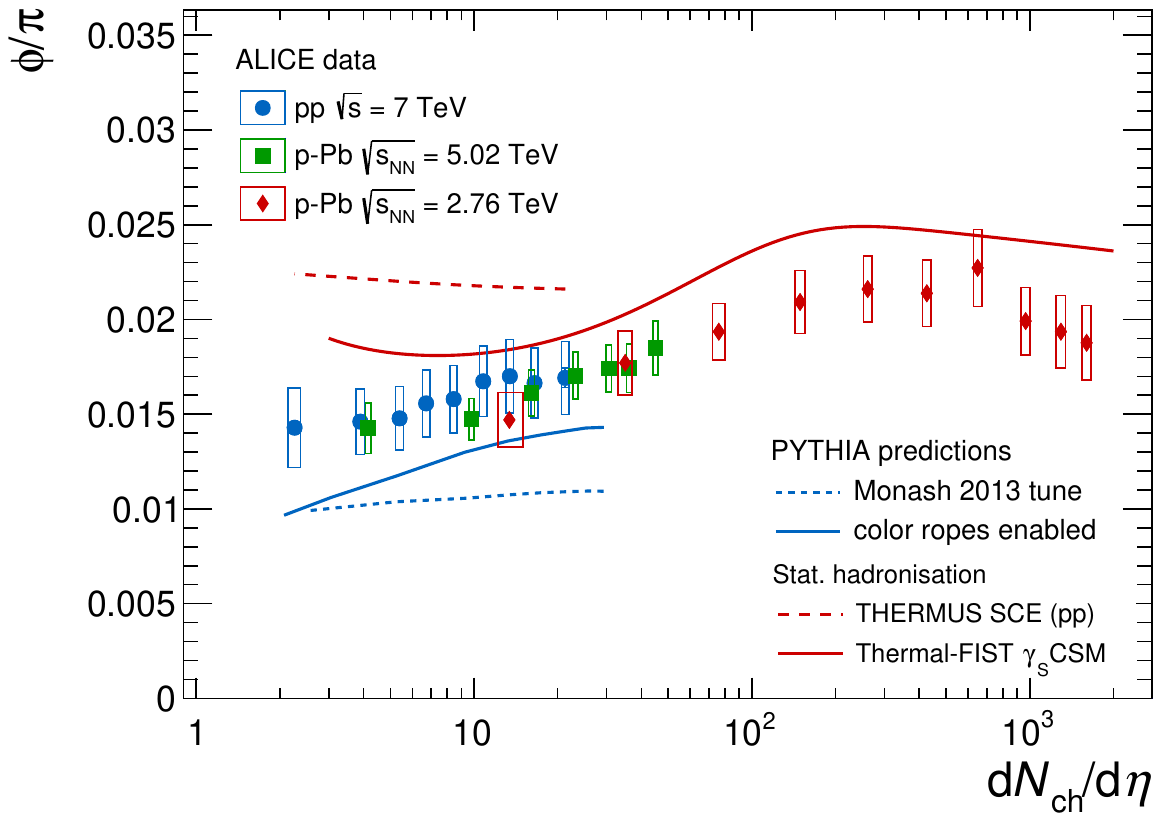}
\end{minipage}\hfill
\begin{minipage}{0.5\textwidth}
\centering
\includegraphics[width=0.99\textwidth]{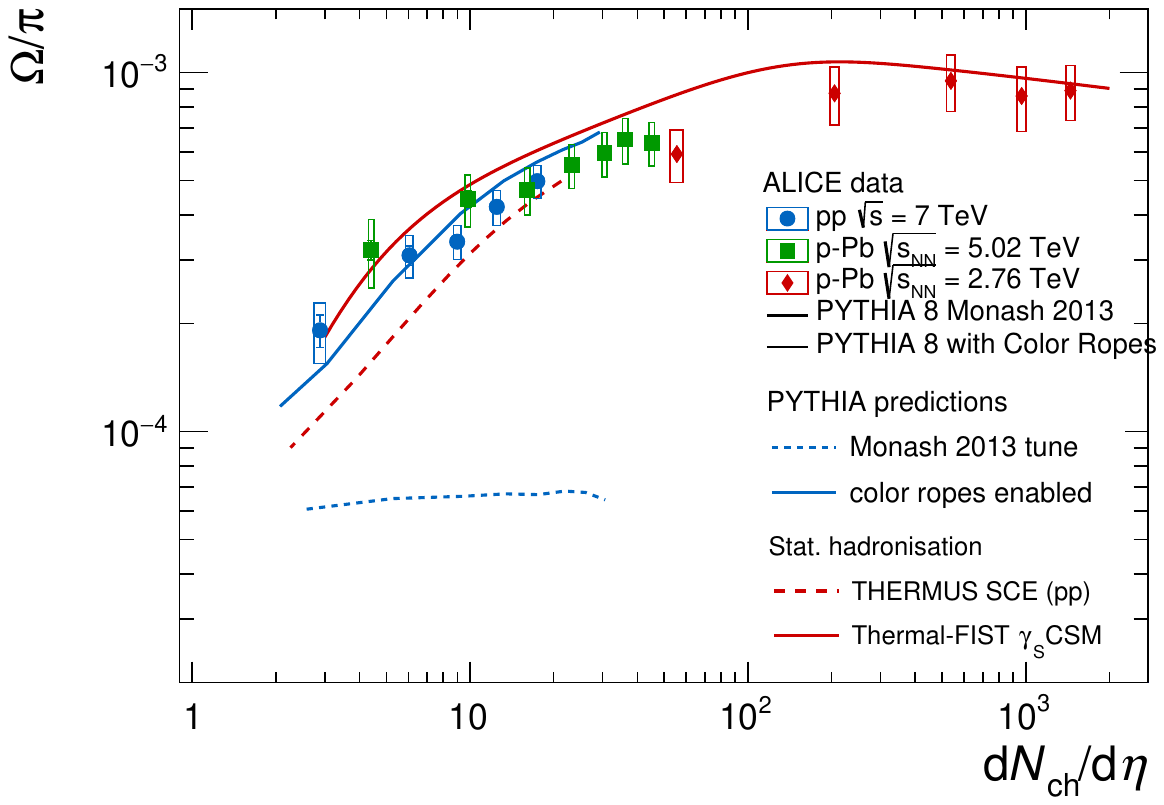}
\end{minipage}\hfill
\begin{minipage}{0.5\textwidth}
\centering
\includegraphics[width=0.99\textwidth]{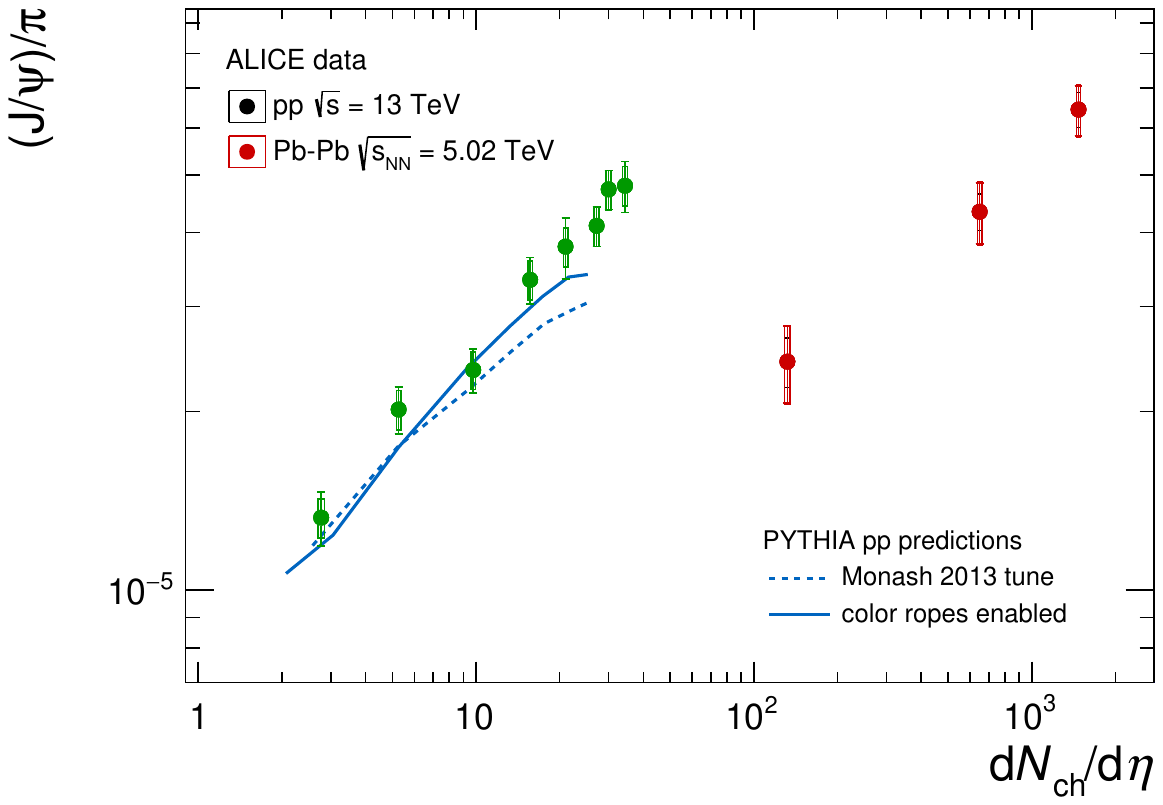}
\end{minipage}\hfill

\caption{Selected measurements from the previous figures compared to different theoretical models. (Top left) $p_{\rm T}$-integrated ratio of p$/\pi$ as a function of $\langle {\rm d}N_{\rm ch}/{\rm d}\eta \rangle$ at midrapidity. (Top right) $p_{\rm T}$-integrated ratio of $\phi/\pi$ as a function of $\langle {\rm d}N_{\rm ch}/{\rm d}\eta \rangle$ at midrapidity.
(Bottom left) $p_\mathrm{T}$-integrated ratio of $\Omega/\pi$ as a function of $\langle {\rm d}N_{\rm ch}/{\rm d}\eta \rangle$ at midrapidity compared to PYTHIA 8.2 (Monash 2013 tune)~\cite{Sjostrand:2014zea} and PYTHIA colour ropes model~\cite{Bierlich:2014xba}. (Bottom right) $p_{\rm T}$-integrated ratio of $(\rm{J}/\psi)/\pi$ as a function of $\langle {\rm d}N_{\rm ch}/{\rm d}\eta \rangle$ at midrapidity. 
}
\label{fig3.3} 
\end{center}
\end{figure}

The first striking observation is that the particle yield ratios with respect to pions increase with multiplicity in pp and p--Pb collisions, while they remain almost unchanged in heavy-ion collisions, as can be seen in Fig.~\ref{fig3.2a}. The only exceptions are for resonances such as the $\rm{K}^{*0}(892)$ and $\Lambda(1520)$ (see Sec.~\ref{sec:HadrPhase}). This evolution is smooth and very similar across collision systems, with measured ratios being consistent within uncertainties for event classes with similar multiplicities in pp, p--Pb, Xe--Xe and Pb--Pb collisions. Moreover, the ratios do not depend on centre-of-mass energy.  These are very important observations, which indicate that hadron yields are mostly related to final-state charged-particle multiplicity density rather than collision system or beam energy. 

The relative production rates of particles containing strange quarks, measured using the $\Lambda/\pi$, $\Xi/\pi$ and $\Omega/\pi$ ratios and shown in the left panel of Fig.~\ref{fig3.2a}, increase faster with multiplicity than those containing u and d quarks only. This effect is commonly referred to as ``strangeness enhancement'' and it is proportional to the strangeness content of the hadron: largest for triply-strange $\Omega^{-}$ and progressively less evident for particles with two and one strange quark, being absent for protons as discussed also in Sec.~\ref{sec:SHM}. The $\phi$ meson, a hadron with hidden strangeness $\rm s\overline s$, exhibits a relative yield increase that is mid-way between the ones measured for $\Xi^{-}$ and $\Lambda$. 
The experimental results concerning strangeness implied not only a major paradigm shift for the heavy-ion community, who had to re-evaluate the 
assumption that pp was a valid reference system, but were 
seen with great interest also by the particle physics community at large. As a consequence, theorists working on QCD-inspired models now consider the 
description of strongly-interacting matter an integral part of their programme. 

A general interpretation of the enhancement could be that the production of (multi-)strange baryons requires sufficiently energetic 
processes that are not favoured in low-multiplicity collisions. However, this mechanism is solely dependent on available energy at hadronisation, 
and therefore it should be captured by e.g.\,the most commonly used QCD-inspired pp event generator PYTHIA. Moreover, PYTHIA itself
predicts a different mixture of event processes at low-multiplicity - e.g. diffractive events - and in fact includes no dynamical component able to reproduce
the steady increase of particle ratios such as $\Xi/\pi$ and $\Omega/\pi$ as a function of charged-particle multiplicity in pp collisions.

In PYTHIA, hadron production occurs via the incoherent break-up of colour flux tubes called `strings', which exhibit constant energy density, leading to the conclusion
that even high-multiplicity events would result in unchanged particle ratios. As a consequence of the observation in~\cite{ALICE:2017jyt}, 
PYTHIA modelling had to resort to conceptually different physical mechanisms
to reproduce experimental data, such as the inclusion of `colour ropes' -- colour flux tubes with increased tension that are created whenever several strings
overlap prior to hadronisation in high-multiplicity pp collisions~\cite{Bierlich:2014xba}. The predictions from PYTHIA with colour ropes can be seen in Fig.~\ref{fig3.3} and describe strangeness enhancement in pp collisions within a 10\% accuracy in high-multiplicity collisions. However, it is important to note that the proton-to-pion ratio is not correctly described in this model, which indicates that further theoretical studies are still required for a proper
description of hadrochemistry using QCD-inspired, density-dependent modelling. 
Furthermore, alternative approaches that select on specific strangeness-related string breakup processes in PYTHIA have also been shown to
reproduce the correct relative strangeness production increase~\cite{Loizides:2021ima}, though a complete dynamical description of 
hadrochemistry employing this particular mechanism is still to be developed.

In addition, statistical hadronisation models also provide a good description 
of the relative increase of strangeness production as a function of charged-particle multiplicity 
density, as can be seen in Fig.~\ref{fig3.3}. A canonical statistical model, incorporating exact conservation of baryon number, electric charge and strangeness, and varying $\gamma_{\rm s}$ and $T_{\rm chem}$ using the Thermal FIST package~\cite{Vovchenko:2019kes, Vovchenko:2019pjl} is able to capture the evolution of most particle ratios quite well qualitatively across systems, with a maximum of 20\% deviation seen to be approximately multiplicity-independent.
An alternative canonical calculation restricted to pp and using a fixed $T_{\rm chem} = 156$~MeV based on~\cite{Cleymans:2020fsc} is also able to capture the strangeness increase rather well, though also in that case, discrepancies of up to 30\% are observed in other particle ratios, 
indicating that a comprehensive description of all particle ratios within experimental 
uncertainties remains a theoretical challenge. Further developments in proton--proton yield modelling 
will also include the extension of the GSI-Heidelberg model employing the S-matrix formalism~\cite{Andronic:2021erx} to 
pp and are expected to appear in the near future.

As seen in the right panel of Fig.~\ref{fig3.2a}, the multiplicity evolution of particle ratios involving resonances and their non-resonant hadronic states, such as $\rm K^{*0}(892)$ and $\rm K^{\pm}$, were also measured by the ALICE Collaboration. As discussed in Sec.~\ref{sec:HadrPhase}, the scope of this study in heavy-ion collisions is to determine the presence and lifetime of the hadronic phase which follows chemical freeze-out in heavy-ion collisions. The same study was performed in small collision systems (pp and p--Pb), showing again a rather smooth trend across multiplicity. Given the statistical and systematic uncertainties in the measurements, there is no evidence of a significant multiplicity evolution for long-lived resonances, while for the $\rm K^{*0}$ a hint of a decreasing trend is observed, pointing to a hadronic stage whose density and energy allow the rescattering of the decay products. The measurement of the $\rho^{0}$ meson still carries uncertainties that prevent any statement regarding the evolution of the $\rho^{0}/\pi$ ratio. 

The yield of light nuclei also evolves with multiplicity, as shown in the left panel of Fig.~\ref{fig3.2b}{, }increasing smoothly across different systems and energies and reaching the values observed in heavy-ion collisions for the highest multiplicities. The increase is more pronounced for $^{3}\rm{He}$ nuclei, which contain three nucleons, than for deuterons. %

Finally, heavy-flavour yields were also measured from proton--proton to central Pb--Pb collisions, 
and the corresponding yield ratios can be seen in Fig.~\ref{fig3.2b} (right), which shows clear evidence for an 
increase in the $(\rm{J}/\psi)/\pi$ ratio in high-multiplicity proton--proton collisions. However, contrary to what is observed for particles containing
only light-flavour quarks, the evolution of the $(\rm{J}/\psi)/\pi$ is not continuous with charged-particle
multiplicity density. This is likely due to the fact that charm quarks are produced in hard scattering processes with a rate that depends on collision system and centre-of-mass energy. 
While the factorisation 
approach can be used to calculate single inclusive production in minimum-bias interactions, the multiplicity dependence of charmed-particle production in small collision systems remains a topic in which further theoretical developments are needed for a better understanding of the role of MPIs in the building of the hadronic multiplicity and the interplay between the soft and hard components of the event. 
More precise measurements in Run 3 will be needed to explore potential dissociation of J$/\psi$ ranging from ground states to excited 
states through ratios of relative yields as a function of the charged-particle multiplicity density.

In heavy-ion collisions, the transverse momentum spectra of identified particles are sensitive to the radial flow imparted by the system expansion. This translates into an evolution of the spectra towards higher $\langle p_{\rm{T}} \rangle$ for more central A--A collisions, as discussed in Sec.~\ref{sec:TG2particlespectra}. 
The same effect has been observed when measuring $p_{\rm{T}}$ spectra as a function of the charged-particle multiplicity in pp and p--Pb interactions.
The low-momentum part of the spectra (up to around 2.5 GeV/$c$) becomes progressively more flat as the multiplicity increases, leading to an increase of the $\langle p_{\rm{T}} \rangle$ for all particle species. This opens the intriguing question: is the origin of the hardening of the spectra the same in different colliding systems? Is it the QGP in high-multiplicity pp interactions causing the increase in $\langle p_{\rm{T}} \rangle$? Can this effect be solely due to radial flow in an expanding system?

\begin{figure}[ht]
\begin{center}
\includegraphics[width=0.99\linewidth]{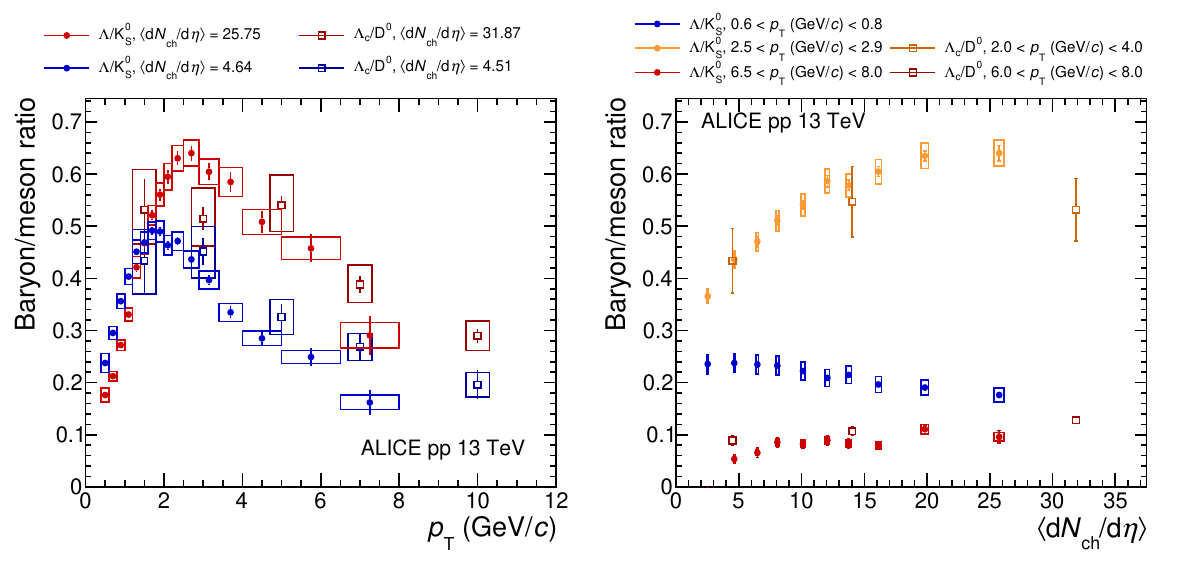}
\caption{(Left) Transverse-momentum dependence of $\Lambda/\rm{K}^{0}_{S}$ at midrapidity for high- and low-multiplicity 
proton--proton collisions at 13 TeV~\cite{ALICE:2019avo} drawn together with $\Lambda_{\rm c}/\rm{D}^{0}$ ratios~\cite{ALICE:2021npz}. (Right) Multiplicity 
dependence of the $\Lambda/\rm{K}^{0}_{S}$ and $\Lambda_{\rm c}/\rm{D}^{0}$ ratios at midrapidity in selected transverse-momentum intervals 
in pp collisions at 13 TeV.}
\label{fig3.4} 
\end{center}
\end{figure}

To try to answer these questions, the ratio between $\Lambda$ and $\rm{K}^{0}_{S}$ $p_{\rm{T}}$ spectra was measured in pp interactions characterised by different final-state multiplicities~\cite{ALICE:2019avo}. The idea behind this observable is that the radial boost of a collectively-expanding system should impact the heavier hadrons more strongly, giving rise to the observation of an enhanced baryon-to-meson ratio at intermediate-$p_{\rm{T}}$. This enhancement is observed in the $\Lambda/\rm{K}^{0}_{S}$ ratio for small collision systems in a qualitatively very similar way as in heavy-ion interactions. The magnitude of the intermediate-$p_{\rm{T}}$ enhancement increases as the multiplicity increases and the peak position moves towards higher values for high-multiplicity collisions, in agreement with the hydrodynamic origin picture. Moreover,
the increase at intermediate momenta is accompanied by a corresponding depletion of the ratio 
at low-momenta, with the integrated $\Lambda/\rm{K}^{0}_{S}$ ratio exhibiting essentially
no multiplicity dependence in pp collisions. This observation also holds for proton-to-pion ratios and is qualitatively reminiscent of what is observed in Pb--Pb collisions, as described in Sec.~\ref{sec:PtDiffRatioAndv2}. 
In order to study the low-$p_{\rm{T}}$ depletion and mid-$p_{\rm{T}}$ enhancement more quantitatively, specific $p_{\rm{T}}$ regions have been selected and the multiplicity 
dependence of the $\Lambda/\rm{K}^{0}_{S}$ ratio in these intervals is reported in the right panel of Fig.~\ref{fig3.4}. 

Similar studies have also been carried out recently for the charmed baryon-to-meson ratio 
$\Lambda_{\rm c}/\rm{D}^{0}$, where similar behaviour has been observed~\cite{ALICE:2021npz}. The corresponding results are
shown in Fig.~\ref{fig3.4} together with the $\Lambda/\rm{K}^{0}_{S}$ ratio and  
exhibit a pattern consistent with the non-charmed baryon-to-meson ratio, although 
future measurements with smaller uncertainties are still needed to come to a firm conclusion. 

Overall, all results discussed here point to the remarkable fact that the hadrochemistry 
depends solely on the final-state charged-particle multiplicity and is independent of the colliding system and collision energy at the LHC. This observation, together with the specific phenomena that appear to vary smoothly with charged-particle multiplicity, such as strangeness enhancement, constitutes a breakthrough result 
of the ALICE experiment. 

Further understanding regarding particle production rates of specific hadron species is expected
to come in the future with studies in which these particles are analysed 
in association with jets and using other correlation techniques such as net-quantum-number fluctuation
measurements. 

\subsection{Collective effects: anisotropic flow}
\label{section:3.3}

Particles emitted from a collective system created in a collision would exhibit global correlations shared among many particles spanning across a large rapidity range, resulting from the response to an initial anisotropy. This phenomenon can be studied with measurements of azimuthal correlations. A clear illustration can be seen in di-hadron correlations as a function of the difference in azimuthal angle, $\Delta\varphi$, and pseudorapidity, $\Delta\eta$, of the particle pairs, in the form of a double-ridge structure (a cosine modulation in $\Delta\varphi$) extended over a long range in $\Delta\eta$. In particular, the ridge signal in the near-side ($|\Delta\varphi| \approx 0$) is understood as manifestation of a collectively-expanding medium, which is distinctively observed in collisions of heavy ions~\cite{Aamodt:2011by,Adam:2017ucq}.

\begin{figure}[htbp!]
\begin{center}
\includegraphics[width=0.6\linewidth]{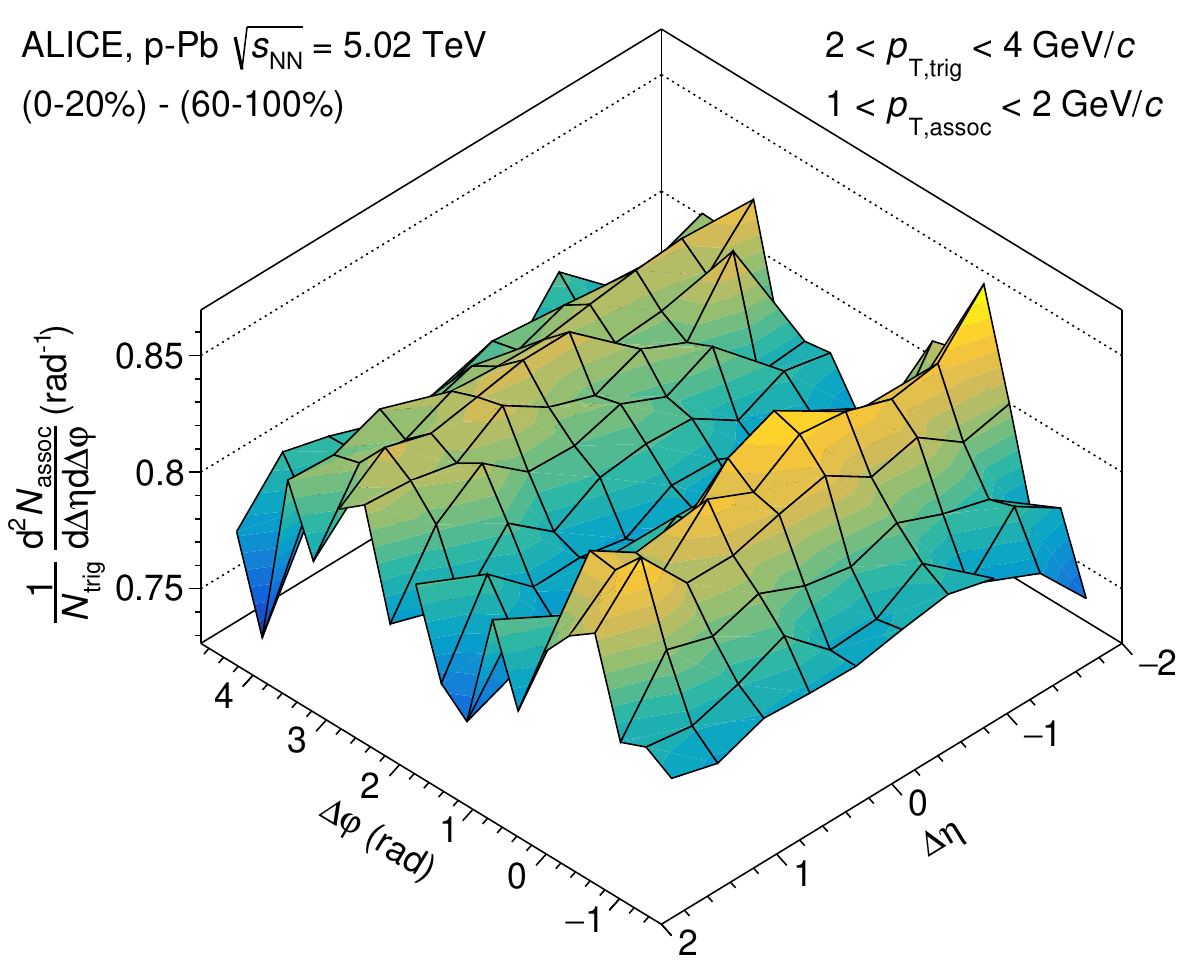}
\caption{Associated yield per trigger particle in $\Delta \varphi$ and $\Delta \eta$ for pairs of charged particles with $2 < p_{\rm T,trig} < 4$~GeV/{\it c} and $1 < p_{\rm T,assoc} < 2$~GeV/{\it c} in p--Pb collisions at $\sqrt{s_{\rm NN}} = 5.02$~TeV for the 0--20\% multiplicity class, after subtraction of the associated yield obtained in the 60--100\% event class under the assumption that the latter multiplicity class is completely dominated by non-flow correlations~\cite{Abelev:2012ola}. A clear double-ridge signal spanning a large range in $\Delta\eta$ can be seen on the near-side $|\Delta\varphi| \approx 0$ and on the away-side $|\Delta\varphi| \approx \pi$, indicative of the possibility of a collective behaviour in p--Pb collisions.}
\label{fig3.5} 
\end{center}
\end{figure}

A significant near-side ridge signal was first reported in high-multiplicity pp collisions by the CMS Collaboration~\cite{CMS:2010ifv}. Shortly after, a double-ridge signal was evidenced in high-multiplicity p--Pb collisions at $\sqrt{s_{\rm NN}} = 5.02$~TeV by the ALICE Collaboration~\cite{Abelev:2012ola} (Fig.~\ref{fig3.5}) and confirmed by the CMS and ATLAS Collaborations~\cite{CMS:2012qk,Aad:2014lta}. Such observations indicated the possibility that small systems could exhibit a collective behaviour. A similar double-ridge signal, though with smaller magnitude than in p--Pb or Pb--Pb collisions, was also later measured in the high-multiplicity pp collisions~\cite{CMS:2016fnw,Aad:2015gqa}. Whether the ridge results from multi-particle global correlations having an origin similar to those in heavy-ion collisions will be further discussed in this section using measurements of anisotropic flow $v_n$ obtained from azimuthal particle correlations.

Effects that do not arise from global correlations relative to common symmetry planes, in particular short-range correlations between few particles originating from jets or resonance decays, are called non-flow. This contamination is dominant in low-multiplicity collisions, which makes measurements in small systems more sensitive to non-flow than in heavy-ion collisions.
In the case of measurements of two-particle correlations, non-flow can be reduced by imposing an $\eta$ gap between the correlated particles or between two sections of a phase space called subevents. The contamination can be further suppressed by subtracting correlations measured in low-multiplicity events, representing the non-flow sample. Different methods are used based on the assumption of whether these events contain only short-range correlations, or whether flow is present there too, yielding to slightly different results. The peripheral subtraction method~\cite{Ajitanand:2005jj} is based on the first assumption and has been widely used in experimental measurements. The results presented in Fig.~\ref{fig3.5} used the low-multiplicity 60--100\% centrality class event sample, selected using the V0A detector, to subtract non-flow from the 0--20\% centrality class containing the high-multiplicity p--Pb collisions. The robustness of the subtraction method was studied by projecting the correlations onto the azimuthal axis and comparing the per-trigger yields in low-multiplicity p--Pb collisions and minimum bias pp collisions; no significant difference was found~\cite{Abelev:2012ola}. The latter assumption of an azimuthal correlations even in low-multiplicity events is used in the template fit method developed by the ATLAS Collaboration~\cite{Aad:2015gqa}. Since the main difference between the methods is the way they treat the non-flow in the baseline sample, the resulting flow harmonics differ mainly in the low-multiplicity region, where the measurements of $v_n$ from the template fit method are approximately constant with multiplicity, while the $v_n$ obtained from the peripheral subtraction method exhibits a decreasing trend with multiplicity~\cite{ATLAS:2017rtr}.
In contrast to two-particle correlations, multi-particle cumulants~\cite{Borghini:2000sa} suppress few-particle correlations by construction and are thus less prone to non-flow contamination. The remaining non-flow effects in multi-particle cumulants are expected to be negligible in large systems, but can remain significant at small multiplicities typical of p--Pb or pp collisions. Thus, the subevent method consisting of an $\eta$ separation between subevents has also been implemented for multi-particle cumulants~\cite{Jia:2017hbm}. 
Given that more particles are used to construct the correlation, three or more subevents may be used for non-flow suppression. While the two-subevent method largely suppresses short-range correlations e.g.\,within a jet cone, a signal from correlated particles between two cones of a dijet separated in $\eta$ (inter-jet correlations) may still remain. In addition, it is not yet clear whether this method is affected by possible flow decorrelation effects in the longitudinal direction, which may further decrease the resulting flow signal when a large $\eta$ gap is introduced. Therefore, using more subevents further reduces the non-flow contamination by also suppressing inter-jet correlations and by being less sensitive to flow decorrelations. Similarly as in the case of the subtraction methods mentioned above, the subevent method in multi-particle correlations may also yield differences when comparing results from narrow (e.g.\,ALICE) and wide acceptance (e.g.\,ATLAS and CMS) measurements due to the varying interplay between non-flow and possible decorrelation effects. 
Proper suppression of non-flow is a crucial element for a correct data-to-model comparison.

\begin{figure}[ht]
\begin{center}
\includegraphics[width=1.0\linewidth]{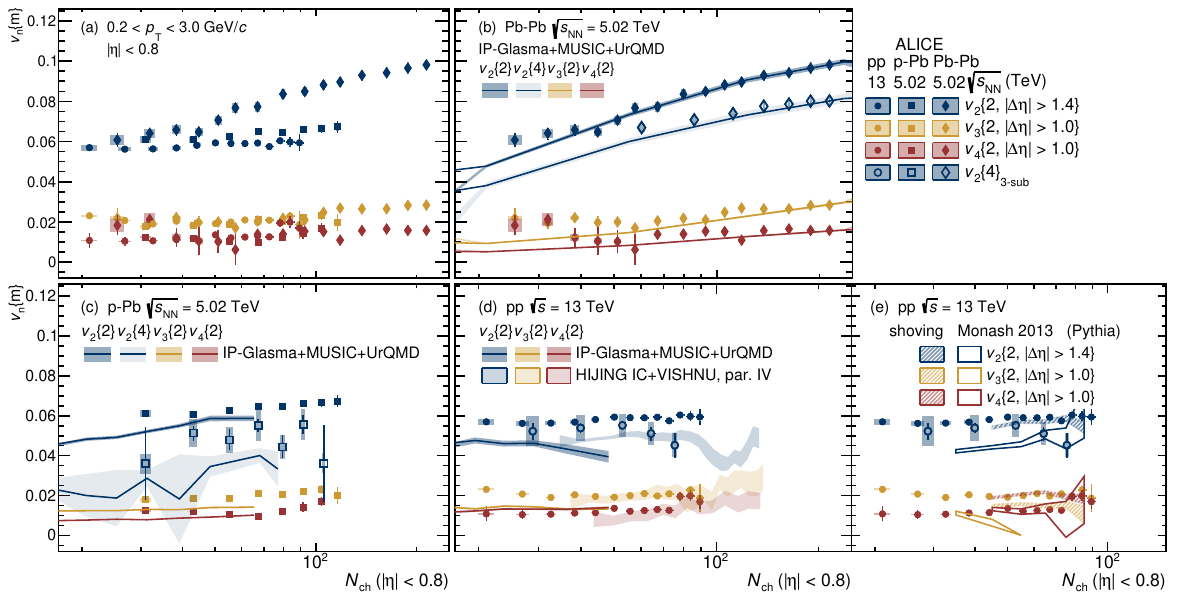}
\caption{Multiplicity dependence of $v_n\{m\}$ in pp, p--Pb and Pb--Pb collisions~\cite{Acharya:2019vdf}.
	Statistical uncertainties are shown as vertical lines and systematic uncertainties as filled boxes.
	The results of two-particle cumulants $v_n\{2\}$ in all collision systems are shown together in panel (a).
	The same results together with the four-particle cumulant $v_2\{4\}$ and comparison to models are reported individually in Pb--Pb (b), p--Pb (c) and pp collisions (d).
	Data are compared with PYTHIA 8.210 (Monash 2013 tune)~\cite{Sjostrand:2014zea} simulations of pp collisions at $\sqrt{s} = 13$ TeV, and the same simulations with string shoving mechanism with the string amplitude of the shoving force $g=10$. Data are further compared to IP-Glasma+MUSIC+UrQMD~\cite{Schenke:2020mbo} calculations of pp collisions at $\sqrt{s} = 13$ TeV and p--Pb and Pb--Pb collisions at $\sqrt{s_{\rm{NN}}} = 5.02$ TeV, and iEBE-VISHNU calculations of pp collisions at $\sqrt{s} = 13$ TeV with HIJING initial conditions and parameter set IV~\cite{Zhao:2020pty}. The width of the bands represent the statistical uncertainty of the model.}
\label{fig3.6} 
\end{center}
\end{figure}

Measurements of $v_n$ in pp and p--Pb collisions as a function of charged-particle multiplicity $N_{\rm{ch}}(|\eta|<0.8)$ are shown in Fig.~\ref{fig3.6} (a) together with measurements from peripheral collisions of Pb nuclei corresponding to centralities higher than 60\%.
A large pseudorapidity gap $|\Delta\eta| > 1.4$ (1.0) in the case of $v_2$ ($v_3$ and $v_4$) is applied to ensure measurements with maximum possible non-flow suppression.
The elliptic flow is found to be compatible between all reported systems at $N_{\rm{ch}} (|\eta|<0.8) \leq 50$,
while at higher multiplicities the rise of $v_2$ for Pb--Pb collisions is in contrast with the weak multiplicity dependence of $v_2$ in pp and p--Pb collisions.
On the contrary, the magnitudes of $v_3$ and $v_4$ are similar and without significant dependence on $N_{\rm ch}$ in both small and large systems over the entire measured multiplicity range.
These observations support the picture of $v_2$ reflecting the system's collective response to the initial geometry and higher harmonics originating from its fluctuations, as discussed for heavy-ion collisions in Sec.~\ref{sec:QGPevolution}.
Within this picture, the weak $N_{\rm ch}$ dependence of $v_n$ in small systems shows the dominance of fluctuations, while in Pb--Pb the results of $v_2$ clearly differentiate between the fluctuation-dominated region at low-$N_{\rm ch}$ and the influence of the large geometrical eccentricity of the nuclear overlap, which is setting in already at $N_{\rm ch} \sim 50$ (roughly corresponding to the centrality interval of 70--80\%). 

Measurements of $m$-particle cumulants for $m>2$ can help to determine whether the measured correlations presented above are of a collective nature shared among many particles. The measurement of the four-particle cumulant $c_n\{4\}$ is of particular interest, since its sign must be negative to result in a real-valued $v_n$, with $v_n\{4\} = \sqrt[4]{-c_n\{4\}}$. This condition is fulfilled in large systems~\cite{Abelev:2014mda,Acharya:2018lmh}, allowing us to calculate the $v_2\{4\}$, shown in Fig.~\ref{fig3.6} (b) together with the same measurements of two-particle cumulants presented in panel (a). Note that the measurement is reported with the 3-subevent method in order to be consistent with the results of small systems shown in the other panels, nevertheless, the method did not yield any significant change with respect to the default measurement of $v_2\{4\}$~\cite{Acharya:2019vdf}.
The relation $v_2\{2\} > v_2\{4\}$ observed in Fig.~\ref{fig3.6} (b) is due to the event-by-event flow fluctuations affecting different orders of cumulants differently, which are discussed in more detail in Chap.~\ref{ch:InitialState}.
In contrast to heavy-ion collisions, measuring a negative $c_2\{4\}$ in small systems is not trivial since the dominating non-flow effects are known to give rise to $c_2\{4\} > 0$~\cite{Abelev:2014mda,Jia:2017hbm}.
A negative $c_2\{4\}$ has been observed in high-multiplicity p--Pb collisions~\cite{Abelev:2014mda}. In pp collisions, this was only possible with the 3-subevent method that further suppresses dominating non-flow correlations, and with a specific high-multiplicity trigger using the V0 detector~\cite{Acharya:2019vdf} which acts as a positive bias towards events with a smaller presence of jets at midrapidity (see Sec.~\ref{section:3.1}). The resulting $v_2\{4\}_{\rm{3-sub}}$ is shown in Fig.~\ref{fig3.6} (c) for p--Pb and (d) and (e) for pp collisions, together with the same measurements of two-particle cumulants presented in panel (a). 
The relation $v_2\{2\} > v_2\{4\}$, apparent in Pb--Pb collisions, is less pronounced in p--Pb collisions, and the measurements are compatible within uncertainties in pp collisions. This may indicate different types of flow fluctuations, larger longitudinal decorrelations, or stronger non-flow effects in the measurements done in small systems. Nevertheless, it should be noted that the region of $N_{\rm ch}$ in which the measurements presented in Fig.~\ref{fig3.6} overlap is small. It should be noted, that a similar comparison was also performed by the ATLAS~\cite{ATLAS:2017rtr} and CMS~\cite{CMS:2016fnw} Collaborations. While ATLAS reports that $v_2\{2\}$ is clearly larger than $v_2\{4\}$ in all collision systems at similar multiplicities, this relation does not hold at low multiplicities in the CMS measurements. 
Compared to the $v_2\{4\}$ measurement (not shown here, see Ref.~\cite{Acharya:2019vdf}), the subevent method allowed us to obtain a real-valued $v_2\{4\}_{\rm{3-sub}}$ down to lower multiplicities in p--Pb collisions, suggesting that the presence of collective effects extends down to regions with multiplicities only few times larger than that of  minimum-bias collisions. In addition, it has been observed that $v_2\{4\} \approx v_2\{6\}$ within uncertainties in both p--Pb and pp collisions~\cite{Acharya:2019vdf}, confirming the collective nature of small systems.

Despite the many similarities with large systems, the origin of the observed collectivity in pp and p--Pb collisions is not clear and further insight can be obtained by comparing the measurements with model calculations. Several approaches are able to explain some features of the experimental data, ranging from purely initial-state gluon momentum correlations~\cite{Mace:2018vwq,Blok:2018xes}, to final state effects where the particle azimuthal anisotropy arises as the system's response to initial spatial anisotropy. The latter is described either with a macroscopic model, such as hydrodynamics~\cite{Weller:2017tsr,Zhao:2020pty,Mantysaari:2017cni}, or a microscopic approach, such as transport models~\cite{Kurkela:2018ygx,Nie:2018xog}, hadronic rescatterings~\cite{Zhou:2015iba,Romatschke:2015dha}, or PYTHIA with the string shoving mechanism~\cite{Bierlich:2017vhg}. 

Hydrodynamic calculations within the IP-Glasma+MUSIC+UrQMD framework~\cite{Schenke:2020mbo} are quantitatively consistent with the $v_n\{m\}$ measurements in Pb--Pb collisions, as shown in Fig.~\ref{fig3.6} (b). The same model shows, in panel (c), a good description of the $v_n\{2\}$ and even a qualitative agreement with $v_2\{4\}$ results in p--Pb collisions, which could only be achieved by assuming sub-nucleon fluctuations, i.e.\ that the proton contains smaller structures modelled as three hot spots~\cite{Mantysaari:2017cni}. 
Instead, the data are not reproduced by the model if the proton is assumed to have a Gaussian profile in the transverse plane. 
While the contribution from fluctuations at such small scales is relatively mild in heavy-ion collisions where the spatial eccentricity is predominantly defined by the colliding nucleons, they become more important in small collision systems.
Despite the successful description of Pb--Pb and p--Pb collisions, the IP-Glasma+MUSIC+UrQMD model~\cite{Schenke:2020mbo} does not reproduce the weak dependence of $v_2$ on multiplicity in pp collisions, and it significantly underestimates the $v_3$ (see Fig.~\ref{fig3.6} (d)). 
In the same panel the iEBE-VISHNU hydrodynamic model calculations with HIJING initial conditions~\cite{Zhao:2017rgg,Zhao:2020pty} are shown. Only simulations with the parameter set denoted as ``IV'', providing the best description of our data, are chosen for comparison.
The model tends to follow the weak trend of $v_n$ with multiplicity. While a quantitative agreement is found with measurements of $v_3$ and $v_4$, this was not achieved in the case of $v_2$ with any of the parameter sets used in the simulations. 
This discrepancy may be due to the fact that the model parameters are tuned to describe the $v_2$ measurements obtained from two-particle correlations with the template fit method used for non-flow subtraction, while our results were performed using two-particle cumulants with a $|\Delta\eta|>1.4$ gap.
Further tuning of the theoretical models could improve the comparison, though it should be noted that non-flow may not be fully removed when only two-particle correlations are used for the measurement. 
In addition, the real-valued $v_2\{4\}$ in pp collisions has not been obtained in any hydrodynamic model so far, possibly because of the large nonlinear hydrodynamic response to initial eccentricities causing a positive $c_{2}\{4\}$~\cite{Schenke:2019pmk,Zhao:2020pty}.

PYTHIA 8 (Monash 2013 tune) model calculations~\cite{Sjostrand:2014zea,Bierlich:2020naj}, in which there is no thermalised medium, are compared to pp data in Fig.~\ref{fig3.6} (e).
The calculations cannot fully reproduce all the aspects of the $v_n$ measurements after applying the subevent method and implementing the selection of high-multiplicity events as in data.
As shown in Fig.~\ref{fig3.6} (e), PYTHIA underestimates the $v_2$ calculated with the two-particle cumulant with an $|\Delta\eta|$ gap. In addition, it should be noted that the PYTHIA results for $v_2\{4\}$ are not shown in the figure due to a positive sign of $c_2\{4\}$, persistent even after the subevent method and the high-multiplicity selection were applies, which had proved to be sufficient in suppressing non-flow effects in data. The higher order harmonics obtained from PYTHIA 8 (Monash 2013 tune) show an opposite trend with respect to the data at high multiplicities, i.e., $v_4 > v_3$. For $N_{\rm ch} (|\eta|<0.8) \geq 75$ the third harmonic could not be calculated in the model due to corresponding negative two-particle cumulants, while the $v_3$ measured in data exhibits a constant multiplicity dependence.
Nevertheless, recent calculations with the additional string shoving mechanism, where overlapping strings repel each other inducing flow-like effects without the presence of a medium, qualitatively reproduce some features of the data. As also shown in Fig.~\ref{fig3.6} (e), the shoving mechanism increases the values of $v_2\{2\}$ leading to a better comparison with the data. It further revealed a positive value of $v_3$ at high-multiplicity in contrast to the non-shoving case. Even though the increase of $v_4$ with the shoving mechanism does not agree with the data, the fact that a model without the presence of a medium is able to reproduce some aspects of the measurements shows the potential for an alternative explanation of the collective effects revealed in the smallest collision systems. An interesting finding was shown in~\cite{Bierlich:2020naj} where a hint for a negative sign of the four-particle cumulant was revealed after applying the high-multiplicity selection, though further investigations are needed to confirm and fully understand such observations.

\begin{figure}[htbp!]
\begin{center}
\centering
\begin{minipage}{0.5\textwidth}
\centering
\includegraphics[width=0.95\linewidth]{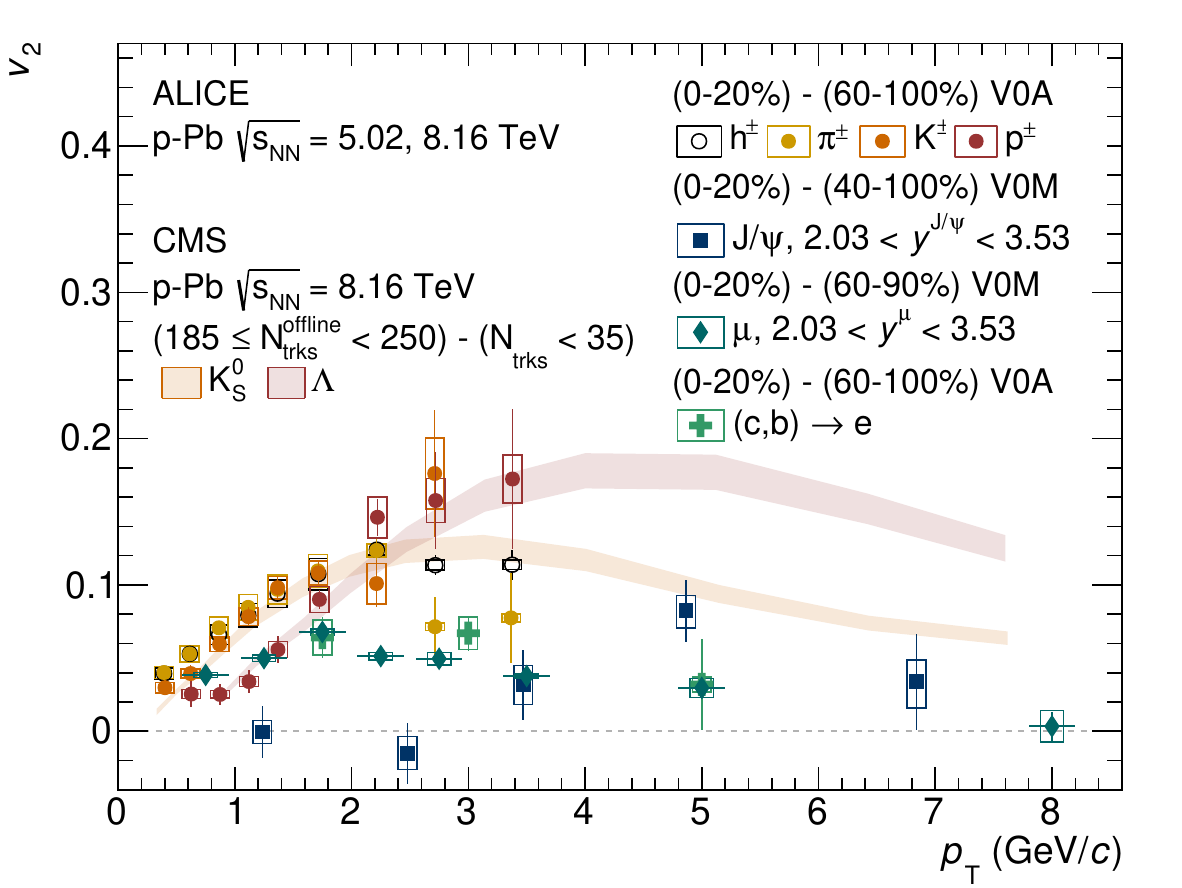}
\end{minipage}\hfill
\begin{minipage}{0.5\textwidth}
\centering
\includegraphics[width=0.95\textwidth]{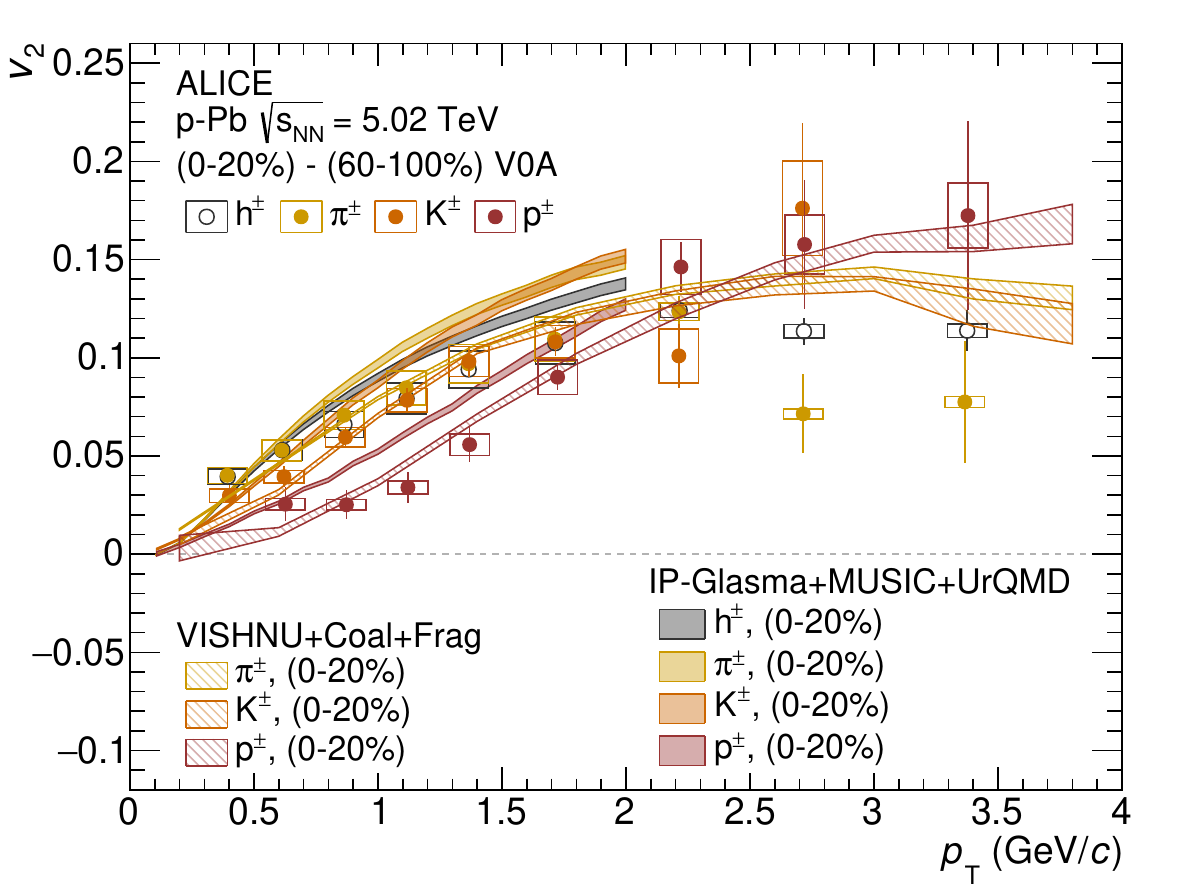}
\end{minipage}\hfill
\caption{(Left) Transverse-momentum dependence of $v_2$ of hadrons, pions, kaons, protons, J$/\psi$, electrons from heavy-flavour hadron decays and inclusive muons in the 0--20\% multiplicity class of p--Pb collisions at $\sqrt{s_{\rm NN}} = 5.02$~TeV and $8.16$~TeV after subtraction of the low-multiplicity class~\cite{ABELEV:2013wsa,Acharya:2018dxy,Acharya:2017tfn,ALICE:2022ruh}. The muons at $p_{\rm T}<1.5$~GeV/$c$ are dominated by decays of light hadrons, while they predominantly originate from heavy-flavour hadron decays at $p_{\rm T}>2$~GeV/$c$. The results of J$/\psi$ are combined results from p--Pb collisions at $\sqrt{s_{\rm NN}} = 5.02$ and 8.16~TeV. The data is plotted at the average-$p_{\rm{T}}$ for each considered $p_{\rm{T}}$ interval and particle species under study. Error bars show statistical uncertainties while boxes denote systematic uncertainties. The bands illustrate the CMS measurements for $\rm K^0_S$ and $\Lambda$~\cite{CMS:2018loe}. (Right) Data of light hadrons are compared to the IP-Glasma+MUSIC+UrQMD~\cite{Schenke:2020mbo} calculations and iEBE-VISHNU calculations with the coalescence and fragmentation model~\cite{Zhao:2020wcd}. The width of the bands represents the statistical uncertainty of the models. }
\label{fig3.7} 
\end{center}
\end{figure}

Studies of azimuthal anisotropies of identified hadron species in small systems, in particular their mass dependence, have a large potential to determine the presence of a partonic medium.
In heavy-ion collisions, the mass ordering of the anisotropic flow of identified hadrons at low-$p_{\rm T}$ is explained as a consequence of a radial expansion of the medium pushing heavier particles to larger $p_{\rm T}$, and is described well by viscous hydrodynamic models (for further discussion see Sec.~\ref{sec:TG2pidflow}). The baryon-meson grouping of $v_n$ at intermediate-$p_{\rm T}$ is understood as an effect of particle production via quark coalescence, hence pointing to the presence of partonic collectivity in heavy-ion collisions. These effects are thus considered as characteristic flow features, discussed in more detail in Sec.~\ref{sec:MicroscopicHadronization}. The analysis of two-particle correlations of identified hadrons in high-multiplicity p--Pb collisions, after the subtraction of correlations from the low-multiplicity data sample, showed the near-side ridge structure~\cite{ABELEV:2013wsa} as observed previously in charged hadron measurements (see Fig.~\ref{fig3.5}). The Fourier coefficiencts of this ridge structure, $v_2$, of pions, kaons and protons as a function of $p_{\rm T}$ shown in Fig.~\ref{fig3.7} (left) reveal a mass dependence similar to the one caused by a collectively expanding medium in heavy-ion collisions. In particular, at low-$p_{\rm T}$ they indicate a mass ordering effect where the $v_2$ of protons is shifted to larger $p_{\rm T}$ with respect to pions and kaons. For comparison, we also show in Fig.~\ref{fig3.7} (left) measurements of $\rm K^0_S$ and $\Lambda$ in p--Pb collisions by the CMS Collaboration~\cite{CMS:2018loe}. Despite the differences in the selection of multiplicity classes (based on the uncorrected number of tracks at midrapidity), wider pseudorapidity acceptance and larger pseudorapidity difference between correlated particles ($|\Delta\eta| > 1.0$), the measurements tend to follow the trend of the mass dependence of $v_2$ seen by ALICE for $\pi$, K, p hadrons.

The IP-Glasma+MUSIC+UrQMD model calculations~\cite{Schenke:2020mbo}, which provide a quantitative description of such measurements in Pb--Pb collisions (see Sec.~\ref{sec:TG2hydrodescription}), overestimate the data in p--Pb collisions at low- $p_{\rm T}$ (see Fig.~\ref{fig3.7} (right)). While this could point to the need for finding a better estimate of the bulk viscosity in p--Pb collisions, owing to its strong impact on the identified particle flow in the model, it should be noted that in the experimental results the choice of the non-flow reference sample in the subtraction technique may influence the model-to-data comparison on a quantitative level.
Besides hydrodynamics, other models based on hadronic rescatterings~\cite{Zhou:2015iba} and on parton transport~\cite{Li:2016ubw} are able to generate the mass ordering effect. In addition, calculations based purely on initial-state momentum correlations in the IP-Glasma model coupled with Lund string fragmentation~\cite{Schenke:2016lrs} revealed this feature too. Thus, the observation of mass ordering alone cannot be taken as evidence of flow-like collectivity in small systems. The measurements of $v_2$ of identified hadrons are further compared to the hydrodynamic model iEBE-VISHNU with the coalescence mechanism~\cite{Zhao:2020wcd}, allowing for comparison to the data to be extended to the intermediate-$p_{\rm T}$ region (shown in Fig.~\ref{fig3.7} (right)). Even though the current statistical uncertainty of ALICE $v_2$ measurements is still considerable, the model provides an excellent description of the $v_2$ of light hadrons over the entire $p_{\rm T}$ range; such an agreement can not be achieved without invoking coalescence. This indicates the importance of partonic degrees of freedom in the system evolution in high-multiplicity p--Pb collisions at the LHC, which in turn supports the hydrodynamic-like origin of the mass ordering at low-$p_{\rm T}$ discussed above.

In Fig.~\ref{fig3.7} (left), $v_2$ measurements of the light hadrons presented above are compared to results of J$/\psi$, electrons from decays of open-heavy-flavour hadrons, and muons~\cite{ABELEV:2013wsa,Acharya:2018dxy,Acharya:2017tfn,ALICE:2022ruh} (the latter are predominantly originating from open-heavy-flavour decays for $p_{\rm T}>2$ GeV/$c$ while being dominated by decays of light hadrons for $p_{\rm T}<1.5$ GeV/$c$). 
The early formation time of heavy quarks suggests that their $v_2$ is mainly developed during the early stages of the collision, namely in hard parton scatterings off colour domains in the collision target. In this picture, the colour fields are locally organised into colour domains. Each parton scatters off these domains independently and receives a momentum kick in the process. If more partons that are in the same colour state scatter off the same domain, they will receive a similar kick, which may create anisotropies in the momentum space. As the orientation of the colour fields fluctuates event-by-event independently of the overall geometry, the initial momentum anisotropies are not directly correlated with the global spatial anisotropy~\cite{Lappi:2015vta}. Several model studies tend to support the idea of heavy-flavour hadron flow being driven by the initial gluon momentum anisotropies. These models are rooted in a purely initial-state calculation based on the colour Glass Condensate (CGC) formalism that was able to describe the measurements of flow of heavy-flavour hadrons~\cite{Zhang:2019dth,Zhang:2020ayy}, and a study showing that final-state interactions alone cannot generate sufficient $v_2$ of J$/\psi$~\cite{Du:2018wsj}. On the other hand, it was found that the momentum anisotropies from the initial state are uncorrelated with the global spatial anisotropy~\cite{Schenke:2015aqa}. Thus, the experimental measurements of the flow of heavy-flavour hadrons performed by correlating them with light hadrons, the flow of which tends to align with the spatial anisotropy (based on the current data and comparison to phenomenological models discussed above), would result in zero flow. This is in contrast to the measurements presented in Fig.~\ref{fig3.7} (left) exhibiting a finite $v_2$ of heavy-flavour hadrons, which could therefore be interpreted as an indication of final-state induced flow of these particles. It should be noted that it is not yet clear what is the contribution of initial-state correlations to flow of light hadrons, thus more investigations in this area are desirable in the future.
While the flow of open heavy-flavour particles is composed of both the heavy and light quarks with poorly known relative contributions in the case of hadronisation via coalescence, the flow of J$/\psi$ may directly reflect the collective behaviour of the charm quark. This is supported by the apparent ordering of $v_2$ in Fig.~\ref{fig3.7} (left) which tends to be followed by the heavy flavour measurements. It should be noted that different samples of low-multiplicity events were used for non-flow subtraction in the case of the $v_2$ of J$/\psi$, which could potentially result in different remaining non-flow contamination in the final measurements and/or different contribution of $v_2$ from the subtracting peripheral bin. In addition, the rapidity intervals differ between the measurements, and the measured $p_{\rm T}$ of heavy-flavour decay leptons does not directly correspond to the original $p_{\rm T}$ of their parents. The data comparison is therefore not straightforward. 
Future data taking in small collision systems will provide larger samples of the rare particle species, potentially allowing for their direct reconstruction and thus measurements of correlations of the original parent particles instead of using the decay products carrying partial kinematics information. In addition, it could be possible to perform the measurements using the state-of-the art methods to suppress non-flow, exploited so far only in the light hadron measurements, which would allow for quantitative comparisons.

Similar identified particle $v_n$ measurements in pp collisions are highly desirable to assess whether these features remain for even smaller collision systems. So far, only $v_2$ of $\Lambda$ and K$^0_S$ were measured in pp collisions by the CMS Collaboration~\cite{CMS:2016fnw}, with a hint of mass dependence at low-$p_{\rm T}$. Measurements of more particle species extended to the intermediate-$p_{\rm T}$ region are therefore necessary to arrive at a conclusive data-model comparison. In addition, a more systematic study of the non-flow subtraction in two-particle correlations, which still results in very large uncertainties on the measurements, should be performed. One of the advanced approaches would be measuring anisotropic flow of identified hadrons using multi-particle cumulants with the subevent method, which has been developed for the ALICE projections of high-energy pp collisions in Run 3 at the LHC~\cite{ALICE-PUBLIC-2020-005} but not yet performed in data due to the limited amount of collected data.

It was shown already in Pb--Pb collisions that measurements of $v_n$ alone are not enough to provide detailed constraints on theoretical models. The symmetric cumulants SC($m,n$), i.e.\,correlations between the second moments of different flow harmonics, possess the ability to constrain the initial conditions~\cite{ALICE:2016kpq}. The most sensitive correlation is between $v_2^2$ and $v_3^2$, i.e.\,SC(3,2), while the correlation between the second moments of the second and fourth harmonics, SC(4,2), is sensitive to the transport properties of the medium in heavy-ion collisions (see more details in Sec.~\ref{sec:TG2symmetriccumulants}). 
In a similar spirit, the determination of the origin of collectivity in small systems can be addressed with other measurements, sensitive to different aspects of the system evolution and their theoretical description, in addition to $v_n$ coefficients. 
Results for symmetric cumulants have shown a similar trend with collision multiplicity in both large and small systems~\cite{Acharya:2019vdf}, indicating that the observed collective correlations share a similar origin, which may vary with multiplicity. 

The measurements of azimuthal particle correlations presented in this section revealed the presence of collective effects in small collision systems. 
Available model comparisons to the results at high multiplicities suggest that the particle correlations originate as a system's response to the initial spatial anisotropy during final state interactions. In particular, a strong indication for the presence of a partonic system was found in high-multiplicity p--Pb collisions. On the other hand, at multiplicities of the order of the average multiplicity, the initial-state (gluon momentum) correlations may become significant~\cite{Schenke:2016lrs,Greif:2017bnr,Nie:2019swk,Kurkela:2019kip}. It is very challenging to confirm this with the present experimental results, but the upcoming Run 3 or new observables could promise to provide additional information to address the relative importance of initial correlations to the measured flow.

%

\newcommand{\pp}           {pp\xspace}
\newcommand{\ppbar}        {\mbox{$\mathrm {p\overline{p}}$}\xspace}
\newcommand{\XeXe}         {\mbox{Xe--Xe}\xspace}
\newcommand{\PbPb}         {\mbox{Pb--Pb}\xspace}
\newcommand{\pA}           {\mbox{pA}\xspace}
\newcommand{\pPb}          {\mbox{p--Pb}\xspace}
\newcommand{\AuAu}         {\mbox{Au--Au}\xspace}
\newcommand{\dAu}          {\mbox{d--Au}\xspace}

\newcommand{\s}            {\ensuremath{\sqrt{s}}\xspace}
\newcommand{\snn}          {\ensuremath{\sqrt{s_{\mathrm{NN}}}}\xspace}
\newcommand{\pt}           {\ensuremath{p_{\rm T}}\xspace}
\newcommand{\meanpt}       {$\langle p_{\mathrm{T}}\rangle$\xspace}
\newcommand{\ycms}         {\ensuremath{y_{\rm CMS}}\xspace}
\newcommand{\ylab}         {\ensuremath{y_{\rm lab}}\xspace}
\newcommand{\etarange}[1]  {\mbox{$\left | \eta \right |~<~#1$}}
\newcommand{\yrange}[1]    {\mbox{$\left | y \right |~<~#1$}}
\newcommand{\dndy}         {\ensuremath{\mathrm{d}N/\mathrm{d}y}\xspace}
\newcommand{\dnchdy}       {\ensuremath{\mathrm{d}N_\mathrm{ch}/\mathrm{d}y}\xspace}
\newcommand{\dndeta}       {\ensuremath{\mathrm{d}N/\mathrm{d}\eta}\xspace}
\newcommand{\dnchdeta}     {\ensuremath{\mathrm{d}N_\mathrm{ch}/\mathrm{d}\eta}\xspace}
\newcommand{\avdndeta}     {\ensuremath{\langle\dndeta\rangle}\xspace}
\newcommand{\dNdy}         {\ensuremath{\mathrm{d}N/\mathrm{d}y}\xspace}
\newcommand{\Npart}        {\ensuremath{N_\mathrm{part}}\xspace}
\newcommand{\meanNpart}    {$\langle N_\mathrm{part}\rangle$\xspace}
\newcommand{\Ncoll}        {\ensuremath{N_\mathrm{coll}}\xspace}
\newcommand{\meanNcoll}    {$\langle N_\mathrm{coll}\rangle$\xspace}
\newcommand{\dEdx}         {\ensuremath{\textrm{d}E/\textrm{d}x}\xspace}
\newcommand{\RpPb}         {\ensuremath{R_{\rm pPb}}\xspace}
\newcommand{\RAA}          {\ensuremath{R_{\rm AA}}\xspace}
\newcommand{\Tc}           {\ensuremath{T_{\rm c}}\xspace}
\newcommand{\Tch}          {\ensuremath{T_{\rm ch}}\xspace}
\newcommand{\Tfo}          {\ensuremath{T_{\rm fo}}\xspace}
\newcommand{\Tkin}         {\ensuremath{T_{\rm kin}}\xspace}
\newcommand{\muB}          {\ensuremath{\mu_{\rm B}}\xspace}

\newcommand{\nineH}        {$\sqrt{s}~=~0.9$~Te\kern-.1emV\xspace}
\newcommand{\seven}        {$\sqrt{s}~=~7$~Te\kern-.1emV\xspace}
\newcommand{\twoH}         {$\sqrt{s}~=~0.2$~Te\kern-.1emV\xspace}
\newcommand{\twosevensix}  {$\sqrt{s}~=~2.76$~Te\kern-.1emV\xspace}
\newcommand{\five}         {$\sqrt{s}~=~5.02$~Te\kern-.1emV\xspace}
\newcommand{\twosevensixnn}{$\sqrt{s_{\mathrm{NN}}}~=~2.76$~Te\kern-.1emV\xspace}
\newcommand{\fivenn}       {$\sqrt{s_{\mathrm{NN}}}~=~5.02$~Te\kern-.1emV\xspace}
\newcommand{\LT}           {L{\'e}vy-Tsallis\xspace}
\newcommand{\GeVc}         {Ge\kern-.1emV/$c$\xspace}
\newcommand{\MeVc}         {Me\kern-.1emV/$c$\xspace}
\newcommand{\TeV}          {Te\kern-.1emV\xspace}
\newcommand{\GeV}          {Ge\kern-.1emV\xspace}
\newcommand{\MeV}          {Me\kern-.1emV\xspace}
\newcommand{\GeVmass}      {Ge\kern-.2emV/$c^2$\xspace}
\newcommand{\MeVmass}      {Me\kern-.2emV/$c^2$\xspace}
\newcommand{\lumi}         {\ensuremath{\mathcal{L}}\xspace}

\newcommand{\ITS}          {\rm{ITS}\xspace}
\newcommand{\TOF}          {\rm{TOF}\xspace}
\newcommand{\ZDC}          {\rm{ZDC}\xspace}
\newcommand{\ZDCs}         {\rm{ZDCs}\xspace}
\newcommand{\ZNA}          {\rm{ZNA}\xspace}
\newcommand{\ZNC}          {\rm{ZNC}\xspace}
\newcommand{\SPD}          {\rm{SPD}\xspace}
\newcommand{\SDD}          {\rm{SDD}\xspace}
\newcommand{\SSD}          {\rm{SSD}\xspace}
\newcommand{\TPC}          {\rm{TPC}\xspace}
\newcommand{\TRD}          {\rm{TRD}\xspace}
\newcommand{\VZERO}        {\rm{V0}\xspace}
\newcommand{\VZEROA}       {\rm{V0A}\xspace}
\newcommand{\VZEROC}       {\rm{V0C}\xspace}
\newcommand{\Vdecay} 	   {\ensuremath{V^{0}}\xspace}

\newcommand{\ee}           {\ensuremath{e^{+}e^{-}}} 
\newcommand{\pip}          {\ensuremath{\pi^{+}}\xspace}
\newcommand{\pim}          {\ensuremath{\pi^{-}}\xspace}
\newcommand{\kap}          {\ensuremath{\rm{K}^{+}}\xspace}
\newcommand{\kam}          {\ensuremath{\rm{K}^{-}}\xspace}
\newcommand{\pbar}         {\ensuremath{\rm\overline{p}}\xspace}
\newcommand{\kzero}        {\ensuremath{{\rm K}^{0}_{\rm{S}}}\xspace}
\newcommand{\lmb}          {\ensuremath{\Lambda}\xspace}
\newcommand{\almb}         {\ensuremath{\overline{\Lambda}}\xspace}
\newcommand{\Om}           {\ensuremath{\Omega^-}\xspace}
\newcommand{\Mo}           {\ensuremath{\overline{\Omega}^+}\xspace}
\newcommand{\X}            {\ensuremath{\Xi^-}\xspace}
\newcommand{\Ix}           {\ensuremath{\overline{\Xi}^+}\xspace}
\newcommand{\Xis}          {\ensuremath{\Xi^{\pm}}\xspace}
\newcommand{\Oms}          {\ensuremath{\Omega^{\pm}}\xspace}
\newcommand{\degree}       {\ensuremath{^{\rm o}}\xspace} 

\subsection{Charmonium and bottomonium suppression in p--Pb collisions}
\label{sec:quarkoniumpA}

Studies of quarkonium production in collision systems involving light ions or protons were carried out since the very beginning of QGP investigations. In particular, the study of ion--ion collisions with systems lighter than Pb--Pb or Au--Au were meant to obtain information about the onset of suppression effects or to better understand scaling properties of the suppression as a function of variables related to the geometry of the system (In--In collisions at SPS energy by NA60~\cite{Arnaldi:2007zz}, Cu--Cu~\cite{Adare:2008sh} and Cu--Au~\cite{Aidala:2014bqx} by PHENIX at RHIC). Studies of p--A (or d--Au at RHIC) collisions were also fundamental for the investigation of the role and the size of cold nuclear matter effects that were found to be responsible for a significant fraction of the observed quarkonium suppression both at SPS and RHIC energies~\cite{Alessandro:2003pi,Arnaldi:2010ky,Alessandro:2006jt,Adare:2013ezl,Adare:2016psx,Acharya:2019zjt,Adamczyk:2016dhc,Adamczyk:2013poh}. Most of these results concerned the \jpsi state, but also the production of $\psi(\rm 2S)$ and $\Upsilon(\rm 1S)$ was studied although with larger uncertainties.

At the LHC, ALICE has obtained a result on the inclusive \jpsi \RAA in \XeXe collisions at $\snn=5.44$ TeV~\cite{Acharya:2018jvc}. The integrated luminosity available was rather low ($L_{\rm int}\sim 0.34$ $\mu{\rm b}^{-1}$), but the observed compatibility between \RAA values for \XeXe and \PbPb at a given $N_{\rm part}$ has shown that the relative contribution of suppression and (re)generation processes is similar for collisions involving a similar number of participant nucleons in different nuclear collision systems.

On the other hand, much more accurate studies of quarkonium production in \pPb collisions were carried out, at $\snn=5.02$ and 8.16 TeV involving \jpsi, $\psi(\rm 2S)$ and the $\Upsilon(\rm 1S,2S,3S)$ states~\cite{Acharya:2018kxc,Acharya:2019lqc,Acharya:2020wwy,Abelev:2013yxa,Abelev:2014zpa,Abelev:2014oea}. These investigations were meant to evaluate the influence of initial state effects, and in particular of nuclear shadowing, on the observed quarkonium yields. Remarkably, final-state cold nuclear matter effects, related to the break-up of the resonances via interactions with the nucleons of the colliding ion(s), are negligible at LHC energies due to the extremely small time spent by the heavy-quark pair in the nuclear matter. In fact, $\tau=\langle  L\rangle/\beta_{\rm z}\gamma$ values ranging from $\sim$ $7\times 10^{-2}$ to 10$^{-4}$ fm/$c$ for low-\pt ${\rm c\overline c}$ pairs are respectively obtained at backward (Pb-going) and forward (p-going) rapidity, with $\langle L\rangle$ being the average length of nuclear matter crossed by the pair, $\beta_{\rm z}$ the velocity of the $c{\overline c}$ along the beam direction in the nucleus rest frame, and $\gamma = E_{\rm c\overline c}/m_{\rm c\overline c}$~\cite{Adam:2016ohd}. In fact, nuclear shadowing was found to be the main effect influencing the production of ground states (\jpsi, $\Upsilon(\rm 1S)$), as shown in Fig.~\ref{fig:fig5} in Sec.~\ref{sec:2.5nonQGP}, where it was demonstrated that shadowing calculations based on the EPPS16 set of nPDFs fairly reproduce the observed \RpPb values.

\begin{figure}[ht!]
\begin{center}
\includegraphics[width=0.7\linewidth]{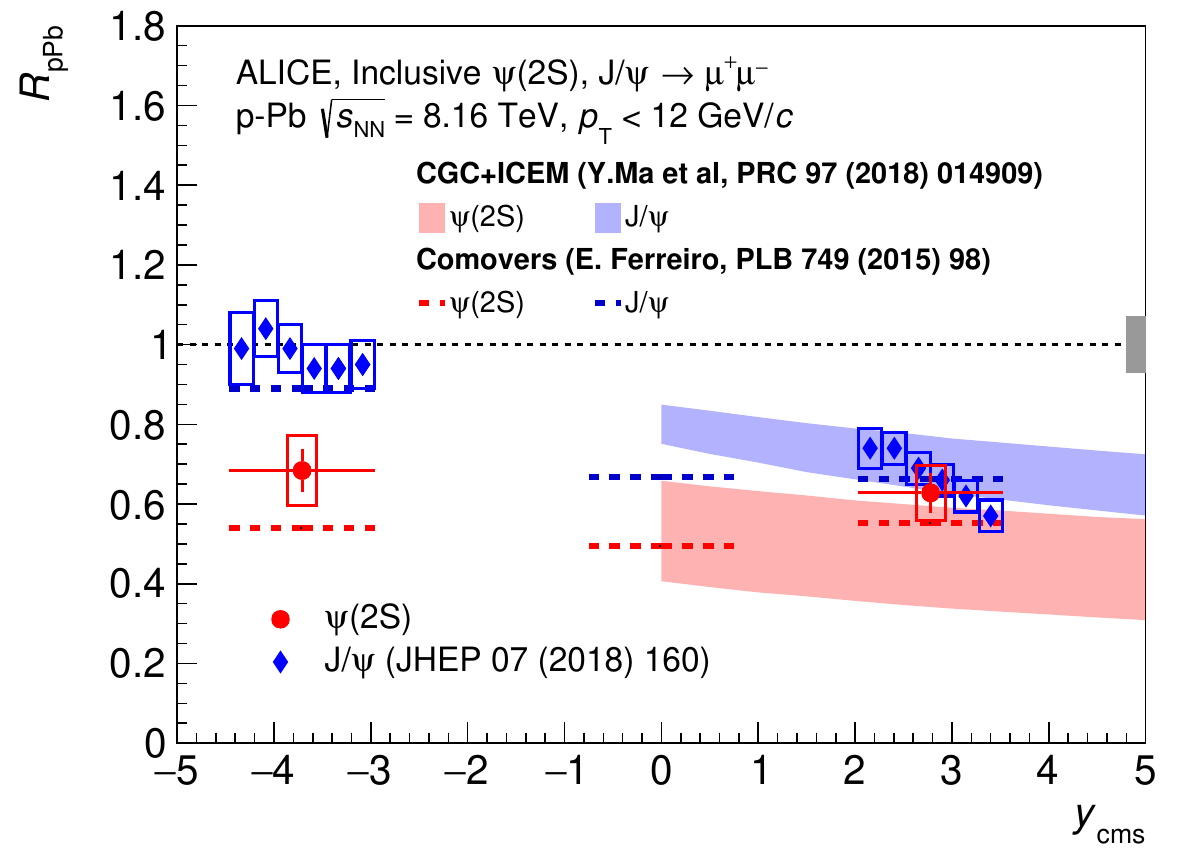}
\caption{The nuclear modification factor for \jpsi and $\psi(\rm 2S)$ production in \pPb collisions at $\snn=8.16$ TeV~\cite{Acharya:2020wwy}. Data are compared with two model calculations that include final state effects~\cite{Ferreiro:2014bia,Ma:2017rsu}.}
\label{fig:psi2svsyppb} 
\end{center}
\end{figure}

However, the situation significantly changes when considering charmonium and bottomonium states with lower binding energy. In the charmonium sector, ALICE studied the rapidity, transverse momentum and centrality dependence of the $\psi(\rm 2S)$ \RpPb~\cite{Adam:2016ohd,Abelev:2014zpa,Acharya:2020wwy,Acharya:2020rvc}. If nuclear shadowing continues to be the main effect at play, similar values of the nuclear modification factors should be expected for \jpsi and $\psi(\rm 2S)$, since the Bjorken-$x$ values of the partons involved in the hard scattering are different by at most a few percent, due to the small mass difference between the two states. This comparison is shown in Fig.~\ref{fig:psi2svsyppb} as a function of rapidity for $\snn=8.16$ TeV. At forward $y$ (p-going), the \jpsi and $\psi(\rm 2S)$ \RpPb are fully compatible, while at backward $y$ (Pb-going) a stronger suppression is observed for $\psi(\rm 2S)$. Considering the correlation of the  uncertainties, the significance of this observation is 3.1\,$\sigma$. This effect is found to have no appreciable dependence on collision energy 
and has no strong dependence as a function of $p_{\rm T}$ (not shown). 

These observations suggest that the suppression seen at backward rapidity for the $\psi(\rm 2S)$ must be due to final state effects and more precisely to the dissociation of the weakly bound ${\rm c\overline c}$ pair in the strongly-interacting system created in the collision. The ALICE measurements are compared to calculations from two theory models which include, in addition to initial state effects, also the final state dissociation of the quarkonia. In the comover model~\cite{Ferreiro:2014bia}, the break-up originates from ``comovers'', i.e.\,particles that travel along with the ${\rm c\overline c}$ pair whose density is constrained by the measured \dndy, which interact with the quarkonia with cross sections tuned on lower energy experimental data. The CGC+ICEM model~\cite{Ma:2017rsu} is based on an improvement of the Colour Evaporation Model~\cite{Fritzsch:1977ay} for the production process and on the Colour Glass Condensate effective theory for the treatment of saturation effects in the small-$x$ regime~\cite{Gelis:2010nm}. In this model, the suppression is here due to parton comovers hadronising on longer time scales than the ${\rm c\overline c}$ pair that have soft colour exchanges with the final-state charmonia. The models are in fair agreement with data (only the comover model can give predictions in the Pb-going rapidity region), advocating that at LHC energies  a dense interacting system is produced, able to selectively dissociate some of the quarkonium states.

\begin{figure}[ht!]
\begin{center}
\includegraphics[width=0.7\linewidth]{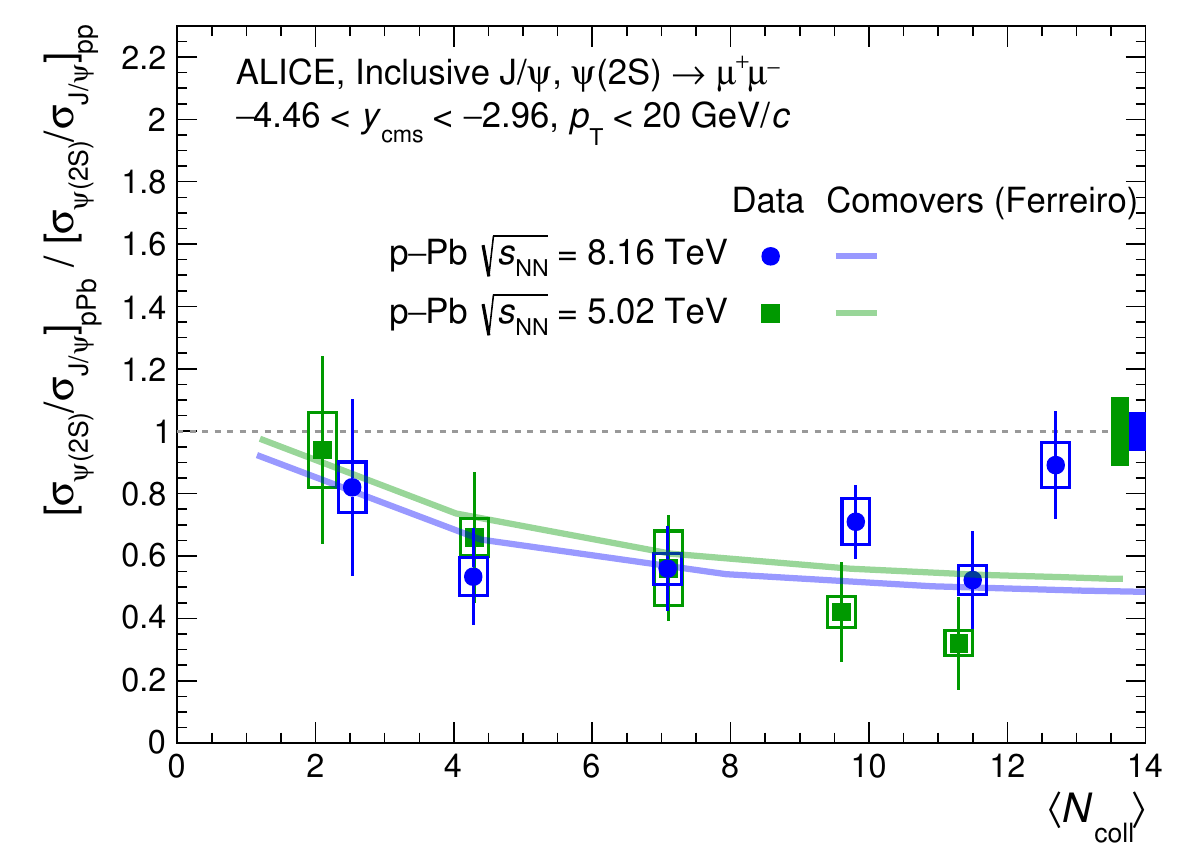}
\caption{The double ratio between the $\psi(\rm 2S)$ and \jpsi cross sections in \pPb and pp collisions at $\snn=5.02$ and 8.16 TeV~\cite{Adam:2016ohd,Acharya:2020rvc}, shown as a function of $\langle N_{\rm coll}\rangle$ for the backward (Pb-going) rapidity region and compared to the comover model results~\cite{Ferreiro:2014bia}. The boxes around unity
correspond to the global systematic uncertainty on the ratio between the $\psi(\rm 2S)$ and \jpsi cross sections in pp collisions.}
\label{fig:psi2svscentppb} 
\end{center}
\end{figure}

A suppression effect was already observed by PHENIX when studying the centrality dependence of the ratio of the $\psi(\rm 2S)$ and \jpsi nuclear modification factors in \dAu collisions at $\snn=0.2$ TeV, where it was shown that for central collisions ($\langle N_{\rm coll}\rangle\sim 15$) the $\psi(\rm 2S)$ suppression was 3 times stronger than the one of \jpsi~\cite{Adare:2013ezl}. The corresponding ALICE result on the centrality dependence of the double-ratio between the $\psi(\rm 2S)$ and \jpsi cross sections in \pPb and pp collisions, shown in Fig.~\ref{fig:psi2svscentppb} for data taken at $\snn=5.02$ and 8.16 TeV, also indicates an increase of the relative suppression of $\psi(\rm 2S)$, which already sets in at smaller values of $\langle N_{\rm coll}\rangle$~\cite{Adam:2016ohd,Acharya:2020rvc}. The comover model fairly reproduces this suppression~\cite{Ferreiro:2014bia}.   

Similar studies were also carried out in the bottomonium sector by comparing the suppression effects between the ground state $\Upsilon(\rm 1S)$ and the 2S and 3S states. In particular, a stronger relative suppression of $\Upsilon(\rm 2S)$ with respect to $\Upsilon(\rm 1S)$ might indicate the presence of final-state effects on the former resonance which has a binding energy of about 500 MeV, similar to that of the \jpsi and significantly larger than that of the $\psi(\rm 2S)$. ALICE measured the double-ratio between the $\Upsilon(\rm 2S)$ and $\Upsilon(\rm 1S)$ in \pPb with respect to pp at $\snn=8.16$ TeV, obtaining $R_{\rm 2S/1S}=0.85\pm0.16\,{\rm (stat.)}\pm0.06\,{\rm (syst.)}$ in the Pb-going rapidity region where final-state effects, based on $\psi(\rm 2S)$ results, should be stronger~\cite{Acharya:2019lqc}. This value is described by comover model calculations, which give $R_{\rm 2S/1S}=0.77\pm0.04$~\cite{Ferreiro:2014bia}. A stronger experimental indication (2.8\,$\sigma$) for a value of $R_{\rm 2S/1S}$ smaller than unity was obtained by ATLAS at $\snn=5.02$ TeV in the rapidity region $-2<y_{\rm cms}<1.5$ ($R_{\rm 2S/1S}=0.765\pm0.069\,{\rm (stat.)}\pm0.048\,{\rm (syst.)}$)~\cite{Aaboud:2017cif}. 
Finally, for the even more weakly bound $\Upsilon(\rm 3S)$, evidence for a stronger suppression with respect to $\Upsilon(\rm 1S)$ was obtained by LHCb ($R_{\rm 3S/1S}=0.44\pm0.15$ at $\sqrt{s_{\rm NN}}=8.16$ TeV for $-4<y_{\rm cms}<-2.5$~\cite{Aaij:2018scz}), with the corresponding ALICE results being $R_{\rm 3S/1S}=0.87\pm 0.27\,{\rm (stat.)}\pm 0.07\,{\rm (syst.)}$~\cite{Acharya:2019lqc}.

\subsection{Searches for jet quenching in small systems}
\label{Sect:JetQuenchSmallSystems}

Measurements of small collision systems explore the question of the limits of QGP formation. Such measurements, at both RHIC and the LHC, have revealed significant effects in the low-\pT\ sector that are associated with collective dynamics of the QGP in large collision systems (Sec.~\ref{section:3.2} and~\ref{section:3.3}). However, interpretations other than QGP formation have been proposed for these small-systems phenomena, based on quantum interference or more conventional hadronic physics (see e.g.\,~\cite{Blok:2017pui,Bierlich:2017vhg}).

Jet quenching is a necessary consequence of the formation of an extended QGP system. Current theoretical approaches for the magnitude of jet quenching in small systems are based on calculations of jet quenching in large systems. Their predictions vary greatly, however, from significant and readily observable signals~\cite{Zakharov} to negligible effects~\cite{Chen:2015qmd}. In addition, there may be significant initial-state effects that can mask signatures of jet quenching~\cite{Kang:2015mta,Huss:2020dwe}. Current experimental searches for jet quenching in small systems have taken two different approaches: yield measurements of inclusive hadrons or jets, analogous to the measurements of \RAA\ in large systems~\cite{Adams:2003im, Adler:2006wg, Adare:2015gla, Aad:2016zif, ALICE:2012mj, Khachatryan:2015xaa, Khachatryan:2016odn, Adam:2015hoa, ATLAS:2014cpa, Khachatryan:2016xdg, Adam:2016jfp} and coincidence measurements~\cite{Chatrchyan:2014hqa, Acharya:2017okq}. To date, no significant jet quenching effects have been observed, within experimental uncertainties. 

In this section we discuss ALICE measurements using both approaches to searching for jet quenching in small systems. The focus of this program in ALICE has been on p--Pb collisions, with measurements carried out both inclusively and selected by Event Activity (EA), as measured by forward charged-particle multiplicity in the V0 detectors or by beam-rapidity neutral energy in the ZDCs.

For inclusive observables, jet quenching is measured using \RpPb, the ratio of the inclusive yield measured in \pPb\ collisions to the inclusive production cross section of the same observable in \pp\ collisions scaled by the geometric factor \TpPb, where $\left<\ldots\right>$ indicates the average over the EA interval used for event selection. Suppression in inclusive hadron production due to jet quenching would correspond to $\RpPb<1$. 

The value of \TpPb\ is determined using Glauber modeling (Sec.~\ref{sec:TG1centrality}), which is based on the assumption that EA is closely correlated with event geometry or ``centrality''. However, as discussed in Sec.~\ref{section:3.1}, the correlation between EA and event geometry in small systems is subject to large fluctuations, and the EA distribution is biased by the presence of a hard process in the event. 

\begin{figure}[htb]
\centering
\includegraphics[width=0.55\textwidth]{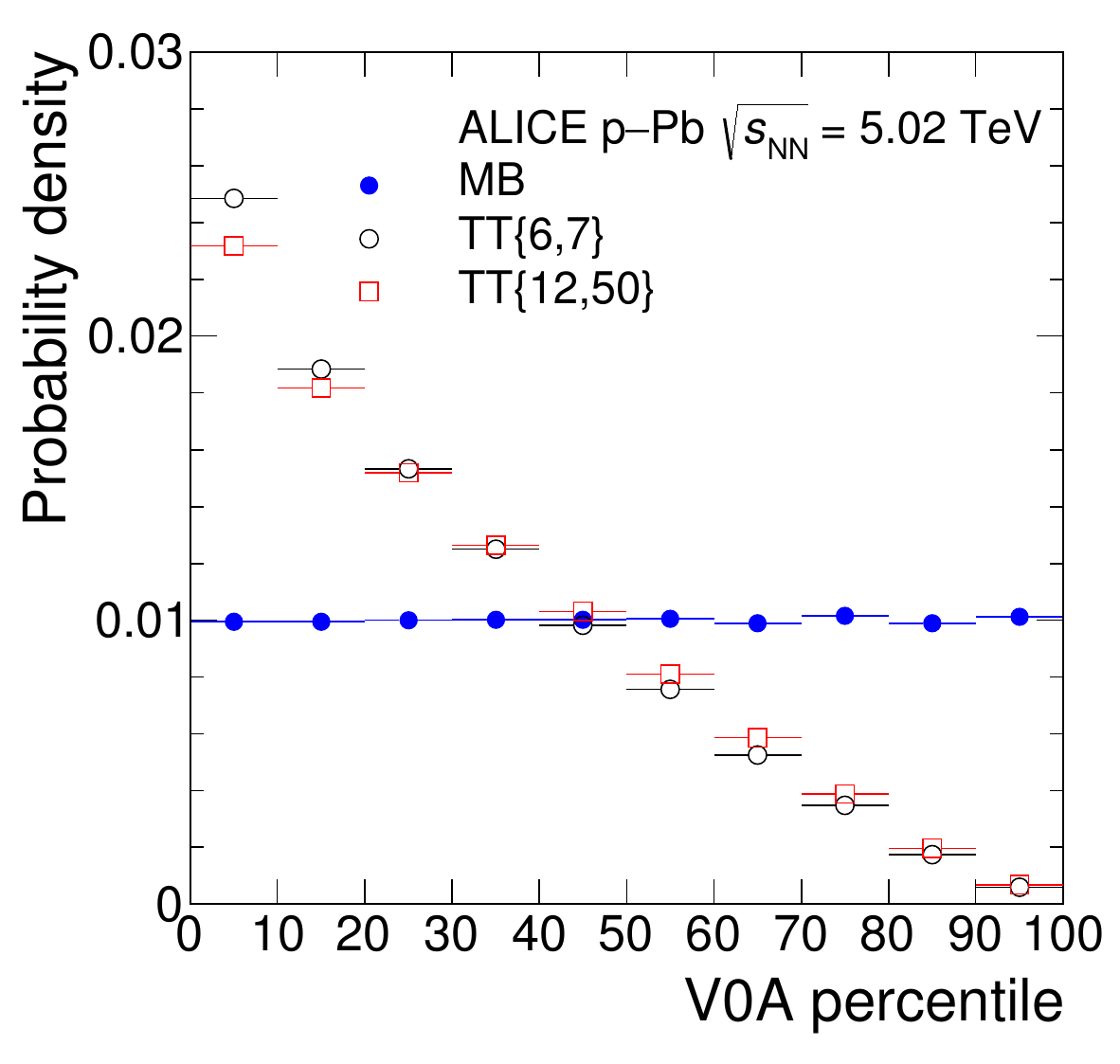}
\caption{Probability distribution of EA (V0A) in \pPb\ events, in decile bins~\cite{Acharya:2017okq}. Blue: MB events; red and black: events selected by presence of a high-\pT\ charged track in the central region. See text for details.} 
\label{fig:hJetpPb_bias} 
\end{figure}

Figure~\ref{fig:hJetpPb_bias} shows the measurement of this hard-process bias in \pPb\ collisions~\cite{Acharya:2017okq}. The blue histogram is the probability distribution in decile bins of the V0A signal (forward charged multiplicity in the Pb-going direction) for the minimum-bias (MB) event population. By construction, this distribution is uniform; indeed, that is the criterion for setting the decile bin boundaries in V0A signal. The black and red histograms show this probability distribution using the same bin boundaries, but for events selected with a high-\pT\ charged track in the central barrel for two \pT\ intervals (``TT\{12,50\}'' denotes $12<\pT<50$ \gev). In this case the probability is biased towards the low percentile bins, corresponding to higher V0A signal, with little to no dependence on track \pT. Such a correlation between soft and hard processes is well-established, and phenomenological approaches have been developed to model it. A recent calculation using the PYTHIA Angantyr model is able to reproduce this measurement, attributing the bias to the increase in the contribution of MPI in the presence of hard processes
~\cite{Adolfsson:2020dhm}.

Such biases are minimised experimentally by utilising forward neutron energy in the ZDC for EA~\cite{Adam:2014qja}, providing the largest possible phase-space separation between the EA measurement and jet quenching observables in the central barrel. However, this approach requires the model assumption that ZDC signal is correlated with centrality through the emission of ``slow neutrons,'' where ``slow'' refers to beam rapidity~\cite{Sikler:2003ef}. Since the modeling of forward slow neutron production is complex, with significant systematic uncertainties, ALICE defines a model-dependent variant of \RpPb, denoted \QpPb, to report jet quenching for inclusive observables using EA-selected \pPb\ data. 

\begin{figure}[htb]
\centering
\includegraphics[width=0.6\textwidth]{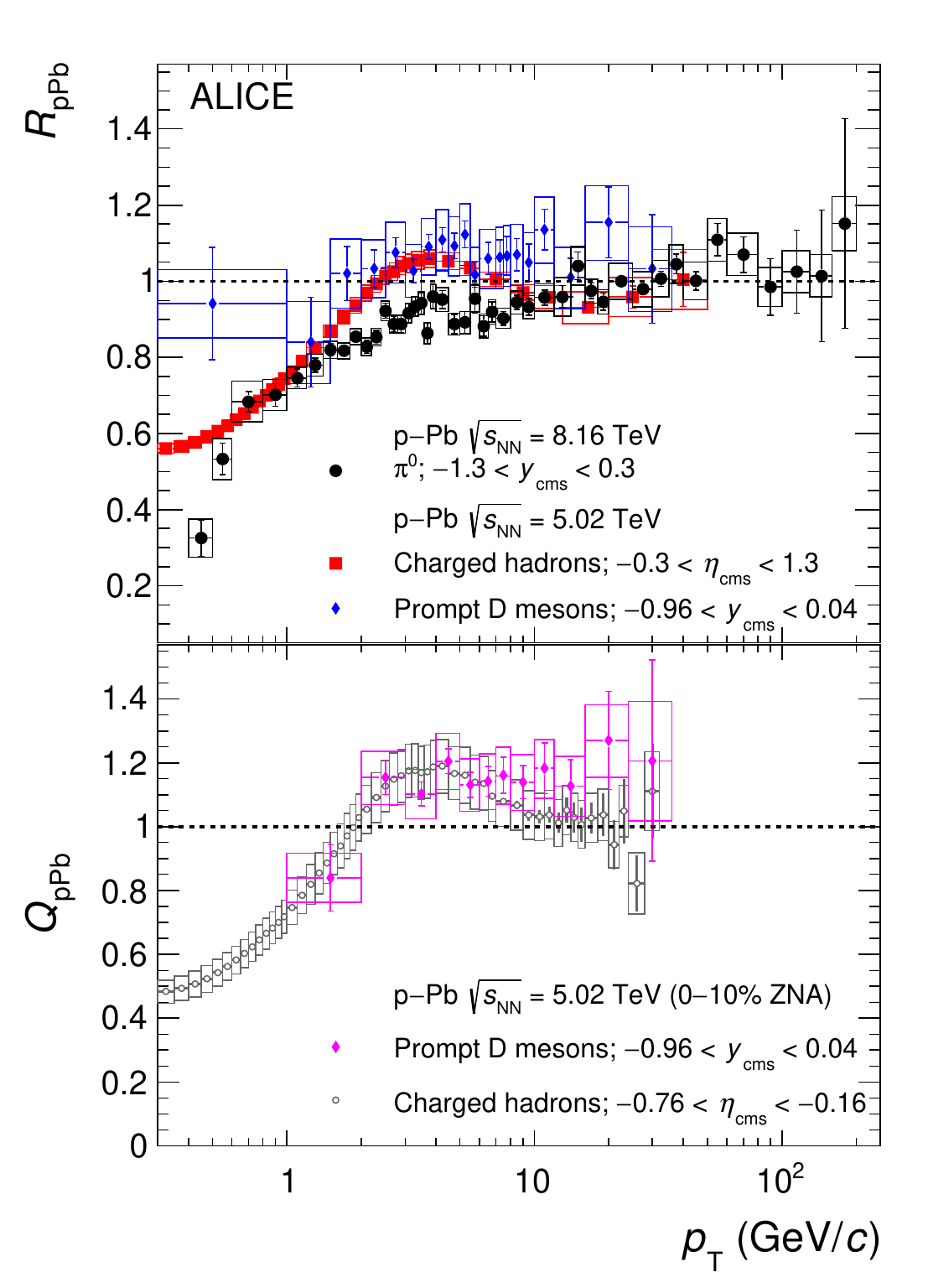}
\caption{Inclusive hadron production in \pPb\ collisions.
(Top) Inclusive yield ratio \RpPb\ for MB \pPb\ collisions, for charged hadrons~\cite{Acharya:2018qsh}, \pizero~\cite{ALICE:2021est}, and prompt D mesons~\cite{ALICE:2019fhe}.  
(Bottom) Inclusive yield ratio \QpPb\ for EA-selected \pPb\ collisions (0--10\% ZNA),  for inclusive charged hadrons~\cite{Adam:2014qja} and  prompt D mesons~\cite{ALICE:2019fhe}. Global normalisation uncertainties of 4\% for \RpPb\ and 7\% for \QpPb\ are not shown in the figure.} 
\label{fig:RpPbSmallSystems} 
\end{figure}

The irreducible modeling uncertainties for correlating \TpPb\ with collision geometry can be avoided by carrying out measurements using MB \pPb\ data, in which there is no EA-based event selection. For MB \pA\ collisions, hard cross sections scale precisely with the nuclear mass $A$~\cite{PhysRevLett.35.407,Alessandro:2003pi} so that $\TpPb=A/\sigma_\mathrm{pPb}^\mathrm{inel}$, where $\sigma_\mathrm{pPb}^\mathrm{inel}$ is the \pPb\ inelastic cross section. Therefore, $R_{\rm pPb}$ is constructed as the ratio of the production cross section in p--Pb divided by the production cross section in pp and by the mass number $A$. While the MB measurement has the advantage of smaller systematic uncertainties, it provides less dynamic range in collision volume than the EA-selected observables, and both should be considered together.

Figure~\ref{fig:RpPbSmallSystems} shows ALICE measurements of inclusive hadron production in \pPb\ collisions. The upper panel shows \RpPb\ for MB \pPb\ collisions, for inclusive charged hadrons, \pizero, and prompt D mesons. The lower panel shows \QpPb\ for EA-selected \pPb\ collisions (0--10\% ZNA), likewise for inclusive charged hadrons and prompt D mesons. For $\pT>5$\,\gev, where the yields of all particle species are dominated by hard-scattering processes, \RpPb\ and \QpPb\ are consistent with unity within the experimental uncertainties for all particle species. Within the sensitivity of these measurements, there is no evidence of jet quenching in \pPb\ collisions.

\begin{figure}[htb]
\centering
\includegraphics[width=0.55\textwidth]{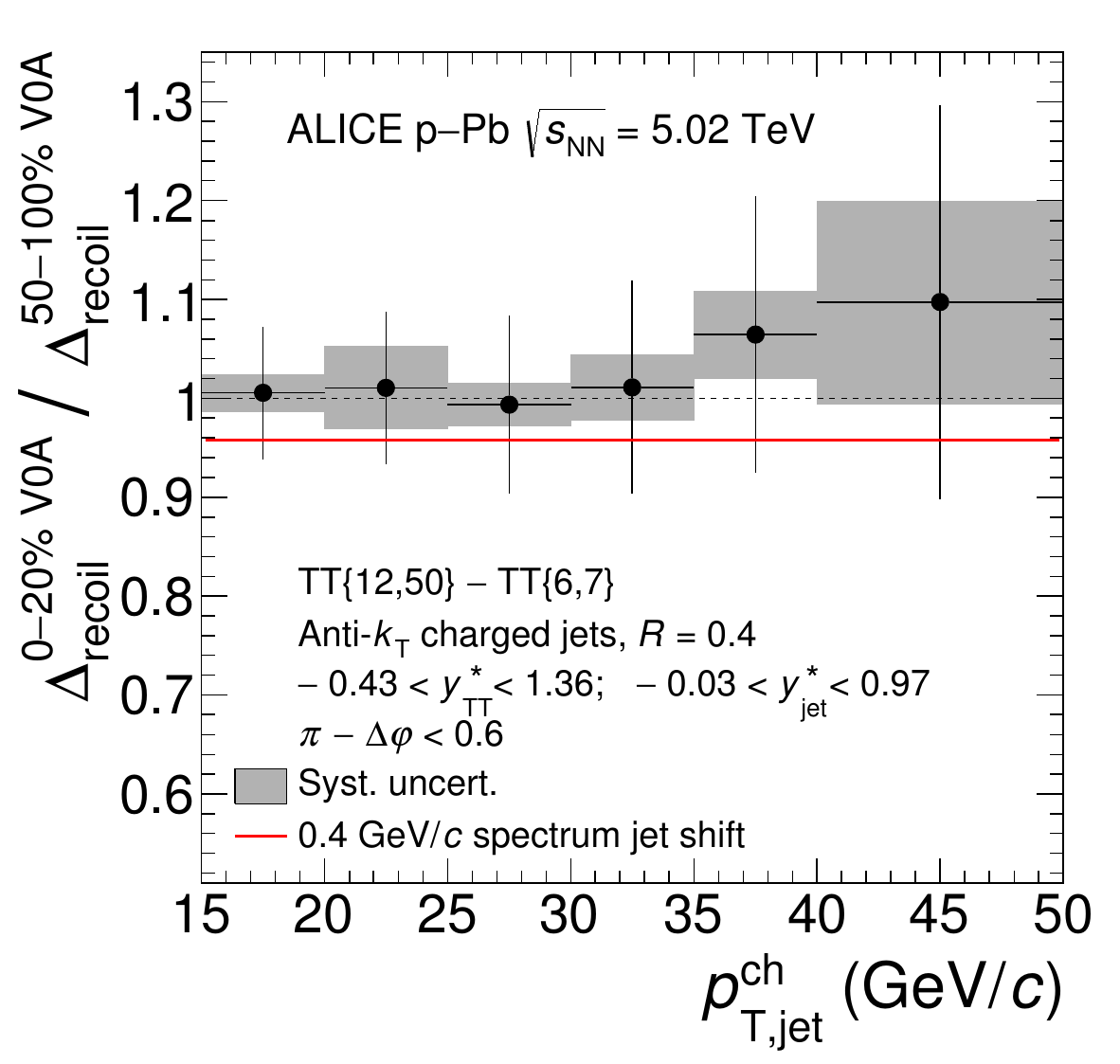}
\caption{Ratio of semi-inclusive recoil jet distributions, \Drecoil, for high-EA and low-EA \pPb\ collisions~\cite{Acharya:2017okq}, using V0A. Charged-particle jets are reconstructed with anti-\kT, $R=0.4$. Vertical lines show the statistical error.
Grey boxes show the systematic uncertainty of the ratio, which accounts for correlated uncertainty in numerator and denominator. The red line indicates the ratio for a \pT-shift of the high-EA distribution of $-0.4$\,\gev, corresponding to the 90\% CL limit in medium-induced energy transport out of the jet cone from this measurement.} 
\label{fig:hJetpPb} 
\end{figure}

In the second approach to the search for jet quenching, based on coincidence observables, ALICE utilises a semi-inclusive observable to measure the trigger-normalised yield of reconstructed jets recoiling from a high-\pT\ charged-hadron trigger~\cite{Adam:2015doa,Acharya:2017okq}. In this approach, the trigger population is sampled, consistently with the inclusive hadron \pT\ distribution, and all recoil jets are counted. This ensures that the resulting trigger-normalised recoil jet distribution is equivalent to the ratio of hard cross sections for hadron+jet and inclusive hadron production. Since the numerator and denominator in this ratio both scale with \TpPb\ for EA-selected data, the geometric factor \TpPb\ cancels exactly~\cite{Acharya:2017okq}. This observable, therefore, enables the measurement of jet quenching in EA-selected data whose interpretation does not require the assumption that EA and collision geometry are correlated, in contrast to \QpPb. This observable therefore has greater systematic precision for searching for small jet quenching signals in EA-selected data from small systems, and does so based on reconstructed jets rather than hadrons.

Jet measurements in high-energy nuclear collisions must contend with large uncorrelated backgrounds (see Sec.~\ref{sec:PartonInteractions}). In order to subtract such backgrounds in a fully data-driven way, ALICE utilises the distribution of \Drecoil, which is the difference of two semi-inclusive recoil jet distributions with high-\pT\ TT and low-\pT\ TT~\cite{Adam:2015doa,Acharya:2017okq}. Figure~\ref{fig:hJetpPb} shows the ratio of \Drecoil\ distributions for high-EA \pPb\ (0--20\% V0A) and low-EA \pPb\ (50--100\% V0A), for charged-particle jets with $R=0.4$. The ratio is compatible with unity, again indicating no evidence of jet quenching in high-EA \pPb\ events, consistent with the inclusive measurements based on \RpPb\ and \QpPb.

In this case, however, the experimental uncertainty is dominated by the statistical error, and it is possible to measure a limit on jet quenching effects. The \Drecoil\ distribution as a function of \pT\ is well-approximated by an exponential function. Assuming that jet quenching corresponds to medium-induced energy transport out of the jet cone whose magnitude is independent of \pTjet, such an energy loss corresponds to a uniform \pT-shift in the \Drecoil\ spectrum, and uniform suppression in the ratio shown in the figure. The red line in the figure corresponds to a \pT-shift of $-0.4$\,\gev\ of the high-EA spectrum, which is the one-sided 90\% CL limit on medium-induced energy transport out of the jet cone $R=0.4$ in high-EA \pPb\ collisions. This limit should be compared to the result of the equivalent measurement for central \PbPb\ collisions, in which significant recoil yield suppression is observed, corresponding to a recoil jet spectrum shift of $(-8\pm2)$\,\gev~\cite{Adam:2015doa}. The limit of $-0.4$\,\gev\ shown in Fig.~\ref{fig:hJetpPb} for high-EA \pPb\ is therefore a factor 20 smaller than the magnitude of jet quenching effects observed in central \PbPb\ collisions. While this measurement provides the most precise limit on jet quenching in small systems currently available, theoretical calculations are needed to determine whether it is still compatible with the mechanisms thought to underlie jet quenching in large systems.

Projections for Runs 3 and 4 at the LHC indicate that the limits on jet quenching from such a semi-inclusive approach will be in the vicinity of $0.1$--$0.2$\,\gev\ for an array of small systems, including high-EA \pp\, high-EA \pPb, and EA-selected O--O collisions. Additional studies are underway to elucidate and quantify the correlations between forward EA and high-$Q^{2}$ processes in the central region.

\subsection{Conclusions}

    \paragraph{Event selection. }The study of event and multiplicity selection in small collision systems has been proven to be
    crucial for a proper understanding of the physics observables across different colliding nuclei. These 
    studies indicate that further measurements in the field of relativistic heavy-ion collisions must
    take proper care to understand such selections carefully, especially in high-multiplicity proton--proton 
    and proton-nucleus collisions. 
    
    \paragraph{Identified particle production.} A comprehensive set of identified particle measurements from low-multiplicity proton--proton 
    to highest multiplicity Pb--Pb collisions has conclusively demonstrated that identified particle ratios 
    such as $\Xi/\pi$ and $\Omega/\pi$ evolve continuously as a function of charged-particle 
    multiplicity density, regardless of collision system or beam energy. Furthermore, specific 
    final-state effects such as the depletion / excess of baryons with respect to mesons at low-/ intermediate-
    transverse momentum have been observed to be present throughout all collisions systems in different 
    magnitudes. Recent measurements have even extended this pattern to the $\Lambda_{\rm c}/{\rm D}^{0}$ ratio. A successful description of all these phenomena for all collision systems remains a theoretical challenge of the field. 
    
    \paragraph{Collective behaviour.} Measurements of azimuthal correlations and anisotropic flow in small collision systems exhibit features of a collectively expanding system, similar to those observed in heavy-ion collisions, where they are believed to originate from the presence of QGP medium. The origin of the collective effects depends on particle multiplicity of a collision. While hydrodynamic-like description seems to be favored by data especially at high multiplicities, the effects of initial state effects from initial gluon momentum correlations may play an important role at low-$N_{\rm ch}$.
    
    \paragraph{Charmonium and Bottomonium. } Charmonium measurements in p--Pb collisions indicate the $\psi$(2S) suppression in the backward rapidity region (Pb-going) to be larger with respect to that of the J/$\psi$. This is not expected from cold nuclear matter effects and is suggestive of final state interactions between the weakly bound c$\overline{\rm c}$ and a strongly interacting partonic or hadronic medium, leading to its dissociation. 
    Similar suppression patterns are observed also in the bottomonium sector for $\Upsilon$(2S) and $\Upsilon$(3S).
    
    \paragraph{Searches for jet quenching.} Measurements in small systems have not observed significant jet quenching effects thus far within the sensitivity of the measurements. These measurements however provide the first quantitative limit on jet quenching in small systems, corresponding to medium-induced energy transport out of a jet cone with radius 0.4 less than 400~MeV at 90\% confidence level in high-event-activity p--Pb collisions.

\newpage

\section{The initial state of the collision}
\label{ch:InitialState}

\newcommand{\jpsi}         {\ensuremath{\rm{J/\psi}}\xspace}
\newcommand{\rhozero}{\ensuremath{\rho^{0}}\xspace}
\newcommand{\eightonesix}         {$\sqrt{s}~=~8.16$~Te\kern-.1emV\xspace}
\newcommand{\fivefourfournn}       {$\sqrt{s_{\mathrm{NN}}}~=~5.44$~Te\kern-.1emV\xspace}

\newcommand{\jgc}[1]{{\color{RoyalBlue}{\colorbox{RoyalBlue}{\color{white}{JGC:}} #1}}}

An accurate description of the initial state of nucleus--nucleus collisions is crucial for the interpretation of all measurements carried out in heavy-ion collisions at the LHC. In particular, the description of hard processes strongly depends on the longitudinal momentum distribution of partons inside the nucleus, while the spatial transverse profile of partons and its fluctuations are important to properly interpret observables related to azimuthal anisotropies and correlations.

Parton distribution functions (PDFs) of protons, describing the probability to find a parton in the proton with a longitudinal momentum fraction $x$ at resolution scale $Q^2$, show a rapid growth towards small values of $x$. This growth is expected to slow down and saturate due to unitarity constraints~\cite{Gribov:1984tu}. However, no compelling evidence for saturation has been found so far. Besides, PDFs are significantly modified inside the nucleus with respect to the free proton PDFs. In particular, nuclear PDFs at $x<10^{-2}$ demonstrate a clear suppression compared to proton PDFs, a phenomenon referred to as nuclear shadowing~\cite{Armesto:2006ph}. The most recent parameterisations of nuclear PDFs, such as nCTEQ15~\cite{Kovarik:2015cma}, EPPS16~\cite{Eskola:2016oht}, or nNNPDF~\cite{AbdulKhalek:2020yuc}, are based on global QCD fits to available data samples of nuclear deep-inelastic scattering including \pPb data on dijets and electroweak bosons. These parameterisations are affected by large uncertainties due to the limited kinematic coverage of the available data and the largely indirect determination of the gluon distributions. Latest results on the heavy-flavour and charmonium nuclear modification factors in \pPb collisions at the LHC provide new stringent constraints on nuclear PDFs~\cite{Lansberg:2016deg,Eskola:2019dui,Eskola:2019bgf}. However, the role of final state effects in these measurements, such as energy loss and radial flow, still needs to be clarified. On the contrary, electroweak-boson measurements are not affected by final-state effects and can serve as particularly clean probes of light quark PDFs in nuclei at large scales $Q^2 \sim M_{\rm W,Z}^2$. ALICE results on $\rm Z^0$ and $\rm W^{\pm}$ boson production in \pPb and \PbPb collisions are summarised in Sec.~\ref{sec:ew}. 

ALICE measurements of  heavy-vector-meson photoproduction  in ultra-peripheral collisions, described in Sec.~\ref{sec:upc}, are sensitive to the gluon distribution in nuclei at hard scales corresponding to the heavy-quark mass~\cite{Guzey:2013xba}, while $\rho^0$ photoproduction also allows for the study of the approach to the black-disk limit in QCD at semi-hard scales~\cite{Frankfurt:2002wc}. 
The rapidity dependence of the coherent cross section of heavy vector mesons gives information on the energy evolution of the gluon distribution, providing one of the cleanest probes of shadowing at  small $x$ down to $10^{-5}$. So far, these results have not been included in the global parton analyses to constrain the nuclear gluon distributions in heavy nuclei, the reason being uncertainties in the choice of scale and implementation of NLO effects. Intense theoretical work, however, is ongoing to resolve these issues~\cite{Flett:2019pux,Eskola:2022vpi}. Furthermore, the transverse-momentum distribution of the coherently-produced vector meson encodes information on the average distribution of colour fields in the impact-parameter plane~\cite{Guzey:2016qwo}, while the
incoherent production of vector mesons is sensitive to  fluctuations of the gluon distribution, particularly to the different possible geometries of the nuclear initial state in the impact-parameter plane~\cite{Mantysaari:2016jaz}.

The characterisation of the spatial distribution of nuclear matter in the initial state is also a crucial task in the study of relativistic heavy-ion collisions. Not only it allows mapping the structure of the nucleus at high-energy, but this knowledge of the initial state also helps constrain calculations of key QGP medium properties, which were discussed in the previous chapters. The initial state models generate energy or density profiles prior to the time of the formation of the QGP in the overlap region of a collision, and this can be accomplished in several ways. Almost all implementations take advantage of the assumptions associated with the Monte Carlo (MC) Glauber approach~\cite{Miller:2007ri}. This approach assumes the nucleons in the nucleus are positioned randomly according to the Woods-Saxon distribution, the nucleons travel in an unperturbed trajectory irrespective of whether they interact with other nucleons, and the criterion for a nucleon--nucleon interaction depends on the inelastic cross section (which can be inferred from experimental measurements in \pp collisions).  In the simplest MC Glauber nucleon model, the density of the overlap region in a collision is the sum of the densities of the participating nuclear matter. The T$_{\rm{R}}$ENTo model~\cite{Moreland:2014oya} also uses nucleons as the relevant degrees of freedom, however in its optimal configuration, it assumes the overlap densities are the square root of the product of the individual nuclear densities. The MC-KLN~\cite{Kharzeev:2000ph,Kharzeev:2002ei} and IP-Glasma~\cite{Schenke:2012wb,Schenke:2012hg} models take a different approach, and assume that the relevant degrees of freedom are gluons in the nucleons. They use the Colour Glass Condensate effective theory coupled with saturation equations to determine the density profiles of gluons in the overlap region. Another initial state model, EKRT~\cite{Eskola:1999fc,Niemi:2015qia}, determines parton densities from next-to-leading-order perturbative QCD using the saturation conjecture. Finally, the MC Glauber constituent quark model~\cite{Loizides:2016djv} is an extension of the MC Glauber nucleon model, however the relevant degrees of freedom are constituent quarks in the nucleons.

Particle multiplicity distributions and the anisotropic flow measurements can be extremely valuable in the understanding of the initial state. After two decades of developments, viscous hydrodynamic calculations using various initial-state models can often describe the multiplicity distributions, particle momentum spectra, and integrated flow measurements simultaneously, which has been demonstrated in earlier chapters. However, around the time of the first LHC collisions, it was stated that modelling uncertainties in the initial state, which in turn provided an extraction of $\eta/s$ from RHIC data, led to uncertainties for $\eta/s$ of a factor of around 2.5~\cite{Song:2010mg}. That factor arose from the ambiguity of whether the MC Glauber nucleon model~\cite{Miller:2007ri}, or the MC-KLN model~\cite{Kharzeev:2000ph,Kharzeev:2002ei}, should be used to determine QGP transport properties based on comparisons to data. Therefore, assessing whether these modelling uncertainties still apply is critical for the determination of QGP transport properties. The situation has significantly improved in the past few years, after new developments arising from measurements on event-by-event flow fluctuations (and the construction of probability density function), namely the fluctuations of flow-vectors, the correlations between different-order flow coefficients via Symmetric Cumulants~\cite{Bilandzic:2013kga} that has been introduced in Sec.~\ref{sec:TG2symmetriccumulants}, and measurements sensitive to the non-linear hydrodynamic response of higher-order flow with the corresponding correlations between different-order symmetry planes. All of these observables have varying sensitivities to the initial-state anisotropy, which is a key feature in all of the initial-state models described, and which will be addressed in Sec.~\ref{sec:mf}.

\subsection{Electroweak-boson measurements}
\label{sec:ew}

\begin{figure}[htb]
\centering
\includegraphics[width = 0.8\textwidth]{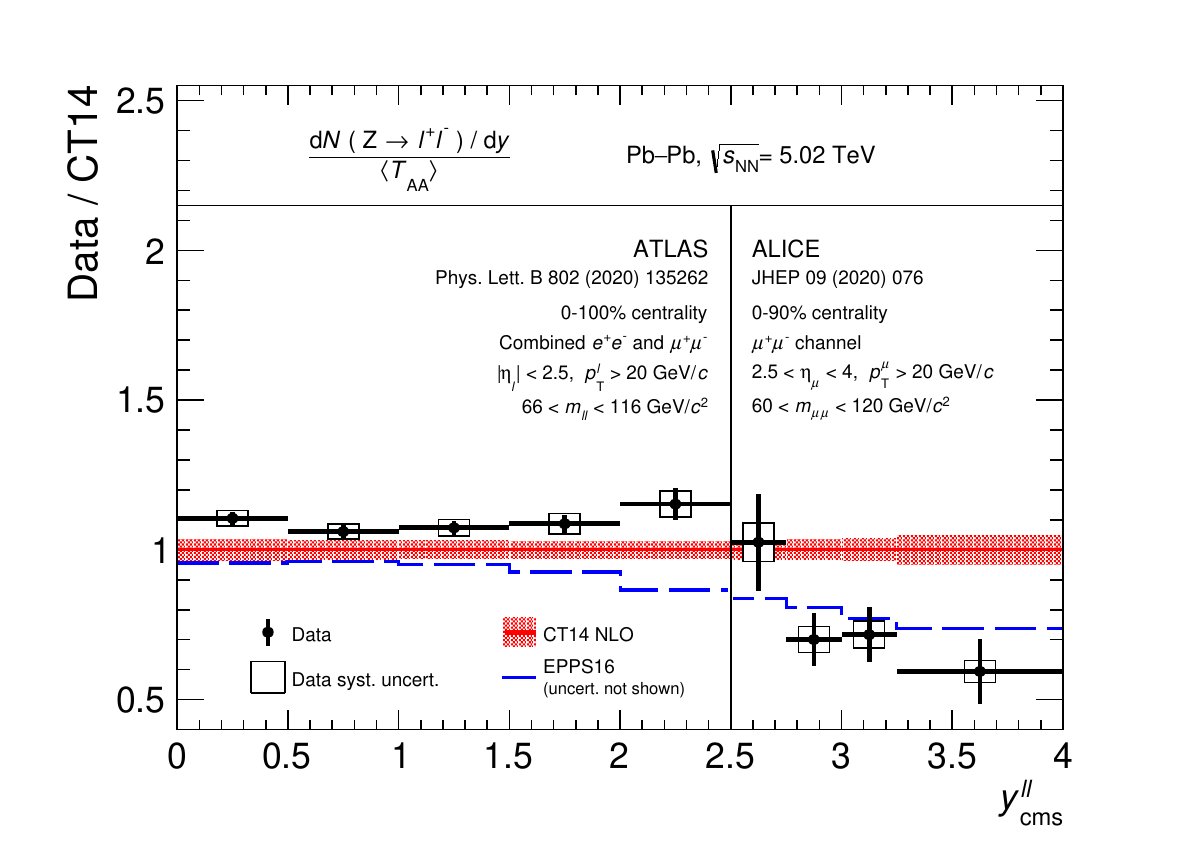}
\caption{The ratio of the $\rm Z^0$-boson yield  measured by ALICE~\cite{Acharya:2020puh} and ATLAS~\cite{Aad:2019lan} in \PbPb collisions at \fivenn and free-nucleon  PDF (CT14) predictions. The ratios are compared to theoretical calculations including EPPS16 parameterisation of the nuclear modification effects.}
\label{fig:ew1}
\end{figure}

Electroweak bosons are predominantly produced via the Drell-Yan process, i.e.\,$\rm q\overline q$ annihilation into a lepton pair, therefore they can be used as particularly clean probes of light-quark PDFs since leptons are not affected by final-state effects. The resolution scale $Q^2$ is determined by the boson mass $Q^2 = M_{\rm Z,W}^2$ while Bjorken-$x$ values of light quarks are directly related to the rapidity $y$ and the transverse mass $m_{\rm T}$ of the dilepton pair as  $x = \frac{m_{\rm T}}{\sqrt{s}} \exp (\pm y)$. ALICE performed detailed studies of $\rm Z^0$ and W$^{\pm}$ boson production in \pPb and \PbPb collisions with the muon spectrometer at forward rapidity. These measurements are sensitive to nuclear effects at typical $x$ values either below $10^{-3}$, in the shadowing region, or above $10^{-1}$ where antishadowing and EMC effects play a role. 

While ALICE measurements of $\rm Z^0$ boson production in \pPb collisions are described by models that include both free-nucleon and nuclear-modified PDFs~\cite{Alice:2016wka,Acharya:2020puh}, the \PbPb results at \fivenn appear to be more sensitive to nuclear shadowing effects, showing a clear deviation from calculations based on free-nucleon PDFs~\cite{Acharya:2020puh}. $\rm Z^0$ boson production is measured in the dimuon channel for muons with $p_{\rm T} > 20 $~GeV/$c$ and $2.5 < y^{\mu\mu} < 4$ in a broad centrality range from 0 to 90\%. The ratio of $\rm Z^0$ boson yield and corresponding free-nucleon PDF (CT14) predictions is shown in Fig.~\ref{fig:ew1} as a function of rapidity together with ATLAS results at midrapidity~\cite{Aad:2019lan} The ratio is compared to theoretical calculations including the EPPS16 parameterisation of the nuclear modification effects. The ALICE data are described by nPDF calculations (CT14 + EPPS16) but the free-nucleon PDF calculations (CT14) overestimate the data. A $3.4\,\sigma$ deviation is found between the rapidity-integrated yield and free-nucleon PDF calculations indicating that nuclear shadowing effects play an important role.

\begin{figure}[htb]
\centering
\includegraphics[width = 0.8\textwidth]{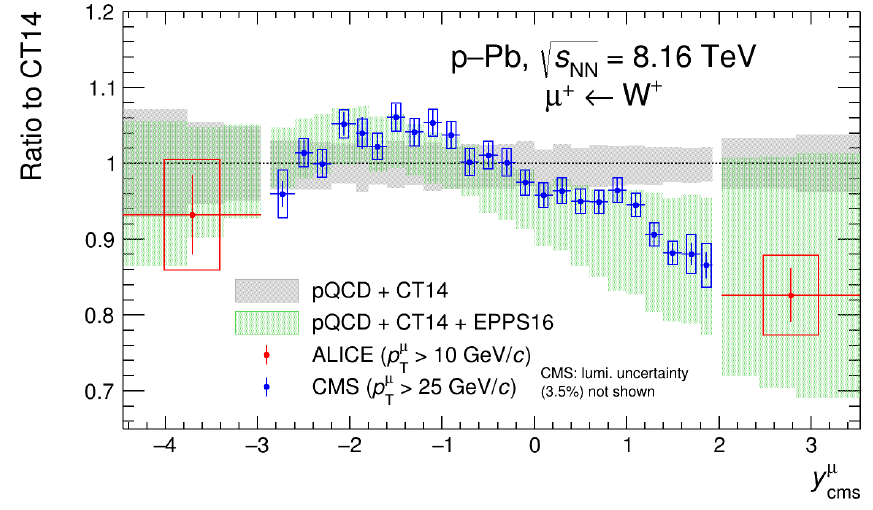}
\caption{The ratio of $\rm W^+$-boson cross section measured in \pPb collisions at \eightonesixnn to the free-nucleon PDF (CT14) predictions for ALICE~\cite{ALICE:2022cxs} and CMS~\cite{Sirunyan:2019dox} data. The ratios are compared to theoretical calculations including  nuclear PDF modification effects (CT14 + EPPS16).} 
\label{fig:ew2}
\end{figure}

The $\rm W^{\pm}$-boson production has been measured by ALICE via muonic decays requiring muon tracks with $p_{\rm T} > 10 $~GeV/$c$ in the muon spectrometer acceptance~\cite{ALICE:2022cxs}. Figure~\ref{fig:ew2} shows the ratio of the cross section of muons from W$^+$ measured by ALICE~\cite{ALICE:2022cxs} and CMS~\cite{Sirunyan:2019dox} in \pPb collisions at \eightonesixnn  to the free-nucleon PDF (CT14) predictions. While the ALICE measurement agrees with predictions based on EPPS16 nPDFs, it deviates by 2.7\,$\sigma$ from the free-nucleon PDF calculation at positive rapidities, corresponding to Bjorken-$x$ below $10^{-3}$ in Pb nuclei. These measurements may provide significant constraints to future nuclear PDF global fits. 

\subsection{Photon-induced processes in heavy-ion collisions}
\label{sec:upc}

Due to the short range of the strong force, collisions where the incoming nuclei pass by each other at a distance larger than the sum of their radii, called ultra-peripheral collisions (UPCs), are dominated by photon-induced processes. The intensity of the photon flux depends on the square of the electric charge of the incoming particle, so Pb nuclei are intense sources of photons at the LHC. Recent reviews of  UPC physics  can be found in Refs.~\cite{Baltz:2007kq,Contreras:2015dqa,Klein:2019qfb}.

One important process accessible in UPC is the diffractive photoproduction of vector mesons off hadronic targets, which is particularly sensitive to the gluon distribution in the target, because it proceeds through a colourless exchange and thus involves at least two gluons. 
ALICE has measured three variations of this process. The exclusive production of a vector meson off a proton target, and the production of the vector meson by the interaction with the full nucleus (coherent production), or with one of the nucleons in the nucleus (incoherent production). Measurements where a $\jpsi$ meson is produced allow for a pQCD treatment of the process at a scale $\approx M_{\jpsi}/2$, while those where a $\rhozero$ meson is produced are an excellent tool to study the approach to the black-disk limit of QCD at a semi-hard scale~\cite{Frankfurt:2002wc}.

\begin{figure}[htb]
\centering
\includegraphics[width = 0.70\textwidth]{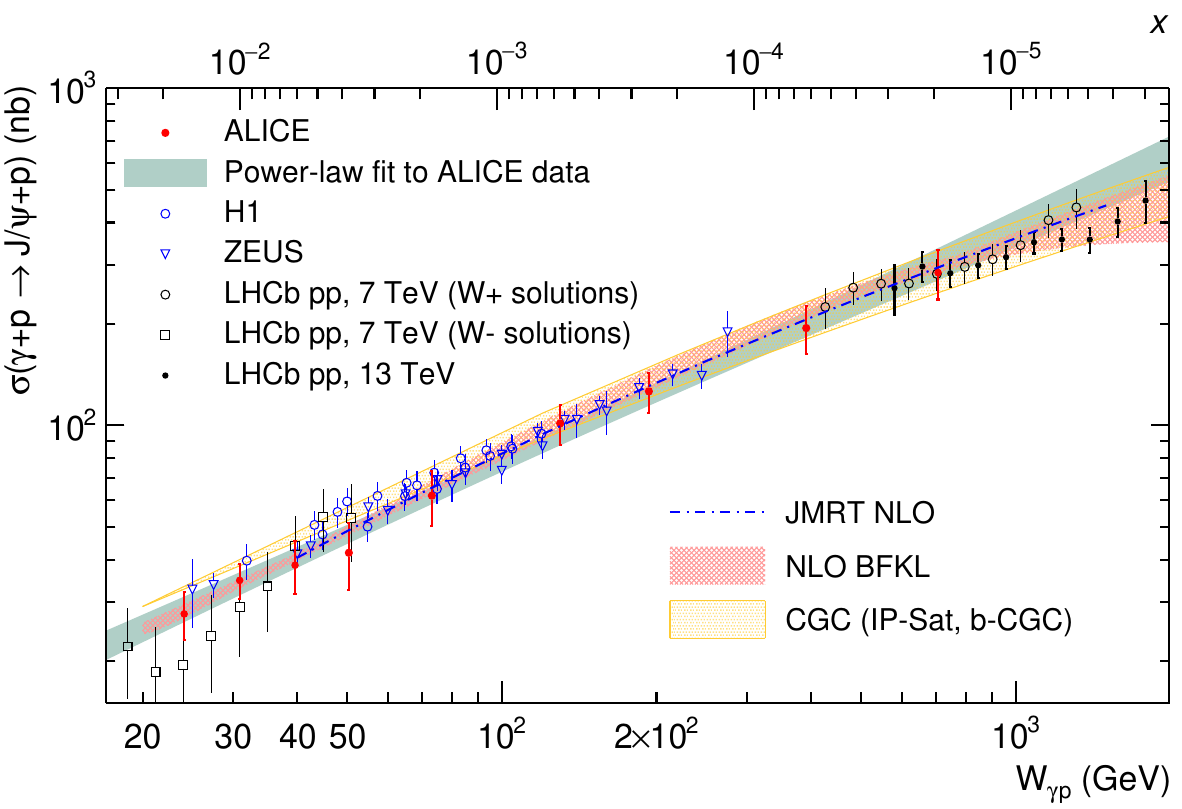}
\caption{Cross section of exclusive $\jpsi$ photoproduction off protons as a function of $W_{\gamma p}$ energy measured by ALICE in \pPb UPC~\cite{TheALICE:2014dwa,Acharya:2018jua}, compared to HERA~\cite{Alexa:2013xxa,Chekanov:2002xi} and LHCb~\cite{Aaij:2018arx,Aaij:2014iea} results and theoretical models~\cite{Jones:2013pga,Armesto:2014sma,Bautista:2016xnp}.} 
\label{fig:upc_jpsi_pPb}
\end{figure}

The exclusive photoproduction of $\jpsi$ off protons has been measured by ALICE in \pPb UPC at \fivenn~\cite{TheALICE:2014dwa,Acharya:2018jua}. The  cross section, shown in Fig.~\ref{fig:upc_jpsi_pPb}, covers a broad range of centre-of-mass energies in the $\gamma$p frame, $W_{\gamma{\rm p}}$, extending from 20 GeV to above 700 GeV, which correspond to three orders of magnitude in Bjorken-$x$ from $2\times10^{-2}$ to $2\times10^{-5}$, where $x=(M/W_{\gamma{\rm p}})^2$ with $M$ being the mass of the vector meson. The behaviour of the cross section in this large kinematic range can be described by a power law in $W_{\gamma{\rm p}}$ with an exponent $0.70 \pm 0.05$. In the leading order pQCD, this behaviour implies that the gluon distribution in the proton keeps rising as a power law with decreasing $x$, without a clear signal of becoming saturated. Data can be described by models based on NLO BFKL equations~\cite{Bautista:2016xnp} or the standard DGLAP pQCD without saturation effects, such as JMRT NLO shown in Fig.~\ref{fig:upc_jpsi_pPb}~\cite{Jones:2013pga}. Models including saturation effects, e.g.\,CGC~\cite{Armesto:2014sma} in the figure, predict a similar cross section to those predicted by models without saturation in the considered $W_{\gamma{\rm p}}$ range. Nonetheless, a recent analysis claims that the data indicate the onset of gluon saturation in the proton~\cite{Garcia:2019tne}. The ongoing analysis of Run 2 data from \pPb UPC at \eightonesixnn, and of future larger samples, will increase the experimental reach to the region above 1 \TeV in $W_{\gamma{\rm p}}$ with smaller uncertainties than in the current results. These measurements may provide a definitive answer about the presence of gluon saturation in the proton.

\begin{figure}[b]
\center
\includegraphics[height = 0.55\textwidth]{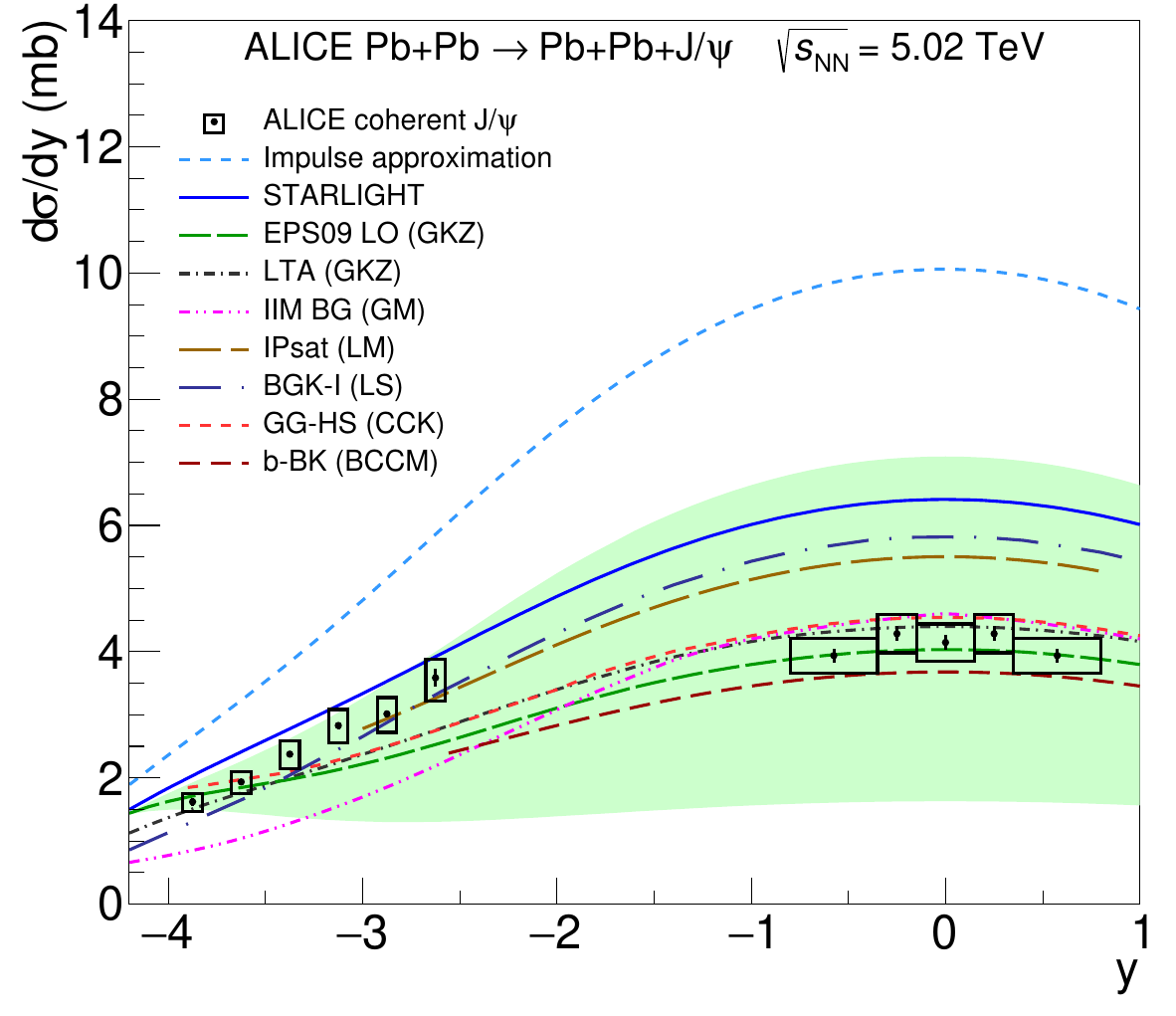}
\hbox{\hspace{0.3cm} \includegraphics[height = 0.5\textwidth]{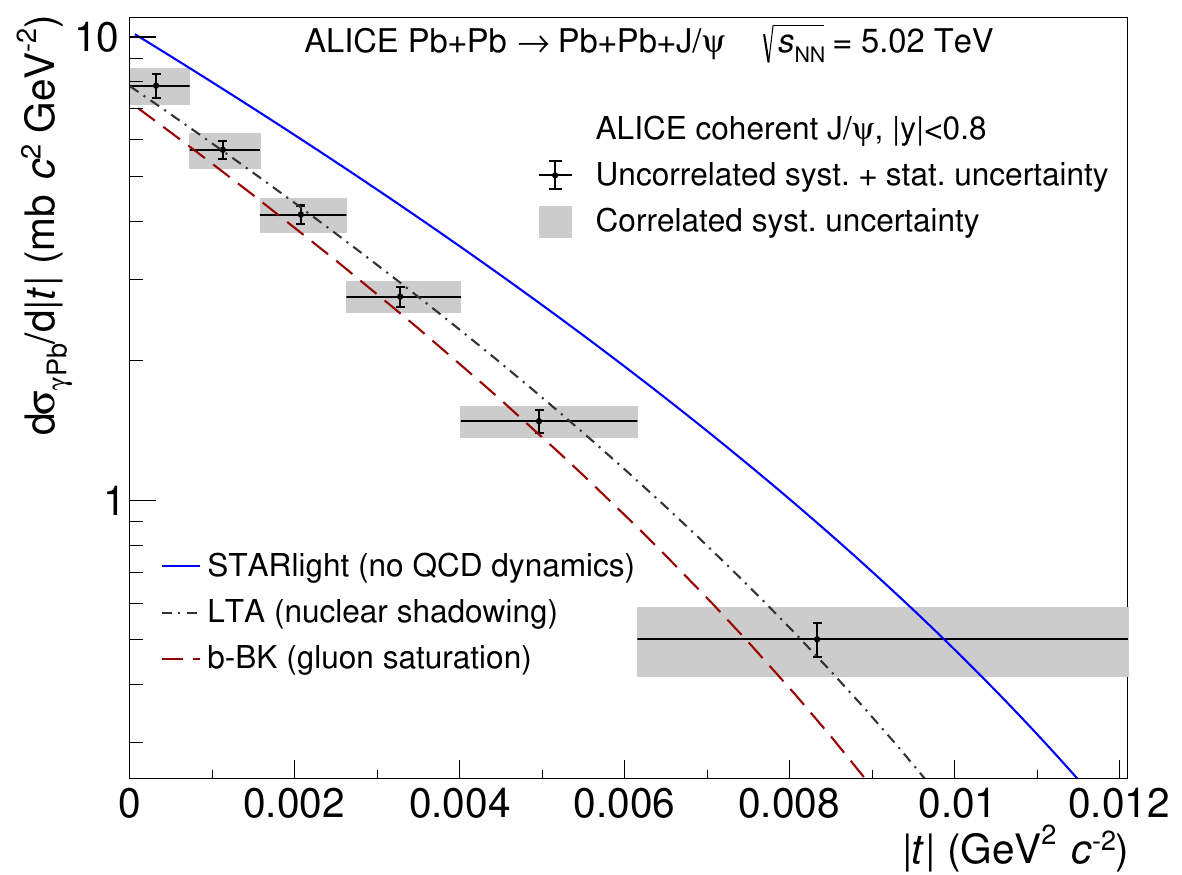}}
\caption{(Top) Cross section of coherent $\jpsi$ production measured by ALICE in \PbPb UPC at \fivenn as a function of rapidity~\cite{ALICE:2021gpt} and (Bottom) $|t|$-differential $\jpsi$ photoproduction cross section extracted from the $p_{\rm T}^2$ distribution of coherent \jpsi at midrapidity (right panel)~\cite{ALICE:2021tyx}. The data are compared to various theoretical models.} 
\label{fig:upc_jpsi_PbPb}
\end{figure}

The photoproduction of $\jpsi$ in \PbPb UPC has been studied by  ALICE at \twosevensixnn~\cite{ALICE:2012yye,Abbas:2013oua} and \fivenn~\cite{Acharya:2019vlb,ALICE:2021gpt,ALICE:2021tyx}. The results reported in Ref.~\cite{ALICE:2012yye} are the first measurement of photon-induced processes at the LHC. They demonstrated that this type of  analysis could be carried out at the LHC with good precision and in a kinematic domain not accessible to other machines. 
ALICE has measured the coherent production of $\jpsi$ in UPC in a wide rapidity range as shown in the top plot of Fig.~\ref{fig:upc_jpsi_PbPb}~\cite{ALICE:2021gpt}. 
The comparison of the rapidity dependence of data with that of the impulse approximation~\cite{Guzey:2013xba}, computed neglecting all nuclear effects under the assumption that the scattering of photons on nuclei is given by the coherent superposition of the scattering on the individual nucleons, provides one of the cleanest, and clearest, signatures of gluon shadowing and its Bjorken-$x$ dependence. 
Data rise quickly towards midrapidity and then seem to saturate. This behaviour is not completely reproduced by models. The STARlight prediction~\cite{Klein:1999qj}, that includes only Glauber-like suppression, is above the data, particularly at midrapidity where the colour-dipole models including saturation,  IPsat~\cite{Lappi:2013am} and BGK-I~\cite{Luszczak:2019vdc} in Fig.~\ref{fig:upc_jpsi_PbPb}, also overshoot the data by a large margin. 

On the other hand, the leading-twist approximation~\cite{Guzey:2016piu}, LTA in the figure, and the energy-dependent hot-spot model GG-HS~\cite{Cepila:2017nef}, which also includes saturation, give the best overall description of the rapidity dependence. However, these models underpredict the data at semi-forward rapidities in the range $2.5 < |y| < 3.5$, indicating that the nuclear shadowing might have a smaller effect at the Bjorken $x \sim 10^{-2}$ or $x \sim 5\times 10^{-5}$ corresponding to this rapidity range. Figure~\ref{fig:upc_jpsi_PbPb} also shows a wide green band corresponding to the uncertainties of the model~\cite{Guzey:2016piu} based on the EPS09 LO parameterisation of nuclear PDFs~\cite{Eskola:2009uj}, illustrating that the ALICE data have a great potential to improve the uncertainties of global nPDF fits.

In \PbPb UPCs there is an ambiguity regarding which of the nuclei is the source of the photon and which the target~\cite{Klein:1999gv}. There are thus two contributions to the cross section, one from a high, the other from a low, energy photon. Only at midrapidity the two contributions are equal and the $\gamma$Pb cross section can be extracted directly from the UPC measurement. Comparing the $\gamma$Pb cross section with the impulse approximation, a gluon shadowing factor can be defined~\cite{Guzey:2013xba} and determined from data to be around 0.6 for $x \approx 10^{-3}$. Future measurements of coherent $\jpsi$ production, accompanied by secondary exchanges of photons and resulting in electromagnetic dissociation of nuclei, can serve as one of the most promising tools to disentangle low- and high-energy contributions at forward rapidities, since coherent $\jpsi$ cross sections with and without nuclear dissociation are sensitive to different impact parameter ranges~\cite{Guzey:2013jaa}.
A proof-of-principle of this approach is provided by the recent measurement of Pb--Pb electromagnetic dissociation cross sections using the detection of forward neutrons in the ALICE Zero Degree Calorimeters~\cite{ALICE:2022hbd}.

The bottom plot in Fig.~\ref{fig:upc_jpsi_PbPb} shows the distribution of the momentum squared transferred at the target vertex, $|t|$, measured by ALICE at midrapidity in \PbPb UPC at \fivenn~\cite{ALICE:2021tyx}. These data correspond to $x=0.6\times10^{-4}$ and are obtained from the measured distribution of the square of the transverse momentum of the $\jpsi$, which, for colliders, is quite close to $|t|$. As the transferred $p_{\rm T}$ is the Fourier conjugate of the impact parameter, this measurement provides information on the gluon distribution in the plane perpendicular to the motion of the nuclei. The shape of the measured $|t|$ distribution deviates from the simple form-factor-based dependence used in STARlight~\cite{Klein:1999qj}, but is well reproduced by models including gluon shadowing according to the leading-twist approximation~\cite{Guzey:2016qwo}, or gluon saturation effects from the impact-parameter-dependent Balitsky-Kovchegov equation~\cite{Bendova:2019psy}.

\begin{figure}[htb]
\centering
\includegraphics[width = 0.75\textwidth]{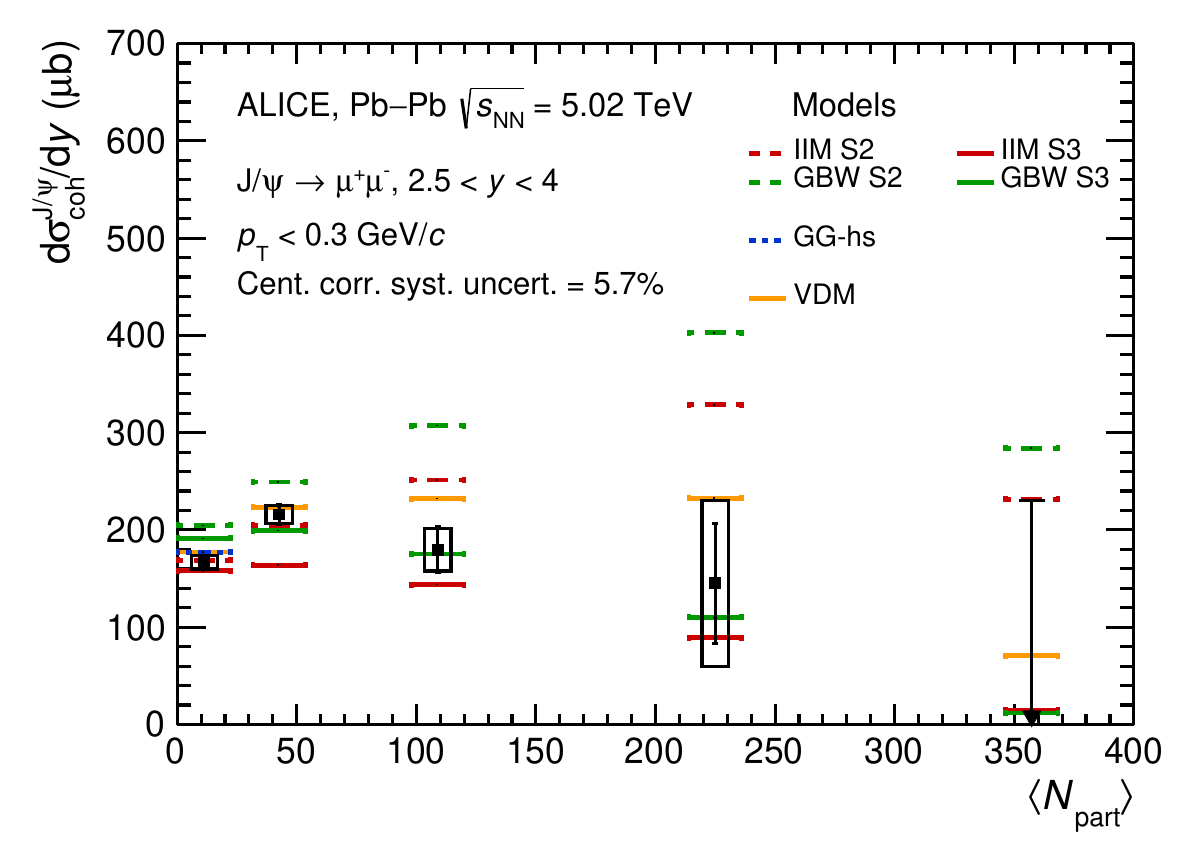}
\caption{Coherent $\jpsi$ photoproduction cross section measured by ALICE in hadronic \PbPb collisions as a function of $N_{\rm part}$~\cite{ALICE:2022zso}.  Results are compared with theoretical calculations from Ref.~\cite{Cepila:2017nef} (GG-hs), Ref.~\cite{GayDucati:2018who} (S2 and S3 scenarios within the IIM and GBW dipole models) and from Ref.~\cite{Klusek-Gawenda:2015hja} (VDM).} 
\label{fig:upc_per}
\end{figure}

ALICE has also discovered an excess of $\jpsi$ at low-transverse momenta, $p_{\rm T}<0.3\, {\rm GeV}/c$, in peripheral and semi-central \PbPb collisions~\cite{Adam:2015gba,ALICE:2022zso}, which was interpreted as a sign of the coherent $\jpsi$ photoproduction in nucleus--nucleus collisions that also include hadronic interactions~\cite{Klusek-Gawenda:2015hja,Zha:2017jch,GayDucati:2018who,Contreras:2016pkc}. Coherent $\jpsi$ cross sections are shown in Fig.~\ref{fig:upc_per} as a function of $N_{\rm part}$ corresponding to the range of centralities from 30\% to 90\% in comparison to various models based on different assumptions on how spectator and non-spectator nucleons participate in the coherent reaction. The IIM and GBW dipole model predictions~\cite{GayDucati:2018who} steadily increase with centrality in the scenario with unmodified photonuclear cross section (S2), while the use of an effective cross section where the overlap region between the two nuclei is assumed not to contribute to coherent photoproduction (S3) results in a reduction of the cross section toward more central collisions, providing a better description of the data.
However, the reduction of the coherent cross section in central collisions has been also obtained in the VDM model~\cite{Klusek-Gawenda:2015hja} with unmodified photonuclear cross sections where the photon flux has been calculated differently compared to the S2 scenario.
Though the mechanism of the coherent photoproduction in hadronic events still needs to be clarified, these measurements are considered as a promising tool to decouple low- and high-energy $\gamma$Pb cross sections in UPC measurements~\cite{Contreras:2016pkc} and a potential probe of the quark--gluon plasma in more central events, see e.g.\,Ref.~\cite{Shi:2017qep}.

\begin{figure}[htb]
\centering
\hbox{\hspace{3.3cm} \includegraphics[width = 0.6\textwidth]{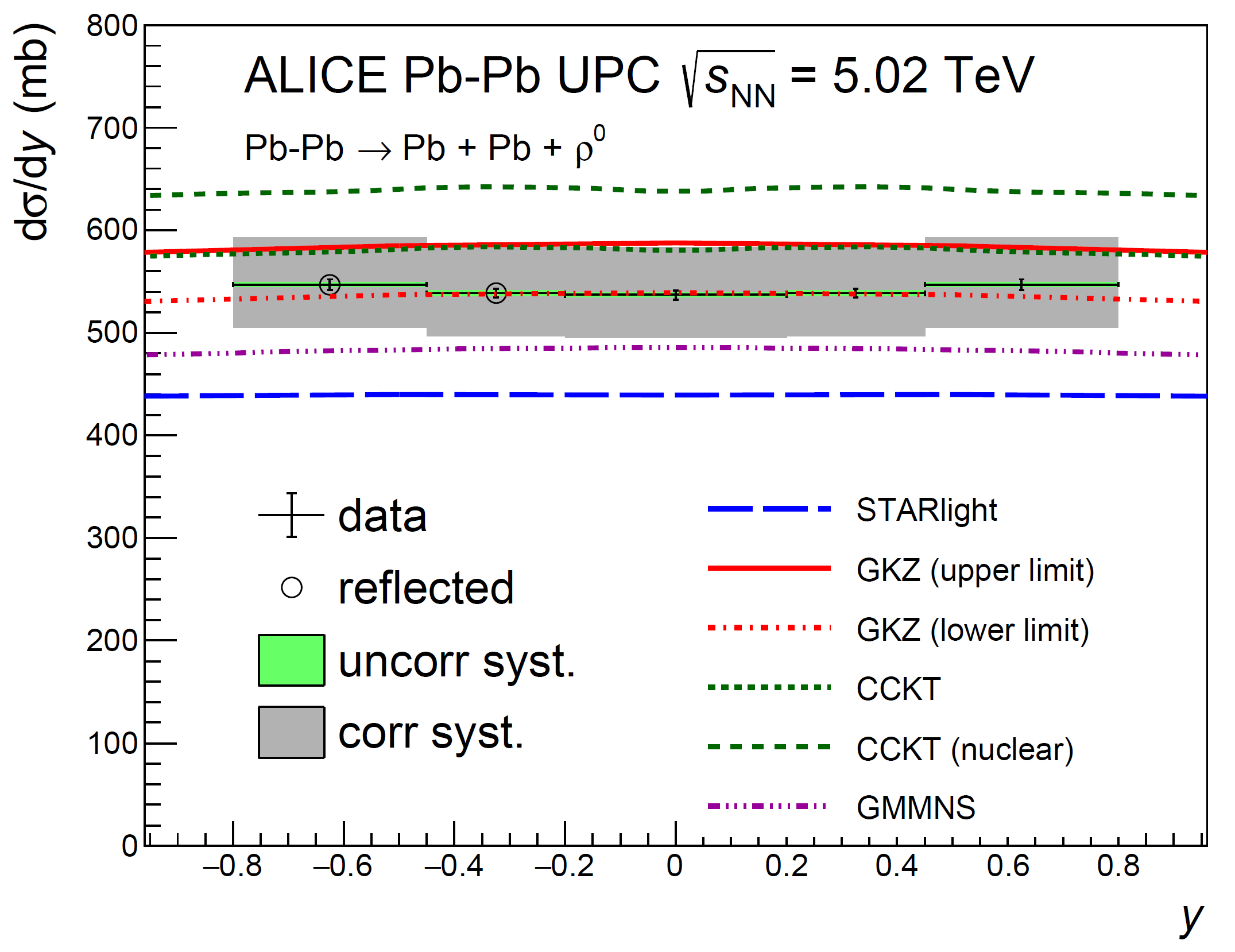}}
\includegraphics[width = 0.6\textwidth]{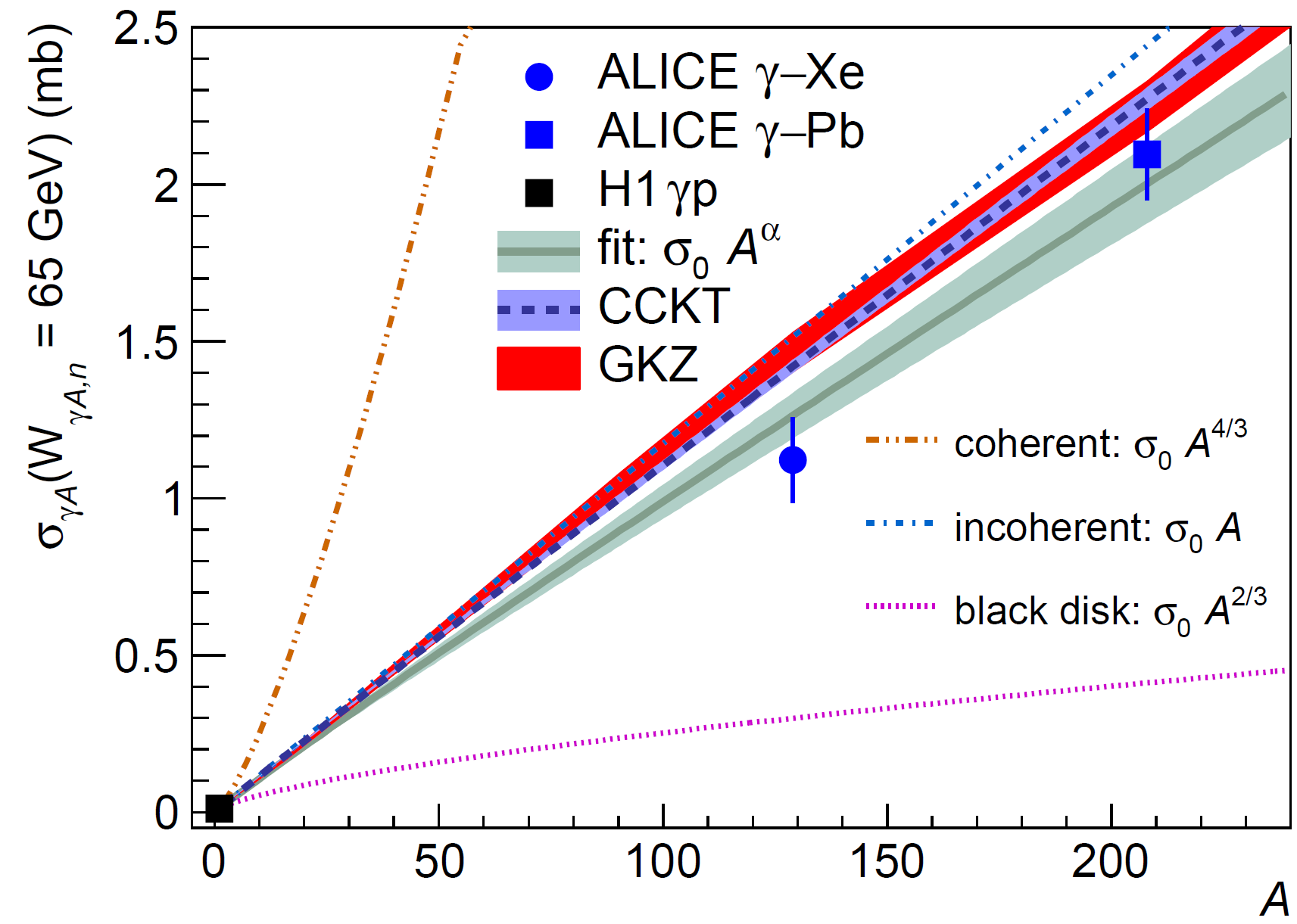}
\caption{(Top) The coherent $\rho$ photoproduction cross section as a function of rapidity compared to theoretical calculations~\cite{Acharya:2020sbc}. (Bottom) $A$-dependence of the coherent $\rho$ photoproduction cross section with a power-law fit shown by the band and general expectations for several extreme cases discussed in the text~\cite{ALICE:2021jnv}.} 
\label{fig:upc_rho}
\end{figure}

The knowledge of the structure of nuclei has been extended to even smaller momentum-exchange scales with the measurements of the coherent photoproduction of $\rhozero$ vector mesons at midrapidity in \PbPb collisions at \twosevensixnn~\cite{Adam:2015gsa} and \fivenn~\cite{Acharya:2020sbc}, and \XeXe collisions at \fivefourfournn~\cite{ALICE:2021jnv}. 
The coherent $\rhozero$ cross section  in \PbPb collisions at \fivenn~\cite{Acharya:2020sbc} without forward-neutron selection is shown in Fig.~\ref{fig:upc_rho}, top, where it is compared with theoretical calculations. 
The GKZ model~\cite{Guzey:2016piu}, based on a modified vector-dominance model, in which the hadronic fluctuations of the photon interact with the nucleons in the nucleus according to the Gribov-Glauber prescriptions, describes correctly the data. A slightly worse, but still quite satisfactory, description of data is obtained with the energy-dependent hot-spot model CCKT~\cite{Cepila:2017nef}. The GMMNS~\cite{Goncalves:2017wgg} and STARlight~\cite{Klein:2016yzr} models underpredict the data. Furthermore, in~\cite{Acharya:2020sbc} the cross section is presented in classes defined by the presence of forward neutrons produced by electromagnetic dissociation of the interacting nuclei due to secondary photon exchanges. The measurements in several neutron-emission classes were proposed as a tool to decouple high- and low-energy contributions to the UPC cross sections~\cite{Guzey:2013jaa}, since the presence of additional photon interactions effectively reduces the range of impact parameters and modifies the flux of photons participating in the coherent photoproduction process. ALICE measurements of the coherent $\rhozero$ production in neutron-emission classes are well described by models implemented in STARlight~\cite{Klein:2016yzr,Baltz:2002pp,Baur:2003ar} and the NOON MC~\cite{Broz:2019kpl}, thus confirming the factorisation of the coherent photoproduction process and the additional photon exchanges assumed in these models. As the measurement at midrapidity can be converted from the UPC to the $\gamma$Pb cross section without ambiguity, the agreement between data and models suggests that the method proposed in~\cite{Guzey:2013jaa} can be applied to disentangle the high- and low-energy photon contribution to the UPC cross section at other rapidities. The comparison of models with the measurement in  \XeXe UPC produces a similar message. Furthermore, as the energy dependence of the $\gamma$Pb cross section is quite mild, the small difference in centre-of-mass energy in the \PbPb and \XeXe systems can be ignored and the measurements can be used to study the $A$-dependence of the cross section as shown in Fig.~\ref{fig:upc_rho}, bottom.
The trend in data is quite different from that expected for coherent production without any other effect, represented in the figure by the model scaling as $A^{4/3}$. Indeed, the data are compatible with a power-law behaviour with exponent $\alpha=0.963\pm0.019$, demonstrating clear shadowing, but still far from the black-disk limit that predicts $A^{2/3}$.

The measurements described in the preceding paragraphs can still be improved substantially with the  data samples from the LHC Run 3 and 4, which are expected to  be at least three orders of magnitude larger than those from Run 2~\cite{Citron:2018lsq}. This huge increase in the available number of events will not only reduce the statistical uncertainties, but will permit a series of studies to reduce the systematic uncertainties. The expected data samples will also allow us to explore multidimensional investigations of the current observables as well as to study new signatures beyond the reach of our current data.

\subsection{Multiplicity and flow measurements}
\label{sec:mf}

\begin{figure}[htb]
\centering
\includegraphics[width = 0.75\textwidth]{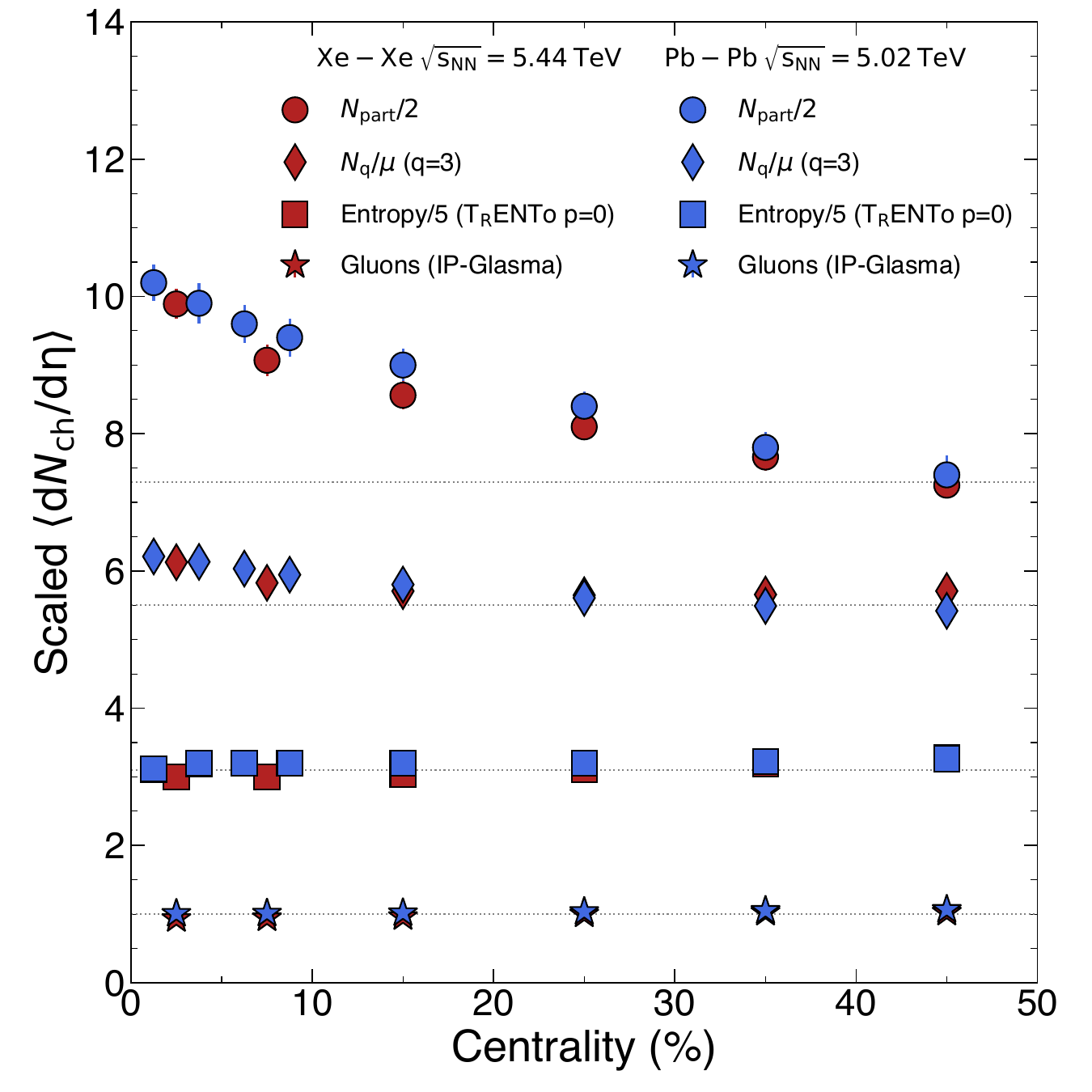}
\caption{Heavy-ion d$N_{\rm ch}/$d$\eta$ data (at midrapidity)~\cite{Adam:2016ddh,Acharya:2018hhy} scaled by the degrees of freedom from a variety of initial state models. The dotted lines are to guide the eye. The circular markers represent d$N_{\rm ch}/$d$\eta$ divided  by the number of participating nucleons in the MC Glauber nucleon model~\cite{Miller:2007ri}, diamonds use the number of constituent quarks from the MC Glauber constituent quark model~\cite{Loizides:2016djv}, squares the entropy from  T$_{\rm{R}}$ENTo~\cite{Moreland:2014oya}, and stars the gluons from IP-Glasma~\cite{Schenke:2012wb,Schenke:2012hg}.} 
\label{fig:Mult_IS}
\end{figure}

In this section, we begin with a discussion of ALICE data from two different heavy-ion collision systems, namely \PbPb and \XeXe, that have yielded various measurements~\cite{Adam:2016ddh,Acharya:2018hhy,Acharya:2018lmh,Acharya:2018ihu}, which can be directly compared to the initial state models described earlier in this review. Figure~\ref{fig:Mult_IS} shows the measured d$N_{\rm ch}/$d$\eta$ divided by the degrees of freedom in each initial state model. It has been demonstrated that entropy is approximately conserved through the evolution of the QGP using hydrodynamic simulations for $\eta/s$ values around $1/4\pi$~\cite{Jeon:2015dfa}. The number of degrees of freedom in the initial state (or initial state entropy) should therefore be proportional to the number of particles produced in the final state. It is clear that this criterion is not met for the MC Glauber nucleon model since the scaled d$N_{\rm ch}/$d$\eta$ increases for more central collisions, and based on the more central data, the MC Glauber constituent quark model also seems to fail. On the other hand, both the T$_{\rm{R}}$ENTo (which provides as an output the initial state entropy directly) and IP-Glasma models show a much better scaling, since the scaled d$N_{\rm ch}/$d$\eta$ appears more constant as a function of centrality. 

\begin{figure}[htb]
\centering
\includegraphics[width = 0.8\textwidth]{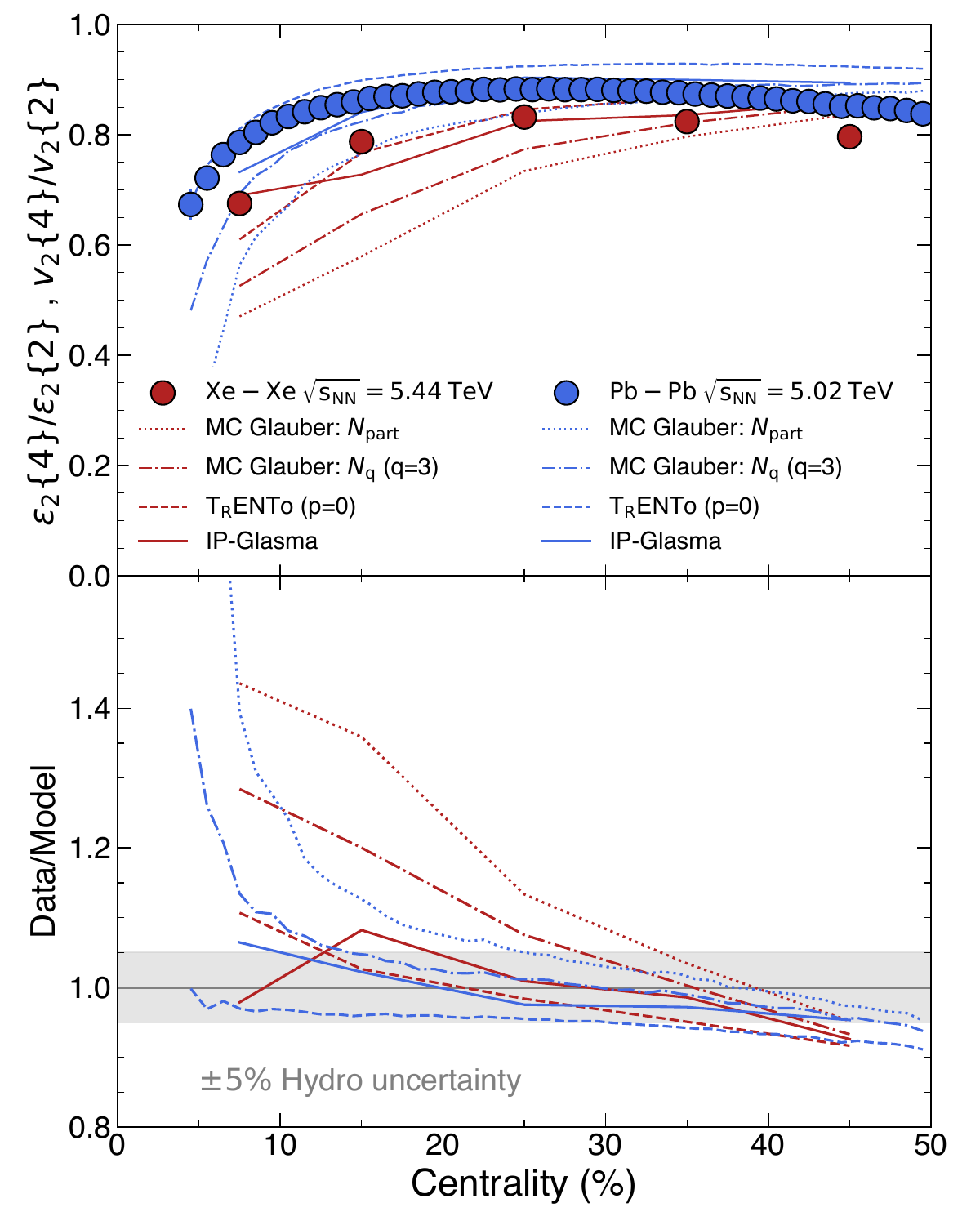}
\caption{Comparisons of ALICE $v_2\{4\}/v_2\{2\}$ results from~\cite{Acharya:2018lmh,Acharya:2018ihu} to $\varepsilon_{2}\{4\}/\varepsilon_{2}\{2\}$ from the same initial state models shown in the previous figure:
the MC Glauber nucleon model~\cite{Miller:2007ri}, the MC Glauber constituent quark model~\cite{Loizides:2016djv},   T$_{\rm{R}}$ENTo~\cite{Moreland:2014oya}, and IP-Glasma~\cite{Schenke:2012wb,Schenke:2012hg}.
} 
\label{fig:vnFluc_IS}
\end{figure}

Figure~\ref{fig:vnFluc_IS} shows ALICE measurements of the ratio of elliptic flow coefficients estimated from 2- and 4-particle correlations $v_{2}\{4\}/v_{2}\{2\}$, compared to the same model predictions for the ratio of eccentricities $\varepsilon_{2}\{4\}/\varepsilon_{2}\{2\}$. The second order eccentricity $\varepsilon_{2}$ in the initial state models is determined based on the configuration of the overlap density. Both $v_{2}\{m\}$ and $\varepsilon_{2}\{m\}$ are sensitive to event-by-event fluctuations of $v_{2}$ and $\varepsilon_{2}$ respectively. Such fluctuations have a positive contribution to $v_{2}\{2\}$ and $\varepsilon_{2}\{2\}$, and have a negative contribution to $v_{2}\{4\}$ and $\varepsilon_{2}\{4\}$. Assuming $v_{2} \propto \varepsilon_{2}$, the ratios $v_{2}\{4\}/v_{2}\{2\}$ and $\varepsilon_{2}\{4\}/\varepsilon_{2}\{2\}$ should therefore be equivalent. Hydrodynamic calculations have indeed shown these two quantities are compatible within 5\% (denoted by the gray area). It is clear here that the T$_{\rm{R}}$ENTo and IP-Glasma models offer the most competitive descriptions, which is also the case for the multiplicity measurements shown in Fig.~\ref{fig:Mult_IS}.

The success of both the T$_{\rm{R}}$ENTo and IP-Glasma models poses some key questions, but also provides reassurances when these models are used to extract QGP properties, as discussed in Sec.~\ref{ch:QGPproperties}, since they appear to model the initial state quite well. As mentioned previously, each model has a fundamentally different approach to modelling the initial state, and the differing ways these models combine nuclear matter can be understood by considering the nuclear matter density projected in the $xy$ plane. In this case, $T_{a}(x,y)$ represents such a distribution for the projectile nucleus, while $T_{b}(x,y)$ for the target nucleus. The IP-Glasma model has its roots in a saturation QCD based approach~\cite{Schenke:2012wb}, where the assumption of a high density of gluons in the initial state leads to the prediction that the overlap density should be characterised by $T_{a}T_{b}$. This is a feature of both weak and strong coupling theories in QCD~\cite{Romatschke:2017ejr}. The T$_{\rm{R}}$ENTo model assumes that this combination should be $\sqrt{T_{a}T_{b}}$, and such an approach has been suggested by a 3D modelling of the initial state~\cite{Shen:2020jwv}. It therefore seems clear that measurements sensitive to both initial state transverse 
and longitudinal effects %
might provide further distinguishing power. 

\begin{figure}[htbp]
\begin{center}
\includegraphics[width=0.8\linewidth]{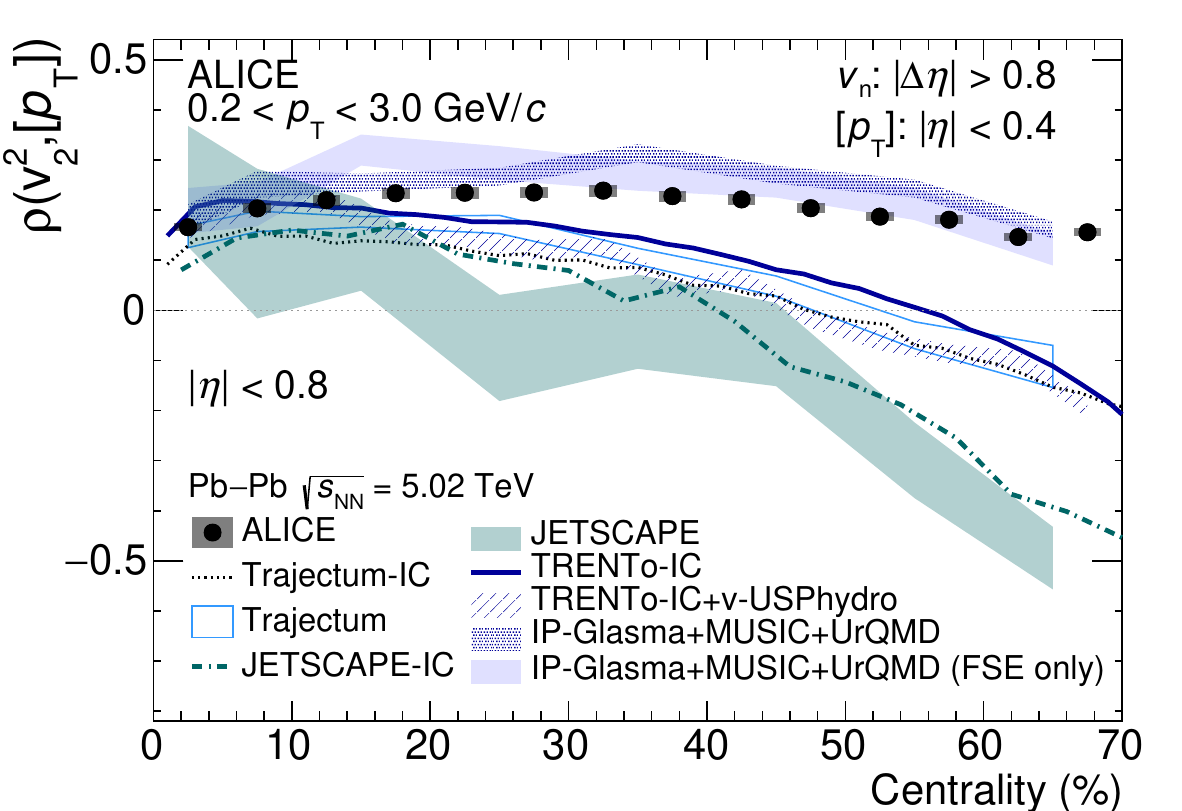}
\caption{Centrality dependence of $\rho \left(v_{\rm 2}^{2}, [ p_{\rm T} ] \right)$ in Pb--Pb collisions at $\sqrt{s_{\rm NN}} = 5.02$~TeV~\cite{ALICE:2021gxt}. The statistical (systematic) uncertainties are shown with vertical bars (filled boxes). The initial state estimations are represented by lines in the figures, while IP+Glasma+MUSIC+UrQMD~\cite{Schenke:2020uqq},  v-USPhydro~\cite{Giacalone:2020lbm}, Trajectum~\cite{Nijs:2020ors}, and JETSCAPE~\cite{JETSCAPE:2020shq} hydrodynamic model calculations are shown with hatched bands. }
\label{fig:rho_2}
\end{center}
\end{figure}

Measurements of correlations between the average transverse momentum of all particles in a single event $[p_{\rm T}]$ and their anisotropic flow coefficients $v_n$, reflecting the correlations between energy density (therefore the size) and the shape of the initial conditions, offer another avenue to resolve this question. They are quantified by a modified Pearson correlation coefficient $\rho(v_{\rm n}^{2}, [p_{\rm T}])$. Because the physical mechanism driving $\rho(v_{\rm n}^{2}, [p_{\rm T}])$ arises from the initial state conditions~\cite{Giacalone:2020dln,Schenke:2020uqq,Giacalone:2020byk}, measurements of $\rho(v_{\rm n}^{2}, [p_{\rm T}])$ can provide valuable information in this regard. Figure~\ref{fig:rho_2} shows the centrality dependence of $\rho(v_{\rm 2}^{2}, [p_{\rm T}])$ in Pb--Pb collisions at $\sqrt{s_{\rm NN}} = 5.02$~TeV measured by ALICE~\cite{ALICE:2021gxt}. It has a weak centrality dependence and is positive for the presented centrality range. This means that $v_2$ and [$p_{\rm T}$], hence the size and shape of the system in the initial stages, are positively correlated. 
Hydrodynamic model calculations from v-USPhydro~\cite{Giacalone:2020lbm}, Trajectum~\cite{Nijs:2020ors}, JETSCAPE~\cite{JETSCAPE:2020shq} based on $\rm T_{R}ENTo$ initial conditions, and IP-Glasma+MUSIC~\cite{Schenke:2020uqq}, are shown for comparison. The IP-Glasma+MUSIC calculation quantitatively reproduces the measured $\rho(v_{\rm 2}^{2}, [p_{\rm T}])$. For the calculations using $\rm T_{R}ENTo$ initial conditions, but with different tuned input parameters, both v-USPhydro and Trajectum show a strong centrality dependence and underestimate the data by more than 50\% for centrality above 30\%. Furthermore, these models predict an opposite sign of $\rho(v_{\rm 2}^{2}, [p_{\rm T}])$ with respect to data for centralities above 40\%. The discrepancies are more pronounced for the JETSCAPE predictions~\cite{JETSCAPE:2020shq}, which become negative for centralities above 20\%. Such discrepancies between IP-Glasma and $\rm T_{R}ENTo$ cannot be attributed to the effect of the initial momentum anisotropy predicted by the CGC framework in IP-Glasma, as its impact is insignificant in the presented centrality ranges. Instead, the discrepancies are expected to arise from different geometric effects in the initial state. In particular, $\rho(v_{\rm 2}^{2}, [p_{\rm T}])$ measurement favours smaller values of the width of the colliding nucleons~\cite{Giacalone:2021clp}. Such smaller values are also supported by a recent study based on the Trajectum framework~\cite{Nijs:2022rme} that uses the ALICE measurement of the hadronic Pb--Pb cross section~\cite{ALICE:2022xir}.  These observations provide significant distinguishing power between different initial state models, which was not possible via the multiplicity or anisotropic flow fluctuations measurements described previously. Moreover, $\rho(v_{\rm 2}^{2}, [p_{\rm T}])$ is found to be sensitive to the nuclear quadrupole deformation~\cite{Giacalone:2020awm} and has the potential to probe nuclear triaxial structure~\cite{Bally:2021qys}. Such measurements therefore open a unique window for the study of nuclear structure in heavy-ion collisions at the LHC. Full exploitation of the LHC as an imaging tool will advance our understanding of strongly-correlated nuclear systems via probes and techniques complementary to those utilized in low-energy applications. Such studies will ultimately yield unique insight into the behavior of quantum chromodynamics (QCD) across systems and energy scales~\cite{Bally:2022vgo}.

\begin{figure}[htb]
\begin{center}
\includegraphics[width = 0.8\textwidth]{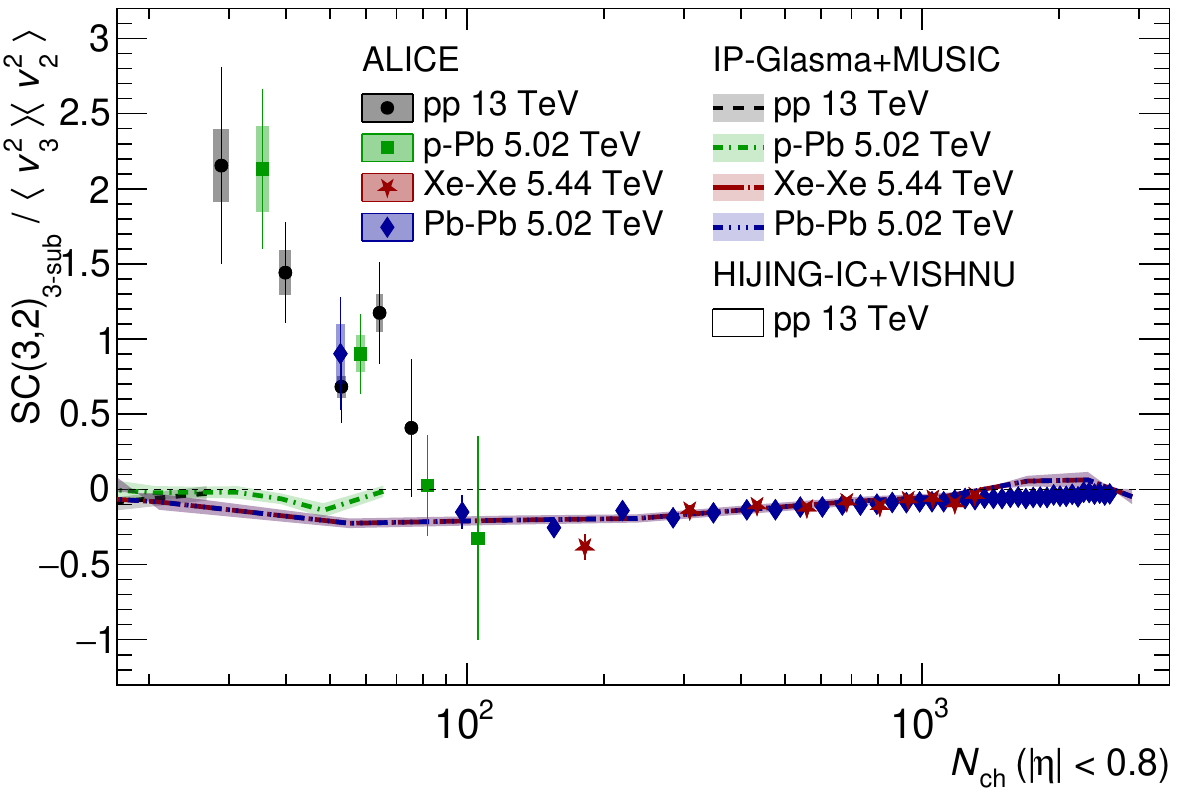}
\caption{Normalised Symmetric Cumulants of the second and third orders from a variety of collisions as a function of the number of charged hadrons~\cite{Acharya:2019vdf}. They are compared to predictions from the IP-Glasma+MUSIC model~\cite{Mantysaari:2016jaz,Schenke:2020mbo}.} 
\label{fig:NSC}
\end{center}
\end{figure}

Finally, assuming that $v_2$ ($v_3$) has a linear response to the initial $\varepsilon_2$ ($\varepsilon_3$), it is expected that the normalised symmetric cumulant $NSC(3,2)$ measured in the final state could reflect the initial correlations between $\varepsilon_2^{2}$ and $\varepsilon_3^{2}$. This has been validated in hydrodynamic calculations~\cite{ALICE:2016kpq,Zhu:2016puf} where a good agreement between initial $NSC^{\varepsilon}(3,2)$ and final state $NSC(3,2)$ is seen, independent of the type of the initial state models or the transport coefficients applied. Most of hydrodynamic model calculations~\cite{Zhu:2016puf,Niemi:2015qia,Zhao:2017yhj,McDonald:2016vlt} can only qualitatively or at best semi-quantitatively describe the ALICE $NSC(m,n)$, as discussed in detail in Sec.~\ref{section:3.3}. If such a linear response of $v_2$ ($v_3$) to $\varepsilon_2$ ($\varepsilon_3$) holds also in small collision systems, then one can use the measured $NSC(3,2)$ to probe the initial $\varepsilon_2^{2}$ and $\varepsilon_3^{2}$ correlation, and thus constrain the initial state for small collision systems, which is still poorly known so far. In Fig.~\ref{fig:NSC}, normalised $SC(3,2)$ measurements are presented in both small and large systems~\cite{Acharya:2019vdf}, where the 3-subevent method is used to largely suppress the non-flow contamination~\cite{Jia:2017hbm,Huo:2017nms}. Negative $NSC(3,2)_{\rm 3-sub}$ is observed in Xe--Xe and Pb--Pb collisions down to multiplicity of $N_{\rm ch} \approx$ 100, which suggests an anticorrelation between $v_{2}^{2}$ and $v_{3}^{2}$. There is a hint of a change to a positive sign of $NSC(3,2)_{\rm 3-sub}$ in Pb--Pb collisions for multiplicity below 100. This tendency is also seen at even lower multiplicities in \pPb and \pp collisions. It is not observed in the measurement using larger $\eta$ acceptances in ATLAS~\cite{Aaboud:2018syf} and CMS~\cite{Sirunyan:2019sef}. This could be due to different contributions of non-flow and longitudinal decorrelations when different kinematic regions in pseudorapidity are used.
To further understand non-flow contamination, calculations from PYTHIA 8 model~\cite{Sjostrand:2014zea} were used, which show that the estimated non-flow effect cannot describe the ALICE data, nor the difference of ALICE with respect to the other measurements~\cite{Aaboud:2018syf,Sirunyan:2019sef}. To improve the understanding of the origin of the measured $NSC(3,2)_{\rm 3-sub}$, several hydrodynamic calculations are presented for comparisons. It is found that the IP-Glasma+ MUSIC+UrQMD~\cite{Mantysaari:2016jaz,Schenke:2020mbo} calculations for Xe--Xe and Pb--Pb collisions reproduce the negative correlation at large multiplicities. This negative sign persists in simulations down to the lowest multiplicities, the change of the sign is not observed even in pp and p--Pb collisions. However, different results are seen in  T$_{\rm{R}}$ENTo+vUSPhydro model calculations~\cite{Sievert:2019zjr},  which predicted sign changes at low-multiplicity ($N_{\rm part}$) for Pb--Pb, Xe--Xe, Ar--Ar and also O--O collisions, although there is no prediction for p--Pb and pp available. For pp collisions, the positive $NSC(3,2)_{\rm 3-sub}$ qualitatively agrees with the iEBE-VISHNU calculations using HIJING initial conditions~\cite{Zhao:2017rgg, Zhao:2020pty}. 
Last but not least, two calculations from the initial state correlation could also reproduce the sign changes in pp collision~\cite{Dusling:2017dqg,Albacete:2017ajt}: one is based on the initial momentum correlations from colour domains~\cite{Dusling:2017dqg} and the other is based on spatial correlations between gluonic hot spots~\cite{Albacete:2017ajt}.

The differences between different experiments and between different theoretical models are not yet fully understood. However, the differences in various model calculations show that the $NSC(3,2)$ observable has a good sensitivity to initial conditions. Thus, the presented ALICE measurements should be very useful to further pin down the uncertainty in the initial state models and significantly improve the overall understanding of the initial conditions in large and small collision systems. This is essential to extract precise information on the properties of the QGP created in heavy-ion collisions at the LHC, as discussed in Sec.~\ref{sec:TG2symmetriccumulants}, and to reveal the origins of the collectivity observed in small collision systems (Sec.~\ref{section:3.3}).

\subsection{Conclusions}

\paragraph{Nuclear shadowing.} $\rm Z^0$ and $\rm W^{\pm}$ boson measurements by ALICE show clear evidence of nuclear shadowing effects at a resolution scale $Q^2 \sim M_{\rm Z,W}^2$ increasing  towards forward rapidity where smaller $x$ values are probed. First measurements of the coherent \jpsi\,photoproduction in \PbPb UPCs provide strong constraints on gluon nPDFs, in particular, a gluon shadowing factor $R_{\rm g} \approx 0.65$ at $x\approx 10^{-3}$ and $Q^2\sim \frac{1}{4}m_{\jpsi}^2$ can be determined from the ALICE data. The first measurement of the $t$-dependence of the the coherent \jpsi\,photoproduction illustrates the potential of $t$-differential measurements for nuclear shadowing constraints in the impact parameter plane.
Coherent $\rho^0$ photoproduction measurements in \PbPb and \XeXe UPCs show an approximately linear dependence of the $\gamma A$ cross section as a function of atomic mass number $A$ revealing an importance of Gribov-Glauber inelastic shadowing effects at LHC energies. ALICE measurements of coherent $\rho^0$ photoproduction in neutron emission classes  confirmed that forward neutron emission can be used to tag different impact parameter ranges in UPCs.

\paragraph{Gluon saturation.} Exclusive \jpsi\ photoproduction cross section off protons, measured by ALICE  in \pPb UPC over 3 orders of magnitude in photon--proton centre-of-mass energy, keeps rising as a power law with energy showing no clear evidence of gluon saturation effects in the proton down to $x \sim 10^{-5}$. On the other hand, suppression of vector meson photoproduction observed by ALICE in \PbPb UPCs can be well described by both calculations with nuclear modified PDFs and by saturation-based models.

\paragraph{Nuclear structure and overlap.} The charged particle multiplicity d$N_{\rm ch}/$d$\eta$ is expected to scale with the number of degrees of freedom and/or the entropy in the initial state. Such scaling fails for the MC Glauber nucleon and constituent quark models, but holds for the T$_{\rm{R}}$ENTo and IP-Glasma models. In this case, the densities in the initial state are assumed to scale as $\sqrt{T_{a}T_{b}}$ and $T_{a}T_{b}$, respectively, where $T_i$ is the nuclear thickness function. The ratio of 2- and 4-particle flow coefficients $v_{n}$ gives direct access to information on the initial eccentricity and its fluctuations. As with the multiplicity measurements, the T$_{\rm{R}}$ENTo and IP-Glasma models offer the most competitive descriptions. On one hand, this cements their role in extracting the $\eta/s$ and $\zeta/s$ QGP transport properties. On the other hand, our measurements of $\rho(v_{\rm n}^{2}),[p_{\rm T}])$ provide further distinguishing power, and significantly favour a small nucleon width for the initial condition, which is consistent with the hadronic nucleus-nucleus cross section study. Such observations have broader implications regarding the structure of a nucleus at high energies in terms of its constituents, and how these constituents combine in a heavy-ion collision to form the QGP. %

\paragraph{Initial state geometry in small and large systems.} The study of correlations between different order flow vectors, especially with the normalised symmetric cumulant $NSC(3,2)$ in large and small collision systems, provides new possibilities to investigate the role of initial state geometry and possible initial momentum correlations. These will eventually help to pin down the origin of the flow signal observed in small systems.

\newpage

\section{Nuclear physics at the LHC: (anti)nuclei formation and hadron--hadron interactions}
\label{ch:NuclPhysLHC}

Besides being at the forefront in the study of the QGP evolution and its properties, in recent years the ALICE experiment has demonstrated to be uniquely sensitive to specific aspects of low-energy nuclear and hadronic physics. The matter--antimatter symmetry that governs hadron production at LHC collision energies allows for the study of production yields, and of the elastic and the inelastic interaction cross sections with the detector materials of light nuclei and antinuclei at the same time~\cite{Acharya:2020cee}. The production of (anti)deuteron, (anti)triton, (anti)\he, (anti)\hefour and even (anti)hypertriton (\hyp) nuclei have been studied in several collision systems~\mbox{\cite{Adam:2015vda,Acharya:2017fvb,ALICE:2017nuf,Acharya:2019rgc,ALICE:2019bnp,Acharya:2020sfy,ALICE:2015oer}}, providing a solid test for microscopic models of (anti)nucleus production and their inelastic interactions. Such capabilities position ALICE as a discovery machine for exotic nuclei.

The residual strong interaction between hadron pairs, which is responsible for the  stability of atomic nuclei, also acts among all hadrons produced in collisions at accelerators. To infer the properties of this residual strong interaction, the ALICE experiment has measured correlations in the momentum space among particle pairs with the femtoscopy technique~\mbox{\cite{Acharya:2018gyz,Acharya:2019ldv,Acharya:2019sms, ALICE:2019eol,Acharya:2019bsa,Acharya:2020oci,Acharya:2019kqn,Acharya:2020asf,ALICE:2021njx,Acharya:2021hwz,ALICE:2021cpv,ALICE:2021cyj}}. 
This method has been applied to several hadron--hadron pairs, including kaon--nucleon pairs, all ground states of hyperons, and to baryon--antibaryons pairs. In particular, the strong interaction has been quantified for several hadron pairs for which it was not previously measured~\cite{Acharya:2019ldv,Acharya:2019sms,Acharya:2019bsa,Acharya:2020oci,Acharya:2019kqn,Acharya:2020asf,ALICE:2021njx,Acharya:2021hwz,ALICE:2021cpv}. 

Information on the strong interaction among hadrons can also be inferred from the measurement of the lifetime and binding energy of hypernuclei. The measurement of the \hyp lifetime achieved by ALICE~\mbox{\cite{ALICE:2015oer, Acharya:2019qcp,ALICE:2022rib}}, which has steadily acquired increasing precision in recent years and represents the most precise measurement ever carried out, provides a value compatible with the free $\Lambda$ lifetime. This has contributed to solving a long-standing puzzle, i.e.\,an observed hypertriton lifetime rather short compared to the free $\Lambda$ lifetime.

This chapter illustrates the main results achieved by ALICE in the field of nuclear physics with and without strangeness, showing how the different observables can be connected to each other. The highlighted connections lay the groundwork for future and more accurate studies which will enable ALICE to contribute with unprecedented precision to nuclear physics within the SU(3) flavour sector.

\subsection{Production of nuclei from small to large collision-system size}\label{ch:nuclei}
A detailed understanding of the formation of light (anti)nuclei and hypernuclei was achieved by investigating their production as a function of the system size. In small collision systems, such as pp or p--Pb, the nucleon emitting source radius is about 1--2~fm~\cite{Acharya:2020dfb}. The size of the source is thus smaller than (or similar to) the size of the (anti)nuclei that are formed from it, as for instance the RMS charge radius of the deuteron is 2.14~fm~\cite{Mohr:2015ccw}. In large collision systems like Pb--Pb the opposite is true as the radius of the homogeneity region of the emitting source (see Sec.~\ref{sec:TG1sizelifetime}) is about 4--9~fm~\cite{Adam:2015vja}. A special role can be attributed to the hypertriton, whose wave function is wider than the fireball produced in central Pb--Pb collisions (\hyp RMS radius $\sqrt{\langle r^{2}_{\mathrm{d}\Lambda} \rangle} \approx 10.6$~fm~\cite{Hildenbrand:2019sgp}). 
 In addition to this, effects of baryon number conservation on the particle production yields are best studied by varying the system size, because the yield of protons per event and unit of rapidity varies from about 0.12 in inelastic pp collisions~\cite{Adam:2015qaa} to about 33 in central Pb--Pb collisions~\cite{Abelev:2013vea}. In practice, such studies are often performed as a function of the event multiplicity expressed in terms of the charged-particle pseudorapidity density (\dnchdeta) and the particle-source radius is found to be proportional to $\langle\dnchdeta\rangle^{1/3}$~\cite{Adam:2015pya} (see Sec.~\ref{sec:TG1sizelifetime}).

\subsubsection{Light-nuclei yield measurements}
\begin{figure}[!hbt]
    \begin{center}
    \includegraphics[width = 0.48\textwidth]{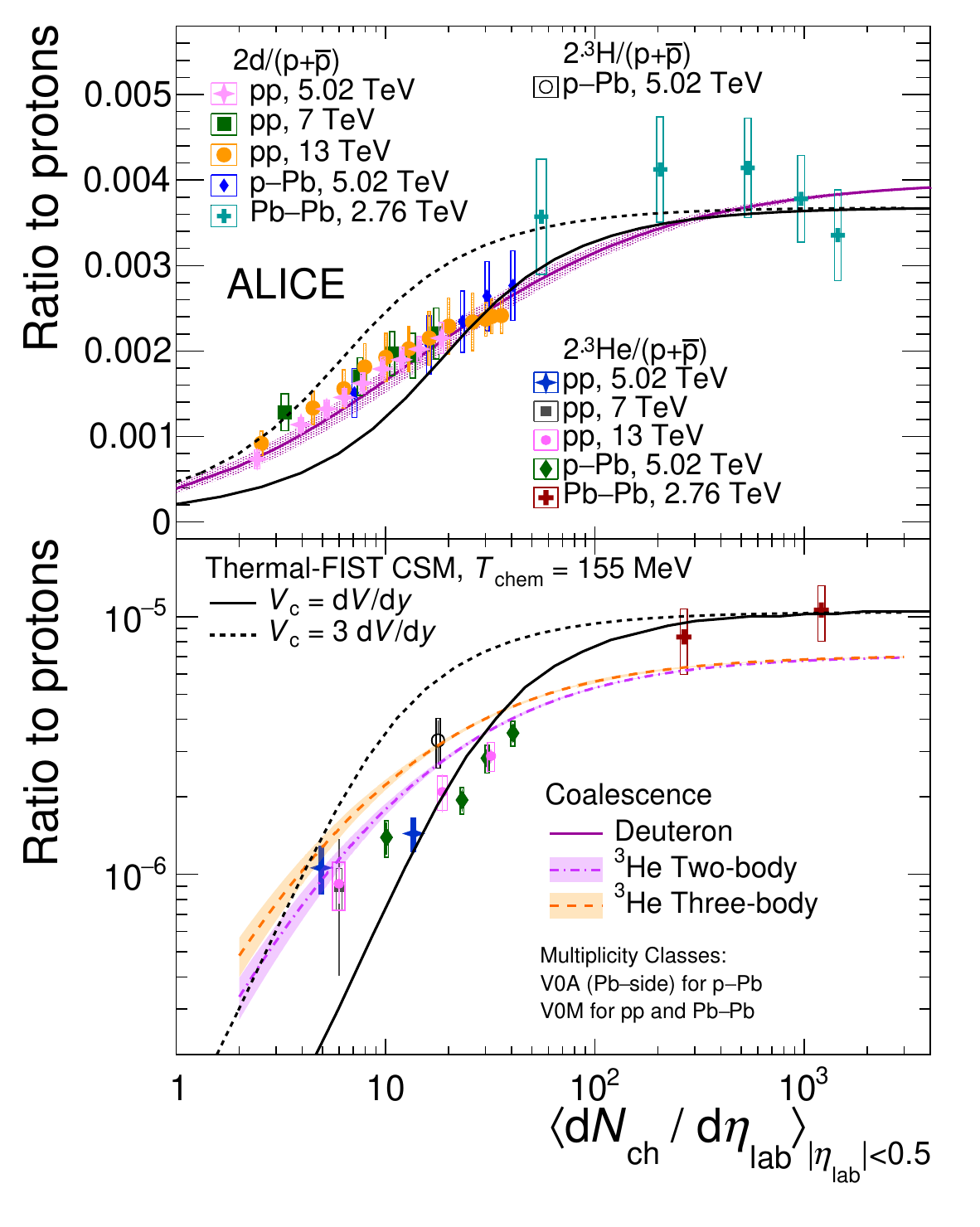}
    \includegraphics[width = 0.48\textwidth]{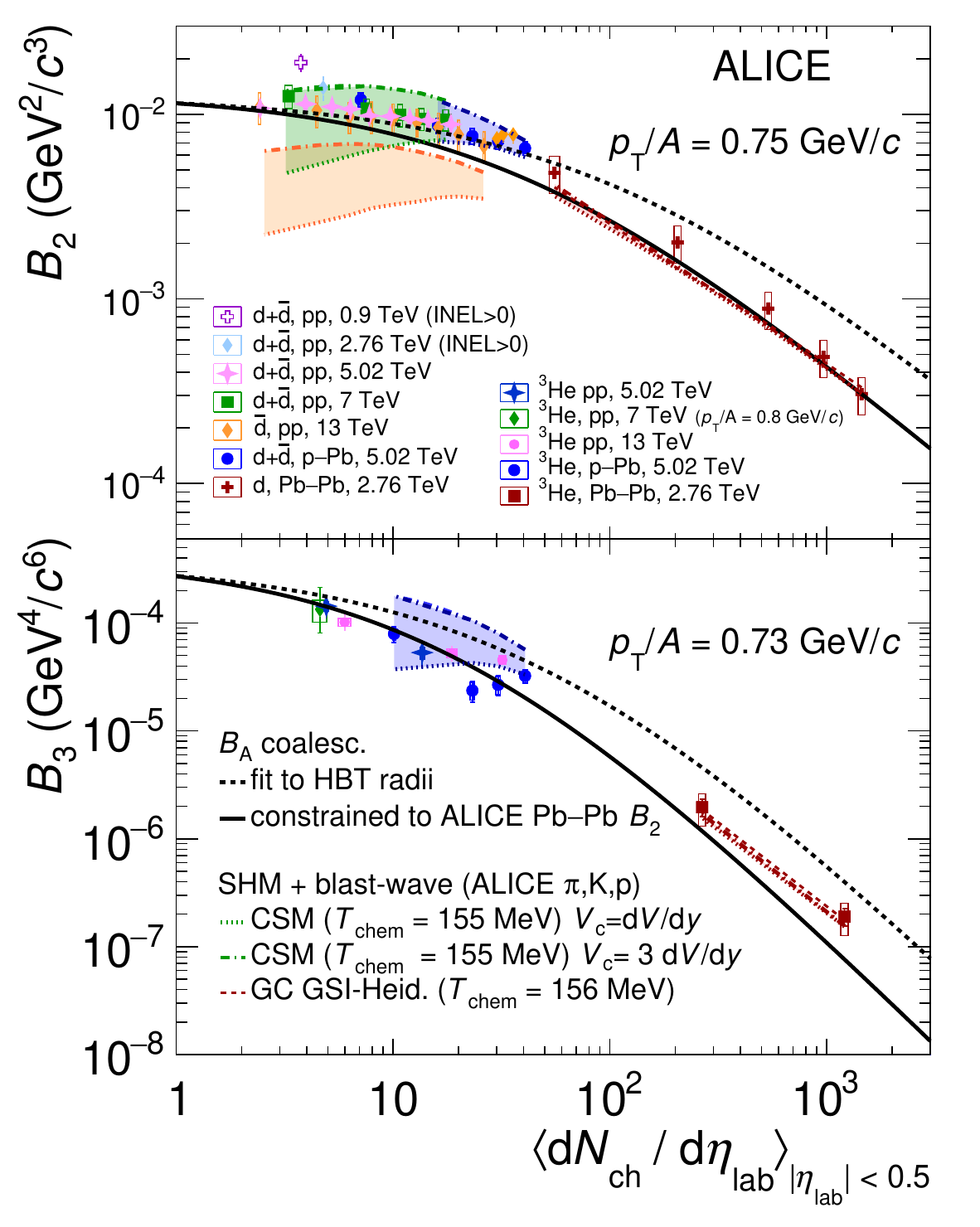}
    
    \end{center}
    \caption{(Left) d/p ratio (Top) and \he/p ratio (Bottom) in pp, p--Pb, and Pb--Pb collisions~\cite{ALICE:2019bnp, Acharya:2019rgc, Acharya:2020sfy, Adam:2015vda, Acharya:2017fvb, Acharya:2019xmu, ALICE:2021ovi, ALICE:2021mfm} as a function of the mean charged-particle pseudorapidity density. %
    (Right) Coalescence parameters $B_2$ (Top) and $B_3$ (Bottom) as a function of the mean charged-particle pseudorapidity density for \pt/$A$~=~0.75~\gmom\ (0.73~\gmom) calculated using the average of particles and antiparticles. 
    In all panels the expectations for the canonical statistical hadronisation model (SHM) evaluated with Thermal-FIST~\cite{Vovchenko:2019pjl} and coalescence approaches~\cite{Vovchenko:2018fiy, Sun:2018mqq, Bellini:2018epz} are shown. For the SHM, two different values of the correlation volume $V_{\rm c}$~\cite{Sun:2018mqq} are displayed. The uncertainties of the coalescence calculations, which are due to the theoretical uncertainties on the emission source radius, are denoted as shaded bands.
    }
\label{fig:NucleiProductionAndBn}
\end{figure}
The measured yields of (anti)nuclei relative to those of protons as a function of $\langle \dnchdeta \rangle$  are summarised in the left panel of Fig.~\ref{fig:NucleiProductionAndBn}. 
Similarly to other light-flavoured hadrons~(see Sec.~\ref{section:3.2}), a smooth trend with $\langle \dnchdeta \rangle$ across different collision systems and independence on the centre-of-mass energy are observed. The nucleus-to-proton ratios exhibit an increasing trend with increasing multiplicity for low-$\langle\dnchdeta \rangle$ and they tend to saturate for $\langle\dnchdeta \rangle\gtrsim$~100. The yield ratios at high-multiplicity are in good agreement with predictions from statistical hadronisation models (SHM) in the grand canonical limit (see Sec.~\ref{sec:SHM}). 
In these models, the production yield per unit of rapidity of a given particle species, ${\rm d}N/{\rm d}y$, is mostly determined by the chemical freeze-out temperature (\Tch)~\cite{Andronic:2017pug} and particle mass $m$ with an approximately-exponential ${\rm d}N/{\rm d}y \propto \exp(-m/\Tch)$ dependence.
The decreasing trend towards smaller multiplicities is also explained in the model as a suppression due to baryon number conservation, i.e.\,a limitation of the available configurations that are compatible with a given canonical ensemble~\cite{Vovchenko:2018fiy}. %
This picture only holds if correlation volumes $V_{\rm c}$ between one and three units of rapidity are allowed. However, these values are in contradiction with those obtained for proton-to-pion ratios~\cite{Vovchenko:2019kes}.

Besides the SHM, the data in the left panel of  Fig.~\ref{fig:NucleiProductionAndBn} are also compared to predictions from the coalescence model, in which nuclei are formed by protons and neutrons that are close in phase space after the kinetic freeze-out~\cite{Scheibl:1998tk}. In this approach, the invariant yield of a (anti)nucleus with mass number $A$ can be written as

\begin{equation}
E_A \frac{\mathrm{d}^3N_A}{\mathrm{d}p_A^3} = B_A \left(E_{\mathrm{p}} \frac{\mathrm{d}^3N_{\mathrm{p}}}{\mathrm{d}p_{\mathrm{p}}^3}\right)^A,
\label{eq:coal}
\end{equation}

\noindent where $p_{\mathrm{p}}$ is the momentum of the nucleon, $p_A = Ap_{\mathrm{p}}$ is the nucleus momentum, and the proportionality factor $B_{A}$ is referred to as the coalescence parameter. The dependence of $B_{A}$ at a given value of \pt/$A$ on the event multiplicity is shown in the right panel of Fig.~\ref{fig:NucleiProductionAndBn} for deuteron and \he. A strong decrease with increasing multiplicity is observed for both $B_2$ and $B_3$ that is explained in the coalescence model by an inverse proportionality to the source volume:  the coalescence probability of two or three nucleons is suppressed if they are largely separated in configuration space. This effect becomes less apparent in small collision systems where the size of the produced nucleus is larger than the source size. %
This effect becomes less apparent in small collision systems where the size of the produced nucleus is larger than the source volume.
In general, the measured yields of light nuclei and their multiplicity dependence across different collision systems can be qualitatively described by the SHM for particle production at the chemical freeze-out and by the coalescence of nucleons at the kinetic freeze-out, even though some open questions still need to be addressed (see Sec.~\ref{section:3.2}).

Due to its wide wave function, hypertriton yield measurements are fundamental to distinguish between the models, because of the relevant role of the size of the produced hypernucleus in the coalescence model predictions~\mbox{\cite{Bellini:2018epz,Sun:2018mqq}}. Recent results on the \hyp/$\Lambda$ yield ratio in \pPb collisions~\cite{ALargeIonColliderExperiment:2021puh} are shown in the left panel of Fig.~\ref{fig:HypandNucleiV2} along with the same measurements in Pb--Pb collisions at \mbox{$\sqrt{s_{\rm NN}}=2.76$~TeV}~\cite{ALICE:2015oer}. The measured ratio excludes with high significance the canonical versions of the SHM with $V_{\rm c} \geq 3\,\mathrm{d} V/\mathrm{d} y$. Nevertheless, it remains to be seen whether advanced versions of the SHM using the S-matrix approach to account for the interactions among hadrons~\cite{Cleymans:2020fsc}  will be able to solve this discrepancy. The \hyp/$\Lambda$ ratio is well described by the two-body coalescence prediction (dark blue band), while the three-body formulation (light blue band) is disfavoured by the measurement.  

The right panel of Fig.~\ref{fig:NucleiProductionAndBn} shows the strong sensitivity of coalescence studies to the source volume, which will be discussed in more detail in Sec.~\ref{sec:sourceSmallCollidigSystems}. The coalescence process and the femtoscopic correlation between nucleons are closely related to each other~\cite{Blum:2019suo,Bellini:2020cbj} and simultaneous measurements of both quantities in future analyses will lead to precise understanding of the formation process of QCD bound states in high-energy particle collisions.

Recently, the measurement of d production fluctuation has been measured~\cite{ALICE:2022amd}. Simple coalescence models fail to fit simultaneously the measurement of the cumulants ratios and the proton deuteron correlation coefficient. The state of-the-art SHM models can describe the data, but with a value of $V_c$ significantly smaller with respect to that that describe p/$\pi$ ratio and cosistent with that obtained from deuteron yield measurement.

\subsubsection{Flow of light nuclei in Pb--Pb collisions}
\begin{figure}[!hbt]

    \begin{minipage}[c]{\textwidth}
    \centering
    \raisebox{-0.5\height}{\includegraphics[height=7.5cm]{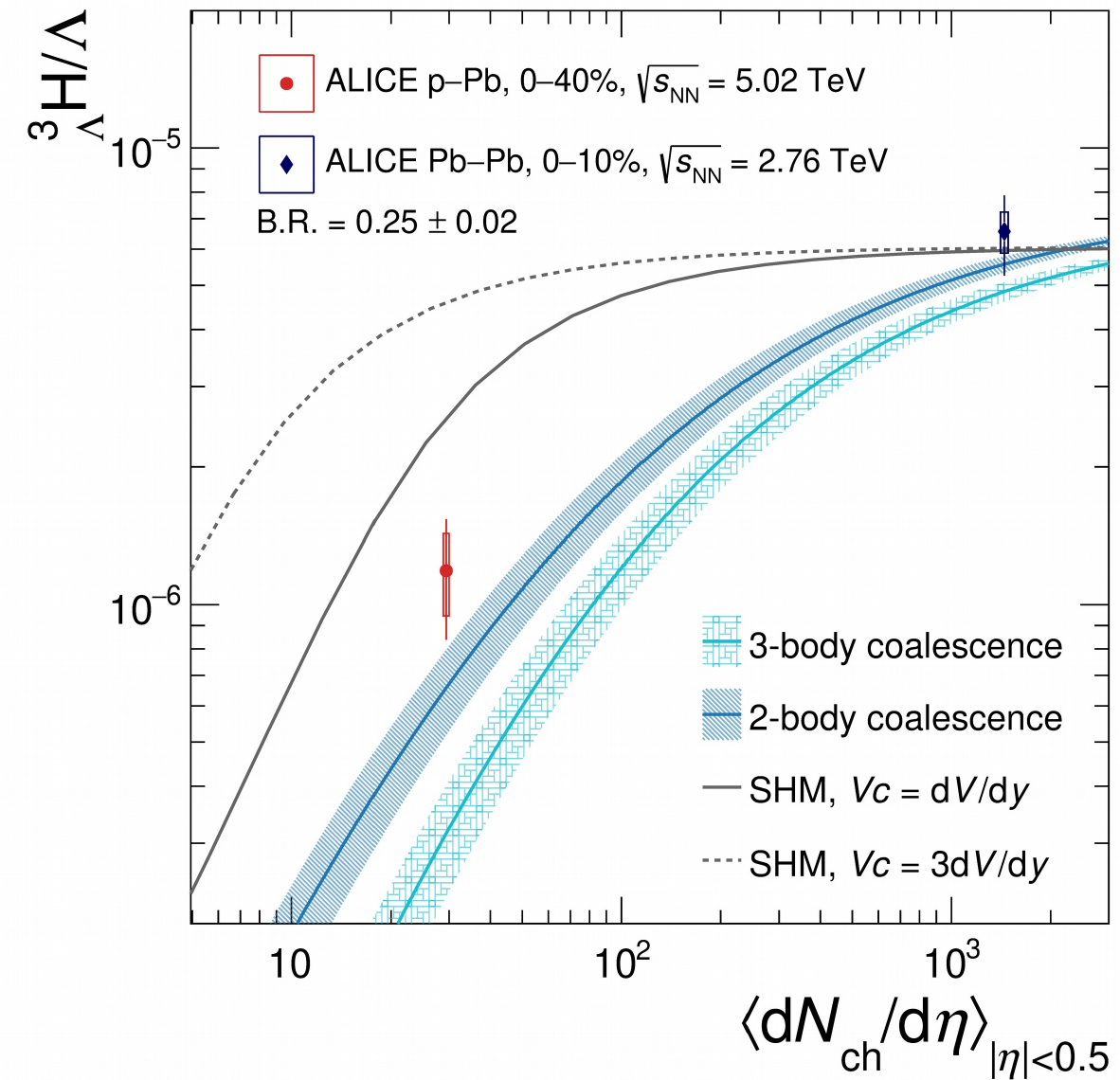}}
    \hspace*{.2in}
    \raisebox{-0.5\height}{\includegraphics[height=8cm]{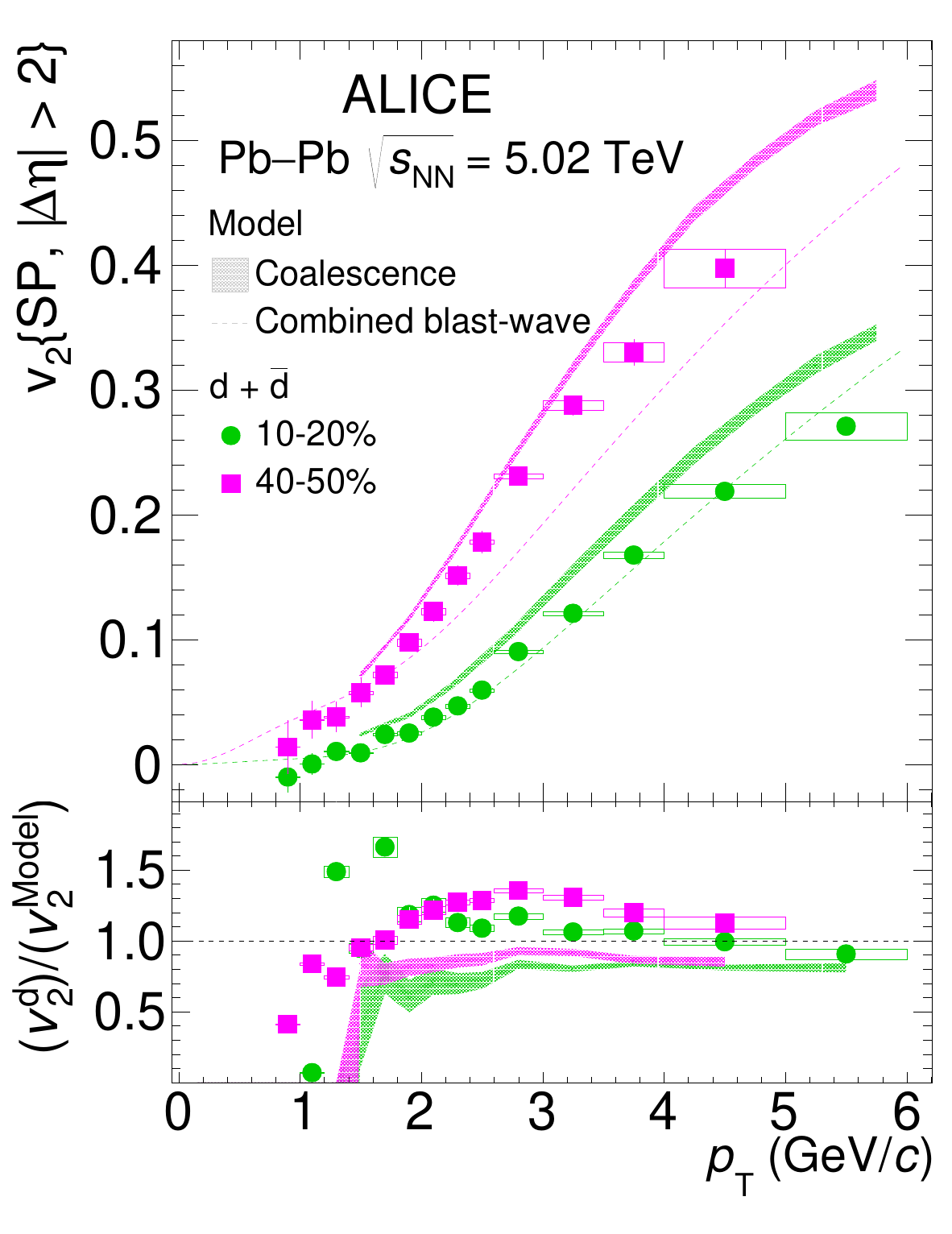}}
    \end{minipage}

    \caption{(Left) $^{3}_\Lambda$H/$\Lambda$ measurements in p--Pb (in red) and Pb--Pb collisions (in blue) as a function of $\langle \dnchdeta \rangle$~\cite{ALargeIonColliderExperiment:2021puh,ALICE:2015oer}. The vertical lines and boxes are the statistical and systematic uncertainties, respectively. The expectations for the SHM and coalescence models are shown as lines and bands, respectively. 
    (Right) (anti)deuteron \vtwo\ as a function of \pt\ for \PbPb at \snn~=~5.02~TeV in two centrality intervals. Results for \mbox{10--20\%} and \mbox{40--50\%} centrality intervals shown as green circles and magenta squares, respectively. Data are compared with the expectations from simple coalescence model and predictions obtained from a blast-wave fit to the \vtwo(\pt) of pions, kaons, and protons~\cite{Acharya:2019yoi}, shown as shaded bands and dashed lines, respectively. 
    In the lower panel, the ratios between the coalescence model predictions are shown as bands, while the ratio to the \mbox{blast-wave} predictions are shown as markers.}
\label{fig:HypandNucleiV2}
\end{figure}

The measurements of flow harmonics can provide additional insights into the 
production mechanism of antinuclei.
The \vtwo\ of (anti)deuteron has been measured in \PbPb collisions at \snn~=~2.76~\cite{ALICE:2017nuf} and \snn~=~5.02~TeV~\cite{Acharya:2020lus}. The measured \vtwo as a function of \pt for two centrality intervals in \PbPb collisions at \snn~=~5.02~TeV is shown in the right panel of Fig.~\ref{fig:HypandNucleiV2}. The \vtwo\ of (anti)deuteron has been evaluated by means of the scalar-product method~\cite{Adler:2002pu,Voloshin:2008dg}, which is a two-particle correlation technique. In the measured \pt\ interval, \vtwo increases with increasing \pt and going from central to more peripheral \PbPb collisions, as expected from the relativistic hydrodynamic description of the collective expansion of a hot and dense medium~\cite{Jeon:2015dfa}. The measurement shows that these effects, related to initial geometrical anisotropy and also observed for most hadron species at LHC energies~\cite{Adam:2016nfo, Acharya:2018zuq}, are also visible for (anti)deuterons.

The elliptic flow of (anti)deuterons is compared with the blast-wave model~{\cite{Schnedermann:1993ws, Adler:2001nb, Siemens:1978pb}}, which is a simplified version of a full relativistic hydrodynamical calculation (see also Sec.~\ref{sec:TG2particlespectra}). 
In the lower right panel, the ratios between the measured (anti)deuteron \vtwo and the one calculated with the blast-wave parameterisation for the two considered centrality intervals are shown as markers. The blast-wave model underestimates the measured (anti)deuteron elliptic flow in semi-peripheral collisions for \pt~$>$~1.4~\gmom, while it describes  the data for central events. 
The (anti)deuteron \vtwo\ is also compared to a coalescence model based on mass-number scaling and isospin symmetry, for which the proton and neutron $v_{2}$ are identical, and the \vtwo of deuteron can thus be calculated from the measured \vtwo of protons.  The width of the band represents the combination of the statistical and systematic uncertainties on the measured proton \vtwo. 
The coalescence model overestimates the (anti)deuteron \vtwo\ by about 20$\%$ to 30$\%$ in central collisions and is closer to the data for semi-peripheral collisions, as illustrated in the lower panel. 

The two simplified models bracket an interval in which the measured deuteron  \vtwo is located and can be used to describe the data in different multiplicity regimes, indicating that currently none of these two models is able to describe the (anti)deuteron anisotropic flow in the low- and high-multiplicity intervals at the same time. %

\subsection{Determination of a universal source for small colliding systems}
\label{sec:sourceSmallCollidigSystems}

\begin{figure}[!ht]
    \centering
    \includegraphics[width=0.49\textwidth]{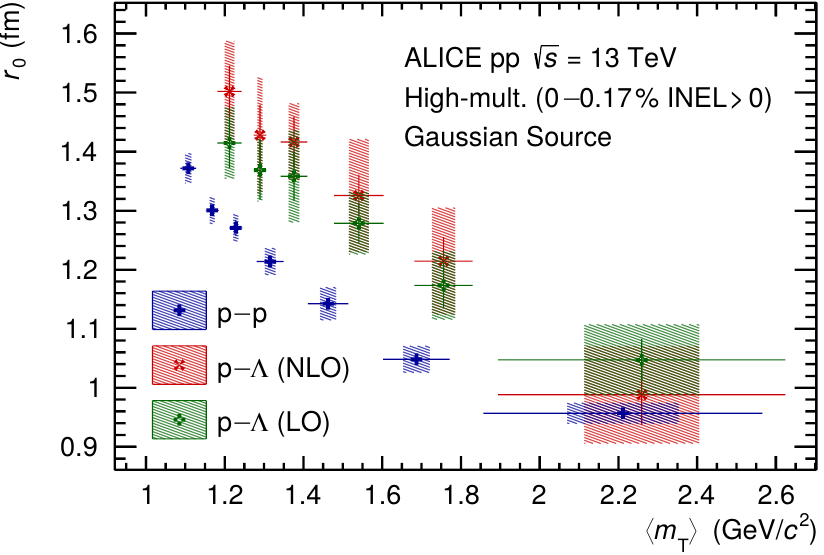}
    \includegraphics[width=0.49\textwidth]{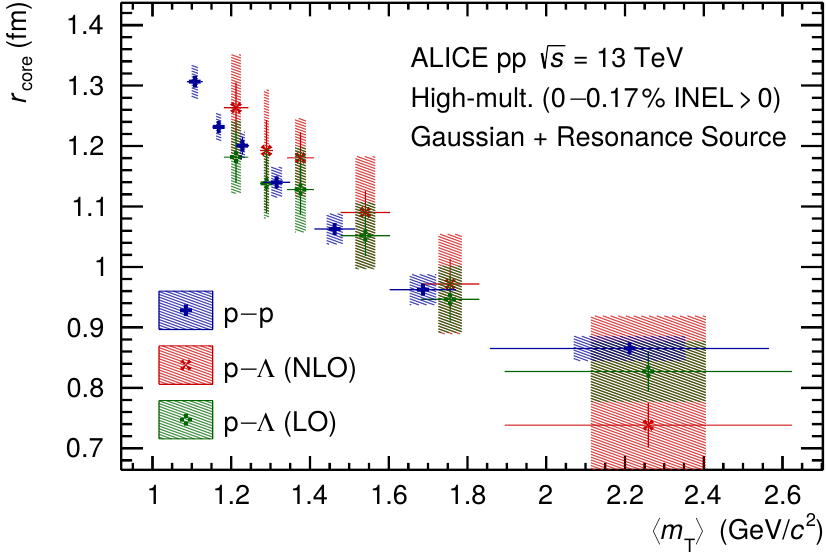}
    \caption{(Left) Source radius $r_\mathrm 0$ as a function of $\langle\mt\rangle$ for the assumption of a purely Gaussian source. (Right) Source radius  $r_{\mathrm{core}}$ as a function of $\langle\mt\rangle$ for the assumption of a Gaussian source with added resonances. The blue crosses result from fitting the \pP correlation function with the strong Argonne $v_{18}$~\cite{Wiringa:1994wb} potential. The green squared crosses (red diagonal crosses) result from fitting the \pL correlation functions with the strong $\chi$EFT LO~\cite{Polinder:2006zh} (NLO~\cite{Haidenbauer:2019boi}) potential. 
    Statistical and systematic uncertainties are indicated by vertical bars and boxes, respectively~\cite{Acharya:2020dfb}.}
    \label{fig:resomT}
\end{figure}

The size of the particle emitting source created in different collision systems plays an important role in the study of (anti)nucleus formation~\cite{Blum:2019suo,Bellini:2020cbj}. It can be measured by employing the femtoscopy techniques (see Sec.~\ref{sec:TG1sizelifetime}). This source size plays also a fundamental role in the study of the residual strong interaction between pairs of hadrons.
The correlation function employed in both kinds of studies reads
\begin{equation}\label{eq:KooninPratt}
C(k^*)=\int S(r^*)\left|\Psi(\vec{k}^*,\vec{r}^*)\right|^2\mathrm{d}^3r^*\,,
\end{equation}
where $r^*$ is the relative distance between the two particles and $k^*$ is half of their relative momentum, both measured in the pair rest frame (*). The source function $S(r^*)$ is, in general, the generic expression for a three dimensional Gaussian source. 
In case of a symmetric source, as considered here, the relevant one-dimensional probability density function of $r^*$ is obtained with a trivial angular phase space factor. The source size $r_0$ obtained from the fit to the correlation function of p--p and p--$\Lambda$ pairs for different $m_\mathrm{T}$ intervals is shown on the left panel of panel of Fig.~\ref{fig:resomT}.  
Theoretical approaches, such as chiral effective field theory ($\chi$EFT)~\cite{Wiringa:1994wb,Polinder:2006zh, Haidenbauer:2019boi} or lattice QCD computations~\cite{Sasaki:2019qnh,Iritani:2018sra}, provide information on the wave function making it possible to compare their predictions directly to the measured $C(k^*)$, given a known source function. %
Small collision systems, pp in particular, are beneficial for study the source function knowing the interaction among particles. It is known that strongly decaying resonances may lead to significant exponential tails of the source distribution, modifying the measured source radii. A new procedure was developed~\cite{Acharya:2020dfb} to quantify these modifications. Within this model, the core radius $r_\mathrm{core}$, which represents the universal Gaussian core source common to all the produced particles, is a free fit parameter. The abundances  of the resonances which modify the source size are taken from the statistical hadronisation model~\cite{Becattini:2009ee} and their branching ratios and lifetimes from the PDG~\cite{Zyla:2020zbs}. The kinematic distributions related to the emission are based on the EPOS transport model~\cite{Pierog:2013ria}. 
For both proton and $\Lambda$ baryons, approximately 2/3 of the total production yield is associated with intermediate resonances with a mean lifetime of 1.7~fm/$c$ and 4.7~fm/$c$, respectively. The fit to the data employing the new source model leads to an identical $m_\mathrm{T}\in(1.1\text{--}2.2)$~GeV/$c^2$ scaling of $r_\mathrm{core}\in(0.85\text{--}1.3)$~fm for p--p and p--$\Lambda$ pairs (right panel of Fig.~\ref{fig:resomT}). This result supports the assumption of a common source parameterisation with $r_\mathrm{core}(m_\mathrm{T})$ for any hadron pair produced in pp collisions. 

\subsection{Hyperon--nucleon and kaon--nucleon interactions}
\subsubsection{Hypertriton lifetime}\label{ch:hypertriton}
The measurement of the properties of hypernuclei gives acccess to the study of interactions among hyperons and nucleons. 
The hypertriton \hyp, for example, is a weakly bound state of a proton, a neutron and a $\Lambda$ hyperon. The binding energy, or more specifically the separation energy of the $\Lambda$ to the pn core, was recently measured by ALICE~\cite{ALICE:2022rib}.
The measurement is consistent with the previous world-average
 \mbox{$B_\Lambda =  (130~\pm~50$)~keV}~\mbox{\cite{Juric:1973zq,Juric:1973zq,Bando:1990yi,Botta:2012xi,Gal:2016boi}}. 
This small separation energy leads to the expectation that the lifetime of the hypertriton is very close to that of the free $\Lambda$ hyperon, $\tau_\Lambda = (263.2 \pm 2.0)$~ps~\cite{Zyla:2020zbs}. Similar conclusions can be expected considering the \hyp\ wave function. 
In a simple quantum-mechanical model, the RMS radius of this hypernucleus is about $\sqrt{\langle r^{2}_{\mathrm{d}\Lambda} \rangle} = 10.6$~fm, if a deuteron-$\Lambda$ bound state is assumed~\cite{Braun-Munzinger:2018hat,Doenigus2020}. Similar values are extracted from more sophisticated theoretical models for the RMS radius of the hypertriton~\cite{Nemura:1999qp,Hildenbrand:2019sgp}. Since this means that the $\Lambda$ is, with a very high probability, several femtometres away from the other nucleons, the lifetime can be expected to be close to the one of the $\Lambda$.

\begin{figure}[!htb]
    \centering
    \hspace{-2.5cm}
    \includegraphics[width=0.8\textwidth]{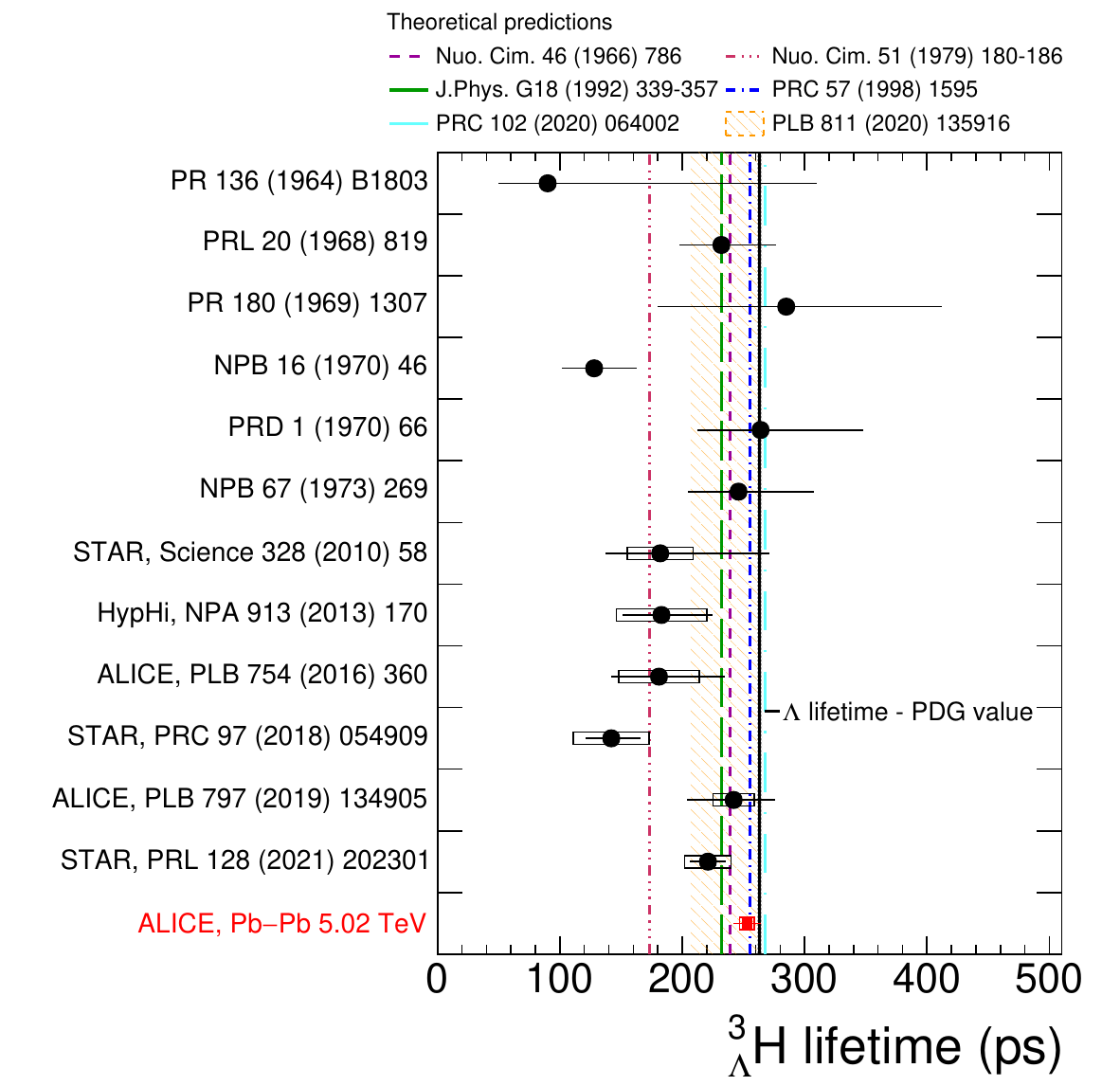}
    \caption{Collection of the \hyp\ lifetime measurements obtained with different experimental techniques. The lowest point corresponds to the latest ALICE measurement~\cite{ALICE:2022rib}.%
    The horizontal lines and boxes are the statistical and systematic uncertainties respectively. The orange band represents the average of the lifetime values and the lines at the edge correspond to 1$\sigma$ uncertainty. The dashed-dotted lines represent five theoretical predictions~\cite{Congleton:1992kk,Kamada:1997rv,Gal:2018bvq,Perez-Obiol:2020qjy}.}
    \label{fig:hypertriton}
\end{figure}

A compilation of the available measurements of the lifetime of hypertriton is displayed in Fig.~\ref{fig:hypertriton}. The data points, shown as markers, were obtained using several experimental techniques and collisions systems. Specifically, emulsions, bubble, and tracking chambers were used to detect \hyp produced using meson beams on different targets~\cite{Prem:1964hyp,Keyes:1968zz,Phillips:1969uy,Bohm:1970se,Keyes:1970ck,Keyes:1974ev} and general purpose detectors were used to detect \hyp produced in heavy-ion collisions at different energies~\cite{Abelev:2010rv,Rappold:2013fic,ALICE:2015oer,Adamczyk:2017buv,Acharya:2019qcp,STAR:2021orx,ALICE:2022rib}.
It should be noted that the data from heavy-ion collisions are characterised by statistical and systematic uncertainties of less than 15\%, smaller by a factor three than bubble chambers experiment results. The early measurements by STAR~\cite{ALICE:2015oer,Adamczyk:2017buv} and ALICE~\cite{Abelev:2010rv} provided values typically lower than the free $\Lambda$ lifetime, which are in contrast with the small value of the $\Lambda$ separation energy. This raised the so-called ``hypertriton lifetime puzzle''. The most recent ALICE measurement (shown as a red marker in Fig.~\ref{fig:hypertriton}), with unprecedented precision, is consistent with the $\Lambda$ lifetime and conclusively settles the puzzle.

Besides the experimental results, the theoretical predictions for the \hyp\ lifetime are reported in Fig.~\ref{fig:hypertriton} for comparison with the data. The calculation performed by Rayet and Dalitz~\cite{Rayet:1966fe} takes into account the phase space factors and the Pauli principle, including also corrections to account for final state pion scatterings and the non-mesonic weak decay channel. 
More recently, Congleton~\cite{Congleton:1992kk} published a prediction for the \hyp\ lifetime, which was obtained using updated values
for nucleon--nucleon and hyperon--nucleon potentials and is expected to be close to one of the free $\Lambda$ hyperon. 
The prediction by Kamada et al.~\cite{Kamada:1997rv} was obtained with a rigorous determination of the hypernucleus wave function and of the three nucleons scattering states, thus finding a value of 256~ps, which is the closest to the free $\Lambda$ lifetime value. Recently, Garcilazo and Gal performed a calculation~\cite{Gal:2018bvq} using the wave function generated by solving three-body Faddeev equations. This calculation is the first one which includes also a final-state interaction effect of the pion. This leads to a reduction of the expected hypertriton lifetime down to 81\% of the free ${\rm{\Lambda}}$ value due to additional attraction from the final-state interaction between the hypertriton decay products.
A more recent approach~\cite{Perez-Obiol:2020qjy} incorporates a hypertriton wave function that is based on chiral effective field theory, distorted waves for the pions and p-wave interactions, while in Ref.~\cite{Gal:2018bvq} only s-waves were included. In addition, in Ref.~\cite{Perez-Obiol:2020qjy} show for the first time that the off-shell $\Sigma \rightarrow {\rm{N}} + \pi$ weak decay 
contribution to the hypertriton decay rate reduces its lifetime by about 10\%. 
Finally, the most recent calculation by Hildenbrand and Hammer~\cite{Hildenbrand:2020kzu} show a value of the hypertriton lifetime very close (about 0.9~$\tau_{\rm{\Lambda}}$) to the one of the free ${\rm{\Lambda}}$. In this theory, a pion-less effective field theory  approach with ${\rm{\Lambda}}$d degrees of freedom and the commonly accepted separation energy of $B_{\rm{\Lambda}}$~=~(130 $\pm$ 50)~keV is used. The dependence of the lifetime on the separation energy was studied:  increasing $B_{\rm{\Lambda}}$ up to 2~MeV, led to a reduction of \hyp\ lifetime down to about 0.9~$\tau_{\rm{\Lambda}}$.

\subsubsection{Nucleon--hyperon and hyperon--hyperon interactions: \texorpdfstring{p--$\Lambda$}{pL}, \texorpdfstring{$\Lambda$--$\Lambda$}{LL} and \texorpdfstring{p--$\Sigma^0$}{psigma}}\label{ch:femto_hyppo}

A direct measurement of the two-body nucleon--hyperon (N--Y) interaction 
was carried out in recent years. The strangeness $|S|=1$ sector, p--$\Lambda$ in particular, is already partially constrained by the measurement of scattering processes~\cite{Alexander:1969cx,SechiZorn:1969hk,Eisele:1971mk}, yet the precision of these data is rather limited.
Theoretical methods allow for the extrapolation of the interaction from the vacuum to large baryonic densities~\cite{Petschauer:2020urh}. Thus, it is of utmost importance to determine the two-body N--Y and three body N--N--Y interactions to study different interacting systems (see Sec.~\ref{ch:neutronstars}).  
The recent measurements of the p--$\Lambda$~\cite{ALICE:2021njx} and \mbox{p--$\Sigma^0$}~\cite{Acharya:2019kqn} interaction in high-multiplicity pp collisions at $\s$~=~13~TeV by ALICE provided new and valuable input to the study of both channels and their strong coupling  N$\Sigma\leftrightarrow$N$\Lambda$.

\begin{figure}[!hbt]
    \centering
    \includegraphics[width = 0.8\textwidth]{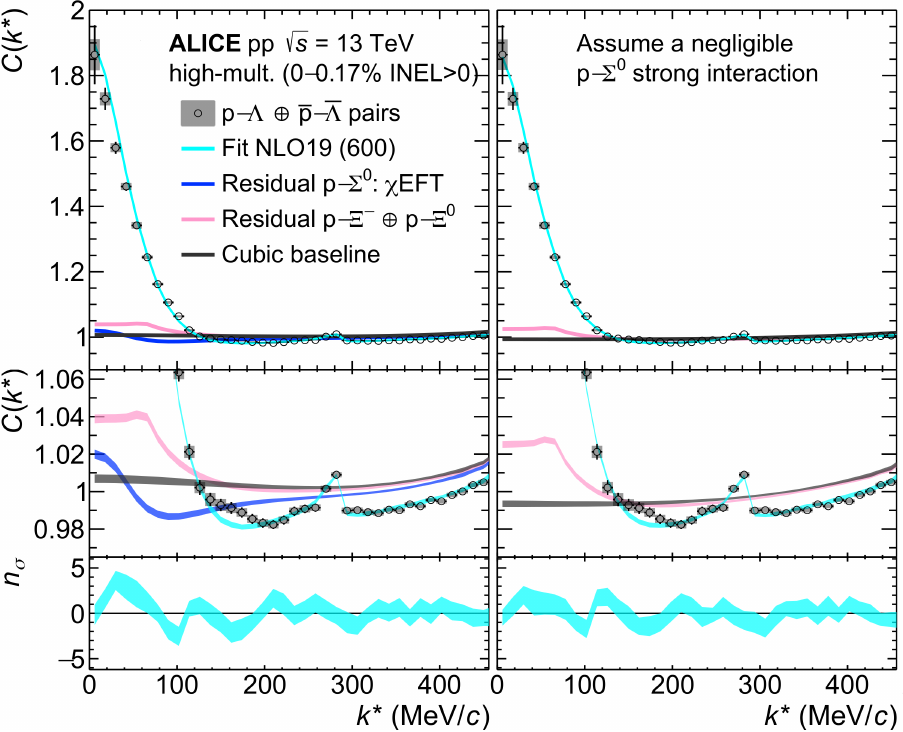}
    \caption{p--$\Lambda$ correlation function measured in pp collisions at $\sqrt{s}$~=~13~TeV~\cite{ALICE:2021njx} compared to theoretical predictions obtained using the NLO19 version of $\chi$EFT. On the right fit is performed assuming a residual p--$\Sigma^0$ interaction as predicted by \chiEFT, while on the right a negligible interaction is assumed. The middle panels show a zoomed version along the $y$-axis, to further investigate the kinematic cusp, around 290~MeV/$c$ corresponding to the N$\Sigma\leftrightarrow$N$\Lambda$ coupling. The bottoms panel show the discrepancy of model to the data. %
    }
\label{fig:pLambda}
\end{figure}

The measurement for the p--$\Lambda$ system, shown in Fig.~\ref{fig:pLambda}, represents the experimental result with highest statistical precision at low relative momentum $k^*$, about a factor of 25 better with respect to the available scattering-experiment results. Moreover, this measurement provided the first direct observation of the kinematic cusp at $k^*\approx 290$~MeV/$c$, corresponding to the N$\Sigma$ threshold. The sensitivity of the data at $k^*<100$~MeV/$c$ is sufficient to probe the correlation signal due to feed-down of 
\mbox{p--$\Xi^{-}\rightarrow$p--$\Lambda$} and \mbox{p--$\Sigma^0\rightarrow$p--$\Lambda$}. The former is modelled using the results from lattice calculations by the HAL QCD collaboration, which are in agreement with ALICE data (see Sec.~\ref{ch:HeavyHyperons}). Due to the coupling $\mathrm{N}\Sigma\leftrightarrow\mathrm{N}\Lambda$, the p--$\Sigma^0\rightarrow$p--$\Lambda$ has to be treated within the same theoretical framework as p--$\Lambda$. Thus, the prediction of $\chi$EFT has been employed.
The fit to the data using the $\chi$EFT predictions is shown in the left panel of Fig.~\ref{fig:pLambda} where the bottom panel represents the deviation between the theory and the data in each $k^*$ interval. The kinematic cusp is well described. However, there is a systematic deviation related to the slope for $k^*<100$~MeV/$c$. The right panel of Fig.~\ref{fig:pLambda} shows the fit result obtained by modifying the p--$\Sigma^0$ contribution to a flat function, mimicking a very weak interaction.
A similar effect can be achieved by slightly reducing the strength of the p--$\Lambda$ attraction. The main conclusion is that the current data call for more precise theoretical calculations.

The N--$\Sigma$ interaction was also directly probed employing correlation studies. The pioneering measurement by ALICE in high-multiplicity \pp collisions~\cite{Acharya:2019kqn} analysed the decay channel $\Sigma^0 \rightarrow \Lambda \gamma$ and the subsequent decay of the $\Lambda$ to p$\pi^-$. The soft photon from the decay is detected via conversions to an electron--positron pair in the ALICE detector material. The resulting purity of the $\Sigma^0$ sample is 34.6\%, thus rather low, and it is owed to the dominant contribution from uncorrelated primordial $\Lambda \gamma$ combinations. The data were compared to state-of-the-art theoretical descriptions of the interaction, such as \fss~\cite{fss2}, two versions of the soft-core Nijmegen models (\ESC~\cite{Nagels:2015lfa}, \NSC ~\cite{NSC97}), and results of \chiEFT at Next-to-Leading Order (NLO)~\cite{Haidenbauer:2013oca}. The current precision of the experimental data does not yet allow for a discrimination among the different models. Nevertheless, these studies demonstrate the feasibility of the measurement and the prospects of the analysis of the larger data samples to be collected during the ongoing LHC Run~3~\cite{ALICE-PUBLIC-2020-005}.

\begin{sloppypar}
The $\Lambda$--$\Lambda$ interaction was  investigated by directly measuring the correlation among the hyperons and by measuring the $\phi$--N correlation. The direct measurement of the $\Lambda$--$\Lambda$ correlation in pp collisions at $\sqrt{s}$~=~13~TeV and p--Pb collisions at $\sqrt{s_{\mathrm{NN}}}$~=~5.02~TeV evidenced a shallow attractive strong interaction potential but could not exclude the presence of a $\Lambda$--$\Lambda$ bound state~\cite{ALICE:2019eol}. The most precise upper limit for the binding energy of this state was extracted and found to be equal to \mbox{$B_{\Lambda\Lambda}= \,[ 3.2\,^{+1.6}_{-2.4}(\mathrm{stat})^{+1.8}_{-1.0}(\mathrm{syst})]$~MeV}.
An alternative method to study the hyperon--hyperon interaction is to measure the coupling between the $\phi$ meson and either nucleons or hyperons~\cite{WEISSENBORN2013421}. The first measurement of the p--$\phi$ correlation in pp collisions at $\sqrt{s}$~=~13~TeV was also carried out and the \Ledn\ fit~\cite{Lednicky:1981su} was used to extract the scattering parameters~\cite{ALICE:2021cpv}. The small scattering parameters indicate a rather shallow p--$\phi$ interaction potential, which is consistent with the observed shallow $\Lambda$--$\Lambda$ interaction potential.
\end{sloppypar}

\subsubsection{Multistrange hyperon--nucleon interactions}\label{ch:HeavyHyperons}

The strangeness $|S| = 2$ and $|S| = 3$ sectors have been also recently studied using phenomenological approaches~\cite{Nagels:2015lfa,Sekihara:2018tsb}
and lattice QCD calculations~\cite{Sasaki:2019qnh,Iritani:2018sra}.
Experimentally little is known about the interaction of $\Xi$ and $\Omega$ baryons with nucleons.
The practical difficulties of creating a beam of such particles prevent the construction of a reliable data base of scattering data, hence the experimental information relies on data on $\Xi$ hypernuclei~\cite{Nakazawa:2015joa,Nagae:2017slp} and on  a pioneer study of p--$\Omega$ two-particle correlation by STAR collaboration using Au--Au interactions~\cite{STAR:2018uho}. Data on $\Xi$ hypernuclei indicate the possibility that $\Xi$ baryons can be bound in a nucleus; data from p--$\Omega$ tend to favour the existence of a p--$\Omega$ bound state. 

\begin{figure}[ht]
\centering
    \includegraphics[width = 0.6\textwidth]{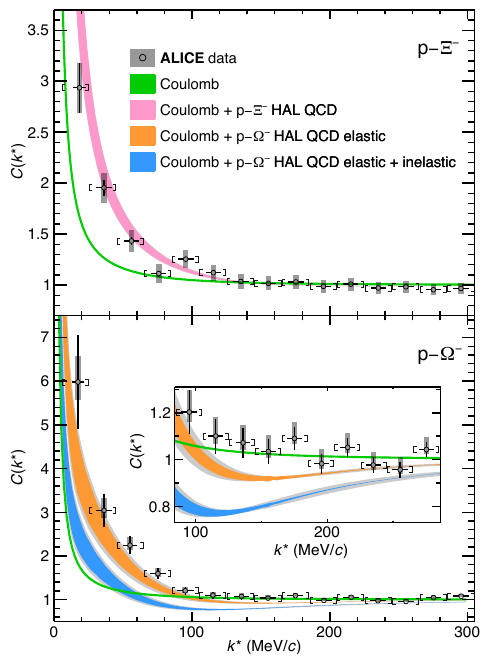}
\caption{%
Measured \pXim (Top) and \pOm (Bottom) momentum correlation functions in \pp collisions at \mbox{$\sqrt{s}=13$~Te\kern-.1emV\xspace}~\cite{Acharya:2020asf}. The experimental data are shown as black symbols. The black vertical bars and the grey boxes represent the statistical and systematic uncertainties, and the square brackets show the $k^*$ bin width. The measurements are compared with theoretical predictions, shown as coloured bands, that assume either Coulomb or Coulomb+strong \HALQCD interactions. The width of the curves including \HALQCD predictions represents the uncertainty associated to the calculation and the grey shaded band represents the uncertainties associated with the determination of the source radius.
} 
 \label{fig:pXipOmega}
\end{figure}

The ALICE collaboration has measured with high precision the p--$\Xi$ correlation function in p--Pb collisions at \snn~=~5.02~TeV~\cite{Acharya:2019sms} and the p--$\Xi$ and p--$\Omega$ correlation functions in pp collisions at $\sqrt{s}$~=~13~TeV using high-multiplicity events~\cite{Acharya:2020asf}, demonstrating the attractive character of both interactions.
The correlation function obtained by ALICE~\cite{Acharya:2020asf} for p--$\Xi^- \oplus \mathrm{\overline{p}}$--$\overline{\Xi}^+$ and p--$\Omega^- \oplus \mathrm{\overline{p}}$--$\overline{\Omega}^+$ using high-multiplicity pp collisions at $\sqrt{s}$~=~13~TeV are shown in Fig.~\ref{fig:pXipOmega}. 
The data are corrected for experimental artefacts and \mbox{feed-down} contributions, displaying the genuine correlation function of the pairs of interest and it can be directly compared to the theoretical  predictions for a given source.

The data shown in Fig.~\ref{fig:pXipOmega} are compared with the correlation function corresponding to the Coulomb-only expectation (green curves) and with predictions that consider also the strong interaction calculated on the lattice by the HAL QCD collaboration~\cite{Sasaki:2019qnh,Iritani:2018sra}.
The lattice QCD calculations predict an attractive interaction for both the \mbox{p--$\Xi$} and the \mbox{p--$\Omega$}  systems. The results obtined in \pp confirms what was observed in the analysis of p--Pb collisions~\cite{Acharya:2019sms}. 
In the \mbox{p--$\Omega$} system, the interaction is attractive at all distances due to the absence of a repulsive core~\cite{Sasaki:2019qnh}. The attractive core can accommodate the presence of a \mbox{p--$\Omega$} bound state with a binding energy of 1.54~MeV~\cite{Iritani:2018sra}. %
The presence of the bound state has a direct effect on the expected correlation function causing a depletion that, for the given radius, reduces the correlation function to a value lower than unity in the $k^*$ range between 100~MeV/$\it{c}$ and 200~MeV/$\it{c}$, as highlighted in the inset in the right panel of Fig.~\ref{fig:pXipOmega}. 
The blue and orange coloured bands corresponding to the HAL QCD prediction for the p--$\Omega$ system in Fig.~\ref{fig:pXipOmega} reflect the current uncertainty of the calculations due to the presence of coupled-channels in the p--$\Omega$ system.

In the case of the \mbox{p--$\Omega$} the lattice QCD predictions lie below the experimental data, and the depletion characteristic of the presence of a bound state is not supported by the ALICE data.
In order to obtain firm conclusions on the possible existence of the p--$\Omega$ bound state, and, if existent, quantify experimentally its binding energy, a differential analysis of the p--$\Omega$ correlations in systems with slightly different source sizes is necessary.
A recent measurement of the $\Lambda\mbox{--}\Xi^{-}$ correlation function in high-multiplicity pp collisions at $\sqrt{s}$~=~13~TeV supports a very small coupling strength for this channel~\cite{ALICE:2022uso} and hence the inelastic contribution to the p--$\Omega$ state should be rather small.
More precise measurement of both the $\Lambda\mbox{--}\Xi^{-}$ and $\Sigma^{0}\mbox{--}\Xi^{-}$ correlations will further reduce the uncertainties of the calculations.
Such measurements will be performed at the LHC by ALICE in pp, \pPb, and peripheral Pb--Pb collisions.

\subsubsection{K--p interaction: evidence of coupled-channel contributions}

Coupled-channel processes are widely present in hadron--hadron interactions whenever pairs of particles, relatively close in mass, share the same quantum numbers: baryonic charge, electric charge and strangeness. The coupling translates into on/off-shell processes from one system to the other. 
The multi-channel dynamics deeply modifies the hadron--hadron interaction and is at the origin of several phenomena, such as bound states and resonances, which crucially depend on the coupling between the inelastic channels. A key example can be found in the $\Lambda(1405)$ resonance, a molecular state arising from the coupling of antikaon-nucleon ($\mathrm{\overline K}$--N) to $\Sigma$--$\pi$~\cite{Lambda1405_lat,Miyahara:2015bya}.

\begin{figure}[!ht]
    \centering
    \includegraphics[width=1\textwidth]{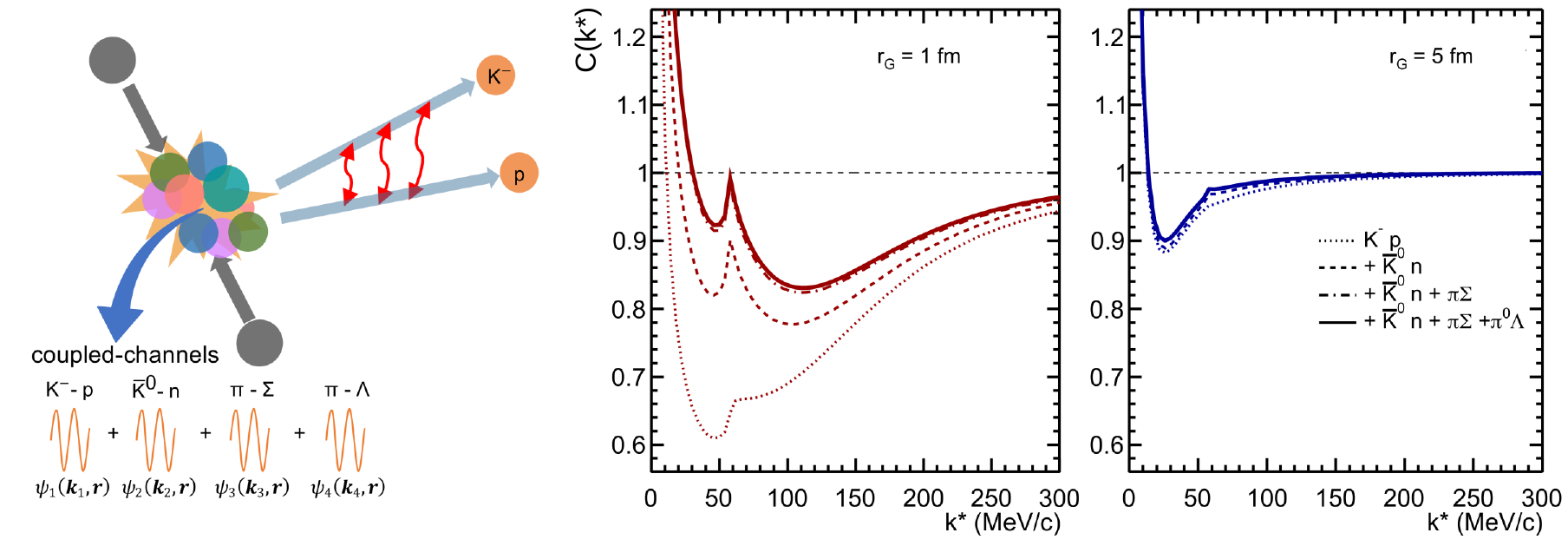}
    \caption{%
    (Left) Sketch of the system configuration in femtoscopic measurements, where only the final $\mathrm{K^-}$--p channel is measured. (Middle and right) Theoretical correlation function for $\mathrm{K^-}$--p, from the pure elastic term (dotted line) to the full \CF (solid line) with all coupled-channels ($\mathrm{\overline{K}^0}$--n, $\pi$--$\Sigma$, $\pi$--$\Lambda$) included. The results are shown for two different radii, typically achieved in pp collisions (1~fm) and in heavy-ion collisions (5~fm). %
    }
    \label{fig:FemtoBox_CC}
\end{figure}

The advantage of the femtoscopic measurement with respect to scattering experiments for the study of coupled channels is that the final state is fixed by the measured particle pair. Hence, the corresponding correlation function represents an inclusive quantity sensitive to all the available initial inelastic channels produced in the  collision~\cite{Haidenbauer:2018jvl,Kamiya:2019uiw}.

Estimates for the weights to be assigned to the different initial channels can be obtained using combined information on yields from statistical hadronisation models~\cite{Vovchenko:2019pjl,Becattini:2001fg, Wheaton:2011rw} and on the kinematics of the produced pairs from transport models~\cite{Pierog:2013ria}. 
A coupled-channel contribution can modify the \CF in two different ways, depending on  whether the minimum energy at which they can be produced occurs below or above the reduced mass of the pair. Inelastic channels opening below threshold do not introduce any shape modification on the \CF, acting as an effective attraction and increasing the signal of the correlation function. On the other hand, channels appearing above threshold lead to a modification of the \CF~in the vicinity of the opening, which is typically translated into a cusp-like structure, whose height is driven by the coupling strength. 
These two main differences are illustrated in Fig.~\ref{fig:FemtoBox_CC} for the $\rm K^-$--p system.   

The K--p system contains couplings to several inelastic channels below threshold such as $\pi$--$\Lambda$, $\pi$--$\Sigma$ and, due to the breaking of isospin symmetry, to charge-conjugated $\rm\overline{K}^0$--n at roughly 4~MeV above threshold. %
In the left panel of Fig.~\ref{fig:FemtoBox_CC}, a schematic representation of the collision is shown.
The correlation of K--p pairs composing the final-state is measured and its decomposition into contributions of different channels contributions is shown for two different source sizes in middle and right panel, respectively. The largest contributions on the \CF\ from coupled-channels occur for a small emitting source with Gaussian radius $\mathrm{r_G}$~=~1~fm as shown in middle panel.  
The \CF\ signal increases as the inelastic contributions are added and the cusp structure, visible when the ${\rm K^0}$--n channel is explicitly added, indicates the opening of this channel above threshold. 
For both source radii, this structure already appears when the mass difference between \kam and $\rm \overline K ^0$ is considered, and it is present also in the elastic $\rm K^-$--p$\rightarrow$ $\rm K^-$--p contribution (dotted line).
The effect of coupled-channels is suppressed when the source size is increased up to $\mathrm{r_G}$~=~5~fm, as in central heavy-ion collisions (right panel in Fig.~\ref{fig:FemtoBox_CC}).

\begin{figure}[!t]
    \centering
    \includegraphics[width=1\textwidth]{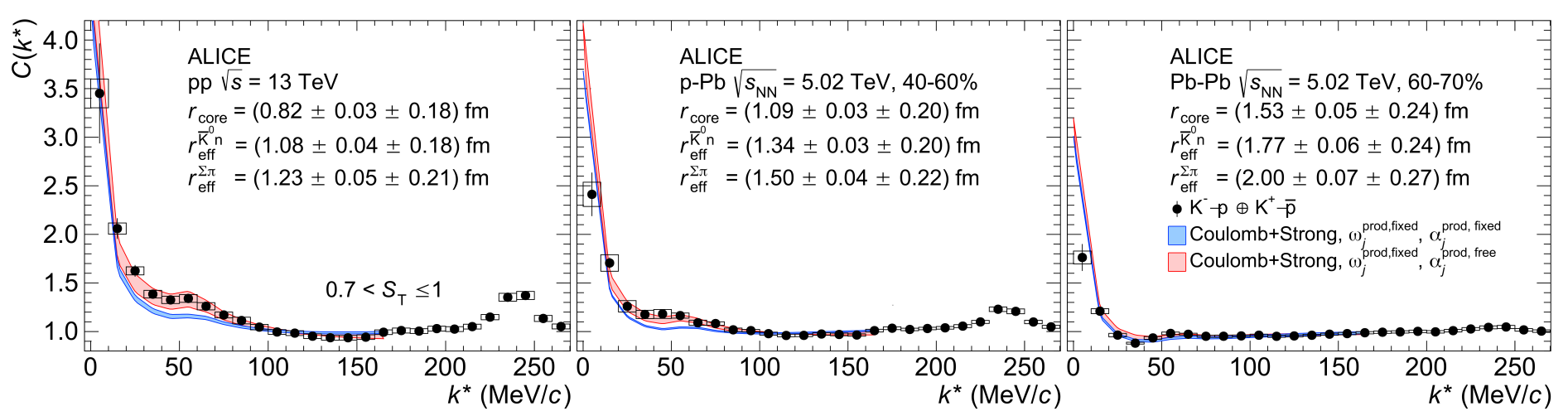}
    \caption{%
    Measured correlation function versus the relative momentum $\ks$ for \mbox{$\kam$--p $\oplus$ $\kap$--$\pbar$} pairs in pp collisions at $\sqrt{s}$~=~13~TeV (left),  in \mbox{40--60\%} centrality interval in \pPb collisions at \snn~=~5.02~TeV (middle) and in \mbox{60--70\%} centrality Pb--Pb collisions at \snn~=~5.02~TeV (right)~\cite{ALICE:2022yyh}. In all the panels, data are compared with the \chiEFT-based potential~\cite{Kamiya:2019uiw} with fixed (blue bands) and free (red bands) coupling weights.
    } %
    \label{fig:Kmp_CC_CollSyst}
\end{figure}

This theoretical scenario and the extreme sensitivity of the correlation function to coupled-channel dynamics have been confirmed by the measurements of the ${\rm K^-}$--p correlation function, measured in different colliding systems, shown in Fig.~\ref{fig:Kmp_CC_CollSyst}. From left to right, the size of the emitting source increases as the correlation function is measured respectively in pp collisions~\cite{Acharya:2019bsa} and Pb--Pb collisions~\cite{Acharya:2021hwz} and the shape of the measured correlation function varies significantly. The cusp signature at $k^*=58$~\MeVc is not present in large systems. As the source size increases, the enhancement at low momenta due to coupled-channel contributions follows the same trend, becoming less pronounced. 
In large collision systems such as Pb--Pb, the asymptotic part of the wave function is probed, hence the effects of coupled-channels on \CF are noticeably suppressed and a partial direct access to the elastic term can be obtained. This leads to a good agreement between the scattering length $f_0$ obtained fitting the Pb--Pb data and the values extracted from kaonic atom measurements, in which coupled-channel effects are typically not considered~\cite{Bazzi:2011zj}.
The small colliding systems on the other hand provide the opportunity to test the coupling strength of different initial channels to the final channel ${\rm K^-}$--p with unprecedented precision~\cite{ALICE:2022yyh}.

Coupled-channel studies have been recently extended to the charm sector by studying the p--D$^-$ system in high-multiplicity pp collisions~\cite{ALICE:2022amd}. This study suggests the possibility of the formation of a N--D bound state, but the current statistical significance of the data is not enough to draw a clear conclusion. Precise measurements of correlation functions between charm and light-flavour or strange hadrons will become accessible with the Run 3 data samples. 

\subsection{Baryon--antibaryon interactions in Pb--Pb collisions: possible existence of bound states}

\begin{figure}[!b]
\centering
\includegraphics[width=0.6\textwidth]{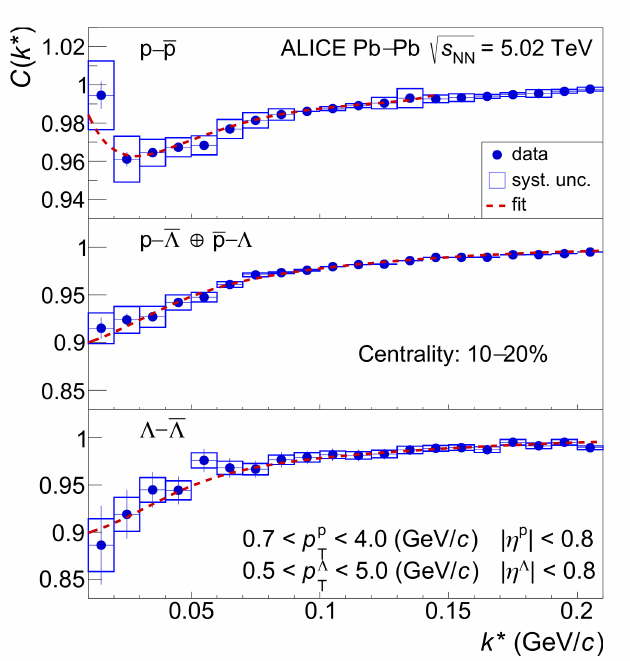}
\caption{Example of correlation functions (points) of $\rm p$--$\rm \overline{p}$, $\rm p$--$\overline{\Lambda}\oplus\overline{\rm p}$--$\Lambda$, and $\Lambda$--$\overline{\Lambda}$ pair for the 10--20\% centrality class~\cite{Acharya:2019ldv}. Dashed lines show a part of the global fit, performed simultaneously to correlation functions of all three pair types in 6 centrality classes.}
\label{fig:data_fit_bab1}
\end{figure}

\begin{figure}[!ht]
  \centering
  \includegraphics[width=0.7\textwidth]{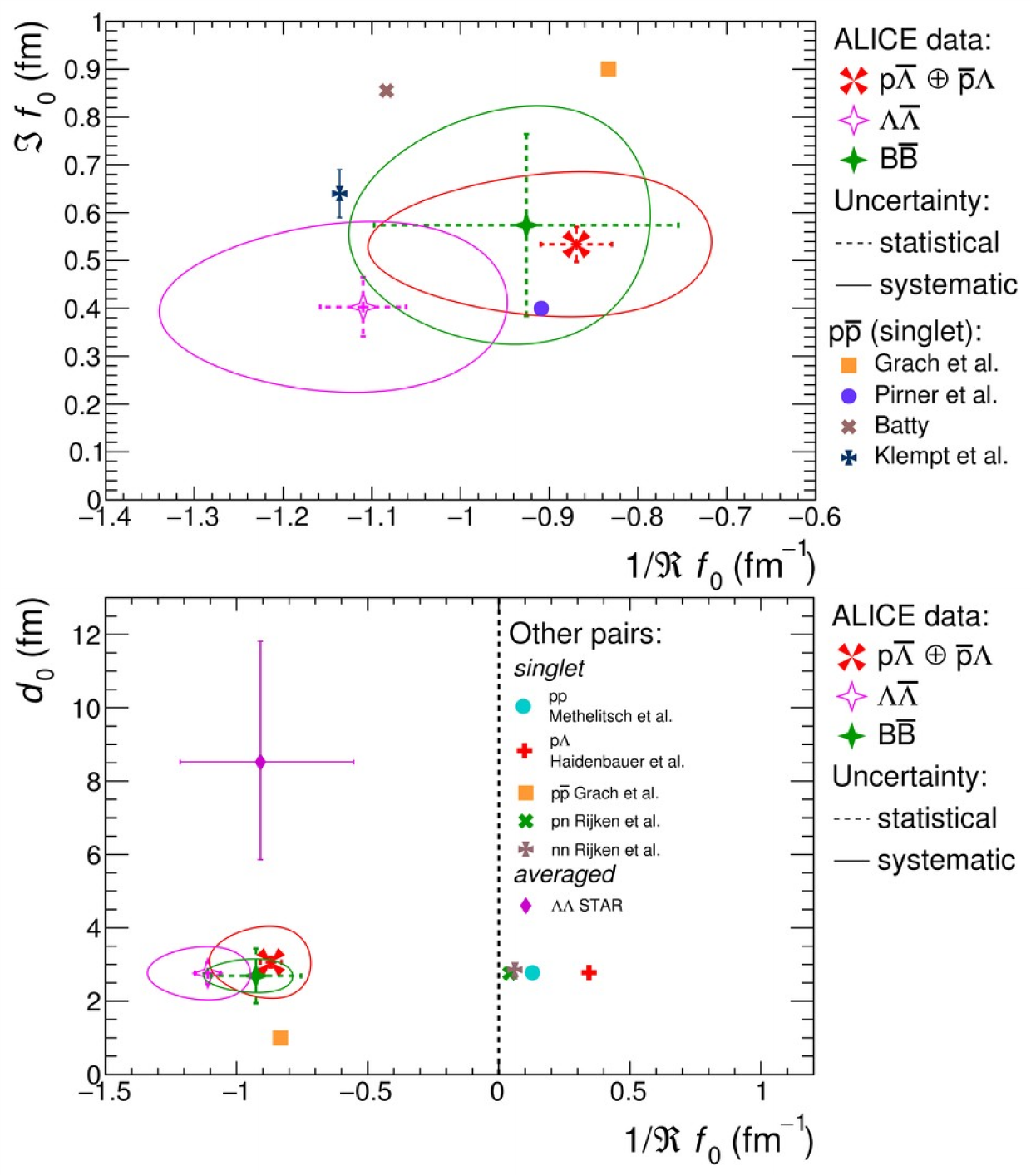}
\caption{Comparison of extracted spin-averaged scattering parameters $\Im f_{0}$ and $\Re f_{0}$ (Top) and $d_{0}$ and $\Re f_{0}$ (Bottom) for $\rm p$--$\overline{\Lambda}\oplus \overline{\mathrm{p}}$--$\Lambda$, $\Lambda$--$\overline{\Lambda}$ pairs and effective parameters accounting for the contribution of heavier $\rm B$--$\rm\overline{B}$ pairs~\cite{Acharya:2019ldv}. Results are compared with theoretical models~\cite{Grach:1987rp,Pirner:1991mn, Klempt:2002ap, Batty:1989gg, Mathelitsch:1984hq,  Rijken:2010zzb, Haidenbauer:2013oca} and with results by the STAR collaboration~\cite{Adamczyk:2014vca}.}
\label{fig:data_fit_bab2}
\end{figure}

The study of the strong interaction in the baryon--antibaryon sector is very challenging, because obtaining beams or targets of antimatter is very difficult. Hence, very little is known about the  baryon--antibaryon interaction. 
From proton--antiproton scattering it is known that the formation of protonium (or antiprotonic hydrogen) occurs~\cite{Batty:1989gg,Klempt:2002ap}. Its 1S and 2P states are  
of particular interest as there is evidence of a contribution from the strong force. Nevertheless, the nature of protonium, whether it can be considered a nuclear bound state or a result of the Coulomb interaction, remains an open question.
Much less is known for baryon--antibaryon pairs with non-zero strangeness, because only few experimental measurements from scattering experiments exist. 

 Two-particle momentum correlations have been also employed to study the strong interaction in the baryon--antibaryon sector
in ultrarelativistic pp and Pb--Pb collisions at the LHC, where approximately the same amount of baryons and antibaryons are produced~\cite{ALICE:2013yba}. 

Correlation functions for $\rm p$--$\rm \overline{p}$, $\rm p$--$\rm \overline{\Lambda}\oplus\overline{p}$--$\Lambda$ and $\Lambda$--$\overline{\Lambda}$ were measured in \mbox{Pb--Pb} collisions at energies of $\sqrt{s_{\rm NN}}=2.76$~TeV and $\sqrt{s_{\rm NN}}=5.02$~TeV~\cite{Acharya:2019ldv} and in pp collisions at $\sqrt{s}=13$~TeV~\cite{ALICE:2021cyj}.
The analysis in \mbox{Pb--Pb} was performed in several centrality intervals; 
an example for the 10--20\% centrality interval is reported in Fig.~\ref{fig:data_fit_bab1}. 
A simultaneous fit of all the correlation functions was performed to extract the scattering parameters of the strong interaction. 
The spin-averaged scattering parameters, i.e.\,$\Re f_0$ the real and $\Im f_0$ imaginary parts of the spin-averaged scattering length, and $d_0$ for the real part of the spin-averaged effective range of the interaction for $\rm \overline{p}$--$\rm \Lambda \oplus p$--$\overline{\Lambda}$ and  $\Lambda$--$\overline{\Lambda}$ was obtained and the scattering parameters for %
\mbox{baryon--antibaryon} pairs,  which were not measured directly, were estimated.  
All of the extracted parameters from the femtoscopic fit are summarised in  Fig.~\ref{fig:data_fit_bab2} and compared with theoretical models and with other measurements.

The real and imaginary parts of the scattering length, and the effective interaction range extracted from the Pb--Pb fits have similar values for all \mbox{baryon--antibaryon} pairs at low-$k^{*}$. 
The significant magnitude of  $\Im f_0$  shows that inelastic processes (annihilation) can occur for baryon--antibaryon systems. This finding was verified by the analysis in pp collisions, where an even larger contribution to the inelastic part of the interaction was found for the p--$\overline{\mathrm{p}}$ and p--$\overline{\Lambda}\oplus \overline{\mathrm p}$--$\Lambda$ pairs, while the same scattering parameters found in \mbox{Pb--Pb} collisions provided a good description of the $\Lambda$--$\overline{\Lambda}$ correlation~\cite{ALICE:2021cyj}. 
In general, the $\Im f_0$ is very different for the three baryon--antibaryon pairs when looking at both colliding systems and it gets smaller for baryon--antibaryon pairs of higher total pair mass. 
The elastic part of the interaction exhibits also the same mass hierarchy and the negative values of $\Re f_0$ show either that the interaction between baryons and antibaryons is repulsive, or that baryon--antibaryon bound states can be formed. 
These findings can not exclude the presence of a bound state in the baryon--antibaryon channels but the presence of sizeable inelastic components makes it very difficult to disentangle the two effects.

\subsection{Conclusions}
    \paragraph{Light nuclei.} The production mechanism of light (anti)nuclei was investigated in small and large collision systems and the measured yields allowed one to test different production models. A tension remains between the coalescence and the statistical hadronisation approaches since a coherent description of all observables in all colliding systems is yet to be achieved. An ingredient recognised as essential for future studies is the accurate measurement of the nucleon source radii in the different colliding systems, which could be achieved with femtoscopy techniques. The measurement of the flow parameter $v_2$ for light (anti)nuclei in Pb--Pb collisions provided a complementary tool to study the production mechanisms. So far, the simplified modelling of such an observable via coalescence or blast--wave approaches describe the data qualitatively. 
    
   \paragraph{Particle emitting source in small collision systems.} The precision of the system size measurement for small colliding systems has been significantly improved by means of a new model that accounts for the contributions of strong decays. A universal source for all hadrons produced in small colliding systems was characterised and its properties hint to the presence of collective effects in these collisions.
    
    \paragraph{Hypertriton lifetime and binding energy.} The most accurate measurement of the hypertriton lifetime was obtained by ALICE
    and
    is consistent with the free $\Lambda$ lifetime, thus solving in a definite way a long-standing puzzle about this observable. The precise determination of the binding energy of the hypertriton
    confirms the weakly-bound nature of \hyp and is in agreement with the binding energy value describing the p--$\Lambda$ momentum correlations measured with the femtoscopy technique.%
    
    \paragraph{Proton--Hyperon interactions.}  Correlations in momentum space measured in pp and p--Pb collisions provided the first high precision data on the strong p--$\Lambda$ interaction (25 times more precise than previous scattering data) and the first assessement of strong p--$\Sigma^0$ and p--$\phi$ interaction. These precise measurements challenge the current versions of effective field theoretical calculations. The $\Lambda$--$\Lambda$ and p--$\phi$ correlations allowed for assess a rather shallow hyperon--hyperon interaction potential. The so far unknown p--$\Xi^-$, $\Lambda$--$\Xi^-$ and p--$\Omega^-$ residual strong interactions have been measured for the first time via the analysis of correlations in the momentum space in pp and p--Pb collisions. These measurements confirmed the attractive interaction predicted by lattice QCD calculations and allowed to search for the existence of di-baryon bound states. So far the precise data measured by ALICE provide no evidence that supports the formation of either a p--$\Xi^-$ or a p--$\Omega^-$ bound state.
    
    \paragraph{Kaon--proton interactions.} The $\rm K^-$--p interaction was also revisited with unprecedented precision and the combined measurements of correlations in small and large colliding systems allowed one to  extract scattering parameters with a precision matching that of measurements with kaonic atoms. A direct measurement of the coupled channels that characterise the $\rm K^-$--p interaction as a function of the system size was also carried out. The measurements in different collisions systems are the only ones available to calibrate the parameters within effective field theories.
    
    \paragraph{Baryon--antibaryon correlations.} Baryon--antibaryon correlations were investigated in pp and  Pb--Pb collisions and the real and imaginary part of the  scattering parameters for p--$\overline{\Lambda}$ and $\Lambda$--$\overline{\Lambda}$ were be extracted for the first time. No evidence of bound states has been observed.

\newpage

\newcommand{\sqrts}{\ensuremath{\sqrt{s_{\mathrm{NN}}}}}
\newcommand{\GeVc}{\ensuremath{\mathrm{GeV}\kern-0.05em/\kern-0.02em c}}

\newcommand{\pTjet}{\ensuremath{p_{\mathrm{T}}^{\mathrm{jet}}}}
\newcommand{\pTchjet}{\ensuremath{p_{\mathrm{T,\;ch\; jet}}}}
\newcommand{\tg}{\ensuremath{\theta_{\mathrm{g}}}}
\newcommand{\rg}{\ensuremath{R_{\mathrm{g}}}}
\newcommand{\zg}{\ensuremath{z_{\mathrm{g}}}}
\newcommand{\zcut}{\ensuremath{z_{\mathrm{cut}}}}
\newcommand{\etajet}{\ensuremath{\eta_{\mathrm{jet}}}}

\newcommand{\ang}{\ensuremath{\lambda_{\alpha}}}
\newcommand{\angNP}{\ensuremath{\lambda_{\alpha}^{\mathrm{NP}}}}
\newcommand{\angsd}{\ensuremath{\lambda_{\alpha,\mathrm{g}}}}
\newcommand{\angsdNP}{\ensuremath{\lambda_{\alpha,\mathrm{g}}^{\mathrm{NP}}}}
\newcommand{\pTi}{\ensuremath{p_{\mathrm{T},i}}}

\newcommand{\s}            {\ensuremath{\sqrt{s}}\xspace}
\newcommand{\pt}{p_{\rm T}} 
\newcommand{\kt}{k_{\rm T}}
\newcommand{\snn}          {\ensuremath{\sqrt{s_{\mathrm{NN}}}}\xspace}
\newcommand{\meanpt}       {$\langle p_{\mathrm{T}}\rangle$\xspace}
\newcommand{\ycms}         {\ensuremath{y_{\rm CMS}}\xspace}
\newcommand{\ylab}         {\ensuremath{y_{\rm lab}}\xspace}
\newcommand{\etarange}[1]  {\mbox{$\left | \eta \right |~<~#1$}}
\newcommand{\yrange}[1]    {\mbox{$\left | y \right |~<~#1$}}
\newcommand{\dndy}         {\ensuremath{\mathrm{d}N_\mathrm{ch}/\mathrm{d}y}\xspace}
\newcommand{\dndeta}       {\ensuremath{\mathrm{d}N_\mathrm{ch}/\mathrm{d}\eta}\xspace}
\newcommand{\avdndeta}     {\ensuremath{\langle\dndeta\rangle}\xspace}
\newcommand{\Npart}        {\ensuremath{N_\mathrm{part}}\xspace}
\newcommand{\Ncoll}        {\ensuremath{N_\mathrm{coll}}\xspace}
\newcommand{\Tfo}         {\ensuremath{T_\mathrm{fo}}\xspace}
\newcommand{\Th}         {\ensuremath{T_\mathrm{h}}\xspace}
\newcommand{\de}          {\ensuremath{\mathrm{d}}\xspace}

\newcommand{\Hc}           {\ensuremath{\mathrm{H_c}}\xspace}
\newcommand{\Dzero}        {\ensuremath{\mathrm{D^0}}\xspace}
\newcommand{\Dplus}        {\ensuremath{\mathrm{D^+}}\xspace}
\newcommand{\Dstar}        {\ensuremath{\mathrm{D^{*+}}}\xspace}
\newcommand{\Ds}           {\ensuremath{\mathrm{D_s^+}}\xspace}
\newcommand{\Lc}           {\ensuremath{\Lambda_\mathrm{c}^+}\xspace}
\newcommand{\XicZero}      {\ensuremath{\Xi_\mathrm{c}^0}\xspace}
\newcommand{\XicPlus}      {\ensuremath{\Xi_\mathrm{c}^+}\xspace}
\newcommand{\XicPlusZero} 
{\ensuremath{\Xi_\mathrm{c}^{+,0}}\xspace}
\newcommand{\Sigmac}{\rm \Sigma_{c}}
\newcommand{\SigmacZeroPlusPlus}{{\rm \Sigma_{c}^{0,++}}}
\newcommand{\Omegac}       {\ensuremath{\Omega_\mathrm{c}^0}\xspace}
\newcommand{\Jpsi}         {\ensuremath{\mathrm{J}/\psi}\xspace}
\newcommand{\Lb}           {\ensuremath{\Lambda_\mathrm{b}^+}\xspace}
\newcommand{\Xib}          {\ensuremath{\Xi_\mathrm{b}^{+,0}}\xspace}
\newcommand{\DzerotoKpi}   {\ensuremath{\mathrm{D^0\to K^-\pi^+}}}

\newcommand{\Qc}           {\ensuremath{\mathrm{c}}}
\newcommand{\AQc}       {\ensuremath{\mathrm{\overline{c}}}}
\newcommand{\Qb}           {\ensuremath{\mathrm{b}}}
\newcommand{\AQb}       {\ensuremath{\mathrm{\overline{b}}}}
\newcommand{\Qq}           {\ensuremath{\mathrm{Q}}}
\newcommand{\AQq}       {\ensuremath{\mathrm{\overline{Q}}}}
\newcommand{\Ap}       {\ensuremath{\mathrm{\overline{p}}}}

\section{QCD studies with high-\texorpdfstring{$Q^2$}{} processes in pp collisions}
\label{ch:QCDpp}

\label{sec:Introduction}

Perturbative QCD (pQCD) is one of the cornerstones of high-energy physics.
A thorough understanding of the phase space in which pQCD is under
precise control provides a baseline for the discovery of new physics, including
searches for physics Beyond the Standard Model (BSM) 
\cite{Nath:2010zj, Larkoski:2017jix, Deur:2016tte, Marzani:2019hun},
and emergent behaviors of the quark--gluon plasma,
as detailed in Chap.~\ref{ch:QGPproperties}.

Theoretical descriptions of high-$Q^2$ processes in QCD employ
various perturbative approximations to compute cross sections. 
Fixed-order expansions in $\alpha_{\rm S}$ include all terms up to a 
given power in $\alpha_{\rm S}$, with the precision of many recent calculations achieving Next-to-Next-to-Leading-Order (NNLO)~\cite{Currie:2016bfm, Czakon:2019tmo}. 
However, for certain observables, terms appearing at all orders in $\alpha_{\rm S}$ 
contribute significantly, due to large logarithms that arise
from characteristic scales in the observable's phase space.
These scale-dependent terms can be computed analytically by all-order
resummations~\cite{Bauer:2001yt, Dasgupta:2013ihk} 
or by matching fixed-order calculations with parton showers 
\cite{Nagy:2007ty, Nagy:2019pjp, Dasgupta:2020fwr, Nason:2004rx, Frixione:2007vw, Alioli:2010xd, Sjostrand:2014zea}.
Other controlled approximations, such as the
leading colour approximation~\cite{tHooft:1973alw, Witten:1979kh} 
or the power suppression 
of other characteristic scales~\cite{Larkoski:2017jix, Marzani:2019hun}, are similarly employed.
In all cases, scale variations are used to estimate
the contributions of higher-order terms. 
General-purpose Monte Carlo (MC) event generators are built
with these first-principles calculations in mind, and attempt to provide 
approximate descriptions of broad classes of observables.

The factorisation approach in QCD enables the connection of perturbative calculations to measurements, by relating perturbatively calculable
cross sections to non-perturbative initial and final states~\cite{Collins:1989gx, Brock:1993sz}.
Observables from high-$Q^2$ processes therefore intrinsically
involve long-distance QCD interactions parameterized by
parton distribution functions (PDFs), fragmentation functions (FFs),
and other non-perturbative objects assumed to be universal across collision systems and observables
\cite{Schuler:1996ku, Chang:2013rca, Jouttenus:2011wh}.
The relevant non-perturbative objects involved depend on the observable
considered; for example, parton-to-hadron fragmentation is 
typically described using FFs, 
whereas parton-to-jet fragmentation is often computed purely perturbatively.
Measurements of high-$Q^2$ processes in \pp\ collisions
test the importance of higher-order terms
in state-of-the-art pQCD calculations,
and constrain these non-perturbative objects
\cite{Deur:2016tte}.

ALICE has made significant contributions to studying high-$Q^2$ processes in \pp{} collisions. While the experimental design of ALICE limits its data-taking rate and high-\pT{} capabilities, and thereby its sensitivity to the rarest processes produced at the LHC, ALICE records collisions in a low-pileup environment ideal for precision QCD studies. The ALICE detector design is complementary to the other LHC detectors; its high precision tracking and particle identification capabilities,
with a focus on low- to moderate-transverse momentum $\pT\lesssim 100$ GeV/$c$, are unique at the LHC (see Ch.~\ref{sec:ALICEExp}).
ALICE also puts special emphasis on measurements in \pp\ collisions of high-$Q^2$ observables that are of importance for the study of the QGP, which provides a complementary focus to the observables
studied by experiments that concentrate more on BSM physics.
In this chapter, we discuss several such measurements
of high-$Q^2$ processes in \pp\ collisions which contribute to 
the fundamental understanding of QCD.
We begin with jet measurements that test the perturbative accuracy 
of state-of-the-art pQCD calculations in Sec.~\ref{sec:pp-jets},
before highlighting identified particle measurements
which inform the understanding of the fragmentation process and its universality in Sec.~\ref{sec:pp-fragmentation}, and finally discuss an innovative jet substructure measurement
used to reveal the QCD dead-cone in Sec.~\ref{sec:pp-deadcone}.

\subsection{Testing perturbative accuracy with jets} \label{sec:pp-jets}

Jets are collimated sprays of hadrons arising from 
the fragmentation of 
energetic quarks and gluons generated in high-$Q^2$ interactions. Jets are fundamental objects in QCD, and are produced copiously at collider energies~\cite{Arnison:1986vk, Appel:1985rm, Aaltonen:2008eq, Abazov:2008ae, Abelev:2006uq, Acharya:2019tku, Abelev:2013fn, Acharya:2019jyg, Khachatryan:2016jfl, Chatrchyan:2014gia, Khachatryan:2016mlc, Khachatryan:2016wdh, Sirunyan:2020uoj, Aad:2013lpa, Aad:2011fc, Aad:2014vwa, Aaboud:2017wsi}.
Inclusive and heavy-flavour jet production cross sections,
as well as jet substructure observables, can be calculated analytically 
in pQCD. ALICE measurements test the perturbative accuracy of these calculations 
and provide insight into the regime of applicability of perturbative 
approximations.
ALICE reconstructs full jets using its tracking system combined with information from the EMCal (MC corrections are applied for missing neutral hadron energy, due to the absence of a hadronic calorimeter), as well as charged-particle jets using the tracking system alone. 
The reconstruction of full jets allows a more direct comparison to theoretical calculations, and are frequently used for measurements of inclusive jet production, whereas charged-particle jets allow better angular resolution, and are often used by ALICE for jet substructure measurements.
The ALICE approach to jet measurements is discussed further in Sec.~\ref{sec:JetsMatter}. 

\subsubsection{Inclusive jet production}

The inclusive jet production cross section in \pp{} collisions is one of the most fundamental quantities in QCD; it was recently calculated to NNLO~\cite{Currie:2016bfm, Czakon:2019tmo}. 
Inclusive jet measurements in elementary collisions are used in global fits to determine PDFs and the strong coupling constant $\alpha_{\mathrm{S}}$~\cite{Britzger:2017maj, Ball:2018iqk, Malaescu:2012ts, Khachatryan:2016mlc, Aad:2013lpa}. 
Historically, jet measurements and calculations with 
large- and moderate-$R$ (resolution parameter), $R \gtrsim 0.4$, have been preferred for high precision studies because they minimise hadronisation effects. Calculation of the inclusive jet production cross section for small-$R$ jets is challenging, due to large logarithms of the jet resolution parameter $R$ that can contribute at all orders in $\alpha_{\mathrm{S}}$. However, this issue was addressed recently by calculations of the inclusive jet cross section with all-order resummation of logarithms of the jet resolution parameter~\cite{Dasgupta:2014yra, Dasgupta:2016bnd, Kang:2016mcy, Kang:2017frl, Liu:2017pbb, Liu:2018ktv}.
Small-$R$ jets are of current interest both because of their importance in measuring jet quenching effects in the high-background environment of heavy-ion collisions (Sec.~\ref{sec:JetsMatter}), and because theoretical techniques for all-order resummation are best tested with small-$R$ jets due to their large resummation effects. 
Furthermore, the jet cross section as a function of $R$ can disentangle the contributions from pQCD, the underlying event, and hadronisation~\cite{Dasgupta:2016bnd}.

\begin{figure}[!t]
\centering{}
\includegraphics[scale=0.8]{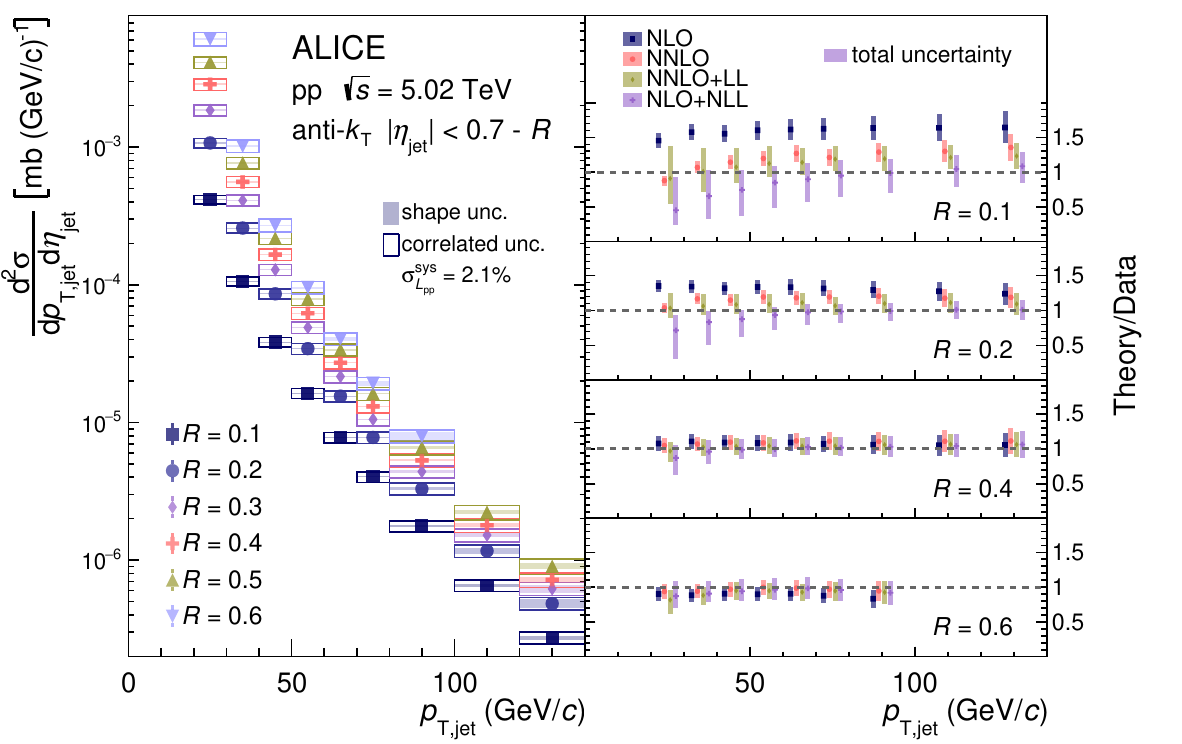}
\caption{(Left) Inclusive full jet cross sections in \pp\ collisions at $\sqrt{s} = 5.02$ TeV for $R=0.1-0.6$, measured by ALICE~\cite{Acharya:2019jyg}. 
(Right) Ratio of various pQCD calculations to data
\cite{Currie:2016bfm,Dasgupta:2014yra, Dasgupta:2016bnd, Liu:2017pbb, Liu:2018ktv, Kang:2016mcy}. The systematic uncertainties in the ratio, shown as boxes, are the quadratic sum of the systematic uncertainties in data and calculations. No systematic uncertainty for non-perturbative corrections are included, except
for the case of NNLO+LL.}
\label{fig:inclusive-jet}
\end{figure}

In order to explore these theoretical developments, ALICE has measured the inclusive jet cross section for a wide range in $R$ in \pp\ collisions at $\sqrt{s} = 2.76$ TeV~\cite{Abelev:2013fn} ($0.2<R<0.4$) and $5.02$ TeV~\cite{Acharya:2019jyg} ($0.1<R<0.6$), with the latter shown in the left panel of Fig.~\ref{fig:inclusive-jet}. 
The right panel of Fig.~\ref{fig:inclusive-jet} shows comparisons of these measurements to four pQCD calculations: NLO and NNLO fixed-order calculations~\cite{Currie:2016bfm},  NNLO fixed-order calculations taken from~\cite{Currie:2016bfm} matched to a leading-logarithmic (LL) resummation of $R$ (NNLO+LL)~\cite{Dasgupta:2014yra, Dasgupta:2016bnd}, and NLO fixed-order calculations with resummation of logarithms of $R$ and threshold logarithms
to next-to-leading-logarithmic (NLL) accuracy (NLO+NLL)~\cite{Liu:2017pbb, Liu:2018ktv, Kang:2016mcy}. 
Non-perturbative corrections due to hadronisation and the underlying event are applied using MC event generators\footnote{In the NNLO+LL calculation, 
the non-perturbative corrections are constructed from an envelope of several MC generators (PYTHIA, SHERPA, HERWIG), and are included in the uncertainty bands.
For all other calculations, PYTHIA8 Monash 2013 alone is used, and no 
uncertainty is included; the 
difference in central values between these approaches is $<6\%$ at all
data points.}.
The non-perturbative corrections are large for small-$R$ and low-\pT{},  reaching approximately a factor 2 for $R=0.1$ for $\pT < 50\; \GeVc$~\cite{Acharya:2019jyg, Dasgupta:2016bnd}; the comparisons at low-\pT{} should be interpreted with this in mind.  

For $R=0.4$ and 0.6, all four theoretical calculations describe the data well, 
suggesting that NNLO and resummation effects are small in these cases. In contrast, for $R=0.1$ and $0.2$, the level of description of the data varies significantly depending on the type of calculation. The NLO fixed-order calculation has poorest description, while the NNLO calculation is closer to the data, demonstrating
the crucial importance of NNLO contributions.
However, at high-\pT\ for $R=0.1$, the NNLO calculations overestimate the 
data, suggesting that resummation effects are important when
$R$ becomes sufficiently small.
Furthermore, the theoretical uncertainties in the 
 NLO and NNLO calculations do not overlap,
corroborating the importance of resummation effects. 
This is shown directly by the NLO+NLL and NNLO+LL calculations, which
describe the data consistently within uncertainties. 
Note that while the NLO and NLO+NLL calculations are dramatically different, 
the NNLO and NNLO+LL calculations differ only slightly, suggesting that further
theoretical work is needed to clarify the magnitude of LL vs. NLL contributions
and differences in the calculational frameworks employed.
Together, these observations show clearly the importance of both NNLO
contributions and resummation contributions for small-$R$ jets, both to ensure a good description of data and to ensure theoretical consistency between calculational orders.

The $R$-dependence of the inclusive jet cross section is determined most precisely by measuring the ratio of experimental cross sections with different $R$, taking account of correlated systematic uncertainties which cancel in the ratio~\cite{Abelev:2013fn,Acharya:2019jyg}. Initial considerations suggested that theoretical uncertainties may also cancel significantly in such ratios~\cite{Soyez:2011np}. Comparison of recent ALICE cross section ratio measurements~\cite{Acharya:2019jyg} with the calculations described above (not shown here) indicate that for the ratio $R=0.2/R=0.4$ all calculations describe the data within uncertainties. For the ratio $R=0.2/R=0.6$, the NLO+NLL and NNLO calculations describe the data, while the NNLO+LL calculation exhibits a tension. Note, however, that scale variations can reach into the non-perturbative regime at low-\pT{} and prevent  theoretical uncertainty cancellation, which is treated differently in each calculation. The experimental data are now precise enough that theoretical uncertainty is the limiting factor in the comparison~\cite{Bellm:2019yyh, Currie:2018xkj}.       %
\subsubsection{Heavy-flavour jet production}

The production of jets containing charm or beauty quarks can be similarly
tested against pQCD calculations~\cite{Dai:2018ywt,Bauer:2013bza,Li:2018xuv}.
Heavy-flavour jets in \pp{} collisions are relevant both to 
understand the flavour-dependence of jet quenching in heavy-ion collisions
(Sec.~\ref{sec:JetsMatter}) and as a Standard Model background to the decay
of massive particles, such as $\rm{H}\rightarrow \rm{b}\overline{\rm{b}}$~\cite{Voutilainen:2015lqa}.

\begin{figure}[!ht]
\includegraphics[scale=0.42]{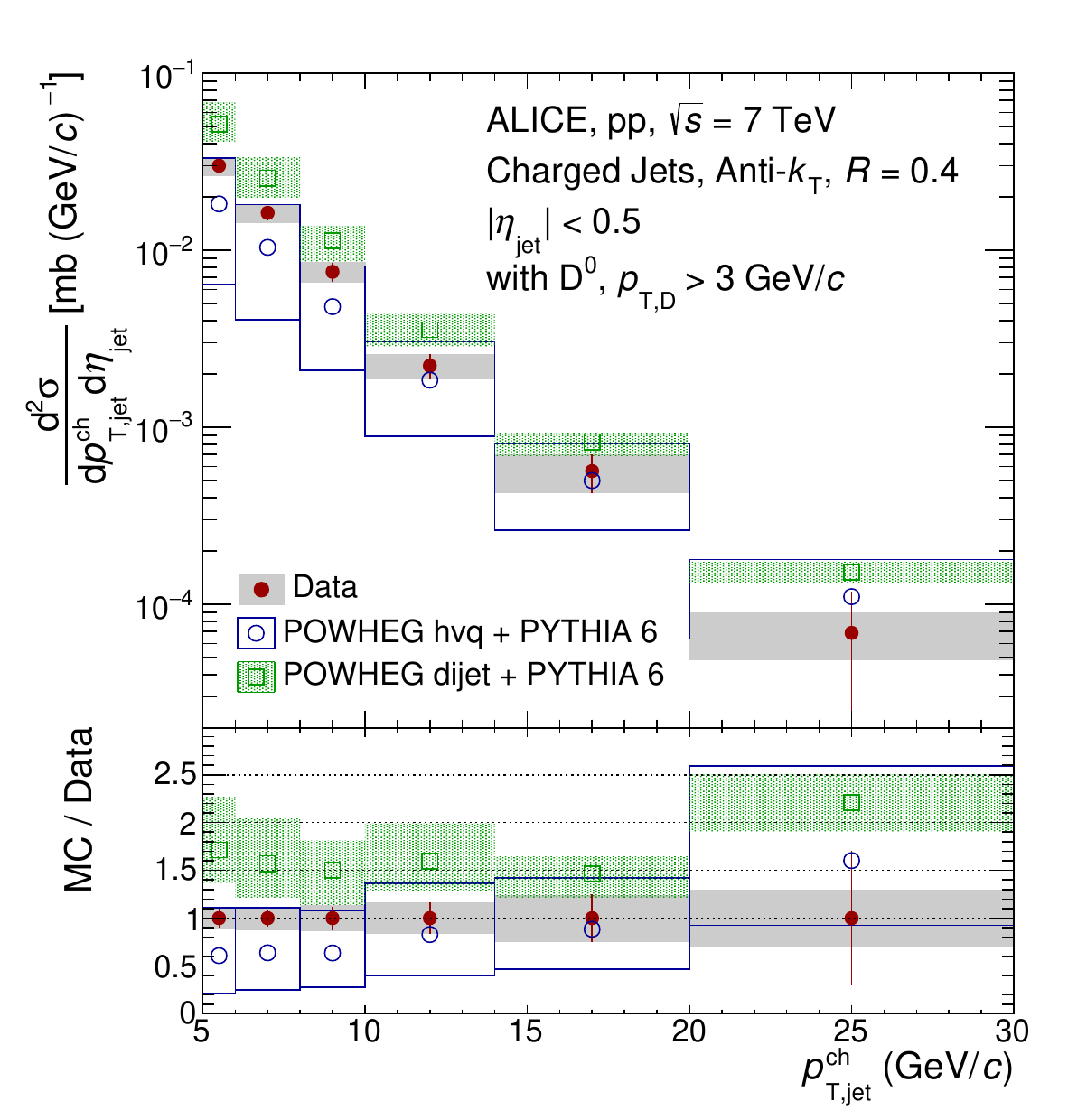}
\includegraphics[scale=0.37]{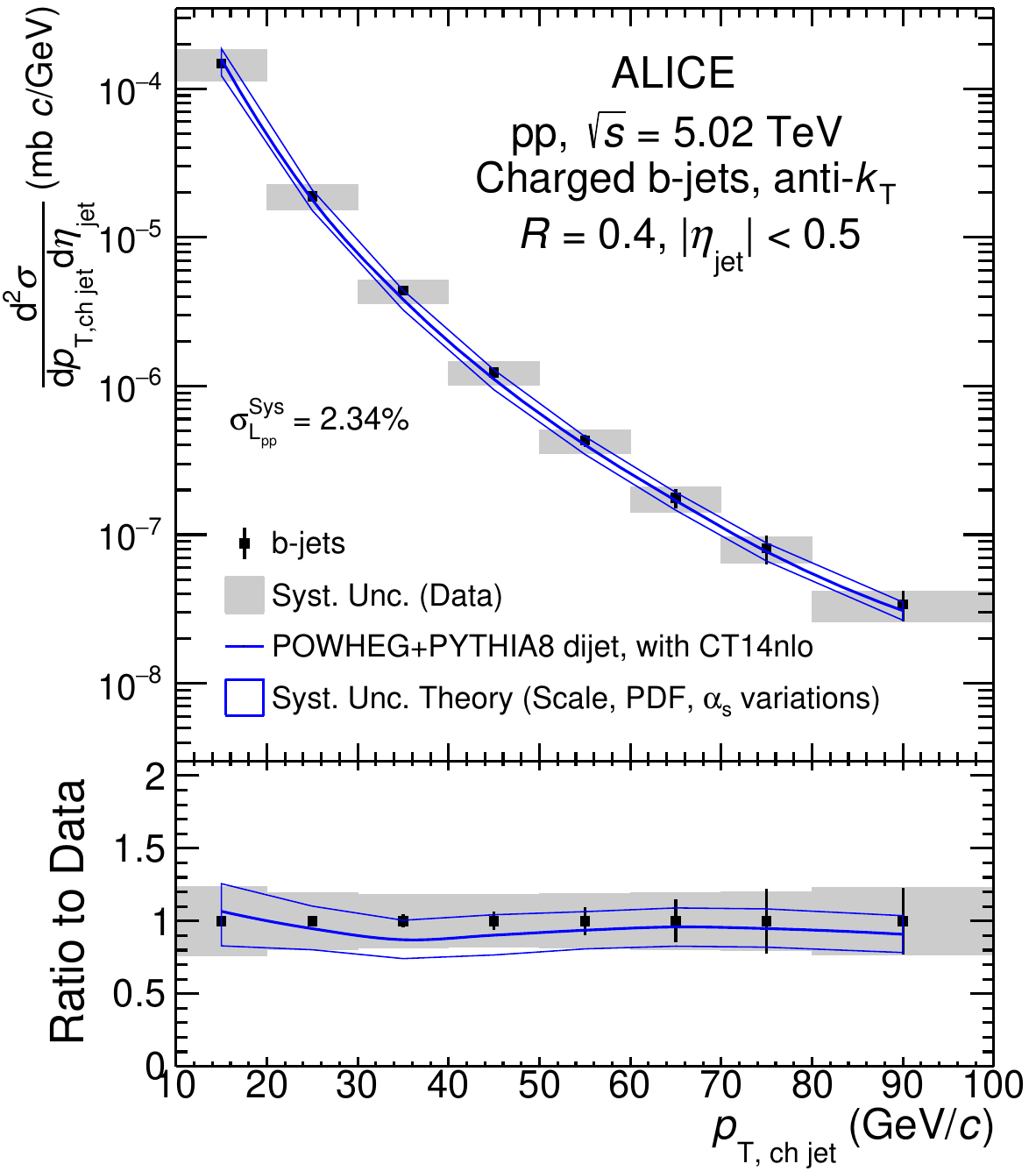}
\caption{(Left) \pT{}-differential cross section of charged-particle jets tagged with $\rm D^0$ mesons~\cite{Acharya:2019zup} in \pp\ collisions at $\sqrt{s}=7$~TeV, compared to the POWHEG heavy-quark and di-jet implementations matched to PYTHIA~6 parton shower.
(Right) \pT{}-differential cross section of b-tagged charged-particle jets  in \pp\ collisions at $\sqrt{s}=5.02$~TeV~\cite{ALICE:2021wct}, 
compared to NLO pQCD prediction by the POWHEG di-jet implementation matched to PYTHIA8 parton shower.}
\label{fig:HF-jet}
\end{figure}

Figure~\ref{fig:HF-jet}, left panel, shows the ALICE measurement of the $\pT$-differential production cross section of charged-particle jets containing a D$^{0}$ meson among its constituents in pp collisions at $\sqrt{s} = 7$ TeV, in the range $5 \leq \pTchjet < 30$ GeV/$c$, at midrapidity~\cite{Acharya:2019zup}. 
The D$^{0}$-tagged jet over inclusive jet fraction was also measured (not shown here), and found to increase with \pTchjet{} from $\approx 4.2$\% at 5 GeV/$c$ 
to $\approx 8.0$\% at 30 GeV/$c$. 
The \pTchjet\ dependence of the D$^{0}$-tagged jet production cross section and of its ratio to inclusive-jet cross section 
are consistent with 
NLO pQCD calculations provided by POWHEG~\cite{Nason:2004rx,Frixione:2007vw} matched with PYTHIA~6~\cite{Sjostrand:2006za,Sjostrand:2007gs} for the generation of the parton showering, hadronisation and particle decays. 
The absolute magnitude of the description is less satisfactory, with discrepancies up to a factor 2, though data and model uncertainties are rather large.
The production cross section of D$^{0}$-tagged jets was also evaluated as a function of the jet-momentum fraction carried by the D$^{0}$ meson in the jet-axis direction ($z^{\rm ch}_{||}$) (not shown here). 
The predictions provide good agreement with the data for the $z^{\rm ch}_{||}$ dependence of the cross section, apart from a small tension for $15 \leq \pTchjet < 30$ GeV/$c$, where data tend to favour a softer charm-quark fragmentation (see 
Sec.~\ref{sec:open-heavy-flavor} for further discussion of charm fragmentation). Measurements of $\rm D^0$-tagged jets were also carried out by ALICE in pp collisions at $\sqrt s=5.02$ and 13~TeV~\cite{ALICE:2022mur}.

Additionally, ALICE has reconstructed charged jets at midrapidity produced by beauty-quark fragmentation in pp collisions at $\sqrt{s} = 5.02$ TeV~\cite{ALICE:2021wct}, as shown in right panel of 
Fig.~\ref{fig:HF-jet}. The jet is tagged using two techniques, each of which exploits the relatively long lifetime of beauty hadrons: the reconstruction of displaced secondary vertices from jet constituent tracks, and the search for jet constituent tracks with a large impact parameter. Both techniques exploit typical features of the decay of long-lived beauty hadrons.
The $\pT$-differential production cross section of beauty jets has been measured in a jet transverse-momentum interval $10 \leq \pTchjet < 100$ GeV/$c$. These are the lowest \pTchjet\ beauty-jet measurements at the LHC. These measurements also demonstrate agreement with ATLAS and CMS in their region of overlapping kinematics~\cite{CMS:2012pgw, ATLAS:2011ac}. The fraction of charged jets originating from beauty quarks has also been measured in the same collision system, and found to increase from $\approx 2$\% at 10 GeV/c to $\approx 3.5$\% at 100 GeV/$c$. Such measurements were compared with NLO pQCD calculations provided by POWHEG, using the dijet process implementation, with subsequent parton showering provided by PYTHIA~8. The resulting predictions are able to correctly reproduce the absolute value and the $\pT$ dependence of both the beauty-jet production cross section and the beauty-jet to inclusive-jet fraction.
These measurements can be compared in the future to analytical calculations of beauty-tagged jet production~\cite{Dai:2018ywt, Bauer:2013bza, Li:2018xuv}.

Overall, the fair description of the measurements indicates that perturbative QCD is capable, albeit within large theoretical uncertainty at the current level of perturbative accuracy, of describing the production of jets obtained from heavy-quark fragmentation.                %
\subsubsection{Jet substructure}

Jet substructure, defined by observables constructed from the distribution
of constituents within a jet, provides the ability to access
specific regions of QCD radiation phase space~\cite{Larkoski:2017jix}.
Jet reclustering techniques enable jet radiations to be organised 
into a phase space diagram known as the Lund plane~\cite{Dreyer:2018nbf}; traditional jet substructure 
observables target specific regions of this phase space, and thereby
emphasise certain features of jets.
The N-subjettiness~\cite{Thaler:2010tr}, for example, characterises the number of prong-like 
structures in the jet, while the jet mass~\cite{Acharya:2017goa} characterises 
the virtuality of the jet-initiating parton.
Jet substructure is therefore used as a tool to study rare event topologies in 
\pp{} collisions, such as boosted objects that decay into jets
\cite{Butterworth:2008iy, Almeida:2008yp}.
A detailed understanding of jet substructure is therefore needed in order 
to distinguish rare signals from the prevailing QCD background.
Studies of jet substructure observables across a wide range of phase
space enable differential tests of our understanding of pQCD.
Jet substructure can also be used to remove pileup
and study non-perturbative effects including hadronisation
\cite{Amoroso:2020lgh, Krohn:2013lba}.

The ALICE tracking system is ideal for jet substructure measurements, for which
fine angular resolution is paramount. 
While track-based jet observables are collinear-unsafe~\cite{Chang:2013rca, Chen:2020vvp, Chien:2020hzh},
they can be measured with greater precision than calorimeter-based jet observables,
and recent measurements have demonstrated that for many substructure observables 
track-based distributions are compatible with the corresponding collinear-safe distributions~\cite{ATLAS:2019mgf}.
With a focus on low-\pT{} jets, the ALICE jet substructure program 
in \pp{} collisions tests the applicability of pQCD approximations 
as \alphaS{} %
increases and the theory becomes increasingly non-perturbative. 
ALICE has measured a variety of jet substructure observables, including
longitudinal and transverse fragmentation distributions, the groomed jet 
momentum fraction and radius, and N-subjettiness
\cite{Acharya:2018eat, ALICE:2022vsz, Acharya:2020oay, ALICE:2021obz, ALICE:2022hyz, Acharya:2018uvf, Acharya:2019djg}.
These results have been compared to 
MC generators, and provide future input for constraints on FFs
and the tuning of MC event generators.

One of the most important classes of jet substructure observables
are the infrared- and collinear- (IRC-) safe jet angularities~\cite{Larkoski:2014pca}, defined as

\begin{equation} \label{eq:1}
\lambda_{\alpha} \equiv \sum_i z_i \theta_i^{\alpha},
\end{equation}

where the sum runs over the jet constituents,
and $\alpha>0$ is a continuous free parameter.
The first factor, $z_i \equiv \pTi /\pTjet$, describes the momentum fraction carried by
the constituent, and the base of the second factor, $\theta_i \equiv \Delta R_i / R$,
denotes the rapidity ($y$) - azimuth ($\varphi$) separation of the constituent from the jet axis,
where $\Delta R_i \equiv \sqrt{\Delta y_i^2 + \Delta \varphi_i^2}$. 
By varying $\alpha$, one can systematically vary
the contribution of collinear radiation within the shower,
providing stringent tests of pQCD calculations.
The IRC-safe angularities
include the case $\lambda_1$, commonly referred to as
the radial moment or ``girth"~\cite{Acharya:2018uvf, Aad:2012meb, Sirunyan:2018asm, Aaltonen:2011pg}, and the case $\lambda_2$,
commonly referred to as the ``thrust'',
which is closely related to the jet mass~\cite{Sirunyan:2018asm, ATLAS:2012am, Aad:2012meb, Aad:2019wdr, ATLAS:2019mgf, Sirunyan:2019rfa, Sirunyan:2018gct, Acharya:2017goa, Aaltonen:2011pg}.

\begin{figure}[!t]
\centering{}
\includegraphics[scale=0.8]{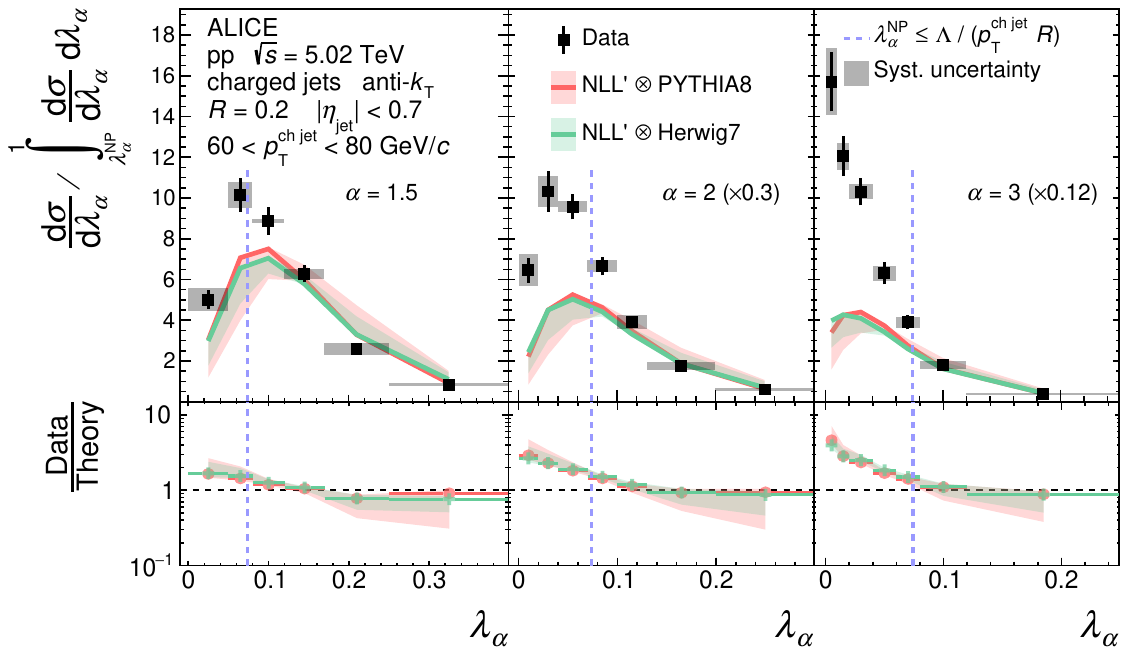}
\caption{Measurements of ungroomed jet angularities \ang{} in \pp{} collisions at $\sqrt{s}=5.02$ TeV for $R=0.2$ charged-particle jets~\cite{ALICE:2021njq}
compared to analytical NLO+NLL predictions~\cite{Kang:2018qra, Kang:2018vgn}. 
The theoretical calculations are corrected with a folding procedure to account for 
hadronisation and underlying event effects as well as charged particle jets,
using PYTHIA~8 Monash 2013 and HERWIG~7.
The distributions are normalised such that the integral of the perturbative region defined 
by $\ang > \Lambda / (\pTchjet R)$ is unity. Note that the transition from perturbative to non-perturbative is smooth,
and this vertical line is merely a visual guide.}
\label{fig:ang}
\end{figure}

ALICE has performed a systematic measurement
of the IRC-safe jet angularities in \pp{} collisions~\cite{ALICE:2021njq}.
Figure~\ref{fig:ang} shows measurements of $\lambda_{1.5},\lambda_2,\lambda_3$ 
for $R=0.2$ jets. 
As $\alpha$ increases, the distributions skew towards small $\ang$, 
since $\theta_i$ is smaller than unity.
The measured angularities are compared to
analytical calculations with all-order resummations of large logarithms to
next-to-leading-logarithmic (NLL$^\prime$) accuracy~\cite{Kang:2018qra, Kang:2018vgn, Kang:2016ehg}.
The angularity distributions are expected to be dominated by pQCD 
only at sufficiently large values of \ang,
with the non-perturbative regime estimated by $\angNP \leq \Lambda / (\pTchjet R)$,
for an energy scale $\Lambda$ at which $\alpha_{\rm S}$ becomes non-perturbative.
A dashed vertical line is included as a rough estimate of the boundary between
perturbative- and non-perturbative-dominated regions, 
with $\Lambda=1\;\GeVc$.
Overall agreement is found between the theoretical calculations and the
measurements in the pertubative regime, while discrepancies are found
in the non-perturbative regime.
For $\alpha=1.5$, the majority of the distributions can
be described perturbatively -- especially for $R=0.4$ (see Ref.~\cite{ALICE:2021njq}) -- while as $\alpha$ increases to $\alpha=3$, the majority of the distributions become strongly non-perturbative.
It was also demonstrated that by applying jet grooming algorithms, 
the perturbative region can be extended and good agreement of perturbative calculations can be achieved to even smaller values of $\lambda$~\cite{ALICE:2021njq, CMS:2021iwu}.

In addition to testing the limits of the applicability of pQCD, these considerations provide critical guidance for jet substructure 
measurements in heavy-ion collisions.
If jet substructure distributions such as those in Fig.~\ref{fig:ang}
are normalised to unity over the full range of the observable, 
as is typically done in heavy-ion collisions, then
disagreement with theory predictions in the non-perturbative region --
which is expected -- will necessarily induce apparent disagreement 
with theory predictions in the perturbative region.
This demands rigorous and careful choice of the observables used to 
quantify jet quenching effects and of the range of values of
the observable that can be described by pQCD, such as the dependence 
of the applicability of pQCD on $\alpha$, $R$, and $\pTchjet$
in the case of jet angularities.
These measurements bring guidance from first-principles pQCD
and illustrate that additional scrutiny is needed 
when interpreting measurements of jet substructure in heavy-ion collisions.
The importance of capturing both perturbative and non-perturbative
processes is evident, as well as the need to discriminate between the two regimes in order
to achieve a well controlled theoretical understanding of observed jet quenching modifications.     %

\subsection{Studying fragmentation with identified particles} \label{sec:pp-fragmentation}

Measurements of hadron cross sections involve not only 
perturbatively calculable production cross sections, as in the case of jets,
but also non-perturbative parton-to-hadron FFs.
The study of identified single particle production in high-$Q^2$\ processes provides a ground for simultaneously testing both of these elements,
as well as QCD factorisation and 
the universality (across collision systems and observables) and evolution of both FFs and PDFs.
In particular, measurements at the LHC extend to unprecedentedly large values of centre-of-mass energy and hadron \pT{}, and are useful for studying the gluon-to-hadron FFs and their universality~\cite{Metz:2016swz, dEnterria:2013sgr}.

\subsubsection{Light flavour hadrons}
ALICE has measured identified hadron production at high-transverse momentum with different %
techniques and detectors:
\begin{itemize}
    \item Charged pions, charged kaons, and protons with the TOF and HMPID detectors (up to \pT{} of about 5~GeV/$c$) and using the relativistic rise of the specific energy loss in the TPC (up to 20~GeV/$c$)~\cite{Abelev:2014laa,Acharya:2019yoi,ALICE:2020nkc,ALICE:2020jsh}
    \item Charged kaons with kink topologies (up to about 7~GeV/$c$)~\cite{Acharya:2019yoi,ALICE:2020nkc,ALICE:2020jsh}
    \item Strange particles (${\rm K^0_S}$, $\Lambda$ and also $\Xi$ and $\Omega$) with topological reconstruction of their weak decays into charged particles (up to about 10--20~GeV/$c$)~\cite{Abelev:2012jp,ALICE:2020jsh}
    \item Hadronic resonances such as ${\rm K^{*0}(892)}$ and $\phi(1020)$ based on invariant mass analyses and relying on particle identification for daughter tracks (up to about 20~GeV/$c$)~\cite{ALICE:2017ban,Acharya:2019wyb}
    \item $\pi^0$ and $\eta$ with the electromagnetic calorimeters (up to about 200~GeV/$c$ and 30~GeV/$c$, respectively)~\cite{Abelev:2012cn,Abelev:2014ypa,Acharya:2017hyu,Acharya:2017tlv}
\end{itemize}
Measurements in these \pT{} ranges are optimal to study the gluon-to-hadron fragmentation, given that the \pT{} is high enough for the production to be already dominated by pQCD processes, but low enough that the fraction of hadrons coming from gluon fragmentation remains dominant. In fact, model calculations~\cite{Chiappetta:1992uh,Sassot:2010bh} indicate that the fraction of pions originating from gluon fragmentation is above 75\% for $p_{\rm T}< 30$  GeV/$c$ at LHC energies --- while the same happens only for $p_{\rm T}<$~5--7~GeV/$c$ at top RHIC energy~\cite{Adare:2010cy} --- and remains dominant ($>50$\%) up to $p_{\rm T}= 100$~GeV/$c$. In the following the measurements of the $\pi^0$ and $\eta$ mesons are discussed %
along with their implications in this context. 

\begin{figure*}[!t]
        \center
        \includegraphics[width=0.49\textwidth]{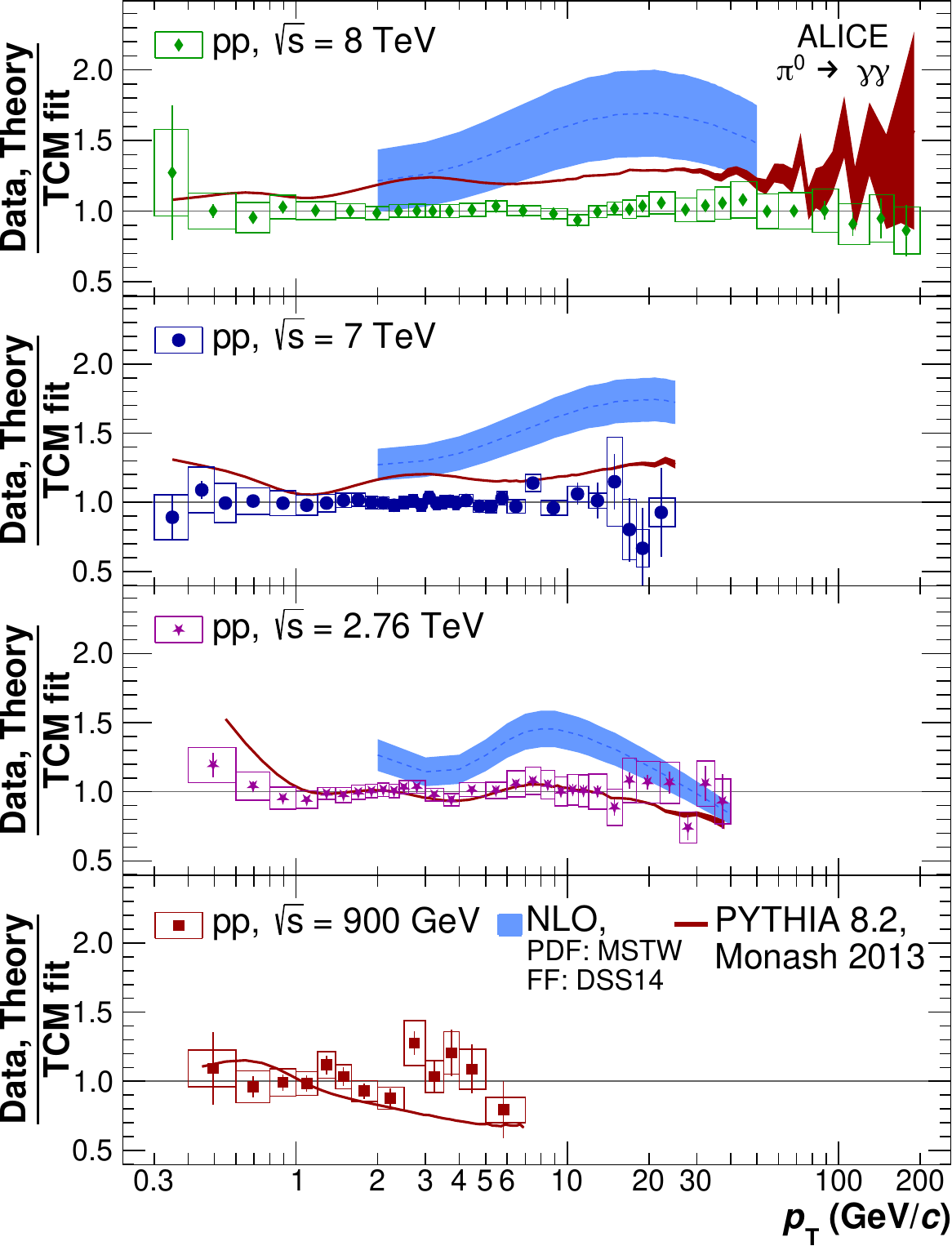}
        \raisebox{0.31\height}{\includegraphics[width=0.49\textwidth]{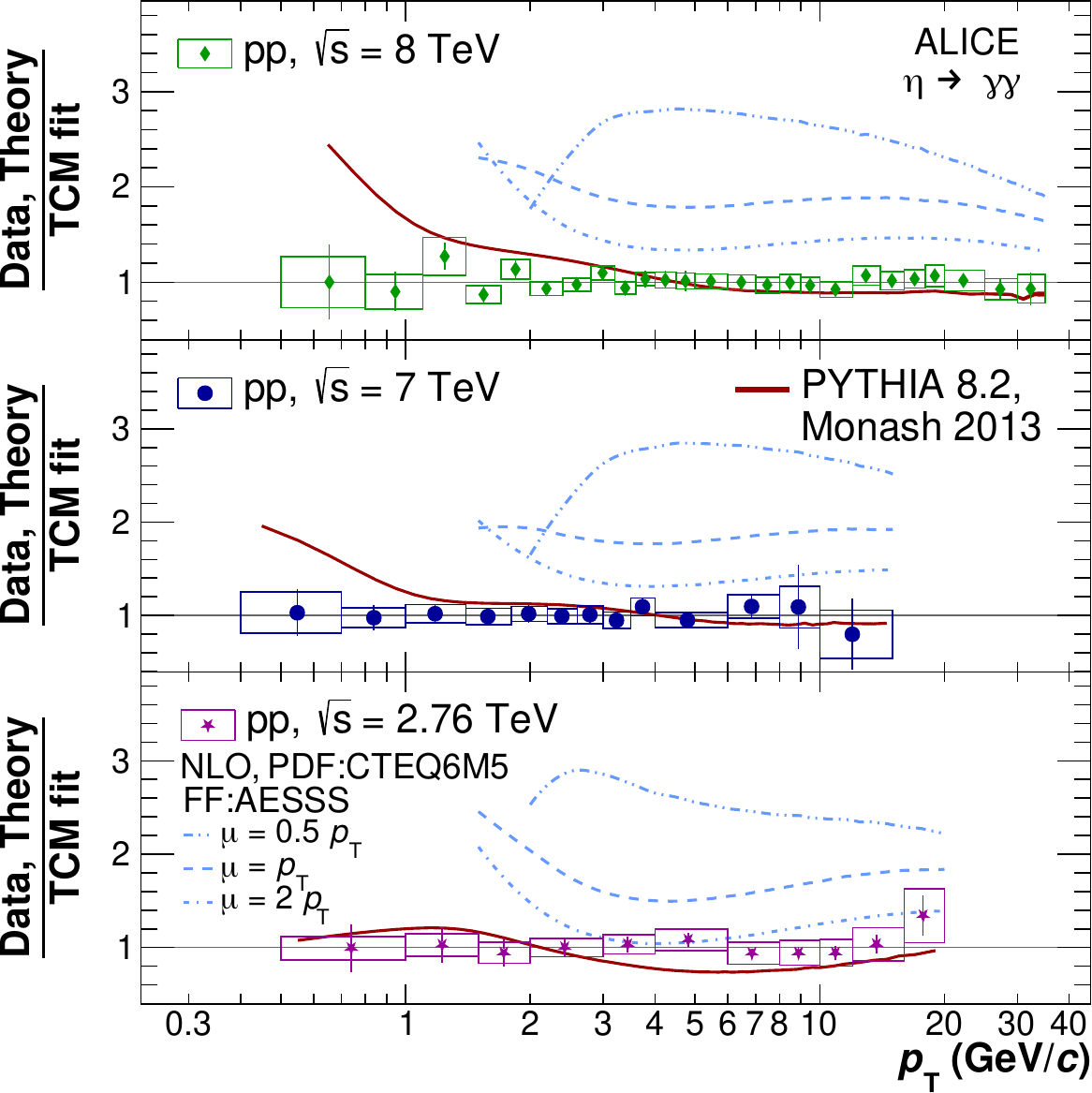}}
        \caption{Ratio of the neutral pion (Left) and eta meson (Right) invariant cross sections to their respective TCM parametrisations~\cite{Abelev:2012cn,Abelev:2014ypa,Acharya:2017hyu,Acharya:2017tlv}. Additionally, the ratios of NLO pQCD calculations and PYTHIA~8 Monash 2013 predictions to the TCM data fit are shown at the respective centre-of-mass energies. The NLO pQCD blue band in the left panel for $\pi^0$ is obtained considering the factorisation scale values $0.5\pT < \mu < 2\pT$.}
        \label{fig:MesonRatioWithNLO}
\end{figure*}

ALICE has measured the invariant differential cross sections of inclusive neutral pion and eta meson production at centre-of-mass energies $\sqrt{s} = 0.9,~2.76,~7$, and $8$~TeV, covering a maximum transverse momentum range of $0.3 <\pT<200$~\GeVc 
~\cite{Abelev:2012cn,Abelev:2014ypa,Acharya:2017hyu,Acharya:2017tlv}.
Figure~\ref{fig:MesonRatioWithNLO} shows the ratio of the measured cross sections
to the fit performed with the two-component model (TCM) proposed in Ref.~\cite{Bylinkin:2015xya}, which describes the spectra of the $\pi^0$ and $\eta$ meson over the full \pT{} range at all energies.
The invariant differential production cross sections are compared with NLO pQCD calculations~\cite{deFlorian:2014xna,Aidala:2010bn}
using MSTW08 (PDF)~\cite{Martin:2009iq} with DSS14 (FF)~\cite{deFlorian:2014xna} for $\pi^{0}$ and
CTEQ6M5 (PDF)~\cite{Tung:2006tb} with AESSS (FF)~\cite{Aidala:2010bn} for the $\eta$ meson.
The same scale value $\mu$ ($0.5\pT < \mu < 2\pT$) is chosen for the factorisation, renormalisation, and fragmentation scales used in the NLO pQCD calculations.
The variations of the $\mu$ values reflect the uncertainty of the NLO pQCD calculation.
 
For the $\pi^{0}$, the combination of the NLO PDF, production cross section, and FF describes the RHIC data~\cite{Adare:2015ozj} and the lowest energy LHC data at $\sqrt{s} = 0.9$~TeV~\cite{Abelev:2012cn} rather well, whereas at $\sqrt{s}=2.76$~TeV pQCD overpredicts the data by 30\% at moderate \pT\ and 
describes the data 
at higher \pT~\cite{Acharya:2017hyu}. 
For all $\mu$ values, these calculations overestimate the measured cross sections for both $\pi^{0}$ and $\eta$ mesons for all centre-of-mass energies above $\sqrt{s} = 2.76$~TeV, although better description of data is achieved for $\mu=2\pT$, for which calculations overestimate the data by $10$--$40\%$,  depending on \pT.
This suggests that higher order corrections are necessary to 
describe the data, as already argued when considering inclusive full jet cross sections for small value of cone radius. 
It has to be noted that FF uncertainties in the NLO pQCD calculations have been considerably reduced after including the published $\pi^{0}$ measurement at $\sqrt{s}=7$~TeV~\cite{Abelev:2012cn} in the calculation of DSS14.
Including precise new data for $\eta$ meson production measured at $\sqrt{s}=2.76$, $7$, and $8$~TeV~\cite{Acharya:2017hyu,Abelev:2012cn} will also help to considerably reduce fragmentation function uncertainties in this  case.

In addition, the neutral meson measurements are compared to PYTHIA~8~\cite{Sjostrand:2014zea} Monash~2013 tune~\cite{Skands:2014pea}.
PYTHIA reproduces the $\pi^{0}$ spectrum well at the two lower centre-of-mass energies, while it overpredicts the data by 10--20\% for $\sqrt{s} = 7$ and $8$~TeV.
Concerning the $\eta$ meson, the Monash~2013 tune reproduces the measured spectra for $\pT>1.5$~\GeVc{} within uncertainties at all energies.
At lower transverse momenta %
it deviates significantly in magnitude and shape from data.
Consequently, at higher transverse momenta, the $\eta/\pi^0$ ratio is underestimated by about 10--15\% in this PYTHIA~8 tune.
The tuning parameters of the soft QCD part of PYTHIA apparently fail to describe the measured $\eta$ meson spectrum below $\pT<1.5$~\GeVc, whereas further tension is observed up to $\pT\approx3.5$~\GeVc~\cite{Acharya:2017tlv}.
This suggests that these data will provide significant constraints 
on fragmentation functions and for modeling of hadronisation effects in MC event generators.

Using the electromagnetic calorimeter, ALICE has also measured the inclusive production cross section of isolated photons at midrapidity in pp collisions at $\sqrt{s}=7$~TeV for the transverse momentum interval $10<p_{\rm T}^{\gamma}<60$~GeV/$c$~\cite{ALICE:2019rtd}. This cross section is dominated by prompt photons from hard-scattering processes (the isolation cut suppresses  photons from parton fragmentation) and it is well described by NLO pQCD calculations. 
The data sample collected during LHC Run 2 at $\sqrt{s}=13$~TeV will enable an extension of the measurement up to a $\pT^{\gamma}$ of about 200~GeV/$c$.         %

\subsubsection{Open heavy flavour hadrons} 
\label{sec:open-heavy-flavor}

The study of the production of hadrons containing heavy quarks, i.e. charm and beauty, in \pp{} collisions at LHC energies is a sensitive test of QCD calculations based on the factorisation approach. 
Factorisation is implemented in terms of the squared momentum transfer $Q^2$ (collinear factorisation)~\cite{Collins:1989gx} or of the partonic transverse momentum $\kt$~\cite{Catani:1990eg}. At LHC energies, calculations based on collinear factorisation are available in the general-mass variable-flavour-number scheme, GM-VFNS~\cite{Kniehl:2004fy,Kniehl:2005mk,Kniehl:2012ti,Helenius:2018uul}, and in the fixed-order plus next-to-leading logarithms approach, FONLL~\cite{Cacciari:1998it,Cacciari:2012ny}, both of them having NLO accuracy with all-order resummation of next-to-leading logarithms. Recently, fixed-order calculations with NNLO accuracy became available for charm-quark~\cite{dEnterria:2016ids} and beauty-quark~\cite{Catani:2020kkl} production cross sections. 
Within the $\kt$-factorisation framework, heavy-flavour production cross section calculations exist only at LO in $\alpha_{\rm s}$~\cite{Luszczak:2008je,Maciula:2013wg,Catani:1990eg}. 
The FONLL, GM-VFNS, and $\kt$-factorisation frameworks utilise different fragmentation functions, all based on dedicated fits to \ee data~\cite{Cacciari:2005uk,Kniehl:2020szu,Peterson:1982ak}. 

Precise measurements of the $\pt$-differential cross sections of charm and beauty hadrons provide a fundamental test for pQCD-based calculations, to validate the factorisation assumption and, indirectly, the ingredients used in the calculations. This is important also to constrain the charm- and beauty-quark $\pt$-differential spectra, which are used as input ingredients in models describing the heavy-quark interaction in the QGP. These models often rely on FONLL calculations.

\begin{figure}[!t]
\centering{}
\includegraphics[scale=0.8]{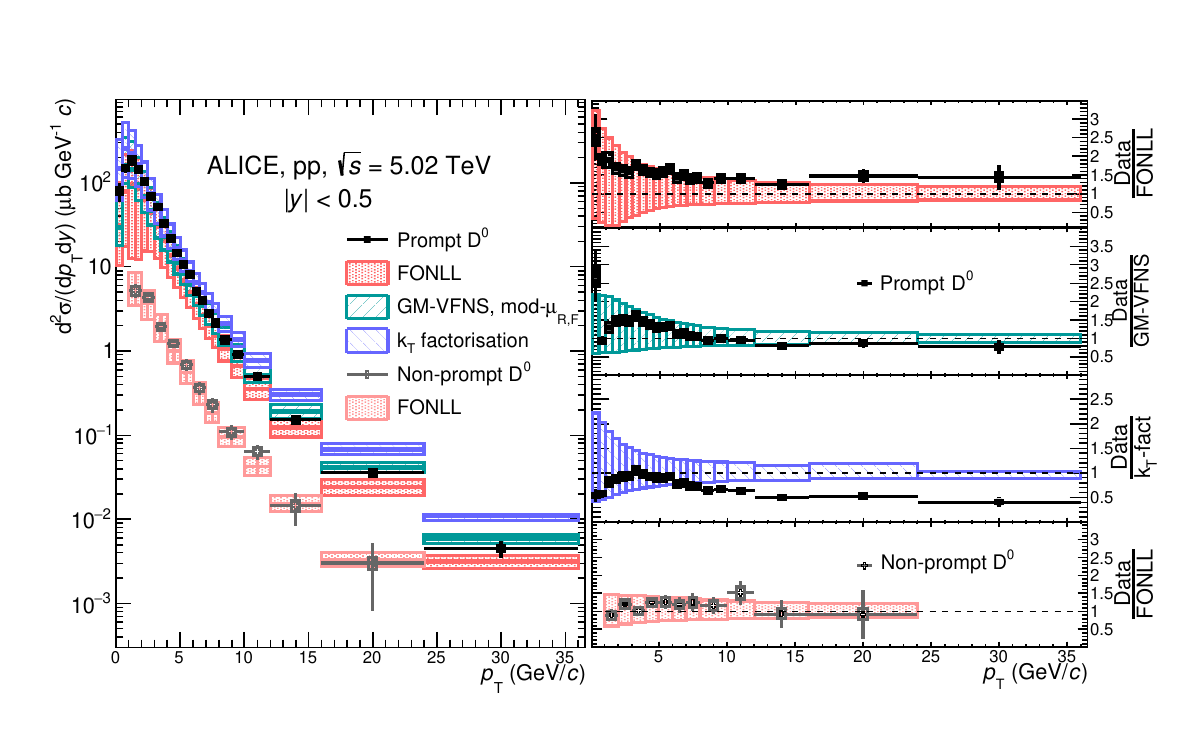}
\caption{(Left) $\pt$-differential cross sections of prompt and non-prompt $\Dzero$ mesons in pp collisions at $\s=5.02$~TeV~\cite{ALICE:2019nxm, ALICE:2021mgk} compared with various pQCD-based theoretical calculations. (Right) Ratios of $\pt$-differential production cross sections of prompt and non-prompt $\Dzero$ measurements to theoretical predictions.}
\label{fig:DnpD}
\end{figure}

In Fig.~\ref{fig:DnpD} the $\pt$-differential cross section of prompt and non-prompt $\Dzero$ mesons, the latter coming from the decays of beauty hadrons, in $|y|<0.5$ in pp collisions at $\sqrt{s}=5.02$~TeV is reported and compared to theoretical predictions. 
For prompt $\Dzero$ mesons~\cite{ALICE:2019nxm}, the data points are systematically higher than the FONLL central values by a factor 1.5--2 but remain  compatible within the theoretical uncertainty, while the measured non-prompt $\Dzero$-meson cross section~\cite{ALICE:2021mgk} is closer to the central value of the prediction. This can be better appreciated from the ratio of the data to the theoretical predictions, shown in the right panel of Fig.~\ref{fig:DnpD}. 
For the GM-VFNS calculation, the central values tend to underestimate the data at low- and intermediate-$\pt$ and to overestimate them at high-$\pt$. 
The $k_{\rm T}$-factorisation predictions describe the data at low- and intermediate-$\pt$ within uncertainties, but overshoot them for $\pt > 7$~GeV/$c$. The data-to-theory ratios indicate that the shape of the $\pt$-differential cross section is better reproduced by the FONLL calculation, which shows a flatter ratio than GM-VFNS and $k_{\rm T}$-factorisation for $\pt>2$~GeV/$c$.
It is worth noting that in the case of charm production, the uncertainties on theoretical predictions, which are dominated by the choice of the factorisation and renormalisation scales, are significantly larger than the uncertainties on the measured data points. 

At forward rapidity (2.5 $< y <$ 4), similar observations are derived from the (\pT,$y$)-double-differential comparison of FONLL predictions with the cross section of muons from semi-leptonic decays of charm and beauty hadrons measured in pp collisions at $\sqrt{s}=5.02$~TeV~\cite{Acharya:2019mky}. The \pT- and $y$-differential cross sections as well as the \pT-differential cross section ratios between different centre-of-mass energies and different rapidity intervals are described, within experimental and theoretical uncertainties, by FONLL. The data, significantly more precise than FONLL calculations, set important constraints for pQCD calculations in a kinematic region important for probing PDFs at low Bjorken-$x$ values, down to about $10^{-5}$.

In the mentioned theoretical calculations, the FFs are typically parametrised from measurements performed in \ee{} or ep collisions~\cite{Braaten:1994bz,Gladilin:2014tba,Lisovyi:2015uqa}, under the assumption that the hadronisation of charm quarks into charm hadrons is a universal process across different colliding systems. 
This assumption may be broken by higher-twist effects or other effects related to the heavy-quark kinematics and the pp underlying event~\cite{Collins:1998rz,Ellis:1982cd,Alekhin:2017kpj}. 
Recent measurements of charmed-baryon production at midrapidity in pp collisions at various collision energies evidence higher baryon-to-meson cross section ratios for $\Lc/\Dzero$~\cite{Acharya:2017kfy,Acharya:2020uqi,Acharya:2020lrg}, $\SigmacZeroPlusPlus/\Dzero$~\cite{ALICE:2021rzj}
, and $\XicPlusZero{}/\Dzero$~\cite{Acharya:2017lwf,ALICE:2021psx,ALICE:2021bli} 
 at low-$\pt$ with respect to \ee{} and ep collisions. It is worth noting that those by ALICE are the first and, to date, unique measurements of $\XicPlusZero$ and $\SigmacZeroPlusPlus$ cross sections in hadronic collisions. The $\Omegac{}/\Dzero$ ratio was recently measured as well~\cite{ALICE:2022cop}, although without an absolute normalisation because of the lack of absolute measurements of the branching ratios of $\Omegac{}$ decays.
 
 \begin{figure}[!t]
\centering{}
\includegraphics[height=8cm,width=8cm,keepaspectratio]{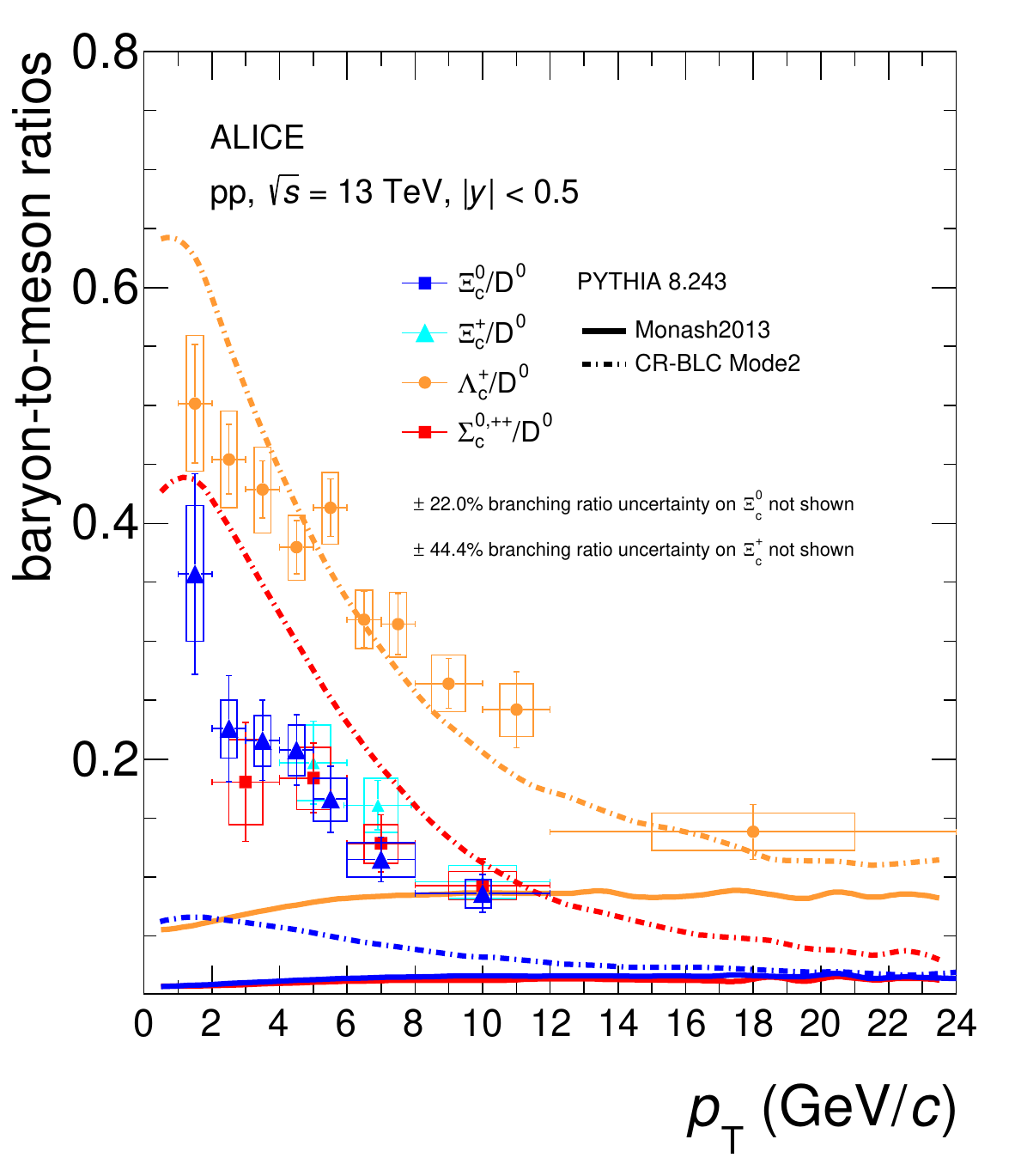}
\includegraphics[height=8cm,width=8cm,keepaspectratio]{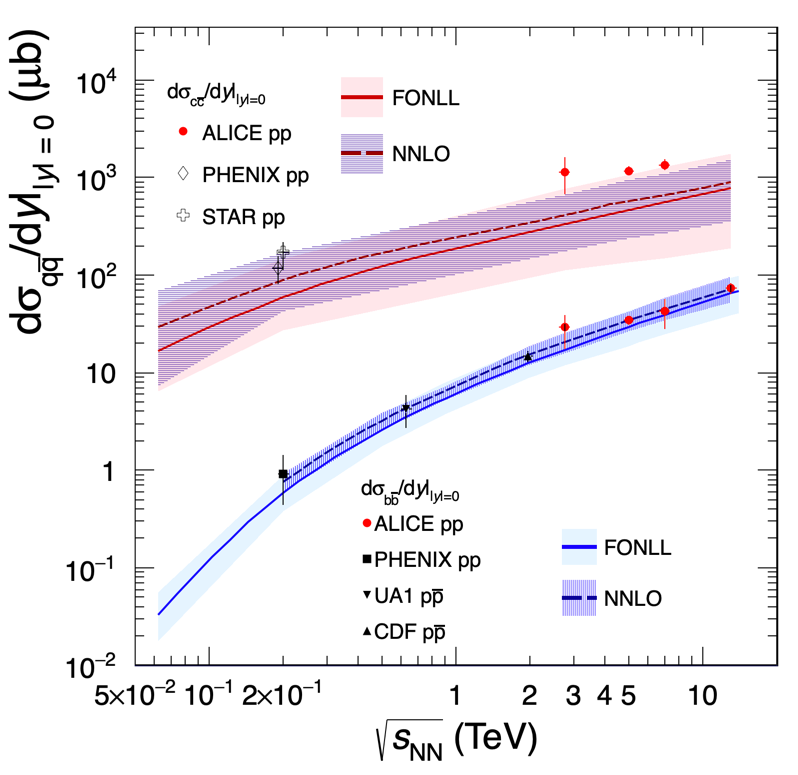}
\caption{(Left) Charmed baryon-to-meson $\Lc/\Dzero$, $\rm \Sigma_{c}^{0,++}/\Dzero$, and $\XicPlusZero{}/\Dzero$ ratios in pp collisions at $\sqrt{s}$ = 13~TeV, in comparison with PYTHIA~8. (Right) Charm and beauty production cross sections per unit of rapidity at midrapidity as a function of the collision energy~\cite{ALICE:2021dhb,Abelev:2014hla,Abelev:2012gx,ALICE:2021edd,ALICE:2021mgk,Adamczyk:2012af, Adare:2010de, Adare:2009ic,Acosta:2004yw,Albajar:1990zu}. STAR and PHENIX measurements in pp collisions at $\sqrt{s}=$ 200~GeV are slightly displaced along the horizontal direction for visibility. Comparison to the FONLL model calculations~\cite{Cacciari:1998it, Cacciari:2012ny} and NNLO calcuations~\cite{dEnterria:2016ids,Catani:2020kkl} are reported. }
\label{fig:ccbarTot}
\end{figure}

 The results at $\sqrt{s}=13$~\tev{} are reported in the left panel of Fig.~\ref{fig:ccbarTot}. The predictions from the default tune of PYTHIA~8.243 (Monash2013), essentially driven by $\rm{e}^+\rm{e}^-$ measurements, severely underestimate the ratios. Hadronisation in PYTHIA is based on the Lund string fragmentation model~\cite{Andersson:1983ia}. In the default PYTHIA tune strings are formed by colour-connecting partons in the Leading-Colour (LC) approximation, which allows to reproduce adequately $\rm{e}^+\rm{e}^-$ data. However, in pp collisions at LHC energies several partons are created via multiple-parton interactions and colour reconnections beyond LC topologies become important~\cite{Christiansen:2015yqa}. Among these, 3-leg or ``Y-shaped junction'' topologies have larger probabilities to fragment into baryons, increasing baryon production.
Indeed, a tune of PYTHIA~8 that implements colour reconnection beyond the leading-colour approximation (CR-BLC Mode2)~\cite{Christiansen:2015yqa} better describes the 
$\Lc/\Dzero$ and $\SigmacZeroPlusPlus/\Dzero$ ratios, though it underestimates the $\XicPlusZero{}/\Dzero$ data.
These measurements suggest that charm hadronisation involves mechanisms in pp collisions at LHC energies that do not play a role in  $\rm{e}^+\rm{e}^-$ collisions at LEP energies. Overall, the hadronisation process is not fully understood. Further investigations are needed to fully elucidate charm-hadron production across different colliding systems. In this respect, a detailed discussion of the $\Lc/\Dzero$ measurements in pp and Pb--Pb collisions is reported in Sec.~\ref{sec:MicroscopicHadronization}.

A consequence of the significant difference between the charmed baryon-to-meson ratios measured in pp and \ee{} and ep collisions is that charm-quark fragmentation fractions,  $f({\rm c} \rightarrow \Hc)$, i.e. the probabilities of a c quark to hadronise as a given charmed hadron species \Hc, estimated from \ee{} and ep  data, cannot be used to calculate the total charm-quark production cross section from the measurement of D-meson production cross section alone.  
Therefore, the \ccbar{} production cross section per unit of rapidity at midrapidity ($\de\sigma^{\ccbar}/\de y|_{|y| < 0.5}$) was calculated by summing the $\pt$-integrated cross sections of all ground-state charmed hadrons (\Dzero{}, \Dplus{}, \Ds{}, \Lc{}, and \XicZero{}) measured in pp collisions at $\sqrt{s}=5.02$~TeV.
The contribution of \Jpsi{} and \Omegac{} is considered negligible with respect to the other hadron species, as specified in Ref.~\cite{ALICE:2021dhb}. A 7\% uncertainty was assigned to account for the possibility that \Omegac{} production may be significantly larger than expectations from $\ee{}$ collisions, as a recent ALICE measurement and the Catania coalescence model suggest~\cite{ALICE:2022cop,Minissale:2020bif}. %

The charm fragmentation fractions $f({\rm c} \rightarrow \Hc)$ are then obtained by dividing the $\pt$-integrated cross section of each measured hadron species by the total charm cross section~\cite{ALICE:2021dhb}.  %
They are listed in Tab.~\ref{tab:FF}.
The fragmentation fractions measured in pp collisions at the LHC are different from those measured in \ee{} and ep collisions~\cite{Gladilin:2014tba}, providing %
evidence that the assumption of universality (colliding-system independence) of parton-to-hadron fragmentation functions is not valid for charm production. 

   \begin{table}[!t]
\caption{Charm fragmentation fractions into charmed hadrons, $f({\rm c} \rightarrow \Hc)$, measured in pp collisions at $\s = 5.02$~TeV~\cite{ALICE:2021dhb}, in comparison with the average of the LEP measurements~\cite{Gladilin:2014tba}. For the former, the statistical and systematic uncertainties are reported separately; for the latter, the first quoted uncertainty is the combination of statistical and systematic errors, while the second uncertainty originates from the limited knowledge of the relevant branching fractions.}
	\begin{center}
	\renewcommand\arraystretch{1.5}
	\begin{tabular}{|c|c|c|}
	\hline
	  $\Hc$  & ALICE $f({\rm c} \rightarrow \Hc$)\%  & average LEP $f({\rm c} \rightarrow \Hc$)\%\\
	  	\hline
	  \Dzero{} &  $39.1 \pm 1.7 ({\rm stat}) _{-3.7}^{+2.5} ({\rm syst}) $ & $54.2 \pm 2.4 (\rm unc.) \pm 0.7 (\rm BR)$ \\
	  \Dplus{} &   $17.3 \pm 1.8 ({\rm stat}) _{-2.1}^{+1.7} ({\rm syst})   $ &  $22.5 \pm 1.0 (\rm unc.) \pm 0.5 (\rm BR)$ \\
	  \Ds{} & $7.3 \pm 1.0 ({\rm stat}) _{-1.1}^{+1.9} ({\rm syst}) $ & $9.2 \pm 0.8 (\rm unc.) \pm 0.5 (\rm BR)$  \\
	  \Lc{} &  $20.4 \pm 1.3 ({\rm stat}) _{-2.2}^{+1.6} ({\rm syst}) $ & $5.7 \pm 0.6 (\rm unc.) \pm 0.3 (\rm BR)$  \\
	  \XicZero{} & $8.0 \pm 1.2 ({\rm stat}) _{-2.4}^{+2.5} ({\rm syst}) $  & - \\
	\hline
	\end{tabular}
	\end{center}
	\label{tab:FF}
\end{table}

In the right panel of Fig.~\ref{fig:ccbarTot}, ALICE measurements of the \ccbar{}~\cite{ALICE:2021dhb} and \bbbar{}~\cite{Abelev:2014hla,Abelev:2012gx,ALICE:2021edd,ALICE:2021mgk} production cross sections in pp collisions %
are shown as a function of the collision energy and compared to FONLL and NNLO~\cite{Catani:2020kkl,dEnterria:2016ids} predictions, as well as to results in pp and \ppbar{} at lower collision energies~\cite{Adamczyk:2012af, Adare:2010de, Adare:2009ic,Acosta:2004yw,Albajar:1990zu}.

For the \ccbar{} measurements at $\sqrt{s}= 2.76$~and~7~\tev{}, the D-meson cross sections and the FF determined at $\sqrt{s} = 5.02$~\tev\ are used to calculate the total cross section, as explained in~\cite{ALICE:2021dhb}.  
The measurements are higher than the upper edge of the  FONLL and NNLO calculation, though compatible within $\sim1\,\sigma$ of the experimental uncertainty. 
The STAR~\cite{Adamczyk:2012af} and PHENIX~\cite{Adare:2010de} results estimated in pp collisions at $\s = 200$~GeV, assuming fragmentation fractions measured at \ee{} collisions, are also shown.   

For the \bbbar{} cross section, the ALICE results at different energies are obtained in different manners: 
at $\sqrt{s} = 2.76$~\tev~ the results are extrapolated from the  production cross section
of electrons from semi-leptonic decays of beauty hadrons, at $\sqrt{s}~=~5.02$~\tev~the results are estimated from the averaged results of non-prompt D-meson and non-prompt \Jpsi measurements, and at $\sqrt{s} = 7$ and 13~\tev{} they are obtained via non-prompt \Jpsi measurements.
The ALICE measurements are shown along with other existing measurements at lower centre-of-mass energies, by PHENIX~\cite{Adare:2010de},  CDF~\cite{Acosta:2004yw} and UA1~\cite{Albajar:1990zu} Collaborations. 
The experimental results are found to be compatible with
FONLL and NNLO calculations.
Though, especially for the charm case, the data uncertainties are significantly smaller than the theory ones, it is worth noting that FONLL as well as NNLO calculations %
can reproduces within uncertainties heavy-quark cross sections varying by 3 orders of magnitude over a factor 65 variation in collision energy. 

These measurements not only provide further constraints to pQCD calculations but are also fundamental as a reference for the investigation of the charm- and beauty-quark interaction with the medium formed in heavy-ion collisions.

\subsubsection{Quarkonium} 

Quarkonia are bound states of either a \Qc\AQc\ (charmonia) or a \Qb\AQb\ (bottomonia) quark pair (indicated generically with \Qq\AQq\ in the following). As for open heavy flavour production, the initial hard parton-parton  scattering process (mostly gluon-gluon scattering at LHC energy) that produces the \Qq\AQq\ pair can be described within pQCD, while the subsequent formation and evolution of the bound state are non-perturbative processes that involve long distances and soft momentum scales. 
Due to the large rest mass of both \Qq\ and \AQq\ valence quarks, their relative motion in bound states is non-relativistic thus making the formation and evolution of the quarkonium a unique test bench to study the properties of the fundamental strong force.   
Different theoretical approaches have been developed to describe the quarkonium production, for comprehensive reviews see e.g. Ref.~\cite{Brambilla:2010cs, Andronic:2015wma}. Among these, in the colour evaporation model~\cite{Fritzsch:1977ay, Amundson:1996qr} it is assumed that all \Qq\AQq\ pairs with an invariant mass above the threshold for heavy-quark pair production and below twice the open heavy flavor (D or B meson) threshold production evolve into quarkonium states. In the Colour Singlet Model (CSM)~\cite{Baier:1981uk} the quantum state of the \Qq\AQq\ pair does not have any transition from its production to its hadronisation: only colour-singlet heavy-quark pairs are thus assumed to be produced. Instead, in the so called ``Non-Relativistic QCD'' (NRQCD) approach~\cite{Bodwin:1994jh} also colour-octet heavy-quark pairs can be initially formed and evolve towards colour-singlet bound states. 
In this approach, so called "Long Distance Matrix Elements" (LDMEs), which are postulated to be universal, i.e. independent of the energy and production process, 
express the probability for a \Qq\AQq\ pair to evolve into a given quarkonium state, and are determined from existing measurements.

Experimentally, prior to the LHC start, charmonium production in p\Ap\ and pp collisions was studied at the Tevatron~\cite{Abachi:1996jq, Abbott:1998sb, Acosta:2004yw,Abulencia:2007us} and RHIC~\cite{Adler:2004kf,Adare:2006kf,Adare:2010pol,Abelev:2010jp} colliders.  %
The available approaches were not able to describe in a consistent way the cross section, the $p_{\rm T}$ distribution and the polarisation measurements. 
The LHC experiments have provided a wealth of results for ground and excited charmonium states  in a 
wide range of phase space, and also for bottomonium states. %
  ALICE contributed with competitive and complementary measurements, thanks to its moderate magnetic field, low material budget of the barrel detectors, and the event selection strategy based on either minimum-bias trigger or %
  on a selection of the muon candidates in the forward muon arm down to transverse momenta as low as 0.5 GeV/$c$.
  In particular, ALICE measured  production cross sections of J/$\psi$ down to $p_{\rm T}=0$ %
   at midrapidity %
in the di-electron channel and of J/$\psi$, $\psi$(2S) and ${\rm \Upsilon(1S,2S,3S)}$ states at  forward rapidity in the dimuon channel~\cite{ALICE:2011zqe,Abelev:2012opp,Abelev:2012jjj,Abelev:2012mmm,Abelev:2014mmm,Adam:2016mmm,Acharya:2017mmm,Acharya:2019mmm, ALICE:2021dtt, ALICE:2021zkd, ALICE:2021edd, ALICE:2021qlw}. %
Additionally, thanks to the very low-$p_{\rm T}$ threshold on single muon in the triggers, ALICE also measured the polarisation of the J/$\psi$ meson reconstructed in the dimuon channel at forward rapidity and down to low-$p_{\rm T}$~\cite{Abelev:2012pol,Acharya:2018pol}.
\begin{figure}[!t]
\centering{}
\includegraphics[scale=0.26]{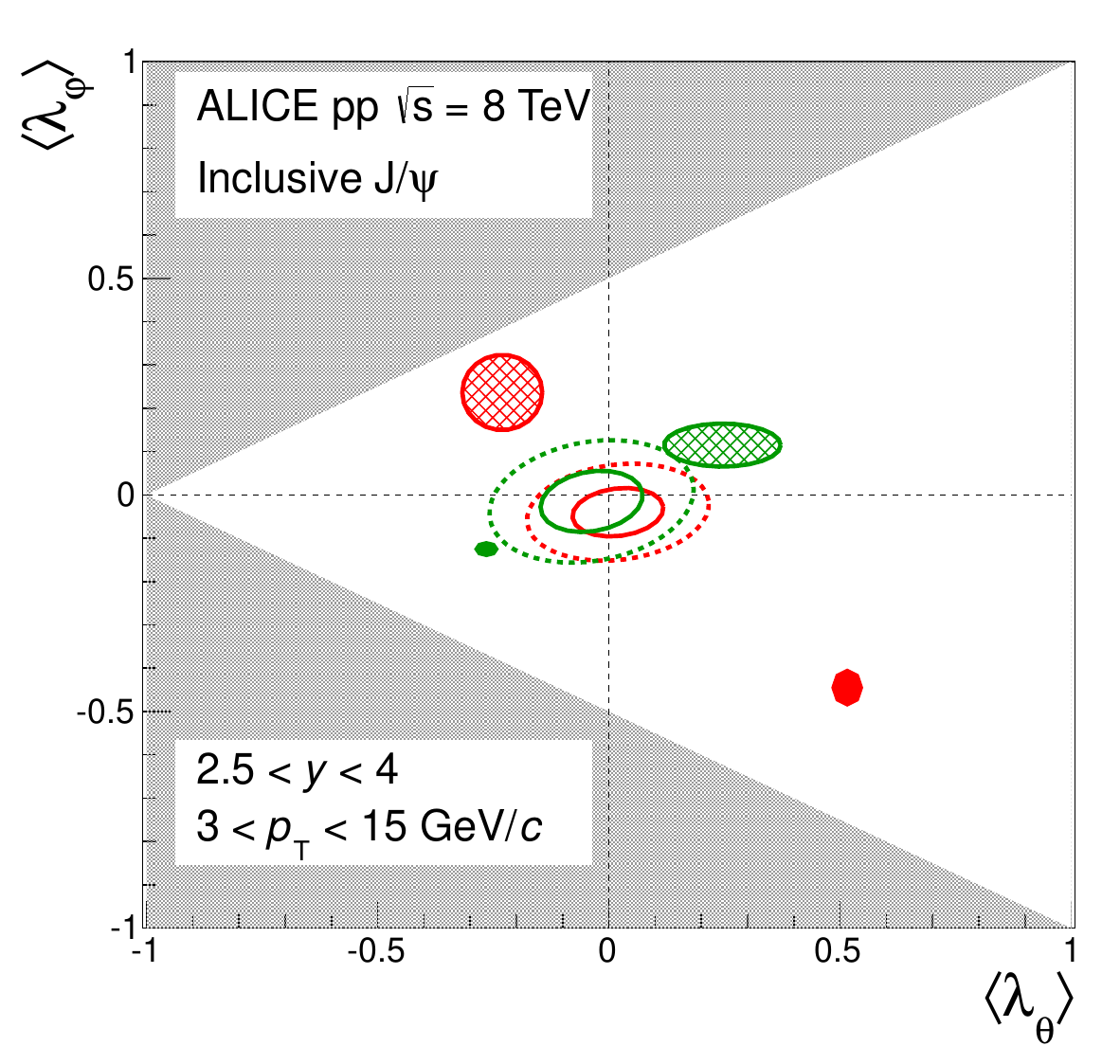}
\includegraphics[scale=0.26]{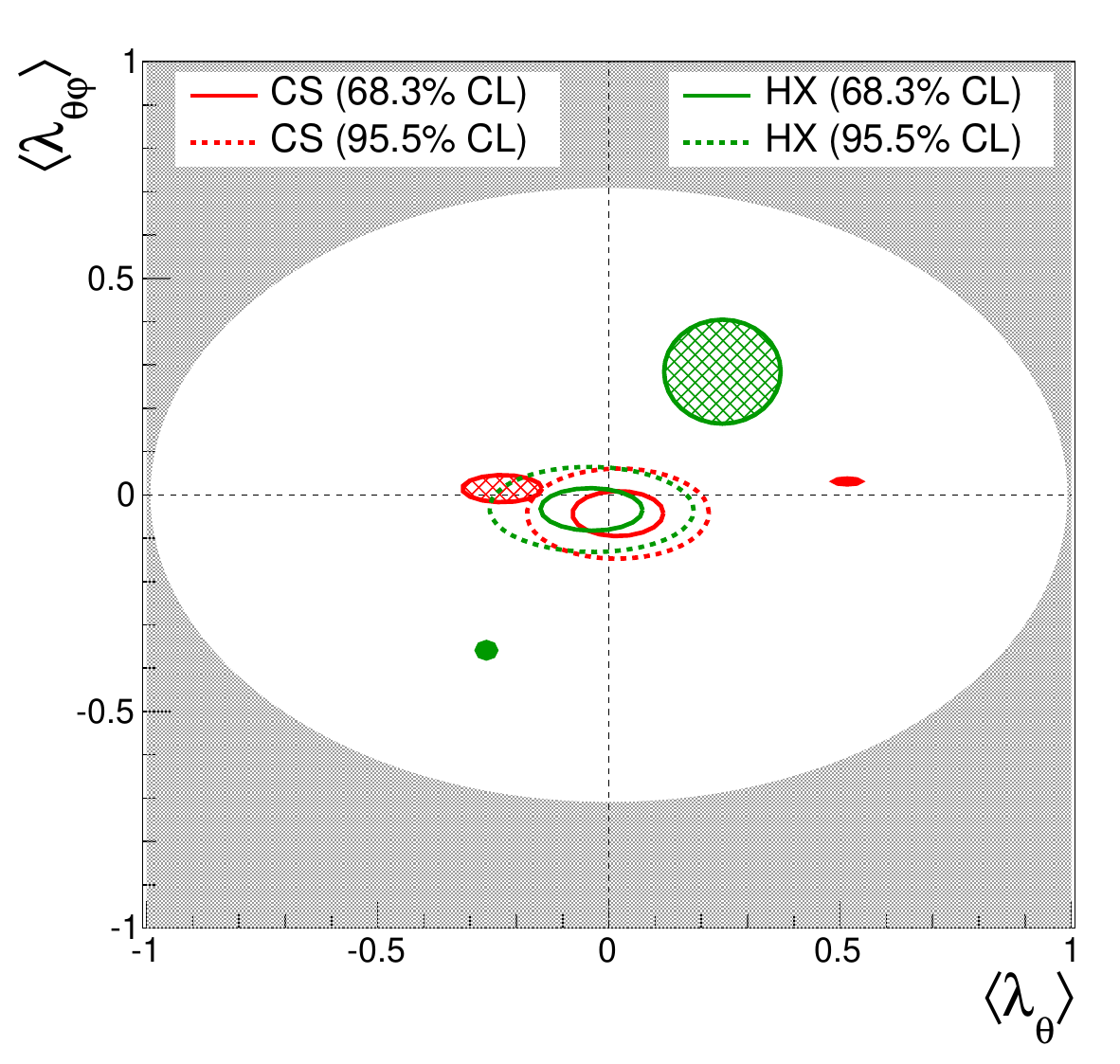}
\includegraphics[scale=0.26]{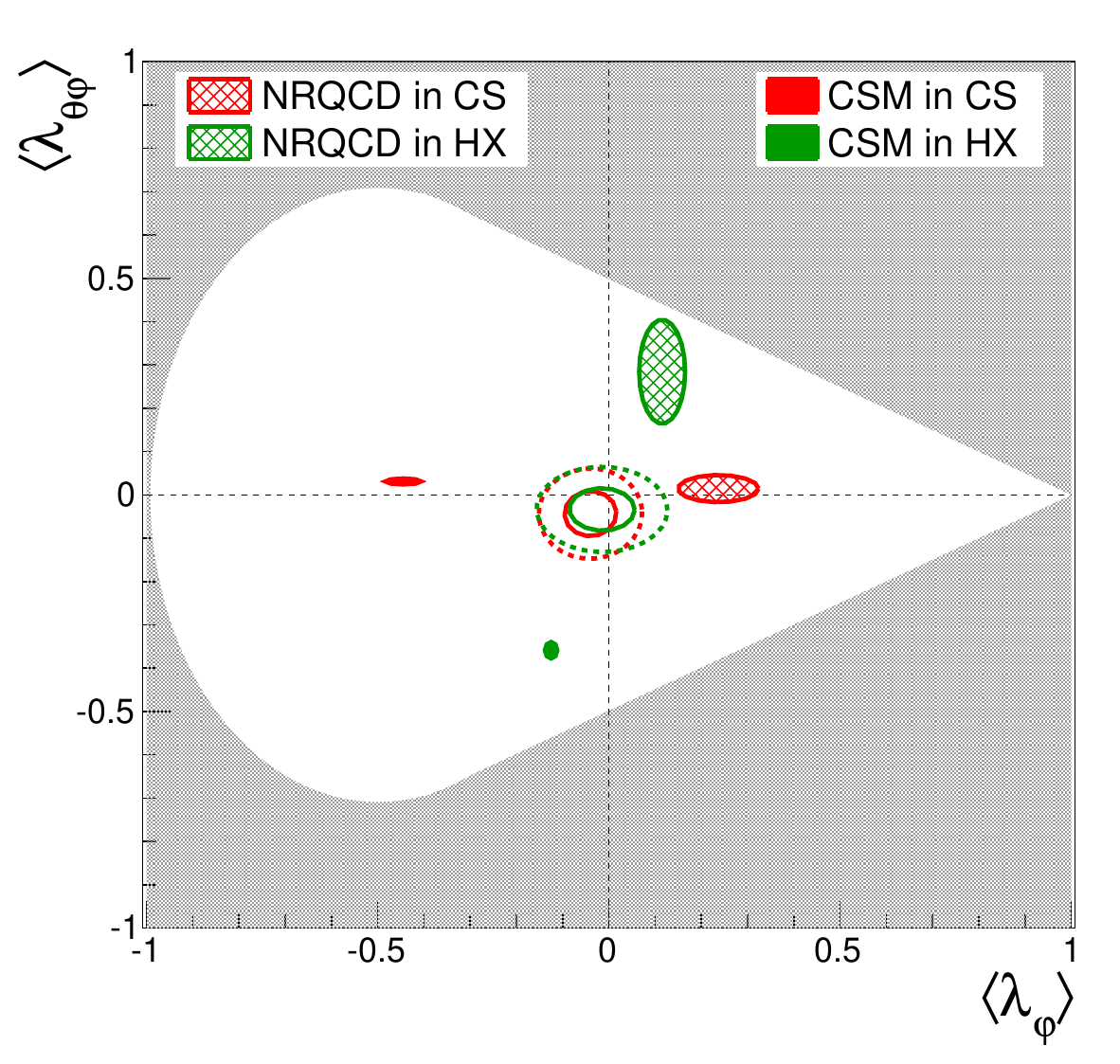}
\caption{ Inclusive J/$\psi$ polarisation parameters measured for 
$3<p_{\rm T}<15$~GeV/$c$ and $2.5<y<4$ in pp collisions at $\sqrt{s}=8$~TeV. Full (dashed) contour lines show the 1$\sigma$ (2$\sigma$) confidence intervals in the Collins-Soper (CS, red contours) and helicity (HX, green contours)) frames~\cite{Acharya:2018pol}. Filled contours correspond to predictions of either the CSM (full filled contours) or the NRQCD model (shaded filled contours) in the corresponding frames (CS in red and HX in green).}
\label{fig:JPSIpol}
\end{figure}

Despite %
significant recent progress on the theory side, see 
e.g.~\cite{Cheung:2018tvq,Cheung:2018upe,Lansberg:2020rft,Campbell:2007ws,Artoisenet:2008fc,Butenschoen:2012qr}, none of the approaches can still provide a satisfactory description of the different observables. 
The case of the NRQCD approach is illustrative: 
while on the one hand a global fit to J/$\psi$ production cross sections measured in hadroproduction, photoproduction, ${\rm e^+ e^-}$ collisions and $\gamma\gamma$ scattering constrained the LO colour octet LDMEs and demonstrated their relevance, the resulting predictions for the J/$\psi$ polarisation in pp at the LHC energy %
are in striking disagreement with measurements. 
Fig.~\ref{fig:JPSIpol} shows the ALICE results on the J/$\psi$ polarisation in pp collisions at $\sqrt{s}=8$~TeV as compared to predictions from the NRQCD model and CSM~\cite{Acharya:2018pol}. The averages of the $p_{\rm T}$-integrated inclusive J/$\psi$    polarisation parameters $(\lambda_\theta,\lambda_\phi,\lambda_{\theta\phi})$ in the allowed 2-D regions (white areas) are displayed as 1$\sigma$ (2$\sigma$) full (dashed) contour ellipses in Collins-Soper (in red) and helicity (in green) frames.     
On the other hand, within the same NRQCD approach, fits to both J/$\psi$ yield and polarisation measurements in hadro-production work reasonable well,  but predictions based on these fits have drastically failed in describing either J/$\psi$ results from other than hadronic collisions or the LHCb results on $\eta_{\rm c}$\ production~\cite{Aaij:2014bga,Butenschoen:2015eta,Han:2015eta,Zhang:2015eta}. 

It is evident that, also in the quarkonium sector, we do not have yet a thorough understanding of the hadronisation mechanism of heavy flavour quarks produced in different colliding systems at high energies.

\subsection{Direct experimental observation of the QCD dead cone} \label{sec:pp-deadcone}

The angular distribution of radiation emitted by a massive moving particle is predicted to depend on its mass-to-energy ratio. In particular, the radiation is to be suppressed for angles $\theta < m/E$. This ``dead cone'' effect known from classical theories conforming to special relativity and gauge theories in general has been worked out in QCD for massive quarks in Ref.~\cite{Dokshitzer:1991fc,Dokshitzer:1991fd}. As an indirect evidence of the effect, a difference in the charged particle multiplicity between events containing jets initiated by beauty quarks and those by light quarks was first reported at the LEP~\cite{DELPHI:2000edu}, but a direct observation was still missing. More recently, new methods traditionally applied to jet substructure measurements have been proposed in order to expose the dead cone of heavy quarks~\cite{Maltoni:2016ays,Cunqueiro:2018jbh}. Characterising the dead-cone effect experimentally is not only critical to the understanding of parton showers in vacuum, but also plays a central role in the studies of the physical properties of the QGP. The assessment of mass and flavour dependence of jet quenching effects is an area of active exploration in ultrarelativistic heavy-ion collisions (see also Sec.~\ref{sec:PartonInteractions}). 

\begin{figure}[!t]
\centering{}
\includegraphics[scale=0.8]{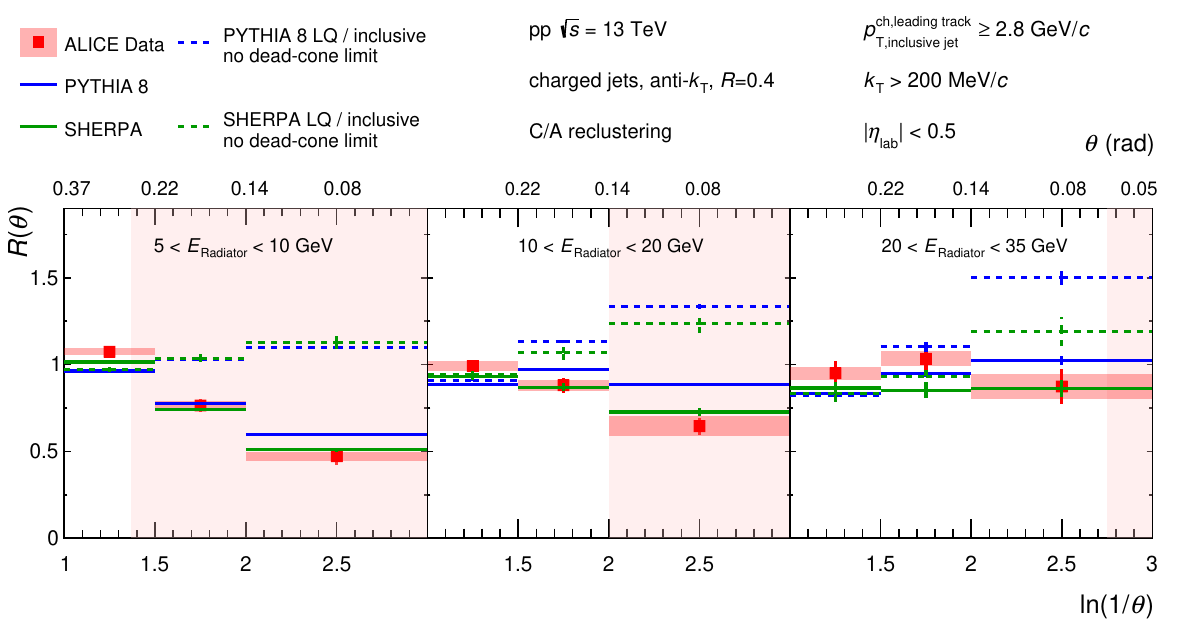}
\caption{The ratio of the angular distributions of emissions from $\rm{D}^{0}$-tagged to inclusive charged-particle jets in \pp{} collisions at $\sqrt{s}=13$ TeV~\cite{ALICE:2021aqk}, shown in three radiator energy intervals of $5 \leq E_{\rm{radiator}} < 10$ GeV/$\it{c}$, $10 \leq E_{\rm{radiator}} < 20$ GeV/$\it{c}$ and $20 \leq E_{\rm{radiator}} < 35$ GeV/$\it{c}$. A selection on the splitting scale of $k_{\rm{T}} > \Lambda_{\rm QCD}$ is applied to suppress hadronisation effects. The shaded regions correspond to the angles within which emissions are suppressed by the dead-cone effect, assuming a charm-quark mass of 1.275~GeV/$c^2$.}
\label{fig:deadcone-jet}
\end{figure}

The ALICE Collaboration has experimentally implemented the method proposed in Ref.~\cite{Cunqueiro:2018jbh}, using \pp\ collisions recorded at $\sqrt{s}=13$ TeV~\cite{ALICE:2021aqk}. Jets were reconstructed from charged particles with the anti-$k_{\rm{T}}$ algorithm and a resolution parameter of $R=0.4$. Charm quark initiated jets were tagged via the presence of a \Dzero{} (or $\mathrm{\overline{D}}^0$) meson among their constituents, reconstructed via its hadronic decay channel into  $\rm{K}^{+}\rm{\pi}^{-}$ ($\rm{K}^{-}\rm{\pi}^{+}$). The jets were then re-clustered using the Cambridge-Aachen~\cite{Cacciari:2008gp} algorithm, giving access to the angular ordered splitting tree. A primary Lund Plane~\cite{Dreyer:2018nbf} was populated by following the branch containing the \Dzero{} meson at each reclustering step and registering the kinematics of the emission. The branch containing the \Dzero{} meson at each splitting step coincided with the hardest branch to percent level accuracy. This allowed for comparison to a reference measurement in a (predominantly gluon-initiated) inclusive jet sample, where the reclustering procedure follows the hardest branch. To mitigate the bias imposed by the presence of a \Dzero{} meson, a minimum $p_{\rm{T}}$ selection on the leading particle of the leading branch was imposed for each accepted splitting. To suppress hadronisation effects a cut on the $k_{\rm{T}}$ of the reconstructed splittings~\cite{Cunqueiro:2018jbh} was applied. The dead-cone effect was then studied through projections of the Lund plane onto the splitting angle axis $\theta$. 

The ratios $R(\theta)$ of these angular distributions in intervals of the radiating prong energy $E_{\rm radiator}$ are shown in Fig.~\ref{fig:deadcone-jet} as a function of $\ln(1/\theta)$. A suppression of emissions at small angles is observed for charm-tagged jets compared to inclusive jets, with the magnitude of suppression increasing with decreasing radiator energy, as predicted by the dead-cone effect. Furthermore, the measurement was compared to MC simulations (PYTHIA~8 and SHERPA) for both the dead-cone and no dead-cone limits. These comparisons corroborate the observation that the suppression seen in data is compatible with the dead-cone effect directly related to the mass of the quarks. This measurement in the charm sector paves the way for future measurements both of the beauty quark dead cone, which will allow the mass dependence of the effect to be explored, as well as in relativistic heavy-ion collisions where a new dimension to studies of the mass dependence of energy loss can be added~\cite{Cunqueiro:2022svx,Dai:2022sjk,Andres:2023ymw}.             %

\subsection{Conclusions} 

In this chapter we have highlighted ALICE measurements
of high-$Q^2$ processes in \pp\ collisions that contribute to 
our fundamental understanding of QCD:

    \paragraph{Precision pQCD tests.} ALICE jet measurements provide stringent tests of state-of-the-art 
    pQCD calculations, demonstrating that NNLO and resummation contributions
    are important for inclusive jet production, and investigating the 
    boundary at which pQCD applies to jet substructure observables.
    \paragraph{Identified particle production.} ALICE measurements of identified particles including neutral mesons, charm and beauty hadrons, and quarkonia test descriptions of %
    hadronisation 
    in both the light and heavy flavour sector.
    Precise measurements of charm hadron production, including the measurements of some of the charmed baryons for the first time in hadronic collisions, allow us to compute the charm cross section at midrapidity and to measure the fragmentation fractions in \pp{} collisions, finding a large increase in charm baryon fractions and demonstrating  the non-universality of charm fragmentation compared to \ee{} and ep collisions. 
    \paragraph{Dead-cone effect.} Using reclustering techniques on charm-tagged jets, ALICE reported the first direct experimental
    observation of the QCD dead-cone effect in \pp{} collisions.

\newpage

\newcommand{\pp}           {pp\xspace}
\newcommand{\ppbar}        {\mbox{$\mathrm {p\overline{p}}$}\xspace}
\newcommand{\XeXe}         {\mbox{Xe--Xe}\xspace}
\newcommand{\PbPb}         {\mbox{Pb--Pb}\xspace}
\newcommand{\pA}           {\mbox{pA}\xspace}
\newcommand{\pPb}          {\mbox{p--Pb}\xspace}
\newcommand{\AuAu}         {\mbox{Au--Au}\xspace}
\newcommand{\dAu}          {\mbox{d--Au}\xspace}

\newcommand{\s}            {\ensuremath{\sqrt{s}}\xspace}
\newcommand{\snn}          {\ensuremath{\sqrt{s_{\mathrm{NN}}}}\xspace}
\newcommand{\pt}           {\ensuremath{p_{\rm T}}\xspace}
\newcommand{\meanpt}       {$\langle p_{\mathrm{T}}\rangle$\xspace}
\newcommand{\ycms}         {\ensuremath{y_{\rm CMS}}\xspace}
\newcommand{\ylab}         {\ensuremath{y_{\rm lab}}\xspace}
\newcommand{\etarange}[1]  {\mbox{$\left | \eta \right |~<~#1$}}
\newcommand{\yrange}[1]    {\mbox{$\left | y \right |~<~#1$}}
\newcommand{\dndy}         {\ensuremath{\mathrm{d}N_\mathrm{ch}/\mathrm{d}y}\xspace}
\newcommand{\dndeta}       {\ensuremath{\mathrm{d}N_\mathrm{ch}/\mathrm{d}\eta}\xspace}
\newcommand{\avdndeta}     {\ensuremath{\langle\dndeta\rangle}\xspace}
\newcommand{\dNdy}         {\ensuremath{\mathrm{d}N_\mathrm{ch}/\mathrm{d}y}\xspace}
\newcommand{\Npart}        {\ensuremath{N_\mathrm{part}}\xspace}
\newcommand{\Ncoll}        {\ensuremath{N_\mathrm{coll}}\xspace}
\newcommand{\dEdx}         {\ensuremath{\textrm{d}E/\textrm{d}x}\xspace}
\newcommand{\RpPb}         {\ensuremath{R_{\rm pPb}}\xspace}

\newcommand{\nineH}        {$\sqrt{s}~=~0.9$~Te\kern-.1emV\xspace}
\newcommand{\seven}        {$\sqrt{s}~=~7$~Te\kern-.1emV\xspace}
\newcommand{\twoH}         {$\sqrt{s}~=~0.2$~Te\kern-.1emV\xspace}
\newcommand{\twosevensix}  {$\sqrt{s}~=~2.76$~Te\kern-.1emV\xspace}
\newcommand{\five}         {$\sqrt{s}~=~5.02$~Te\kern-.1emV\xspace}
\newcommand{\twosevensixnn}{$\sqrt{s_{\mathrm{NN}}}~=~2.76$~Te\kern-.1emV\xspace}
\newcommand{\fivenn}       {$\sqrt{s_{\mathrm{NN}}}~=~5.02$~Te\kern-.1emV\xspace}
\newcommand{\LT}           {L{\'e}vy-Tsallis\xspace}
\newcommand{\GeVc}         {Ge\kern-.1emV/$c$\xspace}
\newcommand{\MeVc}         {Me\kern-.1emV/$c$\xspace}
\newcommand{\TeV}          {Te\kern-.1emV\xspace}
\newcommand{\GeV}          {Ge\kern-.1emV\xspace}
\newcommand{\MeV}          {Me\kern-.1emV\xspace}
\newcommand{\GeVmass}      {Ge\kern-.2emV/$c^2$\xspace}
\newcommand{\MeVmass}      {Me\kern-.2emV/$c^2$\xspace}
\newcommand{\lumi}         {\ensuremath{\mathcal{L}}\xspace}

\newcommand{\ITS}          {\rm{ITS}\xspace}
\newcommand{\TOF}          {\rm{TOF}\xspace}
\newcommand{\ZDC}          {\rm{ZDC}\xspace}
\newcommand{\ZDCs}         {\rm{ZDCs}\xspace}
\newcommand{\ZNA}          {\rm{ZNA}\xspace}
\newcommand{\ZNC}          {\rm{ZNC}\xspace}
\newcommand{\SPD}          {\rm{SPD}\xspace}
\newcommand{\SDD}          {\rm{SDD}\xspace}
\newcommand{\SSD}          {\rm{SSD}\xspace}
\newcommand{\TPC}          {\rm{TPC}\xspace}
\newcommand{\TRD}          {\rm{TRD}\xspace}
\newcommand{\VZERO}        {\rm{V0}\xspace}
\newcommand{\VZEROA}       {\rm{V0A}\xspace}
\newcommand{\VZEROC}       {\rm{V0C}\xspace}
\newcommand{\Vdecay} 	   {\ensuremath{V^{0}}\xspace}

\newcommand{\ee}           {\ensuremath{e^{+}e^{-}}} 
\newcommand{\pip}          {\ensuremath{\pi^{+}}\xspace}
\newcommand{\pim}          {\ensuremath{\pi^{-}}\xspace}
\newcommand{\kap}          {\ensuremath{\rm{K}^{+}}\xspace}
\newcommand{\kam}          {\ensuremath{\rm{K}^{-}}\xspace}
\newcommand{\pbar}         {\ensuremath{\rm\overline{p}}\xspace}
\newcommand{\kzero}        {\ensuremath{{\rm K}^{0}_{\rm{S}}}\xspace}
\newcommand{\lmb}          {\ensuremath{\Lambda}\xspace}
\newcommand{\almb}         {\ensuremath{\overline{\Lambda}}\xspace}
\newcommand{\Om}           {\ensuremath{\Omega^-}\xspace}
\newcommand{\Mo}           {\ensuremath{\overline{\Omega}^+}\xspace}
\newcommand{\X}            {\ensuremath{\Xi^-}\xspace}
\newcommand{\Ix}           {\ensuremath{\overline{\Xi}^+}\xspace}
\newcommand{\Xis}          {\ensuremath{\Xi^{\pm}}\xspace}
\newcommand{\Oms}          {\ensuremath{\Omega^{\pm}}\xspace}
\newcommand{\degree}       {\ensuremath{^{\rm o}}\xspace}
\newcommand{\Lambdac}      {\ensuremath{\Lambda_\textrm{c}^{+}}}

\newcommand{\LN}{\ensuremath{\mbox{$\Lambda$--N}}\xspace}
\newcommand{\SN}{\ensuremath{\mbox{$\Sigma$--N}}\xspace}
\newcommand{\pXim}{\ensuremath{\mbox{p--$\Xi^{-}$}}\xspace}
\newcommand{\chiEFT}       {\ensuremath{\chi}\rm{EFT}\xspace}
\newcommand{\La}{\ensuremath{\Lambda}\,}
\newcommand{\aLa}{\ensuremath{\overline{\Lambda}}\,}
\newcommand{\pL}{\ensuremath{\mbox{p--$\Lambda$}}~}
\newcommand{\NXi}{\ensuremath{\mbox{N--$\Xi$}}~}

\section{ALICE contributions beyond QCD physics and synergies of heavy-ion physics with other fields}
\label{ch:ConnectionsOtherFields}

The ALICE Collaboration has contributed not only to the field of QCD physics but also to some other areas of physics. The corresponding measurements are discussed in the following and show also synergies of heavy-ion physics with other fields. These range from connections of ALICE results to the physics of neutron stars, cosmic particles and exotica searches, to the test of fundamental symmetries, theoretical concepts as the AdS/CFT correspondence, the chiral magnetic effect, ultracold gases, Bose-Einstein condensates, and new developments in machine learning.

\subsection{Neutron stars and the nuclear equation of state}
\label{ch:neutronstars}
The interaction of hyperons (Y) with nucleons (N) is one of the key ingredients needed to understand the composition of the most dense stars in our Universe: neutron stars (NS)~\cite{Ozel:2016oaf,Riley:2019yda}.
Neutron stars are one possible final outcome of supernova explosions and are  characterised by large masses ($M\approx 1.2$--2.2 solar masses $M_\odot$) and small radii ($R\approx 9$--13~km)~\cite{Demorest:2010bx,Antoniadis:2013pzd,Cromartie:2019kug}. 
In the standard scenario, the gravitational pressure is typically counter-balanced by the Fermi pressure of neutrons in the core, which, along with electrons, are the only remnants from the mother-star collapse. 
A large interest in this topic has been triggered also from the recent measurements of gravitational wave signals from NS mergers~\cite{TheLIGOScientific:2017qsa,Monitor:2017mdv}, which opened a new gate to the properties of matter inside NS.

The high-density environment ($\rho\approx4\,\rho_0$, with $\rho_0$ being the nuclear density~\cite{Glendenning:1991es,Schaffner:1995th,Balberg:1997yw,Baldo:1999rq,Vidana:2000ew,Tolos:2020aln}) supposed to occur in the interior of NS leads to an increase in the Fermi energy of the nucleons, translating into the appearance of new degrees of freedom such as hyperons. This energetically-favoured production of strange hadrons induces a softening of the Equation of State (EoS)\footnote{A stiff (or hard) EoS is one where the pressure increases strongly for a given increase in density. Such a material would be harder to compress and offers more support against gravity. A soft equation of state leads to a smaller increase of pressure for a change in density and the material is more easy to compress.}.
The mass as a function of the radius is defined by the EoS through the solution of the Tolman–Oppenheimer–Volkoff equations, hence the mass-radius relation strongly depends on the constituents of the NS and on their interactions.
 
The inclusion of hyperons leads to NS configurations unable to reach the current highest mass limit from experimental observations of close to $3\,M_\odot$~\cite{TheLIGOScientific:2017qsa,Lattimer:2019eez,Lattimer:2021emm}.
For this reason, the presence of hyperons inside the inner cores of NS is still under debate, and this ``hyperon puzzle'' is far from being solved~\cite{Djapo:2008au,Tolos:2020aln}. A possible way out is represented by two-body and three-body repulsive YN and YNN interactions. In both cases, a sufficiently strong YN or YNN repulsive interaction can push the appearance of hyperons to larger densities, limiting the possible presence of these particle species inside NS, stiffening the EoS and leading to larger star masses.

The hyperon--nucleon two-body and three-body interactions in vacuum are used as the input for calculations at finite density, relevant for determining the baryon content inside NS, hence high-precision data are required in order to provide solid constraints on the available theoretical models. 
Currently, ALICE femtoscopic measurements in small systems (\pp, \pPb) for baryon-baryon pairs involving hyperons deliver the most precise data on the residual strong interaction between nucleons and strange hadrons (see Secs.~\ref{ch:femto_hyppo} and~\ref{ch:HeavyHyperons}).

The femtoscopic measurements of the \pL strong interaction represents one of the major achievements in this sector.
The $\Lambda$ baryons are typically the first hyperon species that are produced inside NS, due to their low mass. Their appearance is also theoretically favoured by the overall attractive potential,
$U_\Lambda = -30$ MeV, that a $\Lambda$ feels at the saturation
 density~\cite{Hashimoto:2006aw}. The results obtained by the ALICE Collaboration on this system support recent \chiEFT calculations in which an even more attractive interaction of the \La with the surrounding nucleons, due to the \LN$\leftrightarrow$ \SN dynamics, is predicted. In this case the early appearance of \La hyperons in neutron matter will lead to a too soft EoS and ultimately to a too low mass limit of NS configurations. Such a scenario, in order to be consistent with the astrophysical constraints on NS masses, requires the introduction of repulsive forces that might be  present in other Y--N systems and the inclusion of three-body interactions.

Repulsive hyperon--nucleon--nucleon interactions, such as $\Lambda$--N--N, have already been included in several approaches to obtain a stiffer EoS~\cite{Haidenbauer:2016vfq,Lonardoni:2014bwa}.
At the moment, however, calculations of three-body forces rely on the experimental measurements of hypernuclei ($^4 _\Lambda \mathrm{H}$, $^4 _\Lambda \mathrm{He}$) binding energies, in which the determination of the genuine $\Lambda$--N--N interaction is not straightforward and can be affected by many-body effects.
For this reason, the question of the role of three-body terms in the strangeness $|S|=1$ sector inside NS remains open.
ALICE has started to pioneer the study of the p--p--p and $\Lambda$--p--p interaction with three-body femtoscopic measurements~\cite{ALICE:2022vzr} and this sector will be further investigated with the much larger data samples expected during the LHC Runs 3 and 4.

Neutron star properties can also be studied in heavy-ion collisions at lower energies, e.g. HADES at GSI ($\sqrt{s_{\rm NN}} \approx 2.4$~GeV)~\cite{HADES:2022osk} and STAR at RHIC in the Beam Eenergy Scan~\cite{STAR:2017okv,STAR:2021yiu}. These energies corresponds to high-baryon densities and one directly tests certain regions of the EoS there. In particular, measurements of the direct flow observable $v_1$ can be connected to either momentum-dependent potential models or chiral effective field theory calculations involving protons, $\Lambda$ or even hypernuclei~\cite{Nara:2022kbb,Jinno:2023xjr}.

\begin{figure}[!t]
\centering
\includegraphics[width=\textwidth]{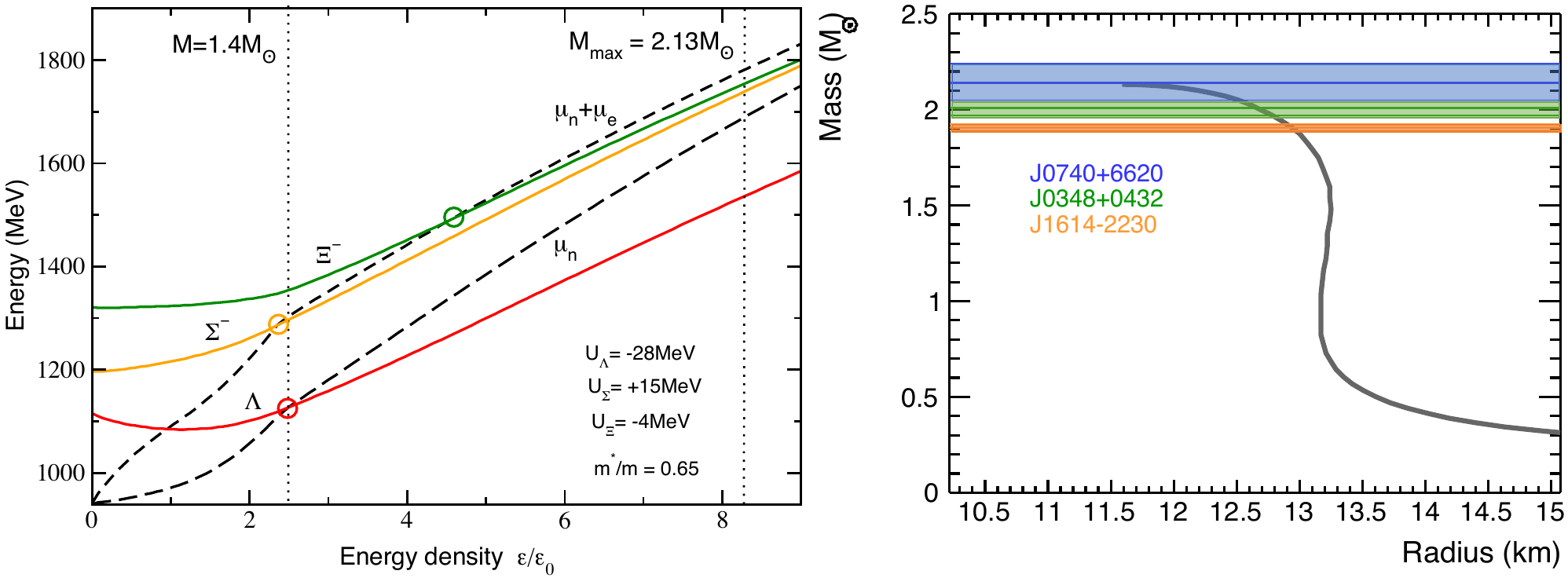}
\caption{(Left) Chemical potential $\mu_i$ of hyperons produced in the inner core of a NS as a function of the energy density, in units of energy density $\epsilon_0$ at the nuclear saturation point. The single-particle potentials depths in symmetric nuclear matter (SNM) for $\Lambda$, $\Sigma$ and $\Xi$ hyperons are displayed.  The vertical dotted lines indicate the central energy densities reached for a standard NS of $1.4\,M_\odot$ and for the maximum mass, $2.13\,M_\odot$, reached within this specific EoS. The mean-field calculations~\cite{Schaffner:1995th,Weissenborn:2011kb,Weissenborn:2011ut,Hornick:2018kfi} have been tuned in order to reproduce the lattice predicted value of $U_\Xi$ in pure neutron matter (PNM) obtained in~\cite{Inoue:2016qxt}, using the in-vacuum results validated by ALICE data in ~\cite{Acharya:2019sms}. The EoS obtained with these constraints provides a stable NS with a maximum mass of $M_{\rm max} = 2.13\,M_\odot$, as seen on the mass-radius plot on the right and is compatible with recent astrophysical measurements of heavy NS, indicated by the orange~\cite{Demorest:2010bx}, green~\cite{Antoniadis:2013pzd} and blue~\cite{Cromartie:2019kug} bands. See also Ref.~\cite{Fabbietti:2020bfg}. }
\label{fig:NS_fractions}
\end{figure}

A major improvement in the understanding of the role played by heavier strange hadrons within NS has been achieved by the validation of lattice QCD predictions for the \NXi interaction.
The ALICE measurements on \pXim pairs~\cite{Acharya:2019sms} (see also Sec.~\ref{ch:HeavyHyperons}) confirmed a strong attractive interaction between these two hadrons and provided a direct confirmation of lattice potentials~\cite{Sasaki:2019qnh}.
Using this two-body interaction to extrapolate to a neutron-rich dense system, a repulsive interaction potential of about $+6$ MeV is obtained~\cite{Inoue:2016qxt}. Currently, models for EoS including $\Xi$ hyperons assume large variations in the values of the single particle potential ($-40,+40$ MeV)~\cite{Weissenborn:2011kb} and hence the
validated lattice predictions impose a much more stringent
constraint.
In Fig.~\ref{fig:NS_fractions}, the chemical potentials $\mu_i$ of different particles, obtained from mean-field calculations~\cite{Schaffner:1995th,Weissenborn:2011kb,Weissenborn:2011ut,Hornick:2018kfi}, produced in the inner part of NS are shown as a function of the energy density.
The isospin coupling strengths of $\Sigma$ and $\Xi$ hyperons to the nucleons have been rescaled  in order to reproduce the current values constrained from scattering data and hypernuclei, and confirmed by ALICE femtoscopic measurements. 
The isovector couplings to the $\Xi$ have been adjusted to reproduce the predicted results in pure neutron matter (PNM) obtained from HAL QCD calculations at finite density~\cite{Inoue:2016qxt}, stemming from the predictions in vacuum validated by ALICE femtoscopic measurements on \pXim~\cite{Acharya:2019sms,Acharya:2020asf}. In the figure the corresponding values for the symmetric nuclear matter case are reported.
The production of neutral hyperons such as \lmb in the NS core becomes energetically favoured when the corresponding chemical potential is equal to the chemical potential of neutrons $\mu_{\rm n}$. In the left panel of Fig.~\ref{fig:NS_fractions}, this is occurring at an energy density $\epsilon \approx 2.5\,\epsilon_0$ where $\epsilon_0$ is the energy density of nuclear matter.
For charged hyperons, as the $\Sigma^-$ and $\Xi^-$, the contribution of the electrons present in the medium has to be taken into account, so that the condition to be fulfilled to start the conversion reads $\mu_i = \mu_{\rm n}+\mu_{\rm e}$.
The slight repulsion acquired by a $\Xi^-$ in pure neutron matter directly translates into the larger energy densities, and hence larger nuclear densities, for the
appearance of this hyperon species. 
In the right panel of Fig.~\ref{fig:NS_fractions}, the resulting mass--radius relation obtained by assuming the predicted HAL QCD $\Xi$ interaction in medium is shown. The production of cascade hyperons occurring at higher densities leads to a maximum NS mass of $2.13\,M_\odot$, which is compatible with the recent measurements, indicated by the coloured bands, of NS close to and above two solar masses~\cite{Demorest:2010bx,Antoniadis:2013pzd,Cromartie:2019kug}.

The results obtained in recent years from femtoscopic measurements in small colliding systems have proven that femtoscopy can play a central role in understanding the dynamics between hyperons and nucleons in vacuum. A comparison between hadronic models and these data is necessary in order to constrain calculations at finite density and to pin down the behaviour of hyperons in a dense matter environment. 
The great possibility to investigate, within the femtoscopy technique, different Y--N interactions and to extend the measurements to three-body forces, can finally provide quantitative input to the long-standing hyperon puzzle.

\subsection{Cosmogenic dark matter}
Thanks to the study of light nucleus and antinucleus production in high-energy collisions and to the measurement of low-energy antinuclei inelastic cross section (see Sec.~\ref{ch:NuclPhysLHC}), ALICE indirectly contributes to the search of dark matter. These measurements providing valuable input for the tuning of the cosmic-ray transport codes used in Monte Carlo simulations to interpret the data collected with experiments for dark matter searches. In addition, the predicted flux of antinuclei from dark-matter annihilation depends on the production mechanism and antinuclei transport properties within the interstellar medium, where ALICE data also is useful as input. The comparison of coalescence models with the ALICE production data on antinucleus production shed light on the results achieved in the indirect search of dark matter, since most of the utilised models are based on a coalescence process. 
Measurements of high-energy antimatter in the cosmic ray spectrum provide a powerful probe of new physics, including the annihilation or decay of dark-matter particles in the halo of the Milky Way~\cite{Bergstrom:1999jc,Hooper:2003ad,Profumo:2004ty,Bringmann:2006im,Pato:2010ih}. An excess of 10–20 GeV cosmic-ray antiprotons~\cite{Hooper:2014ysa,Cirelli:2014lwa,Cuoco:2016eej,Cui:2016ppb} has been identified in data from AMS--02~\cite{Aguilar:2016kjl} (and PAMELA~\cite{Adriani:2010rc}) with characteristics that are consistent with the annihilation of 50–90 GeV dark matter particles. This excess can, in a particular model, also be explained without annihilation of dark matter particles~\cite{Calore:2022stf}. In addition to $\gamma$ rays and antiprotons, dark matter annihilations can produce potentially detectable fluxes of heavier antinuclei, including antideuterons and antihelium~\cite{Donato:1999gy}.   
As kinematic considerations strongly suppress the production of heavy antinuclei in astrophysical processes, the detection of such particles could constitute a smoking-gun for dark matter annihilation. Intriguingly, the AMS-02 Collaboration has reported preliminary evidence of a few candidate antihelium events. Such a rate would be very surprising, as it would significantly exceed predictions from standard astrophysical processes or from annihilating dark matter~\cite{Carlson:2014ssa,Cirelli:2014qia}. A possible additional contribution to a ${}^{3}\overline{\rm He}$ signal observed in near-Earth experiments is the decay of heavy-flavoured hadrons, e.g.\,$\overline{\Lambda}_{\rm b} \rightarrow {}^{3}\overline{\rm He} + X$~\cite{Winkler:2020ltd}.

At the LHC, the ALICE Collaboration has studied d, $^{3}$He and $^{4}$He production in pp, p--Pb and Pb--Pb collisions at centre-of-mass energies per nucleon pair from 0.9 to 13 TeV~\cite{Adam:2015vda,Adam:2015pna,Acharya:2017fvb,Acharya:2017bso,Acharya:2019rgc,ALICE:2019bnp,Acharya:2020sfy} and the measured yields were interpreted by means of coalescence and statistical hadronisation models~\cite{Kapusta:1980zz,Sun:2018mqq,Bellini:2018epz} (see also Sec.~\ref{ch:nuclei}). The ALICE measurements, combined with different coalescence models, have been employed to estimate the antideuteron and antihelium flux from cosmic-ray interactions at the AMS-02 and GAPS experiments~\cite{Korsmeier:2017xzj,Kachelriess:2019taq,Kachelriess:2020uoh,Blum:2017qnn}. In addition, annihilation cross sections for antinuclei are barely known and for antideuterons the inelastic cross sections have been measured for several materials only for two momentum values, $p = 13.3$ GeV/$c$~\cite{Denisov:1971im} and $p = 25$ GeV/$c$ ~\cite{Binon:1970yu}. For antihelium, the inelastic cross section had never been measured. ALICE has contributed in this field by measuring for the first time the antideuteron inelastic cross section in p--Pb data at 5.02 TeV~\cite{Acharya:2020cee}.
The measurement used the ALICE detector material as an effective target, with mean charge number $\langle Z \rangle$ = 8.5 and mass number $\langle A \rangle$ = 17.4 for the antideuteron momentum range $0.3 <p < 0.9$ GeV/$c$, and $\langle Z \rangle$ = 14.8 and $\langle A \rangle$ = 31.8 for $0.9<  p < 4.0$ GeV/$c$.
Recently, these absorption measurements were extended to $^{3}\overline{\mathrm{He}}$ with pp and Pb--Pb data and using two different techniques~\cite{ALICE:2022zuz}: one was the same as antideuterons and the another one used one of the ALICE subdetectors as ``target"  ($\langle A \rangle$ = 34.7 in $1.17 <  p < 10$ GeV/$c$). 

\begin{figure}[htb]
    \begin{center}
    \includegraphics[width = 0.49\textwidth]{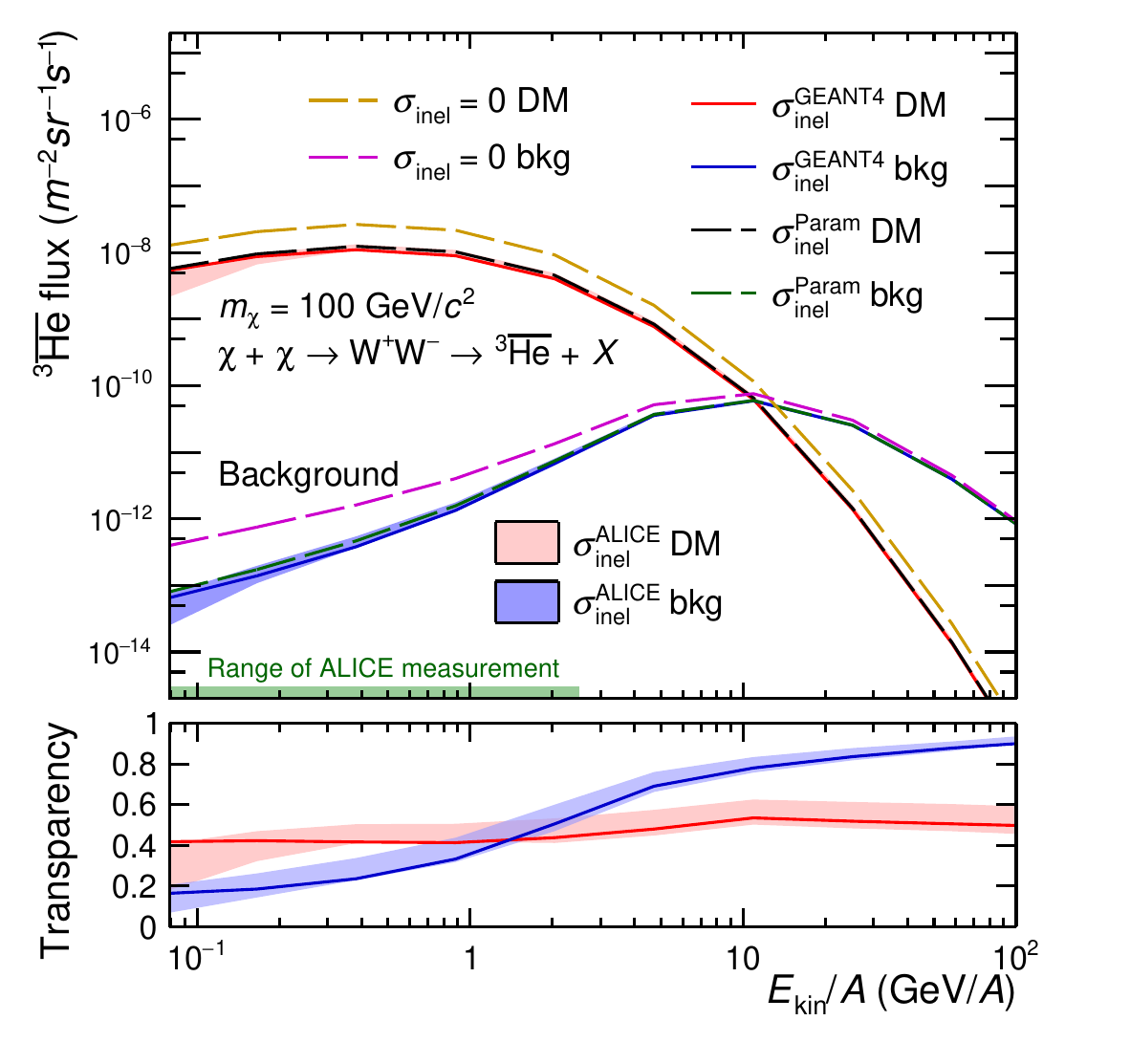}
    \includegraphics[width = 0.49\textwidth]{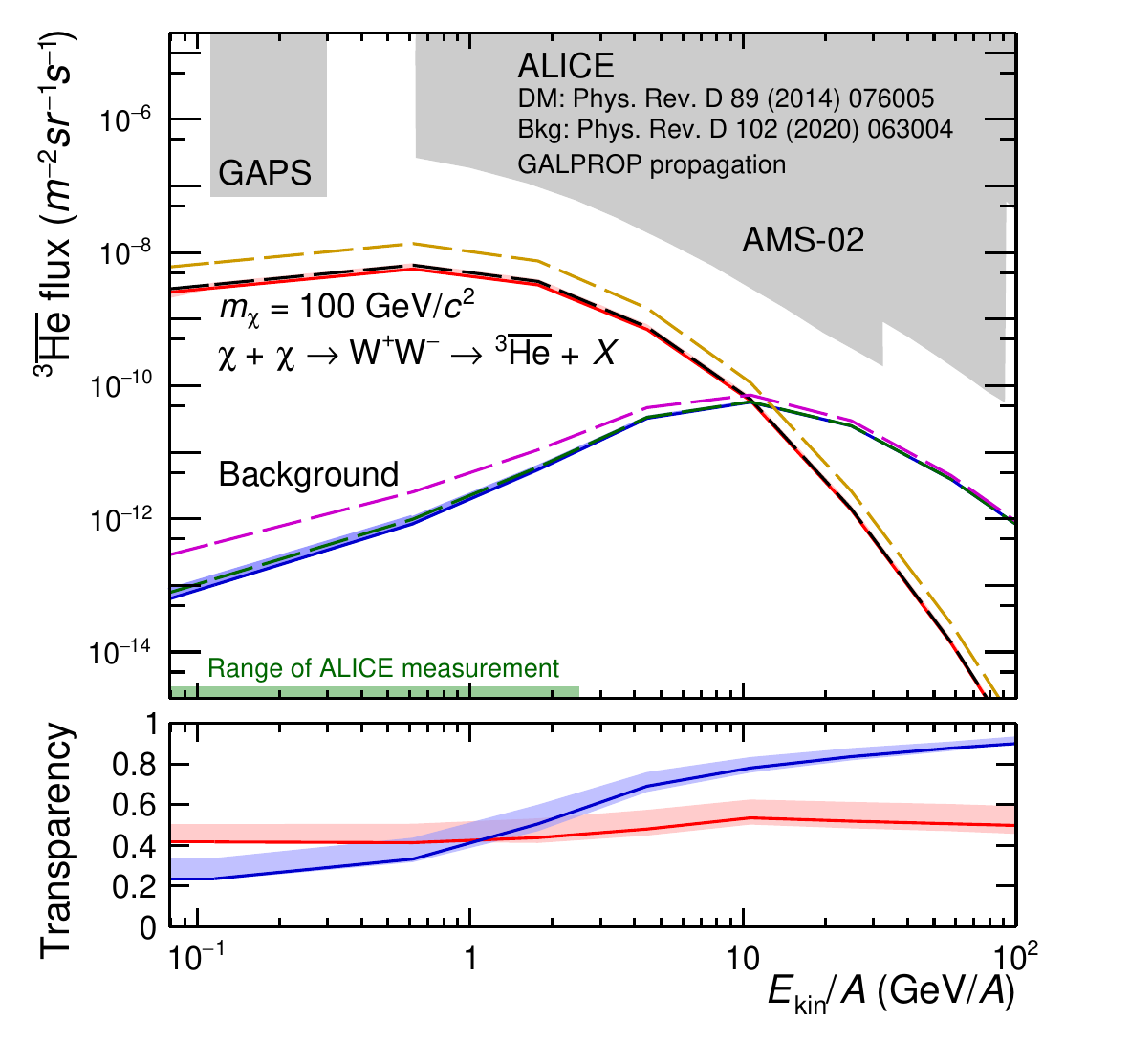}
    \end{center}
    \caption{Expected $^{3}\overline{\mathrm{He}}$ cosmic–ray fluxes near Earth before (Left) and after (Right) solar modulation~\cite{ALICE:2022zuz}. The latter are obtained using Force Field with modulation potential $f = 400$~MV. Upper panels show the fluxes of antihelium nuclei for signal (dark-matter decay, in red) and background (ordinary cosmic rays, in blue) for various cases of inelastic cross section used in the calculations. Bottom panels show the transparency of our galaxy to the propagation of $^{3}\overline{\mathrm{He}}$. Shaded areas on the top-right panel show expected sensitivity of the GAPS~\cite{GAPS:2020axg} and AMS-02~\cite{Korsmeier:2017xzj} experiments. \label{fig:flux}}
\end{figure}

The measured inelastic cross sections and their uncertainties can now be used in propagation models of antinuclei within the interstellar medium for dark-matter searches. Figure~\ref{fig:flux} shows an example of such a calculation using the GALPROP software package~\cite{Strong:1998pw}\footnote{GALPROP version 56 was used, available at https://galprop.stanford.edu.} and showing the comparison between the implemented GEANT4~\cite{GEANT4:2002zbu} absorption cross section~\cite{Uzhinsky:2011zz} and the measurement of ALICE~\cite{ALICE:2022zuz}. The GEANT4 estimation is consistent with the ALICE measurements, but the latter provide for the first time an direct experimental support and experimental uncertainties for this estimation. The obtained transparencies show that $^{3}\overline{\mathrm{He}}$ nuclei can propagate for long distances in our galaxy and can thus be used to study cosmic–ray interactions or possible dark-matter decays.
Future studies with larger data samples during the next LHC runs will represent a unique opportunity to extend these measurements and for the first time measure the inelastic cross sections of heavier antinuclei like ${}_{\overline{\Lambda}}^{3}\overline{\mathrm{He}}$ and $^{4}\overline{\mathrm{He}}$.

\subsection{Cosmic rays}

ALICE contributes to cosmic-ray measurements in two ways.  First, ALICE provides data that can be used to constrain hadronic interaction models (Monte Carlo) that are critical for interpreting measurements on particles reaching the ground, the height of shower maximum, etc., in terms of incident cosmic-ray energy and nuclear composition.  Second, ALICE makes direct measurements of cosmic-ray muon multiplicities and charges. 

\subsubsection{Data for cosmic-ray models}

At energies above about 1 PeV ($10^{15}$ eV), cosmic rays cannot be studied directly; instead, the resulting particle showers are measured in the atmosphere and at ground level.  Relating these observables to the incident cosmic-ray energy and nuclear composition requires a model for the resulting hadronic shower. The shower energy is estimated based on the density of particles reaching the ground (seen by scintillation and Cherenkov detectors), and at high energies via a quasi-calorimetric method, by measuring the
amount of visible light and/or radio waves emitted by the shower. The composition is usually inferred from the height at which the shower contains the maximum number of particles, $X_{\rm max}$ and the number of muons of different energies seen in the shower~\cite{Aab:2017cgk,Abbasi:2018wlq,IceCube:2019hmk}. The muon density, in particular, is sensitive to the hadronic physics as the air shower develops.  

The development of the shower as it travels through the atmosphere depends on the frequency of interactions (i.e.\,the cross section) and the number of particles produced at each interaction. ALICE measurements of multiplicities of several hadron species in pp and p--A collisions are important benchmarks for the air showers models. The baryon-to-meson ratio and the strangeness content also play a role, since baryons have a larger interaction cross section than pions, whereas kaons interact less, and therefore penetrate more deeply. The EPOS code, for example, has been tuned against ALICE data
\cite{Pierog:2013ria}.  %

Most muons and neutrinos in air showers come from the decay of pions and kaons.  Kaons have a shorter lifetime, so are more likely than pions to decay (producing a muon and a neutrino); this also affects the zenith angle distribution~\cite{Gaisser:2014pda}.  Therefore, strangeness enhancement in light systems (as from hadron-oxygen collisions) is critical for air showers. ALICE provides important data on strangeness production (see Secs.~\ref{sec:QGPHadronization} and~\ref{section:3.2}), allowing for tuning the corresponding models.

In a slightly different field, conventional neutrinos $\nu$ (from pion and kaon) and neutrinos produced directly in the primary p--A cosmic collision are an important source of background in astrophysical neutrino studies.  Although conventional neutrinos are fairly well studied~\cite{Aartsen:2015xup}, prompt neutrinos have not been seen~\cite{IceCube:2020acn,IceCube:2021uhz}.  Because of their harder energy spectrum, prompt $\nu$ are easier to confuse with astrophysical neutrinos, and are often included as a nuisance parameter in studies of diffuse astrophysical $\nu$. Measurements of charm production utilising pp collisions (see Sec.~\ref{sec:open-heavy-flavor}) are important in pinning down prompt neutrino calculations~\cite{Bhattacharya:2015jpa}.  Tighter constraints on prompt neutrinos would lead to better estimates of the astrophysical neutrino flux and energy spectrum~\cite{Bhattacharya:2016jce}. These can also improve models of neutrino production in astrophysical sources which include significant charm production~\cite{Enberg:2008jm}.   

High-$p_\textrm{T}$ (here $p_\textrm{T} \geq 2$ GeV/$c$) particle production has also been considered as measure of cosmic-ray composition.  Proton interactions reach higher $\sqrt{s_{\rm NN}}$ than iron nuclei of the same total (not per nucleon) energy, leading to more high-\pt muons. The advantage of this method is that the hadronic interactions can be described in perturbative QCD, rather than the more model-dependent approaches required at lower \pt. The IceCube Collaboration has studied high-\pt~muons (those with large lateral separation) and observed a power-law behaviour at large lateral separations~\cite{Abbasi:2012kza}.  This is as expected from pQCD, but the zenith angle distribution was not in particularly good agreement with any of the models. More precise data on charm production might improve the understanding, and lead to composition measurements. This refers to both, published measurements that have yet to be taken into account in model tuning, and also data with higher precision to be recorded in Run 3 of the LHC.

\subsubsection{ALICE cosmic muon measurements}

\begin{figure}[htb]
    \begin{center}
    \includegraphics[width = 0.9\textwidth]{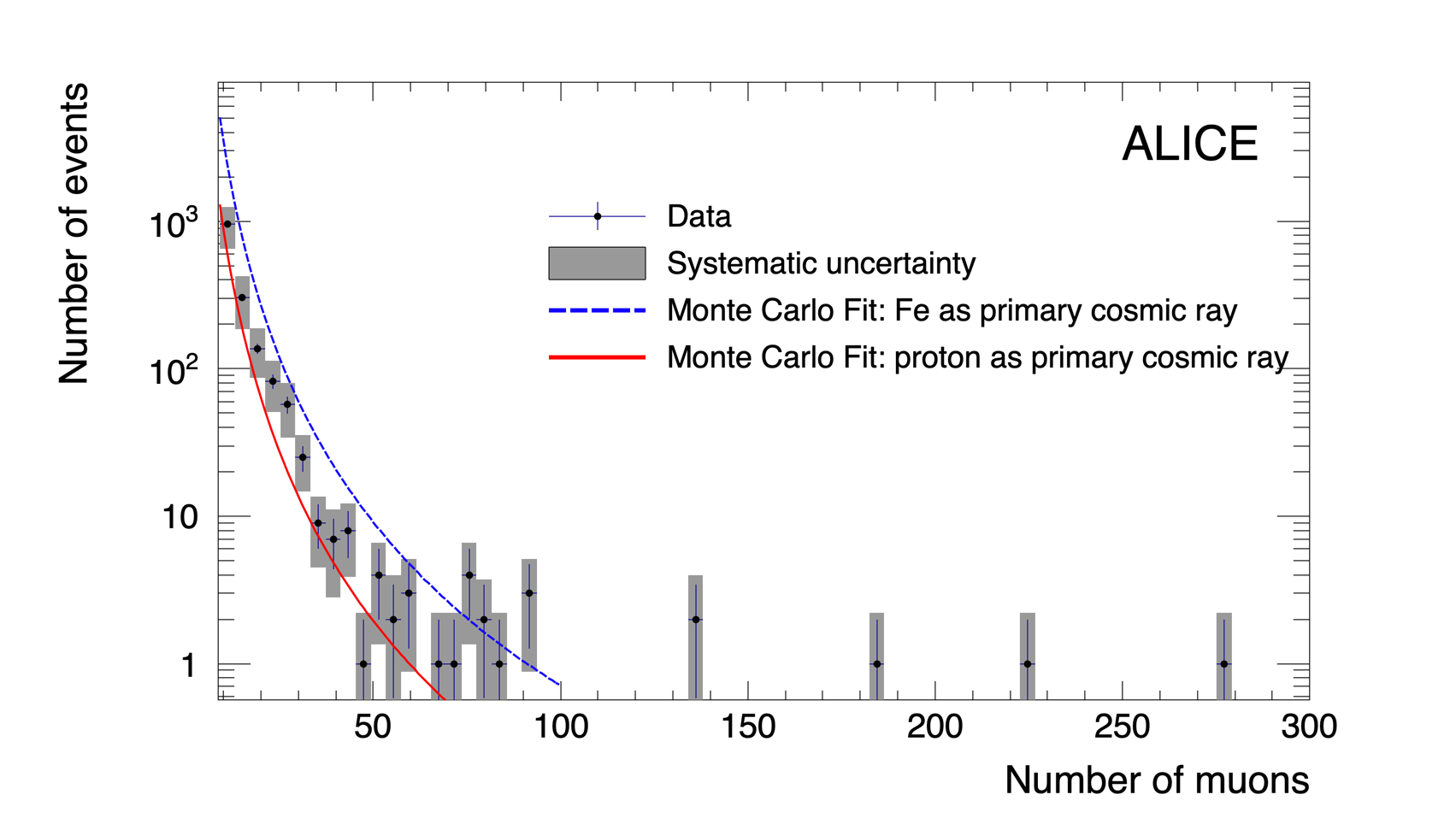}
    \end{center}
    \caption{Number of cosmic-muon events per interval of multiplicity 4 in 30.8 days of data-taking~\cite{ALICE:2015wfa}, compared with a model calculation. \label{fig:mmd}}
\end{figure}

Located 52 meters underground with 28 meters of rock overburden, ALICE has also studied the atmospheric muons coming from Extensive Air Showers (EAS), i.e.\,the shower of particles created by the collisions of cosmic rays with the nuclei of the atmosphere.
The use of collider detectors to study atmospheric muons was pioneered by the LEP experiments ALEPH~\cite{aleph}, DELPHI~\cite{delphi} and L3~\cite{l3}. These experiments concluded that the high-muon-multiplicity events occur with a frequency at almost one order of magnitude above expected.
In this context, ALICE began a cosmic-ray physics program taking 30.8 days of live time of cosmic data in the period 2010--2013, collecting more than 20 million events and more than 7000 multimuon events. The study of the muon multiplicity distribution (MMD) and the measurement of the rate of high muon multiplicity (HMM) events, defined as the events with more than 100 muons ($N_{\mathrm {\mu}}>100$) detected in the Time Projection Chamber (TPC), was described in~\cite{ALICE:2015wfa}. In this work, the experimental results on the MMD and the HMM events were compared with simulations using the CORSIKA\footnote{CORSIKA is a package for detailed simulations of extensive air showers initiated by high-energy cosmic-ray particles, that allows different models for the hadronic interaction to be used with it.} code~\cite{corsika} that utilizes the latest version of QGSJET~\cite{qgsjet_2011} to model hadronic interactions. The data have been compared with pure proton composition, representing the lightest cosmic ray composition that hits the atmosphere and produces the high-multiplicity muon events, and pure iron representing the heaviest composition.
As an example, Fig.~\ref{fig:mmd} shows a comparison of the measured MMD with the simulations in the intermediate range of muon multiplicity. The limited sample size does not allow for a quantitative study of the composition, but the measurement suggests a mixed ion primary cosmic-ray composition with an average mass
increasing with energy in accordance with most of the previous experiments.
The observed rate of HMM events is consistent with the predictions using a pure iron primary composition up to MMD values of about 70 and energies $E > 10^{16}$~eV. 
For the first time the rate of HMM events is well reproduced using
conventional hadronic interaction models and a plausible primary flux extrapolation.
This measurement places significant constraints on alternative, more exotic, production mechanisms.  

\subsection{Atomic physics through lepton-pair production}

Lepton-pair production, $\gamma A\rightarrow \ell^+\ell^- A$ ($AA\rightarrow \ell^+\ell^- AA$ in ultra-peripheral collisions (UPCs)\cite{Baltz:2007kq}) studies atomic physics at very high energies, testing QED in very strong fields.  Because they involve high-$Z$ nuclei at large Lorentz boosts, UPCs are important event classes that enable to study of pair production in ultra-intense electromagnetic fields.  In these high fields, higher-order diagrams should make a substantial contribution.  This was studied intensively in a series of calculations using coupled-channel equations and then all-orders QED calculations.  These calculations found very different cross sections for 
$AA\rightarrow \ell^+\ell^- AA$ before converging on a generally accepted result~\cite{Klein:2004is}. However, theorists have not generally considered higher order corrections to the cross section in a restricted acceptance such as that for ALICE.   Data from STAR 
\cite{Adams:2004zg,STAR:2019wlg}
and ATLAS~\cite{ATLAS:2020mve,Aaboud:2018eph}
are in general agreement with the expectations for lowest-order QED, except for the pair $p_\textrm{T}$ spectrum, which is broader than is expected in the naive calculation.   This is likely due to the limited UPC impact parameter ($b$) range, i.e.\,requiring $b>2R_A$, with $R_A$ being the nucleus radius; the STAR trigger imposes a more sophisticated restriction) for UPCs; when $b$ is limited, $\langle p_\textrm{T}\rangle$ should rise.  However, the calculations are not in complete agreement~\cite{Zha:2018tlq,Klein:2020jom,Brandenburg:2020ozx} and this is still under active study.   Effects such as incoherent photon emission ($A\rightarrow \gamma A^*$) and Sudakov resummation~\cite{Klein:2018fmp} can also increase $\langle p_\textrm{T}\rangle$. 

\begin{figure}[htb]
    \begin{center}
\includegraphics[width = 0.49\textwidth]{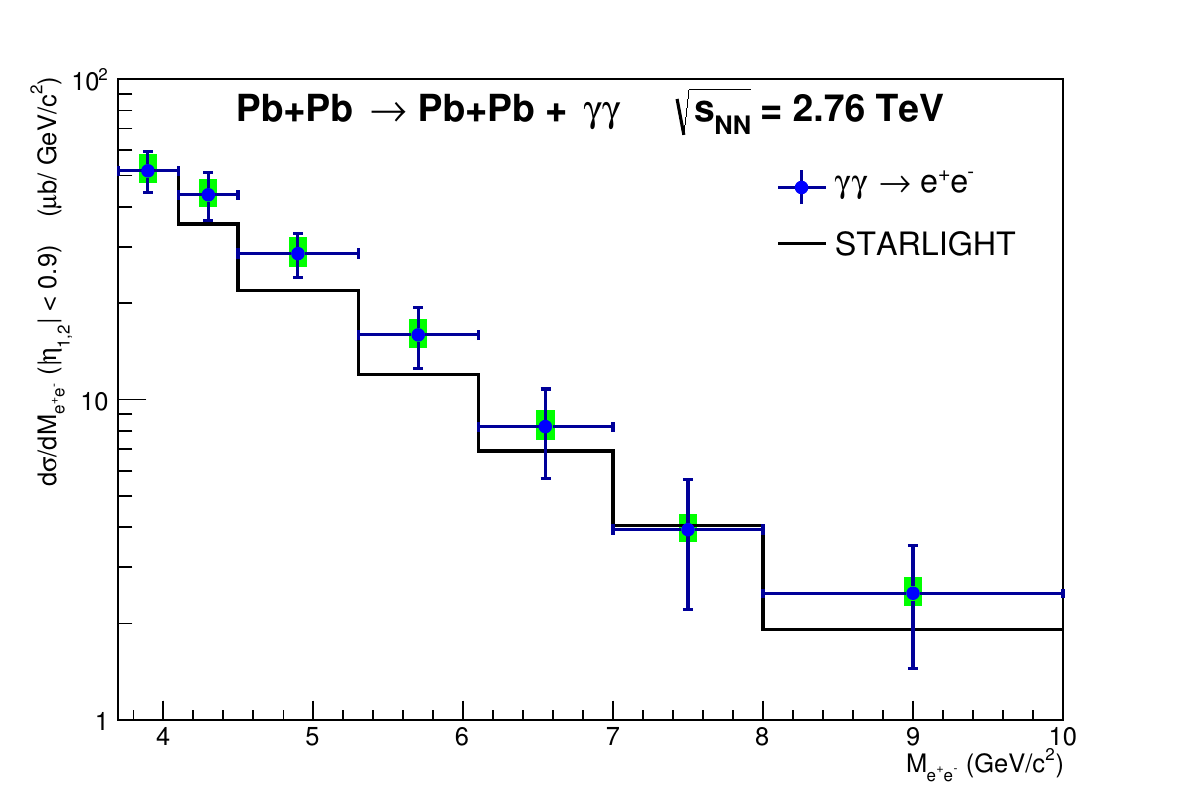}
\includegraphics[width = 0.49\textwidth]{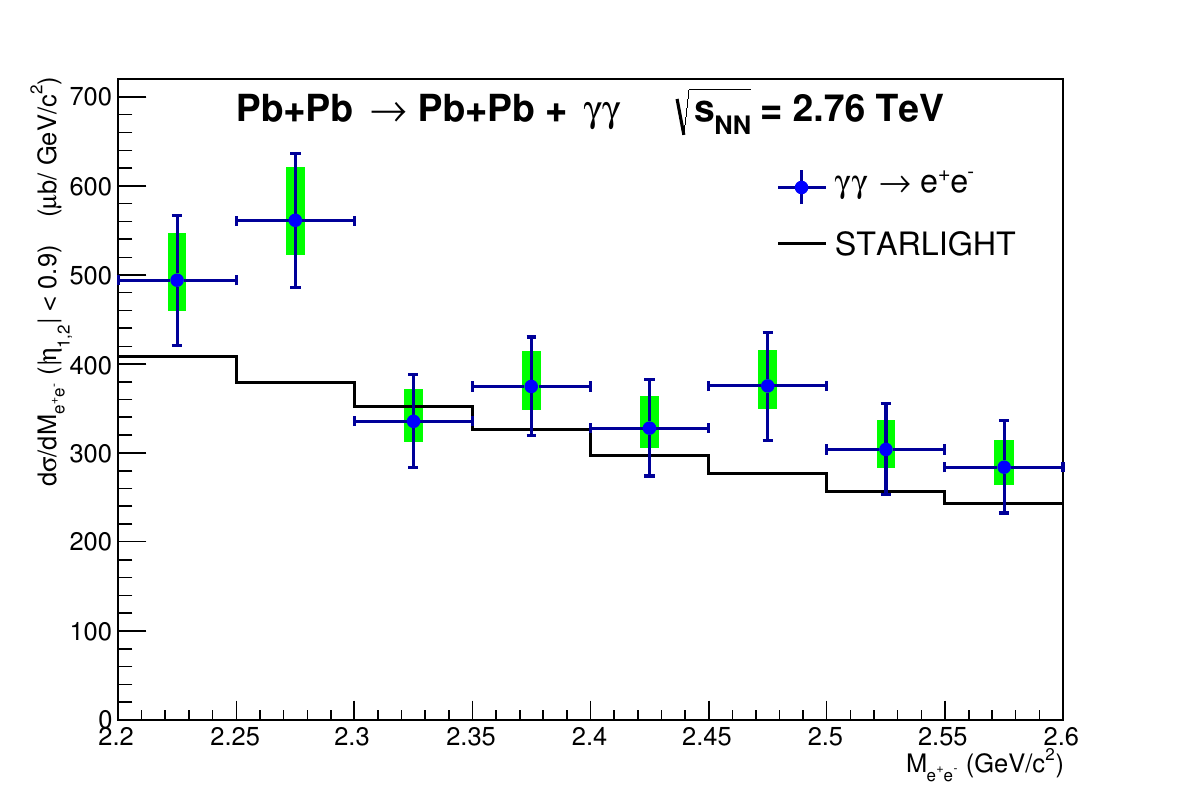}
    \end{center}
    \caption{Invariant-mass distribution for two-photon production of $\rm e^+e^-$ pairs with the leptons within $|\eta|< 0.9$, at $\sqrt{s_{\rm NN}}=2.76$~TeV~\cite{Abbas:2013oua}, compared with the lowest-order QED prediction. %
    \label{fig:leptonpairs}}
\end{figure}

So far, ALICE has mostly studied the production as a background to photoproduction of the J$/\psi\rightarrow \mu^+\mu^-$~\cite{Acharya:2019vlb} and $\rm J/\psi\rightarrow e^+e^-$~\cite{Abbas:2013oua}.  In both cases, the pair mass regions away from the J$/\psi$ mass provide a fairly pure sample of lepton pairs. 
The $\mu^+\mu^-$ mass spectrum was found to be consistent with the expectations from lowest order QED~\cite{ALICE:2012yye}, as implemented in the STARlight event generator~\cite{Klein:2016yzr}. For $\rm e^+e^-$ production at midrapidity (with lepton $|\eta|< 0.9$) at
$\sqrt{s_{\rm NN}}=2.76$ TeV, quantitative results were presented: $154 \pm 11 {\rm (stat.)} ^{+17}_{-11} 
{\rm (syst.)}$ $\mu$b for $2.2 < M_{\rm ee} < 2.6$ GeV/$c$, and 
$91 \pm 10 {\rm (stat.)} ^{+11}_{-8} 
{\rm (syst.)}$ $\mu$b for $3.7 < M_{\rm ee} < 10.$ GeV/$c^2$.  These are in good agreement with the STARlight lowest-order QED predictions of 128 $\mu$b and 77 $\mu$b respectively. Figure~\ref{fig:leptonpairs} shows that the $M_{\rm ee}$ distribution is in good agreement with the predictions.  

Looking ahead, ALICE can study leptons with much lower $p_\textrm{T}$ than ATLAS, and at much smaller $M_{\ell\ell}/\sqrt{s_{\rm NN}}$ than STAR; the latter condition increases the typical impact parameter, limiting the effect of the UPC condition $b>2\,R_A$.  The small $M_{\ell\ell}/\sqrt{s_{\rm NN}}$ also increases the cross section, and so may allow for observation of new phenomena, such as multiple pair production (when one pair of interacting ions produces two pairs), or interference effects in the angular distributions~\cite{Brodsky:1968rd}.  A more detailed study of the pair $p_\textrm{T}$ spectrum with different conditions on neutrons in the ZDCs would also be of interest.  Finally, interesting connections can be made to hadronic interactions (i.e.\,$b < 2\,R_A$); the STAR collaboration has observed apparent $\gamma A\rightarrow \ell^+\ell^- A$ processes in peripheral collisions; they are visible as a large excess of pairs with $p_\textrm{T} < 100$~MeV/$c$~\cite{Adam:2018tdm}.  These pairs may provide some information about the overall reaction plane.

\subsection{Searches for new particle states and new physics}
 ALICE has contributed to the understanding of the antimatter production mechanism in heavy-ion collisions and has also provided insight into the production and lifetime measurement of the hypertriton (i.e.\,the lightest known hypernucleus) (see Secs.~\ref{ch:nuclei} and~\ref{ch:hypertriton}). Besides the nuclei containing one hyperon, more exotic forms of deeply bound states with charm have been proposed as states of matter, either consisting of baryons or quarks. In the last years, this field has attracted much attention with the unexpected observation at electron-positron colliders of the new X, Y and Z states with masses around 4~GeV/$c^2$ (see for instance the reviews on non-$\rm q\overline{q}$ mesons and spectroscopy of mesons containing two heavy quark in Ref.~\cite{Zyla:2020zbs}. These heavy particles show very unusual properties, whose theoretical interpretation is entirely open. Above the D--$\overline{\mathrm{D}}$ threshold the situation is quite different. A number of new states have been recently discovered by BaBar, Belle and CLEO.
One of the most established among the XYZ states is the narrow X(3872) (later denominated $\chi_{\rm c1}(3872)$~\cite{Zyla:2020zbs}) state of less than 2.3~MeV/$c^2$ width. It has been discovered by Belle and
the main hadronic decay modes are $\rm \pi^{+}\pi^{+}J/\psi$ and ${\rm D^{0}}\overline{\mathrm{D}}^{0}\pi^{0}$~\cite{Belle:2003nnu,Olsen:2017bmm}. The small width and the decay properties disfavor a c$\overline{\mathrm{c}}$ structure and the very close
vicinity of the ${\rm D^{0}}\overline{\mathrm{D}}^{0*}$ threshold favors a molecular interpretation with these constituents. Measurements on the production of the $\chi_{\rm c1}(3872)$ at the LHC have been performed by LHCb~\cite{LHCb:2011zzp,LHCb:2013kgk,LHCb:2020fvo,LHCb:2020xds,LHCb:2020sey,LHCb:2021ten}, ATLAS~\cite{ATLAS:2016kwu,ATLAS:2020mve} and CMS~\cite{CMS:2013fpt,CMS:2021znk}. The debate on this topic is still open and ALICE will be able to contribute in the next LHC runs in such a search thanks to measurement of the deuteron production in pp collisions with very high statistics. The comparison with the deuteron (as a loosely bound two-baryon system) transverse momentum spectrum is seen as a possible check between the two possibilities of the $\chi_{\rm c1}(3872)$ being either a compact tetraquark~\cite{Esposito:2015fsa,Esposito:2020ywk,Esposito:2021vhu} or a loosely-bound D-meson (molecular) state~\cite{Guo:2017jvc,Brambilla:2019esw}.

Another field of interest connected to exotic states is the search of the H-dibaryon, a six-quark state (uuddss) predicted by Jaffe in 1977~\cite{Jaffe:1976yi} and later. The discovery of the H-dibaryon or the $\Lambda$n bound state would be a breakthrough in hadron spectroscopy as it would imply the existence of a six-quark state and provide crucial information on the $\Lambda$--nucleon and $\Lambda$--$\Lambda$ interaction.
ALICE has investigated the existence of a possible $\Lambda$n bound state and the H-dibaryon by reconstructing the invariant mass for the $\rm \Lambda p \pi^{-}$ decay channel of the H-dibaryon, under the assumption of a weakly bound state~\cite{ALICE:2015udw}. Since no evidence of a signal for the H-dibaryon was found in the invariant mass distribution, an upper limit ($99\%$ CL) of the production yield was obtained, assuming a $BR = 64$\% and the free $\Lambda$ lifetime. The limit $\approx 3.0 \times 10^{-4}$ is a factor 20 below values predicted by the statistical-thermal models (see Sec.~\ref{sec:QGPHadronization} and ~\ref{ch:NuclPhysLHC}). The $\Lambda$n bound state~\cite{HypHI:2013sxa} was investigated in the decay $\overline{\mathrm{d}}\pi^{+}$, but no signal was found in the invariant mass distribution. Assuming a $BR = 54\%$ and the free $\Lambda$ lifetime, this led to an upper limit ($99\%$ CL) of the production yield  $\approx 2.0 \times 10^{-3}$ that also in this case is a factor 20 below values predicted by thermal models~\cite{ALICE:2015udw}. The possible existence of the H-dibaryon was also investigated through $\Lambda$--$\Lambda$ correlations~\cite{ALICE:2019eol} and this significantly shrank the phase space of the predicted bound state further, but left some space for a resonance structure. A compact six-quark state (uuddss), called S (Sexaquark), was proposed~\cite{Farrar:2022mih} as a possible long-lived or even stable version. It would only be possible to detect it from interaction with (detector) material~\cite{Farrar:2022mih}. Searches for this elusive dark-matter candidate have been started at the LHC. 

Dark matter is a hypothetical form of matter that is responsible for accounting for approximately 85\% of the matter in the Universe. 
Since dark matter cannot be incorporated into the Standard Model, 
new interactions between dark matter particles and the ordinary
Standard Model (SM) particles have to be introduced via unknown dark-sector forces. The dark photon is an extra U(1) gauge boson and 
acts as a messenger of a dark sector with the residual interaction 
to the SM particles.
A number of fixed-target and collider experiments have searched 
for a dark photon, where the dark photon can decay back into charged SM particles with a coupling strength regulated by a `mixing parameter'.
ALICE can contribute to dark photon searches by examining the electron--positron invariant mass in very  large samples of 
$\pi^0$ Dalitz decays ($\pi^0 \rightarrow \gamma A^{\prime}$, $A^{\prime} \rightarrow {\rm e^+e^-}$) for $20 \le M_{\rm ee} \le 100$ MeV/$c^2$. The Run 2 data analysis is ongoing and the data samples to be collected with the upgraded detector in Runs 3--4 are expected to reduce the limit on the mixing parameter down to $10^{-4}$ below 100~MeV/$c^2$~\cite{Citron:2018lsq}. 
 
\subsection{Tests of fundamental symmetries}

After the discovery of the P, CP and T violation, CPT invariance is the only discrete exact symmetry of the Standard Model that still holds, and it is thought to be one of the most fundamental symmetries of nature. The conservation of CPT invariance is theoretically guaranteed by the CPT theorem, which is based on three assumptions: unitarity, locality and Lorentz invariance, formulated in the framework of a quantum field theory.
The experimental search of CPT violating processes thus probes our fundamental understanding of nature.  
Most of the experiments exploit one of the most important consequences of CPT invariance: i.e., that implies that the fundamental properties of antiparticles are identical to those of their matter-conjugates, e.g.\,the inertial mass, the lifetime, the absolute value of the electric charge and the magnetic moment.
The same applies to systems of particles and to their dynamics.
Experimental tests of CPT invariance have been carried out for elementary fermions and bosons, and for QED and QCD systems (a recent compilation can be found in Ref.~\cite{Zyla:2020zbs}).
Despite the different levels of precision reached, which span over several orders of magnitude, each of these distinct tests provides important information.
Indeed, effects in different systems might not be directly comparable. 
The copious production of light antimatter nuclei at LHC energies and the excellent ALICE tracking and PID performance enabled a precise and unique measurement of the mass difference between the deuteron, the $^{3}{\rm He}$ nucleus and their antimatter partners, with relative accuracies of $1.5 \times 10^{-4}$ and $1.3 \times 10^{-3}$, respectively~\cite{Adam:2015pna}.
The first represents the most stringent experimental constraint for CPT invariance in the nuclei sector, probing any matter/antimatter asymmetry in nucleon--nucleon interactions.
In the upcoming LHC Runs, the higher luminosity and the increased ALICE readout rate will allow the sensitivity to be improved and the measurement to be extended to $^3$H and $^4$He nuclei.

\subsection{AdS/CFT correspondence and heavy-ion collisions}

The formation, evolution, and dynamics of a heavy-ion collision are complex (see Sec.~\ref{sec:QGPevolution}). Starting from the initial-state parton distributions in the projectile nuclei, partonic interactions over a wide range in momentum scale drive the system rapidly towards equilibration and formation of the QGP, which then expands and cools. Similarly, jet interactions with the QGP occur over a wide range of momentum scales (see also Sec.~\ref{sec:theotools}). The theoretical description of these dynamical processes is extremely challenging, both because of their non-equilibrium nature, and because the coupling $\alpha_{\rm S}$ runs with momentum scale. Soft, strongly-coupled interactions play a crucial role in both QGP equilibration and jet-medium interactions, but such processes are not amenable to a perturbative treatment. Therefore, other theoretical tools are needed to provide insight into these aspects of the QGP.

A key tool in this program is the AdS/CFT correspondence, which provides a connection between certain conformal super-symmetric gauge theories and classical gravity in curved spacetime with one additional dimension~\cite{Maldacena:1997re}. This remarkable correspondence transforms very difficult out-of-equilibrium calculations in finite-temperature quantum field theory to tractable calculations in the holographic classical gravitational space. While the gauge theories for which this correspondence is known to apply have different symmetries, degrees of freedom, and coupling than QCD, they are nevertheless ``QCD-like’’, and can provide insight into universal features of strongly-coupled non-Abelian gauge theories~\cite{CasalderreySolana:2011us}. Indeed, the landmark observation that the specific shear viscosity, $\eta/s$, has a lower bound at strong coupling (see Sec.~\ref{sec:theotools}), was first made using the AdS/CFT correspondence~\cite{Kovtun:2004de}. 

The AdS/CFT correspondence has been used to study the approach to equilibrium, or ``hydrodynamisation’’, that characterizes the earliest phase of a nuclear collision. Such calculations show that, for strongly-coupled gauge theories, the far-from-equilibrium medium can be described by hydrodynamics after only a short evolution time of order $1/T$, where $T$ is the temperature~\cite{CasalderreySolana:2011us,Chesler:2009cy, Heller:2011ju,Heller:2012km}. These calculations are in good agreement with phenomenological studies comparing flow data to relativistic hydrodynamic calculations (see also Sec.~\ref{sec:theotools}).

The AdS/CFT correspondence has likewise been applied to the process of a far-out-of-equilibrium colour charge (a jet) propagating through the QGP, under the assumption of strong coupling; i.e.\,that the coupling constant is large at all momentum scales relevant to the jet-medium interaction~\cite{Liu:2006ug,CasalderreySolana:2011us}. Such calculations provide detailed pictures of the jet energy loss process, including the generation of a wake~\cite{Chesler:2007an} and the response of the medium (\cite{Casalderrey-Solana:2014bpa} and references therein). However, the strong-coupling assumption is in practice not a good approximation for modelling jet-medium interactions in data. A hybrid model, combining strong coupling using holography and weakly-coupled QCD using PYTHIA, has therefore been developed which incorporates both weak and strong coupling approaches to calculate a broad array of experimental observables~\cite{Casalderrey-Solana:2014bpa,Casalderrey-Solana:2015vaa}. Comparison of Hybrid Model calculations with ALICE jet quenching data is discussed in Sec.~\ref{sec:QGPJets}.

\subsection{Ultracold gases and Bose-Einstein condensates}

Heavy-ion collisions are one of only a few known strongly-interacting, weakly-coupled systems that can be studied experimentally, and it is of particular interest to investigate and compare with other such model systems in order to fully understand strong interactions in their extreme.  Remarkably, aside from the quark--gluon plasma, the most ``perfect'' liquids found in nature are ultracold Bose and Fermi gases, despite having a temperature twenty orders of magnitude lower than the QGP!  Like the QGP, these degenerate atomic gases are also characterised by high occupation numbers, a shear viscosity to entropy density ratio near the proposed quantum lower bound ($\eta/s\geq \hslash/(4\pi k_{\rm B})$), and can be described by fluid dynamic equations over a wide range of length scales.  Similar theoretical approaches, such as within the kinetic theory and holographic duality frameworks, have been used to describe the transport parameters and thermodynamic parameters of both the QGP and ultracold quantum gases (for more details, see the reviews in Refs.~\cite{Schafer:2009dj,Adams:2012th,Berges:2020fwq} and references therein); however, there have been fewer attempts to compare experimental measurements of the two systems.  

Ultracold atomic gases are an important experimental testing ground for strongly-interacting systems, since they can be manipulated in optical traps, and observed at all stages of the system evolution.  Such fine control over the initial conditions and evolution of the system is not possible in ultrarelativistic nucleus-nucleus collisions.  However, roughly analogous tools exist: for example, event shape engineering techniques~\cite{Adam:2015eta} and centrality selection~\cite{ALICE:2018tvk} can give some control over the initial shape and size of the energy-density distribution, and various experimental probes demonstrate sensitivity to different stages of the collision evolution (for example, the azimuthal anisotropy coefficients $v_n$ are self-attenuating probes which give higher sensitivity to the early stages of a heavy-ion collision, while femtoscopic source size measurements are uniquely sensitive to the final state).  While qualitatively similar flow behaviour has been observed in ultracold gases~\cite{O_Hara_2002} as in heavy-ion collisions, in recent years great progress has been made in developing new flow observables which give unique information on the hydrodynamic properties of the QGP, such as the shear ($\eta/s$) and bulk ($\zeta/s$) viscosities~\cite{ALICE:2016kpq,Acharya:2018lmh,Acharya:2017zfg}.  It would be particularly interesting to consider whether some of these measurements of particle correlations could be translated into measurements in ultracold gas systems.  Furthermore, in ultracold dilute Fermi gases, the strength of the interaction can be tuned using Feshbach resonances to span from the weakly attractive BCS regime to the strongly attractive Bose-Einstein condensate (BEC) regime (see for instance Ref.~\cite{RevModPhys.82.1225}.  In the latter, the second-order phase transition occurs near the degeneracy temperature and can be studied in detail. Meanwhile, in heavy-ion collisions, measurements of net-particle fluctuations~\cite{Braun-Munzinger:2020jbk,Acharya:2019izy,Adam:2020unf} are being used to probe the crossover transition between the QGP and hadron gas phases, which coincides with the chemical freeze-out temperature.  Future advancements in understanding the phase structure of strongly-interacting systems may come from considering similar or analogous observables in heavy-ion collisions and ultracold gases.  

\subsection{Chiral magnetic effect and new materials}

The observation of the chiral magnetic effect (CME)~\cite{Kharzeev:2004ey,Kharzeev:2007jp,Fukushima:2008xe} in heavy-ion collisions would be the first direct proof of the topological fluctuations in QCD (ALICE searches are discussed in Sec.~\ref{sec:NovelQCD}). This notion motivated the prediction of the effect in condensed matter systems in Dirac and Weyl semimetals, and eventually led to the discovery of the CME by the BNL–Stony Brook–Princeton–Berkeley group~\cite{Li:2014bha} in a Dirac semimetal ZrTe$_5$. The existence of the CME was later also confirmed in dozens of other materials. The effect also has important potential applications in sensor development and quantum computing (``chiral qubit''). CME is of great interest for both QCD and condensed matter physics in particular as it allows to investigate the role of quantum anomalies.

The the process of baryon asymmetry generation in the early Universe in theory
is presumably also very similar to the process of chirality generation in heavy-ion collisions; both are described by the topological transitions between the vacuum sectors of the non-Abelian gauge theory. 

Another emerging area of cross-disciplinary interest for both the heavy-ion and condensed-matter communities is the role of vorticity in the dynamical evolution of the systems. In heavy-ion collisions,  where large vorticity can be generated in off-central collisions, the vortices are expected to play a very important role in the dynamical evolution of the system. 

\subsection{Machine learning}
Machine learning (ML) represents a powerful tool for uncovering features in large and complex datasets. In the past decade, the number of machine learning applications in high-energy physics has grown tremendously. ML techniques are currently adopted for event classification, event reconstruction and selection, particle identification, detector calibration and monitoring~\cite{Albertsson:2018maf,Feickert:2021ajf}.

\begin{figure}[htb]
    \begin{center}
    \includegraphics[width = 0.51\textwidth]{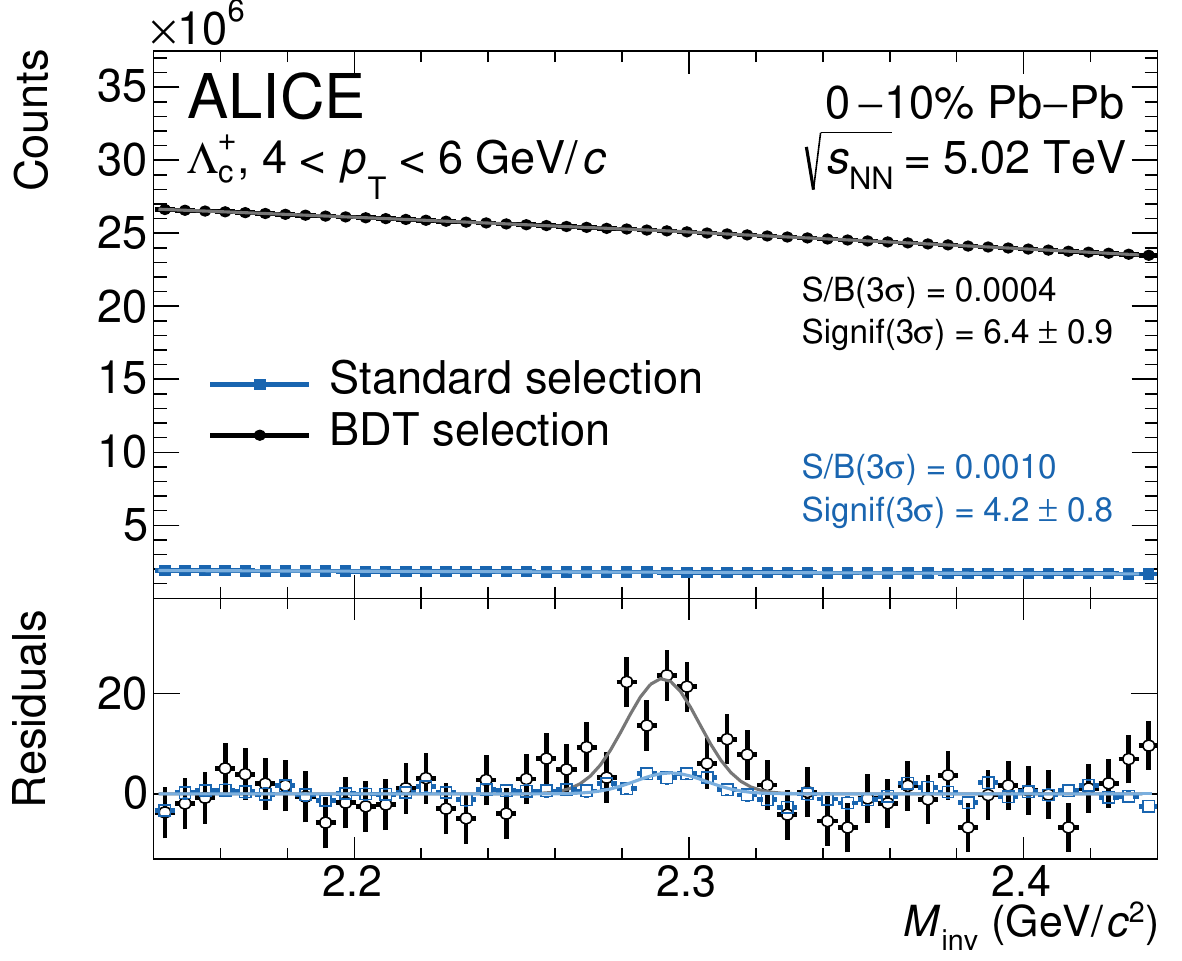}
    \includegraphics[width = 0.44\textwidth]{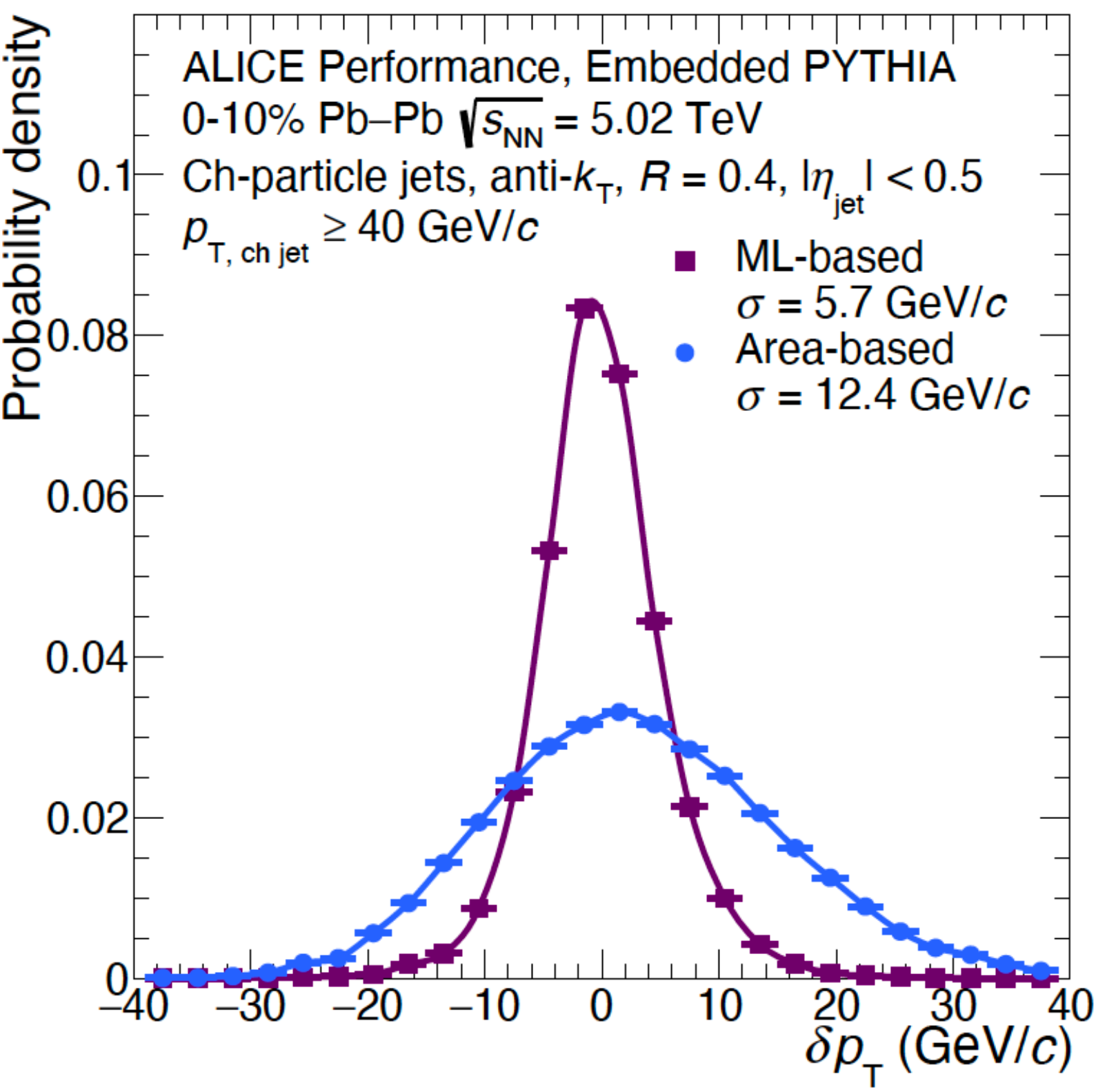}
    \end{center}
    
    \caption{(Left) Invariant mass of \Lambdac candidates using standard (cut-based) and BDT selection techniques in Pb--Pb collisions at $\sqrt{s_{\rm NN}}=5.02$~TeV for an example \pt interval~\cite{ALICE:2021bib}. The upper panel shows the actual counts and the lower panel the residuals after polynomial background subtraction for the two distributions. (Right) $\delta p_{\text{T}}$ distribution for charged jets with $R = 0.4$ in Pb--Pb collisions at 5.02 TeV. Comparison of ML-based and area-based correction.\label{fig:ML}} %
\end{figure}

As described in the previous sections, a large fraction of heavy-ion measurements rely on complex strategies designed to select rare probes like heavy-flavour hadrons down to very low transverse momenta in presence of a large combinatorial background. The ALICE Collaboration has recently published several analyses that rely on ML techniques. As an example, the first measurement of the \Lambdac\ production in central (0--10\%) Pb--Pb collisions at $\sqrt{s_{\rm NN}}= 5.02$~TeV~\cite{ALICE:2021bib} is discussed in the following. The \Lambdac\ selection uses a Boosted Decision Tree (BDT) classifier trained on a sample of signal \Lambdac\ candidates simulated with PYTHIA and background candidates extracted from data in the side-band regions. In the BDT training, both topological and particle-identification selection variables are considered, and the optimised set of BDT (hyper)parameters is identified by using a Bayesian optimisation procedure. The impact of these techniques is highlighted in the left panel of Fig.~\ref{fig:ML}, where the invariant-mass distribution of \Lambdac\ candidates measured in central Pb--Pb collisions obtained with the BDT analysis is compared to the results obtained with a selection based on rectangular cuts (in the upper panel the actual counts and in the lower panel the residuals after polynomial-background-fit subtraction). The BDT analysis achieves a substantial increase of the signal selection efficiency, up to about four times larger, and of the statistical significance, by 50\%  at low-transverse momentum. Similar techniques were adopted in the study of other heavy-flavour hadrons or of hypernuclei in heavy-ion collisions, resulting in similar increases in the statistical significance of the signals (see also Secs.~\ref{sec:open-heavy-flavor} and~\ref{ch:nuclei}). The use of ML classifiers has also been applied to the study of prompt and non-prompt production of D mesons in pp and Pb--Pb collisions~\cite{ALICE:2021mgk,ALICE:2022tji,ALICE:2022xrg}. In these measurements, a BDT classifier was trained to separate three classes of D-meson candidates: prompt D mesons from c-quark hadronisation, non-prompt D mesons from B-meson decays, and combinatorial background. A fit of the total D-meson yield as a function of the BDT score with templates for the prompt and non-prompt components provides an accurate estimation of the fraction of D mesons from B decays. 

ALICE has also made significant strides in using machine learning techniques for jet physics. One such project is the use of machine learning to correct jets for the large fluctuating background in heavy-ion collisions from the underlying event~\cite{Haake:2018hqn}. In this method, the machine learning algorithm is used to build a data-driven mapping from jet properties, including properties of the constituents of the jet, to the background corrected jet $p_{\text{T}}$. The estimator is trained on PYTHIA jets embedded into a realistic Pb--Pb background. A shallow neural network is used with 3 layers and [100, 100, 50] nodes.
The performance of this method is evaluated using the $\delta p_{\text{T}}$ distributions, where a narrowing of the distribution corresponds to a more precise determination of the jet signal. As shown in Fig.~\ref{fig:ML} on the right, the machine learning based method has a much reduced width in $\delta p_{\text{T}}$ as compared to the standard area-based method~\cite{Abelev:2012ej}, corresponding to a reduction in the residual fluctuations remaining after background subtraction. 

The interest in the use of machine learning techniques in heavy-ion collisions and in soft-QCD proton--proton physics is constantly growing. Machine learning techniques are likely to replace standard optimisation routines in the upcoming years for the large majority of the analyses and provide new tools to design new experimental observables. The ALICE Collaboration is currently exploring the impact of ML techniques on a wide range of applications, which go from particle identification of single hadrons and decays to heavy-flavour jet tagging or Monte Carlo reweighting. Deep neural network techniques are also being applied to critical aspects of the calibration and maintenance of the upgraded ALICE detector, from the correction of the electric field distortion fluctuations during data-taking in the ALICE Time Projection Chamber to the data control and quality monitoring of the event reconstruction. 
\newpage

\section{Summary}
\label{ch:Conclusions}

Since the start of LHC collisions in 2009, the ALICE detector has carried out a very successful data taking programme. The experiment is dedicated to exploring QCD at the LHC in the context of the largest collision energies available in the laboratory. Our main focus is the study of many-body QCD interactions at the largest temperatures possible via the formation of the QGP from heavy-ion collisions. We have also extensively explored hadron--hadron strong interactions and several aspects of QCD in pp collisions, using the unique capabilities that the LHC and our detector have provided us. In terms of heavy-ion physics, starting from the BEVALAC to the LHC, this has allowed for some of the most precise explorations of effects discovered previously, and has unraveled its own set of discoveries. Here we summarise the main findings by addressing the questions posed at the end of Chap.~\ref{ch:Introduction}:

\begin{enumerate}[leftmargin=*]
    \item {\it What are the thermodynamic and global properties of the QGP at the LHC?}
    
    Our measurements have shown that heavy-ion collisions at the LHC create conditions that very much exceed those needed to form the QGP. The initial QGP temperature implied by our results in conjunction with theoretical modelling is up to 5 times higher than the predicted QCD deconfinement temperature $T_{\rm pc} = 155$--$158$ MeV, and of the order of a trillion Kelvin, $\sim10^5$ times hotter than the centre of the Sun. An estimate can be made of the largest QGP energy densities from central Pb--Pb collisions, which yields $\sim12$~GeV/fm$^3$ at the early time of 1~fm/$c$. This is about seventy times higher than the energy density of atomic nuclei and about five times higher than the core of the most massive neutron stars.  The lifetime of the system in central Pb--Pb collisions is found to be about 10--13~fm/$c$, which is about 40\% larger than at RHIC. As will be addressed in later questions, such a system reaches equilibrium within this lifetime of the order of $10^{-23}$~s, and no other system in nature has been observed to do so in such a short time. The volume at freeze-out is about $7000$ fm$^3$ and about twice higher than at top RHIC energy.  We therefore conclude that the matter created in heavy-ion collisions at the LHC has the largest temperature and energy density ever observed, and the longer lifetime and larger volume compared to lower energy heavy-ion collisions. These observations provide a critical backdrop for conclusions that will be made in this summary. 
    
    \item {\it What are the hydrodynamic and transport properties of the QGP?}
    
    We have demonstrated that the QGP formed at LHC energies undergoes the most rapid expansion ever observed for a many-body system in the laboratory. The radial flow velocities derived from $p_{\rm T}$ spectra approach about 70\% of the speed of light. Elliptic flow in non-central collisions has been observed for all measured hadron species, including the light nuclei d and $^{3}$He, with the only exception of the $\Upsilon$(1S) bottomonium state. Our measurements of charmed-hadron $v_{2}$ and spectra have revealed, for the first time, that low-momentum heavy quarks participate in the collective motion of the QGP, despite being produced out of equilibrium in the initial stages of the collision. Based on a set of ALICE measurements performed over the last ten years, a suite of new observables has unveiled an extremely rich pattern for the dynamical evolution of the QGP as a whole. These include higher-order anisotropies ($v_{n \geq 3}$), correlations between different order anisotropies, and a translation of the angular momentum of the QGP to the polarisation of its outgoing hadrons. Hydrodynamic calculations, using the QGP equation of state and which assume that the collision system behaves as a liquid about 1~fm/$c$ after the initial collision, describe a wide variety of these observables. This description, which encompasses many more observables compared to hydrodynamic descriptions prior to the LHC, is achieved with the crucial inclusion of small but finite viscous effects, and has shown that hydrodynamics has emerged as a successful effective theory of many-body non-perturbative QCD interactions at high temperature. QGP viscosities explored in these hydrodynamic calculations imply that the system is strongly-coupled at the scale of the QGP temperatures probed at the LHC, with the extracted shear viscosity over entropy density ($\eta/s$) values in the range $1/4\pi < \eta/s < 0.3$, which are at least five times smaller than superfluid helium, and establishes that the QGP is the most perfect liquid ever observed. In the heavy-flavour sector, comparisons of our open-charm hadron measurements with transport models have enabled an estimation of the charm spatial-diffusion coefficient $D_{\rm s}$ in the range $1.5 < 2 \pi D_{\rm s}(T) T < 4.5$ at $T_{\rm pc}$. These values provide evidence that charm quarks couple strongly with the QGP at low momenta, and our measurements in conjunction with the transport model description, indicate how equilibration in the QGP can occur on a microscopic level during the extremely small time scales associated with QGP formation and expansion.
    
    \item {\it How does the QGP affect the formation of hadrons?}
    
    Our measurements of hadron yields over all momenta have provided an extremely extensive mapping of hadro-chemistry in heavy-ion collisions at LHC energies. Models which assume thermalisation at the time of the transition from QGP to a hadron gas (Statistical Hadronisation Models) describe the integrated yields of various light-flavour hadron species over many orders of magnitude. Besides the (u,d,s) sector, ALICE has discovered that, for heavy-ion collisions at the LHC, charmed hadron and quarkonium relative yields follow thermal expectations i.e.\,they can be understood in the context of a common temperature. The light-flavour chemical freeze-out temperature of about 156~MeV is within the predicted range of the deconfinement temperature, suggesting chemical equilibrium cannot be maintained easily after the QGP hadronises. The measurements of identified hadron spectra at intermediate-transverse momentum, where non-equilibrium processes emerge, indicate that an assumption of quark coalescence from the QGP captures the main features of the data for both the light- and heavy-flavour sectors. This is consistent with what was observed prior to the LHC for light flavours, with ALICE measurements revealing that such patterns are also seen in the heavy-flavour sector. At high-transverse momentum, where non-equilibrium processes dominate, we have shown that hadron production remains understood in terms of in-vacuum parton fragmentation. Finally, our investigations into the hadronic phase via measurements of resonance yields and femtoscopic radii have demonstrated complex dynamics at work. They have indicated that this phase is prolonged, and the decoupling of particles from the expanding hadron gas is likely to be a continuous process, rather than a sudden kinetic freeze-out of all particle species at the same temperature. These observations have important implications on how hadrons form and interact at high temperatures, such as those in the early Universe.
    
    \item {\it How does the QGP affect the propagation of energetic partons?}
    
    Energetic, highly-virtual partons generated in hard scatterings in nuclear collisions produce a parton shower that interacts with the surrounding QGP as it expands and cools. The in-medium modification of the jet shower (“jet quenching”) results in several distinct observable effects, enabling a broad multi-messenger exploration of QGP structure and dynamics using jets. ALICE has measured significant yield suppression for a wide range of hadrons and reconstructed jets in both inclusive and coincidence channels, showing that in-medium energy loss occurs at the partonic level and quantifying its magnitude in terms of energy shift. ALICE measurements of heavy-flavour yield suppression provide insight into details of this process: the energy loss is reduced for beauty with respect to charm quarks (e.g.\,by the dead-cone effect) and both radiative and collisional processes are necessary for the models to describe the data. ALICE jet substructure measurements indicate preferential suppression of wide-angle radiation in the jet shower, which is sensitive to colour coherence of the QGP and the space-time structure of jets. Finally, ALICE measurements of di-jet acoplanarity constrain the rate of jet scattering off quasi-particles in the medium, thereby probing the micro-structure of the QGP. A comprehensive description of this rich set of jet measurements in a well-constrained physics picture of the QGP is still work-in-progress. Inclusive jet-suppression measurements are described by a broad range of QGP models incorporating different underlying physics that include both weak and strong-coupling descriptions of the QGP. Overall, ALICE has observed the energy loss of energetic partons in the presence of the QGP and the modification of their shower. 
    \newpage
    \item {\it How does deconfinement in the QGP affect the QCD force?}
    
    Significant modifications of the quarkonium binding in the QGP have been observed, with our results at the LHC showing new and striking features.  For the J/$\psi$ charmonium state, the QGP-induced suppression was shown to be counterbalanced by a strong regeneration effect, taking place during and/or at the fireball chemical freeze-out, that dominates charmonium production at low transverse momenta. This mechanism constitutes a proof of deconfinement, as it implies that coloured partons can move freely over distances much larger than the hadronic scale. At the same time, a significant elliptic flow was observed for the J/$\psi$: an unequivocal signal that charm quarks participate in the medium expansion. The excited $\psi$(2S) state, which has a binding energy lower by one order of magnitude with respect to the J/$\psi$, was found to be more suppressed than the J/$\psi$, by a factor about two. Like for the J/$\psi$, a hint of reduced suppression towards zero transverse momentum supports the presence of the regeneration effect.  
    In the bottomonium sector, a suppression of $\Upsilon$(1S) and $\Upsilon$(2S) with a clear ordering related to binding energy was observed. While the observed suppression of the strongly bound ground state ($\sim$1 GeV binding energy) can be attributed to some extent to cold nuclear matter effects and feed-down from higher mass states, QGP effects are evident for the 2S state. 

    \item {\it Can the QGP lead to discovery of novel QCD effects?}
    
    Our charged-hadron dependent directed flow measurements indicate that, in heavy-ion collisions, extremely large electromagnetic fields are created, strong enough to influence electrically charged degrees of freedom in the QGP. In addition, our $\Lambda$ polarisation measurements provide an upper limit (95\% C.L.) for the magnetic field at freeze-out of $14\times10^{12}$ T at the top LHC heavy-ion collision energy. Multiple searches were carried out to investigate whether parity is violated in the strong interaction, a process which is facilitated by the magnetic fields produced in heavy-ion collisions. The experimental signal is known as the Chiral Magnetic Effect, and has been observed in the QED sector for Type II superconductors at sub-critical temperatures. Our measurements provided upper limits of 15--33\% for the contribution of the CME, consistent with the results at RHIC energies, and constitute state-of-the-art limits.
    
    \item {\it What are the minimal conditions of QGP formation?}
    
    Several of our measurements in high-multiplicity pp and p--Pb collisions exhibit features similar to those observed in heavy-ion collisions, where these are associated with QGP formation. These signatures are observed in events with large charged-particle multiplicity densities and include the enhancement of strange-particle yields, positive values of anisotropic flow determined from multi-particle correlations, a mass ordering of anisotropic flow coefficients as a function of transverse momentum in light-flavour hadrons, and enhanced baryon-to-meson ratios at intermediate-transverse momentum. In p--Pb collisions, another such feature is that $\psi(2S)$ suppression in the backward-rapidity region (Pb-going) is larger with respect to the ground state, which is an observation typically associated to final-state interactions between the weakly-bound c$\overline{\rm c}$ and a strongly-interacting medium. At first sight, the ability of the hydrodynamic framework to describe many of the observed features even at low multiplicities suggests that there is no apparent limit of QGP formation. However, models based on a microscopic description of the system, including even those that do not require a presence of an equilibrated, collectively-expanding medium, can also explain these features and are therefore able to provide an alternative interpretation of our measurements, challenging the idea of creation of a small QGP droplet. This seems to be further supported by the fact that no significant jet quenching has been observed within the sensitivity of the measurements. However, such absence could also be caused by the small spatial extent of a possible QGP system. The wealth of experimental measurements collected until now suggests a smooth evolution of collective effects with particle multiplicity, with further theoretical investigation being required to clarify up to which extent and in which conditions heavy-ion phenomenology and a microscopic description are both able to provide predictions that are consistent with data. 

    \item {\it What is the nature of the initial state of heavy-ion collisions?}
    
    Our vector meson photoproduction measurements in Ultra Peripheral Collisions and electroweak boson studies have provided unique constraints on the gluon parton distribution function for nuclei. They show clear evidence that the partonic structure of nuclei is different compared to free protons, with nuclear shadowing effects increasing with decreasing longitudinal momentum fractions $x$. Using photoproduction measurements in p--Pb collisions, we have probed the gluon densities also in the proton down to $x \sim 10^{-5}$ and found no clear indication of gluon-saturation effects. Advancements in initial-state models that map the spatial distribution of nuclear matter in heavy-ion collisions have been extremely successful in describing multiplicity and flow measurements sensitive to small- and large-scale structures of the nucleus and the nuclear overlap. Such an observation has key and beneficial implications for the extraction of QGP properties. These properties depend on the accuracy of such modelling, with $\eta/s$ being a prime example. Our measurements in both small and large systems have also shed light on the structure of the nuclear matter at high-energy, with indications of nucleon substructure playing an important role. A key question in this regard is whether gluon saturation in the nuclei is a requirement for this success, as competitive descriptions can often be achieved with and without it.

    \item{\it What is the nature of hadron--hadron interactions?}
    
    Collisions at the LHC produce an abundance of short-lived hadrons, with such an  abundance being far beyond previous experimental investigations. Our measurements have studied their properties and interactions with other hadrons at unprecedented levels of accuracy. We have also specifically measured  antimatter nuclear states, which has enabled us to investigate the antimatter properties. These include the first measurements of anti-helium cross sections of interaction with matter, which promise to shed light on the transparency of the interstellar medium in view of Dark Matter searches. For pp and p--Pb collisions, we have observed for the first time that the p--$\Xi$, $\Lambda$--$\Xi$ and p--$\Omega$ interactions are attractive in the strong sector, which is consistent with expectations from Lattice QCD. Although attractive, no bound states for these interactions have been found. Other investigated interactions, such as p--K, p--$\Lambda$, p--$\phi$ and p--$\Sigma$, have challenged Effective Field Theory descriptions. Such observations have implications for the equation of state of the core of neutron stars, where nucleon-hyperon interactions may play a significant role. Our studies on the formation of light nuclei from pp to heavy-ion collisions have explored a tension between the coalescence and the statistical hadronisation approaches for different sizes of the particle-emitting source. Nonetheless it is remarkable that statistical hadronisation models can describe deuteron yields in central heavy-ion collisions, given that the binding energy is approximately 70 times smaller than the medium temperature. On the other hand, the measurement of hypertriton ($^{3}_{\rm \Lambda}{\rm H}$) yields in p--Pb collisions shows the limits of the statistical approach for small colliding systems and is better described by coalescence models. Our hypertriton lifetime and binding energy measurements ($\Lambda$ separation energy) are the most accurate to date and provide stringent constraints for the understanding of the feeble strong interaction that binds hypernuclei. Finally, our studies of matter versus antimatter hadronic interactions have revealed no difference as of yet in the strong sector.

    \item{\it Can ALICE elucidate specific aspects of perturbative QCD and of related ``long distance'' QCD interactions?}
    
    ALICE measurements of high-$Q^2$ processes in pp collisions have provided new insights into the perturbative regime of QCD and, via QCD factorisation, to non-perturbative long distance objects such as fragmentation functions. The importance of high-order perturbative contributions to high-$Q^2$ processes have been demonstrated by ALICE measurements of inclusive jet production and jet substructure, both of which can be calculated in pQCD without non-perturbative fragmentation functions. These measurements provide state-of-the-art tests of the extent to which pQCD can describe low-energy jets, which is crucial as a baseline for interpreting past and future jet quenching measurements. The fragmentation process itself has been studied in detail with a wide range of ALICE measurements of identified particles, including neutral mesons, charm hadrons, and quarkonia. Such identified particle measurements can be described by pQCD in combination with fragmentation functions, allowing the opportunity to constrain fragmentation functions. Precise measurements of charm hadron production, including cross section measurements of several charmed baryons for the first time, have allowed ALICE to compute the total charm cross section and to measure charm fragmentation fractions – demonstrating the non-universality of charm fragmentation in pp collisions compared to ee and ep collisions. By studying jets containing D$^0$ mesons in pp collisions, ALICE has reported the first direct observation of the dead-cone effect in QCD, consisting in a suppression of the gluons radiated by a massive quark (charm, in this case) in a forward cone around its flight direction.

\end{enumerate}

These observations have demonstrated that much has been learnt over the past period of LHC data taking. This has been done in conjunction with other LHC experiments, and with the continuation of the RHIC programme - where huge advances in luminosities and centre-of-mass energy coverage have been achieved. Our studies of high-density QCD have continued to explore emergent behaviour in many-body interactions at highest possible temperatures in the laboratory. These include the hydrodynamic description, that has been tested with unprecedented accuracy in heavy-ion collisions at the LHC, which offer an environment far beyond the usual application in fluid dynamics. Comprehensive efforts at global fitting for precise determination of QGP properties and dynamics are now underway, utilising the powerful approach of Bayesian Inference for rigorous theory-to-experiment comparison. Such analyses have gone through a few iterations in the soft sector, and are starting in the hard-probes sector. We have demonstrated that QGP transitions to thermally and chemically equilibrated hadron yields, with the most extensive set of measurements achieved. QGP-like signatures were discovered in high-multiplicity pp and p--Pb collisions, which probe the thresholds of QGP formation. Such findings have ignited a debate of whether pp and p--Pb collisions at LHC and RHIC energies create the smallest possible QGP droplets. We have also discovered thermalisation effects for charm quarks in the QGP, and these demonstrate microscopically how equilibrium can occur on extremely small time scales via the strong interaction. The discovery of strong regeneration effects in the charmonium sector represents an unambiguous signal for deconfinement of quarks and gluons. Finally, we have investigated two-body interactions to an unprecedented precision, and discovered how rarely-produced hadrons interact, whose behaviour has broad implications for understanding various features of the Universe. However, many open questions remain. The next chapter of this article is dedicated to addressing these questions in the future ALICE programme.

\newpage

\newcommand{\pt}{p_{\rm T}}
\newcommand{\sqrtsNN}{\sqrt{s_{\rm NN}}}
\section{Outlook: ALICE detector and physics for the next two decades}
\label{ch:Outlook}

\subsection{ALICE upgrades and data-taking campaigns for the LHC Runs 3 and 4}
\label{ch:ALICE2}

\subsubsection{Major \texorpdfstring{ALICE\,2}{ALICE 2} detector upgrade}

The ALICE Collaboration has prepared a major upgrade of the experimental apparatus that was installed during the LHC Long Shutdown 2 (2019--2021) and will operate during Runs 3 (2022--2025) and 4 (2029--2032). 
After successful commissioning with injection-energy beams, the upgraded ALICE apparatus has started to record Run~3 proton--proton collisions at $\sqrt s =13.6$~TeV in July 2022. 
The detector upgrade, indicated as `ALICE\,2', was guided by the requirements detailed in Ref.~\cite{Abelevetal:2014cna}.
Track reconstruction performance is improved, in terms of spatial precision and efficiency, in particular for low-momentum particles, in order to select more effectively the decay vertices of heavy-flavour mesons and baryons, and to separate prompt and non-prompt charmonium states.
The sustainable interaction rate is increased up to 50~kHz for Pb--Pb collisions in continuous readout mode (all collisions are collected), that provides full recording efficiency for low-momentum processes. This increase enables recording of a sample of minimum-bias collisions corresponding to the full integrated luminosity of about 13~nb$^{-1}$ that the LHC can deliver during Runs 3 and 4. This sample will be larger by two orders of magnitude than the minimum-bias Run 2 sample. 
The charged-hadron, electron, muon, photon and jet identification capabilities of the apparatus, which are crucial for the selection of heavy-flavour, quarkonium, light nuclei and dilepton signals at low-momentum, is preserved.

The ALICE upgrade, described in detail in Ref.~\cite{Abelevetal:2014cna} and in an upcoming extended review~\cite{ALICE:2023udb}, entails the following main changes to the apparatus.
The ITS has been replaced with the ITS\,2~\cite{Abelevetal:2014dna}, made of seven layers equipped with Monolithic Active Pixel Sensors (MAPS). The innermost layer has a radius of 23~mm, in comparison with 39~mm of the ITS. The hit resolution of the detector is of about 5~$\mu$m and the material budget of each three innermost layers is reduced from the value of 1.15\% (ITS) to 0.35\% (ITS\,2) of the radiation length. These features provide an improvement by a factor of about three for the track impact parameter resolution in the transverse plane and by a factor of about six in the longitudinal direction.
A new Muon Forward Tracker (MFT)~\cite{CERN-LHCC-2015-001} has been installed, made of five double-sided detection disks instrumented with similar MAPS as those used in the ITS\,2 and it provides precise tracking and secondary vertex reconstruction for muon tracks in $-3.6 < \eta < -2.5$.
The TPC readout chambers have been replaced with Gas Electron Multiplier (GEM) chambers~\cite{CERN-LHCC-2013-020} and the readout systems of the TOF detector and Muon Spectrometer have been upgraded~\cite{Antonioli:1603472} to enable continuous readout of Pb--Pb events at an interaction rate up to 50~kHz while keeping the same momentum measurement and particle identification performance for charged hadrons, electrons and muons as in Runs 1--2.
The forward trigger detectors have been replaced with a new Fast Interaction Trigger detector (FIT)~\cite{Antonioli:1603472} based on Cherenkov radiators and scintillator tiles at forward rapidity around the beam pipe.
A new integrated Online-Offline system (O$^2$) has been deployed for data readout, compression and processing on a large GPU-CPU computing facility located at the experimental site~\cite{CERN-LHCC-2015-006}.

\subsubsection{Proposed upgrades for Run 4}

Two additional upgrades are in preparation for the Long Shutdown 3 (2026--2028), with the goal of further enhancing the physics reach of the experiment in Run 4: a new inner barrel for the Inner Tracking System (ITS\,3 project) and a forward calorimeter optimised for photon detection in the range $3.4 < \eta < 5.8$ (FoCal project). 
Both the upgrade projects have been
endorsed by the ALICE Collaboration and are described in Refs.~\cite{Musa:2703140,ALICECollaboration:2719928}. 

The ITS\,3 upgrade~\cite{Musa:2703140} consists in the replacement of the three innermost layers of the ITS\,2 with three truly-cylindrical layers made with curved large-area MAPS sensors~\cite{ALICEITSproject:2021kcd}. Each of the three new layers will have a thickness as low as 0.02--0.04\% of the radiation length. The new layers will be positioned at smaller radii, with the innermost layer having a radius of 18~mm; a new beam pipe with reduced radius (16~mm) and thickness is also part of the upgrade. The ITS\,3 track impact parameter resolution is better than that of the ITS\,2 by a factor of two in both the transverse and longitudinal directions up to a transverse momentum of 5~GeV/$c$, reaching down to about 12~$\mu$m at $\pt=1$~GeV/$c$ and 3~$\mu$m at $\pt=10$~GeV/$c$. The tracking efficiency at very low-transverse momentum is also improved. The projected tracking performance and reduced material budget will strongly enhance the low-mass dielectron, heavy-flavour meson and baryon production measurements.

The FoCal upgrade~\cite{ALICECollaboration:2719928} consists of an electromagnetic calorimeter with high readout granularity for optimal separation of direct-photon showers from those of neutral pions at forward pseudorapidity ($3.4 < \eta < 5.8$), coupled to a hadronic calorimeter for additional hadron rejection. The required granularity is achieved using a combination of MAPS silicon pixel readout planes and standard silicon pad readout planes. The main physics goal of the FoCal is the study of gluon parton distribution functions in the lead nucleus at Bjorken $x$ values down to $10^{-6}$ using the nuclear modification factor $R_{\rm pPb}$ of forward direct photons with transverse momentum $2<\pt<20$~GeV/$c$ in p--Pb collisions at $\sqrtsNN=8.8$~TeV. This kinematic domain is not accessible with other present and planned experiments and it extends to lower $x$ values with respect to the coverage of the Electron-Ion Collider~\cite{Accardi:2012qut}. The gluon densities at $x < 10^{-4}$ are expected to be sensitive to possible nonlinear QCD evolution or saturation effects. The FoCal physics programme also includes forward di-hadron correlations, jet production, as well as measurements in ultra-peripheral p--Pb and Pb--Pb collisions.

\subsubsection{Target data samples and integrated luminosities for LHC Runs 3 and 4}

The future ALICE and LHC physics programme for high-density QCD with small and large colliding systems is extensively discussed in the HL-LHC Physics Yellow Report~\cite{Citron:2018lsq}. The ALICE target samples and integrated luminosities are the following:
Pb--Pb collisions at $\sqrtsNN=5.3$~TeV: $L_{\rm int}=13$~nb$^{-1}$;
p--Pb collisions at $\sqrtsNN=8.5$~TeV: $L_{\rm int}\approx 0.5$~pb$^{-1}$;
pp reference samples at the same centre-of-mass energies as the Pb--Pb and p--Pb samples: $L_{\rm int}\approx 6$~pb$^{-1}$ for each energy;
pp collisions at top LHC energy $\sqrt{s}=13.6$~TeV: $L_{\rm int}=200$~pb$^{-1}$;
$^{16}$O--$^{16}$O collisions at $\sqrtsNN=6.8$~TeV (few days in Run 3 with $L_{\rm int} \approx 0.5$--$1$~nb$^{-1}$).

The Pb--Pb and p--Pb samples will represent an increase by factors of 50--100 and 1000, respectively, in comparison to the samples collected with a minimum-bias trigger during Run 2. A sample of Pb--Pb collisions, corresponding to an integrated luminosity of about 3~nb$^{-1}$, will be collected with the low-field setting of 0.2~T for the ALICE solenoid magnet in order to optimise the performance for low-mass dielectron measurements.
The top-energy pp programme is motivated by studies of high-density QCD effects in high-multiplicity collisions, that complement the p--Pb and O--O programmes, as well as by perturbative QCD and hadronic physics studies to which ALICE can uniquely contribute~\cite{ALICE-PUBLIC-2020-005}. All pp collisions will be inspected in the O$^2$ computing farm and a selection based on event multiplicity or on the presence of specific signals (strange and heavy-flavour hadron decays, quarkonia, light nuclei, jets and photons) will be applied
to record about 1/1000 of the collisions. 
The target integrated luminosity for such studies will be collected mostly in Run 3.

\subsection{\texorpdfstring{ALICE\,3}{ALICE 3} detector proposal for the LHC Run 5 and beyond}
\label{ch:ALICE3}

For the LHC Runs 5 and 6, a completely new setup,
named `ALICE\,3', is proposed~\cite{ALICE:2022wwr}, to enable new measurements in the heavy-flavour sector -- with focus on low-$\pt$, including measurements of multi-charm baryon production and femtoscopic studies of the interaction potentials between heavy mesons -- as well as precise multi-differential measurements of dielectron emission to probe the mechanism of chiral-symmetry restoration and to study the time-evolution of the QGP temperature.

The proposed detector consists of a tracking system with unique pointing resolution over a large pseudorapidity range ($-4<\eta<+4$), complemented by multiple sub-detector systems for particle identification, including silicon time-of-flight layers, a ring-imaging Cherenkov detector with high-resolution readout, a muon identification system and an electromagnetic calorimeter. Unprecedented pointing resolution of 10~$\mu$m at $\pt= 200$~MeV/$c$ at midrapidity in both the transverse and longitudinal directions can be achieved by using thinned silicon sensors with minimal supporting material, similarly to the ITS\,3 proposal, and by placing the first layers as close as possible to the interaction point, on a retractable structure to leave sufficient aperture for the beams at injection energy. The pointing resolution at midrapidity is projected to be about three times better than that of the ITS\,3. In the baseline scenario, the tracking system is placed in a superconducting solenoid with a field of $B= 2$~T to reach a momentum resolution of 1--2\% over a broad pseudorapidity range.  Other options for the magnet are under study. The particle identification systems enable high-purity separation of electrons from $\pt$ as low as 50~MeV/$c$ and up to about 1.5~GeV/$c$, and of hadrons over a broad momentum range. The muon system is optimised for muon identification down to $\pt \approx 1.5$~GeV/$c$, so that J/$\rm \psi$ can be measured with transverse momentum down to zero. This provides access to decays of different $\chi_{\rm c}$ states via their decays to J/$\rm \psi \gamma$ and to the exotic $\chi_{\rm c1}(3872)$ in the J/$\rm\psi\,\pi^+\,\pi^-$ channel.

The programme aims to collect an integrated luminosity of about 35~nb$^{-1}$ with Pb--Pb collisions and 18~fb$^{-1}$ with pp collisions at top LHC energy. The potential to further increase the luminosity for ion running in the LHC by using smaller ions (e.g.\,$^{84}$Kr or $^{128}$Xe), as well as further runs with small collision systems to explore the approach to thermal equilibrium, are being explored.

\subsection{Physics prospects with the \texorpdfstring{ALICE\,2}{ALICE 2} and \texorpdfstring{ALICE\,3}{ALICE 3} detectors}
\label{ch:ALICE23questions}

In the following, the expected advances in addressing the physics questions in the next two decades are illustrated with specific examples from the ALICE~2 and ALICE~3 programmes.

\begin{enumerate}[leftmargin=*]
    \item {\it What are the thermodynamic properties of the QGP at the LHC?}

A major advance in this sector will concern the determination of the QGP temperature and its time evolution. 
The average temperature of the QGP phase will be determined with an accuracy of about 20\% in Run 3 by measuring thermal radiation with dielectron pairs (virtual photons) in the invariant-mass range 1--2~GeV/$c^2$. 
Unlike the effective temperature from the real direct photon spectrum, the temperature obtained from dilepton invariant mass is not affected by the blue shift induced by radial flow.
A temperature measurement with dileptons was not possible in Run~2. A dedicated data-taking period with a reduced magnetic field of 0.2~T will increase the acceptance for low-$p_{\rm T}$ dielectrons and the new Inner Tracking System (ITS\,2) will strongly reduce the background from photon conversions and charm decays. The further upgrade of the inner tracker (ITS\,3) with even lower material thickness will bring the accuracy to about 10\% in Run 4.

The ALICE\,3 detector is specifically designed for ultimate performance for dielectron measurements and will provide novel access to the time-evolution of the temperature of the QGP using differential measurements of dielectrons as function of both transverse momentum and invariant mass, with the higher (lower) mass ranges expected to be dominated by higher (lower) temperature radiation emitted at earlier (later) time. The measurement of the elliptic and higher-harmonic flow coefficients of thermal dielectrons is expected to give access to thermal radiation from the early pre-hydrodynamic QGP phase~\cite{Coquet:2021lca}. This will also provide independent constraints on the transport coefficients of the QGP fluid: the shear viscosity $\eta$ and the bulk viscosity $\zeta$ as a function of time and therefore of temperature~\cite{Vujanovic:2019yih,Kasmaei:2018oag}. In addition, thermal radiation of real photons will be measured via conversions in the tracking system and with the electromagnetic calorimeter. The enhanced low-$\pt$ capabilities of both these systems enable a precise determination of the decay-photon background leading to a reduction of the systematic uncertainties.

In addition, the ALICE\,3 acceptance and performance will allow us to address  the {\it characteristics of the QGP phase transition at vanishing baryochemical potential.} The partial restoration of QCD chiral symmetry is, together with colour deconfinement, one of the fundamental effects that are theoretically associated to the QGP phase transition. Chiral symmetry is expected to  lead to a modification of the dilepton spectrum in the light-vector-meson mass range. Indications of chiral-symmetry restoration were found at SPS energies via measurements of a broadened $\rho^0$-meson mass spectrum by the CERES~\cite{CERES:2006wcq} and, with higher precision, 
NA60~\cite{NA60:2006ymb} experiments. At variance, measurements at high-energy and vanishing baryochemical potential currently have too large uncertainties to allow a concluding experimental study~\cite{STAR:2013pwb,PHENIX:2015vek,Acharya:2018nxm}. 
The vertexing and rate capabilities of the ALICE\,3 detector will be essential for a high precision measurement of the
medium modification of the dilepton spectrum at the LHC 
over a large mass range, in particular in the region around the mass of the $\rho^0$ and $\rm a_1$ mesons. In this mass range, the chiral mixing between the $\rho^0$ and its chiral partner $\rm a_1$ is expected to increase the dielectron yield by about 15\%. Such effect is below the expected accuracy of dielectron measurements with ALICE\,2, while it is within the projected experimental sensitivity of ALICE\,3.

Phase transitions in strongly-interacting matter can be addressed via measurements of fluctuations of conserved charges in heavy-ion collisions~\cite{Ejiri:2005wq,HotQCD:2012fhj,Braun-Munzinger:2016yjz}. These measurements provide information on critical behaviour near the phase boundary between QGP and hadronic matter. The fluctuations, assessed via cumulants of various orders of a given net-charge (e.g.\,baryon number using protons), can be directly related to generalised susceptibilities, which are derivatives of the pressure with respect to the chemical potentials corresponding to the conserved charges and are computed in lattice QCD. 
Critical fluctuations due to the vicinity of the cross-over line to a second-order phase transition of $\mathcal{O}$(4) universality at vanishing u, d quark masses are expected to strongly modify the sixth and higher order cumulants of the net-baryon distribution~\cite{Friman:2011pf}. The target integrated luminosity for ALICE\,2 in Runs 3--4 is expected to be sufficient for observing a possible deviation from unity of the sitxh-order cumulant~\cite{ALICE-PUBLIC-2019-001}. However, only the much larger acceptance of ALICE\,3 (8 units of pseudorapidity, instead of 1.8 of ALICE\,1 and 2) will enable a high-precision measurement and could also give access to lower-order cumulants in the strangeness and charm sectors.

\item  {\it What are the hydrodynamic and transport properties of the QGP?}

A major advance is planned for the determination of the heavy-quark transport parameters of the QGP, in particular the spatial diffusion coefficient $D_{\rm s}$. As discussed in Sec.~\ref{sec:QGPsummary}, unlike the bulk and shear viscosities, the diffusion coefficient can be calculated from first principles in the perturbative QCD and lattice QCD approaches. These approaches are enabled by the large scale given by the masses of charm and beauty quarks. 

The ALICE\,2 upgrade has been designed with precise heavy-flavour measurements as one of the main physics motivations. Nuclear modification factors and azimuthal anisotropy coefficients of several charm-hadron species will be used together with detailed model calculations, for example using Bayesian parameter estimation frameworks~\cite{Xu:2017obm}, to evaluate the charm-quark spatial diffusion coefficient $D_{\rm s}$ with about a factor two better accuracy than the present determinations. Precise measurements of charm-strange mesons ($\rm D_s^+$) and charm baryons ($\rm \Lambda_c^+$) are crucial to constrain the hadronisation mechanisms in nucleus--nucleus collisions for the majority of charm quarks, that have momentum lower than 10 GeV/$c$. This is necessary to remove the large theoretical uncertainty from hadronisation in the estimate of the diffusion coefficient. The ITS\,3 upgrade will further improve the precision on $\rm \Lambda_c$ production measurements in Run 4 by a factor of about four in comparison to Run 3.

The ALICE\,2 measurements of non-prompt D and J/$\psi$ mesons and fully-reconstructed B meson decays from 3~GeV/$c$ in transverse momentum will enable estimates of the diffusion coefficient for beauty quarks. A precise determination of this transport parameter of the QGP will only be enabled by the large acceptance and high pointing resolution of ALICE\,3, as well as the expected recorded integrated luminosity in Runs 5 and 6.
The comparison with first-principle calculations will also become sharper, because the three-fold larger mass of beauty quarks with respect to charm makes the determination of their $D_{\rm s}$ much more accurate in lattice-QCD.
Moreover, the relaxation time of beauty quarks is expected to be about three times larger than for charm quarks (ratio of masses): this aspect will provide additional insight on the temperature-dependence of heavy-quark diffusion in the QGP.

\item  {\it How does the QGP affect the formation of hadrons?}

In Runs 3 and 4, precise measurements of $\rm D_s^+$ and $\rm \Lambda_c^+$ yields over a broad momentum range and the first measurement of $\rm \Lambda_c^+$ elliptic flow will provide strong constraints for the understanding of hadronisation and recombination mechanisms. The ITS\,3 upgrade for Run 4 will enable a first measurement of baryon-to-meson ratios in the beauty sector in Pb--Pb collisions and a precise measurement of beauty + strange hadronisation dynamics using non-prompt $\rm D_s^+$ yields and elliptic flow.

A first, unique, access to the sector of multi-charm baryons (double charm and, possibly, triple charm) is among the main physics goals of the ALICE\,3 experiment. The reconstruction of $\rm \Xi_{\rm cc}$ and $\rm \Omega_{cc}$ cascade chains will be enabled by the unprecedented tracking resolution at the level of a few micrometres, coupled to innovative analysis techniques like strange baryon tracking in the innermost detector layers with radii of a few centimetres. Access to these states allows for a comprehensive study of charm hadronisation from the QGP, including “pure-recombination” hadrons, for which the production probability in string fragmentation is very strongly suppressed by the large charm mass and the conservation of the charm quantum number in the strong interaction~\cite{Becattini:2005hb,Yao:2018zze,Andronic:2021erx}. In addition, the large rapidity acceptance of the ALICE\,3 detector enables studying the dependence of the formation of single- and double-charm baryons on the charm quark rapidity density d$N_{\rm c\overline c}/$d$y$.

\item {\it How does the QGP affect the propagation of energetic partons?}

The ALICE\,2 detector in Runs 3 and 4 will record a fifty-fold larger sample of moderate-$p_{\rm T}$ jets ($\approx 30$--100~GeV/$c$). In conjunction with the excellent tracking performance provided by the ITS\,2 (and ITS\,3 in Run 4), this enables the extension of the studies of the medium-induced modification of internal structure of jets to charm and beauty tagged jets (i.e. tagged parton colour charge and mass) and photon-jet correlations (tagged initial energy of the parton). The large sample of moderate-$p_{\rm T}$ jets will furthermore be used for a high precision search of angular broadening and large-angle Moli\`ere-like deflection with jets that recoil against a high-$p_{\rm T}$ hadron or a photon. This study will provide insight into the existence and nature of scattering centres in the QGP, covering a range of resolution scales (jet energy) and QGP temperatures (related to centre-of-mass energy) that is complementary to studies planned by the sPHENIX Collaboration at RHIC~\cite{PHENIX:2015siv}.

The reconstruction of D mesons in the ALICE\,3 detector with unprecedented large acceptance of eight units of rapidity and large signal-to-background ratio down to almost zero $p_{\rm T}$ will enable the first precise studies of $\rm D$--$\rm \overline D$ azimuthal correlations. This observable provides a direct insight on the interaction mechanisms of charm quarks in the QGP, namely on the contributions of collisional (elastic) and radiative (inelastic) processes~\cite{Cao:2015cba,Nahrgang:2013saa}. Collisional processes are predicted to be dominant for the interaction of low-momentum charm quarks and are considered to be at the origin of the sizeable elliptic flow of hadrons with open and hidden charm. The azimuthal correlation of initially back-to-back charm-anticharm pairs is predicted to be completely washed out by collisional interactions, in analogy to Brownian motion~\cite{Nahrgang:2013saa}.

\item {\it How does deconfinement in the QGP affect the QCD force?}

The ALICE\,2 measurements of charmonium and bottomonium production will benefit from integrated luminosities increased by factors about 15 and 50 at forward and central rapidity, respectively, and from the higher mass resolution and prompt/non-prompt charmonium separation provided by the MFT at forward rapidity. The high precision ratio of the total $\psi$(2S) and  J/$\psi$ yields  is expected to be decisive for the comparison with different models of charmonium formation from deconfined charm and anticharm quarks. This measurement will be possible for the first time also at central rapidity: the rapidity dependence is an important element to discriminate underlying binding mechanisms, because the charm quark density depends on rapidity. At the same time, the charm-quark density will be measured in Pb--Pb collisions using charm mesons and baryons ($\rm \Lambda_c$ and, possibly, $\rm \Xi_c$) at central rapidity and will thus not be a large uncertainty source in models any longer. A high precision J/$\psi$ elliptic flow measurement will provide complementary information on the degree of thermalisation of the deconfined charm quarks in the QGP.

High precision measurements of photoproduction of low-$\pt$ J$/\psi$ mesons in hadronic Pb--Pb collisions will be carried out with ALICE\,2. They will enable studies of possible modification of these states, which are non-relativistic, in interactions with the medium. In addition, they  can be used for an independent estimate of the reaction plane~\cite{Zha:2017jch}.  

The ALICE\,3 design includes both muon detection chambers and electromagnetic calorimetry with acceptance down to low photon energies. These features are instrumental to the measurements in Pb--Pb collisions of P-wave charmonium states ($\chi_{\rm c1,2}$ in the J/$\psi$($\to\mu\mu$)+$\gamma$ channel), which are expected to have different dissociation and recombination dynamics because their binding energy and spatial extent lies in-between those of ground and excited vector states, and because the non-zero orbital angular momentum $L=1$ alters the structure of their wave function. In addition, these measurements will provide for the first time an experimental assessment of the role of feed-down to the S-wave states (in particular J/$\psi$ and $\psi$(2S)). 

\item  {\it Can the QGP lead to discovery of novel QCD effects?}

The search for the Chiral Magnetic Effect induced by parity violation in the strong interaction will be carried out with improved sensitivity using the fifty-fold larger data sample that the ALICE\,2 detector will record in Runs 3 and 4. The search for CME-induced charge separation along the direction of the magnetic field produced by the spectator protons in non-central collisions will reach a sensitivity to a CME fraction as small as 1\% (at 95\% CL)~\cite{ALICE-PUBLIC-2019-001}. 
It is also important to establish direct evidence for the strong magnetic field, which is a key ingredient for the CME, and to determine its strength, which will help significantly constrain theoretical predictions on the magnitude of the CME signal. Measurements of the pseudorapidity-odd component of directed flow will be sensitive to a difference of $5\times 10^{-5}$ between positively and negatively charged particles, that are mainly sensitive to the field value in the late stages of the collisions, and $2\times 10^{-3}$ for D and $\rm \overline D$ mesons that contain positively and negatively charged charm quarks and are sensitive to the field value in the early stages~\cite{ALICE-PUBLIC-2019-001}. 

\item  {\it What are the limits of QGP formation?}

The discovery of collective patterns and strangeness enhancement in particle production in pp and p--Pb collisions with the LHC Run 1 and 2 data questions the view of nucleus--nucleus as the only colliding system in which the QGP can form. This question motivated the decision to pursue increased luminosity goals with pp and p--Pb collisions to carry out a programme of high precision studies of rare probes at high-multiplicity in Runs 3 and 4~\cite{ALICE-PUBLIC-2020-005}. The ten-fold increase of the sample of high-multiplicity pp collisions with respect to Run 2 will open the possibility to study several observables in small and large systems at the same multiplicity. The large integrated luminosity will enable precise measurements of heavy-quark and quarkonium collective flow and a quantitative comparison between pp, p--Pb and Pb--Pb collisions to clarify the origin of the collectivity in small systems and the role of the initial-state of the collisions. The analysis of hadron-jet correlations will allow us to either observe energy loss in small systems for the first time or to put a stringent limit on it (two orders of magnitude smaller than the measured energy loss in central Pb--Pb collisions~\cite{ALICE-PUBLIC-2020-005}). The complementarity of the pp and p--Pb programmes resides in the possibility to separately study the effect of multiplicity and of system size (larger in p--Pb than in pp). In addition, a short $^{16}$O--$^{16}$O run in Run 3 will enable a search for the onset of parton energy loss effects in a system with nucleus--nucleus geometry and with multiplicities similar to those reached in p--Pb collisions~\cite{ALICE-PUBLIC-2021-004}.

The large pseudorapidity acceptance of the ALICE\,3 tracking detectors enables low-bias studies of high-multiplicity pp collisions as a function of global event topology (distribution and dispersion of event activity over eight $\eta$ units, in addition to full azimuth). Furthermore, any future measurement campaigns with lighter ions that may be warranted by the results of Runs 3 and 4 (for instance by oxygen--oxygen studies) will strongly benefit from the much higher rate capability of the ALICE\,3 detector with respect to ALICE\,2.

\item  {\it What is the nature of the initial state of heavy-ion collisions?}

A major advance in the study of the initial state will be achieved in Run 3 with the study of the small Bjorken-$x$ region below $10^{-4}$ where parton phase-space saturation could set in, using quarkonia produced in the photon--lead interactions that occur in ultra-peripheral lead--lead collisions. In particular, the measurements of the J/$\psi$, $\psi$(2S) and $\Upsilon$(1S) photoproduction at central and forward rapidity will allow ALICE\,2 to study the $Q^2$-dependence of the nuclear shadowing in a wide range of Bjorken-$x$ from $10^{-5}$ to $3\times 10^{-2}$. Measurements of both coherent and incoherent production contributions, and of the $\pt^2$ dependence, for these quarkonium states will also provide sensitivity to the variation of the shadowing effect in the transverse area of the nucleus. The production of open charm $\rm D$--$\rm \overline D$ pairs in ultra-peripheral collisions, for which theoretical calculations have smaller uncertainties, are also expected to come into experimental reach.
For Run 4, the proposed dedicated high-granularity forward calorimeter (FoCal) will enable measurements of forward photon production at $3.5< y < 5.8$ and $p_{\rm T}>2$~GeV$/c$ in p--Pb collisions to constrain the gluon densities in the Pb nucleus at very low Bjorken-$x$ ($>10^{-6}$) with unprecedented precision. In addition, FoCal measurements of azimuthal correlations of forward $\pi^0$ will allow us to search for effects of gluon saturation as predicted in the framework of the Colour Glass Condensate.

\item {\it What is the nature of hadron--hadron interactions?}

The large pp and p--Pb samples of Run 3 will allow us to extend the measurements of hyperon--hyperon interactions to states with higher strangeness content, up to the case of $\Omega$--$\Omega$~\cite{ALICE-PUBLIC-2020-005}. Precise measurements of deuteron-hyperon correlation functions will provide constraints to the wave function and production mechanisms of hypernuclei, complementary to direct measurements of hypertriton production in small and large systems. It is also planned to carry out extensive studies of the three-body interaction potential for protons and $\Lambda$ baryons and of the strong interaction between charm mesons and nucleons.

With the large acceptance and signal-to-background ratio for D mesons in ALICE\,3, the measurement of the momentum correlation function between pairs of D mesons (or D--$\rm \overline D$) will provide information on the nature and wave function of exotic hadrons with two charm quarks, such as the $\chi_{\rm c1}$(3872) and the $\rm T_{cc}$, and address the question on their tetraquark or D--D molecular structure. The ALICE\,3 high rate capabilities will also provide access to rare nuclear states, such as light nuclei with $A = 6$, hypernuclei with $A = 4,\,5$. The rate capabilities, combined with the excellent heavy-flavour performance, could possibly give access to the yet undiscovered charm-nuclei~\cite{Andronic:2021erx,ExHIC:2017smd}, in which a nucleon is replaced by a charmed baryon, like c-deuteron (n$\Lambda_{\rm c}^+$ bound state decaying to, e.g., $\rm d+K^- + \pi^+$ or $\rm d+K^0_S$). 

\item {\it Can ALICE elucidate specific aspects of perturbative QCD and of related ``long distance'' QCD interactions?}

As a follow up of the recent observation of the dead-cone effect for charm quarks~\cite{ALICE:2021aqk}, the large pp sample of Run 3 will enable the first direct measurement of this effect also for beauty quarks, using low-$p_{\rm T}$ jets that contain a fully-reconstructed B meson~\cite{ALICE-PUBLIC-2020-005}. This will provide a direct test of the expected quark-mass dependence of the reduction of gluon radiation at small angles. In addition, the studies of charm quark hadronisation in pp collisions will be extended with precise measurements of baryons with charm and strangeness ($\Xi_{\rm c}$ and $\Omega_{\rm c}$), which at present elude a consistent description by string fragmentation models, even when baryon junction effects are included. 
In the next LHC runs, ALICE will study quarkonium production in pp collisions with high precision also at central rapidity, and with prompt and non-prompt component separation also at forward rapidity~\cite{ALICE-PUBLIC-2020-005}. Other and more differential studies will be possible: the production of the $\chi_c$ meson, the associated production of quarkonium with charged particles, to study long-range correlation over up to 5 units in rapidity, and the associated production of J/$\psi$ with a photon or a D meson, to provide a novel insight on the partonic structure of the proton. 

\end{enumerate}

\newpage


\newenvironment{acknowledgement}{\relax}{\relax}
\begin{acknowledgement}
\section*{Acknowledgements}

The ALICE Collaboration would like to thank all its engineers and technicians for their invaluable contributions to the construction of the experiment and the CERN accelerator teams for the outstanding performance of the LHC complex.
The ALICE Collaboration gratefully acknowledges the resources and support provided by all Grid centres and the Worldwide LHC Computing Grid (WLCG) collaboration.
The ALICE Collaboration acknowledges the following funding agencies for their support in building and running the ALICE detector:
A. I. Alikhanyan National Science Laboratory (Yerevan Physics Institute) Foundation (ANSL), State Committee of Science and World Federation of Scientists (WFS), Armenia;
Austrian Academy of Sciences, Austrian Science Fund (FWF): [M 2467-N36] and Nationalstiftung f\"{u}r Forschung, Technologie und Entwicklung, Austria;
Ministry of Communications and High Technologies, National Nuclear Research Center, Azerbaijan;
Conselho Nacional de Desenvolvimento Cient\'{\i}fico e Tecnol\'{o}gico (CNPq), Financiadora de Estudos e Projetos (Finep), Funda\c{c}\~{a}o de Amparo \`{a} Pesquisa do Estado de S\~{a}o Paulo (FAPESP) and Universidade Federal do Rio Grande do Sul (UFRGS), Brazil;
Bulgarian Ministry of Education and Science, within the National Roadmap for Research Infrastructures 2020-2027 (object CERN), Bulgaria;
Ministry of Education of China (MOEC) , Ministry of Science \& Technology of China (MSTC) and National Natural Science Foundation of China (NSFC), China;
Ministry of Science and Education and Croatian Science Foundation, Croatia;
Centro de Aplicaciones Tecnol\'{o}gicas y Desarrollo Nuclear (CEADEN), Cubaenerg\'{\i}a, Cuba;
Ministry of Education, Youth and Sports of the Czech Republic, Czech Republic;
The Danish Council for Independent Research | Natural Sciences, the VILLUM FONDEN and Danish National Research Foundation (DNRF), Denmark;
Helsinki Institute of Physics (HIP), Finland;
Commissariat \`{a} l'Energie Atomique (CEA) and Institut National de Physique Nucl\'{e}aire et de Physique des Particules (IN2P3) and Centre National de la Recherche Scientifique (CNRS), France;
Bundesministerium f\"{u}r Bildung und Forschung (BMBF) and GSI Helmholtzzentrum f\"{u}r Schwerionenforschung GmbH, Germany;
General Secretariat for Research and Technology, Ministry of Education, Research and Religions, Greece;
National Research, Development and Innovation Office, Hungary;
Department of Atomic Energy Government of India (DAE), Department of Science and Technology, Government of India (DST), University Grants Commission, Government of India (UGC) and Council of Scientific and Industrial Research (CSIR), India;
National Research and Innovation Agency - BRIN, Indonesia;
Istituto Nazionale di Fisica Nucleare (INFN), Italy;
Japanese Ministry of Education, Culture, Sports, Science and Technology (MEXT) and Japan Society for the Promotion of Science (JSPS) KAKENHI, Japan;
Consejo Nacional de Ciencia (CONACYT) y Tecnolog\'{i}a, through Fondo de Cooperaci\'{o}n Internacional en Ciencia y Tecnolog\'{i}a (FONCICYT) and Direcci\'{o}n General de Asuntos del Personal Academico (DGAPA), Mexico;
Nederlandse Organisatie voor Wetenschappelijk Onderzoek (NWO), Netherlands;
The Research Council of Norway, Norway;
Commission on Science and Technology for Sustainable Development in the South (COMSATS), Pakistan;
Pontificia Universidad Cat\'{o}lica del Per\'{u}, Peru;
Ministry of Education and Science, National Science Centre and WUT ID-UB, Poland;
Korea Institute of Science and Technology Information and National Research Foundation of Korea (NRF), Republic of Korea;
Ministry of Education and Scientific Research, Institute of Atomic Physics, Ministry of Research and Innovation and Institute of Atomic Physics and Universitatea Nationala de Stiinta si Tehnologie Politehnica Bucuresti, Romania;
Ministry of Education, Science, Research and Sport of the Slovak Republic, Slovakia;
National Research Foundation of South Africa, South Africa;
Swedish Research Council (VR) and Knut \& Alice Wallenberg Foundation (KAW), Sweden;
European Organization for Nuclear Research, Switzerland;
Suranaree University of Technology (SUT), National Science and Technology Development Agency (NSTDA) and National Science, Research and Innovation Fund (NSRF via PMU-B B05F650021), Thailand;
Turkish Energy, Nuclear and Mineral Research Agency (TENMAK), Turkey;
National Academy of  Sciences of Ukraine, Ukraine;
Science and Technology Facilities Council (STFC), United Kingdom;
National Science Foundation of the United States of America (NSF) and United States Department of Energy, Office of Nuclear Physics (DOE NP), United States of America.
In addition, individual groups or members have received support from:
Marie Sk\l{}odowska Curie, European Research Council, Strong 2020 - Horizon 2020 (grant nos. 950692, 824093, 896850), European Union;
Academy of Finland (Center of Excellence in Quark Matter) (grant nos. 346327, 346328), Finland;
Programa de Apoyos para la Superaci\'{o}n del Personal Acad\'{e}mico, UNAM, Mexico.

\end{acknowledgement}
\newpage

\bibliographystyle{utphys}   
\small
\bibliography{Chapter1/bibliography.bib,Chapter2/Chapter2.1/bibliography.bib,Chapter2/Chapter2.2/RefsTG2.bib,Chapter2/Chapter2.3/bibliography-Ch2.3.bib,Chapter2/Chapter2.4/references,Chapter2/Chapter2.5/bibliography2-5,Chapter2/Chapter2.6/RefsTG06,Chapter2/Chapter2.7/bibliography,Chapter3/Chapter3.1-3.3/biblio,Chapter3/Chapter3.4/bibliography,Chapter3/Chapter3.5/references,Chapter4/RefsTG7,Chapter5/bibliography,Chapter6/main.bib,Chapter7/bibliography.bib,Chapter9/bibliography.bib}

\newpage
\appendix

%
%

\section{The ALICE Collaboration}
\label{app:collab}
\begin{flushleft} 
\small

S.~Acharya\,\orcidlink{0000-0002-9213-5329}\,$^{\rm 126}$, 
D.~Adamov\'{a}\,\orcidlink{0000-0002-0504-7428}\,$^{\rm 86}$, 
A.~Adler$^{\rm 70}$, 
G.~Aglieri Rinella\,\orcidlink{0000-0002-9611-3696}\,$^{\rm 32}$, 
M.~Agnello\,\orcidlink{0000-0002-0760-5075}\,$^{\rm 29}$, 
N.~Agrawal\,\orcidlink{0000-0003-0348-9836}\,$^{\rm 51}$, 
Z.~Ahammed\,\orcidlink{0000-0001-5241-7412}\,$^{\rm 134}$, 
S.~Ahmad\,\orcidlink{0000-0003-0497-5705}\,$^{\rm 15}$, 
S.U.~Ahn\,\orcidlink{0000-0001-8847-489X}\,$^{\rm 71}$, 
I.~Ahuja\,\orcidlink{0000-0002-4417-1392}\,$^{\rm 37}$
% ...

\section*{Affiliation Notes}

$^{\rm I}$ Deceased\\
$^{\rm II}$ Also at: Max-Planck-Institut fur Physik, Munich, Germany\\
$^{\rm III}$ Also at: Italian National Agency for New Technologies, Energy and Sustainable Economic Development (ENEA), Bologna, Italy\\
$^{\rm IV}$ Also at: Dipartimento DET del Politecnico di Torino, Turin, Italy\\
$^{\rm V}$ Also at: Department of Applied Physics, Aligarh Muslim University, Aligarh, India\\
$^{\rm VI}$ Also at: Institute of Theoretical Physics, University of Wroclaw, Poland\\
$^{\rm VII}$ Also at: An institution covered by a cooperation agreement with CERN\\

\section*{Collaboration Institutes}

$^{1}$ A.I. Alikhanyan National Science Laboratory (Yerevan Physics Institute) Foundation, Yerevan, Armenia\\
$^{2}$ AGH University of Krakow, Cracow, Poland\\
$^{3}$ Bogolyubov Institute for Theoretical Physics, National Academy of Sciences of Ukraine, Kiev, Ukraine\\
$^{4}$ Bose Institute, Department of Physics  and Centre for Astroparticle Physics and Space Science (CAPSS), Kolkata, India\\
$^{5}$ California Polytechnic State University, San Luis Obispo, California, United States\\
$^{6}$ Central China Normal University, Wuhan, China\\
$^{7}$ Centro de Aplicaciones Tecnol\'{o}gicas y Desarrollo Nuclear (CEADEN), Havana, Cuba\\
$^{8}$ Centro de Investigaci\'{o}n y de Estudios Avanzados (CINVESTAV), Mexico City and M\'{e}rida, Mexico\\
$^{9}$ Chicago State University, Chicago, Illinois, United States\\
$^{10}$ China Institute of Atomic Energy, Beijing, China\\
$^{11}$ Chungbuk National University, Cheongju, Republic of Korea\\
$^{12}$ Comenius University Bratislava, Faculty of Mathematics, Physics and Informatics, Bratislava, Slovak Republic\\
$^{13}$ COMSATS University Islamabad, Islamabad, Pakistan\\
$^{14}$ Creighton University, Omaha, Nebraska, United States\\
$^{15}$ Department of Physics, Aligarh Muslim University, Aligarh, India\\
$^{16}$ Department of Physics, Pusan National University, Pusan, Republic of Korea\\
$^{17}$ Department of Physics, Sejong University, Seoul, Republic of Korea\\
$^{18}$ Department of Physics, University of California, Berkeley, California, United States\\
$^{19}$ Department of Physics, University of Oslo, Oslo, Norway\\
$^{20}$ Department of Physics and Technology, University of Bergen, Bergen, Norway\\
$^{21}$ Dipartimento di Fisica, Universit\`{a} di Pavia, Pavia, Italy\\
$^{22}$ Dipartimento di Fisica dell'Universit\`{a} and Sezione INFN, Cagliari, Italy\\
$^{23}$ Dipartimento di Fisica dell'Universit\`{a} and Sezione INFN, Trieste, Italy\\
$^{24}$ Dipartimento di Fisica dell'Universit\`{a} and Sezione INFN, Turin, Italy\\
$^{25}$ Dipartimento di Fisica e Astronomia dell'Universit\`{a} and Sezione INFN, Bologna, Italy\\
$^{26}$ Dipartimento di Fisica e Astronomia dell'Universit\`{a} and Sezione INFN, Catania, Italy\\
$^{27}$ Dipartimento di Fisica e Astronomia dell'Universit\`{a} and Sezione INFN, Padova, Italy\\
$^{28}$ Dipartimento di Fisica `E.R.~Caianiello' dell'Universit\`{a} and Gruppo Collegato INFN, Salerno, Italy\\
$^{29}$ Dipartimento DISAT del Politecnico and Sezione INFN, Turin, Italy\\
$^{30}$ Dipartimento di Scienze MIFT, Universit\`{a} di Messina, Messina, Italy\\
$^{31}$ Dipartimento Interateneo di Fisica `M.~Merlin' and Sezione INFN, Bari, Italy\\
$^{32}$ European Organization for Nuclear Research (CERN), Geneva, Switzerland\\
$^{33}$ Faculty of Electrical Engineering, Mechanical Engineering and Naval Architecture, University of Split, Split, Croatia\\
$^{34}$ Faculty of Engineering and Science, Western Norway University of Applied Sciences, Bergen, Norway\\
$^{35}$ Faculty of Nuclear Sciences and Physical Engineering, Czech Technical University in Prague, Prague, Czech Republic\\
$^{36}$ Faculty of Physics, Sofia University, Sofia, Bulgaria\\
$^{37}$ Faculty of Science, P.J.~\v{S}af\'{a}rik University, Ko\v{s}ice, Slovak Republic\\
$^{38}$ Frankfurt Institute for Advanced Studies, Johann Wolfgang Goethe-Universit\"{a}t Frankfurt, Frankfurt, Germany\\
$^{39}$ Fudan University, Shanghai, China\\
$^{40}$ Gangneung-Wonju National University, Gangneung, Republic of Korea\\
$^{41}$ Gauhati University, Department of Physics, Guwahati, India\\
$^{42}$ Helmholtz-Institut f\"{u}r Strahlen- und Kernphysik, Rheinische Friedrich-Wilhelms-Universit\"{a}t Bonn, Bonn, Germany\\
$^{43}$ Helsinki Institute of Physics (HIP), Helsinki, Finland\\
$^{44}$ High Energy Physics Group,  Universidad Aut\'{o}noma de Puebla, Puebla, Mexico\\
$^{45}$ Horia Hulubei National Institute of Physics and Nuclear Engineering, Bucharest, Romania\\
$^{46}$ HUN-REN Wigner Research Centre for Physics, Budapest, Hungary\\
$^{47}$ Indian Institute of Technology Bombay (IIT), Mumbai, India\\
$^{48}$ Indian Institute of Technology Indore, Indore, India\\
$^{49}$ INFN, Laboratori Nazionali di Frascati, Frascati, Italy\\
$^{50}$ INFN, Sezione di Bari, Bari, Italy\\
$^{51}$ INFN, Sezione di Bologna, Bologna, Italy\\
$^{52}$ INFN, Sezione di Cagliari, Cagliari, Italy\\
$^{53}$ INFN, Sezione di Catania, Catania, Italy\\
$^{54}$ INFN, Sezione di Padova, Padova, Italy\\
$^{55}$ INFN, Sezione di Pavia, Pavia, Italy\\
$^{56}$ INFN, Sezione di Torino, Turin, Italy\\
$^{57}$ INFN, Sezione di Trieste, Trieste, Italy\\
$^{58}$ Inha University, Incheon, Republic of Korea\\
$^{59}$ Institute for Gravitational and Subatomic Physics (GRASP), Utrecht University/Nikhef, Utrecht, Netherlands\\
$^{60}$ Institute of Experimental Physics, Slovak Academy of Sciences, Ko\v{s}ice, Slovak Republic\\
$^{61}$ Institute of Physics, Homi Bhabha National Institute, Bhubaneswar, India\\
$^{62}$ Institute of Physics of the Czech Academy of Sciences, Prague, Czech Republic\\
$^{63}$ Institute of Space Science (ISS), Bucharest, Romania\\
$^{64}$ Institut f\"{u}r Kernphysik, Johann Wolfgang Goethe-Universit\"{a}t Frankfurt, Frankfurt, Germany\\
$^{65}$ Instituto de Ciencias Nucleares, Universidad Nacional Aut\'{o}noma de M\'{e}xico, Mexico City, Mexico\\
$^{66}$ Instituto de F\'{i}sica, Universidade Federal do Rio Grande do Sul (UFRGS), Porto Alegre, Brazil\\
$^{67}$ Instituto de F\'{\i}sica, Universidad Nacional Aut\'{o}noma de M\'{e}xico, Mexico City, Mexico\\
$^{68}$ iThemba LABS, National Research Foundation, Somerset West, South Africa\\
$^{69}$ Jeonbuk National University, Jeonju, Republic of Korea\\
$^{70}$ Johann-Wolfgang-Goethe Universit\"{a}t Frankfurt Institut f\"{u}r Informatik, Fachbereich Informatik und Mathematik, Frankfurt, Germany\\
$^{71}$ Korea Institute of Science and Technology Information, Daejeon, Republic of Korea\\
$^{72}$ KTO Karatay University, Konya, Turkey\\
$^{73}$ Laboratoire de Physique Subatomique et de Cosmologie, Universit\'{e} Grenoble-Alpes, CNRS-IN2P3, Grenoble, France\\
$^{74}$ Lawrence Berkeley National Laboratory, Berkeley, California, United States\\
$^{75}$ Lund University Department of Physics, Division of Particle Physics, Lund, Sweden\\
$^{76}$ Nagasaki Institute of Applied Science, Nagasaki, Japan\\
$^{77}$ Nara Women{'}s University (NWU), Nara, Japan\\
$^{78}$ National and Kapodistrian University of Athens, School of Science, Department of Physics , Athens, Greece\\
$^{79}$ National Centre for Nuclear Research, Warsaw, Poland\\
$^{80}$ National Institute of Science Education and Research, Homi Bhabha National Institute, Jatni, India\\
$^{81}$ National Nuclear Research Center, Baku, Azerbaijan\\
$^{82}$ National Research and Innovation Agency - BRIN, Jakarta, Indonesia\\
$^{83}$ Niels Bohr Institute, University of Copenhagen, Copenhagen, Denmark\\
$^{84}$ Nikhef, National institute for subatomic physics, Amsterdam, Netherlands\\
$^{85}$ Nuclear Physics Group, STFC Daresbury Laboratory, Daresbury, United Kingdom\\
$^{86}$ Nuclear Physics Institute of the Czech Academy of Sciences, Husinec-\v{R}e\v{z}, Czech Republic\\
$^{87}$ Oak Ridge National Laboratory, Oak Ridge, Tennessee, United States\\
$^{88}$ Ohio State University, Columbus, Ohio, United States\\
$^{89}$ Physics department, Faculty of science, University of Zagreb, Zagreb, Croatia\\
$^{90}$ Physics Department, Panjab University, Chandigarh, India\\
$^{91}$ Physics Department, University of Jammu, Jammu, India\\
$^{92}$ Physics Program and International Institute for Sustainability with Knotted Chiral Meta Matter (SKCM2), Hiroshima University, Hiroshima, Japan\\
$^{93}$ Physikalisches Institut, Eberhard-Karls-Universit\"{a}t T\"{u}bingen, T\"{u}bingen, Germany\\
$^{94}$ Physikalisches Institut, Ruprecht-Karls-Universit\"{a}t Heidelberg, Heidelberg, Germany\\
$^{95}$ Physik Department, Technische Universit\"{a}t M\"{u}nchen, Munich, Germany\\
$^{96}$ Politecnico di Bari and Sezione INFN, Bari, Italy\\
$^{97}$ Research Division and ExtreMe Matter Institute EMMI, GSI Helmholtzzentrum f\"ur Schwerionenforschung GmbH, Darmstadt, Germany\\
$^{98}$ Saga University, Saga, Japan\\
$^{99}$ Saha Institute of Nuclear Physics, Homi Bhabha National Institute, Kolkata, India\\
$^{100}$ School of Physics and Astronomy, University of Birmingham, Birmingham, United Kingdom\\
$^{101}$ Secci\'{o}n F\'{\i}sica, Departamento de Ciencias, Pontificia Universidad Cat\'{o}lica del Per\'{u}, Lima, Peru\\
$^{102}$ Stefan Meyer Institut f\"{u}r Subatomare Physik (SMI), Vienna, Austria\\
$^{103}$ SUBATECH, IMT Atlantique, Nantes Universit\'{e}, CNRS-IN2P3, Nantes, France\\
$^{104}$ Sungkyunkwan University, Suwon City, Republic of Korea\\
$^{105}$ Suranaree University of Technology, Nakhon Ratchasima, Thailand\\
$^{106}$ Technical University of Ko\v{s}ice, Ko\v{s}ice, Slovak Republic\\
$^{107}$ The Henryk Niewodniczanski Institute of Nuclear Physics, Polish Academy of Sciences, Cracow, Poland\\
$^{108}$ The University of Texas at Austin, Austin, Texas, United States\\
$^{109}$ Universidad Aut\'{o}noma de Sinaloa, Culiac\'{a}n, Mexico\\
$^{110}$ Universidade de S\~{a}o Paulo (USP), S\~{a}o Paulo, Brazil\\
$^{111}$ Universidade Estadual de Campinas (UNICAMP), Campinas, Brazil\\
$^{112}$ Universidade Federal do ABC, Santo Andre, Brazil\\
$^{113}$ Universitatea Nationala de Stiinta si Tehnologie Politehnica Bucuresti, Bucharest, Romania\\
$^{114}$ University of Cape Town, Cape Town, South Africa\\
$^{115}$ University of Houston, Houston, Texas, United States\\
$^{116}$ University of Jyv\"{a}skyl\"{a}, Jyv\"{a}skyl\"{a}, Finland\\
$^{117}$ University of Kansas, Lawrence, Kansas, United States\\
$^{118}$ University of Liverpool, Liverpool, United Kingdom\\
$^{119}$ University of Science and Technology of China, Hefei, China\\
$^{120}$ University of South-Eastern Norway, Kongsberg, Norway\\
$^{121}$ University of Tennessee, Knoxville, Tennessee, United States\\
$^{122}$ University of the Witwatersrand, Johannesburg, South Africa\\
$^{123}$ University of Tokyo, Tokyo, Japan\\
$^{124}$ University of Tsukuba, Tsukuba, Japan\\
$^{125}$ Universit\"{a}t M\"{u}nster, Institut f\"{u}r Kernphysik, M\"{u}nster, Germany\\
$^{126}$ Universit\'{e} Clermont Auvergne, CNRS/IN2P3, LPC, Clermont-Ferrand, France\\
$^{127}$ Universit\'{e} de Lyon, CNRS/IN2P3, Institut de Physique des 2 Infinis de Lyon, Lyon, France\\
$^{128}$ Universit\'{e} de Strasbourg, CNRS, IPHC UMR 7178, F-67000 Strasbourg, France, Strasbourg, France\\
$^{129}$ Universit\'{e} Paris-Saclay, Centre d'Etudes de Saclay (CEA), IRFU, D\'{e}partment de Physique Nucl\'{e}aire (DPhN), Saclay, France\\
$^{130}$ Universit\'{e}  Paris-Saclay, CNRS/IN2P3, IJCLab, Orsay, France\\
$^{131}$ Universit\`{a} degli Studi di Foggia, Foggia, Italy\\
$^{132}$ Universit\`{a} del Piemonte Orientale, Vercelli, Italy\\
$^{133}$ Universit\`{a} di Brescia, Brescia, Italy\\
$^{134}$ Variable Energy Cyclotron Centre, Homi Bhabha National Institute, Kolkata, India\\
$^{135}$ Warsaw University of Technology, Warsaw, Poland\\
$^{136}$ Wayne State University, Detroit, Michigan, United States\\
$^{137}$ Yale University, New Haven, Connecticut, United States\\
$^{138}$ Yonsei University, Seoul, Republic of Korea\\
$^{139}$  Zentrum  f\"{u}r Technologie und Transfer (ZTT), Worms, Germany\\
$^{140}$ Affiliated with an institute covered by a cooperation agreement with CERN\\
$^{141}$ Affiliated with an international laboratory covered by a cooperation agreement with CERN.\\

\end{flushleft}

\includepdf{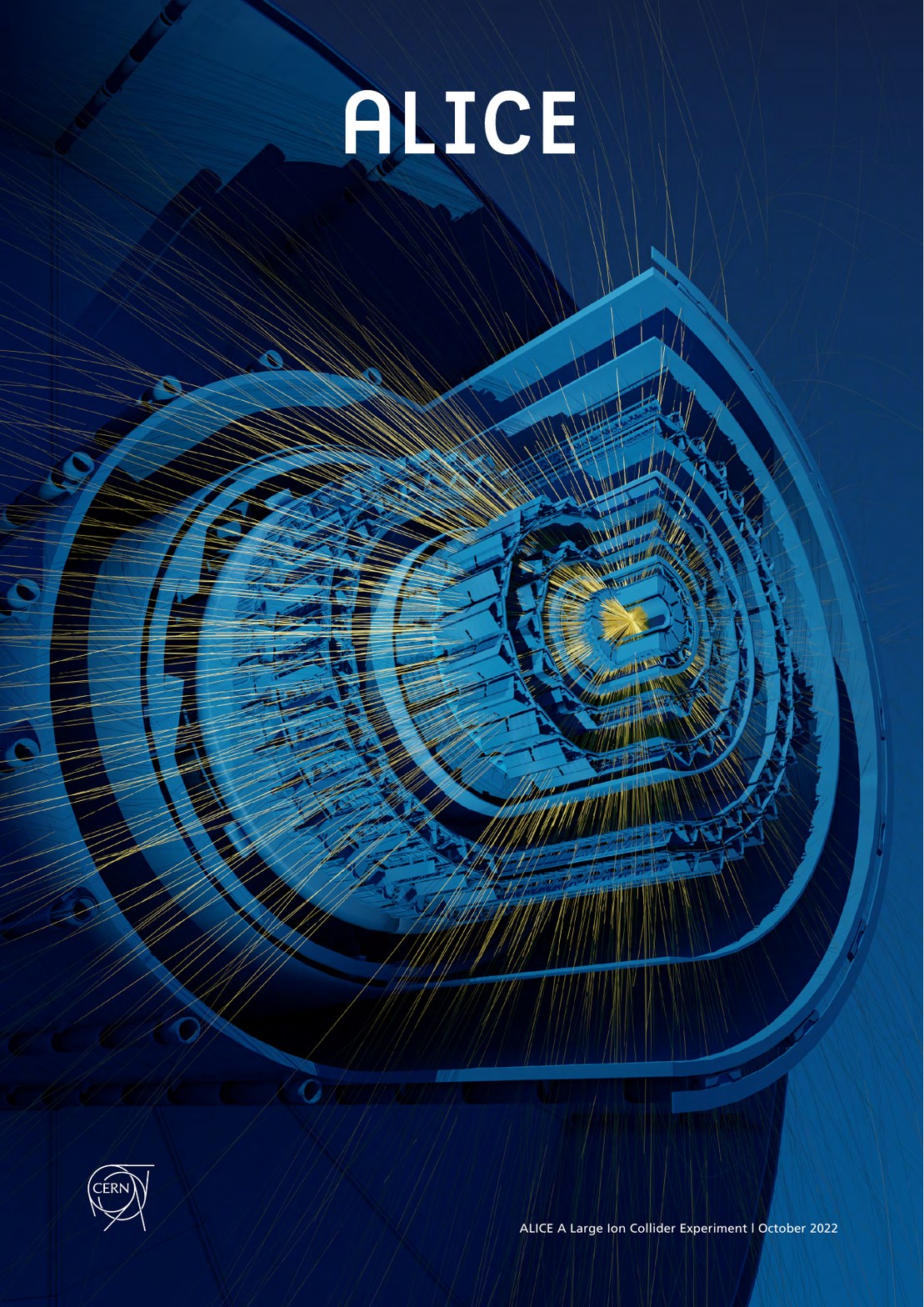}

\end{document}